\documentclass[11pt,twoside,a4paper]{book}

\usepackage{latexsym} 
\usepackage{a4wide}
\usepackage{color} 
\usepackage{amssymb,amsmath} 
\usepackage{fancyhdr} 
\usepackage{epic} 
\usepackage{here}
\usepackage{graphics,epsfig}

\usepackage[french]{babel} 

\newtheorem{theo}{Th\'eor\`eme} 
\newtheorem{defin}{D\'efinition} 
\newtheorem{axiome}{Axiome} 
\newtheorem{remarq}{Remarque} 
\newtheorem{conclu}{Conclusion} 
\newtheorem{coro}{Corollaire} 
\newtheorem{class}{Classification} 
\newtheorem{prop}{Propri\'et\'e}

\newcommand{\bs}[1]{\boldsymbol{#1}} 
\newcommand{\ov}[1]{\overline{#1}} 
\newcommand{\ud}[1]{\underline{#1}} 
\newcommand{\udd}[1]{\underline{\underline{#1}}} 
\newcommand{\id}{\ensuremath{\text{id}}} 
\newcommand{\pv}{\otimes} 
\newcommand{\munite}{\ensuremath{\,\,\mathrm{l}\!\!\!1}} 

\newcommand{\scal}[2]{\langle \; #1 \; ,\; #2 \; \rangle} 
\newcommand{\Mie}[2]{\ensuremath{\text{M}_{#1}^{#2}}} 
\newcommand{\Mlo}[1]{\ensuremath{(M_{#1}(\Lambda^{2}))_0}} 
 
\newcommand{\bmatr}{\begin{pmatrix}} 
\newcommand{\ematr}{\end{pmatrix}} 

\newcommand{\xa}[1]{|\chi_{#1}|^2} 
\newcommand{\xaa}[2]{|\chi_{#1} + \chi_{#2}|^2} 
\newcommand{\ch}[1]{\chi_{#1}} 
\newcommand{\och}[1]{\overline{\chi_{#1}}} 
\newcommand{\xx}[2]{\chi_{#1}.\overline{\chi_{#2}}} 
\newcommand{\hxa}[1]{|\hat{\chi}_{#1}|^2} 
\newcommand{\hxaa}[2]{|\hat{\chi}_{#1}+\hat{\chi}_{#2}|^2} 
\newcommand{\hxaaa}[3]{|\hat{\chi}_{#1}+\hat{\chi}_{#2}+\hat{\chi}_{#3}|^2} 
\newcommand{\hch}[1]{\hat{\chi}_{#1}} 
\newcommand{\hoch}[1]{\overline{\hat{\chi}_{#1}}} 
\newcommand{\hxx}[2]{\hat{\chi}_{#1}.\overline{\hat{\chi}_{#2}}} 
\newcommand{\zd}[2]{\ensuremath{\sigma_{#1}^{#2}}} 
 
\def\otimesdot{\stackrel{\cdot}{\otimes}}

\newcommand{\smalldif}[5]{\unitlength 0.025cm 
\parbox{1cm}{\begin{picture}(40,80) 
\put(20,20){\circle*{6}} 
\put(20,60){\circle*{6}} 
\put(0,0){\line(1,1){20}} 
\put(40,0){\line(-1,1){20}} 
\put(0,80){\line(1,-1){20}} 
\put(40,80){\line(-1,-1){20}} 
\dottedline[$\circ$]{6}(20,20)(20,60) 
\put(7,77){$#1$} 
\put(29,77){$#2$} 
\put(6,-2){$#3$} 
\put(23,-2){$#4$} 
\put(8,36){$#5$} 
\end{picture}}}

\newcommand{\smalldifdual}[5]{\unitlength 0.025cm 
\parbox{2.1cm}{\begin{picture}(80,40) 
\put(0,0){\line(1,1){20}} 
\put(0,40){\line(1,-1){20}} 
\put(80,40){\line(-1,-1){20}} 
\put(80,0){\line(-1,1){20}} 
\dottedline[$\bullet$]{7}(20,20)(60,20) 
\put(-3,30){$#1$} 
\put(78,5){$#4$} 
\put(-3,5){$#3$} 
\put(78,30){$#2$} 
\put(38,27){$#5$} 
\end{picture}}}

\newcommand{\Fun}[6]{ 
\parbox{1cm}{\begin{picture}(40,40) 
\put(0,20){\circle{6}} 
\put(40,20){\circle*{6}} 
\put(20,0){\circle*{6}} 
\put(20,40){\circle*{6}} 
\put(3,20){\line(1,0){34}} 
\put(2,22){\line(1,1){15.8}} 
\put(38,22){\line(-1,1){15.8}} 
\put(2,18){\line(1,-1){15.8}} 
\put(38,18){\line(-1,-1){15.8}} 
\put(20,3){\line(0,1){15}} 
\put(20,37){\line(0,-1){15}} 
\put(30,20){\vector(1,0){0}} 
\put(20,10){\vector(0,-1){0}} 
\put(11,31){\vector(1,1){0}} 
\put(28,8){\vector(-1,-1){0}} 
\put(11,9){\vector(1,-1){0}} 
\put(31,29){\vector(1,-1){0}} 
\put(4,32){$#1$} 
\put(32,32){$#2$} 
\put(4,5){$#3$} 
\put(32,5){$#4$} 
\put(26,22){$#5$} 
\put(13,10){$#6$} 
\end{picture}}}

\newcommand{\Ftrois}[6]{\parbox{40pt}{\begin{picture}(40,40) 
\put(0,20){\circle*{6}} 
\put(40,20){\circle{6}} 
\put(20,0){\circle{6}} 
\put(20,40){\circle{6}} 
\put(3,20){\line(1,0){34}} 
\put(2,22){\line(1,1){15.8}} 
\put(38,22){\line(-1,1){15.8}} 
\put(2,18){\line(1,-1){15.8}} 
\put(38,18){\line(-1,-1){15.8}} 
\put(20,3){\line(0,1){15}} 
\put(20,37){\line(0,-1){15}} 
\put(26,20){\vector(-1,0){0}} 
\put(20,10){\vector(0,-1){0}} 
\put(8,28){\vector(-1,-1){0}} 
\put(28,8){\vector(-1,-1){0}} 
\put(31,29){\vector(1,-1){0}} 
\put(9,11){\vector(-1,1){0}} 
\put(4,32){$#1$} 
\put(32,32){$#2$} 
\put(4,5){$#3$} 
\put(32,5){$#4$} 
\put(26,22){$#5$} 
\put(13,10){$#6$} 
\end{picture}}}

\newcommand{\Fzero}[6]{ 
\parbox{1cm}{\begin{picture}(40,40) 
\put(0,20){\circle*{6}} 
\put(40,20){\circle*{6}} 
\put(20,0){\circle*{6}} 
\put(20,40){\circle*{6}} 
\put(3,20){\line(1,0){34}} 
\put(2,22){\line(1,1){15.8}} 
\put(38,22){\line(-1,1){15.8}} 
\put(2,18){\line(1,-1){15.8}} 
\put(38,18){\line(-1,-1){15.8}} 
\put(20,3){\line(0,1){15}} 
\put(20,37){\line(0,-1){15}} 
\put(30,20){\vector(1,0){0}} 
\put(20,10){\vector(0,-1){0}} 
\put(11,31){\vector(1,1){0}} 
\put(28,8){\vector(-1,-1){0}} 
\put(11,9){\vector(1,-1){0}} 
\put(31,29){\vector(1,-1){0}} 
\put(4,32){$#1$} 
\put(32,32){$#2$} 
\put(4,5){$#3$} 
\put(32,5){$#4$} 
\put(26,22){$#5$} 
\put(13,10){$#6$} 
\end{picture}}}

\newcommand{\Fquatre}[6]{ 
\parbox{1cm}{\begin{picture}(40,40) 
\put(0,20){\circle{6}} 
\put(40,20){\circle{6}} 
\put(20,0){\circle{6}} 
\put(20,40){\circle{6}} 
\put(3,20){\line(1,0){34}} 
\put(2,22){\line(1,1){15.8}} 
\put(38,22){\line(-1,1){15.8}} 
\put(2,18){\line(1,-1){15.8}} 
\put(38,18){\line(-1,-1){15.8}} 
\put(20,3){\line(0,1){15}} 
\put(20,37){\line(0,-1){15}} 
\put(30,20){\vector(1,0){0}} 
\put(20,10){\vector(0,-1){0}} 
\put(11,31){\vector(1,1){0}} 
\put(28,8){\vector(-1,-1){0}} 
\put(11,9){\vector(1,-1){0}} 
\put(31,29){\vector(1,-1){0}} 
\put(4,32){$#1$} 
\put(32,32){$#2$} 
\put(4,5){$#3$} 
\put(32,5){$#4$} 
\put(26,22){$#5$} 
\put(13,10){$#6$} 
\end{picture}}}

\newcommand{\Fdeux}[6]{\parbox{40pt}{\begin{picture}(40,40) 
\put(0,20){\circle{6}} 
\put(40,20){\circle{6}} 
\put(20,0){\circle*{6}} 
\put(20,40){\circle*{6}} 
\put(3,20){\line(1,0){34}} 
\put(2,22){\line(1,1){15.8}} 
\put(38,22){\line(-1,1){15.8}} 
\put(2,18){\line(1,-1){15.8}} 
\put(38,18){\line(-1,-1){15.8}} 
\put(20,3){\line(0,1){15}} 
\put(20,37){\line(0,-1){15}} 
\put(32,20){\vector(1,0){0}} 
\put(20,10){\vector(0,-1){0}} 
\put(11,31){\vector(1,1){0}} 
\put(28,8){\vector(-1,-1){0}} 
\put(11,9){\vector(1,-1){0}} 
\put(29,31){\vector(-1,1){0}} 
\put(4,32){$#1$} 
\put(32,32){$#2$} 
\put(4,5){$#3$} 
\put(32,5){$#4$} 
\put(26,22){$#5$} 
\put(13,10){$#6$} 
\end{picture}}}

\newcommand{\cell}[9]{\unitlength 0.025cm 
\parbox{1.25cm}{\begin{picture}(50,60) 
\put(5,10){\line(1,0){40}} 
\put(5,50){\line(1,0){40}} 
\put(5,10){\line(0,1){40}} 
\put(45,10){\line(0,1){40}} 
\put(28.5,10){\vector(1,0){0}} 
\put(28.5,50){\vector(1,0){0}} 
\put(5,26){\vector(0,-1){0}} 
\put(45,26){\vector(0,-1){0}} 
\put(-2,1){\scriptsize $#4$} 
\put(40,1){\scriptsize $#3$} 
\put(-2,57){\scriptsize $#1$} 
\put(40,57){\scriptsize $#2$} 
\put(21,0){\scriptsize $#6$} 
\put(21,55){\scriptsize $#5$} 
\put(-6,28){\scriptsize $#7$} 
\put(48,28){\scriptsize $#8$} 
\put(25,30){\scriptsize \makebox(0,0){$#9$}} 
\end{picture}}}

\newcommand{\cellbar}[9]{\unitlength 0.025cm 
\parbox{1.25cm}{\begin{picture}(50,60) 
\put(5,10){\line(1,0){40}} 
\put(5,50){\line(1,0){40}} 
\put(5,10){\line(0,1){40}} 
\put(45,10){\line(0,1){40}} 
\put(5,66){\line(1,0){40}} 
\put(28.5,10){\vector(1,0){0}} 
\put(28.5,50){\vector(1,0){0}} 
\put(5,26){\vector(0,-1){0}} 
\put(45,26){\vector(0,-1){0}} 
\put(-2,1){\scriptsize $#4$} 
\put(40,1){\scriptsize $#3$} 
\put(-2,57){\scriptsize $#1$} 
\put(40,57){\scriptsize $#2$} 
\put(21,0){\scriptsize $#6$} 
\put(21,55){\scriptsize $#5$} 
\put(-6,28){\scriptsize $#7$} 
\put(48,28){\scriptsize $#8$} 
\put(25,30){\scriptsize \makebox(0,0){$#9$}} 
\end{picture}}}

\newcommand{\occell}[4]{\unitlength 0.020cm 
\parbox{1.00cm}{\begin{picture}(50,60)
\thinlines
\put(5,10){\line(1,0){40}} 
\put(5,50){\line(1,0){40}} 
\put(5,10){\line(0,1){40}} 
\put(45,10){\line(0,1){40}} 
\put(30,10){\vector(1,0){0}} 
\put(30,50){\vector(1,0){0}} 
\put(5,25){\vector(0,-1){0}} 
\put(45,25){\vector(0,-1){0}} 
\put(2,-2){\scriptsize $#3$} 
\put(41,-2){\scriptsize $#4$} 
\put(2,54){\scriptsize $#1$} 
\put(41,54){\scriptsize $#2$} 
\end{picture}}}

\newcommand{\occelldeux}[6]{\unitlength 0.020cm 
\parbox{1.80cm}{\begin{picture}(90,60)
\thinlines
\put(5,10){\line(1,0){80}} 
\put(5,50){\line(1,0){80}} 
\put(5,10){\line(0,1){40}} 
\put(45,10){\line(0,1){40}} 
\put(85,10){\line(0,1){40}} 
\put(30,10){\vector(1,0){0}} 
\put(30,50){\vector(1,0){0}}
\put(70,10){\vector(1,0){0}} 
\put(70,50){\vector(1,0){0}}  
\put(5,25){\vector(0,-1){0}} 
\put(45,25){\vector(0,-1){0}} 
\put(85,25){\vector(0,-1){0}} 
\put(2,-2){\scriptsize $#4$} 
\put(41,-2){\scriptsize $#5$} 
\put(2,54){\scriptsize $#1$} 
\put(41,54){\scriptsize $#2$} 
\put(80,54){\scriptsize $#3$} 
\put(81,-2){\scriptsize $#6$} 
\end{picture}}}

\newcommand{\occellch}[4]{\unitlength 0.020cm 
\parbox{1.00cm}{\begin{picture}(50,60)
\thinlines
\put(5,10){\line(1,0){40}} 
\put(5,50){\line(1,0){40}} 
\put(5,10){\line(0,1){40}} 
\put(45,10){\line(0,1){40}} 
\put(30,10){\vector(1,0){0}} 
\put(30,50){\vector(1,0){0}} 
\put(5,25){\vector(0,-1){0}} 
\put(45,25){\vector(0,-1){0}}
\put(21,-5){\scriptsize $#2$} 
\put(21,58){\scriptsize $#1$} 
\put(-10,28){\scriptsize $#3$} 
\put(50,28){\scriptsize $#4$}
\end{picture}}}

\begin{document}

\pagenumbering{roman}


\thispagestyle{empty}
\begin{center}

\LARGE{\bf TH\`ESE DE DOCTORAT}\\

\vskip 2.5truecm

\large{pr\'esent\'ee par}\\

\vskip 0.5truecm

\Large{\bf Gil Schieber}\\

\vskip 1.5truecm

\large{pour obtenir le grade de docteur de
l'Universit\'e de Provence, sp\'ecialit\'e}\\
\vskip 0.5truecm
\large{physique des particules, physique
math\'ematique et mod\'elisation.}\\

\vskip 2.0truecm
\large{\bf L'ALG\`EBRE DES SYM\'ETRIES QUANTIQUES D'OCNEANU}
\vskip 0.5truecm
\large{\bf ET  LA CLASSIFICATION DES SYST\`EMES CONFORMES \`A 
2D}\\

\vfill

\large{Soutenue le  16 Septembre 2003,
devant un jury compos\'e de:}
\end{center}

\begin{center}
{\large
\begin{tabular}{l}
R. Amorim (co-directeur de th\`ese, pr\'esident du jury)\\ 
R. Coquereaux (directeur de th\`ese)\\
M. V. Cougo Pinto\\
O. Ogievetsky\\
I. Roditi\\
T. Sch\"ucker\\[0.3truecm]
F. Toppan (rapporteur)\\
R. Trinchero (rapporteur)
\end{tabular}}
\end{center}

\cleardoublepage



\thispagestyle{empty}

\begin{center}
{\huge{\bf Remerciements}}
\end{center}

\vskip 1.5truecm

Je tiens \`a remercier chaleureusement Robert Coquereaux, d'abord pour m'avoir accept\'e comme son \'etudiant de DEA,
et par la suite pour m'avoir orient\'e dans cette th\`ese (m\^eme si souvent \`a distance!). Ses conseils et 
motivations m'ont \'et\'e indispensables pour en arriver l\`a: un grand merci pour ces d\'ej\`a nombreuses 
ann\'ees de collaboration et d'amiti\'e, qui, j'en suis s\^ur, continueront dans le futur.
\newline

Touta ma reconaissance \`a Juan Alberto Mignaco, pour avoir accept\'e d'\^etre mon directeur de th\`ese 
au Br\'esil, et m'avoir accueilli \`a Rio de Janeiro: c'est un grand dommage qu'il n'ait pu \^etre 
l\`a pour en voir la fin. Un grand merci \`a Ricardo Amorim, pour avoir accept\'e de me codiriger 
\`a Rio par la suite.
\newline

Je remercie vivement tous les membres du jury, notamment ceux venant de loin, comme Oleg Ogievetsky; ainsi
que les rapporteurs Francesco Toppan et Roberto Trinchero,
pour avoir accept\'e d'\'evaluer ce travail. 
\newline

Cette th\`ese a \'et\'e effectu\'ee dans deux tr\`es belles villes c\^oti\`eres --
Marseille et Rio de \mbox{Janeiro --}
gr\^ace \`a une convention de cotutelle sign\'ee entre l'Universit\'e de Provence et l'Universit\'e
F\'ed\'erale de Rio de Janeiro. 

Ma gratitude aux membres du secr\'etariat du Centre de Physique Th\'eorique \`a Marseille, notamment
Sylvie, Mich\`ele, Corinne et Dolly; ainsi qu'\`a ceux de l'Instituto de F\'{\i}sica de l'UFRJ, 
Cas\'e et M\'arcia, pour leur tr\`es pr\'ecieuse aide
dans mes embarras administratifs, ainsi que pour leur bonne humeur.   
\newline

Je remercie mes amis et coll\`egues de travail des deux c\^ot\'es de l'Atlantique:
S\'ebastien, Pierre, Sam et David, pour les bons moments et la bonne ambiance au CPT,
ainsi que Pablo et Bernd, pour une non moins bonne ambiance \`a l'UFRJ. 
\newline

Enfin, une mention sp\'eciale pour mes amis belges, que j'ai quitt\'e il y a longtemps d\'ej\`a, 
mais qui ont toujours continu\'e \`a m'encourager et \`a me soutenir dans cette longue entreprise, et \`a
ma famille, pour son soutien moral et financier oh combien n\'ecessaire les derniers mois de cette th\`ese... 
\newline

Ma derni\`ere pens\'ee va \`a Larissa et Irache, qui, \`a des moments diff\'erents de cette th\`ese, ont partag\'e 
ma vie.


\tableofcontents


\chapter*{Introduction} 
\addcontentsline{toc}{chapter}{Introduction} 
\markboth{\uppercase{\bf{Introduction}}} 
{\uppercase{\bf{Introduction}}}

\pagenumbering{arabic}

\thispagestyle{empty}

L'\'epoque des grands savants multidisciplinaires -- qu'ils soient grecs, \'egyptiens ou chinois --
est de nos jours r\'evolue. Le d\'eveloppement naturel du savoir l'am\`ene \`a une ramification
de plus en plus pointue. Toutefois, la math\'ematique et la physique (th\'eorique), m\^eme si pouvant 
\^etre conceptuellement class\'ees comme deux domaines diff\'erents du savoir, ont chemin\'e main dans 
la main de l'Antiquit\'e jusqu'au d\'ebut du vingti\`eme si\`ecle. Archim\`ede, un des plus grands 
math\'ematiciens 
de son \'epoque, \'etait \'egalement un brillant physicien; \`a partir de Galil\'ee et l'av\`enement de la 
physique dite moderne, ces deux derniers domaines \'etaient m\^eme devenus indissociables. Les grands 
bouleversements de paradigmes en
physique ont toujours \'et\'e pr\'ec\'ed\'es ou accompagn\'es de la d\'ecouverte de nouvelles structures
math\'ematiques: m\'ecanique 
classique et calcul diff\'erentiel ont ainsi \'et\'e d\'evelopp\'es conjointement par des 
math\'ematiciens et 
physiciens: Newton, Euler, Lagrange, Hamilton $\ldots$ De m\^eme, les trois grandes r\'evolutions 
physiques du d\'ebut du si\`ecle dernier, \`a savoir la relativit\'e restreinte, la m\'ecanique 
quantique et la th\'eorie de la gravitation d'Einstein, ne peuvent \^etre \'evoqu\'ees sans penser au
groupe de Poincar\'e, \`a l'espace de Hilbert et \`a la g\'eom\'etrie Riemannienne (voir \cite{nahm}). 

L'av\`enement de la th\'eorie quantique des champs (l'unification de la m\'ecanique quantique et de la
relativit\'e restreinte) et ses probl\`emes, notamment ceux inh\'erents \`a l'apparition de divergences, a 
cr\'e\'e un grand \'eloignement entre ces deux communaut\'es. 
La parole \`a Res Jost (cit\'e dans \cite{todorov}): ``$\ldots$ sous l'influence d\'emoralisante de 
la th\'eorie quantique des champs 
perturbative (infest\'ee de divergences), les math\'ematiques n\'ecessaires \`a un physicien 
th\'eoricien
ont \'et\'e r\'eduites \`a la connaissance rudimentaire de l'alphabet Grec et Latin.'' 
Les math\'ematiciens 
n'\'etaient pas en reste, Weil et Dieudonn\'e par exemple affirmant que ``les math\'ematiques du vingti\`eme 
si\`ecle ne souffriront pas l'influence de la physique'' \cite{cartier}. Ainsi Dyson d\'eclara en 1972:
``le mariage entre math\'ematique et physique $\ldots$ a r\'ecemment termin\'e en divorce'' \cite{dyson}.

Ces derni\`eres d\'ecennies, un nouveau rapprochement entre math\'ematique pure et physique th\'eorique
s'est op\'er\'e, b\'en\'efique pour ces deux branches du savoir, car stimulant et mutuellement 
enrichissant. Citons entre autres exemples la formulation de la m\'ethode de {\it scattering} quantique 
inverse introduite par Faddeev, Sklyanin et Takhtajan \cite{QISM} pour les mod\`eles int\'egrables
qui amena \`a la d\'ecouverte des groupes quantiques\footnote{Nous devrions plut\^ot parler de {\sl 
red\'ecouverte} puisque les groupes quantiques sont des cas sp\'eciaux d'alg\`ebres de Hopf 
\cite{sweedler}, structures d\'ej\`a connues des math\'ematiciens.} \cite{Drinfeld}, ou
la d\'ecouverte r\'ecente  d'une structure d'alg\`ebre de Hopf dans la 
combinatoire du programme de renormalisation de la th\'eorie des champs perturbative 
\cite{kreimer,con_krei-1} et ses liens avec un 
probl\`eme de Riemann-Hilbert \cite{con_krei-2}.
Les {\bf th\'eories conformes \`a deux dimensions} offrent un autre exemple marquant d'un terrain d'entente
par sa transparence math\'ematique et ses riches applications 
physiques. Un syst\`eme conforme est invariant sous les transformations de l'espace qui conservent
les angles, donc notamment sous les transformations d'\'echelle. Les applications physiques concernent
les transitions de phase dans les ph\'enom\`enes critiques (car alors aucun param\`etre d'\'echelle 
n'intervient), mais aussi les mod\`eles int\'egrables et principalement la th\'eorie des cordes. 
En math\'ematique, l'\'etude des syst\`emes conformes
est \`a la base de la formulation de nouvelles structures alg\'ebriques introduites par
A. Ocneanu \cite{Oc-Marseille}. De fait, il est assez surprenant et stimulant de savoir que ce 
math\'ematicien, sp\'ecialiste des alg\`ebres d'op\'erateur, a utilis\'e une classification des 
fonctions de partition de th\'eorie des champs conformes pour obtenir ces structures! \\


Depuis l'article fondateur de Belavin, Polyakov et Zamolodchikov \cite{BPZ-Ising}, les syst\`emes 
conformes \`a deux dimensions ont constitu\'e un intense
domaine de recherche. Dans un tel syst\`eme, l'alg\`ebre des transformations (alg\`ebre de 
Virasoro) est de dimension infinie: les contraintes impos\'ees sur les fonctions de corr\'elation du
syst\`eme sont alors telles qu'il est possible dans certains cas de le r\'esoudre explicitement, ouvrant
ainsi la voie \`a des classifications. Un cas important est celui de th\'eories poss\'edant comme 
sym\'etrie \'etendue une alg\`ebre de courant (contenant Virasoro), en particulier les mod\`eles avec 
alg\`ebre affine $\widehat{g}$. 

\`A deux dimensions, le syst\`eme est d\'efini sur un r\'eseau bi-dimensionnel. En d\'efinissant 
des conditions p\'eriodiques selon les deux axes, la g\'eom\'etrie du syst\`eme se ram\`ene \`a celle 
d'un tore. 
Pour les mod\`eles $\widehat{su}(n)$, la classsification des fonctions de partition invariantes
modulaires d\'efinies
sur le tore (de param\`etre modulaire $\tau$) se r\'eduit \`a la classification  des matrices 
$\mathcal{M}$ \`a coefficients entiers non-n\'egatifs qui commutent avec les g\'en\'erateurs $S$ et $T$ 
du groupe modulaire.
$\mathcal{M}$ est appel\'ee l'invariant modulaire, et la fonction de partition s'\'ecrit
alors en termes des caract\`eres $\chi_i$ de l'alg\`ebre $\widehat{su}(n)$:
\begin{equation}
\mathcal{Z}(\tau) = \sum_{i,j}^{ } \chi_i(\tau)\; \mathcal{M}_{ij} \; \overline{\chi}_j(\ov{\tau}).
\end{equation}
La premi\`ere classification des fonctions de partition invariantes modulaires a \'et\'e obtenue 
pour les mod\`eles $\widehat{su}(2)$ en 1987 par Cappelli, Itzykson et Zuber \cite{CIZ-class1,CIZ-class2} 
et est connue sous le nom de  classification $ADE$. \`A chaque fonction de partition $\mathcal{Z}$ 
est associ\'e un graphe tel que son spectre soit cod\'e dans les \'el\'ements diagonaux de 
$\mathcal{M}$. C'est ainsi 
qu'apparaissent les diagrammes de Dynkin de type $ADE$, mais soulignons que cette analogie \'etait \`a 
l'\'epoque myst\'erieuse. Notons que les mod\`eles minimaux (comme par exemple le mod\`ele d'Ising ou 
le mod\`ele de Potts) sont reli\'es aux mod\`eles $\widehat{su}(2)$ par une construction de {\it coset}
\cite{goddard-coset1}: 
la classification des mod\`eles $\widehat{su}(2)$ conduit donc \`a celle des mod\`eles minimaux.
Plus g\'en\'eralement, la classification des mod\`eles affines joue un r\^ole pr\'epond\'erent dans
la classification des th\'eories conformes dites rationnelles.  
La classification des invariants modulaires des mod\`eles $\widehat{su}(3)$ a \'et\'e obtenue en 1994 par 
Gannon \cite{gannon-class}, et \`a cette classification est associ\'ee une liste de graphes appel\'es 
diagrammes de Coxeter-Dynkin g\'en\'eralis\'es.

Quand sont incorpor\'ees des conditions au bord (labell\'ees par $a$ et $b$) ou des lignes de 
d\'efaut (labell\'ees par $x$ et $y$) sur le syst\`eme de mani\`ere compatible avec l'invariance conforme, 
les fonctions de partition s'\'ecrivent:
\begin{eqnarray}
\mathcal{Z}_{a|b}(\tau) &=& \sum_i^{ } \; \mathcal{F}_{ab}^i \; \chi_i(\tau),  \label{introeq1}\\
\mathcal{Z}_{x|y}(\tau) &=& \sum_{i,j}^{ } \; \chi_i (\tau) \; \mathcal{W}_{xy}^{ij} \; 
\overline{\chi}_j(\ov{\tau}), \label{introeq2}
\end{eqnarray}
o\`u $\mathcal{F}_{ab}^i$ et $\mathcal{W}_{xy}^{ij}$ sont des coefficients 
entiers non-n\'egatifs formant des nimreps (``{\it numerical integer valued matrix representation}'')  
de certaines alg\`ebres. Le probl\`eme de la classification des fonctions de partition 
des th\'eories conformes $\widehat{su}(n)$ dans divers environnements se r\'eduit donc \`a la 
d\'etermination de l'ensemble de ces matrices. Or ces coefficients (ou ces matrices)
d\'efinissent les diverses structures d'une nouvelle classe d'alg\`ebres de Hopf, appel\'ees 
``alg\`ebres de Hopf faibles'' \cite{Bohm, Bohm-WHA}.\\


Une alg\`ebre de Hopf faible (WHA) est similaire \`a une alg\`ebre de Hopf usuelle. Elle poss\`ede 
un espace vectoriel muni d'un produit $\circ$ et d'un coproduit $\Delta$, compatibles dans
le sens usuel, et une unit\'e $\munite$, une counit\'e $\epsilon$ et une antipode $S$. Cependant, 
contrairement \`a une alg\`ebre de Hopf usuelle -- pour laquelle $\Delta(\munite) = \munite \otimes 
\munite$ -- dans une alg\`ebre de Hopf faible\footnote{Nous adoptons ici la convention de Sweedler: une sommation sur les indices de type (1) ou (2) est implicite.} $\Delta(\munite) = \munite_{(1)} \otimes \munite_{(2)}$. 
Tous les axiomes reli\'es \`a l'unit\'e doivent alors
\^etre modifi\'es en cons\'equence. Il a \'et\'e montr\'e que toute solution d'un ensemble
d'\'equations connues sous le nom de ``{\it The Big Pentagon Equation}'' fournit un exemple de WHA, dont
un cas particulier de solution provient des diff\'erents coefficients intervenant dans une th\'eorie 
conforme \cite{Oc-Guadeloupe, Bohm}. 

A. Ocneanu associe \`a l'espace des endomorphismes de chemins essentiels\footnote{La notion de chemins
essentiels sera introduite au chapitre {\bf 2}.} d\'efinis sur un graphe $G$ de type $ADE$ une 
{\sl dig\`ebre}, not\'ee $\mathcal{B}(G)$. Cette dig\`ebre 
est un espace vectoriel muni de deux produits $\circ$ et $\odot$. L'existence d'un produit scalaire 
permet de transposer le produit $\odot$ en un coproduit $\Delta$, de mani\`ere \`a ce que $\mathcal{B}(G)$ 
soit techniquement une WHA, m\^eme si aucune v\'erification n'a jamais \'et\'e explicitement men\'ee.
La dig\`ebre $\mathcal{B}(G)$ est semi-simple pour ses deux structures multiplicatives et peut donc \^etre 
diagonalis\'ee pour chacune de ces lois.  $\mathcal{B}(G)$ est isomorphe \`a une somme directe de blocs
matriciels de deux mani\`eres diff\'erentes:
\begin{equation}
\mathcal{B}(G) \, \cong \, \overset{ }{\underset{i}{\oplus}} \; L^i \; \cong \, 
\overset{}{\underset{x}{\oplus}} \; X^x \, . 
\end{equation}  
Les blocs pour la loi $\circ$ sont labell\'es par les vertex d'un graphe not\'e $\mathcal{A}(G)$:
c'est le graphe de la s\'erie $A$ poss\'edant la m\^eme norme que le graphe $G$. Les blocs pour la 
loi $\odot$ sont labell\'es par les vertex d'un autre graphe, appel\'e le graphe d'Ocneanu de $G$, not\'e $Oc(G)$.
L'espace vectoriel engendr\'e par les vertex de chacun de ces deux graphes (relatif \`a une des deux lois) est muni,
vis \`a vis de l'autre loi, d'une structure alg\'ebrique
associative: nous obtenons deux alg\`ebres que nous notons par le m\^eme symbole
que le graphe lui-m\^eme. L'alg\`ebre $\mathcal{A}(G)$ est commutative, mais l'alg\`ebre
$Oc(G)$, aussi appel\'ee l'{\bf alg\`ebre des sym\'etries quantiques} de $G$, ne l'est pas
toujours. 

La connaissance de ces alg\`ebres -- ou la donn\'ee des graphes correspondants -- 
permet de reconstruire l'ensemble des coefficients d\'efinissant 
les fonctions de partition des cas du type $\widehat{su}(2)$. En particulier, \`a un vertex sp\'ecial
du graphe d'Ocneanu (l'unit\'e) est associ\'ee une fonction de partition qui est invariante
modulaire: nous retrouvons ainsi la classification de Cappelli-Itzykson-Zuber.
Mais il est aussi possible d'associer des fonctions de partition aux autres 
points de ce graphe: elles ne sont plus invariantes modulaires, mais sont interpr\'et\'ees en th\'eorie
des champs conformes comme provenant d'un syst\`eme avec une ligne de d\'efauts.
Utilisant l'alg\`ebre des sym\'etries quantiques, il est aussi possible de d\'efinir des fonctions 
de partition provenant d'un syst\`eme avec deux lignes de d\'efauts. Ces fonctions de partition --
\`a une et deux lignes de d\'efauts -- sont appel\'ees {\it twist\'ees} ou g\'en\'eralis\'ees 
\cite{Pet_Zub-gener}. \\


Le travail central de cette th\`ese est la description d'une r\'ealisation de l'alg\`ebre des sym\'etries 
quantiques d'Ocneanu, construite comme un certain quotient du carr\'e tensoriel d'alg\`ebres de graphes d\'ej\`a 
connues. \`A partir de cette r\'ealisation, nous introduisons un algorithme simple permettant 
la d\'etermination de toutes les fonctions de partition (invariante modulaire et g\'en\'eralis\'ees)
pour tous les cas du type $\widehat{su}(2)$ \cite{Coq_Gil-ADE} (voir aussi \cite{Pet_Zub-Oc}, utilisant un formalisme diff\'erent). Par la suite, une caract\'erisation de cette 
r\'ealisation par les propri\'et\'es modulaires du graphe $G$ a permis de construire l'ag\`ebre des 
sym\'etries 
quantiques sans la n\'ecessit\'e de la connaissance pr\'ealable des graphes d'Ocneanu \cite{Coq_Gil-Tmod} 
(toutefois, pour les cas o\`u $Oc(G)$ n'est pas commutative, cette construction n'est pas enti\`erement 
satisfaisante).  

Les graphes d'Ocneanu ne sont connus (publi\'es) que pour les mod\`eles $\widehat{su}(2)$, mais la
liste des diagrammes de Coxeter-Dynkin g\'en\'eralis\'es a \'et\'e obtenue pour
les cas du type $\widehat{su}(3)$ \cite{DiFZub, DiFran_Zub, Oc-Bariloche} et $\widehat{su}(4)$ \cite{Oc-Bariloche}. 
Cependant, l'explicite diagonalisation de la loi $\odot$ pour une (hypoth\'etique?) dig\`ebre 
$\mathcal{B}(G)$ construite sur ces diagrammes 
g\'en\'eralis\'es n'a pas encore \'et\'e effectu\'ee.

 Les fonctions de partition g\'en\'eralis\'ees
des mod\`eles $\widehat{su}(n), n\geq 3$ n'\'etaient donc pas connues.  
Gr\^ace \`a la caract\'erisation introduite pr\'ec\'edemment, notre m\'ethode de construction de $Oc(G)$
se pr\^ete \`a une g\'en\'eralisation aux cas $\widehat{su}(n), n\geq 3$. 
Nous avons \'etudi\'es certains exemples choisis des mod\`eles $\widehat{su}(3)$, et obtenu 
les fonctions de partition associ\'ees. Nous retrouvons les fonctions de partition invariantes
modulaires correspondant \`a la classification de Gannon, confirmant ainsi notre construction,
et nous obtenons les expressions des fonctions de partition \`a une et deux lignes de 
d\'efaut des cas \'etudi\'es \cite{Coq_Gil-Tmod}.

La construction est la suivante. \`A un diagramme de Dynkin $G$ de type $ADE$ (ou possiblement 
g\'en\'eralis\'e) est associ\'e l'espace vectoriel $\mathcal{V}(G)$ engendr\'e par les vertex de ce 
diagramme. Dans certains cas, cet espace vectoriel poss\`ede une structure multiplicative 
associative et commutative avec des coefficients de structure entiers non-n\'egatifs 
(c'est notamment le cas pour les diagrammes de la s\'erie  $\mathcal{A}$) appel\'ee 
{\bf alg\`ebre de graphe}: nous
dirons alors que $G$ poss\`ede {\it self-fusion}. 
M\^eme si $G$ ne poss\`ede pas {\it self-fusion}, $\mathcal{V}(G)$ est toujours un module sous l'action
de l'alg\`ebre du graphe $\mathcal{A}(G)$ ayant la m\^eme norme (de Perron-Frobenius) que $G$. Si nous 
notons $\sigma_a$
les vertex de $G$ et $\tau_i$ les vertex de $\mathcal{A}(G)$, alors:
\begin{equation}
\tau_i \; . \; \sigma_a = \sum_b^{ } \; \mathcal{F}_{ab}^i \; \sigma_b \, , 
\end{equation}
et les coefficients $\mathcal{F}_{ab}^i$ sont les m\^emes que ceux de l'\'equation (\ref{introeq1}).
Dans les cas simples ($A_n$, $E_6$ et $E_8$ pour $\widehat{su}(2)$), l'alg\`ebre $Oc(G)$ est 
isomorphe au carr\'e tensoriel de l'alg\`ebre du graphe $G$, mais o\`u le produit tensoriel est pris 
au-dessus d'une sous-alg\`ebre $J$ de $G$, caract\'eris\'ee par les propri\'et\'es modulaires de $G$:
$Oc(G) \cong G \otimes_J G$.
Un \'el\'ement $x$ de $Oc(G)$ s'\'ecrit alors de la forme $x = \sigma_a \otimes_J \sigma_b$. 
Comme il existe une action de $\mathcal{A}(G)$ sur $\mathcal{V}(G)$, il existe aussi une action 
naturelle \`a droite et \`a gauche de $\mathcal{A}(G)$ sur $Oc(G)$. $Oc(G)$ est donc un bi-module sur 
$\mathcal{A}(G)$ et nous avons:
\begin{equation}
\tau_i \, . \, x \, . \, \tau_j \, = \, \tau_i \, . \, (\sigma_a \otimes_J \sigma_b)\, .\, \tau_j \, = \, 
\sum_{y}^{ } \; \mathcal{W}_{xy}^{ij} \; y \, ,
\end{equation}
o\`u les coefficients $\mathcal{W}_{xy}^{ij}$ sont les m\^emes que ceux de l'\'equation (\ref{introeq2}).
La d\'etermination de ces coefficients permet alors d'obtenir les fonctions de partition 
g\'en\'eralis\'ees du mod\`ele conforme associ\'e au graphe $G$.\\

Soulignons que, bien que les graphes $Oc(G)$ soient {\it a priori} d\'efinis \`a partir de la 
diagonalisation de la loi $\odot$ de $\mathcal{B}(G)$, la construction explicite de ces graphes 
par Ocneanu lui-m\^eme s'est bas\'ee sur la connaissance pr\'ealable de la classification des invariants
modulaires de $\widehat{su}(2)$ de Cappelli-Itzykson-Zuber, ou celle de $\widehat{su}(3)$ par Gannon.
Un autre axe de recherche d\'evelopp\'e dans cette th\`ese est l'\'etude approfondie de la
dig\`ebre $\mathcal{B}(G)$ et de ses structures, notamment \`a travers les cellules d'Ocneanu. L'objectif
est double: d'une part, nous voulons v\'erifier que la dig\`ebre $\mathcal{B}(G)$ est techniquement
une alg\`ebre de Hopf faible, d'autre part nous voulons obtenir la diagonalisation de $\mathcal{B}(G)$
pour la loi $\odot$ dans le but de construire explicitement les graphes d'Ocneanu \cite{Coq_Gil-bigebra}. \\

\noindent Le plan de la th\`ese est le suivant: \\

\noindent $\bullet$ Bien que le travail de recherche \`a proprement parler de cette th\`ese se situe 
plut\^ot au niveau alg\'ebrique, plusieurs r\'esultats obtenus sont interpr\'et\'es dans le langage
de la th\'eorie des champs conformes. Nous avons donc d\'ecid\'e de d\'edier le chapitre {\bf 1}
\`a ces notions. Apr\`es une courte introduction aux th\'eories des champs conformes \`a deux 
dimensions, et notamment aux th\'eories des champs conformes dites {\sl rationelles}, nous pr\'esentons 
les diverses classifications des mod\`eles $\widehat{su}(n)$ et des mod\`eles minimaux, et montrons
comment ces classifications sont naturellement reli\'ees \`a des coefficients formant 
des {\it nimreps} d'un certain ensemble d'alg\`ebres, pouvant \^etre cod\'ees par des graphes.\\

\noindent $\bullet$ Dans le chapitre {\bf 2} est pr\'esent\'ee la construction d'Ocneanu d'une 
dig\`ebre $\mathcal{B}(G)$ associ\'ee \`a un diagramme de Dynkin de type $ADE$. 
Nous montrons comment cette dig\`ebre permet
de d\'efinir les graphes $\mathcal{A}(G)$ et $Oc(G)$ et analysons explicitement l'exemple
du diagramme $A_3$.\\

\noindent $\bullet$ Le chapitre {\bf 3} est consacr\'e \`a la pr\'esentation d'une certaine
r\'ealisation de l'alg\`ebre des sym\'etries quantiques d'Ocneanu. Nous montrons comment cette
r\'ealisation -- qui se pr\^ete 
naturellement \`a une g\'en\'eralisation aux cas $\widehat{su}(n), n\geq 3$ -- permet d'obtenir un 
algorithme simple pour le calcul des divers coefficients entrant dans la d\'efinition des fonctions
de partition du mod\`ele conforme associ\'e. \\

\noindent $\bullet$ Dans le chapitre {\bf 4} nous traiterons explicitement tous les cas
du type $\widehat{su}(2)$ ainsi que trois exemples choisis du type $\widehat{su}(3)$.\\

\noindent $\bullet$ Les diagrammes de Dynkin $ADE$ et leur extension affine $ADE^{(1)}$, une pr\'esentation 
de la correspondance de Mc-Kay (classique et quantique), plusieurs d\'efinitions alg\'ebriques ainsi que 
les expressions des fonctions de partition g\'en\'eralis\'ees pour les exemples 
\'etudi\'es sont donn\'es en Annexe. \\

Les r\'esultats originaux obtenus dans cette th\`ese sont pr\'esent\'es dans le chapitre {\bf 4} (ils
ont en partie \'et\'e publi\'es dans les articles \cite{Coq_Gil-ADE, Coq_Gil-Tmod}) ainsi qu'\`a la fin du 
chapitre {\bf 2} (ils seront publi\'es dans \cite{Coq_Gil-bigebra}).


\chapter{Classification des th\'eories conformes \`a deux dimensions}
\thispagestyle{empty}
Les syst\`emes conformes sont les syst\`emes invariants sous les transformations de l'espace qui 
pr\'eservent les angles. \`A deux dimensions, ces syst\`emes deviennent tr\`es int\'eressants car
l'alg\`ebre des transformations est alors de dimension infinie. Par cons\'equent, il existe des mod\`eles
pour lesquels une classification a pu \^etre \'etablie: c'est notamment le cas pour les mod\`eles affines
$\widehat{su}(2)$ \cite{CIZ-class2}, $\widehat{su}(3)$ \cite{gannon-class} et pour les mod\`eles minimaux
\cite{CIZ-class2}. Dans ce chapitre, nous mentionons
les relations existantes entre ces classifications et un ensemble de coefficients pouvant \^etre cod\'e
par des graphes \cite{Pet_Zub-1996, Pet_Zub-1997, Pet_Zub-CFT}. Ce chapitre est d\'edi\'e \`a des r\'esultats d\'ej\`a connus mais parfois peu divulgu\'es
dans la litt\'erature (principalement en ce qui concerne les syst\`emes avec l'introduction de lignes de d\'efauts
\cite{Pet_Zub-gener}), c'est 
pourquoi nous avons jug\'e utile de les pr\'esenter ici.

\section{Th\'eories conformes: une introduction}

Nous donnons ici une br\`eve introduction aux th\'eories conformes, principalement \`a $2d$. 
Il existe de nos jours plusieurs bons textes traitant du sujet, citons entre 
autres \cite{ginsparg,ketov,FMS-book}.

\subsection{Importance des sym\'etries} 
L'objet principal d'une th\'eorie des champs est l'action $S[\Phi]$, 
fonctionnelle des champs, d\'efinie par: 
\begin{equation} 
S[\Phi] = \int d^d x \; \mathcal{L} (\Phi(x), \partial_{\mu} \Phi(x)), 
\end{equation} 
o\`u $d$ est la dimension de l'espace-temps et $\mathcal{L}$ la densit\'e 
lagrangienne du syst\`eme. 
Ici $\Phi$ est une collection de champs locaux (qui peuvent \^etre de nature 
tr\`es diff\'erente: scalaires, \mbox{spinoriels, \ldots).} 
Au niveau quantique, nous nous int\'eressons plus particuli\`erement aux 
fonctions de corr\'elation entre les champs: 
\begin{equation} 
\langle \Phi(\bs{x_1}) \cdots \Phi(\bs{x_n}) \rangle \doteq \frac{1}{\mathcal{Z}} \int 
[d\Phi] \; \Phi(\bs{x_1}) \cdots \Phi(\bs{x_n})\; e^{-S[\Phi]}, 
\end{equation} 
o\`u $\mathcal{Z}$ est la fonctionnelle g\'en\'eratrice du vide, aussi appel\'ee, en 
analogie avec la physique statistique, fonction de partition. 
\begin{defin} 
Nous dirons qu'un syst\`eme est {\bf soluble} si nous pouvons calculer 
explicitement les fonctions de corr\'elations  
entre tous les champs pr\'esents dans le syst\`eme. 
\end{defin}

De mani\`ere g\'en\'erale, un syst\`eme n'est pas soluble.
Par contre, l'invariance du syst\`eme sous une transformation se pr\'esente 
sous la forme 
de contraintes sur les fonctions de corr\'elations, \`a travers les identit\'es 
de Ward.  
Plus ``grande'' sera la sym\'etrie impos\'ee, plus nombreuses seront ces 
contraintes, pouvant dans certains 
cas nous amener \`a trouver des solutions explicites. Ce sera notamment 
le cas pour des syst\`emes invariants 
sous les transformations conformes \`a deux dimensions, car l'alg\`ebre conforme 
est alors de dimension infinie! 
Analysons tout d'abord une transformation g\'en\'erale sur le syst\`eme.   
 
\subsubsection{Transformation du point de vue actif} 
Une transformation g\'en\'erale sur le syst\`eme est d\'efinie par: 
\begin{equation} 
\begin{array}{rcl} 
\bs{x} &\longmapsto& \bs{x'} \\ 
\Phi (\bs{x}) &\longmapsto& \Phi'(\bs{x'}) = \mathcal{F}(\Phi(\bs{x})) 
\end{array} 
\label{transf} 
\end{equation} 
Nous adoptons ici le point de vue actif: la transformation change le syst\`eme de 
coordonn\'ees ($\bs{x} \mapsto \bs{x'}$), et le 
champ $\Phi$ est lui-m\^eme affect\'e par celle-ci ($\Phi \mapsto 
\Phi'$).
Pour des transformations infinit\'esimales, \`a $n$ param\`etres $w^a$, $a=(1,\ldots,n)$,  
avec $w^a \ll1$: 
\begin{equation} 
\begin{array}{rcl} 
x^{\mu} &\longmapsto& x^{'\mu} = x^{\mu} + \delta x^{\mu} = x^{\mu} + w^{a}  
\displaystyle\frac{\delta x^{\mu}}{\delta w^{a}}\\ 
\Phi(\bs{x}) &\longmapsto& \Phi'(\bs{x'}) = \Phi (\bs{x}) + \delta 
\Phi(\bs{x}) = \Phi(\bs{x}) +  
w^{a}  \displaystyle\frac{\delta \Phi}{\delta w^{a}} (\bs{x})
\end{array} 
\label{transinf} 
\end{equation} 
Le g\'en\'erateur $G_a$ de la transformation infinit\'esimale est d\'efini \`a 
partir de la transformation au {\bf m\^eme point}: 
\begin{equation} 
\Phi'(\bs{x}) = \Phi(\bs{x}) - i w^a G_a \Phi (\bs{x}). 
\label{gener} 
\end{equation} 
Le lien entre g\'en\'erateur et param\`etre est alors donn\'e par: 
\begin{equation} 
i G_a \Phi(\bs{x}) = \frac{\delta x^{\mu}}{\delta w^a} \partial_{\mu} 
\Phi(\bs{x}) - \frac{\delta \Phi}{\delta w^a}(\bs{x}). 
\end{equation} 
Consid\'erons maintenant un syst\`eme invariant sous la transformation 
(\ref{transf}): $S[\mathcal{F}(\Phi)]=S[\Phi]$.

\paragraph{Cons\'equence classique} 
Sous la transformation g\'en\'erale infinit\'esimale (\ref{transinf}), la 
variation de l'action ($S \mapsto S + \delta S$)  
est donn\'ee par: 
\begin{equation} 
\delta S = \int d^d \bs{x} \; \partial_{\mu} j^{\mu}_a(\bs{x}) \; w^a, 
\end{equation} 
o\`u $j^{\mu}_a(\bs{x})$ est le courant associ\'e \`a la transformation: 
\begin{equation} 
j^{\mu}_a (\bs{x}) = \left( \frac{\partial \mathcal{L}}{\partial (\partial_{\mu} 
\Phi)}\partial_{\nu} \Phi -g^{\mu}_{\;\,\nu} \mathcal{L}  
\right) 
\frac{\delta x^{\nu}}{\delta w^a} - \frac{\partial 
\mathcal{L}}{\partial(\partial_{\mu} \Phi)} \frac{\delta \Phi}{\delta w^a}. 
\label{courant} 
\end{equation} 
\begin{theo} 
Si les champs v\'erifient les \'equations classiques du mouvement, alors: 
\begin{equation} 
\delta S =0 \qquad \qquad \Longleftrightarrow \qquad \qquad \partial_{\mu} 
j^{\mu}_a(\bs{x}) =0 
\end{equation} 
\end{theo} 
La cons\'equence classique de l'invariance du syst\`eme sous la transformation 
g\'en\'erale (\ref{transf}) est la loi de conservation du courant associ\'e:  
c'est le {\bf th\'eor\`eme de Noether}.

\paragraph{Cons\'equence quantique} 
Si l'action est invariante sous la transformation (\ref{transf}) et si, de plus, 
nous faisons l'hypoth\`ese que la mesure d'int\'egration  
l'est aussi\footnote{Ce n'est pas toujours le cas, notamment si nous introduisons  
une proc\'edure de r\'egularisation.}, alors 
les fonctions de corr\'elations doivent satisfaire les contraintes suivantes: 
\begin{equation} 
\langle \Phi(\bs{x_1'}) \cdots \Phi(\bs{x_n'}) \rangle = \langle \Phi'(\bs{x_1'}) \cdots 
\Phi'(\bs{x_n'}) \rangle .
\label{fctcorr} 
\end{equation} 
Si nous nous int\'eressons au niveau infinit\'esimal de la transformation, 
l'invariance du syst\`eme se traduit par les  
{\it \bf identit\'es de Ward}: 
\begin{equation} 
\langle \delta \Phi \rangle = \int d^d\bs{x} \partial_{\mu} \langle j^{\mu}_a(\bs{x}) \Phi \rangle w^a. 
\label{ward} 
\end{equation} 
En sp\'ecifiant la variation de $\Phi$ \`a travers les g\'en\'erateurs de la 
transformation (\ref{gener}), nous obtenons la forme 
locale des identit\'es de Ward: 
\begin{equation} 
-i \sum_{i=1}^{n} \delta (\bs{x}-\bs{x_i}) 
\langle \Phi(\bs{x_1}) \cdots G_a \Phi(\bs{x_i}) \cdots \Phi(\bs{x_n}) \rangle = 
\frac{\partial}{\partial x^{\mu}} \langle j^{\mu}_a (\bs{x})\Phi(\bs{x_1'}) \cdots 
\Phi(\bs{x_n'}) \rangle. 
\end{equation} 
Au niveau quantique, l'invariance d'un syst\`eme sous une transformation 
g\'en\'erale se traduit donc par des contraintes 
impos\'ees sur les fonctions de corr\'elations.

\subsection{Transformations conformes} 
Nous consid\'erons un espace-temps \`a $d$ dimensions, avec une m\'etrique 
$g_{\mu \nu}$. Sous un changement de coordonn\'ees 
$\bs{x}\mapsto \bs{x'}$, la m\'etrique se transforme comme: 
\begin{equation} 
g_{\mu \nu} (\bs{x}) \longmapsto g_{\mu \nu}'(\bs{x'}) = \frac{\partial 
x^{\alpha}}{\partial x^{'\mu}}  
\frac{\partial x^{\beta}}{\partial x^{'\nu}} g_{\alpha \beta} (\bs{x}). 
\end{equation} 
\begin{defin} 
Une transformation {\bf conforme} est un \'el\'ement du sous-groupe des transformations de 
coordonn\'ees qui laisse la m\'etrique invariante  
\`a un facteur d'\'echelle pr\`es: 
\begin{equation} 
g_{\mu \nu}' (\bs{x'}) = \Lambda (\bs{x}) g_{\mu \nu} (\bs{x}). 
\label{deftrconf} 
\end{equation} 
\end{defin} 
Les transformations conformes sont les transformations de l'espace qui 
pr\'eservent les angles.\\ 
Consid\'erons maintenant une transformation infinit\'esimale de param\`etres 
$\xi^{\mu}$: $x^{'\mu} = 
x^{\mu} + \varepsilon \; \xi^{\mu}$, avec $\varepsilon \ll 1$. 
En imposant (\ref{deftrconf}), nous obtenons des contraintes sur $\xi^{\mu}$: 
\begin{equation} 
\partial_{\mu} \xi_{\nu} + \partial_{\nu}\xi_{\mu} = \frac{2}{d} g_{\mu\nu}  
\partial_{\rho}\xi^{\rho}.  
\label{defxi} 
\end{equation} 
Ces contraintes nous permettent de sp\'ecifier la transformation conforme, les 
cas \`a $d=2$ et $d>2$ \'etant tr\`es diff\'erents.  
 
\subsubsection{Cas $d>2$} 
Les transformations conformes finies sont: 
$$ 
\begin{array}{lccl} 
\textrm{ $\bullet$ translation:} & \displaystyle x^{'\mu} &=& x^{\mu} + a^{\mu} \\ 
\textrm{ $\bullet$ Lorentz (rotations):} & \displaystyle x^{'\mu} &=& \displaystyle 
\Lambda^{\mu}_{\;\,\nu} x^{\nu} \qquad \qquad \Lambda^{\mu}_{\;\,\nu} = - 
\Lambda^{\nu}_{\;\,\mu} \\ 
\textrm{ $\bullet$ dilatation:} & \displaystyle  x^{'\mu} &=& \alpha \, x^{\mu} \\ 
\textrm{ $\bullet$  SCT:} & \displaystyle   \frac{x^{'\mu}}{\bs{x^{'2}}} &=& 
\displaystyle \frac{x^{\mu}}{\bs{x^2}} - b^{\mu} 
\end{array} 
$$ 
o\`u la derni\`ere transformation est la ``transformation conforme sp\'eciale'' 
(SCT), qui n'est autre qu'une inversion, suivie d'une translation et d'une 
nouvelle inversion. 
Ces transformations forment un groupe \`a un nombre {\bf fini} de param\`etres, 
\'egal \`a: $\frac{1}{2}(d+1)(d+2)$.  
Dans un espace avec signature $(p,q)$ de dimension $d=p+q$, le groupe conforme est isomorphe 
\`a $SO(p+1,q+1)$.   
Consid\'erons un syst\`eme invariant sous le groupe des transformations 
conformes: 
 
\paragraph{Cons\'equence classique} 
Il est bien connu que l'invariance par translation et transformations de Lorentz 
(le groupe de Poincar\'e) conduit \`a  
la conservation du tenseur \'energie-impulsion $T^{\mu\nu}$.
Consid\'erons aussi les dilatations, d\'efinies par: 
\begin{equation} 
x^{\mu} \longmapsto x^{'\mu} = \lambda x^{\mu}, \qquad \qquad \qquad \qquad 
\Phi(\bs{x}) \longmapsto \Phi'(\bs{x'}) =  
\lambda^{-\Delta} \Phi(\bs{x}), 
\end{equation} 
o\`u $\Delta$ est la dimension d'\'echelle du champ $\Phi$.  
Le courant associ\'e \`a la dilatation est: 
\begin{equation} 
j^{\mu}_{D} (\bs{x}) = T^{\mu}_{\;\;\nu} x^{\nu}, 
\end{equation} 
et la conservation de ce courant implique que le tenseur \'energie-impulsion 
est de trace nulle: 
\begin{equation} 
T^{\mu}_{\;\;\mu} =0.  
\label{trnulle} 
\end{equation} 
Le courant conforme associ\'e \`a une transformation g\'en\'erale 
infinit\'esimale (\ref{defxi}), d\'efini par 
$J^{\mu} \doteq T^{\mu\nu} \xi_{\nu}$, est alors conserv\'e si $T^{\mu\nu}$ est 
de trace nulle: 
\begin{equation} 
\partial_{\mu}J^{\mu} = \frac{1}{2} T^{\mu\nu} (\partial_{\mu}\xi_{\nu} + 
\partial_{\nu}\xi_{\mu}) = 0. 
\end{equation} 
De fait, la condition de conservation du tenseur \'energie-impulsion et la 
condition (\ref{trnulle}) impliquent  
l'invariance sous toutes transformations conformes.

\paragraph{Cons\'equence quantique} 
L'invariance par translation et rotation implique par 
(\ref{fctcorr}) que 
la fonction de corr\'elation \`a deux points est de la forme 
suivante:  
\begin{equation} 
\langle \phi_1(\bs{x_1}) \phi_2(\bs{x_2}) \rangle = f(|\bs{x_1}-\bs{x_2}|). 
\end{equation} 
Nous verrons que l'invariance conforme permet de fixer explicitement la forme des fonctions de 
corr\'elations \`a deux et trois points.

\subsubsection{Cas $d=2$} 
Dans un espace euclidien \`a deux dimensions, les contraintes sur $\xi^{\mu}$ 
pour que la transformation soit conforme s'\'ecrivent: 
\begin{equation} 
\partial_0 \xi^1 + \partial_1 \xi^0 =0, \qquad \qquad \partial_0 \xi^0 = 
\partial_1 \xi^1.
\end{equation} 
Ce sont les \'equations de Cauchy-Riemann, qui d\'efinissent une fonction 
holomorphe. 
Il est donc naturel de travailler sur le plan complexe et introduire: 
$$ 
\left\{  
\begin{array}{ccl}  
z &=& x^0 + i x^1 \\ 
\ov{z} &=& x^0 -i x^1 
\end{array} 
\right. 
\qquad \qquad \qquad 
\left\{  
\begin{array}{ccl}  
\xi &=& \xi^0 + i \xi^1 \\ 
\ov{\xi} &=& \xi^0 -i \xi^1 
\end{array} 
\right. 
$$ 
Alors les \'equations de Cauchy-Riemann s'\'ecrivent: 
\begin{equation} 
\frac{\partial}{\partial z} \ov{\xi}(z,\ov{z}) = 0, \qquad \qquad \qquad 
\frac{\partial}{\partial \ov{z}} \xi(z,\ov{z}) = 0,
\end{equation} 
admettant comme solution toute transformation finie analytique (resp. 
anti-analytique): 
\begin{equation} 
z \longmapsto z'=f(z), \qquad \qquad \qquad \ov{z} \longmapsto \ov{z}' = 
\ov{f}(\ov{z}).
\label{transfconf} 
\end{equation} 
Les variables $z$ et $\ov{z}$ se d\'ecouplent, et peuvent \^etre trait\'ees 
comme deux variables complexes  
ind\'ependantes, la condition physique de r\'ealit\'e: $\ov{z}=z^*$, o\`u $z^*$ 
d\'esigne le complexe conjugu\'e de $z$, pouvant 
\^etre impos\'ee \`a tout moment.
Pour une transformation (\ref{transfconf}) infinit\'esimale: 
\begin{equation} 
z \longmapsto z' = z + \varepsilon(z), \qquad \qquad \qquad \qquad \ov{z} 
\longmapsto \ov{z}' = \ov{z} + \ov{\varepsilon}(\ov{z}), 
\end{equation} 
o\`u $\varepsilon(z)$ et $\ov{\varepsilon}(\ov{z})$ peuvent \^etre prises infiniment  
petites dans un disque de rayon fix\'e. Nous pouvons les d\'evelopper en 
s\'erie de Laurent autour de $z=0$: 
\begin{equation} 
\varepsilon(z) = \sum_{n=-\infty}^{+\infty} \varepsilon_n  z^{n+1}, \qquad \qquad 
\qquad \qquad   
\ov{\varepsilon}(\ov{z}) = \sum_{n=-\infty}^{+\infty} \ov{\varepsilon}_n
\ov{z}^{n+1},
\end{equation} 
et les g\'en\'erateurs correspondants sont de la forme: 
\begin{equation} 
\ell_n = - z^{n+1} \partial_z, \qquad \qquad \qquad \qquad \ov{\ell}_n = - \ov{z}^{n+1}
\partial_{\ov{z}}.
\end{equation} 
Le nombre de g\'en\'erateurs des transformations conformes \`a 2d est donc infini! Ces g\'en\'erateurs  
forment l'{\bf alg\`ebre de Witt}, dont les relations de 
commutation sont donn\'ees par: 
\begin{eqnarray} 
\left[ \ell_n,\ell_m \right] &=& (n-m)\; \ell_{n+m} \nonumber\\ 
\label{witt}   
\left[ \ell_n,\ov{\ell}_m \right] &=& 0 \\ 
\left[ \ov{\ell}_n,\ov{\ell}_m \right] &=& (n-m) \; \ov{\ell}_{n+m} \nonumber  
\end{eqnarray} 
Pour former un groupe, les transformations doivent \^etre inversibles et  
d\'efinies en tout point de l'espace, auquel cas nous leur r\'eservons le 
nom de transformations globales. 
L'ensemble des transformations globales forment le groupe conforme $SO(3,1)$ \`a deux dimensions, 
dont les g\'en\'erateurs sont: 
$\{\ell_{-1},\ell_0,\ell_1\} \cup \{\ov{\ell}_{-1},\ov{\ell}_0,\ov{\ell}_1\}$. 
$\ell_{-1}$ et $\ov{\ell}_{-1}$ sont les g\'en\'erateurs des translations, $(\ell_{0} + 
\ov{\ell}_{0})$ et $i(\ell_{0} + \ov{\ell}_{0})$  
sont respectivement les g\'en\'erateurs des dilatations et rotations; $\ell_{1}$ 
et $\ov{\ell}_{1}$ les g\'en\'erateurs des 
transformations sp\'eciales conformes. 
Les autres transformations conformes ne sont pas globales (mais locales), et ne forment 
pas un groupe: c'est pourquoi nous parlerons plus g\'en\'eralement d'alg\`ebre conforme.

\paragraph{Tenseur \'energie-impulsion} 
Dans le plan complexe, les propri\'et\'es de sym\'etrie et de trace nulle du 
tenseur \'energie-impulsion impliquent 
$T^{z\ov{z}}=T^{\ov{z}z}=0$.  
La conservation de ce tenseur s'\'ecrit alors: 
\begin{equation} 
\partial_{\ov{z}}\, T^{zz} =0, \qquad \partial_{z}\, T^{\ov{z}\ov{z}} =0,  
\end{equation} 
et nous introduisons: 
\begin{equation} 
T(z) = - 2 \pi T_{zz}, \qquad \qquad \ov{T}(\ov{z}) =  - 2 \pi T_{\ov{z}\ov{z}}, 
\label{Tmunu} 
\end{equation} 
qui sont des fonctions respectivement holomorphe et anti-holomorphe du plan complexe. 
 
\paragraph{Identit\'es de Ward} 
\`A deux dimensions, nous travaillons dans le plan complexe, ce qui nous permet d'utiliser la 
puissance du calcul analytique.  
Pour $X$ une collection de champs locaux: $X=\phi_1(w_1,\ov{w}_1) \ldots 
\phi_n(w_n,\ov{w}_n)$, les identit\'es de Ward (\ref{ward}) 
provenant de l'invariance conforme -- o\`u le courant conserv\'e est le tenseur  
\'energie-impulsion $T_{\mu\nu}$ --  
s'\'ecrivent: 
\begin{equation} 
\langle \delta_{\varepsilon, \ov{\varepsilon}} X \rangle = \frac{1}{2i\pi} \oint_{\Gamma} dz 
\varepsilon(z) \langle T(z)X \rangle +  
\frac{1}{2i\pi} \oint_{\Gamma} d\ov{z}  \ov{\varepsilon}(\ov{z}) 
\langle \ov{T}(\ov{z})X \rangle 
\label{wardconf} 
\end{equation} 
o\`u le contour $\Gamma$ inclut toutes les positions $(w_i,\ov{w}_i)$ des champs 
contenus dans $X$, et o\`u $T(z)$ est d\'efini en (\ref{Tmunu}).
Pour donner une forme locale pr\'ecise \`a (\ref{wardconf}), il nous faut 
soit conna\^{\i}tre la variation 
des champs sous une transformation conforme (membre de gauche), soit pouvoir 
d\'evelopper le membre de droite et calculer explicitement l'int\'egrale, ce qui nous 
am\`ene \`a introduire  
l'expansion du produit des champs. 
 
\begin{remarq} 
Nous avons vu que les variables $z$ et $\ov{z}$ se d\'ecouplent, nous permettant
de traiter les deux parties s\'eparemment. 
Par la suite, nous allons souvent ne traiter que de la partie holomorphe, la 
partie anti-holomorphe donnant lieu \`a 
des r\'esultats parall\`eles.
\end{remarq}

\subsection{OPE des champs} 
 
\subsubsection{D\'efinition de l'OPE} 
L'expansion en produit d'op\'erateurs (OPE), introduite par Wilson, joue un 
r\^ole important en th\'eorie quantique des champs.  
L'OPE de deux op\'erateurs locaux donne leur comportement \`a courte 
distance ($\bs{x}\rightarrow \bs{y}$): 
\begin{equation} 
A_i(\bs{x}) B_j(\bs{y}) \sim \sum_k C_{ij}^{k}(\bs{x}-\bs{y}) O_k(\bs{y}),
\end{equation} 
o\`u les $C_{ij}^{k}$ sont des coefficients num\'eriques singuliers englobant 
les divergences pour $\bs{x}\rightarrow \bs{y}$, et o\`u 
les $O_k$ forment un ensemble complet d'op\'erateurs locaux.
Cette propri\'et\'e en TQC n'est normalement valable qu'asymptotiquement. Par 
contre, en th\'eorie des champs conformes, 
elle devient une propri\'et\'e exacte, car aucun param\`etre de 
longueur $\ell$ n'appara\^{\i}t 
dans l'expansion compte tenu de l'invariance d'\'echelle.
Traduite en formalisme pour les champs conformes, nous pouvons alors \'ecrire: 
\begin{equation} 
A_i(z) B_j(w) = \sum_{k=-\infty}^{N} \frac{C_{ij}^{k}}{(z-w)^k} O_k (w)   
= \sum_{k=1}^{N} \frac{C_{ij}^{k}}{(z-w)^k} O_k(w) + \textrm{ \bf r\'eg.} 
\end{equation} 
o\`u $N$ est un entier positif. Dans la deuxi\`eme \'equation, nous avons 
s\'epar\'e la partie divergente  
pour $z\rightarrow w$ (terme de gauche) de la partie r\'eguli\`ere (not\'ee {\bf 
r\'eg.}), car seulement cette  
premi\`ere va survivre \`a  l'int\'egration dans les identit\'es de Ward (\ref{wardconf}).
La connaissance de l'OPE des champs pr\'esents dans le syst\`eme est 
d'une grande utilit\'e: elle nous permet 
de ramener le calcul des fonctions de corr\'elations \`a $N$ points 
graduellement \`a celui \`a 2 points.  
L'identit\'e de Ward conforme nous permet d'expliciter l'OPE pour une certaine 
classe de champs dont nous connaissons  
la loi de transformation. 
 
\subsubsection{Champs primaires} 
Il existe des champs, appel\'es {\bf champs primaires}, dont la loi de 
transformation sous une transformation conforme est donn\'ee. 
Soit un champ $\phi$ de spin $s$ et de dimension d'\'echelle $\Delta$, sa dimension 
conforme $h$ (resp. $\ov{h}$) est d\'efinie par: 
\begin{equation} 
h = \frac{1}{2} (\Delta + s), \qquad \qquad \qquad \qquad \ov{h} = \frac{1}{2}
(\Delta - s).
\end{equation} 
\begin{defin} 
Sous une transformation conforme $z \mapsto w(z)$, $\ov{z} \mapsto 
\ov{w}(\ov{z})$, un {\bf champ primaire} est un champ 
qui se transforme comme une $(h,\ov{h})$-forme: 
\begin{equation} 
\phi(z,\ov{z}) \longmapsto \phi'(w,\ov{w}) = \left( \frac{dw}{dz} 
\right)^{-h} \left( \frac{d\ov{w}}{d\ov{z}} \right)^{-\ov{h}}  
\phi(z,\ov{z}).
\end{equation} 
\end{defin} 
En sp\'ecifiant pour une transformation infinit\'esimale $z \mapsto w = z + 
\varepsilon (z)$, la variation du champ primaire 
est donn\'ee par: 
\begin{equation} 
\delta_{\varepsilon}\phi(w) = \phi'(w)-\phi(w)=-\varepsilon(w) \partial_w \phi(w) 
-h \left[ \partial_w \varepsilon(w) \right] \phi(w). 
\label{varprim} 
\end{equation} 
Ceci nous permet de conclure, d'apr\`es l'identit\'e de Ward conforme 
(\ref{wardconf}), que l'OPE d'un champ primaire  
avec le tenseur \'energie-impulsion s'\'ecrit: 
\begin{equation} 
T(z)\phi(w) =  \left\{ \frac{1}{z-w} \partial_{w} \phi(w) + \frac{h}{(z-w)^2} 
\phi(w) \right\} + \text{\bf r\'eg.} 
\end{equation} 
o\`u {\bf r\'eg.} d\'esignent des termes r\'eguliers. En effet, nous pouvons 
v\'erifier qu'en mettant cette expression  
dans (\ref{wardconf}),  
nous retrouvons bien la variation infinit\'esimale du champ primaire donn\'ee en 
(\ref{varprim}). 
Connaissant la loi de transformation des champs primaires, l'invariance 
conforme, par (\ref{fctcorr}), permet de fixer  
la forme des fonctions 
de corr\'elations \`a deux et trois points. Soient $\phi_i(z_i)$ des champs 
primaires, et $z_{ij}=z_i-z_j$, alors: 
\begin{eqnarray*} 
\langle \phi_1(z) \phi_2(w) \rangle &=& \frac{C_{12}}{(z-w)^{2h}}, \qquad \qquad h=h_1=h_2, \\ 
\langle \phi_1(z_1) \phi_2(z_2) \phi_3(z_3) \rangle &=& 
\frac{C_{123}}{z_{12}^{h_1+h_2-h_3}z_{23}^{h_2+h_3-h_1}z_{13}^{h_1+h_3-h_2}}.
\end{eqnarray*} 
Deux champs primaires ne sont corr\'el\'es que s'ils ont la m\^eme dimension 
conforme, et nous pouvons choisir de les normaliser de mani\`ere \`a avoir  
$C_{12}=\delta_{12}$. L'OPE de deux champs primaires peut alors s'\'ecrire 
comme: 
\begin{equation} 
\phi_i(z) \phi_j(w) = \sum_{k} 
\frac{C_{ijk}}{(z-w)^{h_i+h_j-h_k}} \phi_k(w) + \text{\bf r\'eg.} 
\end{equation} 
o\`u les coefficients $C_{ijk}$ sont les m\^emes que ceux apparaissant dans la 
fonction \`a trois points.

\subsubsection{Tenseur d'\'energie-impulsion} 
Il existe un autre champ de la th\'eorie pour lequel les propri\'et\'es de 
transformation sont connues, c'est le tenseur 
\'energie-impulsion:  
\begin{defin} Sous une transformation conforme finie $z \mapsto w(z)$, le 
tenseur \'energie-impulsion se transforme comme: 
\begin{equation} 
T'(w) = \left( \frac{dw}{dz} \right)^{-2} \left[ T(z) - \frac{c}{12} \{w;z\} 
\right],
\label{transf-T} 
\end{equation} 
o\`u $c$ est la charge centrale\footnote{L'interpr\'etation physique de la charge centrale appara\^{\i}t 
lorsque  
nous consid\'erons des g\'eom\'etries restreintes, comme analogue \`a un effet Casimir, 
c.\`a.d. un d\'eplacement fini de l'\'energie libre.} et o\`u nous avons introduit la d\'eriv\'ee 
Schwarzienne: 
\begin{equation} 
\{w;z\} = \frac{(d^3w/dz^3)}{(dw/dz)} - \frac{3}{2} \left( 
\frac{d^2w/dz^2}{dw/dz} \right)^2. 
\end{equation} 
\end{defin} 
Le tenseur \'energie-impulsion se transforme donc comme un champ primaire de 
dimension conforme 2, \`a  
l'anomalie schwarzienne pr\`es, qui s'annule 
pour des transformations conformes globales. 
Donc, sous une transformation conforme globale, le tenseur \'energie-impulsion 
se transforme exactement comme un champ primaire: 
les champs ayant cette propri\'et\'e sont appel\'es {\bf quasi-primaires}.
En sp\'ecifiant pour une transformation infinit\'esimale $z \mapsto w = z + 
\varepsilon (z) $, la variation du tenseur  
\'energie-impulsion est donn\'ee par: 
\begin{equation} 
\delta_{\varepsilon}T(w) \doteq T'(w) - T(w) = -\frac{c}{12} \partial_{w}^3 
\varepsilon(w) - 2\partial_w\varepsilon(w)T(w) -\varepsilon(w) 
\partial_wT(w).
\label{vartmunu} 
\end{equation} 
Ceci nous permet de conclure, d'apr\`es l'identit\'e de Ward conforme 
(\ref{wardconf}), que l'OPE du tenseur   
\'energie-impulsion avec lui-m\^eme s'\'ecrit 
\begin{equation} 
T(z)T(w) = \frac{c/2}{(z-w)^4} + \frac{2T(w)}{(z-w)^2} + \frac{\partial 
T(w)}{(z-w)}  + \text{\bf r\'eg.} 
\label{opetmunu} 
\end{equation} 
Nous pouvons \`a nouveau v\'erifier qu'en mettant cette expression dans 
(\ref{wardconf}),  
nous retrouvons bien la variation infinit\'esimale du tenseur 
\'energie-impulsion  donn\'ee en (\ref{vartmunu}).

\subsection{Alg\`ebre de Virasoro et repr\'esentations $\mathcal{V}_i$} 
 
\subsubsection{Correspondance \'etat-champ} 
Il est toujours utile en TQC d'avoir une vision duale entre champs et \'etats. 
Dans ce but, introduisons la 
proc\'edure de quantification radiale.  
\`A deux dimensions, nous devons faire une distinction entre l'espace et le 
temps: la th\'eorie est initialement 
d\'efinie sur un cylindre infini de diam\`etre L. Le temps $t$ court selon l'axe 
infini du cylindre (t va de  
$-\infty$ \`a +$\infty$) et l'espace est compactifi\'e: $x \in [0,L]$, c.\`a.d. 
$(0,t) \doteq (L,t)$.  
Le cylindre est alors param\'etris\'e par les coordonn\'ees complexes $\xi=t+ix, 
\ov{\xi}=t-ix$, 
et le passage entre le cylindre et le plan complexe se fait \`a travers les 
applications suivantes: 
\begin{equation} 
z \rightarrow  \exp \left( \frac{2\pi \xi}{L} \right), \qquad \qquad \qquad  \xi \rightarrow 
\frac{L}{2 \pi} \ln z.
\label{cylplan} 
\end{equation} 
Le pass\'e lointain ($t\rightarrow -\infty$) sur le cylindre correspond \`a 
l'origine $(z=0)$ du plan, et le futur  
lointain correspond au point \`a l'infini sur le plan complexe (plus exactement 
sur la sph\`ere de Riemann). 
Ceci nous permet de d\'efinir les \'etats entrants et sortants: 
$$ 
|\phi_{\text{in}} \rangle  \doteq \lim_{z,\ov{z}\rightarrow 0} \phi(z,\ov{z}) |0\rangle, \qquad 
\qquad \qquad  
\langle \phi_{\text{out}}| \doteq |\phi_{\text{in}}\rangle^{\dag} = \lim_{z,\ov{z}\rightarrow 
0} \langle0| \phi^{\dag}(z,\ov{z}),
$$ 
o\`u l'adjoint est d\'efini par: 
\begin{equation} 
\phi^{\dag}(z,\ov{z}) \doteq \ov{z}^{-2h}z^{-2\ov{h}} \phi(\frac{1}{\ov{z}},\frac{1}{z}).
\end{equation} 
Un champ conforme $\phi(z,\ov{z})$ de dimension conforme $(h,\ov{h})$ peut 
\^etre d\'evelopp\'e en modes  
(ce qui n'est rien d'autre que sa s\'erie de Laurent autour du point $z=0$) 
selon: 
\begin{eqnarray*} 
\phi(z,\ov{z}) &=& \sum_{m\in\mathbb{Z}} \sum_{n\in\mathbb{Z}} \frac{1}{z^{m+h}} 
\frac{1}{\ov{z}^{n+\ov{h}}} \phi_{m,n}, \\ 
\textrm{o\`u } \quad \phi_{m,n} &=& \frac{1}{2 i \pi} \oint_{0} dz z^{m+h-1}  \frac{1}{2 i \pi} 
\oint_{0} d\ov{z} \ov{z}^{n+\ov{h}-1} \phi(z,\ov{z}), \qquad \qquad \phi^{\dag}_{m,n} = \phi_{-m,-n}. 
\end{eqnarray*}

\subsubsection{Alg\`ebre de Virasoro} 
Le tenseur \'energie-impulsion \'etant un champ quasi-primaire de dimension 
conforme 2, son d\'eveloppement en s\'erie de Laurent est donn\'e par: 
\begin{equation} 
T(z) = \sum_{n\in \mathbb{Z}} \frac{L_n}{z^{n+2}},  \qquad \qquad \ov{T}(\ov{z}) 
= \sum_{n \in \mathbb{Z}} \frac{\ov{L}_n}{\ov{z}^{n+2}}.
\end{equation} 
Les modes $L_n$ seront vus comme des op\'erateurs, agissant sur un espace de 
Hilbert.  
En inversant la relation, nous trouvons: 
\begin{equation} 
L_n = \frac{1}{2\pi i} \oint_{0} dz \; z^{n+1} T(z), \qquad \qquad  \ov{L}_n = 
\frac{1}{2\pi i} \oint_{0} d\ov{z} \;  
\ov{z}^{n+1} \ov{T}(\ov{z}).
\end{equation} 
A partir de l'OPE du tenseur \'energie-impulsion (\ref{opetmunu}), nous 
pouvons en d\'eduire que ces op\'erateurs v\'erifient  
les relations de commutations suivantes: 
\begin{eqnarray} 
\left[ L_n,L_m \right] &=& (n-m) L_{n+m} +\frac{c}{12} n(n^2-1)\delta_{n+m,0} 
\nonumber \\ 
\label{virasoro} 
\left[ L_n,\ov{L}_m \right] &=& 0 \\ 
\left[ \ov{L}_n,\ov{L}_m \right] &=& (n-m) \ov{L}_{n+m} +\frac{c}{12} 
n(n^2-1)\delta_{n+m,0} \nonumber  
\end{eqnarray} 
Elles d\'efinissent l'{\bf alg\`ebre de Virasoro}, qui constitue l'extension 
centrale \cite{garland} de l'alg\`ebre de Witt d\'efinie  
en (\ref{witt}). Les op\'erateurs $L_n,\ov{L}_n$ sont les g\'en\'erateurs des 
transformations conformes, agissant sur un espace de Hilbert.

\subsubsection{Espace de Hilbert et repr\'esentations de Virasoro}
L'Hamiltonien est proportionnel au g\'en\'erateur de translation temporelle sur le  
cylindre: $H \sim L_{-1}^{\textrm{cyl}} + \ov{L}_{-1}^{\textrm{cyl}}$. En passant 
vers le plan complexe, les op\'erateurs de translation deviennent des op\'erateurs 
de dilatation. Sur le plan complexe, nous avons donc: $H \sim L_0+\ov{L}_0$. 
Nous consid\'erons alors les repr\'esentations de Virasoro construites \`a partir de 
l'\'etat de plus haut poids $|h_i \rangle$ -- c'est l'{\bf \'etat primaire} engendr\'e par le
champ primaire $\phi_i(z)$ de dimension conforme $h_i$ -- caract\'eris\'e par:
\begin{equation} 
L_0 |h_i \rangle = h_i |h_i \rangle, \qquad  \qquad \qquad L_n |h_i \rangle = 0, \qquad \qquad \forall n>0. 
\label{h.w.} 
\end{equation} 
Cet \'etat est vecteur propre de l'Hamiltonien du syst\`eme. Les autres \'etats de la repr\'esentation
(les \'etats excit\'es, appel\'es \'etats secondaires), sont construits par application successive
des g\'en\'erateurs $L_{-k}, k>0$:
\begin{equation} 
L_{-k_1} L_{-k_2}\cdots L_{-k_n} |h_i\rangle, \qquad \qquad (k_1<k_2 \cdots < k_n), 
\label{desc}
\end{equation} 
et sont vecteurs propres de $L_0$, de valeur propre $h_i+N$, o\`u  
$N = k_1 + k_2 + \cdots + k_n$ est le niveau de l'\'etat. Les champs correspondants aux \'etats
secondaires sont appel\'es champs secondaires.  
Par exemple, \`a l'\'etat $L_{-n}|h_i \rangle$ correspond le champ secondaire 
$\phi_i^{(-n)}$: 
\begin{equation} 
L_{-n} |h_i \rangle \longrightarrow \phi_i^{(-n)}(w) \doteq \frac{1}{2i \pi} \oint dz 
\frac{1}{(z-w)^{n-1}} T(z)\phi_i(w). 
\end{equation} 
\`A un \'etat primaire $|h_i \rangle$ correspond une infinit\'e 
d'\'etats secondaires de la forme (\ref{desc}): ils 
forment ensemble une famille conforme, not\'ee $[\phi_{h_i}]$. 
Les op\'erateurs $L_n$ sont les g\'en\'erateurs des 
transformations conformes. Sous une transformation conforme, l'\'etat $|h_i \rangle$ et ses 
descendants se transforment donc entre-eux: ils forment une repr\'esentation 
de l'alg\`ebre de Virasoro, appel\'ee {\bf module de Verma}, et not\'ee $V(c,h_i)$
ou plus simplement $\mathcal{V}_i$.  
Nous avons parall\`element le module de Verma associ\'e \`a la partie 
anti-holomorphe $\ov{V}(c,\ov{h}_i)$. L'espace de 
Hilbert est alors d\'efini par: 
\begin{equation} 
\mathcal{H} = \sum_{i,\ov{i}} \mathcal{V}_i \otimes \ov{\mathcal{V}}_{\ov{i}} \;= 
\sum_{h_i,\ov{h}_i} V(c,h_i) \otimes \ov{V}(c,\ov{h}_i) 
\end{equation}  
{\it A priori}, nous n'avons aucune indication sur le nombre de termes apparaissant 
dans la somme, ce nombre pouvant \^etre infini.

\subsubsection{Fonctions de corr\'elations} 
\noindent Il existe deux classes de champs dans une th\'eorie conforme: 
\begin{itemize} 
\item[$\bullet$] les champs primaires $\phi(w)$ de dimension conforme $(h,\ov{h})$;
\item[$\bullet$] les champs secondaires: \`a chaque champ primaire $\phi(w)$ correspond 
une infinit\'e de champs secondaires  
$\phi^{(-k_1,\ldots,-k_n)}(w)$. 
\end{itemize} 
Soit $X$ une collection de champs primaires: $X=\phi_2(w_2) \ldots \phi_n(w_n)$, 
de dimensions conformes $h_i$ $(i=2,\ldots,n)$, 
et $\phi^{(-n)}(w)$ un champ secondaire. La fonction de corr\'elation entre $\phi^{(-n)}(w)$
et $X$ est donn\'ee par: 
\begin{equation} 
\langle \phi^{(-n)}X \rangle = \mathcal{L}_{-n}  \langle \phi(w)X \rangle, 
\end{equation} 
o\`u $\mathcal{L}_{-n}$ est un op\'erateur diff\'erentiel d\'efini par: 
\begin{equation} 
\mathcal{L}_{-n} = \sum_i \left\{ \frac{(n-1)h_i}{(w_i-w)^n} - 
\frac{1}{(w_i-w)^{n-1}}\partial_{w_i} \right\}.
\end{equation} 
Pour un champ secondaire plus g\'en\'eral, de la forme 
$\phi^{(-k_1,-k_2,\ldots,-k_n)}$, nous obtenons de la m\^eme mani\`ere: 
\begin{equation} 
 \langle \phi^{(-k_1,-k_2,\ldots,-k_n)}X \rangle = \mathcal{L}_{-k_1} \ldots 
\mathcal{L}_{-k_n} \langle  \phi(w)X \rangle. 
\label{opdiffn} 
\end{equation} 
Le calcul de fonctions de corr\'elations contenant des champs secondaires se 
r\'eduit donc \`a celui contenant seulement  
des champs primaires, sur lequel nous ferons agir un op\'erateur diff\'erentiel bien 
d\'efini.
Nous sommes donc ramen\'es au seul calcul des fonctions de corr\'elations entre 
champs primaires. Si nous connaissons 
l'OPE des champs primaires, elles nous permettent de passer graduellement du calcul des 
fonctions \`a $N$ points  
\`a celui des fonctions \`a deux points, qui sont explicitement fix\'ees 
par l'invariance conforme.
Nous avons vu que l'OPE de deux champs primaires est donn\'ee par: 
\begin{equation} 
\phi_i(z,\ov{z}) \phi_j(w,\ov{w}) = \sum_{k} 
\frac{C_{ijk}}{(z-w)^{h_i+h_j-h_k}(\ov{z}-\ov{w})^{\ov{h}_i+\ov{h}_j-\ov{h}_k}} 
\phi_k(w,\ov{w}), 
\end{equation} 
o\`u les $\phi_k(w,\ov{w})$ sont des champs primaires ou secondaires. Nous 
pouvons regrouper dans le membre de droite tous les champs
secondaires appartenant \`a la famille conforme $[\phi_p]$ ensemble et diviser 
la sommation selon: 
\begin{equation*} 
\phi_i(z,\ov{z}) \phi_j(w,\ov{w})  = \sum_p 
\sum_{\{k,\ov{k}\}}C_{ijp}^{\{k\ov{k}\}}  \frac{1}{(z-w)^{h_i+h_j-h_k-\sum_l 
k_l}}   
\frac{1}{(\ov{z}-\ov{w})^{\ov{h}_i+\ov{h}_j-\ov{h}_k-\sum_l \ov{k}_l}} 
\phi_p^{\{k,\ov{k}\}}(w,\ov{w}),
\end{equation*} 
o\`u tous les champs descendants du champ primaire $\phi_p$ sont not\'es 
$\phi_p^{\{k,\ov{k}\}}$. 
En utilisant l'\'equation (\ref{opdiffn}), il est possible de montrer que: 
\begin{equation} 
C_{ijk}^{\{k,\ov{k}\}} = \sum_p C_{ijp} \beta_{ij}^{p\{k\}} 
\ov{\beta}_{ij}^{p\{\ov{k}\}},
\end{equation} 
o\`u les $C_{ijp}$ sont les coefficients de l'OPE entre champs primaires 
seulement, et les $\beta_{ij}^{p\{k\}}$,  
$ \ov{\beta}_{ij}^{p\{\ov{k}\}}$ sont des fonctions des quatre param\`etres 
$h_i,h_j,h_p$ et $c$, enti\`erement fix\'ees 
par l'invariance conforme. 
Le calcul des fonctions de corr\'elation entre les champs du syst\`eme est donc ramen\'e  
\`a la connaissance des coefficients $C_{ijk}$. 
Nous sommes donc amen\'e \`a la:
\begin{conclu} 
Toute l'information n\'ecessaire pour compl\`etement sp\'ecifier une th\'eorie 
conforme \`a deux dimensions est la donn\'ee de la charge 
centrale $c$, des dimensions conformes $(h_i,\ov{h}_i)$ des champs primaires   
et des coefficients $C_{ijk}$ provenant de l'OPE de ces champs primaires. Avec ces 
donn\'ees, il est possible de calculer toutes les fonctions de corr\'elations 
du syst\`eme, et par cons\'equent, d'obtenir un syst\`eme soluble!  
\end{conclu} 
Cependant, l'invariance conforme \`a elle-seule ne fixe pas les coefficients $C_{ijk}$, il nous 
faut des informations suppl\'ementaires externes. Nous verrons 
par la suite certaines contraintes permettant de compl\'eter la th\'eorie, 
ouvrant ainsi la voie vers les classifications des th\'eories conformes.


\section{Th\'eories conformes rationelles: RCFT}

\subsection{Alg\`ebre de fusion} 
Nous voulons maintenant transcrire les r\'esultats obtenus jusqu'\`a maintenant 
sous forme alg\'ebrique. L'OPE de 
deux champs quelconques d'une famille conforme est obtenu \`a partir de celle 
des champs primaires correspondants. En consid\'erant 
l'OPE entre deux champs primaires, l'information importante est de savoir quelles 
familles conformes ils vont cr\'eer, \`a travers 
les coefficients $C_{ijk}$. Ceci nous permet d'\'ecrire les r\`egles suivantes: 
\begin{equation} 
[\phi_{h_i}] \times [\phi_{h_j}] = \sum_k \mathcal{N}_{ij}^{k} \; [\phi_{h_k}]. 
\label{Nijk}
\end{equation} 
L'interpr\'etation est la suivante: le membre de gauche repr\'esente l'OPE entre 
un champ conforme de la famille $[\phi_{h_i}]$ 
et un champ conforme de la famille $[\phi_{h_j}]$, le membre de droite indiquant 
quelles familles conformes $[\phi_{h_k}]$ vont appara\^{\i}tre dans cette OPE. 
Les nombres $\mathcal{N}_{ij}^{k}$ sont donc des entiers non-n\'egatifs, reli\'es aux 
coefficients $C_{ijk}$.
Les champs primaires $\phi_{h_i}(z)$ sont en correspondance avec les \'etats de plus haut poids 
$|h_i\rangle$ de la repr\'esentation $\mathcal{V}_i$ de Virasoro. L'\'equation (\ref{Nijk}) peut
donc s'\'ecrire comme la fusion des repr\'esentations:
\begin{equation} 
\mathcal{V}_i \times \mathcal{V}_j = \sum_k \mathcal{N}_{ij}^k \; \mathcal{V}_k \; .
\label{OPE-fusion} 
\end{equation} 
\begin{defin} 
L'{\bf alg\`ebre de fusion} est une alg\`ebre commutative, associative, de 
g\'en\'erateurs $\mathcal{V}_i$, $i=1,\ldots,n$, (n est un entier ou $+\infty$), 
poss\'edant une  
identit\'e $\mathcal{V}_1 = \munite$ (la repr\'esentation identit\'e), 
et un produit not\'e $\times$, dont les r\`egles de multiplication sont 
donn\'ees par (\ref{OPE-fusion}). 
\end{defin} 
D\'efinissons les matrices $N_i$, appel\'ees matrices de fusion, ayant comme \'el\'ements: 
\begin{equation} 
(N_i)_{jk} = \mathcal{N}_{ij}^k. 
\end{equation} 
Alors l'existence de l'identit\'e implique $N_1 = \munite$, 
et la propri\'et\'e d'associativit\'e de l'alg\`ebre de fusion implique la commutativit\'e 
des matrices $N_i$: $N_iN_j = N_jN_i$. L'associativit\'e  peut  
aussi s'\'ecrire comme: 
\begin{equation} 
N_i N_j = \sum_k \mathcal{N}_{ij}^k N_k. 
\end{equation} 
Les matrices $N_{i}$ forment donc une repr\'esentation fid\`ele de l'alg\`ebre de fusion. 
L'information sur les coefficients $C_{ijk}$ est donc ramen\'ee \`a la connaissance de l'alg\`ebre
de fusion -- ou de mani\`ere \'equivalente \`a la connaissance des matrices de fusion $N_i$ -- qui \`a ce
stade reste toutefois \`a d\'eterminer.

\subsection{Unitarit\'e et irr\'eductibilit\'e de $\mathcal{V}_i$} 
La base de la repr\'esentation $\mathcal{V}_i$ est form\'ee par l'\'etat de plus haut poids
$|h_i\rangle$ et tous ses \'etats descendants (\ref{desc}).  
La norme de l'\'etat $L_{-k_1}\cdots L_{-k_n} |h\rangle$ est d\'efinie par: 
\begin{equation} 
\langle h|L_{k_n}\cdots L_{k_1}L_{-k_1}\cdots L_{-k_n}|h\rangle.
\end{equation}

\subsubsection{Unitarit\'e} 
Une repr\'esentation $\mathcal{V}_i$ est dite {\bf unitaire} si elle ne poss\`ede pas d'\'etats de norme 
n\'egative: comme la norme d\'epend de la dimension conforme $h$ et de la charge centrale $c$ 
(\`a travers les relations de commutation  
de l'alg\`ebre), l'unitarit\'e impose donc  
des contraintes sur ces valeurs. L'\'etude de la norme des \'etats des 
repr\'esentation $\mathcal{V}_i$ a \'et\'e effectu\'ee dans \cite{Kac1} et \cite{Fuchs2}: il 
existe 
un vecteur de norme nulle au niveau $\ell=rs$ ($r$ et $s$ entiers) lorsque la dimension conforme de la  
repr\'esentation est donn\'ee par la {\bf formule de Kac}: 
\begin{equation} 
c = 1 - \frac{6}{m(m+1)}, \qquad \qquad h_{r,s}(m) = \frac{\lbrack (m+1)r-ms \rbrack^2-1}{4m(m+1)},  
\qquad \qquad m \in \mathbb{C}.
\end{equation} 
La pr\'esence d'un vecteur de norme nulle permet de d\'elimiter les zones $(c,h)$ d'existence de vecteurs de 
norme n\'egatives (non-unitarit\'e). Les repr\'esentations de Virasoro sont non-unitaires pour $c<0$ et 
pour $h<0$. Pour $h\geq 0$, elles sont 
unitaires si $c \geq 1$. Si $0<c<1$, elles sont unitaires pour les valeurs de la formule de Kac 
avec les contraintes suppl\'ementaires suivantes (spectre fini): 
\begin{equation} 
m \textrm{ entier } >2, \qquad \qquad \qquad 
1 \leq r < m, \qquad \qquad \qquad 1 \leq s \leq r. 
\end{equation} 
 
\subsubsection{Vecteurs singuliers et irr\'eductibilit\'e} 
Lorsque $h = h_{r,s}$, il existe donc un \'etat $|\chi \rangle$ 
au niveau $\ell =rs$ dont la norme est nulle, appel\'e {\bf \'etat singulier}. Cet \'etat satisfait les  
propri\'et\'es (\ref{h.w.}) d'un \'etat de plus haut poids. 
Les \'etats descendants de $|\chi \rangle$ -- aussi de norme nulle -- forment un module  
de Verma not\'e $\mathcal{V}_{\chi}$. L'espace de la repr\'esentation $\mathcal{V}_{h_{r,s}}$ contient un  
sous-espace $\mathcal{V}_{\chi}$ qui est lui-m\^eme une repr\'esentation de Virasoro:  
il est donc {\bf r\'eductible}. Nous contruisons 
des repr\'esentations irr\'eductibles en quotientant par les sous-modules $\mathcal{V}_{\chi}$  
(ce qui \'equivaut \`a identifier les \'etats qui ne diff\`erent que par un \'etat de 
norme nulle). 
\`A l'\'etat singulier $|\chi \rangle$ est associ\'e le champ $\chi(z)$, qui est un champ descendant du 
champ primaire 
$\phi(z)$, mais qui est lui-m\^eme un champ primaire. Le fait que l'\'etat $|\chi \rangle$ soit de 
norme nulle 
(donc orthogonal au module de Verma) se traduit en langage des champs \`a l'annulation des fonctions 
de corr\'elation $\langle \chi(z) X \rangle$, o\`u $X$ est une collection de champs: le champ $\chi(z)$ 
se {\sl d\'ecouple} 
des autres champs. Ceci a comme cons\'equence une \'equation dif\'erentielle (\ref{opdiffn}) pour les 
fonctions 
de corr\'elations $\langle \phi(z) X \rangle$, donnant des contraintes sur l'OPE des champs, 
se traduisant par une troncation de  
l'alg\`ebre de fusion: 
\begin{equation} 
\mathcal{V}_i \times \mathcal{V}_j = \sum_k { }^{'}\; \mathcal{N}_{ij}^k \;\mathcal{V}_k. 
\end{equation}  
Cependant, pour une valeur arbitraire de $c$, le nombre de champs primaires de la th\'eorie peut 
\^etre infini, et il faut d'autres contraintes pour ``fermer'' l'alg\`ebre 
de fusion. 

Les th\'eories conformes pour lesquelles le nombre de champs primaires est fini sont appel\'ees  
rationnelles (RCFT). Ce sont des th\'eories o\`u intervient  
un nombre fini de repr\'esentations $\mathcal{V}_i$ pour lesquelles l'alg\`ebre 
de fusion est ferm\'ee. L'exemple type de RCFT est fourni par les mod\`eles minimaux. 
 
\subsection{Mod\`eles minimaux} 
Pour $p,p'$ deux entiers premiers entre-eux tels que $m = p'/(p-p')$, la formule de Kac s'\'ecrit: 
\begin{equation} 
c = 1 - 6 \frac{(p-p')^2}{pp'}, \qquad \qquad \qquad h_{r,s} = \frac{(pr-p's)^2-(p-p')^2}{4pp'},  
\label{minmod} 
\end{equation} 
et les dimensions conformes sont alors p\'eriodiques $h_{r,s} = h_{r+p',s+p}$. Nous avons notamment: 
\begin{equation} 
h_{r,s} = h_{p'-r,p-s}, 
\end{equation} 
ce qui implique l'existence d'un autre vecteur singulier $|\chi'\rangle$ au niveau $(p'-r)(p-s)$.  
Les dimensions conformes de ces deux \'etats singuliers sont \'egaux \`a: 
\begin{equation} 
h_{\chi} = h_{r,s}+rs, \qquad  \qquad \qquad h_{\chi'} = h_{r,s} + (p'-r)(p-s),  
\end{equation} 
et sont donc aussi donn\'ees par la formule de Kac (elles s'\'ecrivent de la forme $h_{r',s'}$)! 
Ces deux \'etats singuliers engendrent donc des sous-modules de Verma $\mathcal{V}_{\chi}$ et 
$\mathcal{V}_{\chi'}$ r\'eductibles, contenant \`a leur tour des \'etats singuliers engendrant 
des sous-modules r\'eductibles, et ainsi de suite. 
Il existe donc une infinit\'e d'\'etats singuliers dans une repr\'esentation $\mathcal{V}_i$ avec les  
valeurs (\ref{minmod}). 
Chaque \'etat singulier conduit \`a une \'equation diff\'erentielle agissant comme une contrainte 
sur les fonctions de corr\'elation des champs primaires, et donc sur leur OPE.  L'effet global est une 
nouvelle 
troncation de l'alg\`ebre de fusion, qui a comme cons\'equence que seulement un nombre {\sl fini} 
de repr\'esentations $\mathcal{V}_i$ sont \`a consid\'erer: ce sont les mod\`eles 
appel\'es {\bf minimaux}. Les mod\`eles minimaux sont unitaires pour $p'=p+1$ (ou $p'=p-1$) et sont not\'es 
$\mathcal{M}(p',p)$. 
Le premier mod\`ele minimal unitaire non-trivial correspond au cas $p=3$: il a \'et\'e identifi\'e 
comme d\'ecrivant 
le mod\`ele critique d'Ising \cite{BPZ-Ising}. Nous avons les suivantes identifications pour les premiers 
\'el\'ements de la s\'erie unitaire:
$$ 
\begin{array}{ccll} 
\bullet& \mathcal{M}(4,3):& \textrm{mod\`ele critique d'Ising}     & c = 1/2 \\ 
\bullet& \mathcal{M}(5,4):& \textrm{mod\`ele tri-critique d'Ising} & c = 7/10 \\ 
\bullet& \mathcal{M}(6,5):& \textrm{mod\`ele de Potts \`a trois \'etats} & c=4/5 \\ 
\bullet& \mathcal{M}(7,6):& \textrm{mod\`ele tri-critique de Potts \`a trois \'etats} \qquad & c=6/7 
\end{array} 
$$ 
 
\subsection{Mod\`eles $\widehat{g}$-WZWN} 
Une situation fr\'equente en th\'eorie des champs conformes est qu'il existe une alg\`ebre 
``\'etendue'' $\mathcal{A}$ agissant sur les champs  
de la th\'eorie (alg\`ebre de Kac-Moody, supersym\'etrie, alg\`ebre $\mathcal{W}$, $\ldots$), telle que Virasoro soit 
une sous-alg\`ebre de $\mathcal{A}$ 
ou de l'alg\`ebre enveloppante de $\mathcal{A}$. 
Nous nous int\'eressons plus particuli\`erement aux th\'eories conformes 
o\`u l'alg\`ebre \'etendue est une alg\`ebre affine $\widehat{g}$: ce sont des mod\`eles particuliers 
(appel\'es WZWN\footnote{Ils ont \'et\'e introduits par Wess et Zumino, puis compl\'et\'es par Witten et 
Novikov, et sont connus dans la litt\'erature sous le nom de mod\`eles WZWN.})  
en ce sens qu'ils peuvent \^etre formul\'es directement en terme d'une action\footnote{Ce sont des 
mod\`eles construits 
\`a partir d'une action du type  mod\`ele $\sigma$ non-lin\'eaire, avec l'addition d'un terme de 
Wess-Zumino.}.  
Pour ces mod\`eles, les courants additionnels conserv\'es $J^a(z)$ poss\`edent une OPE de la 
forme: 
\begin{equation} 
J^a(z)\, J^b(w) = \frac{{\bf k}\delta_{ab}}{(z-w)} + \sum_{c}\, i \, f_{abc} \frac{J^c(w)}{(z-w)} + 
\textrm{\bf r\'eg.} 
\end{equation} 
et les modes de $J^a(z)=\sum_{n \in \mathbb{Z}} J^a_n z^{-n-1}$ satisfont les relations de commutation 
d'une alg\`ebre affine: 
\begin{equation} 
\lbrack J^a_n J^b_m, \rbrack = \sum_c i\; f_{abc} J^c_{n+m} + {\bf k} \; n \; \delta_{ab} \; \delta_{n+m,0}, 
\end{equation} 
o\`u ${\bf k}$ est un \'el\'ement central. Les repr\'esentations des alg\`ebres affines $\widehat{g}$ 
sont de nos jours bien connues \cite{Kac2,Fuchs1}. 
Une g\'en\'eralisation de la notion de repr\'esentations irr\'eductibles d'une alg\`ebre de Lie simple $g$
est fournie par la notion de repr\'esentations {\bf int\'egrables}. Elles sont labell\'ees par $(\lambda,k)$, o\`u $\lambda$ est le plus haut 
poids et  
$k$ le niveau, et il existe un nombre fini de telles repr\'esentations \`a chaque niveau $k$. 
La dimension conforme et la charge centrale des mod\`eles $\widehat{g}_k$ sont donn\'ees par: 
\begin{equation} 
h_{\lambda} = \frac{(\lambda,\lambda + 2 \rho)}{2(k+\kappa)}, \qquad \qquad \qquad \qquad 
c = \frac{k \dim(g)}{k + \kappa},
\end{equation} 
o\`u $\rho$ est le vecteur de Weyl, et $\kappa$ le nombre (dual) de Coxeter de $g$.  
Les champs primaires sont en correspondance  
avec les plus haut poids $\lambda$ des repr\'esentations int\'egrables: comme il existe un nombre 
fini de telles repr\'esentations \`a chaque niveau $k$, il existe donc un nombre fini de champs primaires. 
Les mod\`eles WZWN fournissent donc un autre exemple de RCFT. De plus, les repr\'esentations int\'egrables 
\'etant unitaires, ces mod\`eles le sont aussi. 
 
Les mod\`eles avec alg\`ebre affine fournissent des exemples non-triviaux de mod\`eles 
quantiques \`a $2d$ exactement solubles, et jouent un r\^ole pr\'edominant 
dans la classification des th\'eories conformes \`a 2d. Nous verrons par exemple que la classification  
des mod\`eles minimaux est reli\'ee, \`a travers une construction de {\it coset}, 
\`a la classification des mod\`eles $\widehat{su}(2)$.

\subsection{Propri\'et\'es modulaires des caract\`eres $\chi_i$ et formule de Verlinde} 
Soit $\mathcal{A}$ l'alg\`ebre d\'ecrivant la sym\'etrie d'une th\'eorie conforme rationelle, et  
$\mathcal{V}_i$ une repr\'esentation de $\mathcal{A}$, pour $i \in \mathcal{I}$, $\mathcal{I}$ 
\'etant un ensemble fini.  
Les repr\'esentations $\mathcal{V}_i$  
sont gradu\'ees par l'action du g\'en\'erateur $L_0$ de Virasoro\footnote{C'est le cas m\^eme si  
$\mathcal{A} \not= Vir$, car les g\'en\'erateurs de Virasoro s'expriment, \`a travers la construction de 
Sugawara, 
en fonction des courants $J$ de $\mathcal{A}$.}. Le spectre de $L_0$ dans $\mathcal{V}_i$ est de la forme  
$\{h_i, h_i +1, h_i+2, \cdots\}$, et nous 
appelons $\#_n$ le nombre d'\'etats lin\'eairement ind\'ependants au niveau $n$  
(donc de valeur propre $h+n$). Nous introduisons 
le caract\`ere $\chi_i$ de la repr\'esentation $\mathcal{V}_i$ comme la fonction g\'en\'eratrice  
des multiplicit\'es $\#_n$, d\'ependant d'une variable complexe $\tau$: 
\begin{equation} 
\chi_i (\tau) = \textrm{Tr}_{\mathcal{V}_i} \; q^{L_0 - \frac{c}{24}} = \sum_{n=0}^{\infty} \, \#_n \; 
q^{n+h-\frac{c}{24}}, 
 \qquad \qquad (q \doteq e^{2i \pi \tau}).  
\label{def-chi} 
\end{equation} 

Les caract\`eres de Virasoro par exemple sont donn\'es par: 
\begin{equation} 
\chi_{(c,h)} (\tau) = (q^{h-\frac{c}{24}}) (\prod_{n=1}^{\infty} \frac{1}{1-q^n}). 
\label{chiVir} 
\end{equation} 
L'expression des caract\`eres des mod\`eles minimaux est plus compliqu\'ee que (\ref{chiVir}), car il  
faut tenir compte de toutes les soustractions des sous-modules: ils sont explicit\'es 
par exemple dans \cite{FMS-book}. Les caract\`eres (\ref{def-chi}) pour 
les alg\`ebres affines $\widehat{g}$ (appel\'es sp\'ecialis\'es car ils comptent les \'etats  
en fonction de la valeur propre de $L_0$ seulement) se trouvent aussi dans \cite{FMS-book}. 
 
\subsubsection{Propri\'et\'es modulaires} 
Une propri\'et\'e remarquable des caract\`eres (de Virasoro, des mod\`eles minimaux ou des alg\`ebres 
affines) est qu'ils satisfont de belles propri\'et\'es de transformation sous l'action du groupe modulaire.  
Le groupe modulaire $SL(2,\mathbb{Z})$ sur une variable $\tau$ est d\'efini par: 
\begin{equation} 
\tau \mapsto \frac{a \tau + b}{c \tau + d}, \qquad  
\left(\begin{array}{cc} a & b \\ c & d \end{array} \right) \in SL(2,\mathbb{Z}), \qquad 
a,b,c,d \in \mathbb{Z}, \qquad ad-bc = 1, 
\label{grmod} 
\end{equation} 
et est engendr\'e par les deux transformations: 
\begin{equation} 
T: \tau \longmapsto \tau + 1, \qquad \qquad S: \tau \longmapsto -\, \frac{1}{\tau}, 
\end{equation} 
satisfaisant les relations $(ST)^3 = S^2=1$.  
Les caract\`eres $\chi_i$ d'une repr\'esentation de l'alg\`ebre $\mathcal{A}$ d'une RCFT forment une 
repr\'esentation 
fini-dimensionelle et unitaire du groupe modulaire: ils se transforment entre-eux sous l'action 
de (\ref{grmod}). Il existe donc deux matrices $S_{ij}$ et $T_{ij}$ telles que: 
\begin{equation} 
\chi_i (\tau) \mapsto \chi_i (\tau +1) = \sum_{j \in \mathcal{I}} T_{ij} \; \chi_j (\tau), \qquad \qquad   
\chi_i (\tau) \mapsto \chi_i (- \, \frac{1}{\tau}) = \sum_{j \in \mathcal{I}} S_{ij} \; \chi_j (\tau). 
\label{transf-chi}   
\end{equation} 
Il est clair d'apr\`es la d\'efinition (\ref{def-chi}) que sous l'action de $T$: 
\begin{equation} 
\chi_i (\tau) \mapsto \chi_i (\tau+1) = e^{2 i \pi (h_i - \frac{c}{24})} \; \chi_i (\tau), 
\end{equation} 
et la matrice $T_{ij}$ est donc une matrice diagonale $(T_{ij} = e^{2 i \pi (h_i - \frac{c}{24})} 
\delta_{ij})$. 
L'expression de la matrice $S$ pour une alg\`ebre affine g\'en\'erale se trouve dans \cite{Kac2}.

\subsubsection{Formule de Verlinde} 
La correspondance entre champs primaires et \'etats de plus haut poids de $\mathcal{V}_i$ permet de 
coder l'OPE de ces champs dans l'alg\`ebre de fusion des repr\'esentations (voir l'\'equation 
(\ref{OPE-fusion})).  
E. Verlinde a montr\'e qu'il existe un lien \'etroit entre les coeficients de fusion $\mathcal{N}_{ij}^k$ 
et la matrice $S_{ij}$ des transformations modulaires des caract\`eres de l'alg\`ebre, donn\'e par 
la {\bf formule de Verlinde} \cite{verlinde}  
\begin{equation} 
\mathcal{N}_{ij}^k = \sum_{\ell \in \mathcal{I}} \frac{S_{i\ell}S_{j\ell}S_{k\ell}^*}{S_{1\ell}}. 
\end{equation} 
Cette relation est hautement non triviale, car elle relie les coefficients de fusion $\mathcal{N}_{ij}^k$, 
qui sont des entiers non-n\'egatifs, aux coefficients de la matrice $S$, qui sont des r\'eels!  
Connaissant les propri\'et\'es de transformation des caract\`eres, nous pouvons donc en d\'eduire les 
r\`egles de fusion; ou r\'eciproquement, la connaissance des r\`egles de fusion nous donne des 
informations sur la matrice $S$.


\section{Invariance modulaire, conditions aux bord et lignes de d\'efaut} 
Nous avons jusqu'\`a pr\'esent utilis\'e implicitement le d\'ecouplage de la th\'eorie en partie 
holomorphe et anti-holomorphe. Les donn\'ees n\'ecesaires pour totalement sp\'ecifier une th\'eorie  
conforme rationelle sont: l'alg\`ebre $\mathcal{A}$, ses repr\'esentations de plus haut poids  
$\mathcal{V}_i$ en nombre fini (ce qui fixe la charge centrale $c$ et les dimensions  
conformes des champs primaires $\phi_{h_i}$), ses caract\`eres $\chi_i$  
et la matrice $S_{ij}$ des transformations modulaires des caract\`eres -- ou de mani\`ere \'equivalente les  
coefficients de fusion $\mathcal{N}_{ij}^k$ des repr\'esentations, obtenus \`a travers la formule  
de Verlinde.   
L'espace de Hilbert du syst\`eme s'\'ecrit: 
\begin{equation} 
\mathcal{H} = \bigoplus_{i, j} \; \mathcal{M}_{ij} \; \mathcal{V}_i \otimes \ov{\mathcal{V}}_j, 
\qquad \qquad \qquad \qquad \mathcal{M}_{ij} \in \mathbb{N},   
\label{hilbert} 
\end{equation} 
o\`u la sommation s'\'etend {\it a priori} sur toutes les dimensions conformes $h_i$, $\ov{h}_j$ 
du syst\`eme. 
Cependant, toute combinaison gauche/droite de repr\'esentations $\mathcal{V}_i$ n'est pas n\'ecessairement
physiquement r\'ealiste. Il s'av\`ere que l'\'etude des th\'eories conformes sur des vari\'et\'es de genre plus 
\'elev\'e (comme le tore) nous fournit de pr\'ecieux renseignements \cite{cardy-invmod,ItzZub-torus}.

\subsection{Mod\`eles d\'efinis sur le tore} 
Un tore est obtenu en sp\'ecifiant deux vecteurs sur le plan -- ou deux nombres complexes $w_1,w_2$ 
(p\'eriodes) 
sur le plan complexe -- d\'efinissant ainsi un r\'eseau,  et en identifiant 
les points qui diff\`erent par une combinaison lin\'eaire enti\`ere de ces vecteurs. Une th\'eorie conforme 
d\'efinie sur le tore ne doit pas d\'ependre de la base choisie sur le r\'eseau pour d\'efinir le tore: elle 
ne d\'ependra que du param\`etre $\tau = w_2 / w_1$, appel\'e {\bf param\`etre 
modulaire}, 
et nous pouvons toujours choisir comme p\'eriodes $1$ et $\tau$. 
 
Une th\'eorie \`a $2d$ est d\'efinie sur un cylindre de diam\`etre $a$, o\`u le temps court selon l'axe du 
cylindre et l'espace est compactifi\'e: 
$x = x + a$,  l'application du cylindre (param\'etrise par $\xi$) vers le plan \'etant $z = \exp 
(\frac{2 i \pi \xi}{a})$. 
L'hamiltonien correspond \`a la translation temporelle sur le cylindre, et est donc proportionnel au 
g\'en\'erateur de translation ($L_{-1} + \ov{L}_{-1}$). Par la loi de transformation (\ref{transf-T}) 
du tenseur \'energie-impulsion,  
l'op\'erateur de translation ($L_{-1}$) devient un op\'erateur de dilatation ($L_0$) sur le plan: 
\begin{equation} 
L_{-1}^{\textrm{cyl}} = -\, \frac{2 i \pi}{a} \left( L_0 - \frac{c}{24}\right), 
\end{equation} 
o\`u le coefficient $c/24$ provient de la d\'eriv\'ee Schwartzienne de l'exponentielle.  
L'op\'erateur d'\'evolution du syst\`eme (l'exponentielle de l'hamiltonien) 
est alors donn\'e par: 
\begin{equation} 
\mathcal{T} \doteq e^{-H a } = e^{2 i \pi (\tau(L_0 - \frac{c}{24}) - \ov{\tau}(\ov{L}_0-\frac{c}{24}))}. 
\label{evol} 
\end{equation} 
La fonction de partition est donn\'ee par la trace de 
l'op\'erateur d'\'evolution $\mathcal{T}$: 
\begin{equation} 
\mathcal{Z}(\tau) = \textrm{Tr}_{\mathcal{H}}\; \mathcal{T} = \textrm{Tr}_{\mathcal{H}}  \;  
q^{(L_0 - \frac{c}{24})}\; \ov{q}^{(\ov{L}_0-\frac{c}{24})}, \qquad \qquad q \doteq e^{2 i \pi \tau}. 
\end{equation} 
En utilisant la d\'ecomposition (\ref{hilbert}) de l'espace de Hilbert et 
l'expression (\ref{def-chi}) des caract\`eres $\chi_i$ de $\mathcal{V}_i$, nous obtenons alors: 
\begin{equation}
\begin{array}{|c|}
\hline 
{ } \\
\displaystyle \quad \mathcal{Z}(\tau) = \sum_{i,j}\; \mathcal{M}_{ij}\; \chi_i (\tau)\; \ov{\chi}_{j} 
(\tau) \quad  \\
{ } \\
\hline
\end{array}
\label{def-Z} 
\end{equation}

\subsection{Fonctions de partition invariante modulaire} 
Physiquement, la fonction de partition d'une th\'eorie conforme d\'efinie sur le tore ne peut d\'ependre 
que du param\`etre  
modulaire $\tau = w_2/w_1$, mais il reste toutefois une redondance. En effet, consid\'erons des p\'eriodes  
$w_1', w_2'$ qui soient des combinaisons lin\'eaires enti\`eres de $w_1$ et $w_2$ (et donc appartenant au 
m\^eme r\'eseau): 
\begin{equation} 
\left( \begin{array}{c} w_1' \\ w_2' \end{array} \right) =   
\left( \begin{array}{cc} a & b \\ c & d \end{array} \right) \;  
\left( \begin{array}{c} w_1 \\ w_2 \end{array} \right), \qquad \qquad a,b,c,d \in \mathbb{Z}, 
\qquad ad-bc=1. 
\label{sl2z} 
\end{equation} 
Par invariance conforme, ces nouvelles p\'eriodes d\'efinissent le m\^eme r\'eseau, et la fonction  
de partition doit donc \^etre invariante sous ces transformations.   
Le groupe engendr\'e par les transformations  
(\ref{sl2z}) est le groupe $SL(2,\mathbb{Z})$, et sous (\ref{sl2z}) le param\`etre modulaire devient: 
\begin{equation} 
\tau \mapsto \frac{a \tau + b}{c \tau + d}. 
\label{transf-tau} 
\end{equation} 
Ce groupe est engendr\'e par les deux transformations $S$ et $T$, la fonction de partition doit donc  
satisfaire\footnote{$\tau$ n'est pas affect\'e par un changement de signe global des param\`etres 
$a,b,c,d$ dans  
(\ref{transf-tau}): la sym\'etrie r\'eelle de la fonction de partition est donc  
$PSL(2,\mathbb{Z}) = SL(2,\mathbb{Z})/\mathbb{Z}_2$.}: 
\begin{equation} 
T: \mathcal{Z} (\tau +1) = \mathcal{Z} (\tau), \qquad \qquad \qquad S: \mathcal{Z} 
\left(-\, \frac{1}{\tau} \right) =  
\mathcal{Z} (\tau).
\label{invmodz}  
\end{equation} 
Utilisant l'expression (\ref{def-Z}) de $\mathcal{Z}$, les propri\'et\'es de transformations (\ref{transf-chi})  
des caract\`eres $\chi_i$ et l'unitarit\'e des  
matrices $S$ et $T$, le probl\`eme de classification des fonctions de partition 
invariantes modulaires se r\'eduit donc \`a la: 
 
\begin{class} 
Trouver toutes les $n \times n$ matrices $\mathcal{M}$, telles que: 
\begin{itemize} 
\item[$\bullet$] $\mathcal{M}_{ij} \in \mathbb{N}$ 
\item[$\bullet$] $\mathcal{M}_{11} = 1$  
\item[$\bullet$] $\mathcal{M}$ commute avec $S$ et $T$: $S\mathcal{M}=\mathcal{M}S$, $T\mathcal{M}=
\mathcal{M}T$ 
\end{itemize}  
\end{class} 
La deuxi\`eme condition impose l'unicit\'e du vide ($\mathcal{V}_1$ est la repr\'esentation identit\'e); 
la troisi\`eme condition exprime sous forme matricielle l'invariance modulaire (\ref{invmodz}) de 
$\mathcal{Z}$. 
Une telle matrice $\mathcal{M}$ est appel\'ee l'{\bf invariant modulaire}, et la fonction de 
partition invariante modulaire s'obtient par (\ref{def-Z}). 
Notons que des solutions \'evidentes de ce probl\`eme sont donn\'ees par les matrices 
$\mathcal{M}=\munite_{n \times n}$.  
Ce sont les th\'eories appel\'ees diagonales. 
 
La classification des fonctions de partition invariantes modulaires a \'et\'e obtenue pour les mod\`eles  
minimaux et les mod\`eles $\widehat{su}(2)$ dans \cite{CIZ-class1,CIZ-class2}, celle des mod\`eles 
$\widehat{su}(3)$ dans \cite{gannon-class}. Nous verrons qu'\`a ces classifications sont naturellement 
associ\'es des graphes (diagrammes de Dynkin de type $ADE$ pour $\widehat{su}(2)$, diagrammes de Di Francesco
- Zuber pour $\widehat{su}(3)$). Ces liens avec des graphes deviennent plus explicites lorsque 
nous consid\'erons des th\'eories conformes avec conditions au bord (BCFT).

\subsection{Conditions au bord} 
Une th\'eorie conforme d\'efinie sur une vari\'et\'e sans bord poss\`ede comme sym\'etrie 
deux alg\`ebres $\mathcal{A}$ et $\ov{\mathcal{A}}$, agissant respectivement (et s\'eparemment) 
sur la d\'ependence holomorphe $(z)$ et anti-holomorphe ($\ov{z}$) des champs de la th\'eorie.  
Il s'av\`ere toutefois n\'ecessaire en physique d'\'etudier des th\'eories d\'efinies sur 
des vari\'et\'es \`a bord
et ses possibles conditions au bord (r\'eseau fini en physique statistique, th\'eorie des cordes, $\ldots$). 
L'\'etude des syst\`emes conformes d\'efinis sur une vari\'et\'e \`a bord a \'et\'e 
initi\'ee  par J. Cardy \cite{cardy-bord1,cardy-bord2}, dont l'exemple type est le  
semi-plan infini $Im(z)>0$, \`a partir duquel par application conforme nous pouvons obtenir 
d'autres exemples 
de g\'eom\'etries. Des conditions sur le bord form\'e par l'axe r\'eel sont impos\'ees, que 
nous labellons de  
mani\`ere g\'en\'erale par $a$ et $b$ sur les domaines $Re(z)>0$ et $Re(z)<0$. Sur une bande 
infinie de largeur 
$L$, cel\`a correspond \`a des conditions sur les bords $x=0$ et $x=L$. 
Les transformations conformes doivent pr\'eserver les conditions aux bords, 
les g\'en\'erateurs de ces transformations ne sont donc plus ind\'ependants sur le bord: 
\begin{equation} 
T(z) = \ov{T}(\ov{z})|_{\textrm{axe r\'eel}}, 
\label{cont} 
\end{equation} 
qui exprime l'absence de flux d'\'energie \`a travers le bord.  
Par cons\'equent, il n'y a plus deux alg\`ebres mais {\bf une seule copie} de l'alg\`ebre 
agissant sur les champs, et 
l'espace de Hilbert se d\'ecompose comme: 
\begin{equation} 
\mathcal{H}_{ab} = \bigoplus_i \; \mathcal{F}_{ab}^i \; \mathcal{V}_i. 
\end{equation} 
o\`u les coefficients $\mathcal{F}_{ab}^i$ sont des nombres entiers non-n\'egatifs d\'ecrivant la 
multiplicit\'e  
de la repr\'esentation $\mathcal{V}_i$ pour un syst\`eme avec des conditions aux bords labell\'ees 
par $a$ et $b$.  
Ces conditions aux bords sont r\'ealis\'ees par des op\'erateurs de bord $\phi_a, \phi_b$, ou par 
des \'etats 
de bord $|a\rangle,|b\rangle$, dont une base compl\`ete est donn\'ee par les $r$ \'etats de Ishibashi 
$|j\rangle\rangle$. 
Les coefficients $\mathcal{F}_{ab}^i$ doivent satisfaire des conditions de compatibilit\'e \`a travers  
l'\'equation de Cardy \cite{cardy-eq}, qui impliquent que les $r \times r$ matrices $F_i$ $((F_i)_{ab}
=\mathcal{F}_{ab}^i)$  
doivent satisfaire l'alg\`ebre de fusion \cite{behrend-BCFT}:  
\begin{equation} 
F_i \;  F_j = \sum_k \mathcal{N}_{ij}^k \; F_k. 
\label{algfus} 
\end{equation} 
L'unicit\'e du vide impose $F_1 = \munite_{r \times r}$, et de mani\`ere g\'en\'erale il 
est seulement n\'ecessaire  
de sp\'ecifier un sous-ensemble de ces matrices qui engendre les autres par fusion, 
\`a travers (\ref{algfus}). 
 
\begin{class} 
La classification des conditions aux bords $(a,b)$ compatibles avec l'invariance conforme se ram\`ene donc  
\`a la classification des matrices $F_i$ de dimension $r$ \`a entr\'ees dans $\mathbb{N}$  
satisfaisant l'alg\`ebre de fusion (\ref{algfus}).  
\end{class} 
Or, comme les matrices \`a entr\'ees dans $\mathbb{N}$ sont associ\'ees \`a des matrices 
d'adjacence 
de graphes, nous voyons que les graphes aparaissent naturellement dans l'\'etude des syst\`emes conformes 
avec 
conditions au bord! Nous verrons que pour les th\'eories $\widehat{su}(2)$, les graphes associ\'es sont 
les diagrammes de Dynkin du type $ADE$.

\subsection{Lignes de d\'efauts et fonctions de partition g\'en\'eralis\'ees} 
La fonction de partition d'une th\'eorie conforme d\'efinie sur le tore s'obtient par la trace de 
l'op\'erateur d'\'evolution $\mathcal{T}$ d\'efini en (\ref{evol}).
Dans \cite{Pet_Zub-gener}, une situation plus g\'en\'erale est consid\'er\'ee o\`u est 
ins\'er\'ee l'action 
d'un op\'erateur $X$ dans la trace de $\mathcal{T}$.
Ceci est interpr\'et\'e comme l'introduction d'une ligne de d\'efaut $x$ dans le syst\`eme, le long 
d'un contour 
non-contractible du cylindre, avant de le fermer en un tore, et dont l'effet est de {\bf \it  twister} les 
conditions 
aux bords. L'op\'erateur $X$ (appel\'e op\'erateur de {\it twist}), n'est pas arbitraire: il doit \^etre 
invariant sous 
une distorsion de la ligne \`a laquelle il est attach\'e, et par cons\'equent doit commuter avec les 
g\'en\'erateurs de Virasoro: 
\begin{equation} 
\lbrack L_n,X \rbrack  = \lbrack \ov{L}_n , X \rbrack = 0. 
\end{equation} 
Deux classes d'op\'erateurs $X$ et $Y$ peuvent \^etre consid\'er\'ees 
(correspondant aux deux contours non-contractibles du tore), et les fonctions de 
partition du mod\`ele -- appel\'ees {\bf g\'en\'eralis\'ees} ou {\bf \it twist\'ees} -- sont
donn\'ees par $\textrm{Tr}_{\,\mathcal{H}} \, X \, Y \, \mathcal{T}$: 
\begin{equation} 
\begin{array}{|c|}
\hline
{ } \\
\displaystyle \quad \mathcal{Z}_{x|y}(\tau) =
\sum_{i,j} \; \chi_i(\tau) \; \mathcal{W}_{xy}^{ij} \;  
\ov{\chi}_j(\tau) \quad \\
{ } \\
\hline
\end{array} 
\end{equation}  
o\`u les coefficients $\mathcal{W}_{xy}^{ij}$ sont des entiers non-n\'egatifs d\'ecrivant 
la multiplicit\'e 
de la repr\'esentation $\mathcal{V}_i \otimes \ov{\mathcal{V}}_j$ dans l'espace de Hilbert 
avec deux lignes de  
d\'efauts (``{\it seams}'') $x$ et $y$. 
Le cas sans lignes de d\'efauts ($x=y=0$) implique que l'on retrouve l'invariant modulaire: 
\begin{equation} 
\mathcal{W}_{00}^{ij} = \mathcal{M}_{ij}. 
\label{inv_mod} 
\end{equation} 
Les coefficients $\mathcal{W}_{xy}^{ij}$ peuvent \^etre cod\'es dans des matrices $\widetilde{W}_{ij}$ 
ou dans des matrices $W_{xy}$:  
\begin{equation} 
(\widetilde{W}_{ij})_{xy} = (W_{xy})_{ij} =\mathcal{W}_{xy}^{ij}. 
\end{equation} 
Des conditions de compatibilit\'e \cite{Pet_Zub-gener,Pet_Zub-Oc} imposent que les matrices  
$\widetilde{W}_{ij}$ 
doivent former une repr\'esentation de l'alg\`ebre carr\'ee de fusion: 
\begin{equation} 
\widetilde{W}_{ij} \; \widetilde{W}_{i'j'} = \sum_{i'',j''} \; \mathcal{N}_{ii'}^{i''} \;  
\mathcal{N}_{jj'}^{j''} \; \widetilde{W}_{i''j''}, 
\label{Wij_xy} 
\end{equation} 
o\`u $\mathcal{N}_{ii'}^{i''}$ sont les coefficients de structure de l'alg\`ebre de fusion.  
L'\'equation (\ref{Wij_xy}) pour $j=j'=1$ (resp. $i=i'=1$) implique ($\mathcal{N}_{11}^{j''} = 
\delta_{j''\,1}$): 
\begin{equation} 
\widetilde{W}_{i1} \; \widetilde{W}_{i'1} = \sum_{i''} \; \mathcal{N}_{ii'}^{i''} \; 
\widetilde{W}_{i''1}, \qquad \qquad \qquad 
\widetilde{W}_{1j} \; \widetilde{W}_{1j'} = \sum_{j''} \; \mathcal{N}_{jj'}^{j''} \;  
\widetilde{W}_{1j''}. 
\label{algfusW} 
\end{equation}  
Les matrices $\widetilde{W}_{i1}$ et $\widetilde{W}_{1j}$ forment donc une repr\'esentation de 
l'alg\`ebre de fusion.  
Leurs coefficients sont des entiers non-n\'egatifs, et \`a leurs matrices correspondantes sont donc naturellement 
associ\'es des graphes.
Il est seulement n\'ecessaire de sp\'ecifier un sous-ensemble de ces matrices qui engendre les 
autres par fusion,  
\`a travers (\ref{algfusW}). Dans le cas des mod\`eles $\widehat{su}(2)$, les matrices  
$\widetilde{W}_{21}$ et $\widetilde{W}_{12}$ sont appel\'ees fondamentales, car elles engendrent les 
autres par fusion.  
Elles correspondent chacune \`a la matrice d'adjacence d'un graphe: les graphes d'Ocneanu sont la  
superposition sur un m\^eme graphe de ces deux graphes.  
Les graphes d'Ocneanu de apparaissent donc naturellement dans la classification des fonctions 
de partition {\it twist\'ees} des mod\`eles $\widehat{su}(2)$. 
 
\paragraph{Matrices toriques} 
D\'efinissons les matrices $W_x \doteq W_{x0}$ (telles que $(W_x)_{ij} =  
\mathcal{W}_{x0}^{ij}$), alors combinant (\ref{inv_mod}) et (\ref{Wij_xy}) prise pour $x=y=0$, nous 
obtenons: 
\begin{equation} 
\sum_x (W_x)_{ij} \; (W_x)_{i'j'} = \sum_{i'',j''} \; \mathcal{N}_{ii'}^{i''} \;  
\mathcal{N}_{jj'}^{j''} \; \mathcal{M}_{i''j''} 
\label{matr_tor} 
\end{equation} 
Les matrices $W_x$ sont appel\'ees {\bf matrices toriques} et sont associ\'ees aux vertex $x$ 
du graphe d'Ocneanu. Elles ont premi\`erement \'et\'e obtenues par Ocneanu \cite{Oc-Marseille}\footnote{Les 
premi\`eres matrices toriques ont \'et\'e publi\'ees dans \cite{Coq-qtetra} pour le mod\`ele $E_6$ de 
$\widehat{su}(2)$.} pour le mod\`ele 
$\widehat{su}(2)$, explicitement calcul\'ees en  
r\'esolvant l'\'equation (\ref{matr_tor}). Sa m\'ethode d'obtention   
des matrices toriques $W_x$ \`a partir de la connaissance de l'invariant modulaire 
$\mathcal{M}$ et 
par la formule (\ref{matr_tor}) est appel\'ee ``{\bf \it modular splitting method}''. Ces matrices toriques 
d\'efinissent les  
fonctions de partition g\'en\'eralis\'ees \`a une ligne de d\'efaut.  
Les matrices toriques g\'en\'eralis\'ees $W_{xy}$ (d\'efinissant les 
fonctions de partition \`a deux  
lignes de d\'efaut) s'obtiennent alors \`a partir de la connaissance des matrices  
toriques $W_x$ et des coefficients de structure $\mathcal{O}_{xy}^z$ (entiers non-n\'egatifs) 
de l'alg\`ebre d'Ocneanu: 
\begin{equation} 
W_{xy} = \sum_z \; \mathcal{O}_{xy}^z \; W_z  
\end{equation} 
L'alg\`ebre d'Ocneanu est aussi appel\'ee alg\`ebre des sym\'etries quantiques, dont une repr\'esentation 
matricielle 
est donn\'ee par les matrices $O_x$ telles que $(O_x)_{yz}=\mathcal{O}_{xy}^z$: 
\begin{equation} 
O_x \; O_y = \sum_z \; \mathcal{O}_{xy}^z \; O_z 
\end{equation} 
 
\begin{conclu} 
Ces d\'efinitions et relations sont {\it a priori} suffisantes pour calculer tous les coefficients 
$\mathcal{W}_{xy}^{ij}$ et ainsi 
obtenir toutes les fonctions de partition (invariante modulaire et g\'en\'eralis\'ees) du mod\`ele 
conforme consid\'er\'e. 
Les donn\'ees initiales indispensables sont les matrices $\widetilde{W}_{21}$ et $\widetilde{W}_{12}$, 
ou de mani\`ere \'equivalente le graphe d'Ocneanu lui-m\^eme, et les coefficients de structure de 
l'alg\`ebre d'Ocneanu.  
\end{conclu} 
Cependant, ces graphes ne sont connus (publi\'es) que pour le cas $\widehat{su}(2)$. Le travail central 
de cette th\`ese est de pr\'esenter une r\'ealisation de l'alg\`ebre des sym\'etries quantiques 
$Oc(G)$. Ceci nous permet 
d'une part 
d'obtenir les coefficients $\mathcal{W}_{xy}^{ij}$ sans faire appel \`a la donn\'ee explicite des 
coefficients 
de structure de l'alg\`ebre $Oc(G)$, obtenant des expressions compactes pour les fonctions de partition
du mod\`ele  $\widehat{su}(2)$. D'autre part, notre m\'ethode permet de g\'en\'eraliser 
cette construction pour les 
mod\`eles $\widehat{su}(n)$, sans faire appel \`a la donn\'ee des graphes d'Ocneanu correspondants. L'unique 
donn\'ee initiale se r\'eduit \`a la connaissance des diagrammes de Coxeter-Dynkin associ\'es. 
 
\subsubsection{Conditions au bord et lignes de d\'efaut} 
Une situation encore plus g\'en\'erale consiste \`a combiner une ligne de d\'efaut $x$ et des conditions 
au bord labell\'ees par $a$ et $b$ \cite{Pet_Zub-CFT}. \`A nouveau, une seule copie de l'alg\`ebre 
intervient dans la th\'eorie, et l'espace de Hilbert se d\'ecompose comme: 
\begin{equation} 
\mathcal{H}_{x;ab} = \bigoplus_i \; (F_i S_x)_{ab} \; \mathcal{V}_i, 
\end{equation} 
o\`u les $S_x$ sont des matrices de m\^eme dimension que les matrices $F_i$, et dont les 
\'el\'ements sont des  
entiers non-n\'egatifs. Des conditions de compatibilit\'e \cite{Pet_Zub-CFT} impliquent qu'elles 
forment une  
repr\'esentation de l'alg\`ebre $Oc(G)$ (g\'en\'eralement de dimension diff\'erente): 
\begin{equation} 
S_x \;  S_y = \sum_z \mathcal{O}_{xy}^z \; S_z, 
\end{equation}  
o\`u les $\mathcal{O}_{xy}^z$ sont les coefficients de structure de l'alg\`ebre $Oc(G)$.


\section{Classifications et graphes} 
 
Dans l'\'etude des th\'eories conformes \`a deux dimensions, nous avons vu que le spectre d'une RCFT dans divers 
environnements (syst\`eme ouvert, avec 
conditions au bord ou avec l'introduction de lignes de d\'efaut) est d\'ecrit par un ensemble de 
coefficients 
qui ont la particularit\'e de former des repr\'esentations matricielles \`a entiers non-n\'egatifs 
d'alg\`ebre 
(appel\'ee dans la litt\'erature ``{\sl nimreps}''): 
\begin{equation} 
\begin{array}{lrcl} 
(N_i)_{jk} \qquad \qquad \qquad & N_i \; N_j &=& \displaystyle \sum_k \; \mathcal{N}_{ij}^k \;N_k \\ 
(F_i)_{ab} & F_i \; F_j &=& \displaystyle \sum_k \; \mathcal{N}_{ij}^k \;F_k \\ 
(\widetilde{W}_{ij})_{xy} & \widetilde{W}_{ij} \; \widetilde{W}_{i'j'} &=& \displaystyle \sum_{i'',j''} 
\mathcal{N}_{ii'}^{i''} \;  
\mathcal{N}_{jj'}^{j''} \; \widetilde{W}_{i''j''} \\ 
(O_x)_{yz} & O_x \; O_y &=& \displaystyle \sum_z \; \mathcal{O}_{xy}^z \; O_z\\ 
(S_x)_{ab} & S_x \; S_y &=& \displaystyle \sum_z \mathcal{O}_{xy}^z \; S_z 
\end{array} 
\end{equation}  
Trois ensembles d'indices  interviennent: $(i,j,k,\ldots)$; $(a,b,c,\ldots)$; $(x,y,z,\ldots)$:  
nous verrons qu'ils sont associ\'es \`a trois types 
de graphes: les graphes $\mathcal{A}(G)$, les graphes $G$ (diagrammes de Dynkin ou g\'en\'eralis\'es) 
et les graphes d'Ocneanu $Oc(G)$.  
Nous verrons comment ces structures apparaissent dans  
l'\'etude de la dig\`ebre $\mathcal{B}(G)$, qui joue donc un r\^ole 
pr\'edominant dans la classification 
des th\'eories conformes rationelles, et qui peut \^etre consid\'er\'ee comme la sym\'etrie 
quantique naturelle associ\'ee au mod\`ele.

\subsection{Mod\`eles $\widehat{su}(2)$} 
La premi\`ere classification des fonctions de partition invariantes modulaires a \'et\'e obtenue en  
1987 par Cappelli, Itzykson et Zuber \cite{CIZ-class2} pour les mod\`eles $\widehat{su}(2)$ et 
est pr\'esent\'ee dans la Tab. \ref{ADE}. Elle consiste en trois s\'eries  
infinies (labell\'ees par $A_n$, $D_{2n}$ et $D_{2n+1}$) 
et trois cas exceptionnels (labell\'es par $E_6$, $E_7$, et $E_8$) et est connue sous le nom de 
classification $ADE$. 
Cette terminologie est utilis\'ee pour mettre en valeur l'analogie existante avec la classification 
de Cartan des alg\`ebres de Lie semi-simples simplement lac\'ees: si nous nous   
concentrons sur les termes diagonaux de $\mathcal{Z}$, leur label $i$ sont \'egaux aux exposants de Coxeter 
des diagrammes de Dynkin $G$ correspondants\footnote{Il y a un {\it shift} global de 1 d\^u \`a notre 
convention de label pour les caract\`eres.}. Pour un tel diagramme $G$, les valeurs propres de sa 
matrice d'adjacence sont de la forme  
$2 \cos \frac{\pi m^C}{\kappa}$, o\`u $\kappa$ est le nombre (dual) de Coxeter de $G$ et $m^C$ 
l'exposant de 
Coxeter de $G$. 
Les valeurs de $\kappa$ et $m^C$ sont illustr\'ees dans la Tab. \ref{coxeter}. 
 
\begin{table}[hhh] 
\small 
$$ 
\begin{array}{|cll|} 
\hline 
{ } & { } & { } \\ 
A_{k+1} & k \geq 0 & \displaystyle \sum_{i=0}^{k} |\chi_i|^2 \\  
{ } & { } & { } \\ 
D_{2\ell+2} & k=4\ell & \displaystyle \sum_{i \textrm{ pair }=0}^{2 \ell - 2} |\chi_i + 
\chi_{4 \ell - i}|^2  
              + 2 \, |\chi_{2 \ell}|^2 \\  
{ } & { } & { } \\ 
D_{2\ell+1} & k=4 \ell - 2& \displaystyle \sum_{i \textrm{ pair }=0}^{4 \ell -2} |\chi_i|^2 + 
|\chi_{2\ell - 1}|^2  
              + \sum_{i \textrm{ impair }=1}^{2 \ell -3}(\chi_i \ov{\chi}_{4 \ell - 2 -i}+ 
\chi_{4 \ell - 2 -i}  \ov{\chi}_i )\\  
{ } & { } & { } \\ 
E_{6} & k=10 &  |\chi_{0} + \chi_{6}|^2 + |\chi_{3} + \chi_{7}|^2 + |\chi_{4} + \chi_{10}|^2 \\  
{ } & { } & { } \\ 
E_{7} & k=16 & |\chi_{0} + \chi_{16}|^2 + |\chi_{4} + \chi_{12}|^2 + |\chi_{6} + \chi_{10}|^2 + |\chi_{8}|^2 
               + \chi_8 (\ov{\chi}_2 + \ov{\chi}_{14}) + (\chi_2 + \chi_{14})\ov{\chi}_{8} \\  
{ } & { } & { } \\ 
E_{8} & k=28 &  |\chi_0 + \chi_{10} + \chi_{18} + \chi_{28}|^2 + |\chi_{6} + \chi_{12} + \chi_{16} + 
\chi_{22}|^2 \\  
{ } & { } & { } \\ 
\hline 
\end{array} 
$$ 
\caption{Classification $ADE$ des fonctions de partition invariantes modulaires pour les mod\`eles 
$\widehat{su}(2)$ 
(en fonction des caract\`eres $\chi_i$ de $\widehat{su}(2)_k$)} 
\label{ADE} 
\normalsize 
\end{table} 
 
\begin{table}[hhh] 
$$ 
\begin{array}{|ccl|} 
\hline 
{ } & { } & { } \\ 
\textrm{Diagramme} &\qquad \kappa \qquad& \textrm{Valeurs de } m^C \\ 
{ } & { } & { } \\ 
A_n  & n+1  & 1,2,3,\ldots,n \\ 
{ } & { } & { } \\ 
D_n & 2n-2 & 1,3,\ldots,2n-3 \text{ et } n-1 \\ 
{ } & { } & { } \\ 
E_6 & 12 & 1,4,5,7,8,11 \\ 
{ } & { } & { } \\ 
E_7 & 18 & 1,5,7,9,11,13,17 \\ 
{ } & { } & { } \\ 
E_8 & 30 & 1,7,11,13,17,19,23,29 \\ 
{ } & { } & { } \\ 
\hline 
\end{array} 
$$ 
\caption{Nombre de Coxeter $\kappa$ et exposants de Coxeter $m^C$ pour les diagrammes $ADE$} 
\label{coxeter} 
\end{table} 
 
Les classifications de type $ADE$ interviennent dans divers domaines des math\'ematiques; en plus des 
alg\`ebres 
de Lie semi-simples, signalons aussi: sous-groupes finis de $SO(3)$, groupes de r\'eflexion en 
cristallographie, 
matrices sym\'etriques de valeur propre comprise entre -2 et +2, $\ldots$. Cependant,  
l'apparition d'une telle classification $ADE$ pour les fonctions de partition invariantes 
modulaires \'etait myst\'erieuse \`a l'\'epoque. Depuis l'av\`enement des \'etudes des conditions au bord 
et des lignes de d\'efaut, une meilleure compr\'ehension de cette analogie a \'et\'e obtenue. 
 
Rappelons qu'au niveau $k$,  
les repr\'esentations int\'egrables de $\widehat{su}(2)_k$ sont labell\'ees par  
$\mathcal{I} = \{0,1,2 \cdots, k\}$, avec $c = 3k/(k+2)$ et $h_i=((i+1)^2-1)/(4(k+2))$. Les matrices 
$T$ et $S$ de transformation modulaire des caract\`eres sont donn\'ees par: 
\begin{eqnarray*} 
T_{ij} &=& \exp \left\lbrack 2 i \pi \left( \frac{j^2}{4(k+2)} - \frac{1}{8} \right) \right\rbrack 
\delta_{ij}, \\  
S_{ij} &=& \sqrt{\frac{2}{k+2}} \sin \left( \frac{\pi \, i \, j }{k+2} \right).  
\end{eqnarray*} 
De la matrice $S$ et par la formule de Verlinde, nous obtenons les coefficients de fusion 
$\mathcal{N}_{ij}^k$. 
Introduisant les matrices de fusion $N_i$ telles que $(N_i)_{jk} = \mathcal{N}_{ij}^k$, elles 
satisfont: 
\begin{equation} 
N_i \; N_2 = N_{i-1} + N_{i+1},  
\label{fusmatrn}
\end{equation} 
avec $N_1 = \munite_{k+1,k+1}$. Les matrices de fusion s'obtiennent donc toutes \`a partir de la 
connaissance de la  
matrice $N_1$, appel\'ee fondamentale: $N_1$ correspond \`a la matrice d'adjacence du diagramme $A_{k+1}$! 
 
En pr\'esence de conditions au bord, le spectre de la th\'eorie est cod\'e par des matrices $F_i$ 
satisfaisant 
la m\^eme alg\`ebre de fusion (\ref{fusmatrn}). La classification des matrices $F_i$ poss\'edant 
cette propri\'et\'e a 
\'et\'e  
compl\'et\'ee.  $F_2$ est aussi appel\'ee fondamentale, car elle engendre les autres par fusion, et 
elle correspond \`a la matrice d'adjacence d'un diagramme $G$ de Dynkin de type $ADE$! Les conditions aux 
bords $a$ et $b$ peuvent \^etre mises en correspondance avec les vertex du diagramme $G$.  
 
Finalement, les graphes d'Ocneanu de $\widehat{su}(2)$ et les alg\`ebres d'Ocneanu correspondantes --- 
\`a partir 
desquels sont d\'efinis les coefficients $\mathcal{W}_{xy}^{ij}$ --- sont connus et 
classifi\'es \cite{Oc-paths}.  
Ils permettent de d\'efinir toutes les fonctions de partition g\'en\'eralis\'ees (\`a une et deux 
lignes de d\'efaut) des mod\`eles $\widehat{su}(2)$ \cite{Coq_Gil-ADE, Pet_Zub-Oc}.

\subsection{Mod\`eles minimaux} 
La classification des mod\`eles minimaux ($c<1$) a \'et\'e compl\'et\'ee \cite{CIZ-class2}, et suit aussi 
une classification  
de type $ADE$: \`a chaque fonction de partition invariante modulaire correspond une paire de diagrammes 
de Dynkin $(A,G)$.  La th\'eorie est unitaire si et seulement si le nombre de Coxeter des deux diagrammes
diff\`ere d'une unit\'e. Cette classification est 
naturellement reli\'ee \`a la classification 
$ADE$ des mod\`eles $\widehat{su}(2)$. La charge centrale pour les modeles minimaux et les mod\`eles 
$\widehat{su}(2)_k$ est respectivement donn\'ee par: 
\begin{equation} 
c=1 - \frac{6}{m(m+1)},  \qquad \qquad \qquad c = \frac{3 k}{k + 2} \geq 1. 
\end{equation} 
Les mod\`eles minimaux ont une charge centrale $c<1$. Pour obtenir ces valeurs \`a partir  d'un syst\`eme avec 
alg\`ebre affine, 
il faut consid\'erer des th\'eories associ\'ees \`a des quotients G/H, avec 
H sous-groupe de G. 
La charge centrale d'une theorie construite sur G/H est donn\'ee par 
$c_{G/H} = c_G - c_H$ \cite{goddard-coset1,goddard-coset2}. En considerant une theorie 
avec des quotients de la forme suivante: 
\begin{equation} 
SU(2)_k \times SU(2)_1 / SU(2)_{k+1}, \qquad \qquad \frac{\widehat{su}(2)_k\oplus \widehat{su}(2)_1}
{\widehat{su}(2)_{k+1}}, 
\end{equation} 
la charge centrale est donn\'ee par: 
\begin{equation} 
c = \frac{3k}{k+2} +1 - \frac{3(k+1)}{k+3} = 1 - \frac{6}{(k+2)(k+3)}, 
\end{equation} 
et nous retrouvons alors exactement la charge centrale des modeles minimaux unitaires (avec $m$ entier 
$ = k+2$)!  
La classification des mod\`eles minimaux unitaires est donc reli\'ee, \`a travers une construction de 
{\it coset},  
\`a la classification des modeles $\widehat{su}(2)$, ce qui explique  
la classification $ADE$ de ces mod\`eles minimaux.

Dans notre approche, la relation entre mod\`eles minimaux et mod\`eles $\widehat{su}(2)$ peut \^etre
reformul\'ee comme suit: la fonction de partition invariante modulaire $\mathcal{Z}$ d'un mod\`ele 
minimal unitaire de type 
$(A,G)$, o\`u $A$ et $G$ sont des diagrammes tels que $\kappa_A = \kappa_G \pm 1$, est d\'efinie par:
\begin{equation}
\mathcal{Z} = \ov{\chi} \; \mathcal{M} \; \chi \; , 
\end{equation}
o\`u la matrice $\mathcal{M}$ est obtenue par le produit tensoriel des invariants modulaires 
des mod\`eles $\widehat{su}(2)$ associ\'es aux graphes $A$ et $G$: 
$\mathcal{M} = \mathcal{M}_A \otimes \mathcal{M}_G$. La matrice $\mathcal{M}$ agit sur
un espace vectoriel dont une base est labell\'ee par les vecteurs $\chi = \chi_i \otimes \chi_j$ (o\`u
$\chi_i$ d\'esigne les caract\`eres de $\widehat{su}(2)$). Toutefois, il faut prendre en compte 
l'action $\sigma$ de $\mathbb{Z}_2$ entre les vecteurs labell\'es par $(i,j)$ et $(\sigma(i),\sigma(j))$
provenant de la sym\'etrie de la table de Kac \cite{Coq_Marina}. La classification des invariants
modulaires $\mathcal{M}$ des mod\`eles $\widehat{su}(2)$ permet alors de retrouver tr\`es simplement
celle des mod\`eles minimaux.

D'autre part, la possibilit\'e de remplacer l'invariant modulaire $\mathcal{M}_G = W_0$ par des matrices
toriques $W_x$ (ou g\'en\'eralis\'ees $W_{xy}$) am\`ene naturellement \`a la classification des 
diff\'erentes fonctions de partition twist\'ees des mod\`eles minimaux:
\begin{equation}
\mathcal{Z}_{xy,x'y'} = \frac{1}{2} \ov{\chi} \left( W_{xy} \otimes W_{x'y'}' \right) \chi
\end{equation} 
Il existe six types diff\'erents de fonctions de partition twist\'ees pour les mod\`eles minimaux, 
obtenues en choisissant 
les indices $(x,y)$ et $(x',y')$ comme suit: $((0,0), (0,0))$; $((x,0), (0,0))$; $((x,y), (0,0))$; 
$((x,0), (x',0))$, $((x,y), (x',0))$, $((x,y), (x',y'))$.

\subsection{Mod\`eles $\widehat{su}(n), n\geq 3$ et mod\`eles minimaux g\'en\'eralis\'es} 
 
\paragraph{Mod\`eles $\widehat{su}(3)$} 
Suivant une d\'emarche combinatoire pour la recherche des invariants modulaires $\mathcal{M}$, 
la classification 
des mod\`eles $\widehat{su}(3)$ a \'et\'e obtenue par Gannon en 1994 \cite{gannon-class}. Elle comporte 
six s\'eries infinies et six cas exceptionnels. De m\^eme que les cas $\widehat{su}(2)$ sont 
reli\'es aux diagrammes de Dynkin $ADE$, \`a chaque fonction de partition $\mathcal{Z}$ 
de $\widehat{su}(3)$ nous pouvons associer un graphe, tel que son spectre (valeurs propres de la matrice 
d'adjacence du graphe) soit cod\'e dans les termes diagonaux de $\mathcal{Z}$, et tel que les conditions 
au bord soient en correspondance avec ses vertex.  
Les diagrammes de $\widehat{su}(3)$ ont premi\`erement \'et\'e d\'etermin\'es de mani\`ere empirique dans 
\cite{DiFZub}, par l'imposition de propri\'et\'es spectrales, et sont 
connus dans la litt\'erature sous le nom de diagrammes de Di Francesco-Zuber. Par la suite, 
ces diagrammes sont apparus dans les travaux d'alg\`ebres d'op\'erateurs 
\cite{Oc-paths,Bock_Evans, Bock_Evans_Kawa}. 
La liste de ces diagrammes a \'et\'e rectifi\'ee par Ocneanu: la liste finale est publi\'ee dans 
\cite{Oc-Bariloche}. 
 
Les graphes d'Ocneanu de $\widehat{su}(3)$ ne sont pas connus (publi\'es), les fonctions de 
partition  
\`a une ou deux lignes de d\'efaut pour ces mod\`eles n'avaient donc pas \'et\'e obtenues.  
Nous verrons que gr\^ace \`a notre r\'ealisation de l'alg\`ebre des sym\'etries quantiques 
(conjectur\'ee en s'inspirant 
de celle de $\widehat{su}(2)$), nous obtenons sur quelques exemples ces fonctions de partition
\cite{Coq_Gil-Tmod}.

\paragraph{Mod\`eles $\widehat{su}(n)$} 
Nous pouvons suivre la m\^eme d\'emarche et associer \`a chaque fonction de partition invariante 
modulaire $\mathcal{Z}$ de $\widehat{su}(n)$ un graphe codant \`a travers ses propri\'et\'es spectrales 
les termes diagonaux $\mathcal{Z}$, et tels que ses vertex classifient les possibles conditions aux 
bords. La liste compl\`ete des graphes a \'et\'e d\'etermin\'ee par Ocneanu pour le cas $\widehat{su}(4)$
(la liste est publi\'ee dans \cite{Oc-Bariloche}), et nous pouvons en d\'eduire une classification 
correspondante compl\`ete des fonctions de partition invariantes modulaires du cas $\widehat{su}(4)$.
Pour des rangs sup\'erieurs \`a 4, il n'existe pas de classification compl\`ete 
des fonctions de partition invariante modulaire, ni de liste compl\`ete des graphes correspondants.

\paragraph{Mod\`eles minimaux g\'en\'eralis\'es}
Les mod\`eles minimaux ``usuels'' font intervenir un nombre fini de repr\'esentations irr\'eductibles
de l'alg\`ebre de Virasoro, et sont labell\'es par une paire de diagrammes de type $su(2)$, c.\`a.d.
les diagrammes $ADE$. Or, l'alg\`ebre de Virasoro est un cas particulier d'alg\`ebres plus g\'en\'erales 
$\mathcal{W}_n$: $Vir=\mathcal{W}_2$. Les mod\`eles minimaux usuels sont de type $\mathcal{W}_2$. Nous
pouvons d\'efinir des mod\`eles minimaux de type $\mathcal{W}_n$, appel\'es g\'en\'eralis\'es. Pour le cas 
$\mathcal{W}_3$ par exemple, ils sont labell\'es par une paire de disgrammes de Di Francesco- Zuber,
et sont unitaires si les nombres de Coxeter (g\'en\'eralis\'es) de ces diagrammes diff\`erent 
d'une unit\'e. 

En ce sens, l'obtention des matrices toriques g\'en\'eralis\'ees $W_{xy}$ des mod\`eles $\widehat{su}(n)$
(dont les expressions pour $\widehat{su}(2)$ et quelques exemple de $\widehat{su}(3)$ sont donn\'ees
dans le chapitre {\bf 4}) est d'une grande utilit\'e pour la classification des mod\`eles minimaux
g\'en\'eralis\'es (d\'etermination de la fonction de partition invariante modulaires et des diff\'erents 
types de fonctions de partition twist\'ees) \cite{Coq_Marina}.


\chapter{L'alg\`ebre des sym\'etries quantiques d'Ocneanu} 
\thispagestyle{empty}
 
La construction d'Ocneanu -- premi\`erement d\'ecrite dans \cite{Oc-Marseille} -- associe une dig\`ebre 
$\mathcal{B}(G)$ \`a chaque diagramme de Dynkin $G$ de type $ADE$. $\mathcal{B}(G)$ est l'espace vectoriel
des endomorphismes de chemins essentiels sur lequel sont d\'efinies deux lois multiplicatives
$\circ$ et $\odot$.
Les divers coefficients provenant des structures 
alg\'ebriques de $\mathcal{B}(G)$ interviennent dans la d\'etermination des fonctions de partition
des mod\`eles conformes $\widehat{su}(2)$. Nous pr\'esentons dans ce chapitre une introduction 
\`a ces structures
alg\'ebriques ainsi que l'\'etude approfondie de la dig\`ebre associ\'ee au diagramme de Dynkin
$A_3$.

\section{Les chemins essentiels sur un graphe $G$} 
La notion de chemins essentiels sur un graphe a \'et\'e introduite par A. Ocneanu dans \cite{Oc-paths}.  
Cet article \'etant tr\`es ``dense'' et les d\'efinitions y \'etant pr\'esent\'ees parfois de mani\`ere  
allusive, nous donnons ici une introduction \`a ces notions (voir aussi \cite{Coq-qtetra,Coq_Gil-Tmod}). 
 
\subsection{Quelques d\'efinitions}

\begin{defin} 
Nous d\'efinissons un {\bf graphe} $G$ par la donn\'ee d'un triplet ($V,E,\psi$) tel 
que: 
\begin{itemize} 
\item[-] $V$ est un ensemble non-vide d'\'el\'ements $v$ appel\'es vertex, 
\item[-] $E$ est un ensemble non vide d'\'el\'ements $\xi$ appel\'es arcs, 
\item[-] $\psi$ est une fonction d'incidence qui associe \`a chaque arc de $G$ une 
paire ordonn\'ee de vertex (non n\'ecessairement distincts) de $G$. 
\end{itemize} 
\end{defin} 
Un graphe est dit {\bf fini} si les ensembles $V$ et $E$ sont finis, et nous notons 
$r$ le nombre de vertex du graphe: $r = Card (V)$. 
Si $\xi$ est un arc, $v_i$ et $v_j$ deux vertex tels que $\psi(\xi)=v_i v_j$, 
alors $\xi$ joint le vertex $v_i$ au vertex $v_j$, et nous notons un tel arc 
$\xi_{ij}$ ou $\vec{ij}$. 
Nous appelons $s(\xi_{ij})=v_i$ la {\bf source} de $\xi_{ij}$ et $r(\xi_{ij})=v_j$ 
l'{\bf extr\'emit\'e} de $\xi_{ij}$. 
Un arc ayant la m\^eme source et extr\'emit\'e est appel\'e une {\bf boucle}.  
Un graphe est dit {\bf simple} s'il ne poss\`ede pas de boucle et si deux arcs 
diff\'erents ne relient pas la m\^eme paire de points. 
Un graphe est dit {\bf bi-orient\'e} 
si pour tout arc $\xi_{ij}$ reliant $v_i$ \`a $v_j$, il existe l'arc inverse 
$\xi_{ji}$ reliant $v_j$ \`a $v_i$, que nous 
notons  $\xi_{ij}^{-1}$. Dans le cas contraire, le graphe est dit {\bf orient\'e}.  
Un {\bf diagramme} est la repr\'esentation picturale du graphe. Dans le diagramme, nous repr\'esentons 
les vertex $v_i$ par des points labell\'es par $\sigma_i$, $(i=1,\ldots,r)$. 
L'arc $\xi_{ij}$ est repr\'esent\'e par un vecteur reliant le vertex $v_i$ au vertex $v_j$. 
Dans le cas d'un graphe bi-orient\'e, les deux arcs $\xi_{ij}$ et $\xi^{-1}_{ij}$ sont repr\'esent\'es 
plus simplement par une seule ligne (non-fl\'ech\'ee) reliant $v_i$ et $v_j$.  
 
Un {\bf chemin \'el\'ementaire} est une s\'equence ($\xi_1,\xi_2,\ldots,\xi_k$) d'arcs telle que 
l'extr\'emit\'e de chaque arc co\"{\i}ncide  
avec la source de l'arc suivant: $r(\xi_i)=s(\xi_{i+1}), 1\leq i \leq k-1$. 
Si le chemin rencontre successivement les vertex $v_1,v_2,\ldots,v_k,v_{k+1}$, 
nous le notons $P(v_1,v_2,\ldots,v_k,v_{k+1})$. 
Un graphe est dit {\bf fortement connect\'e} si pour tout 
couple de vertex $v_i$ et $v_j$, il existe un chemin \'el\'ementaire reliant ces vertex.

Les chemins \'el\'ementaires de longueur $i$ forment donc l'ensemble suivant: 
\begin{equation} 
\mathcal{P}^{(i)}(G) = \{P(v_1,v_2,\cdots,v_i,v_{i+1})\} = \{ \xi^{(i)} = (\xi_1,\xi_2,\cdots,\xi_i) | 
r(\xi_{\ell})=s(\xi_{\ell+1}) \} 
\end{equation} 
et nous consid\'erons les vertex $v$ comme des chemins de longueur 0.  
Nous appelons $Path^{(i)}$ l'espace vectoriel ayant comme base les chemins \'el\'ementaires 
$\xi^{(i)} \in \mathcal{P}^{(i)}(G)$. Nous introduisons un produit scalaire dans cet espace vectoriel, 
not\'e $\langle \, . \, , \, . \, \rangle$, en imposant que les chemins \'el\'ementaires 
soient orthonorm\'es:  
\begin{equation} 
\langle \xi, \eta \rangle = \delta_{\xi\,\eta} \qquad \qquad \qquad \xi, \eta \in Path^{(i)} 
\end{equation} 


\begin{defin} 
Soit $G$ un graphe \`a $r$ vertex. La {\bf matrice d'adjacence} de $G$ est la matrice 
$r \times r$ $\mathcal{G}$ ayant comme entr\'ee  
$\mathcal{G}_{ij}=n$ s'il existe $n$ arcs $\xi_{ij}$ reliant le vertex $v_i$ au 
vertex $v_j$, et $\mathcal{G}_{ij}=0$ sinon. La {\bf norme} du graphe $G$ est d\'efinie comme 
\'etant \'egale \`a la norme de sa matrice d'adjacence $\mathcal{G}$. 
\end{defin} 
Pour un graphe simple, sa matrice d'adjacence v\'erifie donc: $\mathcal{G}_{ii}=0$ et $\mathcal{G}_{ij} 
\in \{0,1\}$. 
Par la suite, nous parlerons de graphes comme un racourci pour graphes simples, finis 
et fortement connect\'es. \\
 
Soit $A=(a_{ij})$ une matrice carr\'ee et appelons $a_{ij}^{(k)}$ l'entr\'ee  
$(i,j)$ de la matrice $A^k$, o\`u $k$ est un entier positif. 
\begin{defin} 
Une matrice $A$ carr\'ee $n \times n$ \`a entr\'ee dans les entiers non-n\'egatifs est dite  
{\bf irr\'eductible} si et seulement si, pour chaque $i,j \in \{1,\ldots,n\}$, il existe 
un entier positif $k$, qui peut d\'ependre de $i$ et $j$, tel que $a_{ij}^{(k)} > 0$.  
\end{defin} 
Nous avons vu plus haut la d\'efinition d'un chemin \'el\'ementaire. Il existe 
une mani\`ere tr\`es simple de compter 
le nombre de tels chemins de longueur $k$ fix\'ee. 
\begin{theo} 
Si $A=(a_{ij})$ est la matrice d'adjacence d'un graphe \`a $n$ vertex, alors le 
nombre de chemins \'el\'ementaires 
distincts de longueur $k$ reliant les vertex $v_i$ et $v_j$ est \'egal \`a 
$a_{ij}^{(k)}$. 
\end{theo} 
Puisque, par d\'efinition, dans un graphe fortement connect\'e il existe au moins  
un chemin reliant tout vertex $v_i$ \`a un vertex $v_j$, sa matrice d'adjacence est donc   
irr\'eductible. 
\begin{theo}[Perron-Frobenius] 
Soit $A$ une matrice carr\'ee irr\'eductible \`a entr\'ees dans les entiers non n\'egatifs. 
Alors, il existe une valeur propre $\beta$ de $A$ telle que: 
\begin{itemize} 
\item[-] $\beta$ est r\'eelle, $\beta >0$; 
\item[-] $\beta$ est la plus grande valeur propre de $A$; 
\item[-] le vecteur propre correspondant \`a $\beta$ est positif\footnote{Toutes ses composantes ont le m\^eme signe.}, et est unique 
\`a une constante multiplicative pr\`es. 
\end{itemize} 
\label{perron} 
\end{theo} 
Ce vecteur-propre est appel\'e vecteur de Perron-Frobenius, et sera not\'e $P$. 
Il satisfait donc \`a l'\'equation suivante: 
\begin{equation} 
\mathcal{G} P = \beta P 
\label{def-beta} 
\end{equation} 
Soit $\{v_1,v_2,\ldots,v_{r}\}$ une base des vertex du graphe $G$: puisque la matrice d'adjacence 
de $G$ est irr\'eductible \`a entr\'ees dans $\mathbb{N}$, le {\bf Th\'eor\`eme} {\bf \ref{perron}} 
s'applique et  
nous d\'efinissons l'application $\mu$ donnant la composante de Perron-Frobenius des vertex: 
\begin{eqnarray*} 
\mu: V(G) &\longmapsto& \mathbb{R}^{+} \\ 
      v_i &\longmapsto& P(i) 
\end{eqnarray*} 
Nous normalisons ce vecteur de telle mani\`ere que $\mu(v_1)=1$, o\`u $v_1$ est choisi 
comme \'etant le vertex\footnote{Dans les cas consid\'er\'es, il n'y aura pas d'ambiguit\'e sur le
choix de ce vertex.} ayant la plus petite composante $P(i)$ (il sera not\'e avec une $\star$ 
sur le diagramme correspondant).

\subsection{Graphes bi-orient\'es et leur classification} 
Dans le cas d'un graphe $G$ bi-orient\'e, sa matrice d'adjacence est 
sym\'etrique, et sa norme est alors donn\'ee par: 
$$ 
||G|| = ||\mathcal{G}|| = \max \{ |\lambda|, \text{o\`u $\lambda$ est valeur 
propre de $\mathcal{G}$}\} 
$$  
Il existe une classification reliant les valeurs possibles de cette norme et son 
graphe correspondant \cite{Jones-book}: 
\begin{theo} 
Soit $\mathcal{G}$ une matrice carr\'ee sym\'etrique \`a entr\'ee dans les entiers non-n\'egatifs. 
Alors: 
\begin{itemize} 
\item $||\mathcal{G}||=2$ si et seulement si $\mathcal{G}$ est la matrice $(\ell+1) \times (\ell+1)$ 
d'adjacence de l'un des graphes suivants: 
$$ 
A_{\ell}^{(1)} (\ell \geq 2),\qquad  D_{\ell}^{(1)} (\ell \geq 4), \qquad E_{\ell}^{(1)} (\ell=6,7,8) 
$$ 
\item $||\mathcal{G}|| <2 $ si et seulement si $\mathcal{G}$ est la matrice $(\ell \times \ell)$ 
d'adjacence de l'un des graphes suivants: 
$$ 
A_{\ell} ({\ell} \geq 2), \qquad D_{\ell} (\ell \geq 4), \qquad E_{\ell} (\ell=6,7,8) 
$$ 
De plus, $ ||\mathcal{G}||=\beta = 2\cos(\frac{\pi}{\kappa})$, o\`u $\kappa$ est par
d\'efinition le nombre (dual) de Coxeter du graphe correspondant. Les autres valeurs propres de
$\mathcal{G}$ sont 
donn\'ees par $\lambda = 2 \cos (\frac{m^c \pi}{\kappa})$ (possiblement avec multiplicit\'e), o\`u  
les $m^c$ sont par d\'efinition les exposants de Coxeter du graphe, prenant $\ell$ valeurs comprises 
entre 1 et $\kappa -1$. 
\end{itemize} 
\end{theo}  
Les graphes d\'efinis ci-dessus sont illustr\'es dans l'Annexe {\bf A}. Ils 
correspondent aux diagrammes de Dynkin des 
alg\`ebres de Lie semi-simples simplement lac\'ees ($A_{\ell},D_{\ell},E_{\ell}$), ou de leur 
extension affine ($A_{\ell}^{(1)},D_{\ell}^{(1)},E_{\ell}^{(1)}$). Insistons sur le fait que nous
n'utiliserons pas la notion d'alg\`ebre de Lie ici (le nombre dual de Coxeter par exemple
est {\sl d\'efini} \`a travers la norme du graphe).

\paragraph{Correspondance de Mc-Kay} 
Les vertex des diagrammes affines ($A_{\ell}^{(1)},D_{\ell}^{(1)},E_{\ell}^{(1)}$) sont en  
correspondance bi-univoque avec les repr\'esentations irr\'eductibles des sous-groupes du groupe  
$SU(2)$, et la composante de Perron-Frobenius de ces vertex est \'egale \`a la dimension 
des irreps: c'est la correspondance de Mc-Kay classique \cite{McKay}. De m\^eme, les vertex 
des diagrammes ($A_{\ell},D_{\ell},E_{\ell}$) peuvent \^etre mis en correspondance avec les 
irreps de ``sous-groupes'' ou ``modules'' associ\'es\footnote{La d\'efinition ici adopt\'ee des mots
``sous-groupe'' et ``module'' diff\`ere de l'acceptation usuelle de ces termes: nous pr\'ecisons dans 
l'Annexe {\bf B} ce que nous entendons par l\`a.} au groupe quantique $U_q(sl(2))$, avec $q$ racine de 
l'unit\'e, les composantes de Perron-Frobenius donnant, par d\'efinition, les dimensions 
quantiques de ces irreps (ce ne sont plus des nombres entiers, mais des $q$-nombres!).  
C'est l'analogie quantique de la correspondance de Mc-Kay. Les correspondances de Mc-Kay (classique  
et quantique) sont pr\'esent\'ees plus en d\'etail dans l'Annexe {\bf B}.

\subsection{Chemins essentiels sur un graphe} 
 
Consid\'erons un graphe $G$ \`a $r$ vertex, et l'espace vectoriel des chemins \'el\'ementaires de 
longueur deux: $Path^{(2)}$. Un \'el\'ement $\xi^{(2)}$ de cet espace sera aussi not\'e 
$\xi \otimes \eta$, o\`u $\xi$ et $\eta$ sont des chemins de longueur un (ce sont des arcs).  
L'op\'erateur d'annihilation d'Ocneanu $C_1$ est d\'efini par: 
\begin{equation} 
\begin{array}{rcl} 
C_1: \qquad \qquad \qquad Path^{(2)} &\longmapsto& Path^{(0)} \\ 
\xi^{(2)} = \xi \otimes \eta &\longmapsto & \displaystyle \delta_{\xi,\eta^{-1}} 
\sqrt{\frac{\mu(r(\xi))}{\mu(s(\xi))}} s(\xi) 
\end{array} 
\label{C1} 
\end{equation}  
L'op\'erateur $C_1$ donne un r\'esultat non-nul si et seulement si le chemin de longueur deux  
sur lequel il agit est un aller-retour: $C_1[P(v_i v_j v_k)] \sim \delta_{i,k} v_i$. 
L'op\'erateur de cr\'eation d'Ocneanu $C^{\dag}_1$ est d\'efini par: 
\begin{equation} 
\begin{array}{rcl} 
C_1^{\dag}: \qquad Path^{(0)} &\longmapsto& Path^{(2)}  \\ 
 v &\longmapsto & \displaystyle \sum_{\xi,s(\xi)=v}  \sqrt{\frac{\mu(r(\xi))}{\mu(s(\xi))}} \xi 
\otimes \xi^{-1} 
\end{array} 
\label{C1dag}  
\end{equation} 
En d'autres termes, l'op\'erateur $C_1^{\dag}$ cr\'ee des allers-retours avec tous les vertex adjacents 
\`a $v$. 
\begin{theo} 
La composition de ces deux op\'erateurs $C_1 \, C_1^{\dag}$ est un scalaire \'egal \`a $\beta$, la plus 
grande valeur 
propre de la matrice d'adjacence du graphe.  
\end{theo} 
{\bf \ud{d\'em.}}: L'\'equation $\mathcal{G}P=\beta P$, o\`u $P_i=\mu(v_i)$, implique: 
$$ 
\sum_j (\mathcal{G})_{ij} \mu(v_j) \; = \beta \,\mu(v_i)\; \Longrightarrow \sum_{\textrm{$v_j$ voisin de 
$v_i$}} \mu(v_j)  
= \beta \, \mu(v_i) \;\Longrightarrow \beta = \sum_{\xi,s(\xi)=v_i} \frac{\mu(r(\xi))}{\mu(s(\xi))}. 
$$ 
En composant (\ref{C1dag}) et (\ref{C1}), la d\'emonstration est alors imm\'ediate. \hfill $\blacksquare$ \\ 
 
\begin{defin} 
Le projecteur de Jones $e_1$ est l'op\'erateur de projection d\'efini par: 
\begin{equation} 
e_1 = \frac{1}{\beta} C_1^{\dag} \, C_1 .
\end{equation} 
\end{defin} 
Il est imm\'ediat de v\'erifier que c'est bien un op\'erateur de projection: $e_1^{\dag}=e_1$, $e_1^2 = 
e_1$.

\begin{figure}[H] 
\unitlength 0.8mm 
\begin{center} 
\begin{picture}(40,13)(0,9) 
\put(5,10){\line(1,0){30}} 
\multiput(5,10)(15,0){3}{\circle*{2}} 
\put(5,3){\makebox(0,0){$\tau_{0}$}} 
\put(20,3){\makebox(0,0){$\tau_{1}$}} 
\put(35,3){\makebox(0,0){$\tau_{2}$}} 
\put(5,17){\makebox(0,0){$[1]$}} 
\put(20,17){\makebox(0,0){$[\sqrt{2}]$}} 
\put(35,17){\makebox(0,0){$[1]$}} 
\end{picture} 
\end{center} 
\caption{Diagramme $A_3$ et dimension quantique de ses vertex} 
\label{grA3chap2}
\end{figure}
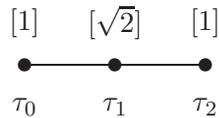

\paragraph{Exemple: diagramme $A_3$} 
Le diagramme $A_3$ poss\`ede 3 vertex not\'es $\tau_0, \tau_1$ et $\tau_2$, et est repr\'esent\'e \`a 
la Fig.\ref{grA3chap2}. 
Nous donnons entre crochets la valeur de la composante du vecteur de Perron-Frobenius (dimension 
quantique) 
de chaque vertex. 
\noindent Les chemins \'el\'ementaires de longueur deux sur le diagramme $A_3$ sont au nombre de 6: 
\begin{eqnarray*} 
P( \tau_0 \tau_1 \tau_0 ) = \vec{01} \otimes \vec{10}   &\qquad \qquad \qquad \qquad&  
P( \tau_1 \tau_2 \tau_1 ) = \vec{12} \otimes \vec{21} \\ 
P( \tau_0 \tau_1 \tau_2 ) = \vec{01} \otimes \vec{12}   &\qquad \qquad \qquad \qquad&  
P( \tau_2 \tau_1 \tau_2 ) = \vec{21} \otimes \vec{12} \\ 
P( \tau_1 \tau_0 \tau_1 ) = \vec{10} \otimes \vec{01}   &\qquad \qquad \qquad \qquad&  
P( \tau_2 \tau_1 \tau_0 ) = \vec{21} \otimes \vec{10}  
\end{eqnarray*} 
L'action de l'op\'erateur de cr\'eation $C_1^{\dag}$ sur les vertex est donn\'ee par: 
\begin{eqnarray*} 
C_1^{\dag} (\tau_0) &=& 2^{1/4} (\vec{01} \otimes \vec{10}) \\ 
C_1^{\dag} (\tau_1) &=& 2^{-1/4} (\, \vec{10} \otimes \vec{01} + \vec{12} \otimes \vec{21} \, ) \\ 
C_1^{\dag} (\tau_2) &=& 2^{1/4} (\vec{21} \otimes \vec{12})  
\end{eqnarray*} 
L'action de l'op\'erateur d'annihilation $C_1$ sur les chemins \'el\'ementaires de longueur deux est 
donn\'ee par: 
$$ 
\begin{array}{rclcrcl} 
C_1(\vec{01} \otimes \vec{10}) &=& 2^{1/4} (\tau_0) &\qquad& C_1(\vec{21} \otimes \vec{12}) &=& 2^{1/4} 
(\tau_2) \\ 
C_1(\vec{10} \otimes \vec{01})  = C_1(\vec{12} \otimes \vec{21}) &=& 2^{-1/4} (\tau_1) &\qquad& 
C_1(\vec{01} \otimes \vec{12})  = C_1(\vec{21} \otimes \vec{10}) &=& 0  
\end{array} 
$$ 
Nous pouvons v\'erifier que la composition des deux op\'erateurs $C_1 C_1^{\dag}$ est bien un scalaire 
\'egal 
\`a la plus grande valeur propre de la matrice d'adjacence ($\beta = \sqrt{2}$): 
$$ 
C_1 C_1^{\dag} (\tau_0) = \sqrt{2} (\tau_0) \qquad \qquad 
C_1 C_1^{\dag} (\tau_1) = \sqrt{2} (\tau_1) \qquad \qquad 
C_1 C_1^{\dag} (\tau_2) = \sqrt{2} (\tau_2)  
$$

Nous pouvons maintenant \'etendre la d\'efinition de ces op\'erateurs \`a des chemins \'el\'ementaires de 
longueur 
quelconque. 
\begin{defin} 
Pour tout entier $n > 1$, l'op\'erateur d'annihilation $C_n$, agissant sur des chemins \'el\'ementaires 
de longueur 
$p$, est d\'efini par: 
$$ 
\begin{array}{crcl} 
\text{si $p \leq n$}:\qquad & C_n (\xi_1 \ldots \xi_p) &=& 0 \; ,\\ 
\text{si $p>n$}:\qquad      &  C_n (\xi_1 \xi_2 \ldots \xi_n \xi_{n+1}\ldots \xi_p) &=& \displaystyle  
\sqrt{\frac{\mu(r(\xi))}{\mu(s(\xi))}} 
\delta_{\xi_n,\xi_{n+1}^{-1}}  (\xi_1 \xi_2 \ldots \widehat{\xi}_n \widehat{\xi}_{n+1} \ldots \xi_p) \; ,
\end{array} 
$$ 
o\`u le symbole $\widehat{\xi}$ signifie que l'on \'elimine l'arc $\xi$ du chemin.
\end{defin} 
L'op\'erateur $C_n$ agissant sur des chemins \'el\'ementaires de longueur $p$ donne donc comme  
r\'esultat soit 0, soit un chemin \'el\'ementaire de longueur $p-2$. 
\begin{defin} 
Pour tout entier $n > 1$, l'op\'erateur de cr\'eation $C_n^{\dag}$, agissant sur des chemins 
\'el\'ementaires de longueur 
$p$, est d\'efini par: 
$$ 
\begin{array}{crcl} 
\text{si $p < n+1$}:   \quad & C_n^{\dag} (\xi_1 \ldots \xi_p) &=& 0 \; , \\ 
\text{si $p \geq n-1$}:\quad & C_n^{\dag} (\xi_1 \ldots \xi_{n-1} \xi_{n}\ldots \xi_p) &=& \displaystyle  
\sum_{\eta ,s(\eta)=r(\xi_{n-1})}\sqrt{\frac{\mu(r(\xi))}{\mu(s(\xi))}} 
(\xi_1  \ldots \xi_{n-1} \,\eta \,\eta^{-1} \,\xi_n  \ldots \xi_p) \, .
\end{array} 
$$ 
\end{defin} 
L'op\'erateur $C_n^{\dag}$ agissant sur des chemins \'el\'ementaires de longueur $p$ donne donc comme  
r\'esultat soit 0, soit une  combinaison lin\'eaire de chemins \'el\'ementaires de longueur $p+2$. 
 
Les projecteurs de Jones $e_k$ sont d\'efinis par: 
\begin{equation} 
e_k \doteq \frac{1}{\beta} C_k^{\dag} C_k,
\end{equation} 
et v\'erifient les relations d\'efinissant une alg\`ebre de Temperley-Lieb 
(voir Annexe {\bf C}).
\\
\begin{defin} 
L'espace des chemins essentiels de longueur $n$ est d\'efini par: 
\begin{eqnarray*} 
{\mathcal{E}ss}^{(n)} = EssPath^{(n)} (G) &=& \left\{ \xi \in Path^{(n)} |\; C_k \, \xi =0  
\qquad \forall k = (1,2,\cdots,n) \right\}\\ 
{ }               &=& \left\{ \xi \in Path^{(n)} |\;  e_k \, \xi =0 \, \qquad \forall k 
\in (1,2,\cdots,n) \right\} 
\end{eqnarray*} 
\end{defin} 
Un chemin est donc essentiel s'il appartient \`a l'intersection du noyau de tous les op\'erateurs 
d'annihilation  
$C_k$ ( ou de tous les projecteurs de Jones $e_k$).  
Tout chemin \'el\'ementaire de longueur 0 et de longueur 1 est aussi un chemin essentiel, car il appartient  
au noyau des op\'erateurs d'annihilations. 
Notons qu'un \'el\'ement de ${\mathcal{E}ss}^{(n)}$  n'est pas toujours un chemin \'el\'ementaire de 
longueur 
$n$, mais possiblement une combinaison lin\'eaire de tels \'el\'ements.

\paragraph{Exemple: diagramme $A_3$} 
Nous avons vu que $C_1(\vec{01} \otimes \vec{12}) = C_1(\vec{21} \otimes \vec{10}) = 0$, donc 
ces deux chemins sont des chemins essentiels. 
Nous avons vu  aussi que $C_1 (\vec{10} \otimes \vec{01}) = C_1 (\vec{12} \otimes \vec{21})$. Soit
le chemin $\gamma = \vec{10} \otimes \vec{01} - \vec{12} \otimes \vec{21}$,  nous avons alors 
$C_1(\gamma) = 0$.  
$\gamma$ est donc aussi un chemin essentiel, bien qu'il ne soit pas \'el\'ementaire mais une combinaison 
lin\'eaire de chemins \'el\'ementaires.\\

Soit ${\mathcal{E}ss}^{(i)}_{a,b}$ l'espace des chemins essentiels de longueur $i$  partant du 
vertex $v_a$ et 
arrivant au vertex $v_b$. Alors: 
\begin{equation} 
{\mathcal{E}ss}^{(i)} = \sum_{v_a,v_b \in V} {\mathcal{E}ss}^{(i)}_{a,b} .
\end{equation} 
\begin{theo}[Ocneanu\cite{Oc-paths}] 
La dimension de l'espace des chemins essentiels ${\mathcal{E}ss}^{(i)}_{a,b}$ est donn\'ee par la formule 
de  r\'ecurrence suivante (loi mod\'er\'ee de Pascal): 
\begin{equation} 
\text{dim}({\mathcal{E}ss}^{(i+1)}_{a,b}) = \sum_{\xi,r(\xi)=v_b} 
\text{dim}({\mathcal{E}ss}^{(i)}_{a,s(\xi)}) -  
\text{dim}({\mathcal{E}ss}^{(i-1)}_{a,b})  .
\label{pascalmod} 
\end{equation} 
\end{theo} 
 
Les chemins essentiels de longueur 0 et 1 sont des chemins \'el\'ementaires (vertex et arcs).
La loi (\ref{pascalmod}) nous permet alors
de calculer la dimension des chemins essentiels de longueur donn\'ee. Ces r\'esultats sont plus facilement
cod\'es dans des matrices.

\paragraph{Matrices $F_i$} D\'efinissons les matrices carr\'ees $r \times r$ $F_i$ telles que la composante  
$(a,b)$ de $F_i$ soit \'egale 
au nombre de chemins essentiels de longueur $i$ reliant le vertex $v_a$ au vertex $v_b$ (donc \'egale \`a la 
dimension de ${\mathcal{E}ss}^{(i)}_{a,b}$). 
La loi mod\'er\'ee de Pascal (\ref{pascalmod}) nous permet d'obtenir une r\'ecurrence simple pour 
calculer ces matrices $F_i$: 
\begin{eqnarray*} 
F_0 &=& \munite_{r \times r}  \\ 
F_1 &=& \mathcal{G} \\ 
F_{i+1} &=& \mathcal{G} F_{i} - F_{i-1} 
\end{eqnarray*} 
La dimension de l'espace vectoriel des chemins essentiels de longueur $i$ est donc donn\'ee par:
\begin{equation}
\dim (\mathcal{E}ss^{(i)}) = \sum_{a,b} (F_i)_{ab}
\end{equation}
{\bf Rappel}: \`A chaque diagramme de type $ADE$ est associ\'e un nombre $\kappa$ (nombre dual de Coxeter), 
d\'efini \`a  
partir de la norme  $\beta$ de la matrice d'adjacence du graphe par la 
relation $\beta = 2 \cos \left( 
\frac{\pi}{\kappa} \right)$. 
\begin{theo}[Ocneanu\cite{Oc-paths}] 
Pour les diagrammes de type $ADE$, il n'existe pas de chemins essentiels de longueur plus grande que 
$\kappa -2$.  
\end{theo}  
Au niveau matriciel, ceci se traduit par le fait que la matrice $F_{\kappa-1}$ est nulle. Au vue de la 
correspondance 
de Mc-Kay, nous verrons au chapitre {\bf 3} que les matrices $F_i$ codent la d\'ecomposition 
du produit tensoriel 
$\tau_i \otimes \sigma_a$ en irreps $\sigma_b$, o\`u les irreps $\sigma_a$ et $\sigma_b$ sont associ\'ees 
aux vertex  
$v_a$ et $v_b$ du graphe $G$ de nombre de Coxeter $\kappa$, 
et l'irrep $\tau_i$ est associ\'ee au vertex $v_i$ du graphe $A_{\kappa-1}$ 
Les graphes $A_n$ correspondent \`a un quotient $\mathcal{H}$ du  groupe 
quantique $U_q(sl(2))$, tandis que les graphes $G$ sont associ\'es 
\`a des ``sous-groupes'' ou ``modules'' de $\mathcal{H}$. 
 
\paragraph{Matrices essentielles $E_a$} 
Il existe une autre mani\`ere de coder ces r\'esultats.  
D\'efinissons $r$ matrices $E_a$ associ\'ees \`a chaque vertex $v_a$ de $G$ et appel\'ees {\bf matrices 
essentielles},  
par: 
\begin{equation} 
E_a[i+1,b] \doteq F_i[a,b] 
\end{equation} 
Alors la composante $(i+1,b)$ de la matrice $E_a$ est le nombre de chemins essentiels de longueur $i$ 
reliant le vertex $v_a$ au vertex $v_b$. Pour le cas de graphes $ADE$, nous obtenons donc $r$ matrices  
$E_a$, de $\kappa-1$ lignes et $r$ colonnes.

\section{La big\`ebre de Hopf faible $\mathcal{B}(G)$} 
Nous consid\'erons \`a partir d'ici seulement les diagrammes $G$ de type $ADE$ (``cas $su(2)$''). 
Toutes ces constructions 
peuvent {\it a priori} \^etre g\'en\'eralis\'ees \`a d'autres diagrammes, notamment les diagrammes 
de Di Francesco-Zuber (``cas $su(3)$''), mais nous nous limiterons ici aux cas $ADE$. 
  
\subsection{Endomorphismes gradu\'es de chemins essentiels} 
Soit $G$ un graphe de type $ADE$, \`a $r$ vertex et nombre (dual) de Coxeter $\kappa$. \`A chaque vertex  
$v_a$ de $G$ est associ\'ee une dimension 
quantique $\mu_a$. Soit $H$ l'espace vectoriel des chemins essentiels sur $G$. $H$ est un espace vectoriel 
fini-dimensionnel, gradu\'e par la longueur: 
\begin{equation} 
H \;  = \; \bigoplus_{i = 0}^{\kappa-1} \; H_i 
\end{equation} 
o\`u $H_i \cong \mathbb{C}^{\,d_i}$ est l'espace vectoriel des chemins essentiels de longueur $i$, de 
dimension $d_i$. 
Soit $\{e_{a,b}^{i,\gamma}\}$ une base de $H_i$ form\'ee des chemins essentiels de longueur $i$, 
partant du vertex $v_a$ et arrivant au vertex $v_b$, de multiplicit\'e $\gamma = 1,2,\cdots, (F_i)_{ab}$. 
La dimension de $H_i$ est $d_i = \sum_{a,b} (F_i)_{a,b}$. Nous choisissons cette base orthonorm\'ee  
par rapport au produit scalaire $\langle \, . \, , \, . \, \rangle$ des chemins \'el\'ementaires: 
\begin{equation} 
\langle \,e_{a,b}^{i,\gamma} \, , \, e_{a',b'}^{i',\gamma'}\, \rangle = 
\delta_{\,a\, a'} \; \delta_{\,b\, b'} \; \delta_{\,i\, i'} \;  \delta_{\,\gamma \, \gamma'} 
\end{equation} 
Pour clarifier les notations, nous omettrons l'indice de multiplicit\'e $\gamma$, et nous noterons 
les \'el\'ements de cette base $\{ |a \stackrel{i}{\longrightarrow} b\rangle \}$, appel\'es chemins 
essentiels normalis\'es.   
Consid\'erons l'espace dual $\widehat{H}$ de $H$. Une base orthonorm\'ee de $\widehat{H}$ est form\'ee 
par les \'el\'ements $\{ \langle d \stackrel{i}{\longleftarrow} c| \}$, tels que: 
\begin{equation} 
\langle \, d \stackrel{\,i}{\longleftarrow} c \,|\, a \stackrel{j\,}{\longrightarrow} b \,\rangle  =  
\; \delta_{\,a\,c} \; \delta_{\,b\,d} \; \delta_{\,i\,j} 
\end{equation}   
 
Soit $\xi$ un \'el\'ement de la base normalis\'ee des chemins essentiels, nous pouvons l'illustrer 
de deux mani\`eres diff\'erentes, avec des triangles bi-color\'es, ou des vertex ``habill\'es'': 
\begin{equation} 
\unitlength 0.04cm 
\xi \qquad = \qquad  |\, a \stackrel{i\,}{\longrightarrow} b \,\rangle \qquad = \qquad 
\parbox{40pt}{\begin{picture}(40,30) 
\put(0,10){\circle*{6}} 
\put(40,10){\circle*{6}} 
\put(20,30){\circle{6}} 
\put(3,10){\line(1,0){34}} 
\put(2,12){\line(1,1){15.8}} 
\put(38,12){\line(-1,1){15.8}} 
\put(21,10){\vector(1,0){0}} 
\put(8,18){\vector(-1,-1){0}} 
\put(32,18){\vector(1,-1){0}} 
\put(20,3){\makebox(0,0){$i$}} 
\put(4,22){\makebox(0,0){$a$}} 
\put(36,23){\makebox(0,0){$b$}} 
\put(20,18){\makebox(0,0){$\xi$}} 
    \end{picture}} 
\qquad 
= \quad 
\parbox{40pt}{\begin{picture}(40,55) 
\put(20,30){\circle*{4}} 
\put(0,50){\line(1,-1){20}} 
\put(40,50){\line(-1,-1){20}} 
\dottedline[$\circ$]{4}(20,30)(20,0) 
\put(8,52){\makebox(0,0){$a$}} 
\put(32,52){\makebox(0,0){$b$}} 
\put(30,30){\makebox(0,0){$\xi$}} 
\put(27,5){\makebox(0,0){$i$}} 
\end{picture}} 
\qquad 
\end{equation} 
 
Consid\'erons l'espace gradu\'e $\mathcal{B} = \oplus_i End(H_i)$ des endomorphismes de l'espace vectoriel 
(gradu\'e) $H$.    
Appelant $\widehat{H}_i$ l'espace vectoriel dual de $H_i$, alors $End(H_i) = H_i \otimes \widehat{H}_i$. 
Les \'el\'ements de la base orthonorm\'ee de $\mathcal{B} = \mathcal{B}(G) = \oplus_i H_i \otimes \widehat{H}_i$ sont 
illustr\'es de la mani\`ere suivante: 
\unitlength 0.04cm 
\begin{equation} 
e_{\xi \eta}(i) \quad = \quad  |\, a \stackrel{i\,}{\longrightarrow} b \,\rangle \;  
\langle \, d \stackrel{\,i}{\longleftarrow} c \,| \quad = \quad 
\parbox{40pt}{\begin{picture}(40,40) 
\put(0,20){\circle*{6}} 
\put(40,20){\circle*{6}} 
\put(20,0){\circle{6}} 
\put(20,40){\circle{6}} 
\put(3,20){\line(1,0){34}} 
\put(2,22){\line(1,1){15.8}} 
\put(38,22){\line(-1,1){15.8}} 
\put(21,20){\vector(1,0){0}} 
\put(8,28){\vector(-1,-1){0}} 
\put(32,28){\vector(1,-1){0}} 
\put(2,18){\line(1,-1){15.8}} 
\put(38,18){\line(-1,-1){15.8}} 
\put(9,11){\vector(-1,1){0}} 
\put(31,11){\vector(1,1){0}} 
\put(20,25){\makebox(0,0){$i$}} 
\put(4,32){\makebox(0,0){$a$}} 
\put(36,33){\makebox(0,0){$b$}} 
\put(4,8){\makebox(0,0){$c$}} 
\put(36,10){\makebox(0,0){$d$}} 
\end{picture}} 
\quad = \quad 
\parbox{40pt}{\begin{picture}(40,80) 
\put(20,20){\circle*{4}} 
\put(20,60){\circle*{4}} 
\put(0,0){\line(1,1){20}} 
\put(40,0){\line(-1,1){20}} 
\put(0,80){\line(1,-1){20}} 
\put(40,80){\line(-1,-1){20}} 
\dottedline[$\circ$]{4}(20,20)(20,60) 
\put(10,79){\makebox(0,0){$a$}} 
\put(30,80){\makebox(0,0){$b$}} 
\put(10,0){\makebox(0,0){$c$}} 
\put(30,1){\makebox(0,0){$d$}} 
\put(30,60){\makebox(0,0){$\xi$}} 
\put(30,20){\makebox(0,0){$\eta$}} 
\put(12,40){\makebox(0,0){$i$}} 
\end{picture}} 
\quad 
\label{baseB} 
\end{equation} 
C'est sur cet espace $\mathcal{B}$ (ou son dual $\widehat{\mathcal{B}}$) que seront d\'efinies diff\'erentes 
structures 
faisant de $\mathcal{B}$ une alg\`ebre de Hopf faible (WHA). Pour une d\'efinition g\'en\'erale des 
propri\'et\'es 
d'une WHA, voir l'Annexe {\bf C}. Nous donnons d'abord une construction g\'en\'erale abstraite de ses 
structures dans la section suivante.

\subsection{Big\`ebre faible $\mathcal{B}(G)$: construction abstraite} 
 
\noindent $\bullet$ $\mathcal{B}$ est un espace vectoriel, $\widehat{B}$ son dual.\\ 
 
\noindent $\bullet$ $\mathcal{B}$ poss\`ede un produit associatif not\'e $\circ$, faisant de 
$(\mathcal{B},\circ)$ 
une alg\`ebre. De mani\`ere duale, $\widehat{\mathcal{B}}$ poss\`ede un coproduit $\widehat{\Delta}$ 
coassociatif: $(\widehat{\mathcal{B}},\widehat{\Delta})$ est une cog\`ebre.\\

\noindent $\bullet$ $\mathcal{B}$ poss\`ede un coproduit $\Delta$ coassociatif, faisant de 
$(\mathcal{B},\Delta)$ 
une cog\`ebre. De mani\`ere duale, $\widehat{\mathcal{B}}$ poss\`ede un produit associatif 
not\'e $\widehat{\odot}$: $(\widehat{\mathcal{B}},\widehat{\odot})$ est une alg\`ebre.\\ 
 
\noindent $\bullet$ Les deux structures sont compatibles, c'est-\`a-dire entre autre que $\Delta$
est un homomorphisme $(\mathcal{B},\circ) \mapsto (\mathcal{B} \otimes \mathcal{B},\circ)$:
\begin{equation}
\Delta (u \circ v) = \Delta(u) \circ \Delta(v), \qquad  \qquad \qquad u,v \in \mathcal{B}.
\end{equation}
Mais les conditions de compatibilit\'e sont prises dans le sens ``faible'', (notamment  
$\Delta(\munite) \not= \munite \otimes \munite$ mais\footnote{Nous utilisons la convention de Sweedler: 
une sommation est implicite.} 
$\Delta(\munite) = \munite_1 \otimes \munite_2$): 
$\mathcal{B}$ est appel\'ee techniquement une big\`ebre de Hopf faible.
Il existe aussi une antipode $S$ dans $\mathcal{B}$ (et $\widehat{S}$ dans $\widehat{\mathcal{B}}$), faisant
de $\mathcal{B}$ (et $\widehat{\mathcal{B}}$) une alg\`ebre de Hopf faible. Tous les axiomes d'une alg\`ebre de Hopf usuelle reli\'es \`a l'unit\'e, \` a la counit\'e et \`a l'antipode sont alors modifi\'es car ils 
contiennent les \'el\'ements $\munite_1$ et $\munite_2$. Les d\'efinitions des structures et propri\'et\'es d'une 
alg\`ebre de Hopf faible sont pr\'esent\'ees dans l'Annexe {\bf C}.  \\ 
 
\noindent $\bullet$  L'alg\`ebre $(\mathcal{B},\circ)$ est semi-simple. Elle peut donc \^etre 
diagonalis\'ee et est isomorphe \`a une somme de blocs matriciels (labell\'es par un index $i$) agissant sur un espace vectoriel 
gradu\'e 
$H = \oplus_i H_i$. Les matrices \'el\'ementaires (unit\'es matricielles) de chaque bloc $i$ sont not\'ees 
$e_{\xi\xi'}(i)$, 
o\`u $\xi,\xi'$ appartiennent \`a l'espace $H_i$. 
Elles sont repr\'esent\'es diagrammatiquement dans (\ref{baseB}). Le produit $\circ$ des \'el\'ements 
de la base de $\mathcal{B}$ correspond alors au produit des unit\'es matricielles:  
$$ 
e_{\xi \xi'}(i) \circ e_{\eta \eta'}(j) = \delta_{\,ij} \; \delta_{\xi'\eta}\; e_{\xi \eta'}(i)  
$$ 
Ce produit est appel\'e {\bf composition} d'endomorphismes, ou aussi {\bf produit vertical}, et est 
repr\'esent\'e  
dans le diagramme suivant: 
$$ 
\parbox{0pt}{\begin{picture}(0,90) \end{picture}} 
\smalldif{a}{b}{c}{d}{i}{}{}{}{} \quad \circ \quad \smalldif{c'}{d'}{e}{f}{j}{}{}{}{} \quad =  
\quad \delta_{c,c'} \, \delta_{d,d'} \, \delta_{i,j} \quad \smalldif{a}{b}{e}{f}{i}{}{}{}{} 
$$ 
 
\noindent $\bullet$ L'alg\`ebre $(\widehat{\mathcal{B}},\widehat{\odot})$ est semi-simple. Elle peut donc \^etre 
diagonalis\'ee et est isomorphe \`a une somme de blocs matriciels (labell\'es par un index $x$) 
agissant sur un espace vectoriel gradu\'e 
$V = \oplus_x V_x$. Les matrices \'el\'ementaires (unit\'es matricielles) de chaque bloc $x$ sont not\'ees 
$E^{\alpha\alpha'}(x)$, o\`u  
$\alpha,\alpha'$ appartiennent \`a l'espace $V_x$, 
et sont repr\'esent\'es diagrammatiquement par: 
\begin{equation} 
\parbox{0pt}{\begin{picture}(0,60) \end{picture}} 
E^{\alpha \beta}(x) \quad = \quad   
\parbox{40pt}{\begin{picture}(40,40) 
\put(0,20){\circle*{6}} 
\put(40,20){\circle*{6}} 
\put(20,0){\circle{6}} 
\put(20,40){\circle{6}} 
\put(20,3){\line(0,1){34}} 
\put(2,22){\line(1,1){15.8}} 
\put(38,22){\line(-1,1){15.8}} 
\put(20,17.5){\vector(0,-1){0}} 
\put(8,28){\vector(-1,-1){0}} 
\put(32,28){\vector(1,-1){0}} 
\put(2,18){\line(1,-1){15.8}} 
\put(38,18){\line(-1,-1){15.8}} 
\put(9,11){\vector(-1,1){0}} 
\put(31,11){\vector(1,1){0}} 
\put(25,20){\makebox(0,0){$x$}} 
\put(4,32){\makebox(0,0){$a$}} 
\put(36,33){\makebox(0,0){$b$}} 
\put(4,8){\makebox(0,0){$c$}} 
\put(36,10){\makebox(0,0){$d$}} 
\end{picture}} 
\quad = \quad 
\parbox{80pt}{\begin{picture}(80,40) 
\put(0,0){\line(1,1){20}} 
\put(0,40){\line(1,-1){20}} 
\put(80,40){\line(-1,-1){20}} 
\put(80,0){\line(-1,1){20}} 
\dottedline[$\bullet$]{4}(20,20)(60,20) 
\put(0,8){\makebox(0,0){$c$}} 
\put(0,33){\makebox(0,0){$a$}} 
\put(80,8){\makebox(0,0){$b$}} 
\put(80,33){\makebox(0,0){$d$}} 
\put(22,14){\makebox(0,0){$\alpha$}} 
\put(56,14){\makebox(0,0){$\beta$}} 
\put(40,27){\makebox(0,0){$x$}} 
\end{picture}} 
\quad 
\end{equation} 
Le produit $\widehat{\odot}$ des \'el\'ements de la base de $\widehat{\mathcal{B}}$ est donn\'e par: 
$$ 
E^{\alpha \beta}(x) \, \widehat{\odot}\, E^{\alpha' \beta'}(y) = \delta_{\,xy} \; \delta_{\,\beta \alpha'} \;
E^{\alpha\beta'}(x)  
$$ 
Ce produit est appel\'e convolution d'endomorphismes, ou aussi {\bf produit horizontal}, et est 
repr\'esent\'e dans le diagramme suivant: 
$$ 
\parbox{0pt}{\begin{picture}(0,50) \end{picture}} 
\smalldifdual{a\;}{b}{c\;}{d}{x}{}{}{}{} \quad \widehat{\odot} \quad \smalldifdual{b'\;}{e}{d'\;}{f}{y}{}{}{}{} 
\quad = \quad \delta_{\,b\,b'} \, \delta_{\,d\,d'} \, \delta_{\,x\,y} \quad \smalldifdual{a\;}{e}{c\;}{f}{x}{}{}{}{}  
$$ 
\noindent $\bullet$ Le coproduit $\Delta$ dans $\mathcal{B}$ est cod\'e par un ensemble de coefficients 
$F_1$: 
\begin{equation} 
\parbox{0pt}{\begin{picture}(0,90) \end{picture}} 
\Delta \quad \smalldif{a}{b}{c}{d}{i} \quad = \quad \sum_{j,k \,; \, e,f} \; F_1\{ {}^{aeb}_{kij}\}  \;  
\overline{F_1\{{}^{cfd}_{kij}\}} 
\quad \smalldif{a}{e}{c}{f}{j} \quad 
\otimes \quad \smalldif{e}{b}{f}{d}{k} \qquad 
\end{equation} 
 
\noindent $\bullet$ Le coproduit $\widehat{\Delta}$ dans $\widehat{\mathcal{B}}$ est cod\'e par un ensemble 
de coefficients $F_3$: 
\begin{equation} 
\parbox{0pt}{\begin{picture}(0,50) \end{picture}} 
 \widehat{\Delta} \quad \smalldifdual{a}{b}{c}{d}{x} \quad = \quad \sum_{y,z;e,f} \; 
F_3\{ {}^{aeb}_{zxy}\}  \;   
\overline{F_3\{{}^{cfd}_{zxy}\}} \quad \smalldifdual{a}{e}{c}{f}{y} 
\quad \otimes \quad \smalldifdual{e}{b}{f}{d}{z} \qquad  
\end{equation} 
 
\noindent $\bullet$ Il existe une forme bilin\'eaire ({\it pairing}) 
$\langle \, , \, \rangle : \widehat{\mathcal{B}} \times 
\mathcal{B} \longrightarrow \mathbb{C}$ de compatibilit\'e entre les deux structures duales, 
appel\'ee {\bf cellules d'Ocneanu}, telle que: 
\begin{equation} 
\langle \quad \smalldifdual{a\;}{b}{c\;}{d}{x} \quad , \quad \smalldif{a'}{b'}{c'}{d'\,}{i} \quad \rangle 
= \delta_{aa'}\, \delta_{bb'}\, \delta_{cc'}\, \delta_{dd'}\; F_2 \{ {}^{abx}_{dci}\} 
\end{equation} 
 
L'ensemble de ces coefficients doit bien entendu satisfaire des conditions pour faire de $\mathcal{B}$ 
(et de son dual $\widehat{\mathcal{B}}$) une big\`ebre de Hopf faible, d\'ecrites par un ensemble 
d'\'equations connu sous le nom de ``{\it The Big Pentagon Equation}''.

\subsection{\'Equations pentagonales} 
Utilisant la diagrammation des \'el\'ements de la base de $\mathcal{B}$  et de $\widehat{\mathcal{B}}$ 
en termes de triangles duaux, nous pouvons illustrer les coefficients $F_1$, $F_2$ et $F_3$ comme 
des t\'etrah\`edres comme suit: 
$$ 
F_1\{ {}^{abc}_{ijk}\} = \; \Fun{a}{k}{c}{i}{b}{j} \qquad \qquad \qquad  
F_2\{ {}^{abx}_{dci}\} = \; \Fdeux{a}{c}{b}{d}{x}{i} \qquad \qquad \quad  
F_3\{ {}^{aeb}_{zxy}\} = \; \Ftrois{a}{x}{e}{z}{b}{y}  
$$ 
La condition de coassociativit\'e du coproduit $\Delta$ est v\'erifi\'ee s'il existe une fonction $F_0$,  
illustr\'ee par un t\'etrah\`edre sans vertex \begin{picture}(6,6) \put(3,3){\circle{6}} \end{picture} 
telle que l'\'equation de type pentagonale suivante soit satisfaite: 
\begin{equation} 
\hspace{4cm} F_0 \; F_1 \; F_1 \; = \; F_1 \; F_1 \hspace{6cm} (P_1) \nonumber 
\end{equation} 
Pour que $F_0$ satisfasse ($P_1$) il est aussi n\'ecessaire que $F_0$ satisfasse une \'equation pentagonale 
du type $F_0\,F_0\,F_0 = F_0\,F_0$, not\'ee $(P_0)$. De mani\`ere duale, l'associativit\'e du coproduit 
dans $\widehat{B}$ impose l'existence d'une fonction $F_4$ illustr\'ee par un t\'etrah\`edre sans vertex  
\begin{picture}(6,6) \put(3,3){\circle*{6}} \end{picture}, 
telle qu'elle satisfasse l'\'equation pentagonale du type $F_4\,F_3\,F_3 = F_3\,F_3$, not\'ee $(P_4)$;  
de m\^eme $F_4$ doit satisfaire une \'equation du type $F_4\,F_4\,F_4 = F_4\,F_4$, not\'ee $(P_5)$.  
Les deux fontions $F_0$ et $F_4$ sont illustr\'ees par: 
$$ 
F_0\{ {}^{ijk}_{lmn}\} = \; \Fzero{i}{n}{k}{l}{j}{m\,} \qquad \qquad \qquad \qquad  
F_4\{ {}^{xyz}_{tuv}\} = \; \Fquatre{x}{v}{z}{t}{y}{u} 
$$ 
Enfin, la condition pour que le {\it pairing} transpose le coproduit de $\mathcal{B}$ en le produit de 
$\widehat{\mathcal{B}}$ 
est aussi une \'equation pentagonale de type $F_2\,F_1\,F_1 = F_2\,F_2$ $(P_2)$. De mani\`ere duale, 
la transposition du produit de $\mathcal{B}$ en le produit de $\widehat{\mathcal{B}}$ implique  
$F_2\,F_3\,F_3 = F_2\,F_2$ $(P_3)$. 
 
\begin{remarq}
Nous choisissons une fois pour toute la nature des indices $\{i,j,k,\cdots\}$; 
$\{a,b,c,\cdots\}$; $\{x,y,z,\cdots\}$.
Ils correspondent respectivement \`a trois types d`arcs (orient\'es):
$$\begin{picture}(40,5)
\thicklines
\put(3,0){\line(1,0){34}}
\put(23,0){\vector(1,0){0}}
\put(0,0){\circle*{6}} 
\put(40,0){\circle*{6}}
\put(20,8){\makebox(0,0){$i$}} 
\end{picture}
\qquad \qquad \qquad
\begin{picture}(40,5)
\thicklines
\put(3,0){\line(1,0){34}}
\put(23,0){\vector(1,0){0}}
\put(0,0){\circle{6}} 
\put(40,0){\circle*{6}} 
\put(20,8){\makebox(0,0){$a$}} 
\end{picture}
\qquad \qquad \qquad
\begin{picture}(40,5)
\thicklines
\put(3,0){\line(1,0){34}}
\put(23,0){\vector(1,0){0}}
\put(0,0){\circle{6}} 
\put(40,0){\circle{6}} 
\put(20,8){\makebox(0,0){$x$}} 
\end{picture}
$$
\noindent Insistons sur le fait qu'il n'y a 
pas d'arc reliant \begin{picture}(6,6) \put(3,3){\circle*{6}} \end{picture} \`a \begin{picture}(6,6) 
\put(3,3){\circle{6}} \end{picture}. Alors la notation $\{{}^{...}_{...}\}$ \`a elle seule suffit sans 
qu'il 
soit n\'ecessaire de pr\'eciser s'il s'agit de $F_0,F_1,F_2,F_3$ ou $F_4$.
\end{remarq}
Les diff\'erents coefficients $F$ apparaisent comme des g\'en\'eralisations des symboles $3j$ et $6j$ 
quantiques.
La connaissance d'un ensemble $F = (F_0,F_1,F_2,F_3,F_4)$ v\'erifiant ``{\it The Big Pentagon Equation}''  
$(P) = (P_0,P_1,P_2,P_3,P_4,P_5)$ permet alors de contruire toutes les structures d'une big\`ebre de 
Hopf faible \cite{Bohm}. Toutefois, la r\'esolution explicite de toutes les \'equations et la 
d\'etermination des diff\'erents coefficients pose des probl\`emes techniques extr\^emement complexes. 
 
Le prototype d'une telle construction est donn\'ee par l'espace vectoriel des endomorphismes gradu\'es  
des chemins essentiels sur un graphe $\mathcal{B}=\mathcal{B}(G) = \oplus_i H_i \otimes \widehat{H}_i$. 
Nous verrons que les diff\'erentes ``fonctions t\'etrah\'edriques'' sont reli\'ees aux divers coefficients  
d\'efinissant les fonctions de partition de syst\`emes conformes \`a $2d$ dans diff\'erents environnements.
Pour \'etablir un lien avec les notations introduites dans le chapitre {\bf 1}, signalons simplement pour
l'instant qu'il existe des relations entre les objets suivants: 
\unitlength 0.035cm 
$$ 
\begin{array}{ccc} 
{ }  &  { }  &  { } \\ 
F_0 = \quad \parbox{1.4cm}{\begin{picture}(40,40) 
\put(0,20){\circle*{6}} 
\put(40,20){\circle*{6}} 
\put(20,0){\circle*{6}} 
\put(20,40){\circle*{6}} 
\put(3,20){\line(1,0){34}} 
\put(2,22){\line(1,1){15.8}} 
\put(38,22){\line(-1,1){15.8}} 
\put(2,18){\line(1,-1){15.8}} 
\put(38,18){\line(-1,-1){15.8}} 
\put(20,3){\line(0,1){15}} 
\put(20,37){\line(0,-1){15}} 
\end{picture}} \quad  \leadsto \mathcal{N}_{ij}^k &  { }  &   
F_4 = \quad \parbox{1.4cm}{\begin{picture}(40,40) 
\put(0,20){\circle{6}} 
\put(40,20){\circle{6}} 
\put(20,0){\circle{6}} 
\put(20,40){\circle{6}} 
\put(3,20){\line(1,0){34}} 
\put(2,22){\line(1,1){15.8}} 
\put(38,22){\line(-1,1){15.8}} 
\put(2,18){\line(1,-1){15.8}} 
\put(38,18){\line(-1,-1){15.8}} 
\put(20,3){\line(0,1){15}} 
\put(20,37){\line(0,-1){15}} 
\end{picture}} \quad \leadsto \mathcal{O}_{xy}^z \\ 
{ }  &  { }  &  { } \\ 
{ }  &  F_2 = \quad \parbox{1.4cm}{\begin{picture}(40,40) 
\put(0,20){\circle{6}} 
\put(40,20){\circle{6}} 
\put(20,0){\circle*{6}} 
\put(20,40){\circle*{6}} 
\put(3,20){\line(1,0){34}} 
\put(2,22){\line(1,1){15.8}} 
\put(38,22){\line(-1,1){15.8}} 
\put(2,18){\line(1,-1){15.8}} 
\put(38,18){\line(-1,-1){15.8}} 
\put(20,3){\line(0,1){15}} 
\put(20,37){\line(0,-1){15}} 
\end{picture}} \quad  &  { } \\ 
{ }  &  { }  &  { } \\ 
F_1 = \quad \parbox{1.4cm}{\begin{picture}(40,40) 
\put(0,20){\circle{6}} 
\put(40,20){\circle*{6}} 
\put(20,0){\circle*{6}} 
\put(20,40){\circle*{6}} 
\put(3,20){\line(1,0){34}} 
\put(2,22){\line(1,1){15.8}} 
\put(38,22){\line(-1,1){15.8}} 
\put(2,18){\line(1,-1){15.8}} 
\put(38,18){\line(-1,-1){15.8}} 
\put(20,3){\line(0,1){15}} 
\put(20,37){\line(0,-1){15}} 
\end{picture}} \quad \leadsto \mathcal{F}_{ab}^i  &  { }  &  
F_3 = \quad \parbox{1.4cm}{\begin{picture}(40,40) 
\put(0,20){\circle*{6}} 
\put(40,20){\circle{6}} 
\put(20,0){\circle{6}} 
\put(20,40){\circle{6}} 
\put(3,20){\line(1,0){34}} 
\put(2,22){\line(1,1){15.8}} 
\put(38,22){\line(-1,1){15.8}} 
\put(2,18){\line(1,-1){15.8}} 
\put(38,18){\line(-1,-1){15.8}} 
\put(20,3){\line(0,1){15}} 
\put(20,37){\line(0,-1){15}} 
\end{picture}} \quad \leadsto \mathcal{S}_{ab}^x \\ 
{ }  &  { }  &  { }  
\end{array} 
$$ 

Notre humble but ici n'est pas de pr\'esenter toute la th\'eorie des alg\`ebres de Hopf faibles, ni de
calculer explicitement tous les coefficients de structure dans le cas des big\`ebres associ\'ees \`a des
diagrammes de Dynkin. Nous montrerons plut\^ot comment, \`a partir
des structures d'une big\`ebre $\mathcal{B}(G)$, sont d\'efinis deux types de graphes, les graphes
$\mathcal{A}(G)$ et les graphes d'Ocneanu de $G$, et comment la simple donn\'ee de ces deux graphes 
nous permet d'obtenir tous les coefficients permettant de d\'eterminer les fonctions de partition 
(g\'en\'eralis\'ees) des mod\`eles conformes \`a deux dimensions.

\subsection{Projecteurs minimaux centraux $\pi_i$ et $\varpi_x$ et graphes $\mathcal{A}(G)$ et $Oc(G)$} 
Soit $\mathcal{B}$ l'espace vectoriel des endomorphismes de chemins essentiels sur le graphe $G$,
et $\widehat{\mathcal{B}}$ son dual. $(\mathcal{B},\circ)$ et $(\widehat{\mathcal{B}},\widehat{\odot})$ sont des
alg\`ebres. Utilisant le 
{\it pairing} $\langle \widehat{\mathcal{B}},\mathcal{B} \rangle \rightarrow \mathbb{C}$, le produit $\widehat{\odot}$ 
induit un coproduit
$\Delta$ dans $\mathcal{B}$: $(\mathcal{B},\circ,\Delta)$ est une big\`ebre (faible). Mais nous pouvons aussi
utiliser un produit scalaire dans $\mathcal{B}$ nous permettant
de passer de $\widehat{\mathcal{B}}$ \`a $\mathcal{B}$, et d\'efinir alors le produit $\widehat{\odot}$ sur l'espace
vectoriel $\mathcal{B}$ lui-m\^eme: nous notons alors ce produit par $\odot$. Sur $\mathcal{B}$ sont donc d\'efinies deux structures multiplicatives 
$\circ$ et $\odot$: nous obtenons une {\bf dig\`ebre}. Notons que les deux lois doivent bien s\^ur satisfaire
des propri\'et\'es de compatibilit\'e. Nous consid\'erons dor\'enavant la dig\`ebre 
$(\mathcal{B},\circ,\odot)$. \\
 
\noindent $\bullet$ $(\mathcal{B},\circ)$ est une alg\`ebre semi-simple et est donc isomorphe 
\`a une somme de blocs matriciels labell\'es par $i$. Appelons $\pi_i$ les projecteurs centraux minimaux de 
$(\mathcal{B},\circ)$. Ils v\'erifient par d\'efinition:
\begin{equation} 
\pi_i \circ \pi_j = \delta_{\, i\,j}\; \pi_j \, ,
\end{equation} 
et sont repr\'esent\'es par des matrices unit\'es dans chaque bloc.  
Ils engendrent un espace vectoriel not\'e $\mathcal{A}(G)$, de dimension $\kappa-1$. Multipliant les
projecteurs $\pi_i$ par le  produit $\odot$, nous obtenons:
\begin{equation} 
\pi_i \odot \pi_j  = \sum_k  \; \mathcal{N}_{ij}^{k} \; \pi_k \; . 
\end{equation} 
Les projecteurs $\pi_i$ sont  ferm\'es pour la loi $\odot$.
Lorsque $G$ est un graphe de type $ADE$, $\mathcal{N}_{ij}^k$ sont les coefficients entiers 
non-n\'egatifs de l'alg\`ebre de fusion (apparaissant dans les th\'eories conformes $\widehat{su}(2)$
\`a deux dimensions). Le produit $\odot$ des projecteurs $\pi_i$ est cod\'e par le graphe 
$\mathcal{A}(G)$ correspondant au diagramme de Dynkin de type $A_n$ ayant la m\^eme norme que $G$.
Donc $\mathcal{A}(G) = A_{\kappa-1}$.\\
 
\noindent $\bullet$ $(\mathcal{B},\odot)$ est aussi une alg\`ebre semi-simple, et est donc  
isomorphe \`a une somme de blocs matriciels labell\'es par $x$. Appelons $\varpi(x)$ les projecteurs  
centraux minimaux de $(\mathcal{B},\odot)$. Ils v\'erifient par d\'efinition:
\begin{equation} 
\varpi_x \odot \varpi_y = \delta_{\, x\,y}\; \varpi_y \; .
\end{equation}
Ils engendrent un espace vectoriel not\'e $Oc(G)$, de dimension $s$, et sont
repr\'esent\'es par des matrices unit\'es dans chaque bloc (sommes d'unit\'es matricielles diagonales).
Les unit\'es matricielles pour les lois $\circ$ et $\odot$ ne sont pas normalis\'ees pour le m\^eme
produit scalaire. Ceci conduit \`a la d\'efinition de nouveaux op\'erateurs $\rho_x$ qui s'\'ecrivent
formellement comme les $\varpi_x$ mais dont les termes constituants diff\`erent par des facteurs
de normalisation. Multipliant les op\'erateurs $\rho_x$ \`a l'aide de la loi $\circ$, nous obtenons:  
\begin{equation} 
\rho_x \circ \rho_y  = \sum_{z} \; \mathcal{O}_{xy}^{z} \; \rho_z \; .  
\end{equation} 
Les op\'erateurs $\rho_x$ (moralement les projecteurs $\varpi_x$) sont ferm\'es pour la loi $\circ$. 
Ils engendrent une alg\`ebre, appel\'ee l'{\bf alg\`ebre des sym\'etries quantiques} du graphe $G$, 
qui est cod\'ee par le graphe d'Ocneanu de $G$, not\'e $Oc(G)$.\\
 
\begin{remarq}
Nous ne donnons ici aucune d\'emonstration, nous contentant de d\'ecrire et adapter les r\'esultats 
\'enonc\'es par Ocneanu dans \cite{Oc-paths} (mais jamais explicitement montr\'es sur des exemples 
concrets).     
\end{remarq}

La connaissance du graphe $G$ permet de calculer les coefficients $F_{ia}^b$ (nombre de chemins 
essentiels de longueur $i$ allant du vertex $v_a$ au vertex $v_b$ sur $G$). La diagonalisation explicite
(obtention des projecteurs $\pi_i$ et $\varpi_x$) des deux lois $\circ$ et $\odot$ permet d'obtenir
les coefficients $\mathcal{N}_{ij}^k$ et $\mathcal{O}_{xy}^z$, et donc les graphes 
$\mathcal{A}(G)$ et $Oc(G)$. Invers\'ement,
la donn\'ee de ces deux graphes permet d'obtenir ces coefficients. Nous verrons 
comment calculer les coefficients $\mathcal{S}_{xa}^b$ (nombre de chemins 
``verticaux'' de label $x$ allant du vertex $v_a$ au vertex $v_b$). Nous pouvons alors d\'eterminer 
toutes les fonctions de partition des mod\`eles conformes associ\'es au graphe $G$. 
 
\subsection{Cellules d'Ocneanu} 
Les syst\`emes de cellules d'Ocneanu ont \'et\'e introduits par Ocneanu dans le contexte  
de paragroupes et d'inclusion de sous-facteurs \cite{Oc-string,Oc-qsym} (voir aussi \cite{kawa}).  
Ces cellules  ont par la suite \'et\'e reli\'es \`a des mod\`eles int\'egrables d\'efinis 
sur le r\'eseau (analogie avec les poids de Boltzmann \cite{Roche-OcCell}). Nous pr\'esentons 
ici les d\'efinitions adapt\'ees au contexte de chemins (essentiels) sur des graphes.  
 
Consid\'erons un graphe de Dynkin $G$, ses vertex not\'es $v_i$ et soit $\mu_i$ la composante de 
Perron-Frobenius 
de $v_i$ (dimension quantique). Soient $\xi, \eta, \alpha, \beta$ des chemins \'el\'ementaires sur $G$, 
et notons $|\xi|$ la longueur du chemin $\xi$. Une cellule d'Ocneanu est un carr\'e dont les arr\^etes 
horizontales sont labell\'ees par des chemins $\xi,\eta$ (avec $|\xi|=|\eta|$) et les arr\^etes verticales  
par des chemins $\alpha,\beta$ (avec $|\alpha| = |\beta|$) 
tels que: $s(\xi)=s(\alpha)=v_1^{ };r(\xi)=s(\beta)=v_2^{ };r(\alpha)=s(\eta)=v_4^{ };r(\eta)=r(\beta)=
v_3^{ }$.
Par convention, nous les repr\'esentons comme suit: 
\begin{equation} 
\textrm{\bf Cellule d'Ocneanu} 
\qquad \qquad \qquad \qquad 
\parbox{1.25cm}{\unitlength 0.025cm 
\begin{picture}(50,60) 
\put(-2,1){\scriptsize $v_4^{ }$} 
\put(40,1){\scriptsize $v_3^{ }$} 
\put(-2,57){\scriptsize $v_1^{ }$} 
\put(40,57){\scriptsize $v_2^{ }$} 
\put(21,0){\scriptsize $\eta$} 
\put(21,55){\scriptsize $\xi$} 
\put(-6,28){\scriptsize $\alpha$} 
\put(48,28){\scriptsize $\beta$} 
\end{picture}} 
\qquad \qquad \qquad \qquad 
\end{equation}   
Une celulle sera dite {\bf basique} si les chemins qui la composent sont tous de longueur un: les chemins 
relient alors 
des vertex adjacents sur le graphe $G$.  
Une {\bf connexion} $x$ est une application qui assigne \`a chaque cellule d'Ocneanu un nombre complexe,
et qui doit v\'erifier certaines conditions rappel\'ees plus loin. Ce nombre complexe co\"{\i}ncide 
avec la fonction pentagonale $F_2$ lorsque $\xi,\eta$ sont des chemins essentiels (de longueur $i$) et 
$\alpha, \beta$ des chemins verticaux (de label $x$), et sera not\'e:
\begin{equation}
F_2 \{{}^{v_1v_2x}_{v_4v_3i}\} \quad = \quad
\parbox{1.25cm}{\unitlength 0.025cm 
\begin{picture}(50,30)(0,15) 
\put(5,10){\line(1,0){40}} 
\put(5,50){\line(1,0){40}} 
\put(5,10){\line(0,1){40}} 
\put(45,10){\line(0,1){40}} 
\put(28.5,10){\vector(1,0){0}} 
\put(28.5,50){\vector(1,0){0}} 
\put(5,26){\vector(0,-1){0}} 
\put(45,26){\vector(0,-1){0}} 
\put(-2,1){\scriptsize $v_4^{ }$} 
\put(40,1){\scriptsize $v_3^{ }$} 
\put(-2,57){\scriptsize $v_1^{ }$} 
\put(40,57){\scriptsize $v_2^{ }$} 
\put(21,0){\scriptsize $\eta$} 
\put(21,55){\scriptsize $\xi$} 
\put(-6,28){\scriptsize $\alpha$} 
\put(48,28){\scriptsize $\beta$} 
\put(25,30){\scriptsize \makebox(0,0){${\bf x}$}} 
\end{picture}} 
\end{equation}

Les conventions 
adopt\'ees dans la litt\'erature diff\`erent selon les auteurs: signalons deux types de convention, 
appel\'ees 
connexions U (U pour unitaires) et connexions S (S pour standard). Elles sont reli\'ees entre elles par: 
\begin{equation} 
\cell{v_1^{ }}{v_2^{ }}{v_3^{ }}{v_4^{ }}{}{}{}{}{U} 
\quad = \; \left ( \frac{\mu_2 \mu_4}{\mu_1 \mu_3}\right)^{\frac{1}{4}} \quad 
\cell{v_1^{ }}{v_2^{ }}{v_3^{ }}{v_4^{ }}{}{}{}{}{S} 
\end{equation} 
Pour faciliter la lecture, nous n'\'ecrivons pas explicitement les indices labellant les chemins. 
Nous choississons d'utiliser les connexions U.  Elles doivent satisfaire les conditions 
suivantes \cite{Oc-string, Coq-private}: 
$$ 
\begin{array}{rclccr} 
\cell{v_1^{ }}{v_2^{ }}{v_3^{ }}{v_4^{ }}{}{}{}{}{} \quad &=& \displaystyle \sqrt{\frac{\mu_2 \mu_4}{\mu_1 
\mu_3}} \quad 
\cellbar{v_4^{ }}{v_3^{ }}{v_2^{ }}{v_1^{ }}{}{}{}{}{} \quad &\quad& \textrm{r\'eflexion horizontale} & 
\qquad  (A_1) \\ 
{} & {} & {} & {} & {} & {} \\ 
\cell{v_1^{ }}{v_2^{ }}{v_3^{ }}{v_4^{ }}{}{}{}{}{} \quad &=& \displaystyle \sqrt{\frac{\mu_2 \mu_4}{\mu_1 
\mu_3}} \quad 
\cellbar{v_2^{ }}{v_1^{ }}{v_3^{ }}{v_4^{ }}{}{}{}{}{} \quad &\quad& \textrm{r\'eflexion verticale} & 
\qquad  (A_2) \\ 
{} & {} & {} & {} & {} & {} \\ 
\displaystyle \sum_{v_2^{ }} \quad \cell{v_1^{ }}{v_2^{ }}{v_3^{ }}{v_4^{ }}{}{}{}{}{} \quad  
\cellbar{v_1^{ }}{v_2^{ }}{v_3^{ }}{v_4'}{}{}{}{}{} \quad &=& \delta_{\, v_4,v_4'} &\quad& \textrm{unitarit\'e} & 
\qquad  (A_3) \\ 
\end{array} 
$$ 
o\`u $\overline{\square}$ est le complexe conjugu\'e de $\square$.
Les conditions pr\'ec\'edentes impliquent aussi: 
$$ 
\begin{array}{rclccr} 
 \cell{v_1^{ }}{v_2^{ }}{v_3^{ }}{v_4^{ }}{}{}{}{}{} \quad &=& \quad  
 \cell{v_3^{ }}{v_4^{ }}{v_1^{ }}{v_2^{ }}{}{}{}{}{} \quad &\quad& \textrm{r\'eflexion diagonale} & 
\qquad  (A_4) \\ 
{} & {} & {} & {} & {} & {} \\ 
\displaystyle \sum_{v_4^{ }} \quad \cell{v_1^{ }}{v_2^{ }}{v_3^{ }}{v_4^{ }}{}{}{}{}{} \quad \cellbar{v_1^{ }}{v_2'}{v_3^{ }}{v_4^{ }}{}{}{}{}{} \quad 
&=& \;\; \delta_{v_2^{ },v_2'} \quad &\quad& \textrm{unitarit\'e} & 
\qquad  (A_5) \\ 
{} & {} & {} & {} & {} & {} \\ 
\displaystyle \sum_{v_3^{ }} \quad \cell{v_1^{ }}{v_2^{ }}{v_3^{ }}{v_4^{ }}{}{}{}{}{} \quad \cellbar{v_1'}{v_2^{ }}{v_3^{ }}{v_4^{ }}{}{}{}{}{} \quad 
&=& \; \displaystyle \sqrt{\frac{\mu_2\mu_4}{\mu_1 \mu_3}}\;\; \delta_{v_1^{ },v_1'} \quad &\quad&  
\textrm{unitarit\'e} & \qquad  (A_6) \\ 
{} & {} & {} & {} & {} & {} \\ 
\displaystyle \sum_{v_1^{ }} \quad \cell{v_1^{ }}{v_2^{ }}{v_3^{ }}{v_4^{ }}{}{}{}{}{} \quad \cellbar{v_1^{ }}{v_2^{ }}{v_3'}{v_4^{ }}{}{}{}{}{} \quad 
&=& \; \displaystyle \sqrt{\frac{\mu_2\mu_4}{\mu_1 \mu_3}}\;\; \delta_{v_3^{ },v_3'} \quad &\quad& \textrm{unitarit\'e} & \qquad  (A_7) \\ 
{} & {} & {} & {} & {} & {} \\ 
\end{array} 
$$

Pour un syst\`eme de cellules, les conditions d'unitarit\'e et de r\'eflexions fixent les valeurs 
possibles des connexions, \`a une libert\'e de choix de jauge pr\`es. Si $X(v_1,v_2,v_3,v_4)$ est la 
valeur d'une cellule -- pour une connexion $x$ donn\'ee --
satisfaisant les propri\'et\'es ($A_1,A_2,A_3$), alors 
\begin{equation}
X'(v_1,v_2,v_3,v_4) = U(v_1,v_4)^* U(v_4,v_3)^* X(v_1,v_2,v_3,v_4) U(v_1,v_2) U(v_2,v_3),
\end{equation} 
avec $U(a,b)^*=U(b,a)$, satisfait aussi ces propri\'et\'es \cite{Roche-OcCell}. \`A un choix de jauge pr\`es,
pour un syst\`eme de cellules basiques, il existe deux solutions complexes conjugu\'ees l'une de l'autre. 
Une formule g\'en\'erale pour ces cellules basiques est donn\'ee dans \cite{Oc-qsym} (obtenue aussi par 
\cite{Roberto-cell}).  
 
La valeur d'une connexion pour des cellules g\'en\'erales (appel\'ees {\bf macro-cellules} dans 
\cite{sunder}) s'obtient \`a partir des valeurs des cellules basiques. Il est plus instructif 
de pr\'esenter un exemple. Consid\'erons une cellule o\`u les chemins horizontaux sont de longueur 3 et 
les chemins verticaux de longueur 2. La valeur de la connexion de cette cellule est donn\'ee par la somme
sur toutes les configurations permises pour les vertex int\'erieurs 
de la valeur de la connexion de la cellule ``remplie'', o\`u la valeur 
d'une connexion pour une cellule ``remplie'' est donn\'ee par le produit de toutes les cellules basiques 
qui la composent. 

$$ 
\unitlength 0.025cm 
\parbox{3.1cm}{\begin{picture}(125,100) 
\put(5,10){\line(1,0){120}} 
\put(5,90){\line(1,0){120}} 
\put(5,10){\line(0,1){80}} 
\put(125,10){\line(0,1){80}} 
\put(28.5,10){\vector(1,0){0}} 
\put(28.5,90){\vector(1,0){0}} 
\put(68.5,10){\vector(1,0){0}} 
\put(68.5,90){\vector(1,0){0}} 
\put(108.5,10){\vector(1,0){0}} 
\put(108.5,90){\vector(1,0){0}} 
\put(5,26){\vector(0,-1){0}} 
\put(125,26){\vector(0,-1){0}} 
\put(5,66){\vector(0,-1){0}} 
\put(125,66){\vector(0,-1){0}} 
 
\put(5,10){\circle*{3}} 
\put(45,10){\circle*{3}} 
\put(85,10){\circle*{3}} 
\put(125,10){\circle*{3}} 
\put(5,90){\circle*{3}} 
\put(45,90){\circle*{3}} 
\put(85,90){\circle*{3}} 
\put(125,90){\circle*{3}} 
\put(5,50){\circle*{3}} 
\put(125,50){\circle*{3}} 
 
\put(0,2){\scriptsize $u_1^{ }$} 
\put(40,2){\scriptsize $u_2^{ }$} 
\put(80,2){\scriptsize $u_3^{ }$} 
\put(120,2){\scriptsize $u_4^{ }$} 
\put(0,96){\scriptsize $v_1^{ }$} 
\put(40,96){\scriptsize $v_2^{ }$} 
\put(80,96){\scriptsize $v_3^{ }$} 
\put(120,96){\scriptsize $v_4^{ }$} 
\put(-8,50){\scriptsize $w$} 
\put(129,50){\scriptsize $w'$} 
\end{picture}} 
\quad \quad=\quad \sum_{t_1^{ }} \sum_{t_2^{ }} \quad \;\; 
\parbox{3.0cm}{\begin{picture}(125,100) 
\put(5,10){\line(1,0){120}} 
\put(5,50){\line(1,0){32}} 
\put(53,50){\line(1,0){24}} 
\put(125,50){\line(-1,0){32}} 
\put(5,90){\line(1,0){120}} 
\put(5,10){\line(0,1){80}} 
\put(45,10){\line(0,1){32}} 
\put(45,90){\line(0,-1){32}} 
\put(85,10){\line(0,1){32}} 
\put(85,90){\line(0,-1){32}} 
\put(125,10){\line(0,1){80}} 
\put(28.5,10){\vector(1,0){0}} 
\put(28.5,50){\vector(1,0){0}} 
\put(28.5,90){\vector(1,0){0}} 
\put(68.5,10){\vector(1,0){0}} 
\put(68.5,50){\vector(1,0){0}} 
\put(68.5,90){\vector(1,0){0}} 
\put(108.5,10){\vector(1,0){0}} 
\put(108.5,50){\vector(1,0){0}} 
\put(108.5,90){\vector(1,0){0}} 
\put(5,26){\vector(0,-1){0}} 
\put(45,26){\vector(0,-1){0}} 
\put(85,26){\vector(0,-1){0}} 
\put(125,26){\vector(0,-1){0}} 
\put(5,66){\vector(0,-1){0}} 
\put(45,66){\vector(0,-1){0}} 
\put(85,66){\vector(0,-1){0}} 
\put(125,66){\vector(0,-1){0}} 
 
\put(5,10){\circle*{3}} 
\put(45,10){\circle*{3}} 
\put(85,10){\circle*{3}} 
\put(125,10){\circle*{3}} 
\put(5,90){\circle*{3}} 
\put(45,90){\circle*{3}} 
\put(85,90){\circle*{3}} 
\put(125,90){\circle*{3}} 
\put(5,50){\circle*{3}} 
\put(125,50){\circle*{3}} 
 
\put(0,2){\scriptsize $u_1^{ }$} 
\put(40,2){\scriptsize $u_2^{ }$} 
\put(80,2){\scriptsize $u_3^{ }$} 
\put(120,2){\scriptsize $u_4^{ }$} 
\put(0,96){\scriptsize $v_1^{ }$} 
\put(40,96){\scriptsize $v_2^{ }$} 
\put(80,96){\scriptsize $v_3^{ }$} 
\put(120,96){\scriptsize $v_4^{ }$} 
\put(-8,50){\scriptsize $w$} 
\put(129,50){\scriptsize $w'$} 
\put(45,50){\scriptsize \makebox(0,0){$t_1^{ }$}} 
\put(85,50){\scriptsize \makebox(0,0){$t_2^{ }$}} 
\end{picture}} 
$$ 
 
Nous allons voir que le calcul des connexions nous permet d'obtenir une \'ecriture matricielle
des endomorphismes de chemins essentiels $\xi \otimes \xi'$ pour la loi $\odot$. 
Consid\'erons la dig\`ebre $(\mathcal{B},\circ,\odot)$, et la base des endomorphismes gradu\'es de chemins 
essentiels $e_{\xi\xi'} = \xi \otimes \xi'$ (qui diagonalise $\mathcal{B}$ pour la loi $\circ$).
$(\mathcal{B},\odot)$ est semi-simple, elle est donc isomorphe \`a une somme directe d'alg\`ebre matricielle
(donnant ses repr\'esentations irr\'eductibles), labell\'ees par $\bf{x}$. L'irrep fondamentale, appel\'ee 
$\bf{x}_1$
est reli\'ee \`a la connexion basique aussi not\'ee $\bf{x}_1$ de la mani\`ere suivante \cite{Oc-paths}.
D\'efinissons l'application $\phi_{\alpha \beta}^{\bf{x}_1} : \mathcal{E}ss \mapsto \mathcal{E}ss$:
\begin{equation}
\phi_{\alpha\beta}^{\bf{x}_1} (\xi) = \sum_{\xi'} \quad \parbox{1.25cm}{\unitlength 0.020cm 
\begin{picture}(50,65)(0,-2.5)
\put(5,10){\circle{3}}
\put(5,50){\circle{3}}
\put(45,10){\circle{3}}
\put(45,50){\circle{3}}
\put(5,10){\line(1,0){40}} 
\put(5,50){\line(1,0){40}} 
\put(5,10){\line(0,1){40}} 
\put(45,10){\line(0,1){40}} 
\put(28.5,10){\vector(1,0){0}} 
\put(28.5,50){\vector(1,0){0}} 
\put(5,26){\vector(0,-1){0}} 
\put(45,26){\vector(0,-1){0}} 
\put(21,-5){\scriptsize $\xi'$} 
\put(21,57){\scriptsize $\xi$} 
\put(-10,28){\scriptsize $\alpha$} 
\put(52,28){\scriptsize $\beta$} 
\put(25,30){\scriptsize \makebox(0,0){$\bf{x}_1$}} 
\end{picture}} 
\quad \xi' \; \; ,
\label{mapphi}
\end{equation}
o\`u $\alpha$ et $\beta$ sont des chemins de longueur un et o\`u dans le membre de droite de 
(\ref{mapphi}) appara\^{\i}t la valeur de la cellule pour la connexion $\bf{x}_1$. D\'efinissons alors 
l'application $\Phi^{\bf{x}_1}$:
\begin{equation}
\Phi_{\alpha \beta}^{\bf{x}_1} (\xi \otimes \xi') = \langle \phi_{\alpha \beta}^{\bf{x}_1} (\xi) \, , \, \xi' \rangle
= \quad \parbox{1.25cm}{\unitlength 0.020cm 
\begin{picture}(50,65)(0,-2.5)
\put(5,10){\circle{3}}
\put(5,50){\circle{3}}
\put(45,10){\circle{3}}
\put(45,50){\circle{3}}
\put(5,10){\line(1,0){40}} 
\put(5,50){\line(1,0){40}} 
\put(5,10){\line(0,1){40}} 
\put(45,10){\line(0,1){40}} 
\put(28.5,10){\vector(1,0){0}} 
\put(28.5,50){\vector(1,0){0}} 
\put(5,26){\vector(0,-1){0}} 
\put(45,26){\vector(0,-1){0}} 
\put(21,-5){\scriptsize $\xi'$} 
\put(21,57){\scriptsize $\xi$} 
\put(-10,28){\scriptsize $\alpha$} 
\put(52,28){\scriptsize $\beta$} 
\put(25,30){\scriptsize \makebox(0,0){$\bf{x}_1$}} 
\end{picture}} \;\; ,
\end{equation} 
qui satisfait la propri\'et\'e d'homomorphisme suivante:
\begin{equation}
\sum_{\beta} \Phi_{\alpha \beta}^{\bf{x}_1} (\xi \otimes \xi') \; \Phi_{\beta \gamma}^{\bf{x}_1} (\eta \otimes \eta')
\;=\; \Phi_{\alpha \gamma}^{\bf{x}_1} (\xi \otimes \xi' \; \odot \eta \otimes \eta')\; . 
\end{equation}
Nous obtenons alors la repr\'esentation 
matricielle d'un \'el\'ement $\xi \otimes \xi'$ pour la loi $\odot$ dans le bloc $\bf{x}_1$. 
Le produit tensoriel de repr\'esentations est donn\'e par:
\begin{equation}
\Phi_{\alpha \cdot \alpha'\,,\, \beta \cdot \beta'}^{\bf{x}_1 \otimes \bf{x}_1} (\xi \otimes \xi')
\;=\; \langle \phi_{\alpha' \beta'}^{\bf{x}_1} (\phi_{\alpha \beta}^{\bf{x}_1}(\xi)) \;,\; \xi'\rangle
\;=\; \sum_{\lambda} \quad
\parbox{1.25cm}{\unitlength 0.020cm 
\begin{picture}(50,65)(0,-2.5)
\put(5,10){\circle{3}}
\put(5,50){\circle{3}}
\put(45,10){\circle{3}}
\put(45,50){\circle{3}}
\put(5,10){\line(1,0){40}} 
\put(5,50){\line(1,0){40}} 
\put(5,10){\line(0,1){40}} 
\put(45,10){\line(0,1){40}} 
\put(28.5,10){\vector(1,0){0}} 
\put(28.5,50){\vector(1,0){0}} 
\put(5,26){\vector(0,-1){0}} 
\put(45,26){\vector(0,-1){0}} 
\put(21,-5){\scriptsize $\lambda$} 
\put(21,57){\scriptsize $\xi$} 
\put(-10,28){\scriptsize $\alpha$} 
\put(52,28){\scriptsize $\beta$} 
\put(25,30){\scriptsize \makebox(0,0){$\bf{x}_1$}} 
\end{picture}}
\quad
\parbox{1.25cm}{\unitlength 0.020cm 
\begin{picture}(50,65)(0,-2.5)
\put(5,10){\circle{3}}
\put(5,50){\circle{3}}
\put(45,10){\circle{3}}
\put(45,50){\circle{3}}
\put(5,10){\line(1,0){40}} 
\put(5,50){\line(1,0){40}} 
\put(5,10){\line(0,1){40}} 
\put(45,10){\line(0,1){40}} 
\put(28.5,10){\vector(1,0){0}} 
\put(28.5,50){\vector(1,0){0}} 
\put(5,26){\vector(0,-1){0}} 
\put(45,26){\vector(0,-1){0}} 
\put(21,-5){\scriptsize $\beta$} 
\put(21,57){\scriptsize $\lambda$} 
\put(-10,28){\scriptsize $\alpha'$} 
\put(52,28){\scriptsize $\beta'$} 
\put(25,30){\scriptsize \makebox(0,0){$\bf{x}_1$}} 
\end{picture}} 
\end{equation}
o\`u $\alpha \cdot \alpha'$ repr\'esente la concat\'enation des chemins $\alpha$ et $\alpha'$. Nous sommes
donc en mesure de trouver la repr\'esentation matricielle de tous les \'el\'ements de la base 
$\xi\otimes \xi'$ de $\mathcal{B}$ pour la loi $\odot$, et ainsi d'en d\'eduire la table de multiplication:
\begin{equation}
(\xi\otimes \xi') \odot (\eta \otimes \eta').
\end{equation}
Il suffit alors de trouver la d\'ecomposition du produit tensoriel des irreps $\bf{x}_1 \otimes \bf{x}_1$
en somme de irreps $\bf{x}_i$, et ainsi trouver la d\'ecomposition en blocs (labell\'es par $\bf{x}$)
de $(\mathcal{B},\odot)$. Il est plus instructif d'\'etudier en d\'etail un exemple, nous traitons
le cas du diagramme $A_3$ dans la section suivante.

\section{Une construction explicite}

\subsection{Cas $A_3$} 
Le graphe $A_3$ poss\`ede trois vertex not\'es $0, 1, 2$ et son nombre de Coxeter $\kappa=4$, il est 
repr\'esent\'e ci-dessous, o\`u sont aussi indiqu\'ees (entre crochets) les dimensions 
quantiques $\mu_i$ de ses vertex.
\begin{figure}[H] 
\unitlength 0.8mm 
\begin{center} 
\begin{picture}(40,13)(0,9) 
\put(5,10){\line(1,0){30}} 
\multiput(5,10)(15,0){3}{\circle*{2}} 
\put(5,3){\makebox(0,0){$0$}} 
\put(20,3){\makebox(0,0){$1$}} 
\put(35,3){\makebox(0,0){$2$}} 
\put(5,17){\makebox(0,0){$[1]$}} 
\put(20,17){\makebox(0,0){$[\sqrt{2}]$}} 
\put(35,17){\makebox(0,0){$[1]$}} 
\end{picture} 
\end{center} 
\end{figure} 
Une base orthonorm\'ee des chemins essentiels $\xi$ est donn\'ee par l'ensemble de 3+4+3=10 vecteurs 
suivants: 
$$v_0 = P(0)\,,\, v_1 = P(1)\,,\, v_2 = P(2)$$ 
$$r_0 = P(0,1)\,,\,\ell_1 = P(1,0)\,,\,r_1 = P(1,2)\,,\,\ell_2 = P(2,1)$$ 
$$d = P(0,1,2)\,,\,\gamma = \frac{1}{\sqrt{2}}(P(1,2,1) - P(1,0,1))\,,\,g = P(2,1,0)$$ 
Les endomorphismes gradu\'es de chemins essentiels $\xi \otimes \xi'$ seront not\'es plus simplement
$\xi\xi'$, o\`u $\xi$ et $\xi'$ sont des chemins essentiels de m\^eme longueur.
Ils forment une base de l'espace vectoriel $\mathcal{B} = \mathcal{B}(A_3)$, de dimension $3^2+4^2+3^2=34$. 
\paragraph{Loi de composition}
Le produit de composition $\circ$ dans $\mathcal{B}$ est d\'efini par: 
$$
\xi\eta \circ \xi'\eta' = \langle \eta,\xi'\rangle \xi\eta' = \delta_{\eta \xi'} \; \xi\eta'.
$$ 
$(\mathcal{B},\circ)$ est isomorphe \`a une somme directe de blocs matriciels labell\'es
par la longueur $i$ des chemins essentiels: 
$\mathcal{B} \cong \underset{i=0}{\overset{2}\oplus} L^i$.  
Nous pouvons \'ecrire de mani\`ere condens\'ee la $(\circ)$-repr\'esentation matricielle 
des \'el\'ements $\xi\xi'$ comme:
\begin{equation}
(\mathcal{B},\circ) \cong 
\left( 
\begin{array}{cccccccccc} 
v_0v_0 & v_0v_1 & v_0v_2 & . & . & . & . & . & . & . \\  
v_1v_0 & v_1v_1 & v_1v_2 & . & . & . & . & . & . & . \\  
v_2v_0 & v_2v_1 & v_2v_2 & . & . & . & . & . & . & . \\  
. & . & . & r_0r_0 & r_0\ell_1 & r_0r_1 & r_0\ell_2 & . & . & . \\  
. & . & . & \ell_1r_0 & \ell_1\ell_1 & \ell_1r_1 & \ell_1\ell_2 & . & . & . \\  
. & . & . & r_1r_0 & r_1\ell_1 & r_1r_1 & r_1\ell_2 & . & . & . \\  
. & . & . & \ell_2r_0 & \ell_2\ell_1 & \ell_2r_1 & \ell_2\ell_2 & . & . & . \\  
. & . & . & . & . & . & . & dd & d\gamma & dg \\  
. & . & . & . & . & . & . & \gamma d & \gamma\gamma & \gamma g \\  
. & . & . & . & . & . & . & gd & g \gamma & gg \\  
\end{array} 
\right) 
\end{equation}
La convention adopt\'ee est la suivante: pour obtenir la ($\circ$)-repr\'esentation d'un 
\'el\'ement $\xi\xi'$,  
on remplace cet \'el\'ement par 1 et tous \'el\'ements $\eta\eta' \neq \xi\xi'$ par 0.  
Chaque \'el\'ement est repr\'esent\'e par une matrice \'el\'ementaire. Par exemple:
$$
\scriptsize
v_0v_1 = 
\left( 
\begin{array}{cccccccccc} 
. & 1 & . & . & . & . & . & . & . & . \\  
. & . & . & . & . & . & . & . & . & . \\  
. & . & . & . & . & . & . & . & . & . \\  
. & . & . & . & . & . & . & . & . & . \\  
. & . & . & . & . & . & . & . & . & . \\  
. & . & . & . & . & . & . & . & . & . \\  
. & . & . & . & . & . & . & . & . & . \\  
. & . & . & . & . & . & . & . & . & . \\  
. & . & . & . & . & . & . & . & . & .  
\end{array} 
\right) 
\qquad 
v_1v_0 = 
\left( 
\begin{array}{cccccccccc} 
. & . & . & . & . & . & . & . & . & . \\  
1 & . & . & . & . & . & . & . & . & . \\  
. & . & . & . & . & . & . & . & . & . \\  
. & . & . & . & . & . & . & . & . & . \\  
. & . & . & . & . & . & . & . & . & . \\  
. & . & . & . & . & . & . & . & . & . \\  
. & . & . & . & . & . & . & . & . & . \\  
. & . & . & . & . & . & . & . & . & . \\  
. & . & . & . & . & . & . & . & . & .  
\end{array} 
\right)
$$
$$
v_0v_0 = 
\left( 
\begin{array}{cccccccccc} 
1 & . & . & . & . & . & . & . & . & . \\  
. & . & . & . & . & . & . & . & . & . \\  
. & . & . & . & . & . & . & . & . & . \\  
. & . & . & . & . & . & . & . & . & . \\  
. & . & . & . & . & . & . & . & . & . \\  
. & . & . & . & . & . & . & . & . & . \\  
. & . & . & . & . & . & . & . & . & . \\  
. & . & . & . & . & . & . & . & . & . \\  
. & . & . & . & . & . & . & . & . & .  
\end{array} 
\right)
$$
\normalsize
et la multiplication matricielle reproduit  bien le produit $\circ$:
$$
v_0 v_1 \circ v_1 v_0 = v_0 v_0.
$$
Le neutre pour la loi $\circ$ (matrice unit\'e) est donn\'e par: 
\begin{equation}
1_{\circ} = v_0v_0 + v_1v_1 + v_2v_2 + r_0r_0 + \ell_1\ell_1 + r_1r_1 + \ell_2\ell_2 + dd + 
\gamma\gamma + gg .
\end{equation}
Les projecteurs minimaux centraux $\pi_i$ pour $(\mathcal{B},\circ)$ sont donn\'es par: 
$$\pi_0 = v_0v_0 + v_1v_1 + v_2v_2 \; ,$$ 
$$\pi_1 = r_0r_0 + \ell_1\ell_1 + r_1r_1 + \ell_2\ell_2\; ,$$ 
$$\pi_2 = dd + \gamma \gamma + gg \; .$$  
Nous v\'erifions qu'ils satisfont bien aux relations suivantes: 
\begin{equation}
\pi_i \circ \pi_j = \delta_{\,i\,j}\; \pi_i \;.
\end{equation}

\paragraph{Cellules d'Ocneanu}
Consid\'erons les cellules d'Ocneanu o\`u les chemins horizontaux et verticaux sont de longueur un (cellules
basiques). Il existe
huit cellules diff\'erentes de ce type:
\begin{equation}
\occell{0}{1}{1}{0} \qquad \occell{0}{1}{1}{2} \qquad \occell{1}{0}{0}{1} \qquad \occell{1}{0}{2}{1} \qquad
\occell{1}{2}{0}{1} \qquad \occell{1}{2}{2}{1} \qquad \occell{2}{1}{1}{0} \qquad \occell{2}{1}{1}{2}  
\label{cellA3}
\end{equation}
La connexion $\bf{x_1}$ est d\'efinie sur des cellules o\`u les chemins verticaux sont de longueur 1 (donc notamment sur les huit cellules basiques ci-dessus). Elle associe \`a chaque cellule un nombre complexe,
devant satisfaire des conditions de r\'eflexion et d'unitarit\'e. Nous codons les r\'esultats dans une 
matrice $4 \times 4$ not\'ee $C$, o\`u les indices de colonne correspondent au chemin horizontal sup\'erieur
et les indices de ligne au chemin horizontal inf\'erieur, dans l'ordre suivant 
$(01 \, , \, 10 \, ,\, 12 \, , \, 21)$. Par exemple, la valeur de la cellule basique \`a 
l'extr\^eme-gauche de (\ref{cellA3}) correspond \`a l'\'el\'ement $C_{21}$. Nous fixons la valeur des trois cellules
suivantes:
$$
\occell{0}{1}{1}{0} = a, \qquad \qquad \occell{0}{1}{1}{2}=b, \qquad \qquad \occell{2}{1}{1}{2} = c, 
\qquad \qquad \qquad 
a,b,c \in \mathbb{C}.
$$
Alors, utilisant les propri\'et\'es de r\'eflexion et d'unitarit\'e, la matrice $C$ la plus g\'en\'erale
s'\'ecrit:
$$
C = \left( 
\begin{array}{cccc}
. & \frac{1}{\sqrt2} a^* & \frac{1}{\sqrt2} b^* & . \\
a & . & . & b \\
b & . & . & c \\
. & \frac{1}{\sqrt2} b^* & \frac{1}{\sqrt2} c^* & . 
\end{array}
\right),
\qquad \qquad
|a|^2 = |b|^2 = |c|^2 = 1 \; , \; a^* b + b^*c =0.
$$
Nous pouvons choisir $a=1,b=1,c=-1$ (libert\'e de choix de jauge), et la matrice $C$ s'\'ecrit:
$$ 
C = \left( 
\begin{array}{cccc} 
. & \frac{1}{\sqrt{2}} & \frac{1}{\sqrt{2}} & . \\ 
1 & . & . & 1 \\ 
1 & . & . & -1 \\ 
. & \frac{1}{\sqrt{2}} & -\frac{1}{\sqrt{2}} & . 
\end{array} 
\right)  
$$ 

\paragraph{Produit de convolution}
La repr\'esentation matricielle du produit de convolution $\odot$ des \'el\'ements $\xi\xi'$ est 
calcul\'ee \`a 
travers le calcul des cellules d'Ocneanu. Consid\'erons d'abord la connexion $\bf{x_1}$. \\

\noindent $\bullet$ $\bf{x_1} \quad$
La repr\'esentation matricielle pour la loi $\odot$ des \'el\'ements $\xi\xi'$ est donn\'ee par:
$$
\Phi_{\alpha, \beta}^{\bf{x_1}} (\xi\xi') = \quad \occellch{\xi}{\xi'}{\alpha}{\beta}\;\;\;,
$$
o\`u $\alpha$ et $\beta$ sont des chemins de longueur un. La matrice $\Phi_{\alpha \beta}^{\bf{x_1}}$
est donc une matrice $4 \times 4$, o\`u l'ordre choisi\footnote{Attention, les indices de lignes et de 
colonnes repr\'esentent maintenant des chemins verticaux.} pour les indices est $(\, \substack{0\\1} \, , \, 
\substack{1\\0} \, ,\, \substack{1\\2} \, , \, \substack{2\\1}\, )$:
$$
\small
\Phi_{\alpha,\beta}^{\bf{x_1}} = 
\begin{array}{cc}
&\hspace*{0.3cm} \begin{array}{cccc}
\quad  \substack{0\\1} \quad  & \quad  \substack{1\\0}  \quad  & \quad  \substack{1\\2}  \quad & 
\quad \substack{2\\1} \; \\
{} & {} & {} & {} 
\end{array} \\
& \begin{array}{c}
 \substack{0\\1} \\ {} \\ \substack{1\\0} \\ {} \\ \substack{0\\1} \\ {} \\ \substack{1\\0} 
\end{array} \left( \begin{array}{cccc} \quad \cdot\quad & \quad \cdot \quad & \quad \cdot \quad & \quad \cdot \quad  \\ {} & {} & {} & {} \\ \quad \cdot \quad & \quad \cdot \quad & \quad \cdot \quad & \quad \cdot \quad \\ {} & {} & {} & {} \\ \quad \cdot \quad & \quad \cdot \quad & \quad \cdot \quad & \quad \cdot \quad \\ {} & {} & {} & {} \\ \quad \cdot \quad & \quad \cdot \quad & \quad \cdot \quad & \quad \cdot \quad
\end{array} \right) \\
& {} 
\end{array}
$$
\normalsize
Pour des endomorphismes $\xi\xi'$ de longueur un, la valeur de la connexion $\bf{x_1}$ est cod\'ee dans la 
matrice $C$ (cellules basiques). Nous avons par exemple:
$$
\Phi_{\substack{0 \\1},\,\substack{1 \\0}}^{\underset{ }{\bf{x_1}}} (r_0 \, \ell_1) = 
\occell{0}{1}{1}{0} = 1\, , \qquad \qquad 
\Phi_{\alpha, \beta}^{\bf{x_1}} (r_0\, \ell_1) = 0, \qquad \textrm{si } (\alpha,\beta) \not=
(\substack{0\\1}, \substack{1\\0}), 
$$
et la repr\'esentation matricielle de $r_0\ell_1$ est donc donn\'ee par:
$$
\Phi^{\bf{x_1}}_{\alpha,\beta} (r_0\ell_1) = 
\left(
\begin{array}{cccc}
. & 1 & . & . \\
. & . & . & . \\
. & . & . & . \\
. & . & . & . 
\end{array}
\right)
$$
Pour les endomorphismes de longueur 0, la valeur de la connexion $\bf{x_1}$ est \'egale par convention \`a 1 
(ceci est reli\'e \`a l'axiome de {\it flatness} de Ocneanu \cite{Oc-paths}). Par exemple:
$$
\Phi^{\bf{x_1}}_{\alpha,\beta} (v_0v_1) = 
\left(
\begin{array}{cccc}
1 & . & . & . \\
. & . & . & . \\
. & . & . & . \\
. & . & . & . 
\end{array}
\right),
\qquad
\Phi^{\bf{x_1}}_{\alpha,\beta} (v_1v_2) = 
\left(
\begin{array}{cccc}
. & . & . & . \\
. & . & . & . \\
. & . & 1 & . \\
. & . & . & . 
\end{array}
\right).
$$
Pour les chemins essentiels de longueur deux, prenons l'exemple de $d\gamma$. Rappelons que 
$\gamma = \frac{1}{\sqrt2} (121 - 101)$, nous avons:
$$
\Phi_{\substack{0\\1},\, \substack{2\\1}}^{\underset{}{\bf{x_1}}} (d\gamma) = \frac{1}{\sqrt2} \left( \;
\occelldeux{0}{1}{2}{1}{2}{1} - \occelldeux{0}{1}{2}{1}{0}{1} \; \right) = 
\frac{1}{\sqrt2} \left( \frac{-1}{\sqrt2} + \frac{-1}{\sqrt2} \right) = -1
$$ 
Nous obtenons ainsi la suivante repr\'esentation matricielle des \'el\'ements $\xi\xi'$ pour la loi $\odot$:
\begin{equation}
\Phi_{\alpha,\beta}^{\bf{x_1}} (\xi\xi') = 
\left(
\begin{array}{cccc}
v_0v_1 & r_0\ell_1 & r_0r_1 & -d\gamma \\ 
\frac{1}{\sqrt{2}} \ell_1r_0 & v_1v_0 & -\gamma d & \frac{1}{\sqrt{2}}r_1r_0 \\ 
\frac{1}{\sqrt{2}}\ell_1\ell_2 & -\gamma g & v_1v_2 & -\frac{1}{\sqrt{2}} r_1\ell_2 \\ 
-g \gamma & \ell_2\ell_1 & -\ell_2r_1 & v_2v_1 \\ 
\end{array}
\right)
\label{x1a3}
\end{equation}
La convention adopt\'ee est la suivante: pour obtenir la ($\odot$)-repr\'esentation de
l'\'el\'ement $\xi\xi'$,  
nous rempla\c{c}ons cet \'el\'ement par 1 et tous \'el\'ements $\eta\eta' \neq \xi\xi'$ par 0. 
De (\ref{x1a3}) nous pouvons lire par exemple:
$$
r_0l_1 \odot l_1r_0 = \frac{1}{\sqrt2} v_0v_1, \qquad \qquad \gamma d \odot \ell_1\ell_2 = - \ell_1 r_0.
$$
\vspace{0.2cm}

\noindent $\bullet$ $\bf{x_1} \otimes \bf{x_1}$ 
La repr\'esentation matricielle pour la loi $\odot$ d'un \'el\'ement $\xi  \xi'$ pour le
produit tensoriel est donn\'ee par:
\begin{equation}
\Phi_{\alpha \cdot \alpha'\,,\, \beta \cdot \beta'}^{\bf{x_1} \otimes \bf{x_1}} (\xi \xi')
\;=\; \langle \phi_{\alpha' \beta'}^{\bf{x_1}} (\phi_{\alpha \beta}^{\bf{x_1}}(\xi)) \;,\; \xi'\rangle
\;=\; \sum_{\lambda} \quad
\parbox{1.25cm}{\unitlength 0.020cm 
\begin{picture}(50,65)(0,-2.5)
\put(5,10){\circle{3}}
\put(5,50){\circle{3}}
\put(45,10){\circle{3}}
\put(45,50){\circle{3}}
\put(5,10){\line(1,0){40}} 
\put(5,50){\line(1,0){40}} 
\put(5,10){\line(0,1){40}} 
\put(45,10){\line(0,1){40}} 
\put(28.5,10){\vector(1,0){0}} 
\put(28.5,50){\vector(1,0){0}} 
\put(5,26){\vector(0,-1){0}} 
\put(45,26){\vector(0,-1){0}} 
\put(21,-5){\scriptsize $\lambda$} 
\put(21,57){\scriptsize $\xi$} 
\put(-10,28){\scriptsize $\alpha$} 
\put(52,28){\scriptsize $\beta$} 
\end{picture}}
\quad
\parbox{1.25cm}{\unitlength 0.020cm 
\begin{picture}(50,65)(0,-2.5)
\put(5,10){\circle{3}}
\put(5,50){\circle{3}}
\put(45,10){\circle{3}}
\put(45,50){\circle{3}}
\put(5,10){\line(1,0){40}} 
\put(5,50){\line(1,0){40}} 
\put(5,10){\line(0,1){40}} 
\put(45,10){\line(0,1){40}} 
\put(28.5,10){\vector(1,0){0}} 
\put(28.5,50){\vector(1,0){0}} 
\put(5,26){\vector(0,-1){0}} 
\put(45,26){\vector(0,-1){0}} 
\put(21,-5){\scriptsize $\xi'$} 
\put(21,57){\scriptsize $\lambda$} 
\put(-10,28){\scriptsize $\alpha'\,\;$} 
\put(52,28){\scriptsize $\beta'$} 
\end{picture}} 
\end{equation}
o\`u les chemins verticaux $\alpha \cdot \alpha'$ et $\beta \cdot \beta'$ sont maintenant des 
chemins de longueur deux. Pour le diagramme $A_3$, la matrice $\Phi_{\alpha\cdot \alpha',\beta\cdot\beta'}$
est une matrice $6 \times 6$, o\`u l'ordre choisi pour les indices est $(\, \substack{0\\1\\0} \, , \, 
\substack{0\\1\\2} \, , \,
\substack{1\\0\\1} \, , \, \substack{1\\2\\1} \, , \, \substack{2\\1\\0} \, , \, \substack{2\\1\\1}
\, , \, \substack{0\\1\\0} \, )$:
$$
\scriptsize
\Phi_{\alpha \cdot \alpha' , \beta \cdot \beta'}^{\bf{x_1 \otimes x_1}} = 
\begin{array}{cc}
&\hspace*{0.3cm} \begin{array}{cccccc}
\quad  \substack{0\\1\\0} \quad  & \quad  \substack{0\\1\\2}  \quad  & \quad  \substack{1\\0\\1}  \quad & 
\quad \substack{1\\2\\1} \quad & \quad  \substack{2\\1\\0}  \quad  & \quad  \substack{2\\1\\2}  \quad  \\
{} & {} & {} & {} & {} & {} 
\end{array} \\
& \begin{array}{c}
 \substack{0\\1\\0} \\ {} \\ \substack{0\\1\\2} \\ {} \\ \substack{1\\0\\1} \\ {} \\ \substack{1\\2\\1} \\
{} \\ \substack{2\\1\\0} \\ {} \\ \substack{2\\1\\2\\{}} 
\end{array} \left( \begin{array}{cccccc} \quad \vspace*{0.15cm} \cdot\quad & \quad \cdot \quad & \quad \cdot \quad & \quad \cdot \quad & \quad \cdot \quad & \quad \cdot \quad \\ {} & {} & {} & {} & {} & {} \\ 
\quad \vspace*{0.15cm} \cdot \quad & \quad \cdot \quad & \quad \cdot \quad & \quad \cdot \quad & \quad \cdot \quad & \quad \cdot \quad \\ {} & {} & {} & {} & {} & {} \\ 
\quad \vspace*{0.15cm} \cdot \quad & \quad \cdot \quad & \quad \cdot \quad & \quad \cdot \quad & \quad \cdot \quad & \quad \cdot \quad \\ {} & {} & {} & {} & {} & {} \\ 
\quad \vspace*{0.15cm}\cdot \quad & \quad \cdot \quad & \quad \cdot \quad & \quad \cdot \quad & \quad \cdot \quad & \quad \cdot \quad \\ {} & {} & {} & {} & {} & {} \\ 
\quad \vspace*{0.15cm} \cdot \quad & \quad \cdot \quad & \quad \cdot \quad & \quad \cdot \quad & \quad \cdot \quad & \quad \cdot \quad \\ {} & {} & {} & {} & {} & {} \\ 
\quad \cdot \quad & \quad \cdot \quad & \quad \cdot \quad & \quad \cdot \quad & \quad \cdot \quad & \quad \cdot \quad \\ {} & {} & {} & {} & {} & {} 
\end{array} \right) \\
& {} 
\end{array}
$$
\normalsize

Par exemple, nous avons:
\begin{eqnarray*}
\Phi_{\, \substack{0\\1\\2} \, , \, \substack{1\\0\\1} \,}^{\underset{}{\bf{x_1}\otimes \bf{x_1}}} 
(r_0 \ell_2) &=& 
\langle \phi_{\substack{1\\2} \, , \, \substack{0\\1}}^{\underset{}{\bf{x_1}}} (\phi_{ 
\substack{0\\1} \, , \, \substack{1\\0}}^{\underset{}{\bf{x_1}}}(r_0))
\; , \; \ell_2 \rangle \\
{ } &=& \occell{0}{1}{1}{0} \langle \phi_{\substack{1\\2},\substack{0\\1}}^{\underset{}{\bf{x_1}}}(\ell_1) \; , \; \ell_2 \rangle 
{ } = \occell{0}{1}{1}{0} \; \occell{1}{0}{2}{1} \langle \ell_2 \; , \; \ell_2 \rangle = 
\frac{1}{\sqrt2}  
\end{eqnarray*}
\'Etendant le calcul aux autres \'el\'ements $\xi\xi'$, nous trouvons:
\begin{equation}
\Phi_{\alpha\cdot\alpha',\beta\cdot\beta'}^{\bf{x_1}\otimes \bf{x_1}}(\xi\xi') = 
\left(
\begin{array}{cccccc}
v_0 v_0 & . & \frac{1}{\sqrt{2}} r_0 r_0 & \frac{1}{\sqrt{2}} r_0 r_0 & . & dd \\
. & v_0 v_2 & \frac{1}{\sqrt{2}}r_0 \ell_2 & - \frac{1}{\sqrt{2}} r_0 \ell_2 & dg & . \\
\frac{1}{\sqrt{2}} \ell_1 \ell_1 & \frac{1}{\sqrt{2}} \ell_1 r_1 & v_1 v_1 & \gamma \gamma & \frac{1}{\sqrt{2}} r_1 \ell_1 & \frac{1}{\sqrt{2}} r_1 r_1 \\
\frac{1}{\sqrt{2}} \ell_1 \ell_1 & - \frac{1}{\sqrt{2}} \ell_1 r_1 & \gamma \gamma & v_1 v_1 & - \frac{1}{\sqrt{2}} r_1 \ell_1 & \frac{1}{\sqrt{2}} r_1 r_1 \\
. & gd & \frac{1}{\sqrt{2}} \ell_2 r_0 & -\frac{1}{\sqrt{2}} \ell_2 r_0 & v_2 v_0 & . \\
gg & . & \frac{1}{\sqrt{2}} \ell_2 \ell_2 & \frac{1}{\sqrt{2}} \ell_2 \ell_2 & . & v_2 v_2
\end{array}
\right)
\end{equation}
Par exemple, la $\odot$-repr\'esentation matricielle des \'el\'ements $v_0v_0$, $r_0l_2$ et $\ell_1\ell_1$
est donn\'ee par:
$$
\tiny
r_0r_0 = 
\left(
\begin{array}{cccccc}
. & . & \frac{1}{\sqrt2} & \frac{1}{\sqrt2}  & . & . \\
. & . & . & . & . & . \\
. & . & . & . & . & . \\
. & . & . & . & . & . \\
. & . & . & . & . & . \\
. & . & . & . & . & . 
\end{array}
\right),
\qquad 
\ell_2\ell_2 = 
\left(
\begin{array}{cccccc}
. & . & . & . & . & . \\
. & . & . & . & . & . \\
\frac{1}{\sqrt2} & . & . & . & . & . \\
\frac{1}{\sqrt2} & . & . & . & . & . \\
. & . & . & . & . & . \\
. & . & . & . & . & . 
\end{array}
\right),
\qquad
v_0v_0 = 
\left(
\begin{array}{cccccc}
1 & . & . & . & . & . \\
. & . & . & . & . & . \\
. & . & . & . & . & . \\
. & . & . & . & . & . \\
. & . & . & . & . & . \\
. & . & . & . & . & . 
\end{array}
\right).
$$
\normalsize
et nous pouvons alors en d\'eduire:
$$
r_0r_0 \odot \ell_2\ell_2 = v_0v_0.
$$
Pour diagonaliser la loi $\odot$, nous effectuons le suivant changement de base des indices de la matrice
$\Phi_{\alpha\cdot \alpha',\beta\cdot\beta'}$:
$$
\begin{array}{rclcrclcrcl}
\ud0 &=& \substack{0\\1\\0} & \quad & \ud1 &=& 
\frac{1}{\sqrt2} \left( \substack{1\\0\\1} + \substack{1\\2\\1} \right), &\quad& 
\ud2 &=& \substack{2\\1\\0}, \\
{} & {} & {} & {} & {} & {} & {} & {} & {} & {} & {}\\
\ud{d} &=& \substack{0\\1\\2}, &\qquad& \ud{\gamma} &=& 
\frac{1}{\sqrt2} \left( \substack{1\\2\\1} - \substack{1\\0\\1}\right), &\qquad&
 \ud{g} &=& \substack{2\\1\\0}.
\end{array}
$$
Alors, dans l'ordre $(\ud0,\ud1,\ud2 \, ; \, \ud{g} , \ud{\gamma} , \ud{d})$, nous obtenons:
$$ 
\Phi_{\ud{\alpha},\ud{\beta}}^{\bf{x_1}\otimes \bf{x_1}}(\xi\xi') =  
\left( 
\begin{array}{cccccc} 
v_0v_0 & r_0r_0 & dd & . & . & . \\ 
\ell_1\ell_1 & v_1v_1+\gamma\gamma & r_1r_1 & . & . & . \\ 
gg & \ell_2\ell_2 & v_2v_2 & . & . & .  \\ 
 . & . & . & v_0v_2 & -r_0\ell_2 & dg \\ 
 . & . & . & -\ell_1r_1 & v_1v_1-\gamma \gamma & -r_1\ell_1 \\ 
 . & . & . & gd & -\ell_2r_0 & v_2v_0 
\end{array} 
\right) 
$$ 
et nous obtenons alors la d\'ecomposition suivante du produit tensoriel de repr\'esentations:
\begin{equation}
\Phi^{\bf{x_1}\otimes\bf{x_1}} = \Phi^{\bf{x_0}} \oplus \Phi^{\bf{x_2}}
\end{equation}
qui d\'efinissent deux autres connexions $\bf{x_0}$ (reli\'ee \`a la repr\'esentation identit\'e) et 
$\bf{x_2}$.

\paragraph{Repr\'esentation matricielle de la loi $\odot$}
$(\mathcal{B},\odot)$ est donc isomorphe \`a une somme d'alg\`ebres matricielles (blocs) labell\'ees
par un indice $\bf{x}$: $(\mathcal{B},\odot) \cong \underset{\bf{x}=0}{\overset{2}\oplus} \Phi^{\bf{x}}$
\begin{equation*}
\small
(\mathcal{B},\odot) \cong  
\left( 
\begin{array}{cccccccccc} 
v_0v_0 & r_0r_0 & dd & . & . & . & . & . & . & . \\ 
\ell_1\ell_1 & v_1v_1+\gamma\gamma & r_1r_1 & . & . & . & . & . & . & . \\ 
gg & \ell_2\ell_2 & v_2v_2 & . & . & . & . & . & . & . \\ 
. & . & . & v_0v_1 & r_0\ell_1 & r_0r_1 & -d\gamma & . & . & . \\ 
. & . & . & \frac{1}{\sqrt{2}} \ell_1r_0 & v_1v_0 & -\gamma d & \frac{1}{\sqrt{2}}r_1r_0 & . & . & . \\ 
. & . & . & \frac{1}{\sqrt{2}}\ell_1\ell_2 & -\gamma g & v_1v_2 & -\frac{1}{\sqrt{2}} r_1\ell_2 & . & . & . \\ 
. & . & . & -g \gamma & \ell_2\ell_1 & -\ell_2r_1 & v_2v_1 & . & . & . \\ 
. & . & . & . & . & . & . & v_0v_2 & -r_0\ell_2 & dg \\ 
. & . & . & . & . & . & . & -\ell_1r_1 & v_1v_1-\gamma \gamma & -r_1\ell_1 \\ 
. & . & . & . & . & . & . & gd & -\ell_2r_0 & v_2v_0 \\ 
\end{array} 
\right) 
\end{equation*}
\normalsize
Rappelons que pour obtenir la $(\odot)$-repr\'esentation matricielle d'un \'el\'ement $\xi \xi'$, nous rempla\c{c}ons
cet \'el\'ement par 1 et tous les \'el\'ements $\eta \eta' \not= \xi \xi'$ par 0. \`A partir de cette repr\'esentation
matricielle\footnote{Cette $(\odot)$-repr\'esentation matricielle
a aussi \'et\'e obtenue, de mani\`ere ind\'ependante, par R. Trinchero \cite{trinchero}.}, nous pouvons alors calculer le produit $\odot$ des \'el\'ements de $\mathcal{B}$. La connaissance du produit $\odot$
d\'efini dans $\mathcal{B}$ -- et par cons\'equent du produit $\widehat{\odot}$ dans $\widehat{\mathcal{B}}$ -- nous permet de calculer
le coproduit $\Delta$ dans $\mathcal{B}$, \`a travers:
\begin{equation}
\langle u \otimes v, \Delta (x) \rangle = \langle u \, \widehat{\odot}\, v , x \rangle \qquad \qquad \qquad \qquad 
x \in \mathcal{B}; \, u,v \in \widehat{\mathcal{B}}  
\end{equation} 
Nous v\'erifions alors que la propri\'et\'e d'homomorphisme $\Delta(x \circ y) = \Delta(x) \circ \Delta(y)$ est
satisfaite. 
Les matrices \'el\'ementaires non-triviales correspondent aux positions $(2,2)$ dans le premier bloc
(not\'ee $a_{22}$) et dans le troisi\`eme bloc (not\'ee $c_{22}$). Elles sont donn\'ees par: 
$$ 
a_{22} = \frac{1}{2}(v_1v_1 + \gamma \gamma), \qquad \qquad c_{22} = \frac{1}{2}(v_1v_1 - \gamma \gamma). 
$$ 
Le neutre pour cette loi (la matrice identit\'e) s'\'ecrit en termes d'endomorphismes: 
$$ 
1_{\odot} = v_0v_0 + v_1v_1 + v_2v_2 + v_0v_1 + v_1v_0 + v_1v_2 + v_2v_1 + v_0v_2 + v_2v_0 
$$  
Les projecteurs minimaux centraux sont donn\'es par:
$$\varpi_0 = v_0v_0 + \frac{1}{2}(v_1v_1 + \gamma\gamma) + v_2v_2 \; , $$
$$\varpi_1 = v_0v_1 + v_1v_0 + v_1v_2 + v_2v_1 \; ,$$
$$\varpi_2 = v_0v_2 + \frac{1}{2}(v_1v_1 - \gamma \gamma) + v_2v_0 \; .$$
Ils v\'erifient les relations:
$$
\varpi_x \odot \varpi_y = \delta_{\,x\,y} \;\varpi_y \; .
$$

\paragraph{Graphes $\mathcal{A}(A_3)$ et $Oc(A_3)$}
Les projecteurs minimaux centraux $\pi_i$ et $\varpi_x$ s'expriment en termes d'endomorphismes. Ils v\'erifient
les relations:
$$
\pi_i \circ \pi_j = \delta_{\,i\,j} \; \pi_j, \qquad \qquad \qquad \varpi_x \odot \varpi_y = 
\delta_{\,x\,y} \; \varpi_y.
$$
Ayant obtenu les d\'ecompositions en blocs de $\mathcal{B}(A_3)$ pour ses deux structures multiplicatives
$\circ$ et $\odot$, nous pouvons alors multiplier les projecteurs d\'efinis pour une loi par l'autre loi.
Nous obtenons:
$$
\begin{array}{rclcrclcrcl}
\pi_0 \odot \pi_0 &=& \pi_0   &\qquad& \pi_1 \odot \pi_0 &=& \pi_1   &\qquad& \pi_2 \odot \pi_0 &=& \pi_2 \\
\pi_0 \odot \pi_1 &=& \pi_1   &\qquad& \pi_1 \odot \pi_1 &=& \pi_0 + \pi_2   &\qquad& \pi_2 \odot \pi_1 &=& \pi_1 \\
\pi_0 \odot \pi_2 &=& \pi_2   &\qquad& \pi_1 \odot \pi_2 &=& \pi_1   &\qquad& \pi_2 \odot \pi_2 &=& \pi_0 
\end{array}
$$ 
Les projecteurs $\pi_i$ pour la loi $\odot$ d\'efinissent l'alg\`ebre $\mathcal{A}_3$. Si nous 
d\'ecidons de coder cette alg\`ebre dans un graphe, tel que ses vertex soient en correspondance avec $\pi_i$
et tel que la multiplication par $\pi_1$ de $\pi_i$ soit \'egale \`a la somme des voisins de $\pi_i$
sur le graphe, nous obtenons le graphe $\mathcal{A}(A_3) = A_3$. 

Consid\'erons maintenant la multiplication $\circ$ des projecteurs $\varpi_x$: ces projecteurs 
ne sont pas ferm\'es pour la loi $\circ$. Par contre, en  d\'efinissant les
projecteurs renormalis\'es $\rho_x$ suivants:
$$ \rho_0 = v_0v_0 + v_1v_1 + \gamma\gamma + v_2v_2 $$
$$ \rho_1 = \varpi_1 $$
$$ \rho_2 = v_0v_2 + v_1v_1 - \gamma\gamma + v_2v_0$$
alors nous obtenons:
$$
\begin{array}{rclcrclcrcl}
\rho_0 \circ \rho_0 &=& \rho_0   &\qquad& \rho_1 \circ \rho_0 &=& \rho_1   &\qquad& \rho_2 \circ \rho_0 &=& \rho_2 \\
\rho_0 \circ \rho_1 &=& \rho_1   &\qquad& \rho_1 \circ \rho_1 &=& \rho_0 + \rho_2   &\qquad& \rho_2 \circ \rho_1 &=& \rho_1 \\
\rho_0 \circ \rho_2 &=& \rho_2   &\qquad& \rho_1 \circ \rho_2 &=& \rho_1   &\qquad& \rho_2 \circ \rho_2 &=& \rho_0 
\end{array}
$$
Ces relations d\'efinissent l'alg\`ebre $Oc(A_3)$, appel\'ee alg\`ebre des sym\'etries quantiques
de $A_3$. La multiplication par le projecteur $\rho_1$ est cod\'ee par le graphe $Oc(A_3)$, et nous
obtenons $Oc(A_3) = A_3$. Le passage des projecteurs
$\varpi_x$ aux projecteurs renormalis\'es $\rho_x$ -- n\'ec\'essaire \`a l'obtention de 
l'alg\`ebre $Oc(A_3)$ -- provient d'un produit scalaire non-trivial (diff\'erent du produit 
scalaire d\'efinissant l'orthonormalit\'e des chemins \'el\'ementaires) pour effectuer le passage 
entre $\mathcal{B}$ et son dual $\widehat{\mathcal{B}}$ \cite{Coq_Gil-bigebra}.
Plus pr\'ecisemment, les unit\'es matricielles associ\'ees \`a la loi $\circ$ constituent une base
orthonorm\'ee pour le produit scalaire hermitien induit de celui des chemins \'el\'ementaires (ceux-ci sont orthonorm\'es), mais les unit\'es matricielles
associ\'ees \`a la loi $\odot$ constituent une base non-orthonorm\'ee pour ce produit scalaire. Nous
pouvons introduire un nouveau produit scalaire hermitien pour lequel elles sont orthonorm\'ees. Cette diff\'erence
de normalisation est \`a l'origine de la diff\'erence existante entre les expressions des op\'erateurs
$\varpi_x$ et des op\'erateurs $\rho_x$.

\subsection{Graphes $G$ du type $ADE$ et g\'en\'eralisations} 
L'\'etude de la dig\`ebre associ\'ee au diagramme de Dynkin $A_3$ montre l'extr\^eme 
richesse de ses structures, mais aussi son extr\^eme complexit\'e, alors que $A_3$ est le cas le plus 
simple -- hormis $A_2$ -- des diagrammes de Dynkin puisqu'il ne comporte que trois points!
Bien que la diagonalisation de la loi de composition $\circ$ et l'obtention des projecteurs $\pi_i$ 
soient presque imm\'ediats, la d\'efinition de la loi $\odot$ -- \`a travers le calcul des cellules 
d'Ocneanu-- sa diagonalisation et l'obtention des projecteurs $\varpi_x$ est nettement plus difficile.

Nous conjecturons que la multiplication des endomorphismes $\xi\xi'$ pour la loi $\odot$ peut
\^etre obtenue \`a partir d'une structure alg\'ebrique $\star$ d\'efinie directement sur l'espace vectoriel
des chemins essentiels:
\begin{equation}
\xi\xi' \odot \eta \eta' \; \stackrel{?}{=} \; \mathcal{P} \, \left\lbrack \, (\xi \star \eta) 
(\xi' \star \eta') \, \right\rbrack \, ,
\end{equation}
o\`u $\mathcal{P}$ est un op\'erateur qui projette sur les endomorphismes gradu\'es par
la longueur. 
La diagonalisation de cette loi $\star$ dans cet espace vectoriel d\'efinirait alors une base des 
chemins verticaux de l'espace $V_x$. L'existence de ce produit $\star$ permettrait alors de d\'efinir
de mani\`ere beaucoup plus directe la loi $\odot$, sans la n\'ec\'essit\'e de faire usage des cellules
d'Ocneanu. Cependant, cette construction n'est pas connue \`a ce jour.

Les graphes d'Ocneanu pour les cas du type $su(2)$ s'obtiennent {\it a priori} \`a travers la 
diagonalisation de la loi $\odot$, et
sont premi\`erement apparus dans la litt\'erature dans \cite{Oc-paths}. Cependant, leur construction
par Ocneanu lui-m\^eme s'est en fait bas\'ee sur la classification des invariants modulaires de 
$\widehat{su}(2)$ de Cappelli-Itzykson-Zuber. La diagonalisation de la loi $\odot$ ne semble
jamais avoir \'et\'e explicitement men\'ee. Nous prendrons donc dans un premier temps ces graphes
d'Ocneanu comme donn\'ee initiale. 
Nous allons \'etablir dans le prochain chapitre
une r\'ealisation de l'alg\`ebre des sym\'etries quantiques d'Ocneanu nous permettant 
de retrouver les graphes d'Ocneanu.  

La dig\`ebre $\mathcal{B}(G)$ est construite \`a partir des diagrammes de Dynkin de type $ADE$. Une
d\'efinition de $\mathcal{B}(G)$ pour des diagrammes de Coxeter-Dynkin g\'en\'eralis\'es est possible,
mais aucune d\'efinition pr\'ecise de cette dig\`ebre g\'en\'eralis\'ee n'est \`a ce jour disponible
dans la litt\'erature. 

Les graphes d'Ocneanu pour les cas du type $su(3)$ et $su(4)$ ont \'et\'e obtenus par Ocneanu (mais jamais
publi\'es): sa construction pour $su(3)$ s'est bas\'ee sur la classification des invariants
modulaires de $\widehat{su}(3)$ de Gannon; pour $su(4)$ une technique originale a \'et\'e d\'evelopp\'ee
\cite{Oc-Bariloche}. Nous verrons que notre r\'ealisation permet aussi de reconstruire les graphes
d'Ocneanu pour certaines classes d'exemples.


\chapter{Des graphes aux fonctions de partition} 
\thispagestyle{empty}
 
Les graphes d'Ocneanu -- et les alg\`ebres des sym\'etries quantiques $Oc(G)$ -- sont connus 
pour les cas du type $su(2)$ \cite{Oc-paths}.   
Dans ce chapitre nous allons pr\'esenter une r\'ealisation de l'alg\`ebre $Oc(G)$ 
comme un certain quotient du carr\'e tensoriel de l'alg\`ebre du graphe $G$, 
o\`u $G$ est un diagramme de Dynkin $ADE$. 
D'une part cette r\'ealisation permet d'obtenir un algorithme tr\`es simple pour le calcul    
des divers coefficients d\'efinissant les fonctions de partition g\'en\'eralis\'ees du mod\`ele  
conforme $\widehat{su}(2)$ associ\'e au graphe $G$. D'autre part, cette r\'ealisation se  
pr\^ete naturellement \`a une 
g\'en\'eralisation aux cas $su(n), n \geq3$, ce qui nous a permis de d\'efinir le graphe 
d'Ocneanu et l'alg\`ebre $Oc(G)$ pour certains exemples choisis du type $su(3)$ et ainsi 
obtenir les fonctions de partition g\'en\'eralis\'ees associ\'ees \`a ces exemples.

\section{Repr\'esentations irr\'eductibles et graphes $\mathcal{A}$}

\subsection{D\'efinitions}  
 
\subsubsection{Cas classique} 
Consid\'erons le groupe $SU(n)$ et ses repr\'esentations irr\'eductibles $(i) \in $ Irr $SU(n)$.   
Ce groupe poss\`ede $n-1$ repr\'esentations fondamentales, en ce sens que les   
autres repr\'esentations irr\'eductibles peuvent \^etre obtenues par tensorialisation puis r\'eduction
\`a partir des fondamentales. 
Chaque repr\'esentation fondamentale donne lieu \`a un graphe infini, dont les vertex sont labell\'es par  
$(i) \in $ Irr $SU(n)$ et dont les arcs correspondent \`a la tensorialisation par la repr\'esentation fondamentale. 
 
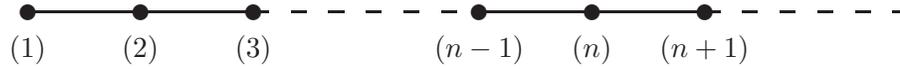
\begin{figure}[hhh]  
\unitlength=0.1cm  
\begin{center}  
\begin{picture}(130,10)(0,5)  
\thinlines  
\multiput(5,10)(15,0){3}{\circle*{2}}  
\multiput(65,10)(15,0){3}{\circle*{2}}  
\thicklines  
\multiput(35,10)(5,0){6}{\line(1,0){2}}  
\put(5,10){\line(1,0){30}}  
\put(65,10){\line(1,0){30}}  
\multiput(95,10)(5,0){6}{\line(1,0){2}}  
\put(5,5){\makebox(0,0){$(1)$}}  
\put(20,5){\makebox(0,0){$(2)$}}  
\put(35,5){\makebox(0,0){$(3)$}}  
\put(65,5){\makebox(0,0){$(n-1)$}}  
\put(80,5){\makebox(0,0){$(n)$}}  
\put(95,5){\makebox(0,0){$(n+1)$}}  
\end{picture}  
\end{center}  
\caption{Graphe $A_{\infty}$ de $SU(2)$.}  
\label{grsu2chap3} 
\end{figure}

Par exemple le groupe $SU(2)$ poss\`ede une repr\'esentation fondamentale,  
de dimension deux, not\'ee $(2)$, et la d\'ecomposition du produit tensoriel d'une irrep $(n) \in $ Irr  
$SU(2)$ par la fondamentale $(2)$ est donn\'ee par la formule suivante (correspondant au couplage  
d'une particule de spin 1/2 avec une particule de spin $j = (n-1)/2$): 
\begin{equation} 
(2) \otimes (n) = (n-1) \oplus (n+1). 
\end{equation}   
Le r\'esultat de cette d\'ecomposition peut \^etre cod\'e dans le graphe $A_{\infty}$ de $SU(2)$ 
illustr\'e \`a la Fig.\ref{grsu2chap3}. 

Les vertex de ce graphe sont labell\'es par les irreps $(i) \in$ Irr $SU(2)$, et  
$(2) \otimes (i)$ se d\'ecompose en la somme directe d'irreps $(j)$, tel que  
$(j)$ soit voisin de $(i)$ sur le graphe. L'exemple de $SU(2)$ est trait\'e de mani\`ere  
plus d\'etaill\'ee dans l'Annexe $\bf B$.

Le groupe $SU(3)$ poss\`ede deux repr\'esentations fondamentales (not\'ees\footnote{Nous avons choisi de  
labeller les irreps de $SU(3)$ par leur dimension, mais nous aurions pu aussi les labeller par 
des diagrammes de Young.} 3 et $\ov{3}$), chacune donnant lieu \`a un graphe   
infini $\mathcal{A}_{\infty}$. Les repr\'esentations 
3 et $\ov{3}$ \'etant conjugu\'ees, le graphe correspondant \`a la tensorialisation par $\ov{3}$ s'obtient   
en inversant la direction des arcs du graphe correspondant \`a 3.  
Le graphe $A_{\infty}$ de $SU(3)$ correspondant \`a la repr\'esentation 3 est illustr\'e \`a la  
Fig. \ref{A_SU3}. Nous lisons sur le graphe,  
par exemple: $3 \otimes 3 = \ov{3} \oplus 6$, car il y a un arc  
reliant $3$ \`a $\ov{3}$ et un arc reliant $3$ \`a $6$.  
  
\unitlength 0.04cm 
\begin{figure}[hhh]  
\begin{center}  
\begin{picture}(200,170)  
 
\put(0,0){\begin{picture}(40,40)(0,-10) 
\put(0,-12){\makebox(0,0){$1$}}  
\put(9,33){\makebox(0,0){$\ov{3}$}}  
\put(0,0){\circle*{4}}  
\put(40,0){\circle*{4}}  
\put(20,30){\circle*{4}}  
\put(0,0){\vector(1,0){21}}  
\put(20,0){\line(1,0){20}}  
\put(40,0){\vector(-2,3){11}}  
\put(30,15){\line(-2,3){10}}  
\put(20,30){\vector(-2,-3){11}}  
\put(10,15){\line(-2,-3){10}}\end{picture}}  
  
\put(20,30){\begin{picture}(40,40)(0,-10)  
\put(9,33){\makebox(0,0){$\ov{6}$}}  
\put(0,0){\circle*{4}}  
\put(40,0){\circle*{4}}  
\put(20,30){\circle*{4}}  
\put(0,0){\vector(1,0){21}}  
\put(20,0){\line(1,0){20}}  
\put(40,0){\vector(-2,3){11}}  
\put(30,15){\line(-2,3){10}}  
\put(20,30){\vector(-2,-3){11}}  
\put(10,15){\line(-2,-3){10}}\end{picture}}  
  
\put(40,0){\begin{picture}(40,40)(0,-10) 
\put(0,-12){\makebox(0,0){$3$}}  
\put(0,0){\circle*{4}}  
\put(40,0){\circle*{4}}  
\put(20,30){\circle*{4}}  
\put(0,0){\vector(1,0){21}}  
\put(20,0){\line(1,0){20}}  
\put(40,0){\vector(-2,3){11}}  
\put(30,15){\line(-2,3){10}}  
\put(20,30){\vector(-2,-3){11}}  
\put(10,15){\line(-2,-3){10}}\end{picture}}  
  
\put(80,0){\begin{picture}(40,40)(0,-10) 
\put(0,-12){\makebox(0,0){$6$}}  
\put(40,-12){\makebox(0,0){$10$}}  
\put(20,40){\makebox(0,0){$15$}}  
\put(0,0){\circle*{4}}  
\put(40,0){\circle*{4}}  
\put(20,30){\circle*{4}}  
\put(0,0){\vector(1,0){21}}  
\put(20,0){\line(1,0){20}}  
\put(40,0){\vector(-2,3){11}}  
\put(30,15){\line(-2,3){10}}  
\put(20,30){\vector(-2,-3){11}}  
\put(10,15){\line(-2,-3){10}}\end{picture}}  
  
\put(60,30){\begin{picture}(40,40)(0,-10)  
\put(0,10){\makebox(0,0){$8$}}  
\put(20,43){\makebox(0,0){$\ov{15}$}}  
\put(0,0){\circle*{4}}  
\put(40,0){\circle*{4}}  
\put(20,30){\circle*{4}}  
\put(0,0){\vector(1,0){21}}  
\put(20,0){\line(1,0){20}}  
\put(40,0){\vector(-2,3){11}}  
\put(30,15){\line(-2,3){10}}  
\put(20,30){\vector(-2,-3){11}}  
\put(10,15){\line(-2,-3){10}}\end{picture}}  
  
\put(40,60){\begin{picture}(40,40)(0,-10) 
\put(8,33){\makebox(0,0){$\ov{10}$}}  
\put(0,0){\circle*{4}}  
\put(40,0){\circle*{4}}  
\put(20,30){\circle*{4}}  
\put(0,0){\vector(1,0){21}}  
\put(20,0){\line(1,0){20}}  
\put(40,0){\vector(-2,3){11}}  
\put(30,15){\line(-2,3){10}}  
\put(20,30){\vector(-2,-3){11}}  
\put(10,15){\line(-2,-3){10}}\end{picture}}  
  
\dashline[10]{4}(120,10)(200,10)  
\dashline[10]{4}(100,40)(180,40)  
\dashline[10]{4}(80,70)(160,70)  
\dashline[10]{4}(60,100)(140,100)  
\dashline[10]{4}(80,130)(120,130)  
  
\dashline[10]{4}(160,10)(80,130)  
\dashline[10]{4}(60,100)(100,160)  
\dashline[10]{4}(80,70)(120,130)  
\dashline[10]{4}(100,40)(140,100)  
\dashline[10]{4}(120,10)(160,70)  
\dashline[10]{4}(160,10)(180,40)  
  
\end{picture}  
\end{center}  
\caption{Graphe $\mathcal{A}_{\infty}$ de $SU(3)$ pour la repr\'esentation $3$.}  
\label{A_SU3}  
\end{figure}
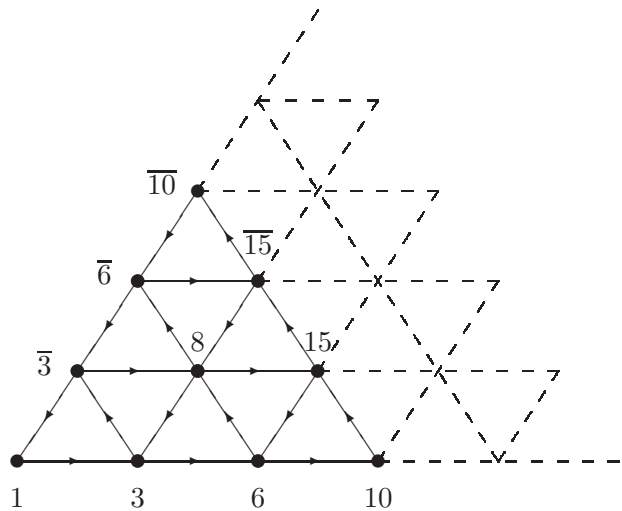

\subsubsection{Cas quantique} 
Les groupes de Lie en g\'en\'eral poss\`edent des d\'eformations quantiques (groupes quantiques),
dont un exemple bien connu est fourni par le groupe quantique $U_q(sl(n))$. Aux racines de l'unit\'e ($q^N=1$),
nous pouvons d\'efinir des quotients de Hopf non semi-simples, de dimension finie -- g\'en\'eriquement 
d\'esign\'es par $u_q(sl(n))$ -- \`a partir du groupe quantique $U_q(sl(n))$: ces groupes quantiques
poss\`edent alors un nombre fini de repr\'esentations irr\'eductibles (nous nous int\'eressons en fait
\`a celles qui sont de q-dimension non-nulle: voir Annexe {\bf B} pour le cas $u_q(sl(2))$). Nous  
d\'ecidons de noter $SU(n)_{\ell} = u_q(sl(n))$ pour $q$ racine de l'unit\'e\footnote{Attention: l'index $q$
de $u_q$ est quelquefois not\'e $q^{1/2}$ dans la litt\'erature.}: 
\begin{equation} 
q^{2(\ell + h)} = q^{2 \kappa} = 1, 
\end{equation} 
o\`u $\ell$ est appel\'e le niveau et $h$ est le nombre (dual) de Coxeter du groupe classique correspondant  
(pour $SU(n)$, $h=n$). $\kappa$ est le nombre de Coxeter g\'en\'eralis\'e (parfois appel\'e altitude)  
et est d\'efini par la relation  $\kappa = \ell + h$.  
De mani\`ere analogue au cas classique, nous pouvons associer   
\`a chaque repr\'esentation fondamentale de $SU(n)_{\ell}$ un graphe: ce graphe  
s'obtient par troncation (au niveau $\ell$) du graphe du cas classique correspondant.

Ainsi, pour $SU(2)_{\ell}$, $h=2$ et \`a la $2(\ell+2)$-i\`eme racine de l'unit\'e, nous obtenons le graphe  
$\mathcal{A}_{\ell} = A_{\ell+1}$\footnote{Pour le cas $SU(2)_{\ell}$, les graphes $\mathcal{A}_{\ell}$  
correspondent aux diagrammes de Dynkin de type $A_{\ell+1}$: nous r\'eservons la notation caligraphique  
pour les graphes labell\'es par le niveau $\ell$, la notation standard pour les graphes labell\'es par 
le nombre de vertex, c.\`a.d. par le rang de  
l'alg\`ebre de Lie correspondante au diagramme de Dynkin.} de $SU(2)_{\ell}$. Pour $SU(3)_{\ell}$,  
$h=3$, et \`a la $2(\ell+3)$-i\`eme racine de l'unit\'e, nous obtenons le graphe  
$\mathcal{A}_{\ell}$ de $SU(3)_{\ell}$.

Pour $SU(2)_4$, nous avons $\mathcal{A}_4 = A_5$, ce graphe poss\`ede 5 vertex not\'es  
$\tau_0, \cdots, \tau_4$.  
La repr\'esentation identit\'e est $\tau_0$ et la fondamentale $\tau_1$.  
Pour $SU(3)_4$, le graphe $\mathcal{A}_4$ poss\`ede 15 vertex. De mani\`ere g\'en\'erale, le graphe  
$\mathcal{A}_{\ell}$ de $SU(3)_{\ell}$ poss\`ede \mbox{$(\ell+1)(\ell+2)/2$} vertex que nous  
labellons\footnote{Notre convention pour les indices des vertex de $SU(3)_{\ell}$ commencent \`a 0 et  
non \`a 1. Certains auteurs adoptent des conventions diff\'erentes.} par $(\lambda_1, \lambda_2)$,  
avec $\lambda_1,\lambda_2 \geq 0$ et $\lambda_1+\lambda_2 \leq \ell$.     
La repr\'esentation identit\'e est $(0,0)$ et les deux repr\'esentations fondamentales (conjugu\'ees)  
sont $(1,0)$ et $(0,1)$: le graphe code la tensorialisation par $(1,0)$. Le graphe de tensorialisation  
par $(0,1)$ est obtenu en inversant la direction des arcs.  
Les graphes $\mathcal{A}_{4}$ de $SU(2)_{4}$ et $SU(3)_{4}$ sont repr\'esent\'es pour  
\`a la Fig. \ref{A_SU2,SU3_4}.  
  
\unitlength 0.035cm 
\begin{figure}[hhh]  
\begin{center}  
\begin{picture}(380,160)  
  
\put(0,75){\line(1,0){160}}  
\put(0,75){\circle*{4}}  
\put(40,75){\circle*{4}}  
\put(80,75){\circle*{4}}  
\put(120,75){\circle*{4}}  
\put(160,75){\circle*{4}}  
\put(0,63){\makebox(0,0){$\tau_0$}}  
\put(40,63){\makebox(0,0){$\tau_1$}}  
\put(80,63){\makebox(0,0){$\tau_2$}}  
\put(120,63){\makebox(0,0){$\tau_3$}}  
\put(160,63){\makebox(0,0){$\tau_4$}}  
  
\put(0,0){\begin{picture}(40,40)(-220,-15)  
\put(0,-12){\makebox(0,0){$(0,0)$}}  
\put(3,33){\makebox(0,0){$(0,1)$}}  
\put(0,0){\circle*{4}}  
\put(40,0){\circle*{4}}  
\put(20,30){\circle*{4}}  
\put(0,0){\vector(1,0){21}}  
\put(20,0){\line(1,0){20}}  
\put(40,0){\vector(-2,3){11}}  
\put(30,15){\line(-2,3){10}}  
\put(20,30){\vector(-2,-3){11}}  
\put(10,15){\line(-2,-3){10}}\end{picture}}

\put(40,0){\begin{picture}(40,40)(-220,-15)  
\put(0,-12){\makebox(0,0){$(1,0)$}}  
\put(0,0){\circle*{4}}  
\put(40,0){\circle*{4}}  
\put(20,30){\circle*{4}}  
\put(0,0){\vector(1,0){21}}  
\put(20,0){\line(1,0){20}}  
\put(40,0){\vector(-2,3){11}}  
\put(30,15){\line(-2,3){10}}  
\put(20,30){\vector(-2,-3){11}}  
\put(10,15){\line(-2,-3){10}}\end{picture}}  
  
\put(80,0){\begin{picture}(40,40)(-220,-15)  
\put(0,0){\circle*{4}}  
\put(40,0){\circle*{4}}  
\put(20,30){\circle*{4}}  
\put(0,0){\vector(1,0){21}}  
\put(20,0){\line(1,0){20}}  
\put(40,0){\vector(-2,3){11}}  
\put(30,15){\line(-2,3){10}}  
\put(20,30){\vector(-2,-3){11}}  
\put(10,15){\line(-2,-3){10}}\end{picture}}

\put(120,0){\begin{picture}(40,40)(-220,-15)  
\put(0,0){\circle*{4}}  
\put(40,0){\circle*{4}}  
\put(20,30){\circle*{4}}  
\put(0,0){\vector(1,0){21}}  
\put(20,0){\line(1,0){20}}  
\put(40,0){\vector(-2,3){11}}  
\put(30,15){\line(-2,3){10}}  
\put(20,30){\vector(-2,-3){11}}  
\put(10,15){\line(-2,-3){10}}  
\put(40,-12){\makebox(0,0){$(4,0)$}}  
\end{picture}}

\put(20,30){\begin{picture}(40,40)(-220,-15)  
\put(0,0){\circle*{4}}  
\put(40,0){\circle*{4}}  
\put(20,30){\circle*{4}}  
\put(0,0){\vector(1,0){21}}  
\put(20,0){\line(1,0){20}}  
\put(40,0){\vector(-2,3){11}}  
\put(30,15){\line(-2,3){10}}  
\put(20,30){\vector(-2,-3){11}}  
\put(10,15){\line(-2,-3){10}}\end{picture}}

\put(60,30){\begin{picture}(40,40)(-220,-15)  
\put(0,0){\circle*{4}}  
\put(40,0){\circle*{4}}  
\put(20,30){\circle*{4}}  
\put(0,0){\vector(1,0){21}}  
\put(20,0){\line(1,0){20}}  
\put(40,0){\vector(-2,3){11}}  
\put(30,15){\line(-2,3){10}}  
\put(20,30){\vector(-2,-3){11}}  
\put(10,15){\line(-2,-3){10}}\end{picture}}

\put(100,30){\begin{picture}(40,40)(-220,-15)  
\put(0,0){\circle*{4}}  
\put(40,0){\circle*{4}}  
\put(20,30){\circle*{4}}  
\put(0,0){\vector(1,0){21}}  
\put(20,0){\line(1,0){20}}  
\put(40,0){\vector(-2,3){11}}  
\put(30,15){\line(-2,3){10}}  
\put(20,30){\vector(-2,-3){11}}  
\put(10,15){\line(-2,-3){10}}\end{picture}}  
  
\put(40,60){\begin{picture}(40,40)(-220,-15)  
\put(0,0){\circle*{4}}  
\put(40,0){\circle*{4}}  
\put(20,30){\circle*{4}}  
\put(0,0){\vector(1,0){21}}  
\put(20,0){\line(1,0){20}}  
\put(40,0){\vector(-2,3){11}}  
\put(30,15){\line(-2,3){10}}  
\put(20,30){\vector(-2,-3){11}}  
\put(10,15){\line(-2,-3){10}}\end{picture}}

\put(80,60){\begin{picture}(40,40)(-220,-15)  
\put(0,0){\circle*{4}}  
\put(40,0){\circle*{4}}  
\put(20,30){\circle*{4}}  
\put(0,0){\vector(1,0){21}}  
\put(20,0){\line(1,0){20}}  
\put(40,0){\vector(-2,3){11}}  
\put(30,15){\line(-2,3){10}}  
\put(20,30){\vector(-2,-3){11}}  
\put(10,15){\line(-2,-3){10}}\end{picture}}

\put(60,90){\begin{picture}(40,40)(-220,-15)  
\put(0,0){\circle*{4}}  
\put(40,0){\circle*{4}}  
\put(20,30){\circle*{4}}  
\put(0,0){\vector(1,0){21}}  
\put(20,0){\line(1,0){20}}  
\put(40,0){\vector(-2,3){11}}  
\put(30,15){\line(-2,3){10}}  
\put(20,30){\vector(-2,-3){11}}  
\put(10,15){\line(-2,-3){10}}  
\put(20,40){\makebox(0,0){$(0,4)$}}  
\end{picture}}

\end{picture}  
\end{center}  
\caption{Les graphes $\mathcal{A}_{4}$ pour $SU(2)_4$ (5 vertex) et $SU(3)_4$ (15 vertex).}  
\label{A_SU2,SU3_4}  
\end{figure}
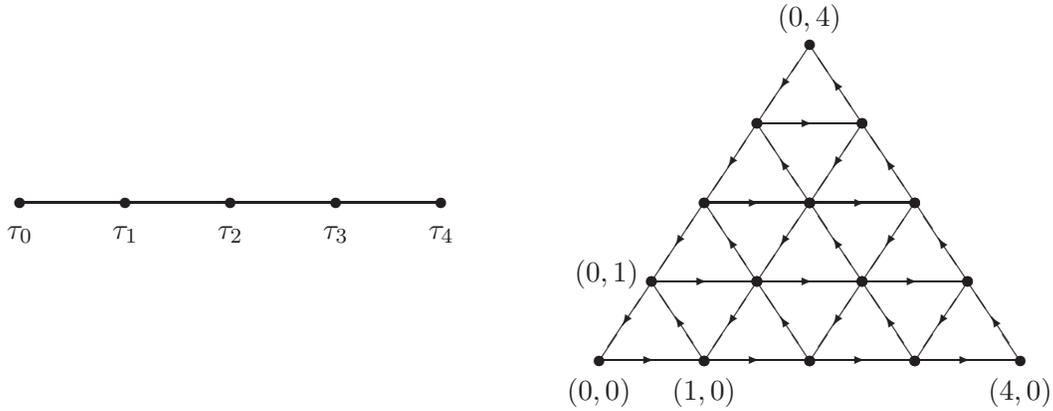

\subsection{Alg\`ebre de graphe et matrices $N_i$}  
Le graphe $\mathcal{A}_{\ell}$ code la d\'ecomposition de la tensorialisation des irreps $(i)$ 
(vertex des graphes)  
par les repr\'esentations fondamentales ($\tau_1$ pour $SU(2)_{\ell}$, $(1,0)$ et $(0,1)$ pour  
$SU(3)_{\ell}$). Cette 
information permet de calculer la fusion de toute irrep:  
\begin{equation} 
(i) . (j) = \sum_k \mathcal{N}_{ij}^k \; (k),  
\end{equation} 
o\`u $\mathcal{N}_{ij}^k$ sont des nombres entiers non-n\'egatifs (multiplicit\'e de $(k)$ dans  
$(i). (j)$). Nous appelons cette alg\`ebre l'{\bf alg\`ebre du graphe} $\mathcal{A}_{\ell}$ ou 
plus simplement l'alg\`ebre $\mathcal{A}_{\ell}$. 
 
\subsubsection{Cas $SU(2)_{\ell}$} 
Les vertex des graphes $A_n$ sont not\'es  
$\tau_i$, pour $i=(0,1,\cdots,\kappa-2)$. Nous savons  
multiplier par l'identit\'e $\tau_0$ et par la fondamentale $\tau_1$ (\`a l'aide du graphe). En imposant   
l'associativit\'e, nous construisons de proche en proche la table de multiplication de l'alg\`ebre  
du graphe $A_n$.  
Par exemple, pour $n>4$, nous avons $\tau_1 . \tau_1 = \tau_0 + \tau_2$, donc nous  
\'ecrivons $\tau_2 = \tau_1 . \tau_1 - \tau_0$, d'o\`u nous d\'eduisons:  
\begin{eqnarray*}  
\tau_2 . \tau_2 = (\tau_1 . \tau_1 - \tau_0). \tau_2 = \tau_1 . \tau_1 . \tau_2 - \tau_0 . \tau_2   
&=& \tau_1 . (\tau_1 + \tau_3) - \tau_2 \\  
{ } &=& \tau_0 + \tau_2 + \tau_2 + \tau_4 - \tau_2 = \tau_0 + \tau_2 + \tau_4   
\end{eqnarray*}  
Par la m\^eme m\'ethode, nous calculons $\tau_2 . \tau_i$, et plus g\'en\'eralement $\tau_i . \tau_j$.  
Nous obtenons ainsi l'alg\`ebre du graphe $A_n$, qui est l'alg\`ebre commutative et associative  
ayant comme \'el\'ements de l'espace vectoriel les combinaisons lin\'eaires des vertex $\tau_i$, et comme multiplication:  
\begin{equation}  
\tau_i . \tau_j = \sum_k \mathcal{N}_{ij}^k \, \tau_k, \qquad \qquad \qquad \mathcal{N}_{ij}^k  
\in \mathbb{N}.  
\label{tau_tau}  
\end{equation}  
Nous associons \`a chaque vertex $\tau_i$ une matrice $(\kappa-1 \times \kappa-1)$  
$N_{i}$, telle que  $(N_i)_{jk} = \mathcal{N}_{ij}^k$. Ces matrices $N_i$ s'obtiennent directement 
par la formule de r\'ecurrence suivante ({\bf formule de r\'ecurrence tronqu\'ee de $SU(2)$}):  
\begin{equation}  
\left.  
\begin{array}{rcl}  
N_0 &=& \munite_{n \times n}  \\  
N_1 &=& \mathcal{G}_{A_n}  \\  
N_1.N_i &=& N_{i-1} + N_{i-2} \qquad \qquad  i=2,\ldots,n-2 \\ 
N_1 . N_{n-1} &=& N_{n-2} 
\end{array}  
\label{SU2_tronquee}  
\right\rbrace  
\end{equation}  
o\`u $\mathcal{G}_{A_n}$ est la matrice d'adjacence du graphe $A_n$. 
Elles forment l'alg\`ebre matricielle du graphe $A_n$ et fournissent une repr\'esentation fid\`ele  
de l'alg\`ebre de graphe $A_n$:  
\begin{equation}   
N_i . N_j = \sum_k \; \mathcal{N}_{ij}^k \, N_k \; = \; \sum_k \; (N_i)_{jk} \, N_k  
\end{equation}

\subsubsection{Cas $SU(3)_{\ell}$} 
Pour $SU(3)_{\ell}$, nous pouvons suivre la m\^eme d\'emarche. Connaissant le graphe $\mathcal{A}_{\ell}$, 
nous obtenons les matrices d'adjacence $N_{(1,0)}$, $N_{(0,1)}$ et 
nous pouvons construire l'alg\`ebre du graphe $\mathcal{A}_{\ell}$, dont une repr\'esentation fid\`ele 
est donn\'ee par les matrices $N_i = N_{\lambda,\mu}$. 
Ces matrices s'obtiennent par la {\bf formule de r\'ecurrence tronqu\'ee de $SU(3)$}:  
\begin{equation}  
\left.  
\begin{array}{lcll}  
N_{\lambda, \mu} &=& 0 & {\text{si} \,   \lambda < 0 \;\, \text{ou} \, \mu <0 } \\  
N_{\lambda, 0} &=& N_{1,0} N_{\lambda-1 , 0} - N_{\lambda-2 , 1} & {} \\  
N_{\lambda, \mu} &=& N_{1,0} N_{\lambda-1 , \mu} -  N_{\lambda-1 ,  
\mu - 1} -  N_{\lambda-2 ,\mu + 1}\qquad & {\text{si} \, \mu \neq 0} \\  
N_{0, \lambda} &=& N_{\lambda, 0}^T & {}  
\end{array}  
\right\rbrace  
\end{equation}  
o\`u $N^T$ d\'esigne la matrice transpos\'ee de $N$.

\subsection{Repr\'esentation du groupe modulaire $SL(2,\mathbb{Z})$ et formule de Verlinde} 
E. Verlinde \cite{verlinde} a montr\'e qu'il existe, dans une th\'eorie conforme rationelle,  
un lien \'etroit entre les coefficients de fusion $\mathcal{N}_{ij}^k$ des champs primaires et  
la matrice $S_{ij}$ des transformations modulaires des caract\`eres de l'alg\`ebre agissant sur  
les champs, donn\'e par la formule de Verlinde:  
\begin{equation}  
\mathcal{N}_{ij}^k = \sum_l \frac{S_{i\ell} S_{j\ell} S_{k\ell}^{\star}}{S_{1\ell}}  
\label{verlindechap3} 
\end{equation}  
Pour les mod\`eles conformes bas\'es sur l'alg\`ebre affine $\widehat{su}(2)_k$ au niveau $k$,  
les matrices $S$ et $T$ de transformation modulaire des caract\`eres sont donn\'ees par: 
\begin{equation} 
\begin{array}{rcl}  
S_{mn} &=& \displaystyle \sqrt{\frac{2}{k+2}} \sin \left( \frac{\pi m n }{k+2} \right) \\  
T_{mn} &=& \displaystyle \exp \left[ 2 i \pi \left( \frac{m^2}{4(k+2)} - \frac{1}{8} \right) \right]  
\end{array} 
\qquad \qquad \qquad m,n=1,\ldots,k+1 
\label{ST-su2} 
\end{equation}  
Un fait remarquable est que les coefficients de fusion $\mathcal{N}_{ij}^k$ calcul\'es par la formule 
de Verlinde \`a partir des matrices donn\'ees en (\ref{ST-su2}) sont identiques aux coefficients 
de structure de l'alg\`ebre de graphe $\mathcal{A}_{k}$ de $SU(2)_{k}$. Les matrices $N_i$ du graphe 
$\mathcal{A}_k$ 
forment donc une repr\'esentation de l'alg\`ebre de fusion de $\widehat{su}(2)_k$: nous  
les appelons par la suite matrices 
de fusion. Cette correspondance est aussi valable pour les cas $SU(n)_{\ell}$. 
Les graphes $\mathcal{A}_{\ell}$ codent donc des informations sur la th\'eorie conforme correspondante. 
Dans une th\'eorie conforme, au lieu de calculer les coefficients de fusion $\mathcal{N}_{ij}^k$ \`a  
partir de la matrice $S$, nous pouvons les obtenir directement \`a partir des matrices de fusion $N_i$ 
calcul\'ees \`a partir du graphe $\mathcal{A}_{\ell}$.  
 
\subsubsection{Matrice $S$} 
Dans le cadre de la correspondance de McKay classique, \`a chaque sous-groupe $\Gamma$ fini de $SU(2)$ 
est associ\'e un ensemble de matrices qui code la tensorialisation des irreps de $\Gamma$. La table 
des caract\`eres de $\Gamma$ peut \^etre reconstruite, sans faire appel \`a la notion de classe 
de conjuguaison, par la diagonalisation simultann\'ee de ces matrices (voir Annexe {\bf B}). 
Dans le cas quantique, nous ne pouvons pas parler de groupe, ni de classe de conjuguaison ni de table 
de caract\`eres. Mais les $N_i$ sont l'analogue des matrices pr\'ec\'edentes. La matrice qui diagonalise 
simultan\'ement les matrices $N_i$ est donc l'analogue quantique de la table des caract\`eres.  
Convenablement normalis\'ee, cette matrice est \'egale \`a la matrice $S$ du groupe modulaire. 
La matrice $S$ peut donc s'obtenir par la diagonalisation des matrices de fusion $N_i$: 
c'est l'inverse de la relation (\ref{verlindechap3}).

\subsubsection{Matrice $T$} 
\`A chaque graphe $\mathcal{A}_k$ de $SU(2)_k$ correspond l'alg\`ebre de  
fusion d'une th\'eorie conforme $\widehat{su}(2)_k$. 
La matrice $T_{ij}$ des transformations des caract\`eres de l'alg\`ebre affine est diagonale. 
Nous pouvons donc associer \`a chaque vertex du graphe $\mathcal{A}_k$ une valeur donn\'ee, 
d\'efinie -- \`a une phase globale pr\`es -- 
par la matrice $T_{ij}$, que nous appelons l'{\bf exposant modulaire}. 
 
Dans le cas de $SU(2)_{k}$, la matrice $T$ est donn\'ee en (\ref{ST-su2}) et l'exposant modulaire 
pour le vertex $\tau_i$ est d\'efini par la quantit\'e $\hat{T}$ donn\'ee par: 
\begin{equation} 
\hat{T} = (i+1) \qquad \mod 4 \kappa \qquad \qquad \qquad \qquad (\kappa=k+2) 
\end{equation} 
Dans le cas de $SU(3)_{k}$, la matrice $T$ est donn\'ee par: 
\begin{equation} 
\left( T^{(k)}\right)_{\lambda \mu} = e_{\kappa} \left[ -(\lambda_1+1)^2 
- (\lambda_1+1).(\lambda_2+1) - (\lambda_2+1)^2 + \kappa \right] 
\delta_{\lambda \mu},  
\end{equation} 
o\`u $\lambda \doteq (\lambda_1,\lambda_2)$, $\mu \doteq (\mu_1,\mu_2)$, $e_{\kappa}[x] \doteq 
\exp \left( \frac{-2 i \pi x}{3\kappa} \right)$, et $\kappa=k +3$. 
L'exposant modulaire est d\'efini par la quantit\'e: 
\begin{equation} 
\hat{T} =  -(\lambda_1+1)^2 
- (\lambda_1+1).(\lambda_2+1) - (\lambda_2+1)^2 + \kappa  \qquad \mod 3 \kappa. 
\end{equation}


\section{Diagrammes de Coxeter-Dynkin g\'en\'eralis\'es $G$}  
\subsection{Historique et d\'efinitions}  
Consid\'erons le groupe $SU(2)$ et l'espace vectoriel dont une base est form\'ee par ses  
irreps $(i)$. Nous pouvons tensorialiser les irreps $(i)$ entre-elles 
et d\'ecomposer le r\'esultat en une somme directe d'irreps: les irreps de $SU(2)$ 
forment une alg\`ebre, dont les coefficients de structure sont des entiers non-n\'egatifs. Soit 
$\Gamma$ un sous-groupe fini de $SU(2)$ et consid\'erons l'espace vectoriel (de dimension finie) 
form\'e par ses irreps $\sigma$. La tensorialisation de l'irrep fondamentale de $SU(2)$ par 
les irreps $\sigma$ de $\Gamma$ est cod\'ee par les diagrammes de Dynkin affine de type $ADE^{(1)}$: 
c'est la correspondance de McKay classique. Nous pouvons reformuler cette correspondance en disant qu'il 
existe une action des irreps de $SU(2)$ sur les irreps de $\Gamma$: nous \'ecrivons que $\widehat{\Gamma}$ 
est un module sur $\widehat{SU}(2)$. Les irreps de $\Gamma$ peuvent aussi \^etre tensorialis\'ees 
entre-elles: elles forment donc aussi une alg\`ebre.  
 
L'analogue quantique de $SU(2)$ sont les groupes quantiques $SU(2)_{\ell}$. Nous pouvons tensorialiser 
les irreps $\tau_i$ de $SU(2)_{\ell}$ entre-elles et d\'ecomposer le r\'esultat en une somme d'irreps. Les 
$\tau_i$ forment donc une alg\`ebre (appel\'ee alg\`ebre de fusion), cod\'ee par les diagrammes 
$\mathcal{A}_{\ell}$: nous dirons que $\mathcal{A}_{\ell}$ poss\`ede {\it self-fusion}. Soit 
un graphe $G$, appelons $\sigma$ ses vertex et consid\'erons l'espace vectoriel $\mathcal{V}(G)$ dont une  
base est form\'ee par l'ensemble des $\sigma$. D\'efinissons l'action du g\'en\'erateur $\tau_1$ de  
$\mathcal{A}_{\ell}$ sur $\sigma$ comme \'etant \'egale \`a la somme des voisins de $\sigma$ sur le graphe $G$. Nous 
pouvons alors rechercher les graphes $G$ de m\^eme norme que $\mathcal{A}_{\ell}$ tels que l'espace  
vectoriel $\mathcal{V}(G)$ soit un module par rapport \`a l'action de $\mathcal{A}_{\ell}$. 
Les graphes poss\'edant ces propri\'et\'es sont\footnote{En fait cette caract\'erisation inclut
\'egalement les diagrammes de type {\it tadpole} qu'il convient d'\'eliminer.} les 
diagrammes de Dynkin de type $ADE$. Les vertex de ces 
diagrammes sont donc consid\'er\'es comme des irreps. Certains de ces diagrammes poss\`edent  
{\it self-fusion}, c.\`a.d. que nous pouvons d\'efinir une structure alg\'ebrique commutative et  
associative dont les coefficients de structure soient des entiers non-n\'egatifs (nous pouvons 
tensorialiser les irreps entre-elles). C'est le cas des diagrammes $A_n, D_{2n}, E_6$ et $E_8$, et nous 
dirons qu'ils forment un ``sous-groupe'' du $SU(2)_{\ell}$ correspondant. Les autres diagrammes ($D_{2n+1}$ 
et $E_7$) seront appel\'es ``modules''. Notons que les diagrammes $D_n$ sont obtenus par une proc\'edure de  
{\it folding} \`a partir des diagrammes $A_n$. Les diagrammes $E_6, E_7$ et $E_8$ sont dits exceptionnels. 
On peut montrer que $E_7$ est obtenu \`a partir d'un {\it twist} du diagramme $D_{10}$.  
Nous avons une correspondance bi-univoque entre les sous-groupes finis de $SU(2)$ (diagrammes 
$ADE^{(1)}$) et leur analogue quantique (diagrammes $ADE$). D'autre part, rappelons 
qu'il existe une correspondance 
bi-univoque entre les fonctions de partition invariante modulaire des th\'eories conformes  
$\widehat{su}(2)$ et les diagrammes $ADE$.\\ 
 
Qu'en est-il pour $SU(3)$? Historiquement, une premi\`ere classification empirique   
des diagrammes de Coxeter-Dynkin $G$ g\'en\'eralis\'es pour $SU(3)_{\ell}$ (appel\'es diagrammes 
de Di Francesco-Zuber) a \'et\'e d\'etermin\'ee par CAF (``{\it Computer Aided Flair}'') par  
Di Francesco et Zuber en 1989 \cite{DiFZub}. Contrairement au cas $SU(2)$, il n'y a pas de correspondance  
bi-univoque entre ces graphes et les graphes codant les sous-groupes finis de $SU(3)$.  
Par la suite, la classification des fonctions de partition invariante modulaire des th\'eories 
conformes $\widehat{su}(3)$ a \'et\'e obtenue par Gannon \cite{gannon-class}, mais la correspondance  
entre les graphes de la liste de Di Francesco-Zuber et la classification de Gannon n'\'etait pas claire.  
Dans un effort de stimulation priv\'ee \`a la recherche, la r\'ecompense propos\'ee par Zuber pour 
la classification des diagrammes g\'en\'eralis\'es pour $SU(3)_{\ell}, SU(4)_{\ell}$ et $SU(5)_{\ell}$ 
\'etait respectivement 1,2 et 3 bouteilles de champagne. 
 
La classification des diagrammes g\'en\'eralis\'es de $SU(3)_{\ell}$ et $SU(4)_{\ell}$ a \'et\'e  
obtenue par A. Ocneanu et pr\'esent\'ee \`a l'Ecole de Bariloche 2000 \cite{Oc-Bariloche}. Pour le cas 
$SU(3)_{\ell}$, il y a la s\'erie $\mathcal{A}_n$, et trois s\'eries qui g\'en\'eralisent les 
deux s\'eries $D_{2n}$ et $D_{2n+1}$ de $SU(2)_{\ell}$, nous pourrions les appeler $\mathcal{D}_{3n}$, 
$\mathcal{D}_{3n+1}$ et $\mathcal{D}_{3n+2}$. Il y a \'egalement deux s\'eries $\mathcal{A}_n^c$ 
et $3\mathcal{A}_n^c$ obtenues \`a partir de la s\'erie $\mathcal{A}_n$ pr\'ec\'edente en utilisant la  
conjuguaison complexe. Sur ces six s\'eries, seulement deux poss\`edent {\it self-fusion}: $\mathcal{A}_n$ 
et $\mathcal{D}_{3n}$. Il y a 7 graphes exceptionnels,  dont trois poss\`edent {\it self-fusion} 
(ils g\'en\'eralisent les graphes $E_6$ et $E_8$ de $SU(2)_{\ell}$): ce sont $\mathcal{E}_5$, 
$\mathcal{E}_9$ et $\mathcal{E}_{21}$. Il y a un graphe $\mathcal{D}_9^t$ obtenu \`a partir de  
$\mathcal{D}_9$ par un twist exceptionnel (comme $E_7$ \`a partir de $D_{10}$). Les trois autres 
exceptionnels ($\mathcal{E}_5^c$, $\mathcal{E}_9^c$ et $\mathcal{D}_9^{tc}$) font usage d'un morphisme 
associ\'e \`a la conjuguaison complexe $c$. La classification de Gannon des fonctions de partition 
invariante modulaire consiste en six s\'eries et six cas exceptionnels. Notons que deux graphes 
distincts (donc deux th\'eories conformes tr\`es diff\'erentes) poss\`edent la m\^eme fonction 
de partition invariante modulaire. 
La liste d'Ocneanu confirme celle trouv\'ee empiriquement par Di Francesco-Zuber, \`a l'exception d'un  
diagramme exceptionnel, exclu pour ne pas satisfaire des conditions locales de nature cohomologique  
(d\'efinition de cellules et de connections internes): c'est le graphe not\'e $\mathcal{E}_3^{(12)}$  
dans \cite{DiFZub}. La liste des diagrammes pour $SU(3)_{\ell}$ ainsi que pour $SU(4)_{\ell}$ est publi\'ee 
dans \cite{Oc-Bariloche}. 
 
Bien qu'utilis\'ee par Ocneanu, insistons sur le fait qu'aucune d\'efinition math\'ematiquement 
rigoureuse des propri\'et\'es 
devant \^etre satisfaites par ces diagrammes (appel\'es ``{\it Higher Coxeter System}'' dans \cite{Oc-MSRI}) 
n'est disponible dans la litt\'erature. Par la suite, 
nous prendrons ces diagrammes comme donn\'ee de d\'epart.

\subsection{Alg\`ebre de graphe} 
Par la suite, nous traiterons en d\'etail le cas $SU(2)_{\ell}$. La g\'en\'eralisation au cas  
$SU(3)_{\ell}$ est souvent imm\'ediate, nous le mentionnerons seulement lorsque cette g\'en\'eralisation 
n'est pas triviale. 
 
\subsubsection{Graphe $G$ et matrice d'adjacence} 
 
Nous prenons comme point de d\'epart les graphes $G$ de type $ADE$, list\'es dans  
l'Annexe {\bf A}, et appelons $r$ le nombre de vertex du graphe consid\'er\'e.  
Les vertex de ces graphes seront not\'es $\sigma_a$, pour $a=(0,1,\cdots,r-1)$. 
\`A chaque graphe $G$ est associ\'ee sa matrice d'adjacence, not\'ee $\mathcal{G}$, qui est une matrice   
carr\'ee $r \times r$, telle que $\mathcal{G}_{ij}=1$ si un arc relie les vertex correspondant \`a $i$ et  
$j$ et $\mathcal{G}_{ij}=0$ sinon.  
Tous les graphes $ADE$ sont bi-orient\'es, leur matrice d'adjacence est donc sym\'etrique.  
La norme du graphe est d\'efinie comme  \'etant la plus grande valeur propre de sa matrice d'adjacence, et  
est not\'ee $\beta$. Pour tous les graphes de type $ADE$, nous avons:  
\begin{equation}  
\beta = 2 \cos \left( \frac{\pi}{\kappa} \right).  
\label{kappaSU2}  
\end{equation}  
Cette relation d\'efinit le nombre $\kappa$, appel\'e le nombre (dual) de Coxeter du  
graphe\footnote{$\kappa$ est ainsi appel\'e car il correspond au nombre (dual) de Coxeter des alg\`ebres  
de Lie semi-simples simplement lac\'ees, d\'ecrites par les diagrammes de Dynkin de type $ADE$  
correspondant: nous le d\'efinissons ici sans faire appel au langage des alg\`ebres de Lie.}.   
Nous avons $1 < \beta < 2$: de fait, les graphes $ADE$ apparaissent aussi dans la classification des  
graphes bi-orient\'es de norme strictement comprise entre un et deux (voir chapitre {\bf 2}).  
Introduisons le nombre quantique $[n]_q$, d\'efini par:  
\begin{equation} 
[n]_{q}=\frac{q^n - q^{-n}}{q - q^{-1}}, \qquad \qquad q = \exp(\frac{i\pi}{\kappa}), \qquad  
\qquad q^{2\kappa}=1.  
\end{equation} 
Alors, la norme $\beta$ du graphe est \'egale au q-nombre deux: $\beta=[2]_q = q+1/q=2 \cos(\pi/\kappa)$.  
Nous rappelons la valeur de $\kappa$ pour chaque graphe $ADE$ dans la Tab. \ref{kappaADE}.  
\begin{table}[hhh]  
$$  
\begin{array}{|c|ccccc|}  
\hline  
\quad {} \quad & \quad A_n \quad & \quad D_{n} \quad &\quad  E_6 \quad  & \quad E_7 \quad & \quad E_8  
\quad \\  
\hline  
\kappa & n+1 & 2(n-1) & 12 & 18 & 30 \\  
\hline  
\end{array}  
$$  
\caption{Nombre (dual) de Coxeter $\kappa$ pour les graphes de type $ADE$.}  
\label{kappaADE} 
\end{table}  
  
Le vecteur propre correspondant \`a $\beta$ est not\'e $P$ (vecteur de Perron-Frobenius).   
Les composantes de ce vecteur correctement normalis\'e nous donnent, par d\'efinition, les  
{\bf dimensions quantiques} des irreps.   
La normalisation est telle que $P(\sigma_0)=[1]_q=1$. Alors, dans tous les cas, la dimension quantique de la  
repr\'esentation fondamentale ($\sigma_1$) est \'egale \`a $[2]_q$.

\subsubsection{Alg\`ebre de graphe}  
Nous consid\'erons l'espace vectoriel $\mathcal{V}(G)$ dont la base est donn\'ee par les  
vertex $\sigma_a$ du graphe $G$, consid\'er\'es comme des \'el\'ements lin\'eairement ind\'ependants.  
Dans tous les cas, $\sigma_0$ est appel\'e l'identit\'e,  et $\sigma_1$ la fondamentale. Dans le cas  
classique $SU(2)$, les graphes $ADE^{(1)}$ codent la d\'ecomposition  
de $f \otimes \sigma_a$ en somme directe d'irreps $\sigma_b$ (o\`u $f$ est la repr\'esentation fondamentale  
de SU(2)). Les graphes $G$ de type $ADE$ sont vus comme codant la d\'ecomposition du produit des  
irreps par la fondamentale $f=\sigma_1$:  
\begin{equation}  
\sigma_1 . \sigma_a = \sigma_a . \sigma_1 = \sum_{b:a} \sigma_b,   
\label{voisins}  
\end{equation}  
o\`u $b:a$ signifie que $\sigma_b$ est voisin de $\sigma_a$ sur le graphe $G$.   
Nous voulons \'etendre la d\'efinition du produit \`a tout l'espace vectoriel $\mathcal{V}(G)$,  
en analogie avec la d\'ecomposition du produit tensoriel du cas classique, de mani\`ere \`a obtenir  
une alg\`ebre commutative et associative:  
\begin{equation}  
\sigma_a . \sigma_b = \sum_c G_{ab}^c \sigma_c, \qquad \qquad \qquad G_{ab}^c \in \mathbb{N},  
\label{prod_ADE}  
\end{equation}  
o\`u les coefficients $G_{ab}^c$ sont vus comme la multiplicit\'e de $\sigma_c$ dans $\sigma_a . \sigma_b$  
(ce sont des entiers non-n\'egatifs). L'obtention de la table de multiplication se fait   
par construction, de proche en proche, comme dans l'exemple de $SU(2)$ trait\'e dans l'Annexe {\bf B}.  
Nous savons multiplier par l'identit\'e $\sigma_0$ et par la fondamentale $\sigma_1$ (\`a l'aide du graphe).  
En imposant l'associativit\'e du produit, nous pouvons essayer de construire la multiplication par les  
autres irreps.  
Le r\'esultat suivant a \'et\'e premi\`erement obtenu par Pasquier\cite{pasquier-ADE}, et les alg\`ebres  
obtenues sont quelquefois appel\'ees alg\`ebres de Pasquier: 
\begin{theo}  
Soit $G$ un graphe de type $ADE$ \`a $r$ vertex, et $\mathcal{V}(G)$ l'espace vectoriel  
dont la   
base est form\'ee par les vertex $\sigma_a$ de $G$. Les graphes pour lesquels il est possible de d\'efinir  
un produit satisfaisant (\ref{voisins}) et (\ref{prod_ADE}) sont les graphes $A_n$, $D_{2n}$, $E_6$ et  
$E_8$.   
\end{theo}  
  
Pour $A_n$, $E_6$ et $E_8$, nous pouvons construire la table de multiplication de mani\`ere unique.  
Pour les cas $D_{2n}$, nous trouvons une famille \`a un param\`etre de solutions, mais en imposant   
que les coefficients de structures soient des entiers non n\'egatifs, la solution devient  
unique \cite{Coq_Gil-ADE} (voir d\'etails dans la section correspondante). Pour ces  
cas, nous obtenons donc une alg\`ebre commutative et associative, dont les coefficients de structure   
sont des entiers non-n\'egatifs, appel\'ee {\bf alg\`ebre de graphe}, et que nous notons   
par le m\^eme symbole $G$ que le graphe lui-m\^eme: nous dirons par la suite que ces graphes   
poss\`edent ``{\bf \it self-fusion}'', ou de mani\`ere \'equivalente, qu'ils sont de {\bf type I}.  
  
Pour $D_{2n+1}$ et $E_7$, il est impossible d'obtenir une alg\`ebre commutative et associative, avec des  
coefficients de structure qui soient des entiers non-n\'egatifs \cite{Pasquier} (pour le cas $E_7$, une  
alg\`ebre existe, mais avec des coefficients de structure n\'egatifs).  
Ces cas ne poss\`edent pas ``{\it self-fusion}'', les graphes sont dits de {\bf type II}.

\subsubsection{Alg\`ebre matricielle de graphe}  
Dans les cas des graphes qui poss\`edent {\it self-fusion}, nous pouvons donner une r\'ealisation  
matricielle de leur alg\`ebre de graphe. \`A chaque irrep $\sigma_a$, nous associons une matrice  
$r \times r$ $G_a$ telle que:  
\begin{equation} 
(G_a)_{bc} = G_{ab}^c.    
\end{equation} 
Ces matrices forment une repr\'esentation fid\`ele de l'alg\`ebre $G$ (lorsqu'elle existe) que nous  
appelons {\bf alg\`ebre matricielle de $G$}, et v\'erifient:  
$$  
G_a . G_b = \sum_c (G_a)_{bc} G_c  
$$  
Comme l'alg\`ebre de graphe est commutative, les matrices $G_a$ commutent \'evidemment entre-elles:  
$$  
G_a.G_b = \sum_c (G_a)_{bc} G_c = \sum_c G_{ab}^c G_c = \sum_c G_{ba}^c G_c = \sum_c (G_b)_{ac} G_c =  
G_b . G_a  
$$  
Nous avons $G_0 = \munite_{r \times r}$ et $G_1 = \mathcal{G}$, la matrice d'adjacence du graphe.  
Dans tous les cas (\`a l'exception de $D_{2n}$), les autres matrices $G_a$ s'expriment comme des  
polyn\^omes de $G_0$ et $G_1$: pour ces cas, nous donnons les expressions polynomiales permettant d'obtenir  
ces matrices.

\subsection{$\mathcal{V}(G)$ comme module sur $\mathcal{A}(G)$: matrices $F_i$ (ou $E_a$)}  
 
\`A chaque graphe $G$ de type $ADE$ est associ\'e un autre graphe, not\'e $\mathcal{A}(G)$,  
appartenant \`a la s\'erie $A_n$ et ayant la m\^eme norme (et donc m\^eme nombre de Coxeter $\kappa$) que $G$.  
Si $\kappa$ est le nombre de Coxeter de $G$, alors $\mathcal{A}(G) = A_{\kappa-1}$.  
Nous avons donc les correspondances suivantes:  
$$  
\begin{array}{cclcl}  
\mathcal{A}(A_n) &=& A_n &\qquad \qquad& \kappa = n+1 \\  
\mathcal{A}(D_n) &=& A_{2n-3} &\qquad \qquad& \kappa = 2n-2 \\  
\mathcal{A}(E_6) &=& A_{11} &\qquad \qquad& \kappa = 12 \\  
\mathcal{A}(E_7) &=& A_{17} &\qquad \qquad& \kappa = 18 \\  
\mathcal{A}(E_8) &=& A_{29} &\qquad \qquad& \kappa = 30   
\end{array}  
$$  
 
Les vertex $\tau_i$ ($i=0,\ldots, \kappa-2$) du graphe $\mathcal{A}(G)$ forment un espace vectoriel, sur  
lequel nous pouvons d\'efinir une multiplication: nous obtenons l'alg\`ebre de graphe $\mathcal{A}(G)$.   
Les vertex $\sigma_a$ ($a=0,\ldots, r-1$) du graphe $G$ forment aussi un espace vectoriel,  
not\'e $\mathcal{V}(G)$.  
Nous voulons d\'efinir une action de $\mathcal{A}(G)$ sur $\mathcal{V}(G)$:  
\begin{eqnarray}  
\mathcal{A}(G) \times \mathcal{V}(G) &\rightarrow & \mathcal{V}(G)  \nonumber \\  
\tau_i . \sigma_a &=& \sum_b  \mathcal{F}_{ia}^b \sigma_b   
\label{tau_sig}  
\end{eqnarray}  
telle que les coefficients $\mathcal{F}_{ia}^b$ soient des entiers non-n\'egatifs  
(ainsi $\mathcal{F}_{ia}^b$ repr\'esente la   
multiplicit\'e de $\sigma_b$ dans $\tau_i . \sigma_a$, en analogie avec le cas classique).  
Pour que cette action soit bien d\'efinie, il faut imposer:  
\begin{equation}  
(\tau_i . \tau_j) . \sigma_a = \tau_i . (\tau_j . \sigma_a). 
\label{EmodsurA}  
\end{equation}  
Utilisant (\ref{tau_tau}) et (\ref{tau_sig}), et du fait que les $\sigma$ sont des \'el\'ements  
lin\'eairement ind\'ependants de $\mathcal{V}(G)$, les coefficients $\mathcal{F}_{ij}^k$ doivent satisfaire  
les relations suivantes:  
\begin{equation}  
\sum_b  \mathcal{F}_{ja}^b \; \mathcal{F}_{ib}^c = \sum_k  \mathcal{N}_{ij}^k \; \mathcal{F}_{ka}^c    
\label{F_N}  
\end{equation}  
Les indices $(a,b,c,\ldots)$ sont r\'eserv\'es pour les vertex $\sigma$ de $G$ $(a,b,c = 0,1,\ldots,r-1)$,  
et les  
indices $(i,j,k,\ldots)$ sont r\'eserv\'es pour les vertex $\tau$ de $\mathcal{A}(G)$  
$(i,j,k = 0,1,\ldots,\kappa-2)$. Nous pouvons parler d'un analogue quantique de la 
correspondance de McKay: le graphe $G$ code la d\'ecomposition des irreps $\sigma$  
de $G$  
par l'irrep fondamentale $\tau_1$ du quotient de $U_q(sl(2))$ correspondant. Il est donc naturel de poser  
$\tau_1 . \sigma_a = \sigma_1 . \sigma_a = \sum_{b:a} \sigma_b$ (la sommation se fait sur les voisins de  
$\sigma_a$ sur le graphe $G$).  
L'action explicite de $\mathcal{A}(G)$ sur $\mathcal{V}(G)$ est d\'efinie par:  
\begin{equation}  
\left.  
\begin{array}{rcl}  
\mathcal{A}(G) \times \mathcal{V}(G) &\rightarrow & \mathcal{V}(G) \\  
\tau_0 . \sigma_a &\doteq& \sigma_0 . \sigma_a = \sigma_a \\  
(\tau_1)^n . \sigma_a &\doteq& (\sigma_1)^n . \sigma_a \qquad \qquad \qquad n>1  
\end{array}  
\right\rbrace  
\label{tau_sig_expl}  
\end{equation}  
\`A noter que m\^eme pour les cas o\`u $G$ ne poss\`ede pas de self-fusion, l'action est bien d\'efinie:  
m\^eme si nous ne pouvons calculer $\sigma_a . \sigma_b$, la multiplication $\sigma_a . \sigma_1$ existe  
toujours.  
$\tau_1$ est l'irrep fondamentale de $\mathcal{A}(G)$: tout irrep $\tau_i$ s'exprime par un polyn\^ome   
sur $\tau_1$ et $\tau_0$. Par exemple:  
$$  
\tau_2 = \tau_1 . \tau_1 - \tau_0 \qquad \qquad \qquad \tau_3 = \tau_1 . \tau_1 . \tau_1 - \tau_1 - \tau_1  
$$  
De mani\`ere g\'en\'erale, nous \'ecrivons: $\tau_i = Pol_i (\tau_0, \tau_1)$.  
Alors, l'action d'un \'el\'ement   
quelconque $\tau_i$ de $\mathcal{A}(G)$ sur $\mathcal{V}(G)$ s'\'ecrit:  
\begin{equation}  
\tau_i . \sigma_a = Pol_i (\tau_0,\tau_1) . \sigma_a  
= Pol_i (\sigma_0, \sigma_1) . \sigma_a  
\label{tau_sig_pol} 
\end{equation}

\subsubsection{Matrices $F_i$}  
  
Nous pouvons coder matriciellement l'action (\ref{tau_sig}). Introduisons $\kappa -1$  
matrices $(r \times r)$ $F_i$ telles que $(F_i)_{ab}= \mathcal{F}_{ia}^b $.   
De la propri\'et\'e de structure de module (\ref{EmodsurA}), nous concluons que les matrices $F_i$  
satisfont:  
\begin{equation}  
F_i\;F_j = \sum_k \mathcal{N}_{ij}^kF_k   
\label{rec_F}  
\end{equation}  
Elles forment donc une repr\'esentation de l'alg\`ebre de fusion de dimension $r$ (les matrices $N_i$  
forment une  repr\'esentation fid\`ele de l'alg\`ebre de fusion, de dimension $\kappa -1$).  
Du fait que $\tau_0 . \sigma_a = \sigma_a$ et que $\tau_1 . \sigma_a = \sigma_1 . \sigma_a$, et par  
(\ref{rec_F}), nous concluons que les matrices $F_i$ s'obtiennent directement par:  
\begin{equation}  
\begin{array}{rcl}  
F_0 &=& \munite_{n \times n}  \\  
F_1 &=& \mathcal{G}_{G}  \\  
F_1.F_i &=& F_{i-1} + F_{i-2} \qquad \qquad  i=2,\ldots,\kappa-2 \\ 
F_1 . F_{n-1} &=& F_{n-2} 
\end{array}   
\label{recur_F} 
\end{equation}

\begin{remarq}  
Par la relation de r\'ecurrence (\ref{recur_F}) d\'efinissant les matrices $F_i$ et comparant avec la  
relation (\ref{pascalmod}), nous concluons que l'\'el\'ement $(F_i)_{ab}$ repr\'esente le nombre de chemins  
essentiels de longueur $i$, partant du vertex  $\sigma_a$ et arrivant au vertex  $\sigma_b$, sur  
le graphe $G$.   
\end{remarq}  
  
\subsubsection{Matrices $E_a$}  
D\'efinissons $r$ matrices   
$(\kappa-1 \times r)$ $E_a$ telles que:  
\begin{equation} 
(E_a)_{ib} = (F_i)_{ab}  
\end{equation} 
Alors l'action (\ref{tau_sig}) s'\'ecrit aussi:  
\begin{equation}  
\tau_i . \sigma_a = \sum_b (E_a)_{ib} \; \sigma_b  
\label{E_a}  
\end{equation}  
Les matrices $E_a$ sont appel\'ees les {\bf matrices essentielles} du graphes $G$. L'une d'entre elles en  
particulier, poss\`ede des propri\'et\'es int\'eressantes: la matrice $E_0$, appel\'ee dans la litt\'erature  
de physique statistique ``{\it intertwiner}''.  
 \begin{theo}  
Soient $G$ un graphe $ADE$ et $\mathcal{A}(G)$ le graphe de la s\'erie $A_n$ ayant le m\^eme   
nombre de Coxeter $\kappa$, et soit $G_1 = \mathcal{G}_G$ et  
$N_1 = \mathcal{G}_{\mathcal{A}(G)}$ leur matrice d'adjacence respective. Alors:  
\begin{equation}  
N_1 \; E_0 = E_0 \; G_1  
\label{sym_E0}  
\end{equation}  
\end{theo}  
{\it \ud{D\'emonstration}}:  
Consid\'erons l'\'equation (\ref{E_a}) pour $a=0$, et appliquons $\tau_1$. Le terme de gauche s'\'ecrit:  
\begin{equation*}  
\tau_1 . \tau_i .  \sigma_0  = (\tau_1 . \tau_i) .  \sigma_0 = \sum_j (N_1)_{ij} \tau_j . \sigma_0   
 = \sum_j \sum_c (N_1)_{ij} (E_0)_{jc} \sigma_c  
\end{equation*}  
Le terme de droite s'\'ecrit:  
\begin{equation*}  
\sum_b (E_0)_{ib} \tau_1 . \sigma_b = \sum_b (E_0)_{ib} \sigma_1 . \sigma_b 
  = \sum_b \sum_c (E_0)_{ib} (G_1)_{bc} \sigma_c  
\end{equation*}  
Comme les $\sigma$ sont lin\'eairement ind\'ependants, l'\'egalit\'e est valable pour tout terme:  
$$  
 \displaystyle \sum_j (N_1)_{ij} (E_0)_{jc} = \displaystyle \sum_b (E_0)_{ib} (G_1)_{bc}   
\;\; \iff \;\;  (N_1 . E_0)_{ic} = (E_0 . G_1)_{ic}     
$$  
\begin{flushright}  
$\blacksquare$  
\end{flushright}  
\begin{coro}  
Soit $\beta$ la plus grande valeur propre de $G$ (donc de $\mathcal{A}(G)$ aussi) et $P$ le vecteur  
propre normalis\'e correspondant. Alors $(E_0 . P)$ est le vecteur propre normalis\'e de $\mathcal{A}(G)$ 
correspondant \`a  $\beta$.   
\end{coro}  
  
Soient deux graphes $G_1$ et $G_2$ tels qu'il existe une matrice card($G_1$)$\times$card($G_2$) $E_0$  
qui relie leur matrice d'adjacence comme dans (\ref{sym_E0}). Si deux mod\`eles statistiques sont d\'ecrits   
par ces deux graphes, alors ils auront certaines propri\'et\'es communes (\'energie libre, charge  
centrale), bien qu'ils diff\`erent au niveau de l'alg\`ebre des op\'erateurs \cite{Roche-OcCell}.   
   
\subsubsection{R\`egles de branchement $\mathcal{A}(G) \hookrightarrow G$}  
Dans le cas o\`u les graphes $G$ poss\`edent {\it self-fusion} (type I), l'action (\ref{tau_sig})  
est compatible avec la structure alg\'ebrique de $G$:  
\begin{equation} 
\tau_i . (\sigma_a . \sigma_b) = (\tau_i . \sigma_a) . \sigma_b  
\end{equation} 
Au niveau matriciel, ceci se traduit par les relations suivantes entre les matrices $F_i$ et $G_a$:  
\begin{equation}  
\left[ F_i, G_a \right] = 0, \qquad \qquad \qquad   
F_i \; G_a = \sum_{b} (F_i)_{ab} G_b.  
\end{equation}  
L'action de $\tau_i$ sur $\sigma_a$ est donn\'ee par (\ref{tau_sig_pol}). Pour $a=0$, nous avons:  
 \begin{equation} 
\tau_i . \sigma_0 = Pol_i (\sigma_0, \sigma_1) = \sum_b (F_i)_{0b} \sigma_b 
\end{equation} 
Pour les cas poss\'edant self-fusion, l'action de $\tau_i$ sur $\sigma_a$ peut donc s'\'ecrire:  
\begin{equation}  
\tau_i . \sigma_a = \sum_b (F_i)_{0b} \; \sigma_b . \sigma_a 
\label{tausig_F}  
\end{equation}  
L'\'equation (\ref{tausig_F}) nous donne les r\`egles de branchement:  
\begin{equation}  
\tau_i \hookrightarrow \sum_b (F_i)_{0b} \; \sigma_b = \sum_b (E_0)_{ib} \; \sigma_b 
\end{equation} 
Les r\`egles de branchement sont donc enti\`erement cod\'ees dans la matrice $E_0$:  
la restriction de l'irrep $\tau_i$ vers les irreps $\sigma$ de $G$ se lit dans la  
ligne correspondante \`a $\tau_i$ de la matrice $E_0$.   
\begin{coro}  
La connaissance de $E_0$ et des matrices $G_a$ nous permet de d\'eterminer les autres matrices $E_a$:  
\begin{equation} 
E_a = E_0 . G_a  
\end{equation} 
\end{coro}  
{\it \ud{D\'emonstration}}:  
D'une part, l'action est donn\'ee par: $\tau_i . \sigma_a = \sum_c (E_a)_{ic} \; \sigma_c$. D'autre part,  
en utilisant les r\`egles de branchement $\tau_i \hookrightarrow \sum_b (E_0)_{ib} \sigma_b$, nous avons:  
\begin{equation*}  
\tau_i . \sigma_a = \sum_b (E_0)_{ib} \sigma_b . \sigma_a = \sum_c \sum_b (E_0)_{ib} (G_a)_{bc}  
\sigma_c  
= \sum_c (E_0 . G_a)_{ic} \sigma_c  
\end{equation*}  
Les $\sigma$ \'etant lin\'eairement ind\'ependants, nous avons finalement:  
$  
(E_a)_{ic} = (E_0.G_a)_{ic}  
$  
\hfill $\blacksquare$  
\vspace{0.15cm}
 
\noindent La correspondance entre $G$ et $\mathcal{A}(G)$ peut \^etre vue de deux mani\`eres  
compl\'ementaires:  
\begin{itemize}  
\item[$\bullet$] Dans le langage de la correspondance de Mc-Kay quantique, le graphe $G$ est reli\'e  
\`a un ``sous-groupe'' (ou ``module'') du quotient de $U_q(sl(2))$ pour $q^{2\kappa=1}$, ce dernier  
\'etant reli\'e au graphe $A_{\kappa-1}$, c.\`a.d. justement \`a $\mathcal{A}(G)$. L'espace
vectoriel $\mathcal{V}(G)$ engendr\'e par les vertex de $G$ est toujours un module sous l'action
de l'alg\`ebre  $\mathcal{A}(G)$. 
\item[$\bullet$] Les chemins essentiels sont d\'efinis sur le graphe $G$, et leur nombre 
est cod\'e dans les matrices $F_i$ ou dans les matrices essentielles $E_a$. $\mathcal{B}(G)$ est la  
dig\`ebre des endomorphismes de chemins essentiels. Les deux lois multiplicatives 
sont la composition $\circ$ et la convolution $\odot$. La diagonalisation de $\mathcal{B}G$ pour  
la loi $\circ$ donne lieu \`a des projecteurs minimaux centraux, dont la multiplication par la loi  
$\odot$ est cod\'ee par le graphe $\mathcal{A}(G)$.  
\end{itemize}  
 
Ces constructions se g\'en\'eralisent pour les cas $SU(n)_{\ell}$: les matrices $F_i$ codant 
l'action de $\mathcal{A}(G)$ sur $\mathcal{V}(G)$ donnent une repr\'esentation de l'alg\`ebre 
de fusion. Cependant, les chemins essentiels n'ont \'et\'e d\'efinis que pour les graphes de type $ADE$ 
(plus exactement pour des graphes bi-partites). L'extension de la d\'efinition des chemins essentiels 
pour des graphes de type $SU(n)_{\ell}$ est possible (la notion de ``longueur'' \'etant remplac\'ee
par un tableau de Young), et nous pouvons \'etablir un lien avec des g\'en\'eralisations
des alg\`ebres de Jones-Temperley-Lieb.


\section{Graphes d'Ocneanu $Oc(G)$}  
 
\subsection{D\'efinition} 
La dig\`ebre $\mathcal{B}(G)$ des endomorphismes de chemins essentiels sur un graphe
$G$ (de type $ADE$) poss\`ede deux lois multiplicatives $\circ$ et $\odot$. La diagonalisation 
de $\mathcal{B}(G)$ pour la loi $\odot$ donne lieu \`a des projecteurs minimaux centraux\footnote{Nous
devons normaliser ces op\'erateurs pour le produit scalaire de l'espace dual: voir la discussion
de l'exemple $A_3$ dans le chapitre {\bf 2}.}, dont la 
multiplication par la loi $\circ$ est cod\'ee par le graphe d'Ocneanu de $G$. C'est ainsi que  
sont formellement introduits les graphes d'Ocneanu de $G$. Soulignons toutefois que la diagonalisation 
explicite de la loi $\odot$ de $\mathcal{B}(G)$ est d'une extr\^eme complexit\'e. Ces graphes ont  
\'et\'e introduits par A. Ocneanu  
\cite{Oc-Marseille}, et sont premi\`erement apparus dans la litt\'erature dans \cite{Oc-paths}. 
Ces graphes et l'alg\`ebre des sym\'etries quantiques jouent un r\^ole important dans l'\'etude  
des syst\`emes conformes \`a deux dimensions.
Ocneanu a utilis\'e la classification des fonctions de partition invariante modulaire des 
mod\`eles $\widehat{su}(2)$ pour construire ces graphes: une explicite diagonalisation  
de la loi $\odot$ ne semble pas avoir \'et\'e vraiment effectu\'ee $\cdots$ 
 
Nous prenons les graphes d'Ocneanu dans un premier temps comme donn\'ee initiale. Nous d\'efinirons 
par la suite une r\'ealisation de l'alg\`ebre d'Ocneanu \`a partir de l'alg\`ebre d'un graphe $G$. 
 
Les graphes d'Ocneanu ne sont d\'efinis et publi\'es que pour les cas $ADE$, donc pour les mod\`eles 
$SU(2)_{\ell}$. Les d\'efinitions de chemins essentiels  n'ont pas \`a nos jours \'et\'e \'etendues 
\`a des diagrammes de Coxeter-Dynkin g\'en\'eralis\'es. La dig\`ebre que l'on pourrait associer 
\`a ces diagrammes reste donc un objet ``virtuel''. M\^eme si une construction de cette big\`ebre 
\'etait d\'efinie, l'explicite diagonalisation des deux lois $\circ$ et $\odot$ et l'obtention 
des projecteurs minimaux centraux pour ces lois, conduisant \`a la construction des graphes 
d'Ocneanu correspondants, serait une t\^ache ardue $\ldots$ 
 
Toutefois, notre r\'ealisation de l'alg\`ebre d'Ocneanu se pr\^ete naturellement 
\`a une g\'en\'eralisation aux cas $SU(n)_{\ell}$. Nous pouvons ainsi obtenir des {\it ``graphes 
d'Ocneanu''} pour ces mod\`eles: nous verrons notamment que les fonctions de partition invariantes 
modulaires qui y sont associ\'ees pour certains exemples trait\'es du cas $SU(3)_{\ell}$ 
sont identiques \`a la classification de Gannon, confirmant ainsi notre construction.

\subsubsection{Alg\`ebre du graphe $Oc(G)$}  
Nous prenons comme donn\'ee initiale les graphes d'Ocneanu $Oc(G)$ du mod\`ele $ADE$ 
publi\'es dans \cite{Oc-paths}. Les vertex de ces graphes sont not\'es   
$(\ud{x},\ud{y},\ud{z},\ldots)$, et nous appelons $s$ le nombre de vertex du graphe. 
Consid\'erons l'espace vectoriel de dimension $s$ form\'e par les vertex du graphe $Oc(G)$.  
Les \'el\'ements $\ud1$ et $\ud{1'}$ sont appel\'es  
respectivement les g\'en\'erateurs chiraux gauche et droit. Le graphe d'Ocneanu est la superposition 
de deux graphes de Cayley de multiplication par les g\'en\'erateurs.  
La multiplication par le g\'en\'erateur gauche $\ud1$  (resp. droit $\ud{1'}$) est donn\'ee par la  
somme des voisins sur le graphe reli\'es \`a $\ud1$  
(resp. \`a $\ud{1'}$) par une ligne continue (resp. discontinue). De plus, l'\'el\'ement  
$\ud0$ est consid\'er\'e comme l'identit\'e pour cette multiplication. Nous avons donc:  
\begin{equation}  
\ud0 \;.\; \ud{x} = \ud{x}, \qquad \;\; 
\ud1 \;.\; \ud{x} = \sum_{\ud{y}-\ud1} \; \ud{y} = \sum_{y} \; (\mathcal{G}_{\ud1})_{xy}\; \ud{y},  
\qquad \;\; 
\ud{1'} \;.\; \ud{x} = \sum_{\ud{y} \cdots \ud{1'}}\; \ud{y} =  
\sum_{y} \; (\mathcal{G}_{\ud{1}'})_{xy}\; \ud{y}. 
\end{equation}  
o\`u $\mathcal{G}_{\ud1}$ et $\mathcal{G}_{\ud{1}'}$ sont les matrices d'adjacence correspondantes 
aux g\'en\'erateurs $\ud1$ et $\ud{1}'$. 
\`A partir de ces donn\'ees et en imposant l'associativit\'e, il est possible d'\'etendre la  
multiplication \`a  tous les vertex de $Oc(G)$.  Nous obtenons alors l'alg\`ebre de graphe $Oc(G)$,  
qui sera aussi not\'ee $Oc(G)$ et dont la multiplication est donn\'ee par:  
\begin{equation}  
\ud{x} \;.\; \ud{y} = \sum_z \mathcal{O}_{xy}^z \; \ud{z}   
\label{Ocprod}  
\end{equation}  
o\`u $\mathcal{O}_{xy}^z \in \{ 0,1, \ldots \}$ est vu comme la multiplicit\'e de $\ud{z}$  
dans  $\ud{x} . \ud{y}$. Cette alg\`ebre est appel\'ee l'{\bf alg\`ebre des sym\'etries quantiques} de $G$. 
Contrairement aux alg\`ebres de graphes $G$, les alg\`ebres $Oc(G)$  
ne sont pas forc\'ement commutatives (pour le cas $SU(2)_{\ell}$,  
$Oc(D_{2n})$ est non-commutative).  
  
\`A chaque vertex $\ud{x}$ du graphe $Oc(G)$  
nous associons la matrice $(s \times s)$ $O_x$ telle que $(O_x)_{yz} =  
\mathcal{O}_{xy}^{z}$. Nous avons notamment:  
$$  
O_0 = \munite_{s \times s}, \qquad \qquad \qquad O_1 = \mathcal{G}_{\ud1}, \qquad \qquad  
\qquad   O_{1'} = \mathcal{G}_{\ud{1}'}  
$$   
Les matrices $O_x$ donnent une repr\'esentation fid\`ele de l'alg\`ebre $Oc(G)$:  
\begin{equation} 
O_x \; O_y = \sum_z \mathcal{O}_{xy}^z \; O_z =  
\sum_z (O_x)_{yz} O_z  \; .
\end{equation}

\subsection{R\'ealisation alg\'ebrique de $Oc(G)$}  
Historiquement, la premi\`ere r\'ealisation d'une alg\`ebre $Oc(G)$ a \'et\'e effectu\'ee  
dans \cite{Coq-qtetra} pour le cas $E_6$. Il y est observ\'e que l'alg\`ebre $Oc(E_6)$ s'obtient comme  
le carr\'e tensoriel de l'alg\`ebre du graphe $E_6$, mais o\`u le produit tensoriel est pris au-dessus  
d'une sous-alg\`ebre de $E_6$, isomorphe \`a l'alg\`ebre de graphe $A_3$:   
\begin{equation} 
Oc(E_6) = \frac{E_6 \otimes E_6}{A_3} = E_6 \otimes_{A_3} E_6  \;\, .
\end{equation} 
Le produit tensoriel pris au-dessus de $A_3$ signifie que l'on identifie les termes $a.u \otimes b$  
et $a \otimes u.b$,  
pour $a,b \in E_6$ et  $u \in A_3 \subset E_6$.   
Un premier travail de cette th\`ese a consist\'e \`a g\'en\'eraliser cette r\'ealisation \`a tous 
les cas de  
type $ADE$ de $SU(2)_{\ell}$. Il est difficile de pr\'esenter une m\'ethode g\'en\'erale,
chaque cas pos\'edant ces particularit\'es. Les diff\'erents cas sont 
trait\'es explicitement dans le chapitre {\bf 4}. Notre r\'ealisation des alg\`ebres d'Ocneanu 
permet d'obtenir de mani\`ere simple les matrices toriques (d\'efinissant les fonctions  
de partition  \`a une ligne de d\'efauts) et les matrices toriques g\'en\'eralis\'ees 
(deux lignes de d\'efauts). Les r\'esultats obtenus sont en partie publi\'es dans \cite{Coq_Gil-ADE}. 
Par la suite, nous avons montr\'e que la sous-alg\`ebre au-dessus de laquelle  
le produit tensoriel est pris est d\'etermin\'ee par les propri\'et\'es modulaires du graphe $G$   
\cite{Coq_Gil-Tmod}: cette caract\'erisation de $Oc(G)$ par les  
propri\'et\'es modulaires de $G$ permet de d\'efinir la r\'ealisation de $Oc(G)$ sans la connaissance 
pr\'ealable du graphe d'Ocneanu (ceci n'est pas tout \`a fait valable pour les cas o\`u l'alg\`ebre 
d'Ocneanu est non-commutative), et nous pouvons l'\'etendre de mani\`ere naturelle aux diagrammes 
de Coxeter-Dynkin g\'en\'eralis\'es. Ceci nous a permis d'obtenir les alg\`ebres d'Ocneanu 
pour certains cas choisis de $SU(3)_{\ell}$, pour lesquels nous obtenons les fonctions de  
partition g\'en\'eralis\'ees.   
 
Nous donnons ici un aper\c{c}u de la caract\'erisation des propri\'et\'es modulaires de $G$ 
et de la r\'ealisation de $Oc(G)$. Tous les cas $ADE$ de $SU(2)_{\ell}$ et les trois cas 
exceptionnels poss\`edant {\it self-fusion} de $SU(3)_{\ell}$ sont trait\'es de mani\`ere d\'etaill\'ee 
dans le chapitre {\bf 4}.

\subsubsection{Cas poss\'edant self-fusion: type I}  
Nous rappelons ici, pour les graphes poss\'edant self-fusion ($A_n, E_6, E_8, D_{2n}$), les r\`egles de  
branchement $\mathcal{A}(G) \hookrightarrow G$:  
\begin{equation}  
\tau_i \hookrightarrow \sum_b \, (E_0)_{ib} \; \sigma_b.  
\end{equation} 
Elles sont enti\`erement cod\'ees dans la matrice essentielle $E_0$, qui est une matrice rectangulaire,  
\`a $\kappa-1$ lignes et $r$ colonnes. Les lignes de $E_0$ sont index\'ees par les irreps $\tau_i$ de  
$\mathcal{A}(G)$, et les colonnes par les irreps $\sigma_a$ de $G$. Pour conna\^{\i}tre la restriction  
$\mathcal{A}(G) \hookrightarrow G$, il suffit donc de lire les \'el\'ements non-nuls de la ligne de  
$E_0$ correspondant \`a $\tau_i$.

Consid\'erons maintenant l'induction $G \hookleftarrow \mathcal{A}(G)$: pour savoir de quelles  
irreps $\tau_i$ provient l'irrep $\sigma_a$ de $G$ (les irreps $\tau_i$ pour lesquelles \ 
$\sigma_a$ appara\^{\i}t dans leur restriction), il suffit de lire les \'el\'ements non-nuls de la  
colonne de $E_0$ correspondant \`a $\sigma_a$.   
 
Les graphes $\mathcal{A}$ sont toujours modulaires, en ce sens que nous  pouvons d\'efinir une  
repr\'esentation de $SL(2,\mathbb{Z})$ sur l'espace vectoriel form\'e par ses irreps $\tau_i$  
(vus ici comme labellant les caract\`eres $\chi_i$ de l'alg\`ebre de fusion).  
En particulier, l'op\'erateur $T$ est diagonal (voir formule (\ref{ST-su2})), et nous pouvons assigner  
une valeur de l'exposant modulaire $\hat{T}$ fix\'ee \`a chaque irrep $\tau_i$ 
de $\mathcal{A}$.  
 
Nous voudrions d\'efinir une valeur de $\hat{T}$ sur l'espace vectoriel   
form\'e par les irreps $\sigma_a$ de $G$, de mani\`ere compatible avec l'induction-restriction entre  
$\mathcal{A}(G)$ et $G$. Supposons que l'irrep $\sigma_a$ apparaisse dans les r\`egles de branchement  
des irreps $\tau_i$ et $\tau_j$: nous pourrions d\'efinir $\hat{T}(\sigma_a)$ par $\hat{T}(\tau_i)$  
ou par $\hat{T}(\tau_j)$, mais si ces valeurs sont diff\'erentes, alors la d\'efinition est 
ambig\"{u}e. De mani\`ere g\'en\'erale, il existe un sous-ensemble $J$ des irreps $\sigma_a$ sur  
lequel $T$ est bien d\'efini, et ce sous-ensemble $J$ est une sous-alg\`ebre de l'alg\`ebre de graphe $G$.

\begin{defin}  
Soit $G$ un graphe de type $ADE$ poss\'edant ``{\it self-fusion}''. Alors, l'ensemble $J$ est form\'e par  
le sous-espace des irreps $\sigma_a$ de $G$ pour lesquels l'exposant modulaire $\hat{T}$ est bien d\'efini.  
$\sigma_a \in J$ si $\hat{T}$ poss\`ede la m\^eme valeur sur les irreps $\tau_i$ de $\mathcal{A}(G)$ dont  
la restriction \`a $G$ contient $\sigma_a$.   
\end{defin}    
Cette d\'efinition donne une caract\'erisation de l'ensemble $J$ pour les graphes de type I. 
Pour les cas o\`u l'alg\`ebre d'Ocneanu est commutative ($A_n,E_6,E_8$) nous obtenons la suivante
r\'ealisation de $Oc(G)$:  
\begin{equation}  
Oc(G) = \frac{G \otimes G}{J} = G \otimes_J G  
\label{iso_OcG}  
\end{equation}  
L'alg\`ebre d'Ocneanu des cas $D_{2n}$ est non-commutative. La r\'ealisation pr\'ec\'edente n'est donc pas  
valable, car elle d\'efinit une alg\`ebre commutative: il faut utiliser la sym\'etrie $\mathbb{Z}_2$ 
du diagramme pour d\'efinir $Oc(D_{2n})$ \`a l'aide d'un produit semi-direct avec $\mathbb{Z}_2$: 
le cas $D_4$ est trait\'e en d\'etail dans le chapitre {\bf 4}.

\subsubsection{Cas ne poss\'edant pas self-fusion: type II}  
Les cas $D_{2n+1}$ et $E_7$ ne poss\`edent pas {\it self-fusion} (type II).
L'alg\`ebre d'Ocneanu d'un graphe de type II est construite \`a partir de l'alg\`ebre du graphe $H$,  
o\`u $H$ est un graphe de type I appel\'e le  ``{\it parent graph}'' de $G$ \cite{Pet_Zub-CFT}. 
Nous obtenons $Oc(E_7) = D_{10} \otimes_{\rho}D_{10} $, o\`u le {\it twist} exceptionnel $\rho$ 
est d\'etermin\'e par les propri\'et\'es modulaires du graphe  $A_{17}$. L'alg\`ebre $Oc(D_{2n+1})$ 
est obtenue comme un quotient (utilisant une application $\rho$) du produit tensoriel d'alg\`ebre 
$\mathcal{A}$ (de m\^eme nombre de Coxeter). Par exemple $Oc(D_5) = A_7 \otimes_{\rho(A_7)} A_7$. 
Les d\'etails sont pr\'esent\'es dans le chapitre {\bf 4}.

\subsection{$\mathcal{V}(G)$ comme module sur $Oc(G)$: matrices $S_x$}  
Tous les graphes d'Ocneanu d\'efinissent une alg\`ebre de graphe $Oc(G)$.   
Nous voulons d\'efinir une action de $Oc(G)$ sur l'espace vectoriel $\mathcal{V}(G)$ des irreps $\sigma$ de  
$G$:  
\begin{eqnarray}  
Oc(G) \times \mathcal{V}(G) &\rightarrow & \mathcal{V}(G)  \nonumber \\  
\ud{x} . \sigma_a &=& \sum_b  \mathcal{S}_{xa}^b \sigma_b   
\label{x_sig}  
\end{eqnarray}  
telle que les coefficients $\mathcal{S}_{xa}^b$ soient des entiers non-n\'egatifs (ainsi  
$\mathcal{S}_{xa}^b$ repr\'esente la  multiplicit\'e de $\sigma_b$ dans $\ud{x} . \sigma_a$).  
Pour que cette action soit bien d\'efinie, il faut imposer:  
\begin{equation}  
(\ud{x} . \ud{y}) . \sigma_a = \ud{x} . (\ud{y} . \sigma_a)  
\label{modsurOc}  
\end{equation}  
Utilisant (\ref{Ocprod}) et (\ref{x_sig}), et du fait que les $\sigma$ sont des \'el\'ements lin\'eairement   
ind\'ependants de $\mathcal{V}(G)$, les coefficients $\mathcal{S}_{xa}^b$ doivent satisfaire les relations  
suivantes:  
\begin{equation}  
\sum_b  \mathcal{S}_{xa}^b \mathcal{S}_{yb}^c = \sum_z  \mathcal{O}_{xy}^z \mathcal{S}_{za}^c    
\label{S_O}  
\end{equation}  
Les indices $(a,b,c,\ldots)$ sont r\'eserv\'es pour les vertex $\sigma$ de $G$ $(a,b,c = 0,1,\ldots,r-1)$,  
et les indices $(x,y,z,\ldots)$ sont r\'eserv\'es pour les $s$ vertex de $Oc(G)$. 
Du fait que $\ud1$ et $\ud{1'}$ sont les g\'en\'erateurs de l'alg\`ebre $Oc(G)$ (l'unit\'e $\ud{0}$ peut aussi 
s'exprimer \`a l'aide de $\ud{1}$ et $\ud{1'}$), tout \'el\'ement $\ud{x}$  
de $Oc(G)$ s'exprime comme un polyn\^ome sur ($\ud0, \ud1, \ud{1'}$):  
$$  
\ud{x} = Pol_x (\ud0, \ud1, \ud{1'})  
$$   
Il est donc suffisant de d\'efinir l'action de $Oc(G)$ par ses g\'en\'erateurs. L'action explicite de 
$Oc(G)$ sur $\mathcal{V}(G)$ est d\'efinie par:  
\begin{equation}  
\left. 
\begin{array}{rcl} 
Oc(G) \times \mathcal{V}(G)  &\rightarrow& \mathcal{V}(G) \\  
\ud0 . \sigma_a &=& \sigma_a \\  
(\ud1)^n . \sigma_a = (\ud{1'})^n . \sigma_a &=& (\sigma_1)^n . \sigma_a \\  
(\ud1 . \ud{1'})^n . \sigma_a &=& (\sigma_1.\sigma_1)^n . \sigma_a   
\end{array}  
\right\rbrace  
\label{x_sig_expl}  
\end{equation}  
\`A noter que m\^eme pour les cas o\`u $G$ ne poss\`ede pas {\it self-fusion}, l'action est bien d\'efinie: la   
multiplication $\sigma_1 . \sigma_a$ existe toujours (cod\'ee par le graphe $G$).  
L'action d'un \'el\'ement quelconque $\ud{x}$ de $Oc(G)$ sur $\mathcal{V}(G)$ s'\'ecrit:  
\begin{equation}  
\ud{x} . \sigma_a = Pol_x (\ud0,\ud1,\ud{1'}) . \sigma_a  
 = Pol_x (\sigma_0, \sigma_1, \sigma_1) . \sigma_a  
\label{x_sig_pol}  
\end{equation}

\subsubsection{Matrices $S_x$}  
Introduisons $s$ matrices  
$(r \times r)$ $S_{x}$ telles que $(S_x)_{ab} = \mathcal{S}_{xa}^b$. Par la propri\'et\'e  
de module de (\ref{modsurOc}), les matrices $S_{x}$ satisfont les relations suivantes:  
\begin{equation} 
S_{x} . S_{y} = \sum_z \mathcal{O}_{xy}^z \; S_{z}  
\end{equation} 
Par la d\'efinition explicite (\ref{x_sig_expl}) de l'action de $Oc(G)$ sur $G$ , nous avons:   
$S_{0} = \munite_{r \times r}$, $S_{1} = \mathcal{G}_{\ud1}$ et $S_{1'} = \mathcal{G}_{\ud{1}'}$.  
Les autres matrices $S_{x}$ peuvent \^etre obtenues \`a partir de la connaissance de la r\'ealisation  
de $Oc(G)$. Si $Oc(G)$ est isomorphe \`a l'alg\`ebre $G \otimes_J G$, nous \'ecrirons un  
\'el\'ement $\ud{x}$ de $Oc(G)$ comme $\ud{x} = \sigma_a \otimesdot  
\sigma_b$\footnote{Selon les cas, $\ud{x}$ peut parfois \^etre une   
combinaison lin\'eaire de tels \'el\'ements.}. Alors l'action de $Oc(G)$ sur $G$ s'\'ecrit:  
\begin{equation} 
\ud{x} . \sigma_c = (\sigma_a \otimesdot \sigma_b) . \sigma_c \doteq \sigma_a . \sigma_b . \sigma_c  
\end{equation} 
auquel cas les matrices $S_{x}$ sont donn\'ees par:  
\begin{equation} 
S_{x} = G_a.G_b \qquad \qquad \qquad \qquad \textrm{pour  }\; \ud{x} = \sigma_a \otimesdot \sigma_b  
\end{equation}


\section{Relations entre $\mathcal{A}(G)$ et $Oc(G)$}

\subsection{Fonctions de partition}

Consid\'erons des th\'eories conformes \`a deux dimensions avec alg\`ebre affine $\widehat{su}(2)$.  
Nous avons vu dans le chapitre {\bf 1} que les fonctions de partition d'un syst\`eme avec deux 
lignes de d\'efauts $x$ et $y$ -- appel\'ees g\'en\'eralis\'ees \cite{Pet_Zub-gener} -- d\'efinies 
sur le tore (de param\`etre modulaire $\tau$) s'\'ecrivent: 
\begin{equation}  
\mathcal{Z}_{x|y}(\tau) =\sum_{i,j} \; \chi_i(\tau)\; \mathcal{W}_{xy}^{ij}\; \ov{\chi}_j(\tau) \; ,  
\end{equation}   
o\`u les $\chi_i(\tau)$ sont les caract\`eres de  
l'alg\`ebre $\widehat{su}(2)$, et les coefficients $\mathcal{W}_{xy}^{ij}$ sont des entiers  
non-n\'egatifs.  
Le cas sans ligne de d\'efauts ($x$=$y$=0) permet d'obtenir l'invariant modulaire $\mathcal{M}$ 
qui comute avec les g\'en\'erateurs $S$ et $T$ du groupe modulaire  
($\mathcal{M}_{ij} = \mathcal{W}_{00}^{ij} $), et la fonction de partition invariante 
modulaire s'\'ecrit: 
\begin{equation} 
\mathcal{Z}(\tau) =  \; \ov{\chi} \; \mathcal{M} \; \chi 
\end{equation} 
D\'efinissant les matrices $(s \times s)$ $\widetilde{W}_{ij}$ telles 
que $(\widetilde{W}_{ij})_{xy} = \mathcal{W}_{xy}^{ij}$, des conditions de compatibilit\'e 
\cite{Pet_Zub-gener, Pet_Zub-Oc} imposent que ces matrices doivent satisfaire l'alg\`ebre carr\'ee de fusion: 
\begin{equation}  
\widetilde{W}_{ij} \; \widetilde{W}_{i'j'} = \sum_{i'',j''} \; \mathcal{N}_{ii'}^{i''}  \; 
\mathcal{N}_{jj'}^{j''} \; \widetilde{W}_{i''j''} \; , 
\label{algfuscarre} 
\end{equation} 
alors que les matrices $\widetilde{W}_{i1}$ et $\widetilde{W}_{1j}$ forment une repr\'esentation de  
l'alg\`ebre de fusion: 
\begin{equation}  
\widetilde{W}_{i1} \; \widetilde{W}_{i'1} = \sum_{i''} \; \mathcal{N}_{ii'}^{i''} \;  
\widetilde{W}_{i''1}, \qquad \qquad \qquad  
\widetilde{W}_{1j} \; \widetilde{W}_{1j'} = \sum_{j''} \; \mathcal{N}_{jj'}^{j''} \;   
\widetilde{W}_{1j''}.  
\label{algfusW3}  
\end{equation}   
La matrice $\widetilde{W}_{11}$ est l'identit\'e, tandis que  
les matrices $\widetilde{W}_{21}$ et $\widetilde{W}_{12}$ correspondent  
respectivement aux deux matrices d'adjacence d'un graphe d'Ocneanu $Oc(G)$: 
\begin{equation} 
\widetilde{W}_{11} =\munite_{s \times s} \;, \qquad \qquad \qquad \widetilde{W}_{21} = O_{1}\; ,   
\qquad \qquad \qquad \widetilde{W}_{12} = O_{1'} \; . 
\end{equation} 
Ainsi la donn\'ee du graphe d'Ocneanu (des matrices $O_1$ et $O_{1'}$) permet d'obtenir les 
matrices $\widetilde{W}_{21}$ et $\widetilde{W}_{12}$, et en utilisant (\ref{algfusW3}) 
puis (\ref{algfuscarre})  
d'obtenir les matrices $\widetilde{W}_{ij}$, et donc les coefficients $\mathcal{W}_{xy}^{ij}$ 
qui d\'efinissent les fonctions de partition g\'en\'eralis\'ees. 
C'est la d\'emarche suivie par Zuber {\it et al } \cite{Pet_Zub-gener, Pet_Zub-Oc}, cependant, dans  
cette approche les graphes d'Ocneanu sont une donn\'ee initiale externe et indispensable. 
Ces graphes n'\'etant pas connus (publi\'es) pour des syst\`emes $su(n), n \geq3$, cette m\'ethode ne 
peut pas s'\'etendre \`a ces cas-l\`a.

Il a \'et\'e par la suite remarqu\'e que pour des th\'eories de  
type I, l'invariant modulaire peut s'\'ecrire sous la forme suivante:  
\begin{equation}  
\mathcal{M}_{ij} = \sum_{c \in J} (F_i)_{1c} (F_j)_{1c}  
\end{equation}   
o\`u les $F_i$ sont des matrices \`a coefficients entiers non-n\'egatifs satisfaisant  
l'alg\`ebre de fusion, et o\`u la sommation s'effectue sur un sous-ensemble $J$.   
Cette formule a originellement \'et\'e obtenue de mani\`ere empirique \cite{DiFZub} et par la suite  
g\'en\'eralis\'ee aux mod\`eles de type II \cite{behrend, behrend-BCFT,Pet_Zub-1996, Pet_Zub-1997}.  
L'\'el\'ement $F_1$ est le g\'en\'erateur de l'alg\`ebre  
de fusion, et correspond \`a la matrice d'adjacence d'un graphe: c'est ainsi que les graphes  
$ADE$ sont formellement apparus dans la classification des fonctions de partition des mod\`eles  
$\widehat{su}(2)$.   
 
Introduisant les caract\`eres \'etendus $\widehat{\chi}_a$, d\'efinis par:  
\begin{equation}  
\widehat{\chi}_a = \sum_i (F_i)_{1a} \; \chi_i \; , 
\end{equation}  
la fonction de partition invariante modulaire des mod\`eles de type I s'\'ecrit:  
\begin{equation}  
\mathcal{Z} = \sum_{a \in J} |\widehat{\chi}_a|^2,  
\end{equation}  
et est donc diagonale par rapport \`a ces caract\`eres. Ceux-ci sont interpr\'et\'es comme des  
caract\`eres d'une alg\`ebre chirale \'etendue \cite{zuber-CFT}. Les fonctions de partition des  
mod\`eles de type II sont obtenues \`a partir de celles de type I par une proc\'edure de  
twist \cite{dijkgraaf,moore1,moore2}.

Notons que les expressions des fonctions de partition invariantes modulaires ou  
g\'en\'eralis\'ees \'etaient obtenues de mani\`ere empirique ou par la donn\'ee du graphe d'Ocneanu. 
Nous allons voir que gr\^ace \`a notre r\'ealisation de l'alg\`ebre d'Ocneanu, nous pouvons d\'eterminer 
ces expressions de mani\`ere naturelle par l'action de $\mathcal{A}(G)$ sur $Oc(G)$, et d'autre part 
que cette m\'ethode se pr\^ete \`a une g\'en\'eralisation aux cas $\widehat{su}(n), n\ \geq 3$.

\subsection{$Oc(G)$ comme bi-module sur $A(G)$: matrices $W_{ij}$ et $W_{xy}$}  
 
Soit $G$ un graphe de type $ADE$ (ou g\'en\'eralis\'e), $\mathcal{V}(G)$ l'espace vectoriel 
form\'e par ses vertex et $\mathcal{A}(G)$ l'alg\`ebre du graphe $A_n$ de m\^eme norme. 
Il existe une action de $\mathcal{A}(G)$ sur $\mathcal{V}(G)$, cod\'ee par les matrices $F_i$. 
Une r\'ealisation de l'alg\`ebre $Oc(G)$ est construite \`a partir du carr\'e tensoriel de l'alg\`ebre 
du graphe $G$: il est donc naturel d'avoir  
une action (\`a gauche et \`a droite) de $\mathcal{A}(G)$ sur $Oc(G)$. Nous avons:  
\begin{equation}  
\tau_i . \ud{x} . \tau_j = \sum_y \mathcal{W}_{xy}^{ij} \ud{y}  
\label{tau_x_tau}  
\end{equation}  
o\`u les coeficients $\mathcal{W}_{xy}^{ij}$ sont des nombres entiers non-n\'egatifs. Introduisons  
alors des matrices   
$s \times s$ $W_{ij}$ telles que $(W_{ij})_{xy} = \mathcal{W}_{xy}^{ij}$.   
\begin{prop}  
Ces matrices $W_{ij}$ satisfont les relations suivantes:  
\begin{enumerate}  
\item $ W_{ij} \; W_{i'j'} = \displaystyle \sum_{i'',j''} \mathcal{N}_{ii'}^{i''}\;   
\mathcal{N}_{jj'}^{j''} \; W_{i''j''} $  
\item   
$W_{i1} \; W_{i'1} = \displaystyle \sum_{i''} \mathcal{N}_{ii'}^{i''} \; W_{i''1}$ \\  
$W_{1j} \; W_{1j'} = \displaystyle \sum_{j''} \mathcal{N}_{jj'}^{j''} \; W_{1j''}$    
\item $O_x \; W_{ij} = W_{ij} \; O_x = \displaystyle \sum_y (W_{ij})_{xy} \; O_y$  
\item $W_{ij} = \displaystyle \sum_y (W_{ij})_{0y} \;  O_y$  
\end{enumerate}  
\end{prop}   
{\bf \ud{D\'emonstration}:}  
La relation (1) provient de l'\'egalit\'e $\tau_{i'}.(\tau_i . x . \tau_j). \tau_{j'} =   
(\tau_{i'}.\tau_i) . x . (\tau_j. \tau_{j'})$. La relation (2) est obtenue \`a partir de (1) pour  
$i=i'=0$ (j=j'=0). La relation (3) est obtenue en multipliant \`a gauche et \`a droite l'\'equation  
(\ref{tau_x_tau}) par $\ud{z}$. Et enfin (4) est une cons\'equence imm\'ediate de (3).  
\hfill $\blacksquare$

\begin{conclu}  
Les matrices $W_{ij}$ d\'efinies \`a partir de la structure de bi-module de $Oc(G)$ sur  
$A(G)$ co\"{\i}ncident avec les matrices $\widetilde{W}_{ij}$ introduites pr\'ec\'edemment.  
Les fonctions de partition g\'en\'eralis\'ees des mod\`eles $\widehat{su}(2)$ s'\'ecrivent donc:  
\begin{equation}  
\mathcal{Z}_{x|y}(\tau) =\sum_{i,j} \; \chi_i(\tau) \; (W_{ij})_{xy} \ov{\chi}_j(\tau).  
\end{equation} 
o\`u les coefficients $W_{ij}^{xy}$ sont calcul\'es en explicitant l'action (\ref{tau_x_tau}) 
de $\mathcal{A}(G)$ sur $Oc(G)$  
\end{conclu}

L'obtention de ces coefficients d\'epend de chaque cas sp\'ecifique.  
Par exemple, pour $x=\sigma_a \otimes_J \sigma_b \in Oc(G)$, avec $\sigma_a, \sigma_b \in G$, nous avons: 
\begin{equation} 
\tau_i \, . \, x \, . \, \tau_j \, = \, (\tau_i . \sigma_a) \, \otimes_J \, (\sigma_b . \tau_j) 
\, = \, \sum_{c,d} \, (F_i)_{ac} (F_j)_{bd} \; \sigma_c \otimes_J \sigma_d \; ,   
\end{equation} 
et il faut alors identifier les \'el\'ements $\sigma_c \otimes_J \sigma_d$ sur la base  
des \'el\'ements de la base $\{y\}$ de $Oc(G)$. Pour un \'el\'ement $y$, nous obtenons alors 
les coefficients $\mathcal{W}_{xy}^{ij}$ \`a partir desquels sont d\'efinies les matrices toriques  
g\'en\'eralis\'ees $W_{xy}$. Pour l'\'el\'ement $y=0$ nous obtenons les matrices toriques 
$W_x = W_{x0}$.   
L'action est bien d\'efinie et permet d'obtenir des formules compactes 
pour les expressions des fonctions de partition invariantes modulaires et g\'en\'eralis\'ees. 
Nous retrouvons ainsi la classification  
de Cappelli-Itzykson-Zuber et donnons des formules pour les expressions des fonctions  
de partition \`a une et deux lignes de d\'efauts de tous les mod\`eles $\widehat{su}(2)$.

Notre r\'ealisation de $Oc(G)$ permet de g\'en\'eraliser les m\'ethodes aux cas $SU(n)_{\ell}$. 
Nous avons \'etudi\'e les trois cas exceptionnels poss\'edant {\it self-fusion} du mod\`ele  
$\widehat{su}(3)$ et d\'etermin\'e
leurs fonctions de partition. Les expressions obtenues pour la fonction de partition 
invariante modulaire co\"{\i}ncident avec celles provenant de la classification de Gannon, 
et nous avons obtenu des expressions g\'en\'erales pour les fonctions de partition 
\`a une et deux lignes de d\'efauts.

Les calculs explicites ainsi que les expressions des fonctions de partition des mod\`eles 
\'etudi\'ees sont pr\'esent\'es dans le chapitre {\bf 4}.

\subsection{Relations de compatibilit\'e alg\'ebrique} 
Soit $G$ un graphe de type $ADE$ et $\mathcal{B}(G)$ l'alg\`ebre gradu\'ee des endomorphismes de chemins  
essentiels sur $G$: $\mathcal{B}(G)$ est techniquement une alg\`ebre de Hopf faible \cite{Oc-paths}. 
Il existe deux produits distincts $\circ$ (composition) et $\odot$ (convolution) dans $\mathcal{B}(G)$,  
v\'erifiant des conditions de compatibilit\'e.

\subsubsection{La dig\`ebre $\mathcal{B}(G)$: r\`egles de somme (quadratique et lin\'eaire)}  
$\mathcal{B}(G)$ est semi-simple pour ses deux  
structures alg\'ebriques,  
nous pouvons donc la diagonaliser pour ses deux lois. L'espace vectoriel $\mathcal{B}(G)$ poss\`ede donc 
deux structures d'alg\`ebre, qu'on peut \'ecrire matriciellement: 
\begin{equation}  
\mathcal{B}(G) \cong \underset{i}{\oplus} \;\, L^i \cong \underset{x}{\oplus}\;\, X^x  
\end{equation}  
\paragraph{Premi\`ere loi: composition $\circ$}  
Pour la loi de composition $\circ$, les blocs $L^i$ sont index\'es par la longueur $i$ des chemins  
essentiels. Les projecteurs minimaux centraux dans chaque bloc sont labell\'es par les vertex du graphe  
$\mathcal{A}(G)$. Rappelons que l'\'el\'ement $(a,b)$ de la matrice $F_i$, pour $i \in \mathcal{A}(G)$,  
est \'egal au nombre de chemins essentiels de longueur $i$ partant du vertex $\sigma_a$ et arrivant  
au vertex $\sigma_b$ de $G$.   
La dimension de chaque bloc $L^i$ est donn\'ee par:  
\begin{equation}  
d_i = \sum_{a,b=0}^{r-1} (F_i)_{ab}  
\end{equation}  
Comme $\mathcal{B}(G) \cong \oplus_i L^i$,  
la dimension de $\mathcal{B}(G)$ est donn\'ee par:  
\begin{equation}  
\dim(\mathcal{B}(G)) = \sum_{i=0}^{\kappa-2} \; d_i^2  
\end{equation}

\paragraph{Deuxi\`eme loi: convolution $\odot$}  
Pour la loi de convolution $\odot$, les blocs $X^x$ sont index\'es par le label $x$. Les  
projecteurs minimaux centraux de chaque bloc sont labell\'es par les vertex du graphe $Oc(G)$. 
En analogie avec le cas pr\'ec\'edent, la dimension de chaque bloc $X^x$ est donn\'ee par  
\begin{equation}  
d_x = \sum_{a,b=0}^{r-1} (S_x)_{ab}  
\end{equation}  
L'\'el\'ement $(a,b)$ de $S_x$, pour $x \in Oc(G)$, est vu comme le nombre de chemins verticaux partant du 
vertex $\sigma_a$ et arrivant au vertex $\sigma_b$ sur $G$, les chemins verticaux \'etant une base 
des chemins qui diagonalisent $\mathcal{B}(G)$ pour la loi $\odot$. 
Comme $\mathcal{B}(G) \cong \oplus_x X^x$,  
la dimension de $\mathcal{B}(G)$ est donn\'ee par:  
\begin{equation}  
\dim(\mathcal{B}(G)) = \sum_{x=0}^{s-1} \; d_x^2  
\end{equation}

\paragraph{R\`egles de somme}  
Nous avons l'\'egalit\'e suivante ({\bf r\`egle de somme quadratique}):  
\begin{equation}  
\dim(\mathcal{B}(G)) = \sum_{l=0}^{\kappa-2} d_l^2 = \sum_{x=0}^{s-1} d_x^2  
\end{equation}  
Une autre relation peut aussi \^etre v\'erifi\'ee dans la plupart des cas\footnote{Dans les cas o\`u cette
relation n'est pas satisfaite, on sait la corriger.}, la {\bf r\`egle de somme lin\'eaire}:  
\begin{equation}  
\sum_{l=0}^{\kappa-2} d_l = \sum_{x=0}^{s-1} d_x  
\end{equation}  
{\it A priori}, il n'existe pas de raison d'obtenir une telle relation pour une dig\`ebre 
semi-simple pour ses deux structures multiplicatives. Son interpr\'etation reste encore myst\'erieuse.  
Elle proviendrait d'un changement de base entre les chemins essentiels (index\'es par  
la longueur $i$) et les chemins verticaux (index\'es par le label $x$), mais une d\'efinition  
pr\'ecise des chemins verticaux et de cette transformation demeure incompl\`ete. \\

\subsubsection{Une autre r\`egle de somme} 
Consid\'erons un graphe $G$ de type $ADE$ et de nombre de Coxeter $\kappa$, et les alg\`ebres 
$\mathcal{A}(G)$ et $Oc(G)$ associ\'ees. Une repr\'esentation de $\mathcal{A}(G)$ et $Oc(G)$ 
est donn\'ee par les matrices de fusion $N_i$ (de dimension $\kappa-1$) et $O_x$ (de dimension $s$). 
D\'efinissons: 
\begin{equation} 
\widehat{d}_i = \sum_{j,k} \; (N_i)_{jk} \;, \qquad \qquad \qquad \widehat{d}_x = \sum_{y,z} \; 
(O_x)_{yz}\;.
\end{equation} 
 
Soient $W_x$ les matrices toriques associ\'ees au mod\`ele. Nous avons v\'erifi\'e que les relations  
suivantes sont satisfaites: 
\begin{equation} 
\sum_{y,z} \; (O_x)_{yz} \, = \, \sum_{i,j} \; (W_x)_{ij}\;. 
\label{doublesum} 
\end{equation} 
Alors, la r\`egle de somme suivante est satisfaite: 
\begin{equation} 
\sum_{i,j} \; \mathcal{M}_{ij} \; \widehat{d}_i \; \widehat{d}_j \, = \, \sum_{x} 
\widehat{d}_{\,x}^{\;2} \,. 
\end{equation} 
{\bf \ud{d\'em}}: Nous partons de l'\'equation (\ref{matr_tor}) 
\begin{equation}  
\sum_x (W_x)_{ij} \; (W_x)_{i'j'} = \sum_{i'',j''} \; \mathcal{N}_{ii'}^{i''} \;   
\mathcal{N}_{jj'}^{j''} \; \mathcal{M}_{i''j''}  
\end{equation} 
Du fait que $(N_i)_{jk} = \mathcal{N}_{ij}^k$, en sommant sur les $i,i',j,j'$, nous obtenons: 
\begin{equation} 
\sum_{i'',j''} \mathcal{M}_{i''j''} \; \widehat{d}_{i''} \; \widehat{d}_{j''} = \sum_x \sum_{ij} \sum_{i'j'} 
\; (W_x)_{ij} \; (W_x)_{i'j'} 
\end{equation} 
et utilisant (\ref{doublesum}) nous arrivons au r\'esultat. \hfill  $\blacksquare$  \\ 
Cette r\`egle de somme est importante car elle relie les nombres caract\'eristiques de l'alg\`ebre  
$Oc(G)$ avec les nombres caract\'eristiques de l'alg\`ebre $\mathcal{A}(G)$. Elle permet donc une  
v\'erification de nos constructions des structures alg\'ebriques construites \`a partir 
de diagrammes $G$ g\'en\'eralis\'es.

\subsubsection{Masses quantiques}  
Pour un graphe $G$, les composantes du vecteur normalis\'e de Peron-Frobenius d\'efinissent les  
dimensions quantiques des vertex $\sigma$ de $G$: ce sont des nombres quantiques   
$[n]_q = qdim(\sigma)$. Rappelons que le nombre quantique $[n]_q$ est d\'efini par:  
\begin{equation} 
[n]_{q}=\frac{q^n - q^{-n}}{q - q^{-1}}, \qquad \qquad q = \exp(\frac{i\pi}{\kappa}), \qquad \qquad q^{2\kappa}=1.  
\end{equation} 
Ces nombres s'\'ecrivent explicitement:  
$$  
\begin{array}{lrcl}  
\textrm{n pair} \qquad  & [n]_q &=& q + q^{-1} + q^3 + q^{-3} + q^5 + q^{-5} + \ldots + q^{n-1} +  
q^{-(n-1)} \\  
{ }             & { }   &=&  2 \cos \left(\frac{ \pi}{\kappa}\right) + 2 \cos  
\left(\frac{3 \pi}{\kappa}\right)   
+ 2 \cos \left(\frac{5 \pi}{\kappa}\right) + 2 \cos \left(\frac{(n-1) \pi}{\kappa}\right)               \\ 
{} & {} & {} & {} \\ 
\textrm{n impair} & [n]_q &=& 1 + q^2 + q^{-2} + q^4 + q^{-4} + q^6 + q^{-6} + \ldots + q^{n-1} +  
q^{-(n-1)} \\  
{ }             & { }   &=& 1 +  2 \cos \left(\frac{2 \pi}{\kappa}\right) + 2 \cos  
\left(\frac{4 \pi}{\kappa}\right)   
+ 2 \cos \left(\frac{6 \pi}{\kappa}\right) + 2 \cos \left(\frac{(n-1) \pi}{\kappa}\right)              
\end{array}  
$$  
\begin{defin}  
Pour um graphe $G$ de type $ADE$ \`a $r$ vertex $\sigma_a$, sa {\bf masse quantique} $m(G)$ est d\'efinie par:  
\begin{equation}  
m(G) = \sum_{a=0}^{r-1} (\textrm{qdim}(\sigma_a))^2  
\end{equation} 
\end{defin}  
Si $Oc(G)$ est de la forme $Oc(G) = G \otimes_J G$, alors nous d\'efinissons la masse quantique  
de $Oc(G)$ par:  
\begin{equation}  
m(Oc(G)) = \frac{m(G) . m(G)}{m(J)} \end{equation} 
\begin{prop}  
Soit  un graphe $G$ tel que la diagonalisation des deux lois $\circ$ et $\odot$ de la dig\`ebre  
$\mathcal{B}G$  soit d\'ecrite respectivement par $\mathcal{A}(G)$ et par $Oc(G)$. Alors les  
masses quantiques de $A(G)$ et de $Oc(G)$ sont \'egales:  
\begin{equation} 
m(A(G)) = m(Oc(G))  
\end{equation}   
\end{prop}  
Cette observation, proprement g\'en\'eralis\'ee\footnote{La d\'efinition de la masse quantique  
de $Oc(G)$ pour les graphes $D_{2n}$ et les graphes de type II doit \^etre adapt\'ee.}, est aussi  
valable pour les graphes de type $ADE$. 
Nous ne connaissons pas de d\'emonstration de cette relation, ni m\^eme son origine. Nous verrons  
qu'elle est aussi v\'erifi\'ee pour les graphes g\'en\'eralis\'es des cas $SU(3)_{\ell}$  
\'etudi\'es. \\

Gr\^ace \`a notre r\'ealisation de l'alg\`ebre d'Ocneanu, nous avons pu d\'efinir les alg\`ebres 
d'Ocneanu pour certains exemples des diagrammes de Coxeter-Dynkin g\'en\'eralis\'es. {\it A priori} 
cette construction 
est une conjecture, car une d\'efinition de la dig\`ebre $\mathcal{B}(G)$ pour des graphes 
g\'en\'eralis\'es $G$ n'est pas encore disponible dans la litt\'erature. 
Pour les cas \'etudi\'es de $SU(3)_{\ell}$, les diverses r\`egles de somme (lin\'eaire, quadratique,
de masse quantique) sont satisfaites, donnant ainsi une confirmation  
de nos constructions.


\chapter{Calculs explicites} 
\thispagestyle{empty}

Dans ce chapitre nous traitons explicitement l'ensemble des cas du type $su(2)$ et trois exemples  
du type $su(3)$, en donnant une r\'ealisation de leur alg\`ebre d'Ocneanu.
Par cette r\'ealisation,  nous d\'eterminons toutes 
les fonctions de partition invariantes modulaires ainsi que  les formules g\'en\'erales permetttant 
d'obtenir les fonctions de partition g\'en\'eralis\'ees, interpr\'et\'ees 
dans le langage de la th\'eorie des champs conformes comme d\'ecrivant des 
syst\`emes d\'efinis sur un tore, avec l'introduction de une ou deux lignes de 
d\'efauts.

\section{Rappels des notations} 
\noindent Nous donnons un rappel des notations introduites pour les diff\'erentes structures 
rencontr\'ees.\\

\noindent $\bullet$ Graphes $G$ 
 
\begin{itemize} 
 
\item  $G$ est un graphe correspondant \`a un diagramme de Dynkin de type $ADE$ ou g\'en\'eralis\'e. 
Nous apelons aussi $G$ l'alg\`ebre du graphe de $G$, lorsque $G$ poss\`ede {\it self-fusion} (type I). 
 
\item  $r$ est le nombre de vertex de $G$. 
 
\item  $\sigma_a$ sont les vertex du graphe $G$ ($a=0,1,\ldots, r-1$ pour un diagramme $ADE$). 
 
\item $\mathcal{V}(G)$ est l'espace vectoriel, de dimension $r$, dont une base est form\'ee par les vertex
$\sigma_a$ de $G$.

\item  $\mathcal{G}$ est la matrice d'adjacence de $G$. 
 
\item  $\beta$ est la plus grande valeur propre de $\mathcal{G}$ (norme de $G$). 
\item  $P$ est le vecteur propre de $\mathcal{G}$ correspondant \`a $\beta$ (vecteur de Perron-Frobenius). 
Les composantes de P d\'efinissent les dimensions quantiques des vertex $\sigma$ de $G$. 
\item  $\kappa$ est le nombre (dual) de Coxeter de $G$. 
\item  $G_{a}=(G_{a})_{bc}$ sont les matrices donnant une repr\'esentation fid\`ele de l'alg\`ebre 
du graphe $G$, lorsqu'elle existe.\\ 
 
\end{itemize} 
\pagebreak
 
\noindent $\bullet$ Graphes $\mathcal{A}(G)$ 
 
\begin{itemize} 
\item  $\mathcal{A}(G)$ est le graphe de la s\'erie $\mathcal{A}$ ayant le m\^eme nombre 
de Coxeter $\kappa$ que $G$. $\mathcal{A}(G)$ d\'esigne aussi l'alg\`ebre du graphe (qui 
existe toujours). 
\item  $\tau_i$ sont les vertex du graphe $\mathcal{A}(G)$ ($i=0,1,\ldots,\kappa-2$ si $G$ est du type 
$ADE$.  
\item  $N_{i}=(N_{i})_{jk}$ sont les matrices donnant une repr\'esentation de l'alg\`ebre du graphe 
$\mathcal{A}(G)$. Elles forment une repr\'esentation de l'alg\`ebre de fusion.  
\item  $F_{i} = (F_{i})_{ab} \doteq (E_{a})_{ib}$ sont les matrices qui codent l'action 
(multiplication externe) de $\mathcal{A}(G)$ sur $\mathcal{V}(G)$. 
Elles forment une repr\'esentation de l'alg\`ebre de fusion, de dimension $r$.  
\item $d_i = \underset{a,b}{\sum} (F_i)_{ab}$ est la dimension des blocs de la dig\`ebre $\mathcal{B}(G)$ 
pour la loi $\circ$.  
\item  $E_{a} = (E_{a})_{ib}$ sont les matrices essentielles du graphe $G$. 
\item  $E_0$ est la matrice essentielle correspondant \`a $\sigma_0$, apel\'ee {\it intertwiner}. 
Si $G$ poss\`ede {\it self-fusion}, elle code les r\`egles de 
branchement $\tau_i \hookrightarrow \sum_b (E_0)_{ib}\; \sigma_b$ de $\mathcal{A}(G)$ vers $G$.\\
\end{itemize} 
\noindent $\bullet$ Graphes $Oc(G)$ 
 
\begin{itemize} 
\item  $Oc(G)$ est le graphe d'Ocneanu associ\'e \`a $G$. $Oc(G)$ d\'esigne aussi l'alg\`ebre du graphe 
$Oc(G)$, appel\'ee alg\`ebre des sym\'etries quantiques (qui existe toujours, mais n'est 
pas forc\'ement commutative). 
\item  $s$ est le nombre de vertex de $Oc(G)$. 
\item  $\ud{x},\ud{y},\ud{z},\ldots$ sont les vertex de $Oc(G)$. 
\item  $O_x=(O_x)_{yz}$ sont les matrices donnant une repr\'esentation de l'alg\`ebre $Oc(G)$. 
\item  $S_x = (S_x)_{ab}$ sont les  matrices qui codent l'action (multiplication externe) 
de $Oc(G)$ sur $\mathcal{V}(G)$. 
\item $d_x = \underset{a,b}{\sum} (S_x)_{ab}$ est la dimension des blocs de la dig\`ebre $\mathcal{B}(G)$ 
pour la loi $\odot$. \\ 
\end{itemize} 
 
\noindent $\bullet$ Fonctions de partition 
 
\begin{itemize} 
\item $W_{xy} = (W_{xy})_{ij}$ sont les matrices toriques g\'en\'eralis\'ees qui codent l'action (\`a gauche et \`a  
droite) de $\mathcal{A}(G)$ sur $Oc(G)$.  
\item $\mathcal{Z}_{x|y} = \sum_{i,j} \chi_i(q) (W_{xy})_{ij} \ov{\chi}_j(q)$ sont les fonctions de  
partition g\'en\'eralis\'ees ({\it twist\'ees}) du mod\`ele consid\'er\'e.  
\item $\chi_i(q)$ sont les caract\`eres de l'alg\`ebre affine du mod\`ele ($\widehat{su}(2)$ ou 
$\widehat{su}(3)$).  
\item $\hat{\chi}_a(q)$ sont les caract\`eres \'etendus associ\'es au graphe $G$. 
\item $\mathcal{M} = W_{00}$ est l'invariant modulaire, qui commute avec les g\'en\'erateurs 
$S$ et $T$ du groupe modulaire. 
\item $\mathcal{Z}_G = \mathcal{Z}_{0|0}$ est la fonction de partition invariante modulaire associ\'ee 
au graphe $G$. 
\end{itemize}

\noindent Les diff\'erentes matrices rencontr\'ees satisfont des relations, que nous rappelons ici: 
\small 
$$ 
\begin{array}{rclcrcl} 
\tau_i . \tau_j &=& \displaystyle \sum_k \mathcal{N}_{ij}^k\; \tau_k  &\rightarrow& N_i . N_j &=& \displaystyle \sum_k (N_i)_{jk} \; N_k \\ 
\sigma_a . \sigma_b &=& \displaystyle \sum_b G_{ab}^c \; \sigma_c &\rightarrow& G_a . G_b &=& \displaystyle \sum_b (G_a)_{bc} \; G_c \\ 
\ud{x} . \ud{y} &=& \displaystyle \sum_z \mathcal{O}_{xy}^z \; \ud{z} &\rightarrow& O_{x} . O_{y} &=& \displaystyle   
\sum_z (O_x)_{yz} \; O_{z} \\ 
\tau_i . \sigma_a &=& \displaystyle \sum_b \mathcal{F}_{ia}^b \; \sigma_b &\rightarrow& F_i . G_a &=&  \displaystyle \sum_b (F_i)_{ab} \; G_b \\ 
(\tau_i . \tau_j) . \sigma_a &=& \tau_i . (\tau_j . \sigma_a)  &\rightarrow &  F_i . F_j &=&  \displaystyle \sum_k (N_i)_{jk} \; F_k \\ 
\ud{x} . \sigma_a &=&  \displaystyle \sum_b \mathcal{S}_{xa}^b \; \sigma_b &\rightarrow & S_x . G_a &=&  \displaystyle \sum_b (S_x)_{ab} \; G_b \\ 
(\ud{x} . \ud{y}) . \sigma_a &=&  \ud{x} . (\ud{y} . \sigma_a) &\rightarrow& S_y . S_x &=& \displaystyle  
\sum_z (O_x)_{yz} \; S_z \\ 
\tau_i . \ud{x} . \tau_j &=& \displaystyle \sum_{y} (W_{ij})_{xy} \; \ud{y} &\rightarrow & W_{ij} . O_x &=& \displaystyle \sum_y (W_{ij})_{xy} \; O_y \\   
\tau_{i'} . (\tau_i . x . \tau_j) . \tau_{j'} &=& (\tau_{i'} . \tau_{i}) . x . (\tau_{j} . \tau_{j'}) &\rightarrow & 
W_{ij} .  W_{i'j'} &=&\displaystyle  \sum_{i''} \sum_{j''} (N_{i})_{i'i''}\, (N_{j})_{j'j''}\, 
W_{i''j''} 
\end{array} 
$$ 
\normalsize


\section{Calculs des cas $\widehat{su}(2)$} 
 
Les fonctions de partition \`a une ligne de d\'efauts des cas du type $su(2)$ ont premi\`erement 
\'et\'e publi\'ees dans \cite{Coq_Gil-ADE} et \cite{Pet_Zub-gener} (obtenues par des m\'ethodes  
de calculs tr\`es diff\'erentes), et des r\'esultats partiels  
sont aussi connus pour les fonctions de partition \`a deux lignes de d\'efauts \cite{Pet_Zub-Oc}. 
Elles y sont donn\'ees en fonction des caract\`eres 
de l'alg\`ebre $\widehat{su}(2)$. Par l'introduction des caract\`eres \'etendus $\hat{\chi}_a$ 
\cite{dijkgraaf,moore1,moore2}, nous 
donnons des formules g\'en\'erales permettant l'obtention de toutes ces fonctions  
de partition de mani\`ere simple et compacte.  
 
Il est difficile d'avoir un traitement unifi\'e pour les diff\'erents cas $ADE$, chaque cas 
ayant sa particularit\'e. Les cas $A_n$ sont 
trop simples, dans le sens o\`u diverses structures co\"{\i}ncident. Les cas $E_6$ et $E_8$ 
(cas exceptionnels) sont assez analogues. Les cas $D_{2n}$ sont les seuls qui conduisent 
\`a une alg\`ebre d'Ocneanu  
non-commutative, et forment avec les cas $A_n$, $E_6$ et $E_8$ les mod\`eles de type I. 
Enfin, les cas $D_{2n+1}$ et $E_7$ sont les mod\`eles de type II:  
leur alg\`ebre d'Ocneanu est d\'efinie \`a partir de l'alg\`ebre d'un graphe de type I.


\subsection{Les cas $A_n$} 
Les cas $A_n$ sont les plus simples des cas $ADE$. Nous traiterons ici l'exemple  
du cas $A_4$. La g\'en\'eralisation aux $A_n$ est presque imm\'ediate: 
nous donnerons un aper\c{c}u des calculs.

\subsubsection{Le cas $A_4$} 
\begin{figure}[hhh] 
\unitlength 0.8mm 
\begin{center} 
\begin{picture}(55,11)(0,10) 
\put(5,13){\line(1,0){45}} 
\multiput(5,13)(15,0){4}{\circle*{2}} 
\put(5,6){\makebox(0,0){$\tau_{0}$}} 
\put(20,6){\makebox(0,0){$\tau_{1}$}} 
\put(35,6){\makebox(0,0){$\tau_{2}$}} 
\put(50,6){\makebox(0,0){$\tau_{3}$}} 
\end{picture} 
\qquad \qquad 
$ 
{\cal G}_{A_4} = 
\left( \begin{array}{cccc} 
     0 & 1 & 0 & 0 \\ 
     1 & 0 & 1 & 0  \\ 
     0 & 1 & 0 & 1   \\ 
     0 & 0 & 1 & 0   \\ 
\end{array} 
\right) 
$ 
\caption{Le graphe $A_4$ et sa matrice d'adjacence.} 
\label{grA4} 
\end{center} 
\end{figure}
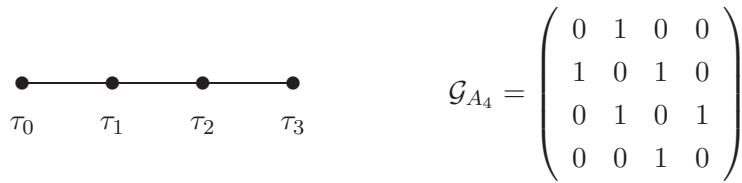

Le graphe $A_4$ et sa matrice d'adjacence sont illustr\'es \`a la Fig. \ref{grA4}, o\`u nous 
avons choisi l'ordre suivant de la base des vertex:  
$\{ \tau_0, \tau_1, \tau_2, \tau_3 \}$. 
 
Pour $A_4$, $\kappa = 5$, et la norme du graphe est \'egale au nombre d'or $\beta = 2 \cos 
(\frac{\pi}{5}) = \frac{1+\sqrt 5}{2}$, solution de l'\'equation $\beta^2 = 1 + \beta$. 
Les composantes du vecteur normalis\'e de Perron-Frobenius (dimensions quantiques des vertex) sont 
donn\'ees par: 
$P = \left( [1]_q, [2]_q, [2]_q, [1]_q \right) $, pour $q=\exp(i \pi / 5)$. 
Le graphe $A_4$ d\'etermine de mani\`ere unique l'alg\`ebre de graphe $A_4$. Nous connaissons dej\`a 
la multiplication par $\tau_0$ (l'identit\'e) et par $\tau_1$ (\`a l'aide du graphe): nous pouvons 
remplir les deux premi\`eres lignes et colonnes de la table.  En \'ecrivant 
$\tau_2 = \tau_1 . \tau_1 - \tau_0$ et $\tau_3 = \tau_1 . \tau_2 - \tau_1 = \tau_1 . \tau_1 . \tau_1 - 2 \tau_0$, nous pouvons alors compl\'eter toute la table, illustr\'ee ci-dessous: 
 
\begin{table}[hhh] 
$$ 
\begin{array}{|c||c|c|c|c|} 
\hline 
{} & \tau_0 & \tau_1 & \tau_2 & \tau_3 \\ 
\hline 
\hline 
\tau_0 & \tau_0 & \tau_1 & \tau_2 & \tau_3 \\ 
\tau_1 & \tau_1 & \tau_0+\tau_2 & \tau_1+\tau_3 & \tau_2 \\ 
\tau_2 & \tau_2 & \tau_1+\tau_3 & \tau_0+\tau_2 & \tau_1 \\ 
\tau_3 & \tau_3 & \tau_2 & \tau_1 & \tau_0 \\ 
\hline 
\end{array} 
$$ 
\caption{Table de multiplication de l'alg\`ebre de graphe $A_4$.} 
\end{table} 
 
Les matrices de fusion $N_i$, dont les \'el\'ements sont les constantes de structure de l'alg\`ebre 
de graphe $A_4$, peuvent se lire de cette table. Elles s'obtiennent plus facilement comme des polynomes 
de la matrice d'adjacence du graphe $A_4$: 
$$ 
\begin{array}{ll} 
N_0  = \munite_{4 \times 4}  \qquad \qquad& 
N_2  = N_1 . N_1 - N_0 \\ 
N_1  = {\cal G}_{A_4} & 
N_3  = N_1 . N_1 . N_1 - 2 . N_1 \\ 
\end{array} 
$$ 
ou bien directement par la formule de r\'ecurrence tronqu\'ee de $SU(2)$. Elles forment une 
repr\'esentation matricielle de l'alg\`ebre de graphe $A_4$. Dans la base choisie, 
elles sont donn\'ees par: 
\small 
$$ 
N_0 = \left( \begin{array}{cccc} 
     1 & 0 & 0 & 0 \\ 
     0 & 1 & 0 & 0 \\ 
     0 & 0 & 1 & 0 \\ 
     0 & 0 & 0 & 1 \\ 
\end{array} 
\right) 
\quad 
N_1 = \left( \begin{array}{cccc} 
     0 & 1 & 0 & 0 \\ 
     1 & 0 & 1 & 0 \\ 
     0 & 1 & 0 & 1 \\ 
     0 & 0 & 1 & 0 \\ 
\end{array} 
\right) 
\quad 
N_2 = \left( \begin{array}{cccc} 
     0 & 0 & 1 & 0 \\ 
     0 & 1 & 0 & 1 \\ 
     1 & 0 & 1 & 0 \\ 
     0 & 1 & 0 & 0 \\ 
\end{array} 
\right) 
\quad 
N_3 = \left( \begin{array}{cccc} 
     0 & 0 & 0 & 1 \\ 
     0 & 0 & 1 & 0 \\ 
     0 & 1 & 0 & 0 \\ 
     1 & 0 & 0 & 0 \\ 
\end{array} 
\right) 
$$ 
\normalsize 
et v\'erifient: 
\begin{equation} 
(N_i)_{jk} = (N_i)_{kj} = (N_k)_{ij} 
\end{equation} 
Le graphe de la s\'erie $A_n$ correspondant \`a $A_4$ est  $\ldots$ $A_4$ lui-m\^eme. 
Par cons\'equent, 
les matrices $F_i$ sont \'egales aux matrices de fusion $N_i$. L'alg\`ebre d'Ocneanu de $A_4$ est 
d\'efinie par: 
\begin{equation} 
Oc(A_4) = A_4 \otimes_{A_4} A_4 = A_4 \otimesdot A_4. 
\end{equation} 
o\`u nous identifions les \'el\'ements $\tau_i \otimesdot \tau_j$ avec $\tau_i . \tau_j \otimesdot \tau_0$.  
C'est une alg\`ebre de dimension 4 engendr\'ee par les quatre \'el\'ements suivants: 
$$ 
\ud0 = 0 \otimesdot 0, \qquad \qquad  \ud1 = 1 \otimesdot 0, \qquad \qquad 
\ud2 = 2 \otimesdot 0, \qquad \qquad  \ud3 = 3 \otimesdot 0, \qquad \qquad 
$$ 
et est donc isomorphe \`a l'alg\`ebre de graphe de $A_4$. Le graphe d'Ocneanu de $A_4$ est identique  
au graphe de $A_4$ lui-m\^eme. Les matrices $O_x$ codant la multiplication dans $Oc(A_4)$ et les matrices  
$S_x$ codant l'action de $Oc(A_4)$ sur $A_4$ sont \'egales aux matrices de fusion $N_i$. La dimension des blocs de 
la big\`ebre $\mathcal{B}(A_4)$ pour ses deux lois multiplicatives est donn\'ee par la somme des \'el\'ements  
des matrices $F_i$ et $S_x$, \'egales \`a $N_i$. Nous avons $d_i=d_x= (4,6,6,4)$ et les r\`egles de somme 
lin\'eaire et quadratique sont \'evidemment v\'erifi\'ees.  
$$ 
\sum_i d_i = \sum_x d_x = 20 \qquad \qquad \dim (\mathcal{B}A_4) = \sum_i d_i^2 = \sum_x d_x^2 = 104. 
$$ 
Les matrices toriques g\'en\'eralis\'ees sont d\'efinies par l'action de $A_4$ sur $Oc(A_4)$. Nous avons: 
\begin{eqnarray*} 
\tau_i . x . \tau_j = \tau_i . (\tau_x \otimesdot \tau_0) . \tau_j &=& \sum_m \sum_n (F_x)_{im} (F_0)_{jn}\; ( \tau_m \otimes \tau_n) \\ 
{ } &=& \sum_y \sum_m \sum_n (N_x)_{im} (N_0)_{jn} (N_m)_{ny} (\tau_y \otimes \tau_0)\\ 
{ } &=& \sum_m (N_x)_{im} (N_y)_{mj} \ud{y} = (N_x . N_y)_{ij} \ud{y} 
\end{eqnarray*} 
Les matrices toriques $W_{xy}$ sont donc \'egales \`a: 
\begin{equation} 
(W_{xy})_{ij} = (N_x . N_y)_{ij} 
\end{equation} 
Les fonction de partition g\'en\'eralis\'ees du mod\`ele $A_4$ sont d\'efinies par: 
\begin{equation} 
\mathcal{Z}_{x|y} = \sum_i \sum_j \chi_i(q) (N_x . N_y)_{ij} \ov{\chi}_j (q) 
\end{equation} 
et la fonction de partition invariante modulaire s'\'ecrit: 
\begin{equation} 
\mathcal{Z}_{A_4} = \sum_{i=0}^{3} |\chi_i(q)|^2. 
\end{equation} 
Nous donnons toutes les fonctions de partition du mod\`ele $A_4$ dans l'Annexe {\bf D}. 
 
\subsubsection{Formules g\'en\'erales pour $A_n$} 
Nous illustrons ci-dessous le graphe $A_n$, pour $n > 4$, et sa matrice d'adjacence, avec comme 
ordre de la base $\{ \tau_0, \tau_1, \tau_2, \ldots, \tau_{n-2}, \tau_{n-1} \}$. 
 
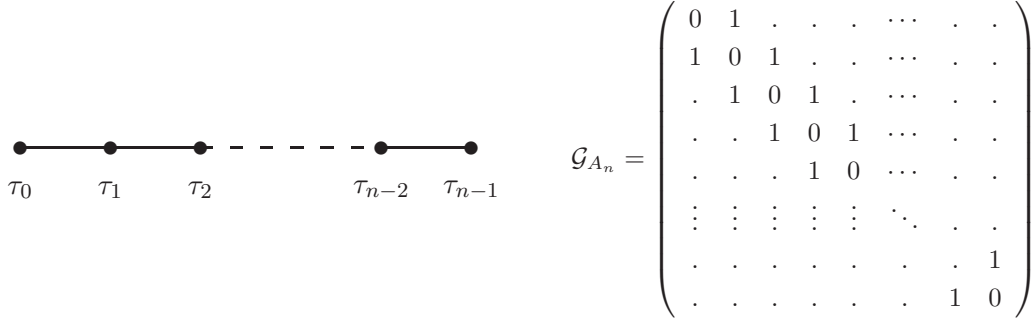
\begin{figure}[H] 
\unitlength 0.8mm 
\begin{center} 
\begin{picture}(90,11)(0,10) 
\thinlines 
\multiput(0,13)(15,0){3}{\circle*{2}} 
\multiput(60,13)(15,0){2}{\circle*{2}} 
\thicklines 
\multiput(30,13)(5,0){6}{\line(1,0){2}} 
\put(0,13){\line(1,0){30}} 
\put(60,13){\line(1,0){15}} 
\put(0,6){\makebox(0,0){$\tau_0$}} 
\put(15,6){\makebox(0,0){$\tau_1$}} 
\put(30,6){\makebox(0,0){$\tau_2$}} 
\put(60,6){\makebox(0,0){$\tau_{n-2}$}} 
\put(75,6){\makebox(0,0){$\tau_{n-1}$}} 
\end{picture} 
\small 
$ 
{\cal G}_{A_n} = 
\left( \begin{array}{cccccccc} 
     0 & 1 & . & . & . & \cdots & . & .  \\ 
     1 & 0 & 1 & . & . & \cdots & . & .  \\ 
     . & 1 & 0 & 1 & . & \cdots & . & .  \\ 
     . & . & 1 & 0 & 1 & \cdots & . & .  \\ 
     . & . & . & 1 & 0 & \cdots & . & .  \\ 
     \vdots & \vdots & \vdots & \vdots & \vdots & \ddots & . & .  \\ 
     . & . & . & . & . & . & . & 1  \\ 
     . & . & . & . & . & . & 1 & 0  \\ 
\end{array} 
\right) 
$ 
\normalsize 
\caption{Le graphe $A_n$ et sa matrice d'adjacence.} 
\label{grAn} 
\end{center} 
\end{figure} 
 
Pour les graphes $A_n$, $\kappa = n+1$ et $\beta = 2 \cos \frac{\pi}{n+1}$. Les composantes du vecteur normalis\'e de 
Perron-Frobenius d\'efinissant les dimensions quantiques des vertex sont donn\'ees par: 
\begin{eqnarray*} 
\textrm{n pair: }P &=& \left( [1]_q, [2]_q, [3_q], \ldots , \left[\frac{n}{2} -1\right]_q , \left[\frac{n}{2}\right]_q,  
\left[\frac{n}{2}\right]_q, \left[\frac{n}{2} -1\right]_q, \ldots, [3]_q, 
[2]_q, [1]_q \right) \\ 
\textrm{n impair: }P &=& \left( [1]_q, [2]_q, [3_q], \ldots , \left[\frac{n-1}{2}\right]_q\, , \,\left[\frac{n+1}{2}\right]_q\,,\,  
\left[\frac{n-1}{2}\right]_q\,,\, \ldots, [3]_q, 
[2]_q, [1]_q\right) 
\end{eqnarray*} 
Pour tous les cas $A_n$, l'alg\`ebre de graphe est enti\`erement d\'etermin\'ee par les donn\'ees du  graphe.  
Les matrices de fusion $N_i$ formant la repr\'esentation  matricielle de l'alg\`ebre de graphe $A_n$ 
s'obtiennent par la formule de r\'ecurrence tronqu\'ee de $SU(2)$.   
L'alg\`ebre d'Ocneanu de $A_n$ est d\'efinie par: 
\begin{equation} 
Oc(A_n) = A_n \otimes_{A_n} A_n = A_n \otimesdot A_n 
\end{equation} 
et co\"{\i}ncide avec l'alg\`ebre du graphe $A_n$. 
Pour les cas $A_n$, nous avons donc $\mathcal{A}(A_n) = Oc(A_n) = A_n$, les matrices $F_i$, $O_x$ et $S_x$ sont donc toutes 
\'egales \`a $N_i$. La dimension des blocs est donn\'ee pour les deux structures par la somme des \'el\'ements 
des matrice $N_i$.\\ 
Les matrices toriques g\'en\'eralis\'ees sont donn\'ees par: 
\begin{equation} 
(W_{xy})_{ij} = (N_x.N_y)_{ij} 
\end{equation} 
et les fonctions de partition g\'en\'eralis\'ees et l'invariante modulaire sont donn\'ees par: 
\begin{equation} 
\mathcal{Z}_{x | y} = \sum_{i=1}^{n-1}\, \sum_{j=1}^{n-1}\, \chi_i(q) \,(N_x . N_y)_{ij} \,\ov{\chi}_j(q) \qquad \qquad 
\mathcal{Z}_{A_n} = \sum_{i=1}^{n-1} \,|\chi_i(q)|^2 
\end{equation}


\subsection{Le cas $E_6$} 
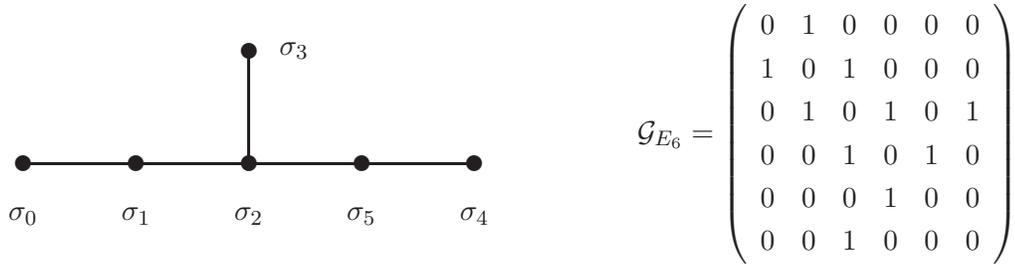
\begin{figure}[H] 
\unitlength 1.0mm 
\begin{center} 
\begin{picture}(80,18)(0,13) 
\thinlines 
\multiput(15,10)(15,0){5}{\circle*{2}} 
\put(45,25){\circle*{2}} 
\thicklines 
\put(15,10){\line(1,0){60}} 
\put(45,10){\line(0,1){15}} 
\put(15,3){\makebox(0,0){$\sigma_0$}} 
\put(30,3){\makebox(0,0){$\sigma_1$}} 
\put(45,3){\makebox(0,0){$\sigma_2$}} 
\put(60,3){\makebox(0,0){$\sigma_5$}} 
\put(75,3){\makebox(0,0){$\sigma_4$}} 
\put(51,25){\makebox(0,0){$\sigma_3$}} 
\end{picture} 
\qquad \qquad 
$ 
{\cal G}_{E_6} = 
\left( \begin{array}{cccccc} 
      0 & 1 & 0 & 0 & 0 & 0  \\ 
      1 & 0 & 1 & 0 & 0 & 0  \\ 
      0 & 1 & 0 & 1 & 0 & 1  \\ 
      0 & 0 & 1 & 0 & 1 & 0  \\ 
      0 & 0 & 0 & 1 & 0 & 0  \\ 
      0 & 0 & 1 & 0 & 0 & 0  \\ 
\end{array} 
\right) 
$ 
\label{grE6} 
\end{center} 
\caption{Le graphe $E_6$ et sa matrice d'adjacence.} 
\end{figure}

\paragraph{Graphe $E_6$ et matrices de fusion} 
Le graphe $E_6$ et sa matrice d'adjacence sont illustr\'es \`a la Fig. \ref{grE6}. Nous choisissons 
l'ordre suivant pour 
les vertex: $\{\sigma_0, \sigma_1, \sigma_2, \sigma_5, \sigma_4, \sigma_3\}$. 
 
Pour $E_6$, $\kappa = 12$, la norme du graphe est $\beta = 2 \cos 
(\frac{\pi}{12}) = \frac{1 + \sqrt 3 }{\sqrt 2}$ 
et les composantes du vecteur de Perron-Frobenius, sont: 
$P = \left( [1]_q, [2]_q, [3]_q, [2]_q, [1]_q, \frac{[3]_q}{[2]_q} 
\right)$, avec $q= \exp (\frac{i \pi }{12})$. 
Le graphe $E_6$ d\'etermine de mani\`ere unique la table de multiplication de l'alg\`ebre de graphe $E_6$, 
illustr\'ee ci-dessous:

\begin{table}[hhh] 
$$ 
\begin{array}{||c||c|c|c|c|c|c||} 
\hline 
{}& \sigma_0 & \sigma_1  & \sigma_2 & \sigma_5 & \sigma_4 & \sigma_3  \\ 
\hline 
\hline 
\sigma_0 & \sigma_0  & \sigma_1      & \sigma_2       & \sigma_5     & \sigma_4 & \sigma_3    \\ 
\sigma_1 & \sigma_1  & \sigma_0+\sigma_2    & \sigma_1+\sigma_3+\sigma_5   & \sigma_2+\sigma_4   & \sigma_5 & \sigma_2    \\ 
\sigma_2 & \sigma_2  & \sigma_1+\sigma_3+\sigma_5  & \sigma_0+\sigma_2+\sigma_2+\sigma_4 & \sigma_1+\sigma_3+\sigma_5 &\sigma_2 & \sigma_1+\sigma_5  \\ 
\sigma_5 & \sigma_5  & \sigma_2+\sigma_4    & \sigma_1+\sigma_3+\sigma_5   & \sigma_0+\sigma_2   & \sigma_1 & \sigma_2    \\ 
\sigma_4 & \sigma_4  & \sigma_5      & \sigma_2       & \sigma_1     & \sigma_0 & \sigma_3    \\ 
\sigma_3 & \sigma_3  & \sigma_2      & \sigma_1+\sigma_5     & \sigma_2     & \sigma_3 & \sigma_0+\sigma_4  \\ 
\hline 
\end{array} 
$$ 
\caption{Table de multiplication de l'alg\`ebre de graphe $E_6$.} 
\end{table} 
Les matrices de fusion $G_a$ de $E_6$ sont donn\'ees par les 
expressions suivantes: 
$$ 
\begin{array}{ll} 
G_0 = \munite_{6 \times 6} \qquad \qquad \qquad& 
G_4 = G_1.G_1.G_1.G_1 - 4 G_1.G_1 + 2 G_0 \\ 
G_1 = {\cal G}_{E_6} & 
G_5 = G_1.G_4 \\ 
G_2 = G_1.G_1 - G_0 & 
G_3 =  - G_1.(G_4 - G_1.G_1 + 2 G_0) 
\end{array} 
$$ 
\paragraph{Induction-restriction} 
Le graphe de la s\'erie $A_n$ poss\'edant le m\^eme nombre de Coxeter que $E_6$ est $A_{11}$. Les  
matrices de fusion $N_i$ de $A_{11}$ et les matrices $F_i$ codant l'action de $A_{11}$ sur $E_6$
s'obtiennent par: 
$$ 
\begin{array}{rclcrcl} 
N_0 &=& \munite_{11 \times 11} &\qquad& F_0 &=& \munite_{6 \times 6} \\ 
N_1 &=& \mathcal{G}_{A_{11}} &\qquad& F_1 &=& \mathcal{G}_{E_6} \\ 
N_i &=& N_1 . N_{i-1} - N_{i-2} &\qquad& F_i &=& F_1 . F_{i-1} - F_{i-2}  \qquad \qquad \qquad 2 \leq i \leq 10 
\end{array} 
$$ 
Les matrices essentielles 
$E_a$ ($(E_a)_{ib} = (F_i)_{ab}$) poss\`edent 11 lignes (labell\'ees par les vertex $\tau_i$ de $A_{11}$) et 6 colonnes 
(labell\'ees par les vertex $\sigma_b$ de $E_6$). La matrice essentielle $E_0$ ({\it intertwiner}) est illustr\'ee \`a la  
Fig. \ref{E0(E6)} et d\'efinit les r\`egles de branchement $A_{11} \hookrightarrow E_6: \tau_i \hookrightarrow  
\sum_a (E_0)_{ia}\sigma_a$. 
 
\begin{figure}[hhh] 
\unitlength 0.7mm 
\begin{center} 
$ 
E_0 = 
\left( \begin{array}{cccccc} 
        1 & . & . & . & . & . \cr . & 
        1 & . & . & . & . \cr . & . & 
        1 & . & . & . \cr . & . & . & 
        1 & . & 1 \cr . & . & 1 & . & 
        1 & . \cr . & 1 & . & 1 & . & 
        . \cr 1 & . & 1 & . & . & . \cr 
        . & 1 & . & . & . & 1 \cr . & 
        . & 1 & . & . & . \cr . & . & 
        . & 1 & . & . \cr . & . & . & 
        . & 1 & . \end{array} \right) 
\qquad \qquad \qquad 
\begin{array}{ccl} 
\tau_0 &\hookrightarrow& \sigma_0 \\  
\tau_1 &\hookrightarrow& \sigma_1 \\ 
\tau_2 &\hookrightarrow& \sigma_2 \\ 
\tau_3 &\hookrightarrow& \sigma_3 + \sigma_5 \\ 
\tau_4 &\hookrightarrow& \sigma_2 + \sigma_4 \\  
\tau_5 &\hookrightarrow& \sigma_1 + \sigma_5 \\ 
\tau_6 &\hookrightarrow& \sigma_0 + \sigma_2 \\ 
\tau_7 &\hookrightarrow& \sigma_1 + \sigma_3 \\ 
\tau_8 &\hookrightarrow& \sigma_2 \\ 
\tau_9 &\hookrightarrow& \sigma_5 \\ 
\tau_{10} &\hookrightarrow& \sigma_4  
\end{array} 
$ 
\caption{Matrice essentielle $E_0$ de $E_6$ et r\`egles de branchement $A_{11} \hookrightarrow E_6$.} 
\label{E0(E6)} 
\end{center} 
\end{figure}
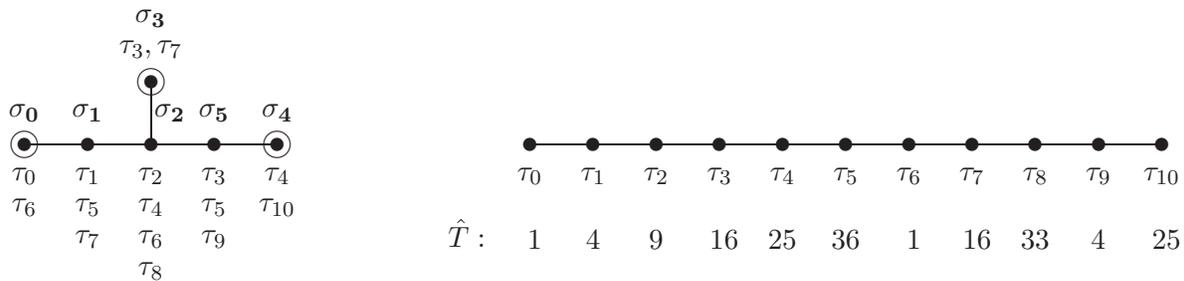
\begin{figure}[hhh] 
 
\unitlength 0.84mm 
 
\begin{center} 
\begin{picture}(180,40) 
\multiput(0,20)(10,0){5}{\circle*{2}} 
\put(20,30){\circle*{2}} 
\put(0,20){\line(1,0){40}} 
\put(20,20){\line(0,1){10}} 
\put(0,25){\makebox(0,0){$\bf{\sigma_{0}}$}} 
\put(0,15){\makebox(0,0){$\tau_{0}$}} 
\put(0,10){\makebox(0,0){$\tau_{6}$}} 
\put(10,25){\makebox(0,0){$\bf{\sigma_{1}}$}} 
\put(10,15){\makebox(0,0){$\tau_{1}$}} 
\put(10,10){\makebox(0,0){$\tau_{5}$}} 
\put(10,5){\makebox(0,0){$\tau_{7}$}} 
\put(23,25){\makebox(0,0){$\bf{\sigma_{2}}$}} 
\put(20,15){\makebox(0,0){$\tau_{2}$}} 
\put(20,10){\makebox(0,0){$\tau_{4}$}} 
\put(20,5){\makebox(0,0){$\tau_{6}$}} 
\put(20,0){\makebox(0,0){$\tau_{8}$}} 
\put(30,25){\makebox(0,0){$\bf{\sigma_{5}}$}} 
\put(30,15){\makebox(0,0){$\tau_{3}$}} 
\put(30,10){\makebox(0,0){$\tau_{5}$}} 
\put(30,5){\makebox(0,0){$\tau_{9}$}} 
\put(40,25){\makebox(0,0){$\bf{\sigma_{4}}$}} 
\put(40,15){\makebox(0,0){$\tau_{4}$}} 
\put(40,10){\makebox(0,0){$\tau_{10}$}} 
\put(20,40){\makebox(0,0){$\bf{\sigma_{3}}$}} 
\put(20,35){\makebox(0,0){$\tau_{3},\tau_{7}$}} 
 
\put(0,20){\circle{4}} 
\put(20,30){\circle{4}} 
\put(40,20){\circle{4}}

\multiput(80,20)(10,0){11}{\circle*{2}} 
\put(80,20){\line(1,0){100}} 
\put(80,15){\makebox(0,0){$\tau_0$}} 
\put(90,15){\makebox(0,0){$\tau_1$}} 
\put(100,15){\makebox(0,0){$\tau_2$}} 
\put(110,15){\makebox(0,0){$\tau_3$}} 
\put(120,15){\makebox(0,0){$\tau_4$}} 
\put(130,15){\makebox(0,0){$\tau_5$}} 
\put(140,15){\makebox(0,0){$\tau_6$}} 
\put(150,15){\makebox(0,0){$\tau_7$}} 
\put(160,15){\makebox(0,0){$\tau_8$}} 
\put(170,15){\makebox(0,0){$\tau_9$}} 
\put(180,15){\makebox(0,0){$\tau_{10}$}}

\put(70,6){\makebox(0,0){$\hat{T}:$}} 
 
\put(80,5){\makebox(0,0){ 1}} 
\put(90,5){\makebox(0,0){4}} 
\put(100,5){\makebox(0,0){9}} 
\put(110,5){\makebox(0,0){ 16}} 
\put(120,5){\makebox(0,0){25}} 
\put(130,5){\makebox(0,0){36}} 
\put(140,5){\makebox(0,0){ 1}} 
\put(150,5){\makebox(0,0){ 16}} 
\put(160,5){\makebox(0,0){33}} 
\put(170,5){\makebox(0,0){4}} 
\put(180,5){\makebox(0,0){ 25}} 
\end{picture} 
\end{center} 
\caption{Graphe d'induction $E_6 \hookleftarrow A_{11}$ et valeurs de l'exposant modulaire $\hat T$  
sur les vertex de $A_{11}$.} 
\label{E6/A11} 
\end{figure}

 
Pour connaitre l'induction $E_6 \hookleftarrow A_{11}$, il suffit de consid\'erer les r\`egles 
de branchement dans la direction 
oppos\'ee. Par exemple, $\sigma_3$ {\sl provient} de $\tau_3$ et de $\tau_7$ ($\sigma_3$ apparait dans la restriction 
de $\tau_3$ et de $\tau_7$ vers $E_6$), que nous \'ecrivons $\sigma_3 \hookleftarrow (\tau_3,\tau_7)$. 
Nous obtenons ainsi le graphe d'induction $E_6 \hookleftarrow A_{11}$ illustr\'e \`a la Fig. \ref{E6/A11}. L'op\'erateur $T$ du groupe 
modulaire est diagonal sur les vertex $\tau$ de $A_{11}$: \`a chaque vertex $\tau_i$ correspond une valeur 
de l'exposant modulaire $\hat{T}$ d\'efinie par $\hat{T}(\tau_i) = (i+1)^2 \textrm{mod } 48$. Les valeurs de 
$\hat{T}$ sur les vertex du graphe $A_{11}$ sont aussi pr\'esent\'ees \`a la Fig. \ref{E6/A11}.

Du graphe d'induction $E_6/A_{11}$, nous lisons par exemple $\sigma_0 \hookleftarrow (\tau_0,\tau_6)$. La valeur 
de $\hat{T}$ sur $\tau_0$ et $\tau_6$ est la m\^eme: ceci permet de d\'efinir une valeur de $\hat{T}$ de mani\`ere 
unique au vertex $\sigma_0$ de $E_6$: $\hat{T}(\sigma_0) = \hat{T}(\tau_0) = \hat{T}(\tau_6)$. Ceci est aussi  
valable pour les vertex $\sigma_3$ et $\sigma_4$ de $E_6$: $\hat{T}(\sigma_3) = \hat{T}(\tau_3) = \hat{T}(\tau_7)$ 
et $\hat{T}(\sigma_4) = \hat{T}(\tau_0) = \hat{T}(\tau_4)$. Pour les autres vertex de $E_6$, nous ne
pouvons pas d\'efinir de valeur fixe de $\hat{T}$. Par exemple:  
$\sigma_1 \hookleftarrow (\tau_1, \tau_5,\tau_7)$, mais $\hat{T}(\tau_1) \neq \hat{T}(\tau_5) \neq \hat{T}(\tau_7)$. 
Les vertex de $E_6$ pour lesquels une valeur de $\hat{T}$ est d\'efinie de mani\`ere unique par le m\'ecanisme 
d'induction forment le sous-espace $J = \{\sigma_0,\sigma_3,\sigma_4\}$, qui est aussi une sous-alg\`ebre 
de $E_6$ (les \'el\'ements de $J$ sont encercl\'es  
sur le graphe d'induction). 
Appelons $\ov{J}$ le sous-espace engendr\'e par les \'el\'ements $\{\sigma_1,\sigma_2,\sigma_5\}$. Alors, nous avons: 
$$ 
E_6 = J \oplus \ov{J}, \qquad \qquad J \; J \subset J, \qquad \qquad J\; \ov{J} \subset J. 
$$ 
$J$ fournit une partition de $E_6$ en classes d'\'equivalence: $\sigma_a \sim \sigma_b$ s'il existe un \'el\'ement 
$\sigma_c \in J$ tel que $(G_c)_{ab} \not= 0$ \cite{Bannai_Ito,DiFran,DiFran_Zub,Pet_Zub-Oc}. Pour $E_6$, 
nous avons par exemple: $\sigma_0.\sigma_1=\sigma_1, \sigma_3.\sigma_1=\sigma_2,\sigma_4.\sigma_1=\sigma_5$. 
Nous avons deux classes d'\'equivalence: 
\begin{eqnarray*} 
\sigma_0 \sim \sigma_3 \sim \sigma_4 &:& \tilde{\sigma_0} = \tilde{\sigma_3} = \tilde{\sigma_4} = \{\sigma_0,\sigma_3,\sigma_4\} = J\\ 
\sigma_1 \sim \sigma_2 \sim \sigma_5 &:& \tilde{\sigma_1} = \tilde{\sigma_2} = \tilde{\sigma_5} =\{\sigma_1,\sigma_2,\sigma_5\} = \ov{J} 
\end{eqnarray*} 
Un \'el\'ement $\sigma_b \notin J$ peut s'\'ecrire $(\rho(\sigma_b) . \sigma_1)$, o\`u $\rho(\sigma_b)$ 
est un \'el\'ement de $J$. Nous avons: 
$$ 
\begin{array}{rclcrcl} 
\sigma_1 &=& \sigma_0 . \sigma_1   &\qquad \qquad& \rho(\sigma_1) &=& \sigma_0 \\ 
\sigma_2 &=& \sigma_3 . \sigma_1   &\qquad \qquad& \rho(\sigma_2) &=& \sigma_3 \\ 
\sigma_5 &=& \sigma_4 . \sigma_1   &\qquad \qquad& \rho(\sigma_5) &=& \sigma_4 \\ 
\end{array} 
$$ 
\paragraph{Alg\`ebre d'Ocneanu} 
L'alg\`ebre d'Ocneanu de $E_6$ est d\'efinie par: 
\begin{equation} 
Oc(E_6) = \frac{E_6 \otimes E_6}{J} = E_6 \otimes_J E_6 = E_6 \otimesdot E_6 
\end{equation} 
Les \'el\'ements de $Oc(E_6)$ sont de la forme $\ud{x} = \sigma_a \otimesdot \sigma_b$, o\`u nous identifions 
les \'el\'ements $(\sigma_a \otimesdot \sigma_b . \sigma_c)$ avec  $(\sigma_a . \sigma_b \otimesdot \sigma_c)$ pour 
$\sigma_b \in J$. Une base de $Oc(E_6)$ est donn\'ee  par les 12 \'el\'ements lin\'eairement ind\'ependants $\sigma_a \otimesdot \sigma_0$ et $\sigma_a \otimesdot \sigma_1$, 
not\'es: 
$$ 
\begin{array}{lclcrcr} 
\ud0 = \sigma_0 \otimesdot \sigma_0, & \qquad \quad  & \ud3 = \sigma_3 \otimesdot \sigma_0, & 
\qquad \qquad & 
\ud{1^{'}} = \sigma_0 \otimesdot \sigma_1,  & \qquad \quad & \ud{31^{'}}=  \sigma_3 
\otimesdot \sigma_1, \\ 
\ud1 = \sigma_1 \otimesdot \sigma_0, & \qquad & \ud4 = \sigma_4 \otimesdot \sigma_0, & \qquad & 
\ud{11^{'}}=  \sigma_1 \otimesdot \sigma_1, & \qquad & \ud{41^{'}}=  \sigma_4 \otimesdot \sigma_1, \\ 
\ud2 = \sigma_2 \otimesdot \sigma_0, & \qquad & \ud5 = \sigma_5 \otimesdot \sigma_0, & \qquad & 
\ud{21^{'}}=  \sigma_2 \otimesdot \sigma_1, & \qquad & \ud{51^{'}}=  \sigma_5 \otimesdot \sigma_1. 
\end{array} 
$$ 
Nous avons les suivantes identifications dans $Oc(E_6)$: 
\begin{equation} 
\begin{array}{rcr} 
\sigma_a \otimesdot \sigma_b &=& \sigma_a .\sigma_b \otimesdot \sigma_0 \qquad \qquad \sigma_b \in J \\ 
\sigma_a \otimesdot \sigma_b &=& \sigma_a . \rho(\sigma_b) \otimesdot \sigma_1 \qquad \qquad \sigma_b \in \ov{J} 
\end{array} 
\label{identif_E6} 
\end{equation} 
La multiplication de l'alg\`ebre $Oc(E_6)$ est d\'efinie par: 
\begin{equation} 
(\sigma_a \otimesdot \sigma_b ) . (\sigma_c \otimesdot \sigma_d ) = \sigma_a . \sigma_c \otimesdot \sigma_b . \sigma_d 
\end{equation} 
L'\'el\'ement $ \ud0$ est l'identit\'e. Les \'el\'ements $\ud1$ et $\ud{1}'$ sont respectivement  
les g\'en\'erateurs chiraux gauche et  
droit: ils engendrent s\'eparemment deux sous-alg\`ebres $E_{6}\otimesdot 0$ et 
$0 \otimesdot E_{6}$, chacune isomorphe \`a l'alg\`ebre de graphe $E_6$. 
La partie ambichirale est par d\'efinition l'intersection de ces deux sous-alg\`ebres,  
elle est engendr\'ee par les \'el\'ements $\{\ud0, \ud3, \ud4 \}$. 
Nous pouvons v\'erifier que la multiplication par les g\'en\'erateurs de $Oc(E_6)$ 
est cod\'ee par le graphe d'Ocneanu de $E_6$, illustr\'e \`a la Fig. \ref{grocE6}.  
La multiplication par le g\'en\'erateur gauche $\ud{1}$ (resp. droit $\ud{1}'$) est donn\'ee 
par la somme des \'el\'ements du graphe reli\'es \`a $\ud{1}$ (resp. $\ud{1}'$) par une ligne 
continue (resp. discontinue).  
Par exemple, $\ud1 . \ud2 = \ud1 + \ud3 + \ud5$ et les vertex  $\ud1, \ud3$ et $\ud5$  
sont reli\'es au vertex $\ud2$ par une ligne continue;  
$\ud{1^{'}} . \ud4 = \ud{41^{'}}$ et les vertex $\ud4$ et $\ud{41^{'}}$ sont 
reli\'es par une ligne discontinue.

\begin{figure}[hhh] 
\unitlength 0.87mm 
\begin{center} 
\begin{picture}(50,70) 
\multiput(25,5)(0,10){3}{\circle*{2}} 
\multiput(25,45)(0,10){3}{\circle{2}} 
\multiput(5,25)(0,10){3}{\circle*{2}} 
\multiput(45,25)(0,10){3}{\circle*{2}} 
 
\thicklines 
\put(5,45){\line(1,1){20}} 
\put(5,35){\line(1,1){20}} 
\put(5,25){\line(1,1){20}} 
\put(5,25){\line(0,1){20}} 
 
\thinlines 
\put(45,45){\line(-1,-1){20}} 
\put(45,35){\line(-1,-1){20}} 
\put(45,25){\line(-1,-1){20}} 
\put(25.3,5){\line(0,1){20}} 
 
\thicklines 
\dashline[50]{1}(45,45)(25,65) 
\dashline[50]{1}(45,35)(25,55) 
\dashline[50]{1}(45,25)(25,45) 
\dashline[50]{1}(45,25)(45,45) 
 
\thinlines 
\dashline[50]{1}(5,45)(25,25) 
\dashline[50]{1}(5,35)(25,15) 
\dashline[50]{1}(5,25)(25,5) 
\dashline[50]{1}(24.7,5)(24.7,25) 
 
\small 
\put(25,68){\makebox(0,0){\ud0}} 
\put(25,58){\makebox(0,0){\ud3}} 
\put(25,48){\makebox(0,0){\ud4}} 
\put(1,45){\makebox(0,0){\ud1}} 
\put(1,35){\makebox(0,0){\ud2}} 
\put(1,25){\makebox(0,0){\ud5}} 
\put(49,45){\makebox(0,0){$\ud{1^{'}}$}} 
\put(49,35){\makebox(0,0){$\ud{31^{'}}$}} 
\put(49,25){\makebox(0,0){$\ud{41^{'}}$}} 
\put(21,25){\makebox(0,0){$\ud{11^{'}}$}} 
\put(21,15){\makebox(0,0){$\ud{21^{'}}$}} 
\put(21,5){\makebox(0,0){$\ud{51^{'}}$}} 
\normalsize 
 
\end{picture} 
\caption{Le graphe d'Ocneanu de $E_6$} 
\label{grocE6} 
\end{center} 
\end{figure} 
Explicitement, la multiplication des \'el\'ements de la base de $Oc(E_6)$ est donn\'ee par: 
\begin{eqnarray*} 
(\sigma_a \otimesdot \sigma_0 ) . (\sigma_c \otimesdot \sigma_0 ) &=& \sum_e (G_a)_{ce}\; (\sigma_e \otimesdot \sigma_0) \\ 
(\sigma_a \otimesdot \sigma_0 ) . (\sigma_c \otimesdot \sigma_1 ) = (\sigma_a \otimesdot \sigma_1 ) . (\sigma_c \otimesdot \sigma_0 )  
&=&  \sum_e (G_a)_{ce}\; (\sigma_e \otimesdot \sigma_1) \\ 
(\sigma_a \otimesdot \sigma_1 ) . (\sigma_c \otimesdot \sigma_1 ) &=& \sum_e (G_a)_{ce} \;(\sigma_e \otimesdot (\sigma_0+ 
\sigma_2))  
\end{eqnarray*} 
Or, $\sigma_2 = \sigma_3.\sigma_1$, donc $(\sigma_e \otimesdot \sigma_2) = \sum_{f} (G_3)_{ef}\; (\sigma_f \otimesdot \sigma_1)$, alors: 
$$ 
(\sigma_a \otimesdot \sigma_1 ) . (\sigma_c \otimesdot \sigma_1 ) = \sum_e (G_a)_{ce}\; (\sigma_e \otimesdot \sigma_0) 
+ \sum_e (G_a.G_3)_{ce} \; (\sigma_e \otimesdot \sigma_1)  
$$ 
Les matrices $(12 \times 12)$ $O_x$ forment une repr\'esentation de l'alg\`ebre $Oc(E_6)$. Les matrices  
correspondants aux g\'en\'erateurs $\ud{1}$ et $\ud{1'}$ sont les matrices d'adjacence du graphe d'Ocneanu.  
En choisissant comme ordre de la base des vertex l'ordre suivant: $\{\ud0,\ud1,\ud2,\ud5,\ud4,\ud3,\ud{1}',\ud{11}', 
\ud{21}',\ud{51}',\ud{41}',\ud{31}'\}$, ces matrices sont donn\'ees par: 
$$ 
\tiny 
O_1 = \left(  
\begin{array}{cccccccccccc} 
. & 1 & . & . & . & . & . & . & . & . & . & .  \\  
1 & . & 1 & . & . & . & . & . & . & . & . & .  \\  
. & 1 & . & 1 & . & 1 & . & . & . & . & . & .  \\  
. & . & 1 & . & 1 & . & . & . & . & . & . & .  \\  
. & . & . & 1 & . & . & . & . & . & . & . & .  \\  
. & . & 1 & . & . & . & . & . & . & . & . & .  \\  
. & . & . & . & . & . & . & 1 & . & . & . & .  \\  
. & . & . & . & . & . & 1 & . & 1 & . & . & .  \\  
. & . & . & . & . & . & . & 1 & . & 1 & . & 1  \\  
. & . & . & . & . & . & . & . & 1 & . & 1 & .  \\  
. & . & . & . & . & . & . & . & . & 1 & . & .  \\  
. & . & . & . & . & . & . & . & 1 & . & . & .   
\end{array} 
\right) 
= 
\Large 
\left(  
\begin{array}{c|c} 
G_1 & 0 \\ 
\hline 
0 & G_1 
\end{array} 
\right) 
$$ 
$$ 
\tiny 
O_{1'} = \left(  
\begin{array}{cccccccccccc} 
. & . & . & . & . & . & 1 & . & . & . & . & .  \\  
. & . & . & . & . & . & . & 1 & . & . & . & .  \\  
. & . & . & . & . & . & . & . & 1 & . & . & .  \\  
. & . & . & . & . & . & . & . & . & 1 & . & .  \\  
. & . & . & . & . & . & . & . & . & . & 1 & .  \\  
. & . & . & . & . & . & . & . & . & . & . & 1  \\  
1 & . & . & . & . & . & . & . & 1 & . & . & 1  \\  
. & 1 & . & . & . & . & . & 1 & . & 1 & . & .  \\  
. & . & 1 & . & . & . & . & . & 1 & . & . & .  \\  
. & . & . & 1 & . & . & . & . & . & . & . & .  \\  
. & . & . & . & 1 & . & . & . & . & . & . & 1  \\  
. & . & . & . & . & 1 & 1 & . & . & . & 1 & .   
\end{array} 
\right)  
= 
\Large 
\left(  
\begin{array}{c|c} 
0 & \munite  \\ 
\hline 
\munite & G_3 
\end{array} 
\right) 
$$ 
\normalsize 
D'apr\`es la multiplication de l'alg\`ebre $Oc(E_6)$, les matrices $O_x$ sont donn\'ees 
par\footnote{Les matrices $O_x$ sont explicitement calcul\'ees ici d'apr\`es la multiplication de notre r\'ealisation alg\'ebrique de $Oc(E_6)$. Dans \cite{Pet_Zub-Oc}, elles sont obtenues de mani\`ere empirique
d'apr\`es le graphe d'Ocneanu.}: 
$$ 
O_x = \left\lbrace 
\begin{array}{cc} 
\left( \begin{array}{c|c} G_a & 0 \\ \hline  0 & G_a \end{array} \right)  &  
\mathrm{pour \ } x=\sigma_a \otimesdot \sigma_0 \\ 
{ } & { } \\ 
\left( \begin{array}{c|c} 0 & G_a \\ \hline  G_a & G_3.G_a \end{array} \right)  & 
\mathrm{pour \ } x=\sigma_a \otimesdot \sigma_1  \\ 
\end{array} 
\right. 
$$ 
L'action de $Oc(E_6)$ sur $E_6$ est cod\'ee par les matrices $S_x$: 
$$ 
\ud{x} . \sigma_c = \sum_d (S_x)_{cd} \sigma_d  
$$ 
Elles sont explicitement donn\'ees par: 
$$S_{x} = \left\lbrace  
\begin{array}{rcl} 
G_a  \qquad \qquad \textrm{pour } \ud{x} &=& \sigma_a \otimesdot \sigma_0 \\ 
G_a.G_1  \qquad \qquad \textrm{pour } \ud{x} &=& \sigma_a \otimesdot \sigma_1  
\end{array} 
\right. 
$$

\paragraph{Dimensions des blocs} 
La dig\`ebre $\mathcal{B}(E_6)$ s'\'ecrit comme une somme de blocs (diagonalisation) pour ses 
deux structures 
multiplicatives. Pour la premi\`ere loi (convolution $\circ$), les blocs sont labell\'es par les onze vertex 
du graphe $A_{11}$. La dimension $d_{i}$, avec $i \in (0,1,2,\ldots 10)$, 
pour ces onze blocs est donn\'ee par la somme des \'el\'ements de la matrice $F_i$: $d_i = \sum_{a,b}(F_i)_{ab}$: 
$$ 
d_i: \qquad (6,10,14,18,20,20,20,18,14,10,6)  
$$ 
Pour la deuxi\`eme loi (convolution $\odot$), la dimension 
$d_{x}$ des douze blocs, labell\'es par $\ud{x}$ dans l'ordre 
$(\ud0, \ud1, \ud2, \ud5, \ud4, \ud3 ; \ud{1}',\ud{11}',\ud{21}',\ud{51}',\ud{41}',\ud{31}')$ est donn\'ee par  
la somme des \'el\'ements de la matrice $S_x$: $d_x = \sum_{a,b} (S_x)_{ab}$. Nous obtenons:
$$ 
d_x: \qquad (6,10,14,10,6,8,10,20,28,20,10,14)  
$$ 
Les r\`egles de somme quadratique et lin\'eaire sont v\'erifi\'ees: 
$$ 
\dim({\cal B}{E}_6) = \sum_{i \in A_{11}} d_i^2 = \sum_{x \in Oc(E_6)} d_x^2 = 2512, \qquad \qquad \qquad 
\sum_{i} {d_{i}} = \sum_{x} {d_{x}} = 156. 
$$ 
La masse quantique de $E_6$ et de $\mathcal{A}(E_6)=A_{11}$ sont d\'efinies par la somme du carr\'e des dimensions 
quantiques de leur irreps (composantes du vecteur de Perron-Frobenius). Nous avons: 
$m(E_6) = 4(3+\sqrt{3}), m(A_{11})=24(2+\sqrt{3})$. L'alg\`ebre d'Ocneanu de $E_6$ est $E_6 \otimes_J E_6$, 
o\`u $J=\{\sigma_0,\sigma_3,\sigma_4\}$. La dimension quantique de $J$ est  
$m(J) = qdim^2(\sigma_0)+qdim^2(\sigma_3)+qdim^2(\sigma_4) = 4$. Alors, la relation de masse quantique entre 
$Oc(E_6)$ et $A_{11}$ est v\'erifi\'ee: 
$$ 
m(Oc(E_6)) = \frac{m(E_6).m(E_6)}{m(J)} = m(A_{11}) = 24(2+\sqrt{3}) 
$$

\paragraph{Matrices toriques g\'en\'eralis\'ees} 
L'action de $A_{11}$ sur $Oc(E_6)$ est cod\'ee par les matrices $W_{xy}$. Pour calculer 
explicitement ces matrices, calculons 
l'action (\`a droite et \`a gauche) de $A_{11}$ sur un \'el\'ement $x = \sigma_a \otimesdot \sigma_b$ de  
$Oc(E_6)$, en utilisant les identifications (\ref{identif_E6}): 
\begin{eqnarray*} 
\tau_i . (\sigma_a \otimesdot \sigma_b) . \tau_j &=& 
 \sum_{(c,d) \in E_6} (F_i)_{ac} (F_j)_{bd}\; (\sigma_c \otimesdot \sigma_d) \\ 
{ } &=& \sum_{c} \sum_{d \in J} (F_i)_{ac} (F_j)_{bd}\; (\sigma_c.\sigma_d \otimesdot \sigma_0)  
+ \sum_{c} \sum_{d \in \ov{J}} (F_i)_{ac} (F_j)_{bd}\; (\sigma_c.\rho(\sigma_d) \otimesdot \sigma_1) \\ 
{ } &=& \sum_e \sum_{c} \sum_{d \in J} (F_i)_{ac} (F_j)_{bd} (G_c)_{de} \; (\sigma_e \otimesdot \sigma_0) \\ 
{ } &+& \sum_e \sum_{c} \sum_{d \in \ov{J}} (F_i)_{ac} (F_j)_{bd} (G_c)_{\rho(d)e}\; (\sigma_e \otimesdot \sigma_1)  
\end{eqnarray*} 
Les matrices de fusion $G$ de $E_6$ commutent entre-elles  et sont sym\'etriques, donc nous avons:  
$(G_a)_{bc} = (G_c)_{ab}$. Alors: 
\begin{eqnarray*} 
\tau_i . (\sigma_a \otimesdot \sigma_b) . \tau_j &=& 
\sum_e \sum_{c} \sum_{d \in J} (F_i)_{ac} (F_j)_{bd} (G_e)_{cd}\; (\sigma_e \otimesdot \sigma_0) \\ 
{ } &+& \sum_e \sum_{c} \sum_{d \in \ov{J}} (F_i)_{ac} (F_j)_{bd} (G_e)_{c\rho(d)} \; (\sigma_e \otimesdot \sigma_1)  
\end{eqnarray*} 
Les matrices toriques g\'en\'eralis\'ees $W_{xy}$ seront not\'ees $W_{ab,ef}$, pour  
$\ud{x} = \sigma_a \otimesdot \sigma_b$ et $\ud{y} = \sigma_e \otimesdot \sigma_f$ ($b$ et $f$ =0,1). Elles  
s'\'ecrivent donc sous la forme compacte suivante: 
\begin{equation} 
\begin{array}{|c|} 
\hline 
{ } \\ 
W_{ab,ef} = \left\lbrace 
\begin{array}{lcl} 
\displaystyle \sum_{c \in E_6} \sum_{d \in J} (F_i)_{ac} (F_j)_{bd} (G_e)_{cd} &\qquad \qquad& f=0 \\ 
\displaystyle \sum_{c \in E_6} \sum_{d \in \ov{J}} (F_i)_{ac} (F_j)_{bd} (G_e)_{c\rho(d)} &\qquad \qquad& f=1 \\ 
\end{array} 
\right. \\ 
{ } \\ 
\hline 
\end{array} 
\label{Wxy_E6} 
\end{equation}  
Nous pouvons  
alors v\'erifier que les matrices $(\widetilde{W}_{ij})_{xy} = (W_{xy})_{ij}$ satisfont bien l'alg\`ebre 
carr\'ee de fusion: 
$$ 
\widetilde{W}_{ij} \; \widetilde{W}_{i'j'} = \sum_{i''} \sum_{j''} \mathcal{N}_{ii'}^{i''} 
\mathcal{N}_{jj'}^{j''} \widetilde{W}_{jj'} \; .
$$  
L'invariant modulaire $\mathcal{M}$ correspond \`a la matrice $W_{00,00}$. Par la formule 
(\ref{Wxy_E6}), et utilisant le 
fait que $(G_0)_{cd}= \munite_{cd} = \delta_{c,d}$, $\mathcal{M}$ est donc \'egal \`a: 
$$ 
\mathcal{M}_{ij} = (W_{00,00})_{ij} = \sum_{d \in J} (F_i)_{0d} (F_j)_{0d} 
$$  
et nous pouvons v\'erifier qu'il commute avec les g\'en\'erateurs $S$ et $T$ du groupe modulaire.

\paragraph{Fonctions de partition g\'en\'eralis\'ees} 
Elles sont d\'efinies \`a partir des matrices toriques g\'en\'eralis\'ees par: 
\begin{equation} 
\mathcal{Z}_{x|y} = \sum_{i \in A_{11}} \sum_{j \in A_{11}} \chi_i (q) (W_{xy})_{ij} \ov{\chi}_j (q) 
\end{equation} 
o\`u les $\chi_i(q)$ sont les caract\`eres de l'alg\`ebre
$\widehat{su}(2)$.  
Introduisons les caract\`eres \'etendus $\hat{\chi}_a(q)$, d\'efinis \`a partir de la matrice essentielle 
$E_0$ par: 
\begin{equation} 
\hat{\chi}_a(q) = \sum_{i \in A_{11}} (E_0)_{ia} \chi_i(q) = \sum_{i \in A_{11}} (F_i)_{0a} \chi_i(q) 
\end{equation} 
Ils sont explicitement donn\'es par: 
$$ 
\begin{array}{rclcrcl} 
\hat{\chi}_0 &=& \chi_0 + \chi_6  &\qquad& \hat{\chi}_3 &=& \chi_3 + \chi_7 \\ 
\hat{\chi}_1 &=& \chi_1 + \chi_5 + \chi_7    &\qquad& \hat{\chi}_4 &=& \chi_4 + \chi_{10} \\ 
\hat{\chi}_2 &=& \chi_2 + \chi_4 + \chi_6 + \chi_8    &\qquad& \hat{\chi}_5 &=& \chi_3 + \chi_5 + \chi_9  
\end{array} 
$$ 
Introduisons aussi les caract\`eres \'etendus g\'en\'eralis\'es $\hat{\chi}_{ab}$, qui sont des combinaisons 
lin\'eaires des caract\`eres \'etendus $\hat{\chi}_a$ ou des caract\`eres $\chi_i$ (utilisant la propri\'et\'e  
$E_a = E_0 . G_a$): 
\begin{equation} 
\hat{\chi}_{ab} = \sum_{c \in E_6} (G_a)_{bc} \; \hat{\chi}_c \; = \sum_{i \in A_{11}} (F_i)_{ab} \chi_i (q) 
\end{equation} 
Alors, toutes les fonctions de partition g\'en\'eralis\'ees du mod\`ele $E_6$ s'\'ecrivent  
sous la forme compacte suivante: 
\begin{equation} 
\begin{array}{|c|} 
\hline 
{ }\\ 
\mathcal{Z}_{ab,ef} = \left\lbrace 
\begin{array}{lcl} 
\displaystyle \sum_c \sum_{d \in J} \hat{\chi}_{ac} (q) \, (G_e)_{cd}\, \ov{\hat{\chi}}_{bd} (q)      &\qquad \qquad& f=0 \\ 
\displaystyle \sum_c \sum_{d \in \ov{J}} \hat{\chi}_{ac} (q)\, (G_e)_{c\rho(d)}\, \ov{\hat{\chi}}_{bd} (q)      &\qquad \qquad& f=1  
\end{array} 
\right. \\ 
{ } \\ 
\hline 
\end{array} 
\end{equation} 
Pour $y = \sigma_0 \otimesdot \sigma_0$, les matrices toriques $W_{x}=W_{x0}$ et les fonctions de partition 
$\mathcal{Z}_{x}$ (une ligne de d\'efauts) sont donn\'ees, pour $\ud{x} = \sigma_a \otimesdot \sigma_b$, par: 
\begin{equation} 
(W_{ab})_{ij} = \sum_{d \in J} (F_i)_{ad} (F_j)_{bd} \qquad \qquad  
\mathcal{Z}_{ab} = \sum_{d \in J} \hat{\chi}_{ad}(q) \ov{\hat{\chi}}_{bd}(q) 
\end{equation} 
Les matrices toriques $W_{x}$ du mod\`ele $E_6$ sont publi\'ees dans \cite{Coq-qtetra}. Les fonctions  
de partition correspondantes $\mathcal{Z}_{x}$ sont donn\'ees dans \cite{Coq_Gil-ADE} en fonction 
des caract\`eres 
de $A_{11}$. Nous les \'ecrivons sous forme compacte en fonction des caract\`eres 
\'etendus $\hat{\chi}_a$ de $E_6$  
dans l'Annexe {\bf D}.
La fonction de partition invariante modulaire de $E_6$ est diagonale en fonction des 
caract\`eres \'etendus $\hat{\chi}_a$: 
\begin{eqnarray*} 
\mathcal{Z}_{E_6} = \sum_{d \in J} \hat{\chi}_d (q) \ov{\hat{\chi}}_d (q)  
&=& |\hat{\chi}_0|^2 + |\hat{\chi}_3|^2 + |\hat{\chi}_4|^2 \\ 
{ } &=& |\chi_0 + \chi_6|^2 + |\chi_3 + \chi_7|^2 + |\chi_4 + \chi_{10}|^2   
\end{eqnarray*} 
et nous retrouvons la fonction de partition invariante modulaire de la classification de  
Cappelli, Itzykson et Zuber\cite{CIZ-class2} labell\'ee par $E_6$.

\paragraph{Propri\'et\'es modulaires} 
Les propri\'et\'es modulaires des fonctions de partition s'\'etudient \`a travers les matrices toriques  
g\'en\'eralis\'ees obtenues par (\ref{Wxy_E6}).  
La fonction de partition $\mathcal{Z}_{E_6}$ est invariante modulaire car $\mathcal{M}$ commute 
avec les g\'en\'erateurs  
$T$ et $S$ du groupe modulaire. Les autres fonctions de partition ne sont pas invariantes 
modulaires. N\'eanmoins, elles satisfont les propri\'et\'es remarquables suivantes \cite{Coq_Marina}: 
\begin{itemize} 
\item Aucune des matrices $W_{xy}$ (autre que $W_{00}$\footnote{La matrice $W_{4,4}$ commute avec $T$ et $S$, mais cela 
provient du fait que $W_{44}=W_{00}$.}) ne commute avec $T$ et $S$. 
\item Toutes les matrices $W_{xy}$ commutent avec l'op\'erateur $ST^{-1}S$. 
\item Les matrices $W_{xy}$ commutent avec une certaine puissance de l'op\'erateur $T$. 
\end{itemize}


\subsection{Le cas $E_8$} 
\paragraph{Graphe $E_8$ et matrices de fusion} 
Le graphe $E_8$ et sa matrice d'adjacence sont illustr\'es \`a la Fig. \ref{grE8}, o\`u l'ordre choisi pour  
repr\'esenter les vertex est: 
$\{\sigma_0, \sigma_1, \sigma_2, \sigma_3, \sigma_4, \sigma_7, \sigma_6, \sigma_5 \}$. 
 
\begin{figure}[hhh] 
\unitlength 0.85mm 
\begin{center} 
\begin{picture}(115,30)(5,10) 
\thinlines 
\multiput(15,10)(15,0){7}{\circle*{2}} 
\put(75,25){\circle*{2}} 
\thicklines 
\put(15,10){\line(1,0){90}} 
\put(75,10){\line(0,1){15}} 
\put(15,3){\makebox(0,0){$\sigma_0$}} 
\put(30,3){\makebox(0,0){$\sigma_1$}} 
\put(45,3){\makebox(0,0){$\sigma_2$}} 
\put(60,3){\makebox(0,0){$\sigma_3$}} 
\put(75,3){\makebox(0,0){$\sigma_4$}} 
\put(90,3){\makebox(0,0){$\sigma_7$}} 
\put(105,3){\makebox(0,0){$\sigma_6$}} 
\put(81,25){\makebox(0,0){$\sigma_5$}} 
\end{picture} 
\footnotesize$ 
{\cal G}_{E_8} = 
\left( \begin{array}{cccccccc} 
      0 & 1 & 0 & 0 & 0 & 0 & 0 & 0 \\ 
      1 & 0 & 1 & 0 & 0 & 0 & 0 & 0 \\ 
      0 & 1 & 0 & 1 & 0 & 0 & 0 & 0 \\ 
      0 & 0 & 1 & 0 & 1 & 0 & 0 & 0 \\ 
      0 & 0 & 0 & 1 & 0 & 1 & 0 & 1 \\ 
      0 & 0 & 0 & 0 & 1 & 0 & 1 & 0 \\ 
      0 & 0 & 0 & 0 & 0 & 1 & 0 & 0 \\ 
      0 & 0 & 0 & 0 & 1 & 0 & 0 & 0 \\ 
\end{array} 
\right) 
$ 
\normalsize 
\caption{Le graphe $E_8$ et sa matrice d'adjacence.} 
\label{grE8} 
\end{center} 
\end{figure} 
 
Pour $E_8$, $\kappa = 30$, la norme du graphe est $\beta = 2 \cos (\frac{\pi}{30})$ et les composantes du vecteur de Perron-Frobenius sont donn\'ees par 
$P = \left( [1]_q, [2]_q, [3]_q, [4]_q, [5]_q, \frac{[7]_q}{[2]_q} , 
\frac{[5]_q}{[3]_q}, 
\frac{[5]_q}{[2]_q} \right)$, avec $q = \exp (\frac{i \pi}{30})$. 
Le graphe $E_8$ d\'etermine de mani\`ere unique l'alg\`ebre de graphe $E_8$, dont la table de  
multiplication est illustr\'ee ci-dessous\footnote{Pour une meilleure visibilit\'e, les vertex $\sigma_a$  
de $E_8$ sont d\'esign\'es uniquement par leur indice $a$.}: 
 
\begin{table}[hhh] 
\tiny 
$$ 
\begin{array}{||c||c|c|c|c|c|c|c|c||} 
\hline 
{}& 0 & 1  & 2 & 3 & 4 & 7 & 6 & 5 \\ 
\hline 
\hline 
0 & 0 & 1     & 2       & 3           &   4         &  7        & 6 
& 5       \\ 
1 & 1 & 0+2   & 1+3     & 2+4         & 3+5+7       & 4+6       & 7 
& 4       \\ 
2 & 2 & 1+3   & 0+2+4   & 1+3+5+7     & 2+4+4+6     & 3+5+7     & 4 
& 3+7     \\ 
3 & 3 & 2+4   & 1+3+5+7 & 0+2+4+4+6   & 1+3+3+5+7+7 & 2+4+4     & 3+5 
& 2+4+6   \\ 
4 & 4 & 3+5+7 & 2+4+4+6 & 1+3+3+5+7+7 & 0+2+2+4+4+4+6 & 1+3+3+5+7 & 2+4 
& 1+3+5+7 \\ 
7 & 7 & 4+6   & 3+5+7   & 2+4+4       & 1+3+3+5+7   & 0+2+4+6   & 1+7 
& 2+4     \\ 
6 & 6 & 7     & 4       & 3+5         & 2+4         & 1+7       & 0+6 
& 3       \\ 
5 & 5 & 4     & 3+7     & 2+4+6       & 1+3+5+7     & 2+4       & 3 
& 0+4     \\ 
\hline 
\end{array} 
$$ 
\normalsize 
\caption{Table de multiplication de l'alg\`ebre de graphe $E_8$.} 
\end{table} 
Les matrices de fusion $G_a$  de  $E_8$ sont donn\'ees par les 
expressions suivantes: 
$$ 
\begin{array}{ll} 
G_0 = \munite_{8 \times 8} \qquad \qquad \qquad& 
G_4 = G_1.G_3 - G_2 \\ 
G_1 = {\cal G}_{E_8} & 
G_6 = G_2.G_4 - G_2 - G_4 - G_4 \\ 
G_2 = G_1.G_1 - G_0 & 
G_7 = G_1.G_6 \\ 
G_3 = G_1.G_2 - G_1 & 
G_5 = G_6.G_3 - G_3 
\end{array} 
$$ 
\paragraph{Induction-restriction} 
Le graphe de la s\'erie $A_n$ poss\'edant le m\^eme nombre de Coxeter que $E_8$ est $A_{29}$. Les 
matrices de fusion $N_i$ de $A_{29}$ et les matrices $F_i$ codant l'action de $A_{29}$ sur $E_8$ 
s'obtiennent 
par la formule de r\'ecurrence tronqu\'ee de $SU(2)$. Les matrices essentielles 
$E_a$ ($(E_a)_{ib} = (F_i)_{ab}$) poss\`edent 29 lignes (label\'ees par les vertex $\tau$ de $A_{29}$) et 8 colonnes 
(label\'ees par les vertex $\sigma$ de $E_8$). La matrice essentielle $E_0$ ({\it intertwiner}) est illustr\'ee \`a la  
Fig. \ref{E0(E8)} et d\'efinit les r\`egles de branchement $A_{29} \hookrightarrow E_8: \tau_i \hookrightarrow  
\sum_a (E_0)_{ia}\sigma_a$. Nous obtenons alors le graphe d'induction $E_8 \hookleftarrow A_{29}$, illustr\'e aussi \`a la  
Fig. \ref{E0(E8)}. 
 
\begin{figure}[hhh] 
\unitlength 0.7mm 
\begin{center} 
\scriptsize 
$ 
E_0 = 
\left( 
\begin{array}{cccccccc} 
1& .& .& .& .& .& .& .\\ 
.& 1& .& .& .& .& .& .\\ 
.& .& 1& .& .& .& .& .\\ 
.& .& .& 1& .& .& .& .\\ 
.& .& .& .& 1& .& .& .\\ 
.& .& .& .& .& 1& .& 1\\ 
.& .& .& .& 1& .& 1& .\\ 
.& .& .& 1& .& 1& .& .\\ 
.& .& 1& .& 1& .& .& .\\ 
.& 1& .& 1& .& .& .& 1\\ 
1& .& 1& .& 1& .& .& .\\ 
.& 1& .& 1& .& 1& .& .\\ 
.& .& 1& .& 1& .& 1& .\\ 
.& .& .& 1& .& 1& .& 1\\ 
.& .& .& .& 2& .& .& .\\ 
.& .& .& 1& .& 1& .& 1\\ 
.& .& 1& .& 1& .& 1& .\\ 
.& 1& .& 1& .& 1& .& .\\ 
1& .& 1& .& 1& .& .& .\\ 
.& 1& .& 1& .& .& .& 1\\ 
.& .& 1& .& 1& .& .& .\\ 
.& .& .& 1& .& 1& .& .\\ 
.& .& .& .& 1& .& 1& .\\ 
.& .& .& .& .& 1& .& 1\\ 
.& .& .& .& 1& .& .& .\\ 
.& .& .& 1& .& .& .& .\\ 
.& .& 1& .& .& .& .& .\\ 
.& 1& .& .& .& .& .& .\\ 
1& .& .& .& .& .& .& . 
\end{array} 
\right) 
\normalsize 
\qquad 
$ 
\begin{picture}(90,0)(0,55) 
\thinlines 
\multiput(0,55)(15,0){7}{\circle*{2}} 
\put(60,70){\circle*{2}} 
\put(0,55){\circle{4}} 
\put(90,55){\circle{4}} 
\thicklines 
\put(0,55){\line(1,0){90}} 
\put(60,55){\line(0,1){15}} 
 
\put(0,60){\makebox(0,0){$\bf{\sigma_{0}}$}} 
\put(15,60){\makebox(0,0){$\bf{\sigma_{1}}$}} 
\put(30,60){\makebox(0,0){$\bf{\sigma_{2}}$}} 
\put(45,60){\makebox(0,0){$\bf{\sigma_{3}}$}} 
\put(64,60){\makebox(0,0){$\bf{\sigma_{4}}$}} 
\put(75,60){\makebox(0,0){$\bf{\sigma_{7}}$}} 
\put(90,60){\makebox(0,0){$\bf{\sigma_{6}}$}} 
\put(60,80){\makebox(0,0){$\bf{\sigma_{5}}$}} 
 
\put(0,50){\makebox(0,0){$\tau_{0}$}} 
\put(0,45){\makebox(0,0){$\tau_{10}$}} 
\put(0,40){\makebox(0,0){$\tau_{18}$}} 
\put(0,35){\makebox(0,0){$\tau_{28}$}} 
 
\put(15,50){\makebox(0,0){$\tau_{1}$}} 
\put(15,45){\makebox(0,0){$\tau_{9}$}} 
\put(15,40){\makebox(0,0){$\tau_{11}$}} 
 
\put(15,35){\makebox(0,0){$\tau_{17}$}} 
\put(15,30){\makebox(0,0){$\tau_{19}$}} 
\put(15,25){\makebox(0,0){$\tau_{27}$}} 
 
\put(30,50){\makebox(0,0){$\tau_{2}$}} 
\put(30,45){\makebox(0,0){$\tau_{8}$}} 
\put(30,40){\makebox(0,0){$\tau_{10}$}} 
\put(30,35){\makebox(0,0){$\tau_{12}$}} 
\put(30,30){\makebox(0,0){$\tau_{16}$}} 
\put(30,25){\makebox(0,0){$\tau_{18}$}} 
 
\put(30,20){\makebox(0,0){$\tau_{20}$}} 
\put(30,15){\makebox(0,0){$\tau_{26}$}} 
 
\put(45,50){\makebox(0,0){$\tau_{3}$}} 
\put(45,45){\makebox(0,0){$\tau_{7}$}} 
\put(45,40){\makebox(0,0){$\tau_{9}$}} 
\put(45,35){\makebox(0,0){$\tau_{11}$}} 
\put(45,30){\makebox(0,0){$\tau_{13}$}} 
\put(45,25){\makebox(0,0){$\tau_{15}$}} 
\put(45,20){\makebox(0,0){$\tau_{17}$}} 
\put(45,15){\makebox(0,0){$\tau_{19}$}} 
\put(45,10){\makebox(0,0){$\tau_{21}$}} 
\put(45,5){\makebox(0,0){$\tau_{25}$}} 
 
\put(60,50){\makebox(0,0){$\tau_{4}$}} 
\put(60,45){\makebox(0,0){$\tau_{6}$}} 
\put(60,40){\makebox(0,0){$\tau_{8}$}} 
\put(60,35){\makebox(0,0){$\tau_{10}$}} 
\put(60,30){\makebox(0,0){$\tau_{12}$}} 
 
\put(60,25){\makebox(0,0){$2\tau_{14}$}} 
\put(60,20){\makebox(0,0){$\tau_{16}$}} 
\put(60,15){\makebox(0,0){$\tau_{18}$}} 
\put(60,10){\makebox(0,0){$\tau_{20}$}} 
\put(60,5){\makebox(0,0){$\tau_{22}$}} 
\put(60,0){\makebox(0,0){$\tau_{24}$}} 
 
\put(75,50){\makebox(0,0){$\tau_{5}$}} 
\put(75,45){\makebox(0,0){$\tau_{7}$}} 
\put(75,40){\makebox(0,0){$\tau_{11}$}} 
\put(75,35){\makebox(0,0){$\tau_{13}$}} 
\put(75,30){\makebox(0,0){$\tau_{15}$}} 
\put(75,25){\makebox(0,0){$\tau_{17}$}} 
\put(75,20){\makebox(0,0){$\tau_{21}$}} 
\put(75,15){\makebox(0,0){$\tau_{23}$}} 
 
\put(90,50){\makebox(0,0){$\tau_{6}$}} 
\put(90,45){\makebox(0,0){$\tau_{12}$}} 
\put(90,40){\makebox(0,0){$\tau_{16}$}} 
\put(90,35){\makebox(0,0){$\tau_{22}$}} 
 
\put(60,75){\makebox(0,0){$\tau_{5},\tau_{9},\tau_{13},\tau_{15},\tau_{19}, 
\tau_{23}$}} 
 
\end{picture} 
\bigskip 
\caption{Matrice essentielle $E_0$ de $E_8$ et graphe d'induction $E_8 \hookleftarrow A_{29}$.} 
\label{E0(E8)} 
\end{center} 
\end{figure} 
 
La valeur de $\hat{T}$ sur les vertex 
$(\tau_0,\tau_1,\tau_2, \cdots,\tau_{28})$ de $A_{29}$ \mbox{(\'egale \`a  $(j+1)^2$ mod 120 pour $\tau_j$)} est 
donn\'ee par la liste suivante: 
$$ 
(\ud{1},4,9,16,25,36,\udd{49},64,81,100, \ud{1},22,\udd{49},76,105,16,\udd{49},84,\ud{1},40, 
81,4,\udd{49},96,25,76,9,64,\ud{1}) 
$$ 
Les seuls vertex $\sigma$ de $E_8$ pour lesquels une valeur de $\hat{T}$ est bien d\'efinie par le m\'ecanisme 
d'induction sont $\sigma_0$ et $\sigma_6$: $\hat{T}(\sigma_0) = 1$, $\hat{T}(\sigma_6)=49$. Ils engendrent le 
sous-espace $J$, qui est une sous-alg\`ebre de $E_8$.  
$J$ fournit une partition de $E_8$ en quatre classes d'\'equivalence:  
$$ 
\begin{array}{rclcc} 
\sigma_0 \sim \sigma_6 &:& \tilde{\sigma_0} = \tilde{\sigma_6}  = \{\sigma_0,\sigma_6\} = J  
& \qquad \qquad & \phi(\sigma_0) = \phi(\sigma_6) = \sigma_0 \\ 
\sigma_1 \sim \sigma_7 &:& \tilde{\sigma_1} = \tilde{\sigma_7}  = \{\sigma_1,\sigma_7\}  
& \qquad & \phi(\sigma_1) = \phi(\sigma_7) = \sigma_1 \\ 
\sigma_2 \sim \sigma_4 &:& \tilde{\sigma_2} = \tilde{\sigma_4}  = \{\sigma_2,\sigma_4\}  
& \qquad & \phi(\sigma_2) = \phi(\sigma_4) = \sigma_2 \\ 
\sigma_5 \sim \sigma_3 &:& \tilde{\sigma_5} = \tilde{\sigma_3}  = \{\sigma_3,\sigma_5\}  
& \qquad & \phi(\sigma_5) = \phi(\sigma_3) = \sigma_5 \\ 
\end{array} 
$$ 
o\`u l'application $\phi$ choisit un repr\'esentant dans chaque classe d'\'equivalence: l'ensemble $\Phi$ 
est donn\'e par $\Phi = \{\phi(\sigma_a)\} = \{\sigma_0,\sigma_1,\sigma_2,\sigma_5 \}$.  Notons  
qu'un \'el\'ement $\sigma_b \notin \Phi$ peut s'\'ecrire $\sigma_b = \sigma_6 . \phi(\sigma_b)$. En effet: 
$$ 
\sigma_6 = \sigma_6 . \sigma_0, \qquad  \sigma_7 = \sigma_6 . \sigma_1, \qquad  
\sigma_4 = \sigma_6 . \sigma_2, \qquad  \sigma_3 = \sigma_6 . \sigma_5 
$$ 
Introduisons alors l'application $\rho$ d\'efinie par: 
$$ 
\begin{array}{rcl} 
\rho(\sigma_a) = \sigma_0 & \qquad &\textrm{si } \sigma_a \in \Phi : \sigma_a \in \{\sigma_0, \sigma_1, \sigma_2,  
\sigma_5 \} \\ 
\rho(\sigma_a) = \sigma_6 & \qquad &\textrm{si } \sigma_a \notin \Phi : \sigma_a \in \{\sigma_6, \sigma_7, \sigma_4,  
\sigma_3 \}  
\end{array} 
$$ 
 
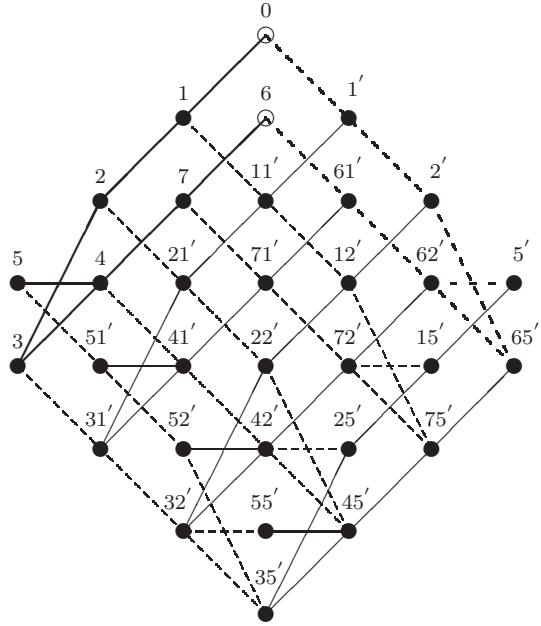
\begin{figure}[hhh] 
\unitlength 1.1mm 
\begin{center} 
\begin{picture}(70,80) 
\multiput(5,35)(0,10){2}{\circle*{2}} 
\multiput(15,25)(0,10){4}{\circle*{2}} 
\multiput(25,15)(0,10){6}{\circle*{2}} 
\multiput(35,5)(0,10){6}{\circle*{2}} 
\multiput(35,65)(0,10){2}{\circle{2}} 
\multiput(45,15)(0,10){6}{\circle*{2}} 
\multiput(55,25)(0,10){4}{\circle*{2}} 
\multiput(65,35)(0,10){2}{\circle*{2}}

\thicklines 
\put(35,75){\line(-1,-1){20}} 
\put(5,35){\line(1,1){30}} 
\put(5,35){\line(1,2){10}} 
\put(5,45){\line(1,0){10}} 
 
\thinlines 
\put(45,65){\line(-1,-1){20}} 
\put(15,25){\line(1,1){30}} 
\put(15,25){\line(1,2){10}} 
\put(15,35){\line(1,0){10}}

\put(55,55){\line(-1,-1){20}} 
\put(25,15){\line(1,1){30}} 
\put(25,15){\line(1,2){10}} 
\put(25,25){\line(1,0){10}} 
 
\put(65,45){\line(-1,-1){20}} 
\put(35,5){\line(1,1){30}} 
\put(35,5){\line(1,2){10}} 
\put(35,15){\line(1,0){10}}

\thicklines 
\dashline[50]{1}(35,75)(55,55) 
\dashline[50]{1}(65,35)(35,65) 
\dashline[50]{1}(65,35)(55,55) 
\dashline[50]{1}(65,45)(55,45)

\thinlines 
\dashline[50]{1}(25,65)(45,45) 
\dashline[50]{1}(55,25)(25,55) 
\dashline[50]{1}(55,25)(45,45) 
\dashline[50]{1}(55,35)(45,35)

\dashline[50]{1}(15,55)(35,35) 
\dashline[50]{1}(45,15)(15,45) 
\dashline[50]{1}(45,15)(35,35) 
\dashline[50]{1}(45,25)(35,25)

\dashline[50]{1}(5,45)(25,25) 
\dashline[50]{1}(35,5)(5,35) 
\dashline[50]{1}(35,5)(25,25) 
\dashline[50]{1}(35,15)(25,15)

\scriptsize 
\put(35,78){\makebox(0,0){0}} 
\put(35,68){\makebox(0,0){6}} 
\put(35,59){\makebox(0,0){$11^{'}$}} 
\put(35,49){\makebox(0,0){$71^{'}$}} 
\put(35,39){\makebox(0,0){$22^{'}$}} 
\put(35,29){\makebox(0,0){$42^{'}$}} 
\put(35,19){\makebox(0,0){$55^{'}$}} 
\put(35.5,10){\makebox(0,0){$35^{'}$}} 
 
\put(25,68){\makebox(0,0){1}} 
\put(25,58){\makebox(0,0){7}} 
\put(25,49){\makebox(0,0){$21^{'}$}} 
\put(25,39){\makebox(0,0){$41^{'}$}} 
\put(25,29){\makebox(0,0){$52^{'}$}} 
\put(24.5,19){\makebox(0,0){$32^{'}$}}

\put(46,69){\makebox(0,0){$1^{'}$}} 
 
\put(45,59){\makebox(0,0){$61^{'}$}} 
 
\put(45,49){\makebox(0,0){$12^{'}$}} 
 
\put(45,39){\makebox(0,0){$72^{'}$}} 
\put(45,29){\makebox(0,0){$25^{'}$}} 
\put(46,19){\makebox(0,0){$45^{'}$}} 
 
\put(15,58){\makebox(0,0){2}} 
\put(15,48){\makebox(0,0){4}} 
\put(15,39){\makebox(0,0){$51^{'}$}} 
\put(15,29){\makebox(0,0){$31^{'}$}} 
 
\put(56,59){\makebox(0,0){$2^{'}$}} 
\put(55,49){\makebox(0,0){$62^{'}$}} 
\put(55,39){\makebox(0,0){$15^{'}$}} 
\put(56,29){\makebox(0,0){$75^{'}$}} 
 
\put(5,48){\makebox(0,0){5}} 
\put(5,38){\makebox(0,0){3}} 
 
\put(66,49){\makebox(0,0){$5^{'}$}} 
\put(66.5,39){\makebox(0,0){$65^{'}$}} 
\normalsize 
\end{picture} 
\caption{Le graphe d'Ocneanu de $E_8$.} 
\label{grocE8} 
\end{center} 
\end{figure}

\paragraph{Alg\`ebre d'Ocneanu} 
L'alg\`ebre d'Ocneanu $Oc(E_8)$ est d\'efinie par: 
\begin{equation} 
Oc(E_8) = \frac{E_8 \otimes E_8}{J} = E_8 \otimes_J E_8 = E_8 \otimesdot E_8 
\end{equation} 
Les \'el\'ements de $Oc(E_8)$ sont de la forme $\ud{x} = \sigma_a \otimesdot \sigma_b$, o\`u nous identifions 
les \'el\'ements $(\sigma_a  \otimesdot \sigma_b . \sigma_c)$ avec  $(\sigma_a . \sigma_b \otimesdot \sigma_c)$ pour 
$\sigma_b \in J$. L'alg\`ebre $Oc(E_8)$ est de dimension $6 . 8 /2 =32$. 
Une base de $Oc(E_8)$ est donn\'ee par les 32 \'el\'ements lin\'eairement ind\'ependants suivants: 
$$ 
\ud{a} = \sigma_a \otimesdot \sigma_0, \qquad \ud{a1'} = \sigma_a \otimesdot \sigma_1, \qquad  
\ud{a2'} = \sigma_a \otimesdot \sigma_2, \qquad \ud{a5'} = \sigma_a \otimesdot \sigma_5.  
$$ 
et utilisant les applications $\rho$ et $\phi$ introduites, nous avons les identifications  
suivantes dans l'alg\`ebre $Oc(E_8)$: 
\begin{equation} 
\sigma_a \otimesdot \sigma_b = \sigma_a . \rho(\sigma_b) \otimesdot \phi(\sigma_b) 
\label{identif_E8} 
\end{equation} 
La multiplication dans $Oc(E_8)$ est d\'efinie par: 
\begin{equation} 
(\sigma_a \otimesdot \sigma_b) . (\sigma_c \otimesdot \sigma_d) = \sigma_a . \sigma_c \otimesdot \sigma_b . \sigma_d 
\end{equation} 
L'\'el\'ement $ \ud0$ est l'identit\'e. Les \'el\'ements $\ud1$ et $\ud{1}'$ sont respectivement  
les g\'en\'erateurs chiraux gauche et  
droit: ils engendrent s\'eparemment deux sous-alg\`ebres $E_{8}\otimesdot 0$ et 
$0 \otimesdot E_{8}$, chacune isomorphe \`a l'alg\`ebre de graphe $E_8$. 
Les seuls \'el\'ements ambichiraux sont $\ud0$  et $\ud6$. 
Nous pouvons v\'erifier que la multiplication par les g\'en\'erateurs de $Oc(E_8)$ 
est cod\'ee par le graphe d'Ocneanu de $E_8$, illustr\'e \`a la Fig. \ref{grocE8}.  
La multiplication par le g\'en\'erateur gauche $\ud{1}$ (resp. droit $\ud{1}'$) est donn\'ee 
par la somme des \'el\'ements du graphe reli\'es \`a $\ud{1}$ (resp. $\ud{1}'$) par une ligne 
continue (resp. discontinue).

 
 
Les matrices $O_x$ qui codent la multiplication dans $Oc(E_8)$ sont explicitement donn\'ees dans 
la base $\{ \{\sigma_a \otimesdot \sigma_0\}, \{\sigma_a \otimesdot \sigma_0\}, \{\sigma_a \otimesdot \sigma_0\},\{\sigma_a \otimesdot \sigma_0\} \}$ par: 
$$ 
\small
O_x = \left\lbrace 
\begin{array}{cc} 
\left( \begin{array}{c|c|c|c}  
G_a & 0 & 0 & 0 \\  
0 & G_a & 0 & 0 \\ 
0 & 0 & G_a & 0  \\ 
0 & 0 & 0 & G_a  
\end{array} \right)  &  
\mathrm{pour \ } x=\sigma_a \otimesdot \sigma_0 \\ 
{ } & { } \\ 
\left( \begin{array}{cccc}  
0 & G_a & 0 & 0 \\  
G_a & 0 & G_a & 0 \\ 
0 & G_a & 0 &  G_a.G_6   \\ 
0 & 0 & G_a.G_6 & 0   
\end{array} \right)  &  
\mathrm{pour \ } x=\sigma_a \otimesdot \sigma_1 \\ 
{ } & { } \\ 
\left( \begin{array}{c|c|c|c}  
0 & 0 & G_a & 0 \\  
0 & G_a & 0 & G_a.G_6 \\ 
G_a & 0 & G_a+G_a.G_6 & 0  \\ 
0 & G_a.G_6 & 0 & G_a.G_6 
\end{array} \right)  &  
\mathrm{pour \ } x=\sigma_a \otimesdot \sigma_2 \\ 
{ } & { } \\ 
\left( \begin{array}{c|c|c|c}  
0 & 0 & 0 & G_a \\  
0 & 0 & G_a.G_6 & 0 \\ 
0 & G_a.G_6 & 0 & G_a.G_6  \\ 
G_a & 0 & G_a.G_6 & 0  
\end{array} \right)  &  
\mathrm{pour \ } x=\sigma_a \otimesdot \sigma_5 \\ 
{ } & { }  
\end{array} 
\right. 
\normalsize
$$ 
Nous d\'eterminons ainsi par notre r\'ealisation de $Oc(E_8)$ les matrices $O_x$, qui co\"{\i}ncident 
avec celles publi\'ees dans \cite{Pet_Zub-Oc}, o\`u elles sont d\'etermin\'ees de mani\`ere empirique.
Les matrices $S_x$ qui codent l'action de $Oc(E_8)$ sur $E_8$ sont explicitement donn\'ees par: 
$$ 
S_x = G_a . G_b \qquad \qquad \textrm{pour } \ud{x} = \sigma_a \otimesdot \sigma_b 
$$ 
\paragraph{Dimension des blocs} 
Les dimensions $d_i$ des 29 blocs de la dig\`ebre $\mathcal{B}(E_8)$ pour la loi $\circ$ 
sont donn\'ees par $\sum_{a,b}(F_i)_{ab}$. Nous avons $F_{i-28}=F_i$. 
Donc, pour $i$ dans $(0,1,\ldots,14)$:  
$$ 
d_i = d_{28-i} = (8,14,20,26,32,38,44,48,52,56,60,62,64,64,64)  
$$ 
Les dimensions $d_x$ des 32 blocs de la dig\`ebre $\mathcal{B}(E_8)$ pour la loi $\odot$ 
sont donn\'ees, pour ($\ud{x} = (\sigma_a \otimesdot \sigma_0), (\sigma_a \otimesdot \sigma_1),  
(\sigma_a \otimesdot \sigma_2),(\sigma_a \otimesdot \sigma_5)$), par: 
\scriptsize 
$$ 
d_x = (8,14,20,26,32,16,12,22),(14,28,40,52,64,32,22,44),(20,40,60,78,96,48,32,64),(16,32,48,64,78,40,26,52) 
$$ 
\normalsize 
Nous v\'erifions la r\`egle de somme quadratique et lin\'eaire: 
$$ 
dim(\mathcal{B}E_8) = \sum_{i \in A_{29}} d_i^2 = \sum_{x \in Oc(E_8)} d_x^2 = 63136 \; , \qquad \qquad 
\sum_i d_i   = \sum_x d_x = 1240 \; .
$$  
D\'efinissant la masse quantique de $Oc(E_8)$ \`a travers sa r\'ealisation $Oc(E_8)=E_8 \otimes_J E_8$, avec  
$m(J)=qdim^2(\sigma_0)+qdim^2(\sigma_6)$, alors la relation de masse entre $Oc(E_8)$ et $A_{29}$ est satisfaite: 
$$ 
m(Oc(E_8)) = \frac{m(E_8).m(E_8)}{m(J)} = m(A_{29}) 
$$   
 
\paragraph{Matrices toriques g\'en\'eralis\'ees} 
Nous calculons l'action de $A_{29}$ sur $Oc(E_8)$, en utilisant l'\'equation (\ref{identif_E8}) qui donne 
les identifications des \'el\'ements de l'alg\`ebre $Oc(E_8)$: 
\begin{eqnarray*} 
\tau_i . (\sigma_a \otimesdot \sigma_b) . \tau_j &=& \sum_c \sum_d (F_i)_{ac} (F_j)_{bd}\; (\sigma_c \otimesdot \sigma_d) \\ 
{ } &=& \sum_c \sum_d (F_i)_{ac} (F_j)_{bd}\; (\sigma_c . \rho(\sigma_d)  \otimesdot \phi(\sigma_d) ) \\ 
{ } &=& \sum_c \sum_d \sum_e (F_i)_{ac} (F_j)_{bd} (G_e)_{c\rho(d)}\; (\sigma_e \otimesdot \phi(\sigma_d)) 
\end{eqnarray*} 
Divisant alors la sommation sur $d$ en une sommation sur chaque classe d'\'equivalence, nous obtenons la formule 
compacte suivante pour les matrices toriques g\'en\'eralis\'ees: 
\begin{equation} 
\begin{array}{|c|} 
\hline 
{ } \\ 
W_{ab,ef} = \displaystyle \sum_c \sum_{d \in \tilde{f}}  (F_i)_{ac} (F_j)_{bd} (G_e)_{c \rho(d)}\\ 
{ } \\ 
\hline 
\end{array}  
\label{Wxy_E8} 
\end{equation} 
o\`u la sommation sur $d$ se fait sur les \'el\'ements de la classe d'\'equivalence de $\sigma_f$.  
Pour $f=0$, la sommation sur $d$ est sur $\tilde{0} = J$, et pour $\sigma_d \in J$, $\rho(d)=d$. Alors, 
les matrices toriques $W_{x} = W_{x0}$ sont donn\'ees par: 
\begin{equation} 
W_x = \sum_{c \in J} \; (F_i)_{ac} \; (F_j)_{bc} \qquad \qquad x = \sigma_a \otimesdot \sigma_b 
\end{equation} 
et nous pouvons v\'erifier que l'invariant modulaire $\mathcal{M} = W_0$ commute avec les g\'en\'erateurs $S$ et 
$T$ du groupe modulaire. 
 
\paragraph{Fonctions de partition g\'en\'eralis\'ees} 
Elles sont d\'efinies \`a partir des matrices toriques g\'en\'eralis\'ees par: 
\begin{equation} 
\mathcal{Z}_{x|y} = \sum_{i \in A_{29}} \sum_{j \in A_{29}} \chi_i (q) (W_{xy})_{ij} \ov{\chi}_j (q) 
\end{equation} 
o\`u les $\chi_i(q)$ sont les caract\`eres de l'alg\`ebre $\widehat{su}(2)$.  
Introduisons les caract\`eres \'etendus $\hat{\chi}_a(q)$ du mod\`ele $E_8$, d\'efinis \`a partir  
de la matrice essentielle $E_0$ par: 
\begin{equation} 
\hat{\chi}_a(q) = \sum_{i \in A_{11}} (E_0)_{ia} \chi_i(q) = \sum_{i \in A_{11}} (F_i)_{0a} \chi_i(q) 
\end{equation} 
et les caract\`eres \'etendus g\'en\'eralis\'es $\hat{\chi}_{ab}$, qui sont des combinaisons 
lin\'eaires des caract\`eres \'etendus $\hat{\chi}_a$ ou des caract\`eres $\chi_i$ (utilisant la propri\'et\'e  
$E_a = E_0 . G_a$): 
\begin{equation} 
\hat{\chi}_{ab} = \sum_{c \in E_8} (G_a)_{bc} \; \hat{\chi}_c \; = \sum_{i \in A_{29}} (F_i)_{ab} \chi_i (q) 
\end{equation} 
Les caract\`eres \'etendus du mod\`ele $E_8$ sont donn\'es dans l'Annexe {\bf D}.  
Les fonctions de partition g\'en\'eralis\'ees du mod\`ele $E_8$ s'\'ecrivent  
sous la forme compacte suivante: 
\begin{equation} 
\begin{array}{|c|} 
\hline 
{ } \\ 
\mathcal{Z}_{ab,ef} = \displaystyle \sum_c \sum_{d \in \tilde{f}} \, \hat{\chi}_{ac} \;(G_e)_{c \rho(d)}\; \ov{\hat{\chi}}_{bd}\\ 
{ } \\ 
\hline 
\end{array}  
\label{Zxy_E8} 
\end{equation} 
Les fonctions de partition $\mathcal{Z}_{x}$ (une ligne de d\'efauts) sont donn\'ees, pour $\ud{x} = \sigma_a \otimesdot \sigma_b$, par: 
\begin{equation} 
\mathcal{Z}_{ab} = \sum_{d \in J} \hat{\chi}_{ad}(q) \ov{\hat{\chi}}_{bd}(q) 
\end{equation} 
Elles sont publi\'ees dans \cite{Coq_Gil-ADE} en fonction des caract\`eres 
de $A_{11}$, nous les \'ecrivons sous forme compacte en fonction des caract\`eres \'etendus  
$\hat{\chi}_a$ de $E_8$ dans l'Annexe {\bf D}.
La fonction de partition invariante modulaire de $E_8$ est diagonale en fonction des 
caract\`eres \'etendus $\hat{\chi}_a$: 
\begin{eqnarray*} 
\mathcal{Z}_{E_8} = \sum_{d \in J} \hat{\chi}_d (q) \ov{\hat{\chi}}_d (q)  
&=& |\hat{\chi}_0|^2 + |\hat{\chi}_6|^2 \\ 
{ } &=& | \chi_0 + \chi_{10} + \chi_{18} + \chi_{28} |^2 + | \chi_6 + \chi_{12} + \chi_{16} + \chi_{22} |^2    
\end{eqnarray*} 
et nous retrouvons la fonction de partition invariante modulaire de la classification de  
Cappelli, Itzykson et Zuber\cite{CIZ-class2} labell\'ee par $E_8$.


\subsection{Les cas $D_{2n}$} 
 
\subsubsection{Le cas $D_4$} 
\begin{figure}[H] 
\unitlength 0.9mm 
\begin{center} 
\begin{picture}(45,8)(0,9) 
\thinlines 
\multiput(5,10)(15,0){2}{\circle*{2}} 
\put(35,17,5){\circle*{2}} 
\put(35,2,5){\circle*{2}} 
\thicklines 
\put(5,10){\line(1,0){15}} 
\put(20,10){\line(2,1){15}} 
\put(20,10){\line(2,-1){15}} 
\put(5,5){\makebox(0,0){$\sigma_0$}} 
\put(20,5){\makebox(0,0){$\sigma_1$}} 
\put(40,18){\makebox(0,0){$\sigma_2$}} 
\put(41,3){\makebox(0,0){$\sigma_{2^{'}}$}} 
\end{picture} 
\qquad \qquad \qquad 
$ 
{\cal G}_{D_4} = 
\left( \begin{array}{cccc} 
      0 & 1 & 0 & 0   \\ 
      1 & 0 & 1 & 1   \\ 
      0 & 1 & 0 & 0   \\ 
      0 & 1 & 0 & 0   \\ 
\end{array} 
\right) 
$ 
\caption{Le graphe $D_4$ et sa matrice d'adjacence.} 
\label{grD4} 
\end{center} 
\end{figure}
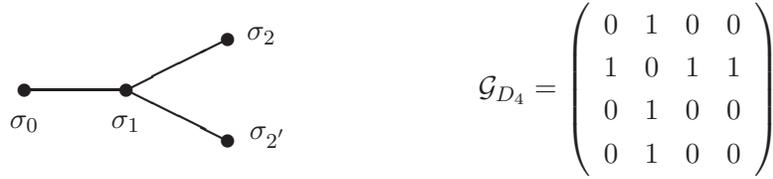

\paragraph{Graphe $D_4$ et matrices de fusion} 
Le graphe $D_4$ et sa matrice d'adjacence sont illustr\'es \`a la Fig. \ref{grD4}, avec l'ordre suivant pour les vertex:  
$\{\sigma_0, \sigma_1, \sigma_2, \sigma_{2^{' }} \}$.

Pour $D_4$, $\kappa = 6$, la norme est $\beta = 2 \cos 
(\frac{\pi}{6}) = \sqrt 3$ et le vecteur de Perron-Frobenius
$P = \left( [1]_q, [2]_q, \frac{[2]_q}{[2]_q}, 
\frac{[2]_q}{[2]_q} \right)$, avec $q = \exp (\frac{i \pi}{6})$.\
Le graphe $D_4$ ne d\'etermine pas de mani\`ere unique la table de multiplication  
de l'alg\`ebre 
de graphe $D_4$: en effet, de l'\'equation $\sigma_1 . \sigma_1 = \sigma_0 + \sigma_2 + \sigma_2'$, nous savons 
comment multiplier par $(\sigma_2 + \sigma_2')$, mais les deux vertex $\sigma_2$ et $\sigma_2'$ de la fourche  
sont indistingables. Nous pouvons remplir toute la table de multiplication, \`a l'exception   
de la multiplication des vertex $\sigma_2$ et $\sigma_2'$ entre-eux.   
N\'eanmoins, en {\sl imposant} que les coefficients de structures de l'alg\`ebre de graphe $D_4$ 
soient des entiers non n\'egatifs, nous trouvons une solution {\sl unique}.
 
\begin{table}[hhh] 
$$ 
\begin{array}{||c||c|c|c|c||} 
\hline 
{}& \sigma_0 & \sigma_1  & \sigma_2 & \sigma_2^{'}   \\ 
\hline 
\hline 
\sigma_0     & \sigma_0     & \sigma_1         & \sigma_2     & \sigma_2^{'}  \\ 
\sigma_1     & \sigma_1     & \sigma_0+\sigma_2+\sigma_2^{'} & \sigma_1     & \sigma_1      \\ 
\sigma_2     & \sigma_2     & \sigma_1         & \sigma_2^{'} & \sigma_0      \\ 
\sigma_2^{'} & \sigma_2^{'} & \sigma_1         & \sigma_0     & \sigma_2      \\ 
\hline 
\end{array} 
$$ 
\caption{Table de multiplication de l'alg\`ebre de graphe $D_4$.} 
\end{table} 
Les matrices de fusion $G_{a}$ sont donn\'ees par: 
$$ 
G_0 = \munite_{4 \times 4} 
\qquad \qquad \qquad 
G_1 = {\cal G}_{D_4} \qquad \qquad \qquad 
G_2 + G_{2^{'}} = G_1.G_1 - G_0 
$$ 
et les matrices des vertex de la fourche s'\'ecrivent:
$$ 
G_2 = \left( \begin{array}{cccc} 
0 & 0 & 1 & 0 \\ 
0 & 1 & 0 & 0 \\ 
0 & 0 & 0 & 1 \\ 
1 & 0 & 0 & 0 \\ 
\end{array} 
\right) 
\qquad \qquad \qquad 
G_{2^{'}} = \left( \begin{array}{cccc} 
0 & 0 & 0 & 1 \\ 
0 & 1 & 0 & 0 \\ 
1 & 0 & 0 & 0 \\ 
0 & 0 & 1 & 0 \\ 
\end{array} 
\right) 
$$ 
Contrairement aux cas $E_6$ et $E_8$, les matrices de fusion $G_a$ de $D_4$  ne sont pas sym\'etriques 
($G_{2'}=G_2^T$). 
\paragraph{Induction-restriction} 
Le graphe de la s\'erie $A_n$ de m\^eme norme que $D_4$ est $A_5$. Les matrices de fusion $N_i$ 
de $A_5$ et les matrices $F_i$ codant l'action de $A_5$ sur $D_4$ sont calcul\'ees par la formule 
de r\'ecurrence tronqu\'ee de $SU(2)$: 
$$ 
\begin{array}{lcl} 
N_0  = \munite_{5 \times 5}    & \qquad \qquad \qquad \qquad & F_0  = \munite_{4 \times 4}\\ 
N_1  = \mathcal{G}_{A_5} & { } & F_1 = \mathcal{G}_{D_4}  \\ 
N_i  = N_1.N_{i-1} - N_{i-2}  & { } & F_i = F_1.F_{i-1} - F_{i-2}  \qquad \qquad 2\leq i \leq 5 
\end{array} 
$$ 
Les matrices essentielles $E_a$ de $D_4$ poss\`edent 5 lignes (label\'ees par les vertex $\tau$ de $A_5$) et 4 
colonnes (label\'ees par les vertex $\sigma$ de $D_4$). L'{\it intertwiner} ($E_0$) 
d\'efinissant les r\`egles 
de branchement $A_5 \hookrightarrow D_4$ et le graphe d'induction $D_4 \hookleftarrow A_5$ sont pr\'esent\'es \`a la  
Fig. \ref{E0_D4}. 
 
\begin{figure}[hhh] 
\unitlength 0.7mm 
\begin{center} 
$ 
E_0 = 
\left( 
\begin{array}{cccc} 
1 & . & . & . \cr 
. & 1 & . & . \cr 
. & . & 1 & 1 \cr 
. & 1 & . & . \cr 
1 & . & . & . \cr 
\end{array} 
\right) 
$ 
\qquad \qquad 
\unitlength 1.0mm 
\begin{picture}(45,20)(0,10) 
\thinlines 
\multiput(5,10)(15,0){2}{\circle*{1.5}} 
\put(35,17,5){\circle*{1.5}} 
\put(35,2,5){\circle*{1.5}} 
\put(0,11){$\ast$} 
\thinlines 
\put(5,10){\line(1,0){15}} 
\put(20,10){\line(2,1){15}} 
\put(20,10){\line(2,-1){15}} 
\put(5,14){\makebox(0,0){{\bf $\sigma_0$}}} 
\put(5,5){\makebox(0,0){$\tau_0$}} 
\put(5,0){\makebox(0,0){$\tau_4$}} 
\put(20,14){\makebox(0,0){{\bf $\sigma_1$}}} 
\put(20,5){\makebox(0,0){$\tau_1$}} 
\put(20,0){\makebox(0,0){$\tau_3$}} 
\put(35,20){\makebox(0,0){{\bf $\sigma_2$}}} 
\put(35,13){\makebox(0,0){$\tau_2$}} 
\put(35,6){\makebox(0,0){{\bf $\sigma_{2'}$}}} 
\put(35,-2){\makebox(0,0){$\tau_{2}$}} 
\end{picture} 
\bigskip 
\caption{Matrice essentielle $E_0$ de $D_4$ et graphe d'induction $D_4 \hookleftarrow A_5$.} 
\label{E0_D4} 
\end{center} 
\end{figure} 
La valeur de $\hat{T}$ sur les vertex $\tau_i$ de $A_5$ est donn\'ee par $(i+1)^2$ mod 24: 
(1, 4, 9, 16, 1). Les vertex $\sigma$ de $D_4$ pour lesquels una valeur de $\hat{T}$ est bien d\'efinie 
sont $\{\sigma_0, \sigma_2, \sigma_{2'}\}$: ils forment le sous-espace $J$, qui est une sous-alg\`ebre 
de l'alg\`ebre de graphe de $D_4$.  
 
\paragraph{Alg\`ebre d'Ocneanu} 
Nous serions tent\'es de d\'efinir l'alg\`ebre d'Ocneanu de $D_4$ 
par $D_4 \otimes_J D_4$. Cette alg\`ebre est engendr\'ee par les six \'el\'ements lin\'eairement ind\'ependants 
suivants: 
$$ 
\sigma_0 \otimesdot \sigma_0, \qquad \quad \sigma_1 \otimesdot \sigma_0, \qquad \quad  \sigma_2 \otimesdot  
\sigma_0, 
\qquad \quad  \sigma_2^{'} \otimesdot \sigma_0, \qquad \quad \sigma_0 \otimesdot \sigma_1, \qquad \quad 
\sigma_1 \otimesdot \sigma_1, 
$$ 
avec $\sigma_1 \otimesdot \sigma_0$ et $\sigma_0 \otimesdot \sigma_1$ comme g\'en\'erateurs chiraux gauche et droit. Maintenant, si nous voulons coder cette alg\`ebre dans un graphe, il existe une obstruction: 
\begin{eqnarray*} 
(\sigma_1 \otimesdot \sigma_0 ).(\sigma_0 \otimesdot \sigma_1) &=& \sigma_1 \otimesdot \sigma_1 \\ 
(\sigma_0 \otimesdot \sigma_1 ).(\sigma_1 \otimesdot \sigma_1) &=& \sigma_1 \otimesdot (\sigma_0 + \sigma_2  + \sigma_{2'}) 
=  \sigma_1 \otimesdot \sigma_0  + \sigma_1 \otimesdot \sigma_0  + \sigma_1 \otimesdot \sigma_0 
\end{eqnarray*}  
Il y aurait donc une ligne reliant $\sigma_1 \otimesdot \sigma_1$ \`a $\sigma_0 \otimesdot \sigma_1 $ mais 
trois lignes dans la direction oppos\'ee \cite{Chui}. Ce probl\`eme est contourn\'e en ``splittant'' 
le vertex $\sigma_1 \otimesdot \sigma_1$ en trois\footnote{Le fait de devoir splitter le point en trois est particulier au cas $D_4$, qui poss\`ede la sym\'etrie $\mathbb{Z}_3$. Pour les cas $D_{2n}, n>2$, il faut splitter le point en deux, car ces graphes poss\`edent seulement la sym\'etrie $\mathbb{Z}_2$.}, en introduisant une extension de l'alg\`ebre 
$D_4 \otimes_J D_4$ par des matrices $2 \times 2$ (voir \cite{Coq_Gil-ADE}). La cons\'equence est que l'alg\`ebre  
d'Ocneanu de $D_4$ 
est non-commutative. Nous allons voir qu'il existe une autre r\'ealisation de $Oc(D_4)$. 
Les graphes $D_{2n}$ sont des {\sl orbifolds} des graphes $A_{4n-3}$ par  
la sym\'etrie 
$\mathbb{Z}_2$. Consid\'erons le graphe $A_5$ poss\'edant 5 vertex $\tau_i, i \in (0,1,2,3,4)$, avec la sym\'etrie classique $\mathbb{Z}_2$ par rapport au vertex central $\tau_2$. Alors le graphe $\mathbb{Z}_2$-orbifold  est obtenu 
en identifiant les vertex sym\'etriques ($\tau_0$ avec $\tau_4$, $\tau_1$ avec $\tau_3$), et en splittant 
le vertex $\tau_2$ (point fixe par rapport \`a la sym\'etrie) en deux composantes \cite{Fendley}: 
nous obtenons 
le graphe $D_4$, qui est un  
$\mathbb{Z}_2$-orbifold du graphe $A_5$. 
Nous d\'efinissons l'alg\`ebre d'Ocneanu de $D_4$ par: 
\begin{equation} 
Oc(D_4) = D_4 \times_{\rho} \mathbb{Z}_2 
\end{equation} 
o\`u $\rho$ est une application qui \'echange les vertex de la fourche et laisse les autres invariants: 
$$ 
\rho(\sigma_0) = \sigma_0, \qquad \rho(\sigma_1) = \sigma_1, \qquad \rho(\sigma_2) = \sigma_{2}', \qquad \rho(\sigma_{2}') = \sigma_2. 
$$ 
Les \'el\'ements de $Oc(D_4)$ sont de la forme $(\sigma_a,+)$ et $(\sigma_a,-)$, la dimension  
de l'alg\`ebre $Oc(D_4)$ est $4 \times 2 = 8$ et la multiplication dans $Oc(D_4)$  
est d\'efinie par: 
\begin{equation} 
\begin{array}{rcl} 
(\sigma_a, +) . (\sigma_b, \pm) &=& (\sigma_a . \sigma_b, \pm)\\ 
(\sigma_a, -) . (\sigma_b, \pm) &=& (\sigma_a . \rho(\sigma_b), \mp) 
\end{array} 
\label{mult_D4} 
\end{equation} 
La table de multiplication de l'alg\`ebre d'Ocneanu de $D_4$ est pr\'esent\'ee \`a la Tab. \ref{D4mult}:

\begin{table}[hhh] 
$$ 
\begin{array}{|c||c|c|c|c||c|c|c|c|} 
\hline 
{ } & \zd{0}{+} & \zd{1}{+} & \zd{2}{+} & \zd{2'}{+} & \zd{0}{-} & \zd{1}{-} & \zd{2}{-} & \zd{2'}{-} \\ 
\hline 
\hline 
\zd{0}{+} & \zd{0}{+} & \zd{1}{+} & \zd{2}{+} & \zd{2'}{+} & \zd{0}{-} & \zd{1}{-} & \zd{2}{-} & \zd{2'}{-} \\ 
\zd{1}{+} & \zd{1}{+} & \zd{0}{+} + \zd{2'}{+} + \zd{2'}{+} & \zd{1}{+} & \zd{1}{+} & \zd{1}{-} & \zd{0}{-} + \zd{2}{-} + \zd{2'}{-}& \zd{1}{-} & \zd{1}{-} \\ 
\zd{2}{+} & \zd{2}{+} & \zd{1}{+} & \zd{2'}{+} & \zd{0}{+} & \zd{2}{-} & \zd{1}{-} & \zd{2'}{-} & \zd{0}{-} \\ 
\zd{2'}{+} & \zd{2'}{+} & \zd{1}{+} & \zd{0}{+} & \zd{2}{+} & \zd{2'}{-} & \zd{1}{-} & \zd{0}{-} & \zd{2}{-} \\ 
\hline 
\hline 
\zd{0}{-} & \zd{0}{-} & \zd{1}{-} & \zd{2'}{-} & \zd{2}{-} & \zd{0}{+} & \zd{1}{+} & \zd{2'}{+} & \zd{2}{+} \\ 
\zd{1}{-} & \zd{1}{-} & \zd{0}{-} + \zd{2}{-} + \zd{2'}{-} & \zd{1}{-} & \zd{1}{-} & \zd{1}{+} & \zd{0}{+} + \zd{2'}{+} + \zd{2'}{+} & \zd{1}{+} & \zd{1}{+} \\ 
\zd{2}{-} & \zd{2}{-} & \zd{1}{-} & \zd{0}{-} & \zd{2'}{-} & \zd{2}{+} & \zd{1}{+} & \zd{0}{+} & \zd{2'}{+} \\ 
\zd{2'}{-} & \zd{2'}{-} & \zd{1}{-} & \zd{2}{-} & \zd{0}{-} & \zd{2'}{+} & \zd{1}{+} & \zd{2}{+} & \zd{0}{+} \\ 
\hline 
\end{array} 
$$ 
\caption{Table de multiplication de l'alg\`ebre d'Ocneanu de $D_4$.}
\label{D4mult} 
\end{table}

\begin{figure}[hhh] 
\unitlength 1.0mm 
\begin{center} 
\begin{picture}(50,70) 
\multiput(25,5)(0,10){2}{\circle*{2}} 
\multiput(25,25)(0,10){2}{\circle{2}} 
\put(25,45){\circle*{2}} 
\put(5,35){\circle*{2}} 
\put(25,65){\circle{2}} 
\put(45,35){\circle*{2}} 
 
\thicklines 
\put(5,35){\line(1,0){20}} 
\put(5,35){\line(2,-1){20}} 
\put(5,35){\line(2,3){20}} 
 
\thinlines 
\put(45,35){\line(-1,-1){20}} 
\put(45,35){\line(-2,-3){20}} 
\put(45,35){\line(-2,1){20}} 
 
\thicklines 
\dashline[50]{1}(45,35)(25,35) 
\dashline[50]{1}(45,35)(25,65) 
\dashline[50]{1}(45,35)(25,25) 
 
\thinlines 
\dashline[50]{1}(5,35)(25,45) 
\dashline[50]{1}(5,35)(25,15) 
\dashline[50]{1}(5,35)(25,5) 
 
\scriptsize 
\put(25,69){\makebox(0,0){$\sigma_0^+$}} 
\put(25,48){\makebox(0,0){$\sigma_0^-$}} 
\put(25,39){\makebox(0,0){$\sigma_2^+$}} 
\put(25,29){\makebox(0,0){$\sigma_{2'}^+$}} 
\put(25,19){\makebox(0,0){$\sigma_2^-$}} 
\put(25,1){\makebox(0,0){$\sigma_{2'}^-$}} 
\put(49,35){\makebox(0,0){$\sigma_1^-$}} 
\put(0,35){\makebox(0,0){$\sigma_1^+$}} 
\normalsize 
\end{picture} 
\caption{Le graphe d'Ocneanu de $D_4$.} 
\label{grOc_D4} 
\end{center} 
\end{figure}
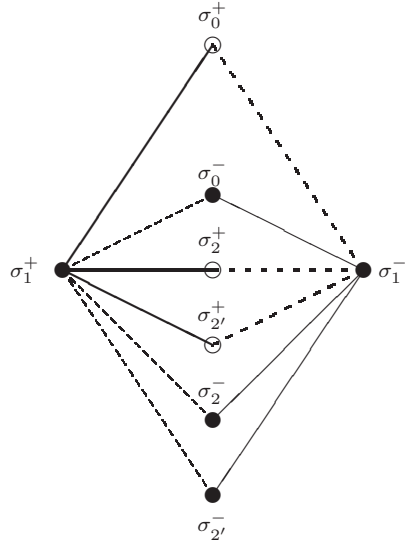

Cette alg\`ebre est non-commutative. Par exemple, de (\ref{mult_D4}) nous avons: 
$$ 
(\sigma_2,+). (\sigma_0,-) = (\sigma_2,-) \quad \not= \quad   
(\sigma_0,-). (\sigma_2,+) = (\sigma_{2'},-) 
$$ 
Les g\'en\'erateurs chiraux gauche et droit de $Oc(D_4)$ sont $(\sigma_1,+)$ et $(\sigma_1,-)$, et 
nous pouvons v\'erifier que la multiplication par les g\'en\'erateurs 
est effectivement cod\'ee dans le graphe d'Ocneanu de $D_4$, pr\'esent\'e \`a la Fig. \ref{grOc_D4}.  
 
D'apr\`es la d\'efinition (\ref{mult_D4}) de la multiplication dans $Oc(D_4)$, il est facile de voir que  
les matrices $O_x$ qui codent cette multiplication sont donn\'ees, dans la base $\{(\sigma_a,+),(\sigma_a,-)\}$, par: 
$$ 
O_{a,+} = \left(  
\begin{array}{cc} 
G_a & 0 \\ 
0 & G_a 
\end{array} 
\right) 
\qquad \qquad \qquad  
O_{a,-} = \left(  
\begin{array}{cc} 
0 & G_a^{\rho} \\ 
G_a^{\rho} & 0 
\end{array} 
\right) 
$$ 
o\`u les matrices $G_a^{\rho}$ sont d\'efinies \`a partir des matrices $G_a$ par: 
$$ 
\sigma_a . \rho(\sigma_b) = \sum_c (G_a^{\rho})_{bc} \qquad \qquad (G_a^{\rho})_{bc} = (G_a)_{\rho(b)c} 
$$ 
Elles sont obtenues \`a partir des matrices $G_a$ en permutant les lignes associ\'ees \`a $\sigma_2$ et $\sigma_{2}'$.\\ 
L'action de $Oc(D_4)$ sur $D_4$ est d\'efinie par: 
\begin{equation} 
\begin{array}{rcl} 
(\sigma_a,+) . \sigma_b &=& \sigma_a . \sigma_b \\ 
(\sigma_a,-) . \sigma_b &=& \sigma_a . \rho(\sigma_b)  
\end{array} 
\end{equation} 
et les matrices $S_x$ ($x = (\sigma_a, \pm)$) qui codent cette action sont donn\'ees par: 
\begin{equation} 
S_x = \left\lbrace \begin{array}{rcl} G_a \qquad x=(\sigma_a,+) \\ 
                                  G^{\rho}_a \qquad x=(\sigma_a,-) \end{array} 
\right. 
\end{equation} 
Nous pouvons v\'erifier que cette action v\'erifie bien $(y.x).\sigma_a = y.(x.\sigma_a)$: 
$$ 
S_x . S_y = \sum_z (O_y)_{xz} S_z 
$$ 
 
Notons que nous avons les suivantes projections $\psi$ des \'el\'ements de $D_4 \otimes D_4$ 
vers $Oc(D_4)$: 
\begin{equation} 
\begin{array}{rcl} 
\psi(\sigma_a \otimes \sigma_b) &=& (\sigma_a . \sigma_b, +) \qquad \textrm{pour } \sigma_b \in J \\ 
\psi(\sigma_a \otimes \sigma_b) &=& (\sigma_a . \sigma_b, -) \qquad \textrm{pour } \sigma_b \notin J  
\end{array} 
\label{proj} 
\end{equation} 
Sous cette projection $\psi$, la multiplication commutative $\mu_1$ dans  
$D_4 \otimes D_4$ et la multiplication non-commutative $\mu_2$ dans  
$Oc(D_4)$ commutent: 
\begin{equation} 
\psi \circ \mu_1 = (\psi \times \psi) \circ \mu_2 
\end{equation} 
Ceci est possible du fait que la projection $\psi$ n'est pas bijective. Par exemple: 
\begin{eqnarray*} 
\psi (\sigma_1 \otimes \sigma_1) &=& (\sigma_0,-) + (\sigma_2,-) + (\sigma_{2'},-) \\ 
\psi (\sigma_0 \otimes \sigma_1) = \psi(\sigma_2 \otimes \sigma_1) =   
\psi(\sigma_{2'} \otimes \sigma_1) &=& (\sigma_1,-) 
\end{eqnarray*}

\paragraph{Dimension des blocs} 
La dimension $d_i$ des blocs de la big\`ebre $\mathcal{B}D_4$ pour la loi de composition 
$(d_i = \sum_{a,b} (F_i)_{ab})$ est donn\'ee, pour $i$ dans $(0,1,2,3,4)$, par: 
$$ 
d_i : (4,6,8,6,4) 
$$  
La dimension $d_x$ des blocs de la big\`ebre $\mathcal{B}D_4$ pour la loi de convolution 
$(d_x = \sum_{a,b} (S_x)_{ab})$ est donn\'ee, pour $x$ dans $((\sigma_a,+),(\sigma_a,-))$,  
par: 
$$ 
d_x : (4,6,4,4 \, ;\, 4,6,4,4) 
$$ 
La r\`egle de somme quadratique, d\'efinissant la dimension de la big\`ebre $\mathcal{B}D_4$ 
est v\'erifi\'ee: 
$$ 
\dim \mathcal{B}D_4 = \sum_{i \in A_5} d_i^2 = \sum_{x \in Oc(D_4)} d_x^2 = 168. 
$$  
Par contre, la r\`egle de somme lin\'eaire (dont l'interpr\'etation est encore myst\'erieuse) 
na\"ive ne l'est pas: $\sum_i d_i = 28, \sum_x d_x = 36$. Il existe un double comptage  
dans la sommation sur les \'el\'ements $x$ de $Oc(D_4)$, d\^u \`a la sym\'etrie originelle 
de la fourche dans $D_4$. En introduisant un facteur $1/2$ pour les termes $x = (\sigma_2,\pm)$ 
et $x =(\sigma_{2'},\pm)$, ou en excluant de la sommation les termes $x =(\sigma_{2'},\pm)$, 
alors $\sum_x^{'} d_x = 28$, et la r\`egle de somme lin\'eaire est  
v\'erifi\'ee\cite{Pet_Zub-Oc}.\\ 
Les masses quantiques de $D_4$ et $A_5$ sont: $m(D_4)=6$, $m(A_5)=12$. Du fait que l'alg\`ebre 
d'Ocneanu de $D_4$ est d\'efinie par: $Oc(D_4)=D_4 \times_\rho \mathbb{Z}_2$, il est naturel de 
d\'efinir la masse quantique de $Oc(D_4)$ par $m(Oc(D_4)) = m(D_4).m(\mathbb{Z}_2)$.  
L'alg\`ebre du graphe $A_2$ est isomorphe \`a $\mathbb{Z}_2$, donc nous d\'efinissons 
$m(\mathbb{Z}_2) = m(A_2) = 2$. Alors la relation de masse quantique est v\'erifi\'ee: 
$$ 
m(Oc(D_4)) = m(D_4).m(A_2) = m(A_5) = 12  
$$

\paragraph{Matrices toriques g\'en\'eralis\'ees} 
L'action de $A_5$ sur les \'el\'ements de type pair de $Oc(D_4)$ est d\'efinie en utilisant  
les projections (\ref{proj}) de $D_4 \otimes D_4$ sur $Oc(D_4)$. Un \'el\'ement  
pair $(\sigma_a,+)$ de $Oc(D_4)$ a comme pr\'e-image ($\sigma_a \otimes \sigma_0$) 
dans $D_4 \otimes D_4$. Alors l'action de $A_5$ sur un tel \'el\'ement est d\'efinie, 
utilisant (\ref{proj}), par: 
\begin{eqnarray*} 
\tau_i . (\sigma_a, +) . \tau_j &=&  \tau_i . (\sigma_a \otimes \sigma_0) . \tau_j \\ 
{ } &=& \sum_b \sum_c (F_i)_{ab} (F_j)_{0c} (\sigma_b \otimes \sigma_c) \\ 
{ } &=& \sum_b \sum_{c \in J} (F_i)_{ab} (F_j)_{0c} (\sigma_b.\sigma_c, +)  
+ \sum_b \sum_{c \notin J} (F_i)_{ab} (F_j)_{0c} (\sigma_b.\sigma_c, -)   \\ 
{ } &=& \sum_b \sum_{c \in J}\sum_d (F_i)_{ab} (F_j)_{0c} (G^{'}_d)_{bc} (\sigma_d, +)  
+ \sum_b \sum_{c \notin J} \sum_d (F_i)_{ab} (F_j)_{0c} (G^{'}_d)_{bc} (\sigma_d, -) \\ 
{ } &=& \qquad \quad \, \sum_d (W_{a+,d+})_{ij} (\sigma_d, +) \qquad \quad \, + \qquad  
\quad \, \sum_d (W_{a+,d-})_{ij} (\sigma_d, -)  
\end{eqnarray*} 
o\`u les matrices $G_d^{'}$ sont d\'efinies par: 
$$ 
(G^{'}_d)_{bc} = (G_b)_{cd} 
$$ 
Notons le fait que l'alg\`ebre d'un graphe $G$ soit commutative implique  
$(G_b)_{cd} = (G_c)_{bd}$, mais  
ces matrices ne sont pas forc\'ement sym\'etriques (donc en g\'en\'eral  
$G^{'}_d \not= G_d$).\\ 
L'action (\`a gauche et \`a droite) de $A_5$ sur les \'el\'ements de type impair de $Oc(D_4)$ est d\'efinie en 
utilisant le fait que $(\sigma_a,-) = (\sigma_a,+) . (\sigma_0,-)$. Alors: 
$$ 
\tau_i .(\sigma_a,-) . \tau_j = \tau_i .(\sigma_a,+).(\sigma_0,-) . \tau_j  = (\tau_i .(\sigma_a,+).\tau_j).(\sigma_0,-),   
$$ 
Les matrices toriques $W_{xy}$ du mod\`ele $D_{4}$ s'obtiennent donc \`a partir de  
la formule g\'en\'erale: 
\begin{equation} 
\begin{array}{|c|} 
\hline 
{ } \\ 
\begin{array}{rclcl} 
W_{a+,d+} &=& \displaystyle \sum_b \sum_{c \in J} (F_i)_{ab} (F_j)_{0c} (G_d^{'})_{bc}  
&\quad=\quad& W_{a-,d-} \\ 
W_{a+,d-} &=& \displaystyle \sum_b \sum_{c \notin J} (F_i)_{ab} (F_j)_{0c} (G_d^{'})_{bc} 
&\quad=\quad& W_{a-,d+}  
\end{array} \\ 
{ } \\ 
\hline 
\end{array} 
\label{Wxy_D4} 
\end{equation} 
Nous pouvons v\'erifier qu'elles satisfont l'alg\`ebre carr\'ee de Verlinde.\\ 
Pour $y =(\sigma_0,+)$, $(G^{'}_0)_{bc} = (G_{b})_{c0} = \delta_{b,\rho(c)}$, et les matrices toriques 
$W_{x,0+}$ s'\'ecrivent donc: 
$$ 
W_{a+,0+} = \sum_{c \in J}    (F_i)_{a\rho(c)} (F_j)_{0c} \qquad \qquad  
W_{a-,0+} = \sum_{c \notin J} (F_i)_{a\rho(c)} (F_j)_{0c} 
$$  
et l'invariant modulaire $\mathcal{M}=W_{0+,0+}$ est donc \'egal \`a: 
$$ 
\mathcal{M}_{ij} = \sum_{c \in J}    (F_i)_{0\rho(c)} (F_j)_{0c} \qquad \Longrightarrow \qquad  
\mathcal{M} = \left( \begin{array}{ccccc}  
1 & . & . & . & 1 \\ 
. & . & . & . & . \\ 
. & . & 2 & . & . \\ 
. & . & . & . & . \\ 
1 & . & . & . & 1  
\end{array} \right) 
$$ 
La non-commutativit\'e de l'alg\`ebre d'Ocneanu de $D_4$ 
se manifeste par la pr\'esence d'un coefficient 2 dans l'invariant modulaire $\mathcal{M}$ (et donc 
dans la fonction de partition correspondante).  
\begin{remarq} 
L'\'el\'ement de matrice $(F_i)_{ab}$ donne le nombre de chemins essentiels de longueur $i$ partant du vertex  
$\sigma_a$ et arrivant au vertex $\sigma_b$ sur le graphe $D_4$. Consid\'erant les chemins partant de $\sigma_0$, 
d\^u \`a la sym\'etrie entre les vertex $\sigma_2$ et $\sigma_{2'}$ de la fourche, alors \mbox{$(F_i)_{0\rho(c)} =  
(F_i)_{0c}$}, et l'invariant modulaire s'\'ecrit donc aussi: 
$$ 
\mathcal{M} = \sum_{c \in J}    (F_i)_{0c} (F_j)_{0c} 
$$  
qui est la formule utilis\'ee notamment dans \cite{Coq_Gil-ADE} et \cite{Pet_Zub-Oc}. 
\end{remarq}

\paragraph{Fonctions de partition g\'en\'eralis\'ees} 
Elles sont d\'efinies \`a partir des matrices toriques g\'en\'eralis\'ees (\ref{Wxy_D4}) 
par: 
\begin{equation} 
\mathcal{Z}_{x|y} = \sum_{i \in A	_5}\, \sum_{j \in A_5} \, \chi_i(q) (W_{xy})_{ij} \,  
\ov{\chi}_j(q) 
\end{equation} 
o\`u les $\chi_i$ sont les caract\`eres de l'alg\`ebre $\widehat{su}(2)$. Introduisons 
les caract\`eres \'etendus $\hat{\chi}_a(q)$, d\'efinis \`a partir de la matrice essentielle 
$E_0$, et les caract\`eres \'etendus g\'en\'eralis\'es $\hat{\chi}_{ab}(q)$ du mod\`ele  
$D_4$: 
\begin{eqnarray*} 
\hat{\chi}_a(q) &=& \sum_{i \in A_5} (E_0)_{ia} \chi_i (q) = \sum_{i \in A_5} (F_i)_{0a}  
\chi_i (q) \\ 
\hat{\chi}_{ab}(q) &=& \sum_{c \in D_4} (G_a)_{cb} \hat{\chi}_c (q) = \sum_{i \in A_5} (F_i)_{ab}  
\chi_i (q)  
\end{eqnarray*} 
Les caract\`eres \'etendus $\hat{\chi}_a(q)$ du mod\`ele $D_4$ sont donn\'es par:  
$$ 
\begin{array}{ccc} 
\hat{\chi}_0 = \chi_0 + \chi_4        &\qquad&      \hat{\chi}_2 = \chi_2 \\ 
\hat{\chi}_1 = \chi_1 + \chi_3        &\qquad&      \hat{\chi}_{2'} = \chi_2  
\end{array} 
$$ 
Les fonctions de partition g\'en\'eralis\'ees du mod\`ele $D_4$ s'\'ecrivent sous la forme 
compacte suivante: 
\begin{equation} 
\begin{array}{|c|} 
\hline 
{ } \\ 
\begin{array}{rclcl} 
\mathcal{Z}_{a+,d+} &=& \displaystyle \sum_b \sum_{c \in J} \hat{\chi}_{ab} \ov{\hat{\chi}}_c (G_d^{'})_{bc}  
&\quad=\quad& \mathcal{Z}_{a-,d-} \\ 
\mathcal{Z}_{a+,d-} &=& \displaystyle \sum_b \sum_{c \notin J} \hat{\chi}_{ab} \ov{\hat{\chi}}_c (G_d^{'})_{bc} 
&\quad=\quad& \mathcal{Z}_{a-,d+}  
\end{array} \\ 
{ } \\ 
\hline 
\end{array} 
\label{Zxy_D4} 
\end{equation} 
et les fonctions de partition \`a une ligne de d\'efauts $\mathcal{Z}_{x} = \mathcal{Z}_{x0}$ 
s'\'ecrivent en fonction des caract\`eres \'etendus: 
$$ 
\mathcal{Z}_{a+} = \sum_b \sum_{c \in J} (G_a)_{cb} \hat{\chi}_b \ov{\hat{\chi}}_c \qquad \qquad 
\mathcal{Z}_{a-} = \sum_b \sum_{c \notin J} (G_a)_{cb} \hat{\chi}_b \ov{\hat{\chi}}_c 
$$  
Elles sont donn\'ees dans \cite{Coq_Gil-ADE} en fonction des caract\`eres de 
$\widehat{su}(2)$, nous les pr\'esentons 
dans l'Annexe {\bf D} en fonction des caract\`eres \'etendus $\hat{\chi}_a$ de $D_4$.  
La fonction de partition invariante modulaire du mod\`ele $D_4$ s'\'ecrit sous 
forme diagonale en fonction des caract\`eres \'etendus:  
\begin{equation} 
\mathcal{Z}_{D_4} = \sum_{c \in J} \, |\hat{\chi}_c |^2 = 2\, |\chi_0|^2 \,+\, |\chi_0 + \chi_4|  
\end{equation} 
et correspond bien \`a celle de la classification de Cappelli-Itzykson-Zuber \cite{CIZ-class2}.

\subsubsection{Formules g\'en\'erales pour $D_{2n}$}

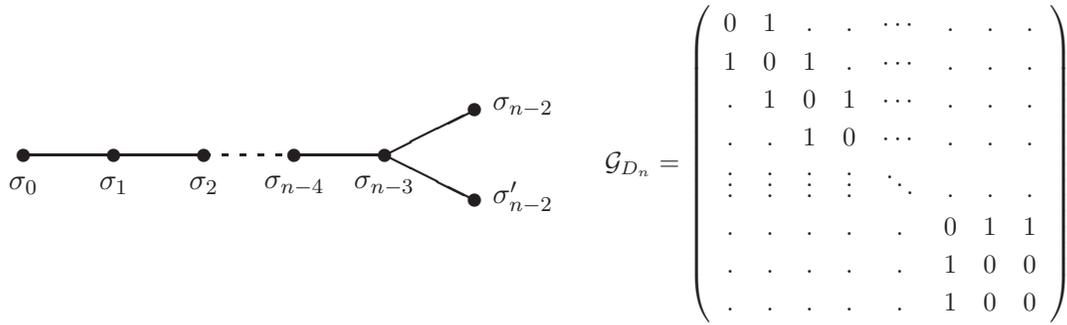
\begin{figure}[hhh] 
\unitlength 0.8mm 
\begin{center} 
\begin{picture}(100,25)(0,10) 
\thinlines 
\multiput(5,12.5)(15,0){3}{\circle*{2}} 
\multiput(50,12.5)(15,0){2}{\circle*{2}} 
\put(80,20){\circle*{2}} 
\put(80,5){\circle*{2}} 
\thicklines 
\put(5,12.5){\line(1,0){30}} 
\put(50,12.5){\line(1,0){15}} 
\put(65,12.5){\line(2,1){15}} 
\put(65,12.5){\line(2,-1){15}} 
\multiput(35,12.5)(3,0){6}{\line(1,0){1}} 
\put(5,7.5){\makebox(0,0){$\sigma_0$}} 
\put(20,7.5){\makebox(0,0){$\sigma_1$}} 
\put(35,7.5){\makebox(0,0){$\sigma_2$}} 
\put(50,7.5){\makebox(0,0){$\sigma_{n-4}$}} 
\put(65,7.5){\makebox(0,0){$\sigma_{n-3}$}} 
\put(88,20.5){\makebox(0,0){$\sigma_{n-2}$}} 
\put(88,5.5){\makebox(0,0){$\sigma_{n-2}'$}} 
\end{picture} 
\small 
$ 
{\cal G}_{D_n} = 
\left( \begin{array}{cccccccc} 
     0 & 1 & . & . & \cdots & . & . & .  \\ 
     1 & 0 & 1 & . & \cdots & . & . & .  \\ 
     . & 1 & 0 & 1 & \cdots & . & . & .  \\ 
     . & . & 1 & 0 & \cdots & . & . & .  \\ 
     \vdots & \vdots & \vdots & \vdots & \ddots & . & . & .  \\ 
     . & . & . & . & . & 0 & 1 & 1  \\ 
     . & . & . & . & . & 1 & 0 & 0  \\ 
     . & . & . & . & . & 1 & 0 & 0  \\ 
\end{array} 
\right) 
$ 
\normalsize 
\caption{Le graphe $D_n$ et sa matrice d'adjacence} 
\label{grDn} 
\end{center} 
\end{figure}

\paragraph{Graphe $D_{2n}$ et matrices de graphe} 
Le graphe $D_n$, pour $n$ pair $>4$, est illustr\'e \`a la Fig. \ref{grDn} avec sa matrice d'adjacence, en adoptant comme ordre de la base des vertex l'ordre suivant: $\{ \sigma_0, \sigma_1, \sigma_2, \cdots, \sigma_{n-3}, \sigma_{n-2},  
\sigma_{n-2}' \}$.

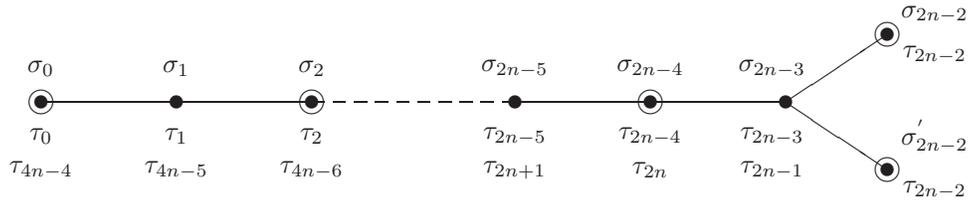
\begin{figure}[hhh] 
\unitlength 0.9mm 
\begin{center} 
\begin{picture}(130,25) 
\put(0,10){\line(1,0){40}} 
\put(70,10){\line(1,0){40}} 
\put(110,10){\line(3,2){15}} 
\put(110,10){\line(3,-2){15}} 
\multiput(0,10)(20,0){3}{\circle*{2}} 
\multiput(70,10)(20,0){3}{\circle*{2}} 
\put(125,20){\circle*{2}} 
\put(125,0){\circle*{2}} 
 
\put(0,10){\circle{3.5}} 
\put(40,10){\circle{3.5}} 
\put(90,10){\circle{3.5}} 
\put(125,20){\circle{3.5}} 
\put(125,0){\circle{3.5}} 
 
\dashline[30]{2}(40,10)(70,10) 
 
\small 
\put(0,5){\makebox(0,0){$\tau_{0}$}} 
\put(0,0){\makebox(0,0){$\tau_{4n-4}$}} 
\put(0,15){\makebox(0,0){$\sigma_{0}$}} 
\put(20,5){\makebox(0,0){$\tau_{1}$}} 
\put(20,0){\makebox(0,0){$\tau_{4n-5}$}} 
\put(20,15){\makebox(0,0){$\sigma_{1}$}} 
\put(40,5){\makebox(0,0){$\tau_{2}$}} 
\put(40,0){\makebox(0,0){$\tau_{4n-6}$}} 
\put(40,15){\makebox(0,0){$\sigma_{2}$}} 
\put(70,5){\makebox(0,0){$\tau_{2n-5}$}} 
\put(70,0){\makebox(0,0){$\tau_{2n+1}$}} 
\put(70,15){\makebox(0,0){$\sigma_{2n-5}$}} 
\put(90,5){\makebox(0,0){$\tau_{2n-4}$}} 
\put(90,0){\makebox(0,0){$\tau_{2n}$}}

\put(90,15){\makebox(0,0){$\sigma_{2n-4}$}} 
\put(108,5){\makebox(0,0){$\tau_{2n-3}$}} 
\put(108,0){\makebox(0,0){$\tau_{2n-1}$}} 
\put(108,15){\makebox(0,0){$\sigma_{2n-3}$}} 
\put(132,23){\makebox(0,0){$\sigma_{2n-2}$}} 
\put(132,17){\makebox(0,0){$\tau_{2n-2}$}} 
\put(132,5){\makebox(0,0){$\sigma_{2n-2}^{'}$}} 
\put(132,-3){\makebox(0,0){$\tau_{2n-2}$}}

\normalsize 
\end{picture} 
\end{center} 
\caption{Le graphe d'induction $D_{2n}$-$A_{4n-3}$.} 
\label{A/D2n} 
\end{figure}

Pour $D_n$, $\kappa = 2n-2$, la norme du graphe est $\beta = 2 \cos (\frac{\pi}{2n-2})$, et les composantes  
du vecteur normalis\'e de Perron-Frobenius sont donn\'ees, par: 
$$ 
P = \left( [1]_q, [2]_q, [3]_q, \cdots, [n-3]_q, [n-2]_q, \frac{[n-1]_q}{[2]_q}, \frac{[n-1]_q}{[2]_q} \right) 
$$ 
Le graphe $D_n$ ne d\'etermine pas de mani\`ere unique l'alg\`ebre de graphe: 
nous pouvons remplir de mani\`ere unique la table de multiplication, \`a l'exception des \'el\'ements 
correspondants \`a la multiplication des vertex $\sigma_{n-2}, \sigma_{n-2}'$ de la fourche. 
Mais en {\sl imposant} que les coefficients de structures soient des entiers non n\'egatifs, la solution devient 
unique: la multiplication des vertex $\sigma_{n-2}, \sigma_{n-2}'$ est donn\'ee par: 
\vspace{0.4cm} 
\begin{itemize} 
\item $\frac{n}{2}$ pair: \ \ \,  \qquad \qquad 
$ 
\begin{array}{c|cc} 
{ } & \sigma_{n-2} & \sigma_{n-2}' \\ 
\hline 
\sigma_{n-2} & \sigma_2+\sigma_6+ \cdots + \sigma_{n-2}' \qquad & \sigma_0+\sigma_4+ \cdots + \sigma_{n-4} \\ 
\sigma_{n-2}' & \sigma_0+\sigma_4+ \cdots + \sigma_{n-4} \qquad & \sigma_2+\sigma_6+ \cdots + \sigma_{n-2} 
\end{array}   
$ 
\vspace{0.4cm} 
\item $\frac{n}{2}$ impair: \qquad \qquad 
$ 
\begin{array}{c|cc} 
{ } & \sigma_{n-2} & \sigma_{n-2}' \\ 
\hline 
\sigma_{n-2} & \sigma_0+\sigma_4+ \cdots + \sigma_{n-2} \qquad & \sigma_2+\sigma_6+ \cdots + \sigma_{n-4} \\ 
\sigma_{n-2}' & \sigma_2+\sigma_6+ \cdots + \sigma_{n-4} \qquad & \sigma_0+\sigma_4+ \cdots + \sigma_{n-2}' 
\end{array}   
$ 
\end{itemize} 
\vspace{0.6cm} 
Les matrices $G_a$ s'obtiennent par la formule de r\'ecurrence tronqu\'ee de $SU(2)$ jusqu'au vertex 
$\sigma_{n-3}$, 
et les matrices correspondantes aux vertex de la fourche s'obtiennent gr\^ace aux relations ci-dessus. Nous d\'efinissons 
aussi les matrices $G^{'}_a$ et $G^{\rho}_a$: 
$$ 
(G_a^{'})_{bc} = (G_c)_{ab} \qquad \qquad (G_a^{\rho})_{bc} = (G_a)_{\rho(b)c} 
$$ 
o\`u $\rho$ est l'application qui permute les vertex $\sigma_{n-2}, \sigma_{n-2}'$ de la fourche et laissent les 
autres invariants. Notons que nous avons: 
$$ 
G_a^{\rho} = G_a \quad \textrm{ si } \frac{n}{2} \textrm{pair}, \qquad \qquad G_a^{\rho} = G^{'}_a \quad \textrm{ si }  
\frac{n}{2}. 
\textrm{impair}  
$$ 
\paragraph{Induction-restriction} 
Le graphe de la s\'erie $A_n$ correspondant \`a $D_{2n}$ est $A_{4n-3}$. Les matrices de fusion $N_i$ 
et les matrices $F_i$  s'obtiennent comme toujours. De 
la matrice essentielle $E_0$ nous obtenons les r\`egles de branchement $A_{4n-3} \hookrightarrow D_{2n}$, 
qui d\'efinissent le graphe d'induction repr\'esent\'e \`a la Fig. \ref{A/D2n}.

La valeur de $\hat T$ sur les irreps $\{ \tau_0, \tau_1, 
\cdots, \tau_{2n-1}, \tau_{2n}, \tau_{2n+1}, \cdots, 
\tau_{4n-5}, \tau_{4n-4} \}$ de $A_{4n-3}$ est donn\'ee par: 
$$ 
\hat{T}(\tau_0) = \hat{T}(\tau_{4n-4}) \qquad 
\hat{T}(\tau_2) = \hat{T}(\tau_{4n-6}) \qquad \cdots \qquad 
\hat{T}(\tau_{2n-4}) = \hat{T}(\tau_{2n}) 
$$ 
Ces valeurs sont sym\'etriques par rapport au vertex central $\tau_{2n}$.  
Par le m\'ecanisme d'induction, nous pouvons assigner une valeur d\'etermin\'ee  
de $\hat{T}$ seulement pour les vertex $( \sigma_0, 
\sigma_2, \cdots, \sigma_{2n-4}, \sigma_{2n-2}, 
\sigma_{2n-2}^{'} )$ de $D_{2n}$ (ils sont entour\'es par un cercle sur le diagramme d'induction). 
Ces vertex engendrent la sous-alg\`ebre $J$. Notons que les vertex de la fourche $\sigma_{2n-2}, 
\sigma_{2n-2}^{'}$ ne sont pas distingu\'es par les valeurs de $\hat{T}$. Comme $\hat{T}$ pourrait 
\^etre d\'efini sur une combinaison lin\'eaire de ces vertex, il est naturel de s'attendre \`a ce  
que cet arbitraire conduise, au niveau de l'alg\`ebre d'Ocneanu de $D_{2n}$, \`a la non-commutativit\'e. 
\paragraph{Alg\`ebre d'Ocneanu} 
De fait, les graphes $D_{2n}$ sont des $\mathbb{Z}_2$-orbifolds des graphes $A_{4n-3}$. Une r\'ealisation 
de l'alg\`ebre d'Ocneanu de $D_{2n}$ est donn\'ee par un produit semi-direct: c'est le quotient, par $\rho$, 
du produit de l'alg\`ebre du graphe $D_{2n}$ par $\mathbb{Z}_2$: 
\begin{equation} 
Oc(D_{2n}) = D_{2n} \times_{\rho} \mathbb{Z}_2 
\end{equation}  
Cette alg\`ebre est de dimension $2 \times 2n$, et ses \'el\'ements sont not\'es: 
$$ 
(\sigma_a,+) \qquad \qquad \qquad (\sigma_a,-) 
$$ 
La multiplication de l'alg\`ebre $Oc(D_{2n})$ est d\'efinie par: 
\begin{equation} 
\begin{array}{rcl} 
(\sigma_a,+ ) . (\sigma_b,{\pm}) &=& (\sigma_a.\sigma_b, \pm) \\ 
(\sigma_a,- ) . (\sigma_b,{\pm}) &=& (\sigma_a.\rho(\sigma_b), \mp)  
\end{array} 
\end{equation} 
Cette alg\`ebre est bien non-commutative, et la multiplication par les g\'en\'erateurs chiraux gauche et droit 
(resp. $(\sigma_1,+ )$ et ($\sigma_1,- )$ est cod\'ee par le graphe d'Ocneanu correspondant. 
Les matrices $O_x$ qui codent la multiplication dans $Oc(D_{2n})$ sont donn\'ees, dans la base  
$\{(\sigma_a,+),(\sigma_a,-)\}$, par: 
\begin{equation} 
O_{a,+} = \left( \begin{array}{cc} G_a & 0 \\ 0 & G_a \end{array} \right) \qquad \qquad  
O_{a,-} = \left( \begin{array}{cc} 0 & G_a^{\rho} \\ G_a^{\rho} & a \end{array} \right)  
\end{equation} 
L'action de $Oc(D_{2n})$ sur $D_{2n}$ est d\'efinie par: 
\begin{equation} 
(\sigma_a,+) . \sigma_b = \sigma_a . \sigma_b \qquad \qquad  
(\sigma_a,-) . \sigma_b = \sigma_a . \rho(\sigma_b)  
\end{equation} 
et les matrices $S_x$ qui codent cette action sont donn\'ees par: 
\begin{equation} 
S_{a,+} = G_a \qquad \qquad S_{a,-} = G_a^{\rho} 
\end{equation} 
Nous avons les projections $\psi$ suivantes entre les \'el\'ements de $D_{2n} \otimes D_{2n}$ et $Oc(D_{2n})$: 
\begin{eqnarray*} 
\psi (\sigma_a \otimes \sigma_b) &=& (\sigma_a.\sigma_b, +) \qquad \textrm{si } \sigma_b \in J \\  
\psi (\sigma_a \otimes \sigma_b) &=& (\sigma_a.\sigma_b, -) \qquad \textrm{si } \sigma_b \notin J   
\end{eqnarray*} 
de mani\`ere \`a ce que les structures multiplicatives $\mu_1$ de $D_{2n} \otimes D_{2n}$ et $\mu_2$ 
de $Oc(D_{2n})$ commutent:  
\begin{equation} 
\psi \circ \mu_1 = (\psi \times \psi) \circ \mu_2 
\end{equation} 
Les \'el\'ements pairs $(\sigma_a,+)$ de $Oc(D_{2n})$ sont repr\'esent\'es dans $D_{2n} \otimes D_{2n}$ 
par les \'el\'ements $\sigma_a \otimes \sigma_0$.

\paragraph{Dimensions des blocs} 
Les dimension $d_i$ et $d_x$ des blocs de la dig\`ebre $\mathcal{B}(D_{2n})$ pour les lois $\circ$
et $\odot$ sont donn\'ees par: 
$$ 
d_i = \sum_{a,b} (F_i)_{ab} \qquad \qquad d_x = \sum_{a,b} (S_x)_{ab} 
$$ 
et la r\`egle de somme quadratique, d\'efinissant la dimension de la dig\`ebre $\mathcal{B}(D_{2n})$ 
est v\'erifi\'ee: 
$$ 
\sum_{i \in A_{4n-3}} d_i^2 = \sum_{x \in Oc(D_{2n})} d_x^2 = \dim \mathcal{B}(D_{2n}) 
$$ 
Par contre, pour satisfaire la r\`egle de somme lin\'eaire, la sommation sur les \'el\'ement  
$x$ de $Oc(D_{2n})$ ne doit se faire que pour l'un des vertex de la fourche: il faut exclure de la somme 
les termes $x =(\sigma_{2n-2}^{'},\pm)$ \cite{Pet_Zub-Oc}. 
La relation de masse quantique entre les graphes $Oc(D_{2n})$ et $\mathcal{A}(D_{2n}) = A_{4n-3}$ est  
v\'erifi\'ee: 
$$ 
m(Oc(D_{2n})) = m(D_{2n}) . m(\mathbb{Z}_2)  = m(D_{2n}) . 2 = m(A_{4n-3}) 
$$

\begin{figure}[H] 
\unitlength 1.0mm 
\begin{center} 
\begin{picture}(160,110)(-15,0) 
\multiput(25,5)(0,10){2}{\circle*{2}} 
\multiput(25,25)(0,10){2}{\circle{2}} 
\put(25,45){\circle*{2}} 
\put(5,35){\circle*{2}} 
\put(45,35){\circle*{2}}

\put(5,75){\circle*{2}} 
\put(45,75){\circle*{2}} 
\put(25,65){\circle{2}} 
\put(25,85){\circle*{2}} 
\put(25,105){\circle{2}}

\thicklines 
\put(5,35){\line(1,0){20}} 
\put(5,35){\line(2,-1){20}} 
\put(5,35){\line(2,3){20}} 
\put(5,75){\line(2,-1){20}} 
\put(5,75){\line(2,3){20}} 
\thinlines 
\put(45,35){\line(-1,-1){20}} 
\put(45,35){\line(-2,-3){20}} 
\put(45,35){\line(-2,1){20}} 
\put(45,75){\line(-2,1){20}} 
\put(45,75){\line(-2,-3){20}}

\thicklines 
\dashline[50]{1}(45,35)(25,35) 
\dashline[50]{1}(45,35)(25,65) 
\dashline[50]{1}(45,35)(25,25) 
\dashline[50]{1}(45,75)(25,65) 
\dashline[50]{1}(45,75)(25,105)

\thinlines 
\dashline[50]{1}(5,35)(25,45) 
\dashline[50]{1}(5,35)(25,15) 
\dashline[50]{1}(5,35)(25,5) 
\dashline[50]{1}(5,75)(25,85) 
\dashline[50]{1}(5,75)(25,45) 
 
 
\put(25,109){\makebox(0,0){$\sigma_0^+$}} 
\put(25,89){\makebox(0,0){$\sigma_0^-$}} 
\put(1,76){\makebox(0,0){$\sigma_2^+$}} 
\put(50,75){\makebox(0,0){$\sigma_1^-$}} 
\put(25,69){\makebox(0,0){$\sigma_2^+$}} 
\put(25,50){\makebox(0,0){$\sigma_2^-$}} 
\put(25,39){\makebox(0,0){$\sigma_4^+$}} 
\put(25,29){\makebox(0,0){$\sigma_{4'}^+$}} 
\put(25,12){\makebox(0,0){$\sigma_4^-$}} 
\put(25,1){\makebox(0,0){$\sigma_{4'}^-$}} 
\put(50,35){\makebox(0,0){$\sigma_3^-$}} 
\put(1,35){\makebox(0,0){$\sigma_3^+$}} 
\normalsize 
 
\put(95,55){\makebox(0,0){ 
$ 
\mathcal{M}=\left( \begin{array}{ccccccccc} 
1 & . & . & . & . & . & . & . & 1 \\ 
. & . & . & . & . & . & . & . & . \\ 
. & . & 1 & . & . & . & 1 & . & . \\ 
. & . & . & . & . & . & . & . & . \\ 
. & . & . & . & 2 & . & . & . & . \\ 
. & . & . & . & . & . & . & . & . \\ 
. & . & 1 & . & . & . & 1 & . & . \\ 
. & . & . & . & . & . & . & . & . \\ 
1 & . & . & . & . & . & . & . & 1 \\ 
\end{array} 
\right) 
$}} 
\end{picture} 
\caption{Le graphe d'Ocneanu de $D_6$ et son invariant modulaire.} 
\label{grOc_D6} 
\end{center} 
\end{figure}
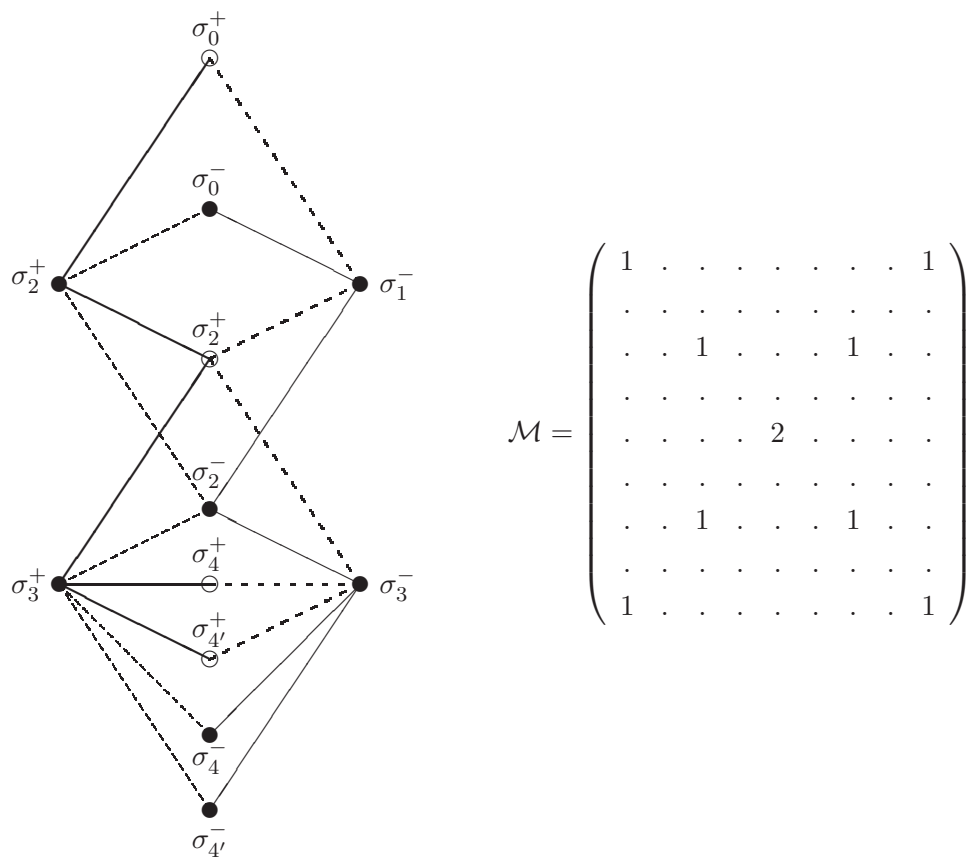

\paragraph{Matrices toriques et fonctions de partition g\'en\'eralis\'ees} 
L'action \`a gauche et \`a droite de $A_{4n-3}$ sur $Oc(D_{2n})$ est d\'efinie sur les \'el\'ements 
de $D_{2n} \otimes D_{2n}$, projet\'es ensuite sur $Oc(D_{2n})$ par la projection $\psi$. Les 
matrices $W_{xy}$ s'obtiennent par les formules suivantes:  
\begin{equation} 
\begin{array}{|c|} 
\hline 
{ } \\ 
\begin{array}{rclcl} 
W_{a+,d+} &=& \displaystyle \sum_b \sum_{c \in J} (F_i)_{ab} (F_j)_{0c} (G_d^{'})_{bc}  
&\quad=\quad& W_{a-,d-} \\ 
W_{a+,d-} &=& \displaystyle \sum_b \sum_{c \notin J} (F_i)_{ab} (F_j)_{0c} (G_d^{'})_{bc} 
&\quad=\quad& W_{a-,d+}  
\end{array} \\ 
{ } \\ 
\hline 
\end{array} 
\label{Wxy_D2n} 
\end{equation} 
et l'invariant modulaire $\mathcal{M}$ des mod\`eles $D_{2n}$ est donn\'e par: 
\begin{equation} 
\mathcal{M} = \sum_{c \in J}\; (F_i)_{0c} \;(F_j)_{0c} 
\end{equation} 
Les fonctions de partition g\'en\'eralis\'ees des mod\`ele $D_{2n}$ s'obtiennent alors par: 
\begin{equation} 
\mathcal{Z}_{xy} = \sum_{i \in A_{4n-3}} \sum_{j \in A_{4n-3}} \chi_i \; (W_{xy})_{ij} \; \ov{\chi}_j 
\end{equation} 
en fonction des caract\`eres $\chi_i(q)$ de l'alg\`ebre $\widehat{su}(2)$.  Elles s'\'ecrivent de mani\`ere 
plus compacte en fonction des caract\`eres \'etendus $\hat{\chi}_a(q)$ de $D_{2n}$ d\'efinis par: 
\begin{equation} 
\hat{\chi}_a(q) = \sum_{i \in A_{4n-3}} \; (E_0)_{ia}\; \chi_i (q) = \sum_{i \in A_{4n-3}} \; (F_i)_{0a}\; \chi_i (q) 
\end{equation} 
Nous donnons \`a la Fig. \ref{grOc_D6} le graphe d'Ocneanu de $D_6$ ainsi que l'invariant modulaire correspondant. 
Notons que la non-commutativit\'e de l'alg\`ebre d'Ocneanu de $D_{2n}$ se manifeste par la pr\'esence  
d'un facteur 2 dans l'invariant modulaire (provenant de la sym\'etrie des vertex de la fourche du graphe $D_{2n}$), 
et par cons\'equent aussi dans la fonction de partition invariante modulaire. 
Les fonctions de partition \`a une ligne de d\'efauts $\mathcal{Z}_{x} = \mathcal{Z}_{x0}$ sont donn\'ees dans 
\cite{Coq_Gil-ADE} en fonction des caract\`eres $\chi_i$ de $\widehat{su}(2)$, nous les rappelons dans 
l'Annexe {\bf D} \'ecrites de mani\`ere plus compacte en fonction des caract\`eres \'etendus $\hat{\chi}_a$. 
 
\subsection{Les cas $D_{2n+1}$}

\subsubsection{Le cas $D_5$} 
\begin{figure}[hhh] 
\unitlength 0.9mm 
\begin{center} 
\begin{picture}(60,20)(0,7.5) 
\thinlines 
\multiput(5,10)(15,0){3}{\circle*{2}} 
\put(50,17,5){\circle*{2}} 
\put(50,2,5){\circle*{2}} 
\thicklines\put(5,10){\line(1,0){30}} 
\put(35,10){\line(2,1){15}} 
\put(35,10){\line(2,-1){15}} 
\put(5,5){\makebox(0,0){$\sigma_0$}} 
\put(20,5){\makebox(0,0){$\sigma_1$}} 
\put(35,5){\makebox(0,0){$\sigma_2$}} 
\put(55,18){\makebox(0,0){$\sigma_3$}} 
\put(56,3){\makebox(0,0){$\sigma_3^{'}$}} 
\end{picture} 
\qquad \qquad 
$ 
{\cal G}_{D_5} = 
\left( \begin{array}{ccccc} 
      0 & 1 & 0 & 0 & 0   \\ 
      1 & 0 & 1 & 0 & 0   \\ 
      0 & 1 & 0 & 1 & 1   \\ 
      0 & 0 & 1 & 0 & 0   \\ 
      0 & 0 & 1 & 0 & 0   \\ 
\end{array} 
\right) 
$ 
\caption{Le graphe $D_5$ et sa matrice d'adjacence.} 
\label{grD5} 
\end{center} 
\end{figure}
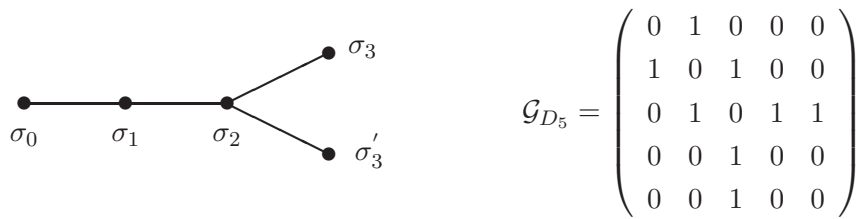

\paragraph{Graphe $D_5$} 
Le graphe $D_5$ et sa matrice d'adjacence sont illustr\'es \`a la Fig. \ref{grD5}, o\`u l'ordre choisi pour les vertex est: 
$\{\sigma_0, \sigma_1, \sigma_2, \sigma_3, \sigma_{3^{'}} \}$.

Pour $D_5$, $\kappa = 8$, la norme du graphe est $\beta = [2]_q = 2 \cos 
(\frac{\pi}{8}) = \sqrt{2 + \sqrt 2}$ et les composantes du vecteur de 
Perron-Frobenius sont donn\'ees par: 
$P = \left( [1]_q, [2]_q, [3]_q, \frac{[3]_q}{[2]_q}, \frac{[3]_q}{[2]_q} 
     \right)$. Il n'est pas possible de d\'efinir una alg\`ebre associative et commutative 
poss\'edant des coefficients de structure entiers non-n\'egatifs pour le cas $D_5$: plus g\'en\'eralement,
on montre que les graphes $D_{2n+1}$ ne poss\`edent pas {\it self-fusion} \cite{pasquier-ADE}.  
 
\begin{figure}[hhh] 
\unitlength 0.85mm 
\begin{center} 
\begin{picture}(90,20)(0,2) 
\put(0,15){\line(1,0){90}} 
\multiput(0,15)(15,0){7}{\circle*{2}} 
\put(0,7.5){\makebox(0,0){$\tau_{0}$}} 
\put(15,7.5){\makebox(0,0){$\tau_{1}$}} 
\put(30,7.5){\makebox(0,0){$\tau_{2}$}} 
\put(45,7.5){\makebox(0,0){$\tau_{3}$}} 
\put(60,7.5){\makebox(0,0){$\tau_{4}$}} 
\put(75,7.5){\makebox(0,0){$\tau_{5}$}} 
\put(90,7.5){\makebox(0,0){$\tau_{6}$}} 
\put(-10,1){\makebox(0,0){$\hat{T}:$}} 
 
\put(0,0){\makebox(0,0){1}} 
\put(15,0){\makebox(0,0){4}} 
\put(30,0){\makebox(0,0){9}} 
\put(45,0){\makebox(0,0){16}} 
\put(60,0){\makebox(0,0){25}} 
\put(75,0){\makebox(0,0){4}} 
\put(90,0){\makebox(0,0){17}} 
 
\end{picture} 
\end{center} 
\caption{Le graphe $A_7$ et les valeurs $\hat T$ sur les vertex $\tau_i$.} 
\label{grA7} 
\end{figure}
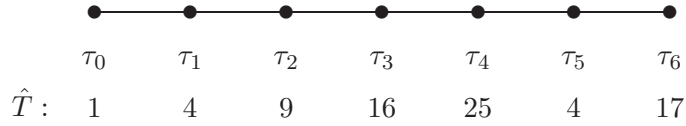

\paragraph{Induction-restriction} 
Comme les graphes $D_{2n}$, les graphes $D_{2n+1}$ sont 
des $\mathbb{Z}_2$-orbifold des graphes $A_{4n-1}$. Le graphe de la s\'erie $A_n$ correspondant \`a $D_5$ 
est $A_7$. M\^eme si $D_{2n+1}$ ne poss\`ede pas de structure alg\'ebrique, il est n\'eanmoins un {\sl module} 
sous l'action de $A_7$. Les matrices de fusion $N_i$ de $A_7$ et les matrices $F_i$ codant l'action de  
$A_7$ sur $D_5$ sont obtenues comme  
d'habitude par la formule de r\'ecurrence tronqu\'ee 
de $su(2)$, avec $N_1 = \mathcal{G}_{A_7}$ et $F_1= \mathcal{G}_{D_5}$. Le graphe $A_7$ est pr\'esent\'e \`a la Fig. \ref{grA7} ainsi que les  
valeurs de $\hat{T}$ sur ses vertex. 
 
\begin{figure}[H] 
\unitlength 0.9mm 
\begin{center} 
\begin{picture}(120,70) 
\multiput(20,5)(0,15){5}{\circle{2}} 
\put(5,35){\circle{2}} 
\put(35,35){\circle{2}} 
 
\thicklines 
\put(5,35){\line(1,1){15}} 
\put(5,35){\line(1,2){15}} 
\put(35,35){\line(-1,-1){15}} 
\put(35,35){\line(-1,-2){15}} 
\put(19.5,50){\line(0,-1){15}} 
\put(20.5,20){\line(0,1){15}} 
 
\thicklines 
\dashline[50]{1}(5,35)(20,20) 
\dashline[50]{1}(5,35)(20,5) 
\dashline[50]{1}(35,35)(20,50) 
\dashline[50]{1}(35,35)(20,65) 
\dashline[80]{1}(20.5,50)(20.5,35) 
\dashline[80]{1}(19.5,20)(19.5,35)

\small 
\put(20,69){\makebox(0,0){$\ud{0}$}} 
\put(20,54){\makebox(0,0){$\ud{2}$}} 
\put(23,35){\makebox(0,0){$\ud{3}$}} 
\put(20,16){\makebox(0,0){$\ud{4}$}} 
\put(1,35){\makebox(0,0){$\ud{1}$}} 
\put(39,35){\makebox(0,0){$\ud{5}$}} 
\put(20,1){\makebox(0,0){$\ud{6}$}} 
\normalsize

\put(100,35){\makebox(0,0){$\mathcal{M}=\left( \begin{array}{ccccccc} 
1 & . & . & . & . & . & . \\
. & . & . & . & . & 1 & . \\
. & . & 1 & . & . & . & . \\
. & . & . & 1 & . & . & . \\
. & . & . & . & 1 & . & . \\
. & 1 & . & . & . & . & . \\
. & . & . & . & . & . & 1 
\end{array} 
\right)
$}} 
\end{picture} 

\caption{Le graphe d'Ocneanu de $D_5$ et son invariant modulaire.} 
\label{grOc_D5} 
\end{center} 
\end{figure}
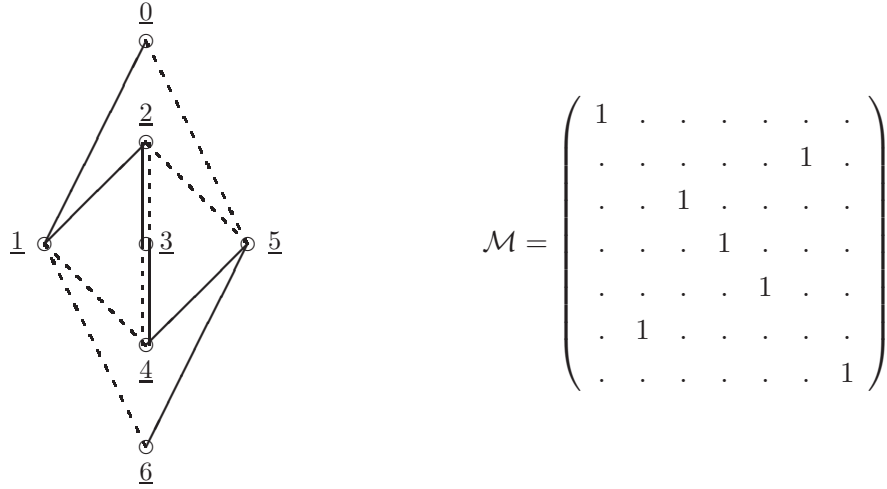

\paragraph{Alg\`ebre d'Ocneanu} 
L'alg\`ebre d'Ocneanu de $A_7$ est d\'efinie par $A_7 \otimes_{A_7} A_7$. Cependant, il est possible 
de d\'efinir un autre quotient sur le produit tensoriel $A_7 \otimes A_7$ pour lequel  
$\hat{T}$ est bien d\'efini. En effet, les 
valeurs de $\hat{T}$ sont les m\^emes ($\hat{T}=4$) sur les vertex $\tau_1$ et $\tau_5$ de $A_7$. D\'efinissons l'application $\rho$ ({\sl twist}) telle que: 
$$ 
\rho(\tau_i) = \tau_i \qquad \text{pour} \;\, i = \{0,2,3,4,6 \} 
\qquad \text{et} \quad 
\rho(\tau_1) = \tau_5, \quad \rho(\tau_5) = \tau_1. 
$$ 
Alors, l'alg\`ebre d'Ocneanu de $D_5$ est d\'efinie par: 
\begin{equation} 
Oc(D_5) = A_7 \otimes_{\rho} A_7 
\end{equation}  
o\`u nous avons les identifications suivantes entre les \'el\'ements de $Oc(D_5)$: 
\begin{equation} 
\tau_a \otimesdot \tau_b = \tau_a . \rho(\tau_b) \otimesdot \tau_0 
\end{equation} 
Une base de $Oc(D_5)$ est donn\'ee par les 7 \'el\'ements lin\'eairement ind\'ependants suivants: 
$$ 
\begin{array}{lcc} 
\ud0 = \tau_0 \otimesdot \tau_0, & {} & \ud4 = \tau_4 \otimesdot \tau_0 = \tau_0 \otimesdot \tau_4, \\ 
\ud1 = \tau_1 \otimesdot \tau_0 = \tau_0 \otimesdot \tau_5, & \qquad \qquad \qquad & \ud5 = \tau_5 
\otimesdot \tau_0 = \tau_0 \otimesdot \tau_1, \\ 
\ud2 = \tau_2 \otimesdot \tau_0 = \tau_0 \otimesdot \tau_2, & {} & \ud6 = \tau_6 \otimesdot \tau_0 
= \tau_0 \otimesdot \tau_6. \\ 
\ud3 = \tau_3 \otimesdot \tau_0 = \tau_0 \otimesdot \tau_3, & {} & {} \\ 
\end{array} 
$$ 
$(\tau_1 \otimesdot \tau_0)$ et $(\tau_0 \otimesdot \tau_1)$ ($= (\tau_5 \otimesdot \tau_0)$) sont respectivement les  
g\'en\'erateurs chiraux gauche et droit. La multiplication par ces g\'en\'erateurs est cod\'ee par le graphe 
d'Ocneanu de $D_5$, repr\'esent\'e \`a la Fig. \ref{grOc_D5}. 
La multiplication dans $Oc(D_5)$ et l'action de $Oc(D_5)$ sur $D_5$ sont cod\'ees par les matrices $O_x$  
et $S_x$, qui, pour $\ud{x} = \tau_i \otimes \tau_0$, sont donn\'ees par: 
\begin{equation} 
O_x = N_i \qquad \qquad S_x = F_i. 
\end{equation} 
 
\paragraph{Dimensions des blocs} La dimension des blocs  
$d_i$ et $d_x$ est donn\'ee par la somme des \'el\'ements des matrices $F_i$ et $S_x$, et comme ces matrices 
sont identiques, les r\`egles de somme lin\'eaire et quadratiques sont \'evidemment v\'erifi\'ees. Nous avons: 
\begin{equation} 
\dim (\mathcal{B}(D_5)) = \sum_{i \in A_7} d_i^2 = \sum_{x \in Oc(D_5)} d_x^2 = 564. 
\end{equation} 
 
\paragraph{Matrices toriques et fonctions de partition g\'en\'eralis\'ees} 
Elles sont obtenues par l'action \`a gauche et \`a droite de $A_7$ sur un \'el\'ement $\ud{x} = \tau_x \otimesdot \tau_0$  
de $Oc(D_5)$: 
\begin{eqnarray*} 
\tau_i . (\ud{x}) . \tau_j =  \tau_i . ( \tau_x \otimesdot \tau_0) . \tau_j &=&  
\sum_k \sum_l (N_i)_{xk}  (N_0)_{jl} \; (\tau_k \otimesdot \tau_l) \\  
{ } &=& \sum_k  (N_x)_{ik} (\tau_k . \rho(\tau_j) \; \otimesdot \tau_0) \\ 
{ } &=& \sum_k \sum_y (N_x)_{ik} (N_k)_{\rho(j)y} \; (\tau_y \otimesdot \tau_0) 
= \sum_k \sum_y (N_x)_{ik} (N_y)_{k\rho(j)} (\ud{y}) 
\end{eqnarray*} 
Donc les matrices toriques g\'en\'eralis\'ees $W_{xy}$, pour $\ud{x} = \tau_x \otimesdot \tau_0$ et 
$\ud{y} = \tau_y \otimesdot \tau_0 $ du mod\`ele $D_5$ s'\'ecrivent de mani\`ere compacte sous: 
\begin{equation} 
\begin{array}{|c|} 
\hline 
{ } \\ 
W_{xy} = (N_x . N_y)_{i \rho(j)} \\ 
{ } \\ 
\hline 
\end{array} 
\end{equation} 
et l'invariant modulaire qui commute avec les g\'en\'erateurs $S$ et $T$ du groupe modulaire s'\'ecrit: 
$$ 
(\mathcal{M})_{ij} = (\munite)_{i \rho(j)} 
$$ 
Les fonctions de partition g\'en\'eralis\'ees du mod\`ele $D_5$ s'obtiennent alors par: 
\begin{equation} 
\mathcal{Z}_{xy} = \sum_{i \in A_7} \sum_{j \in A_7} \chi_i (q) \; (W_{xy})_{ij} \; \ov{\chi}_j(q) 
\end{equation} 
et la fonction de partition invariante modulaire s'\'ecrit: 
\begin{equation} 
\mathcal{Z}_{D_5} = \xa{0} + \xa{2} + \xa{3} + \xa{4} + \xa{6} + (\chi_1 . \ov{\chi}_5 + \textrm{h.c.})  
\end{equation} 
Les fonctions de partition \`a une ligne de d\'efauts sont pr\'esent\'ees dans l'Annex {\bf D} 
en fonction des caract\`eres de l'alg\`ebre $\widehat{su}(2)$.

\subsubsection{Formules g\'en\'erales pour $D_{2n+1}$} 
\begin{figure}[H] 
\unitlength 0.7mm 
\begin{center} 
\begin{picture}(190,100) 
\multiput(35,5)(0,15){7}{\circle{2}} 
\multiput(5,50)(15,0){5}{\circle{2}} 
 
\thicklines 
\put(20,50){\line(1,1){15}} 
\put(20,50){\line(1,2){15}} 
\put(50,50){\line(-1,-1){15}} 
\put(50,50){\line(-1,-2){15}} 
\put(34.5,65){\line(0,-1){15}} 
\put(35.5,35){\line(0,1){15}} 
 
\put(5,50){\line(1,1){30}} 
\put(5,50){\line(2,3){30}} 
\put(65,50){\line(-1,-1){30}} 
\put(65,50){\line(-2,-3){30}} 
 
\thicklines 
\dashline[50]{1}(20,50)(35,35) 
\dashline[50]{1}(20,50)(35,20) 
\dashline[50]{1}(50,50)(35,65) 
\dashline[50]{1}(50,50)(35,80) 
\dashline[80]{1}(35.5,65)(35.5,50) 
\dashline[80]{1}(34.5,35)(34.5,50) 
 
\dashline[50]{1}(65,50)(35,80) 
\dashline[50]{1}(65,50)(35,95) 
\dashline[50]{1}(5,50)(35,15) 
\dashline[50]{1}(5,50)(35,5) 
 
\small 
\put(35,99){\makebox(0,0){$\ud{0}$}} 
\put(35,84){\makebox(0,0){$\ud{2}$}} 
\put(35,69){\makebox(0,0){$\ud{4}$}} 
\put(38,50){\makebox(0,0){$\ud{5}$}} 
\put(35,31){\makebox(0,0){$\ud{6}$}} 
\put(35,16){\makebox(0,0){$\ud{8}$}} 
\put(35,1){\makebox(0,0){$\ud{10}$}} 
\put(1,50){\makebox(0,0){$\ud{1}$}} 
\put(54,50){\makebox(0,0){$\ud{7}$}} 
\put(69,50){\makebox(0,0){$\ud{9}$}} 
\put(16,50){\makebox(0,0){$\ud{3}$}} 
\normalsize 
\put(140,50){\makebox(0,0){$\mathcal{M}=\left( \begin{array}{ccccccccccc} 
1 & . & . & . & . & . & . & . & . & . & . \\ 
. & . & . & . & . & . & . & . & . & 1 & . \\ 
. & . & 1 & . & . & . & . & . & . & . & . \\ 
. & . & . & . & . & . & . & 1 & . & . & . \\ 
. & . & . & . & 1 & . & . & . & . & . & . \\ 
. & . & . & . & . & 1 & . & . & . & . & . \\ 
. & . & . & . & . & . & 1 & . & . & . & . \\ 
. & . & . & 1 & . & . & . & . & . & . & . \\ 
. & . & . & . & . & . & . & . & 1 & . & . \\ 
. & 1 & . & . & . & . & . & . & . & . & . \\ 
. & . & . & . & . & . & . & . & . & . & 1 \\ 
\end{array} 
\right) 
$}} 
 
\end{picture} 
\caption{Le graphe d'Ocneanu de $D_7$ et son invariant modulaire.} 
\label{grOc_D7} 
\end{center} 
\end{figure}
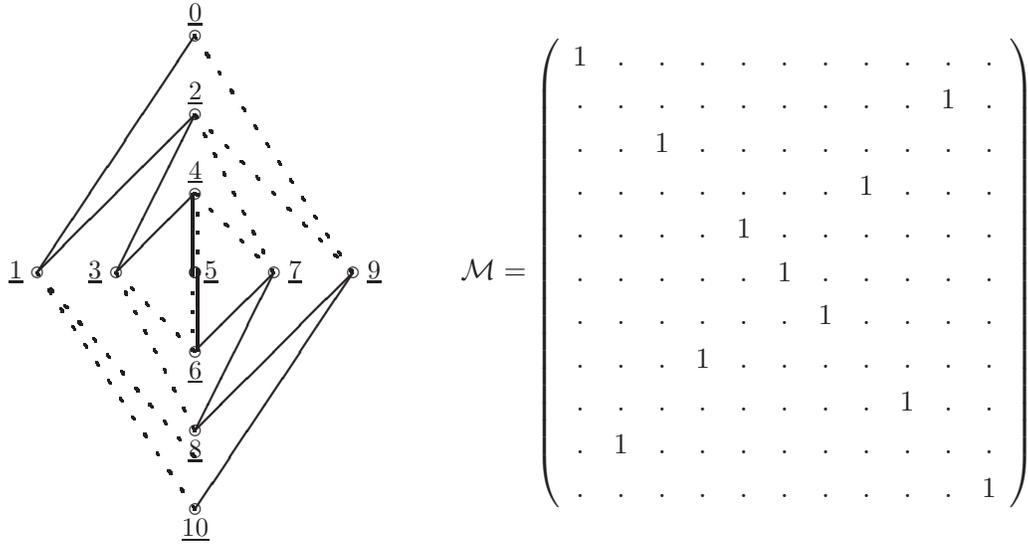

Le graphe $D_n$ g\'en\'eral est illustr\'e \`a la figure (\ref{grDn}) avec sa matrice d'adjacence. 
Il n'est pas possible de 
d\'efinir une structure alg\'ebrique: les graphes  
$D_{2n+1}$ ne poss\`edent pas {\it self-fusion} (type II). L'espace vectoriel form\'e par les vertex $\sigma_a$ de $D_{2n+1}$ 
est cependant un {\sl module} sous l'action de $A_{4n-1}$. 
Les matrices de fusion $N_i$ et les matrices $F_i$ codant l'action de $A_{4n-1}$ sur $D_{2n+1}$ 
s'obtiennent comme d'habitude. La valeur de $\hat{T}$ sur les vertex de $A_{4n-1}$ est la m\^eme pour les vertex 
impairs sym\'etriques: $\sigma_1$ et $\sigma_{4n-3}$, $\sigma_3$ et $\sigma_{4n-5}$, $\ldots$. Ceci permet 
de d\'efinir un quotient de l'alg\`ebre $A_{4n-1} \otimes A_{4n-1}$ par l'application $\rho$, d\'efinie par: 
$$ 
\begin{array}{rclcl} 
\rho(\sigma_a) &=& \sigma_a &\qquad& \textrm{si a pair} \\ 
\rho(\sigma_a) &=& \sigma_{4n-2-a} &\qquad& \textrm{si a impair}  
\end{array} 
$$  
sur lequel l'op\'erateur $\hat{T}$ est bien d\'efini. L'alg\`ebre d'Ocneanu de $D_{2n+1}$ est d\'efinie par: 
\begin{equation} 
Oc(D_{2n+1}) = A_{4n-1} \otimes_{\rho} A_{4n-1} = A_{4n-1} \otimesdot A_{4n-1}  
\end{equation} 
o\`u nous identifions les \'el\'ements $\sigma_a \otimesdot \sigma_b$ et $\sigma_a . \rho(\sigma_b) \otimesdot \sigma_0$. 
Cette alg\`ebre est de dimension $4n-1$. La multiplication dans $Oc(D_{2n+1})$ et l'action de $Oc(D_{2n+1})$ sur $D_{2n+1}$ sont cod\'ees par les matrices $O_x$  
et $S_x$, qui, pour $\ud{x} = \tau_i \otimes \tau_0$, sont donn\'ees par: 
\begin{equation} 
O_x = N_i \qquad \qquad S_x = F_i. 
\end{equation}   
Les matrices $F_i$ et $S_x$ \'etant \'egales, les r\`egles de somme lin\'eaire et quadratique sont bien sur 
satisfaites. 
Les matrices toriques g\'en\'eralis\'ees $W_{xy}$ codant l'action de $Oc(D_{2n+1})$ sur $D_{2n-1}$ sont donn\'ees 
par: 
\begin{equation} 
\begin{array}{|c|} 
\hline 
{ } \\ 
W_{xy} = (N_x . N_y)_{i \rho(j)} \\ 
{ } \\ 
\hline 
\end{array} 
\label{Wxy_D2n+1} 
\end{equation} 
et les fonctions de partition g\'en\'eralis\'ees s'obtiennent par: 
\begin{equation} 
\mathcal{Z}_{xy} = \sum_{i \in A_{4n-1}} \sum_{j \in A_{4n-1}} \chi_i (q) \; (W_{xy})_{ij} \; \ov{\chi}_j(q) 
\label{Zxy_D2n+1} 
\end{equation} 
Pour le mod\`ele $D_7$, le graphe $A_n$ correspondant est $A_{11}$, et le twist  
$\rho: A_{11} \rightarrow A_{11}$ est d\'efini par: 
$$ 
\rho(\tau_i) = \tau_i \quad \text{pour} \quad i \in \{0,2,4,5,6,8,10\}  \quad 
\text{et} \quad 
\rho(\tau_1) = \tau_9, \quad \rho(\tau_3) =\tau_7, \quad \rho(\tau_7) = \tau_3, \quad \rho(\tau_9) =\tau_1. 
$$ 
Le graphe d'Ocneanu de $D_7$ est illustr\'e \`a la Fig \ref{grOc_D7} avec l'invariant modulaire correspondant. 
Les fonctions de partition \`a une ligne de d\'efauts sont publi\'ees dans \cite{Coq_Gil-ADE}, elles s'obtiennent 
tr\`es facilement par (\ref{Wxy_D2n+1}) et (\ref{Zxy_D2n+1}). 
 
\subsection{Le cas $E_7$} 
\paragraph{Graphe $E_7$} 
Le graphe $E_7$ et sa matrice d'adjacence sont illustr\'es ci-dessous. Nous choisissons l'ordre suivant pour 
les vertex: $\{\sigma_0, \sigma_1, \sigma_2, \sigma_3, \sigma_6, \sigma_5, \sigma_4 \}$.

\begin{figure}[hhh] 
\unitlength 0.9mm 
\begin{center} 
\begin{picture}(100,20)(5,10) 
\thinlines 
\multiput(15,10)(15,0){6}{\circle*{2}} 
\put(60,25){\circle*{2}} 
\thicklines 
\put(15,10){\line(1,0){75}} 
\put(60,10){\line(0,1){15}} 
\put(15,3){\makebox(0,0){$\sigma_0$}} 
\put(30,3){\makebox(0,0){$\sigma_1$}} 
\put(45,3){\makebox(0,0){$\sigma_2$}} 
\put(60,3){\makebox(0,0){$\sigma_3$}} 
\put(75,3){\makebox(0,0){$\sigma_6$}} 
\put(90,3){\makebox(0,0){$\sigma_5$}} 
\put(68,25){\makebox(0,0){$\sigma_4$}} 
\end{picture} 
$ 
{\cal G}_{E_7} = 
\left( \begin{array}{ccccccc} 
      0 & 1 & 0 & 0 & 0 & 0 & 0 \\ 
      1 & 0 & 1 & 0 & 0 & 0 & 0 \\ 
      0 & 1 & 0 & 1 & 0 & 0 & 0 \\ 
      0 & 0 & 1 & 0 & 1 & 0 & 1 \\ 
      0 & 0 & 0 & 1 & 0 & 1 & 0 \\ 
      0 & 0 & 0 & 0 & 1 & 0 & 0 \\ 
      0 & 0 & 0 & 1 & 0 & 0 & 0 \\ 
\end{array} 
\right) 
$ 
\caption{Le graphe $E_7$ et sa matrice d'adjacence.} 
\label{grE7} 
\end{center} 
\end{figure}
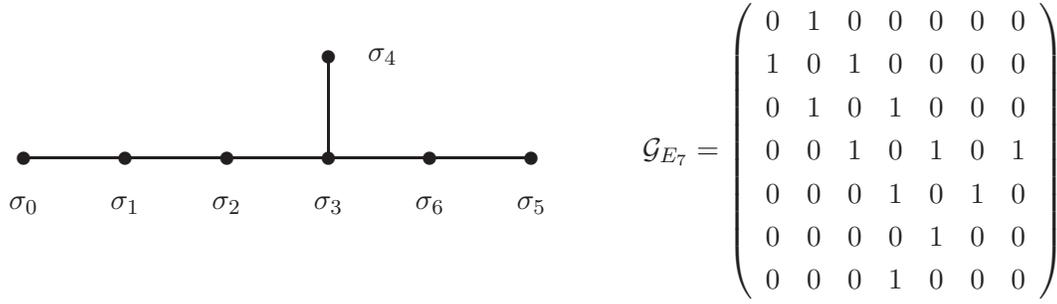 
 
Pour $E_7$, $\kappa = 18$, la norme du graphe est $\beta = [2]_q = 2 \cos (\frac{\pi}{18})$  
et les composantes du vecteur de Perron-Frobenius  sont 
donn\'ees par: 
$P = \left( [1]_q, [2]_q, [3]_q, [4]_q, \frac{[6]_q}{[2]_q}, 
\frac{[4]_q}{[3]_q}, \frac{[4]_q}{[2]_q} \right) $, avec $q = \exp \left( \frac{i \pi }{18} \right)$.   
Le graphe $E_7$ ne d\'efinit pas une alg\`ebre de graphe \`a coefficients entiers non-n\'egatifs: il est 
possible de d\'efinir une alg\`ebre mais dont certains coefficients de structure sont n\'egatifs \cite{Coq_Gil-ADE}:  
cette alg\`ebre ne code donc pas en soi une structure de fusion de ``repr\'esentations irr\'eductibles''
(voir Tab. \ref{multE7}).
 
\scriptsize 
\begin{table}[H] 
$$ 
\begin{array}{||c||c|c|c|c|c|c|c||} 
\hline 
E_7& \sigma_0 & \sigma_1  & \sigma_2 & \sigma_3 & \sigma_6 & \sigma_5 & \sigma_4  \\ 
\hline 
\hline 
\sigma_0 & \sigma_0 & \sigma_1     & \sigma_2       & \sigma_3           & \sigma_6     & \sigma_5     & \sigma_4     \\ 
\sigma_1 & \sigma_1 & \sigma_0+\sigma_2   & \sigma_1+\sigma_3     & \sigma_2+\sigma_4+\sigma_6       & \sigma_3+\sigma_5   & \sigma_6     & \sigma_3     \\ 
\sigma_2 & \sigma_2 & \sigma_1+\sigma_3   & \sigma_0+\sigma_2+\sigma_4+\sigma_6 & \sigma_1+2\,\sigma_3+\sigma_5     & \sigma_2+\sigma_4+\sigma_6 & \sigma_3     & \sigma_2+\sigma_6   \\ 
\sigma_3 & \sigma_3 & \sigma_2+\sigma_4+\sigma_6 & \sigma_1+2\, \sigma_3+\sigma_5 & \sigma_0+2\, \sigma_2+\sigma_4+2\, \sigma_6 & \sigma_1+2\, \sigma_3 & \sigma_2+\sigma_4   & \sigma_1+\sigma_3+\sigma_5 \\ 
\sigma_6 & \sigma_6 & \sigma_3+\sigma_5   & \sigma_2+\sigma_4+\sigma_6   & \sigma_1+2\, \sigma_3       & \sigma_0+\sigma_2+\sigma_6 & \sigma_1+\sigma_5   & \sigma_2+\sigma_4   \\ 
\sigma_5 & \sigma_5 & \sigma_6     & \sigma_3       & \sigma_2+\sigma_4         & \sigma_1+\sigma_5   & \sigma_0-\sigma_4+\sigma_6 & \sigma_3-\sigma_5   \\ 
\sigma_4 & \sigma_4 & \sigma_3     & \sigma_2+\sigma_6     & \sigma_1+\sigma_3+\sigma_5       & \sigma_2+\sigma_4   & \sigma_3-\sigma_5   & \sigma_0+\sigma_6   \\ 
\hline 
\end{array} 
$$ 
\label{multE7}
\caption{Table de multiplication de l'alg\`ebre de graphe $E_7$.} 
\end{table} 
\normalsize 
Les matrices de ``{\sl fusion}''$ G_{a}^{E_7}$ codant cette structure alg\'ebrique sont donn\'ees par: 
$$ 
\begin{array}{ll} 
G_0^{E_7} = \munite_{7 \times 7}  \qquad \qquad \qquad \qquad \qquad \qquad & 
G_5^{E_7} = G_3^{E_7}.G_2^{E_7} - G_1^{E_7} - 2.G_3^{E_7} \\ 
G_1^{E_7} = \mathcal{G}_{E_7} & 
G_4^{E_7} = G_5^{E_7}.G_3^{E_7} - G_2^{E_7} \\ 
G_2^{E_7} = G_1^{E_7}.G_1^{E_7} - G_0^{E_7} & 
G_6^{E_7} = G_2^{E_7}.G_4^{E_7} - G_2^{E_7} \\ 
G_3^{E_7} = G_1^{E_7}.G_2^{E_7} - G_1^{E_7} & {} 
\end{array} 
$$

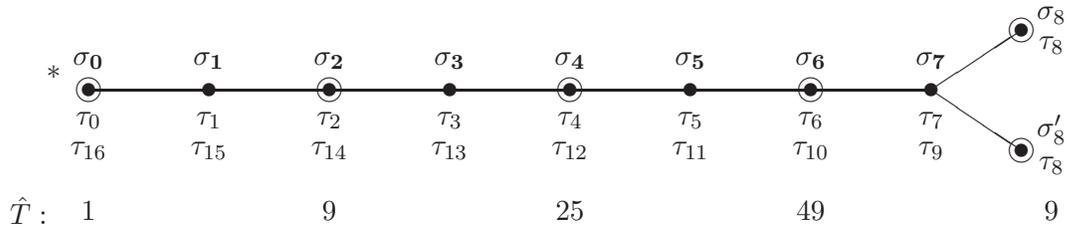
\begin{figure}[hhh] 
\unitlength 0.8mm 
\begin{center} 
\begin{picture}(160,35)(0,-5) 
\put(0,15){\line(1,0){140}} 
\put(140,15){\line(3,2){15}} 
\put(140,15){\line(3,-2){15}} 
\multiput(0,15)(20,0){8}{\circle*{2}} 
\put(155,25){\circle*{2}} 
 
\put(155,5){\circle*{2}} 
\put(155,25){\circle{4}} 
\put(155,5){\circle{4}} 
\multiput(0,15)(40,0){4}{\circle{4}} 
 
\put(0,10){\makebox(0,0){$\tau_{0}$}} 
\put(0,5){\makebox(0,0){$\tau_{16}$}} 
\put(20,10){\makebox(0,0){$\tau_{1}$}} 
\put(20,5){\makebox(0,0){$\tau_{15}$}} 
\put(40,10){\makebox(0,0){$\tau_{2}$}} 
\put(40,5){\makebox(0,0){$\tau_{14}$}} 
\put(60,10){\makebox(0,0){$\tau_{3}$}} 
\put(60,5){\makebox(0,0){$\tau_{13}$}} 
 
\put(80,10){\makebox(0,0){$\tau_{4}$}} 
\put(80,5){\makebox(0,0){$\tau_{12}$}} 
\put(100,10){\makebox(0,0){$\tau_{5}$}} 
\put(100,5){\makebox(0,0){$\tau_{11}$}} 
\put(120,10){\makebox(0,0){$\tau_{6}$}} 
\put(120,5){\makebox(0,0){$\tau_{10}$}} 
\put(140,10){\makebox(0,0){$\tau_{7}$}} 
\put(140,5){\makebox(0,0){$\tau_{9}$}}

\put(0,20){\makebox(0,0){$\bf{\sigma_{0}}$}} 
\put(20,20){\makebox(0,0){$\bf{\sigma_{1}}$}} 
\put(40,20){\makebox(0,0){$\bf{\sigma_{2}}$}} 
\put(60,20){\makebox(0,0){$\bf{\sigma_{3}}$}} 
\put(80,20){\makebox(0,0){$\bf{\sigma_{4}}$}} 
\put(100,20){\makebox(0,0){$\bf{\sigma_{5}}$}} 
\put(120,20){\makebox(0,0){$\bf{\sigma_{6}}$}} 
\put(140,20){\makebox(0,0){$\bf{\sigma_{7}}$}} 
 
\put(160,27.5){\makebox(0,0){$\sigma_{8}$}} 
\put(160,22.5){\makebox(0,0){$\tau_{8}$}} 
 
\put(160,8.5){\makebox(0,0){$\sigma_{8}'$}} 
\put(160,2.5){\makebox(0,0){$\tau_{8}$}} 
 
\put(-10,-5){\makebox(0,0){$\hat{T}:$}} 
\put(0,-5){\makebox(0,0){$1$}} 
\put(40,-5){\makebox(0,0){$9$}} 
\put(80,-5){\makebox(0,0){$25$}} 
\put(120,-5){\makebox(0,0){$49$}} 
\put(160,-5){\makebox(0,0){$9$}} 
 
\put(-7,17){$\ast$} 
 
\end{picture} 
\end{center} 
\caption{Le graphe d'induction $D_{10}$-$A_{17}$ et les valeurs $\hat{T}$.} 
\label{grapheinductionAD} 
\end{figure}

\paragraph{Induction-restriction} 
Le graphe de la s\'erie $A_n$ ayant m\^eme norme que $E_7$ est $A_{17}$. Les matrices de fusion $N_i$ et les  
matrices $F_i^{E_7}$ qui codent l'action de $A_{17}$ sur $E_7$ s'obtiennent comme d'habitude par la formule 
de r\'ecurrence tronqu\'ee de $SU(2)$, avec $F_1^{E_7} = \mathcal{G}_{E_7}$.  
Le graphe $D_{10}$ poss\`ede aussi la m\^eme norme que $A_{17}$ et $E_7$: nous allons voir que l'alg\`ebre  
d'Ocneanu de $E_7$ est construite \`a partir de celle du graphe $D_{10}$ \cite{Oc-paths, Coq_Gil-ADE}.  
Les matrices de fusion de l'alg\`ebre de graphe $D_{10}$ sont not\'ees $G_a^{D_{10}}$, et les matrices 
qui codent l'action de $A_{17}$ sur $D_{10}$ seront not\'ees $F_i^{D_{10}}$, avec $F_1^{D_{10}} = \mathcal{G}_{D_{10}}$. 
Elles permettent --\`a travers la matrice essentielle $E_0^{D_{10}}$-- de d\'efinir l'induction-restriction entre  
$A_{17}$ et $D_{10}$: ce graphe d'induction est illustr\'e \`a la Fig. \ref{grapheinductionAD}.

Par le m\'ecanisme d'induction-restriction, le 
sous-espace $J$ des vertex de $D_{10}$ pour lesquels une valeur de $\hat{T}$ est bien d\'efinie 
est $J = \{\sigma_0, \sigma_2, \sigma_4, \sigma_6, \sigma_8, \sigma_{8'} \}$. Pour $A_{17}$,  
la valeur de $\hat{T}$ sur le vertex central $\tau_8$ 
est \'egale \`a la valeur de $\hat{T}$ sur les vertex: $\tau_2$ et $\tau_{14}$: ainsi, la valeur de $\hat{T}$ 
est la m\^eme pour les vertex $\sigma_2$ et ceux constituant la fourche de $D_{10}$. Ceci nous m\`ene \`a pouvoir 
d\'efinir un {\sl twist} $\rho$ agissant sur les vertex de $D_{10}$ (ce {\it twist} est ``le'' {\it twist} exceptionnel du mod\`ele 
$su(2)$; l'existence de ce {\it twist} n'est pas nouvelle, mais nous montrons ici sa relation avec les propri\'et\'es de l'op\'erateur modulaire): 
$$ 
\begin{array}{ccc} 
\rho(\sigma_0)=\sigma_0, & \qquad \qquad \rho(\sigma_4)=\sigma_4, & 
\qquad \qquad \rho(\sigma_8)=\sigma_2,  \\ 
\rho(\sigma_2)=\sigma_8, & \qquad \qquad \rho(\sigma_6)=\sigma_6, & 
\qquad \qquad \rho(\sigma_8^{'})=\sigma_8^{'}. 
\end{array} 
$$  
 
\paragraph{Alg\`ebre d'Ocneanu} 
L'alg\`ebre d'Ocneanu de $E_7$ est d\'efinie par: 
\begin{equation} 
Oc(E_7) = D_{10} \otimes_{\rho} D_{10} = D_{10} \otimesdot D_{10} 
\end{equation}  
o\`u  nous avons les suivantes identifications: 
\begin{equation} 
\sigma_a \otimesdot \sigma_b.\sigma_c = \sigma_a . \rho(\sigma_b) \otimesdot \sigma_c  \qquad \textrm{pour }  
\sigma_b \in J 
\end{equation}   
Appelons $\ov{J}$ le sous-espace compl\'ementaire de $J$ dans $D_{10}$. Nous avons: 
$$ 
D_{10} = J \oplus \ov{J} \qquad \qquad J.J \subset J \qquad \qquad J.\ov{J} = \ov{J}.J \subset \ov{J} \qquad \qquad \ov{J}.\ov{J} \subset J  
$$ 
et un \'el\'ement $\sigma_a \in \ov{J}$ peut s'\'ecrire comme $\sigma_a^{'}.\sigma_b$, avec $\sigma_a^{'}\in J$ 
et $\sigma_b \in \ov{J}$ (ou une combinaison lin\'eaire de tels  \'el\'ements).  
L'alg\`ebre $D_{10} \otimes D_{10}$ poss\`ede $10 \times 10 = 100$ \'el\'ements lin\'eairement 
ind\'ependants.  
Nous avons les identifications suivantes dans l'alg\`ebre $D_{10} \otimesdot D_{10}$: 
\begin{equation} 
\sigma_a \otimesdot \sigma_b = \sigma_a . \rho(\sigma_b) \otimesdot \sigma_0 \qquad \qquad  
\textrm{pour } \sigma_a \in D_{10} \; ,  \sigma_b \in J 
\label{identif1} 
\end{equation} 
et  
\begin{equation} 
\begin{array}{rcl} 
\sigma_a \otimesdot \sigma_b &=& \sigma_0 \otimesdot \rho(\sigma_a) . \sigma_b \qquad \quad \;\; \textrm{pour } \sigma_a \in J,   
\sigma_b \notin J \\  
\sigma_a \otimesdot \sigma_b  &=& \sum_c \sum_d \sigma_c \otimesdot \sigma_c \qquad \;\;\,\, \textrm{pour }  
(\sigma_a, \sigma_b) \notin J, \textrm{avec } (\sigma_c, \sigma_d) \notin J, 
\end{array} 
\label{identif2} 
\end{equation} 
Les identifications (\ref{identif1}) d\'efinissent 10 \'el\'ements lin\'eairement ind\'ependants, 
not\'es \mbox{$\ud{a} = \sigma_a \otimesdot \sigma_0$}. Les identifications (\ref{identif2}) d\'efinissent 7 \'el\'ements lin\'eairement ind\'ependants, \mbox{not\'es $(\ud{a})$}. 
Une base de l'alg\`ebre d'Ocneanu de $E_7$ est donn\'ee par les 17 \'el\'ements suivants: 
$$ 
\begin{array}{cclcccl} 
\ud0 &=& 0 \otimesdot 0  & \qquad \qquad & \ud{(0)} &=& 0 \otimesdot 1   \\ 
\ud1 &=& 1 \otimesdot 0  & \qquad \qquad & \ud{(1)} &=& 1 \otimesdot 1   \\ 
\ud2 &=& 2 \otimesdot 0 = 0 \otimesdot 8 & \qquad \qquad & \ud{(2)} &=& 2 
\otimesdot 1 = 0 \otimesdot 7  \\ 
\ud3 &=& 3 \otimesdot 0  & \qquad \qquad & \ud{(3)} &=& 3 \otimesdot 1 = 1 
\otimesdot 3  \\ 
\ud4 &=& 4 \otimesdot 0 = 0 \otimesdot 4 & \qquad \qquad & \ud{(4)} &=& 0 
\otimesdot 3   \\ 
\ud5 &=& 5 \otimesdot 0  & \qquad \qquad & \ud{(5)} &=& 5 \otimesdot 1 - 3 
\otimesdot 1  \\ 
\ud6 &=& 6 \otimesdot 0 = 0 \otimesdot 6 & \qquad \qquad & &=&1 \otimesdot 5 - 
1 \otimesdot 3   \\ 
\ud7 &=& 7 \otimesdot 0  & \qquad \qquad & \ud{(6)} &=& 0 \otimesdot 5   \\ 
\ud8 &=& 8 \otimesdot 0 = 0 \otimesdot 2 & \qquad \qquad & & &   \\ 
\ud{8^{'}} &=& 8^{'} \otimesdot 0 = 0 \otimesdot 8^{'} & \qquad 
\qquad & & &   \\ 
\end{array} 
$$ 
$\ud1$ et $\ud{(0)}$ sont respectivement les g\'en\'erateurs chiraux gauche et droit. 
La partie ambichirale est engendr\'ee par  
$\{\ud{0},\ud{2},\ud{4},\ud{6},\ud{8},\ud{8^{'}}\}$. Les \'el\'ements $\ud{a}$ engendrent une sous-alg\`ebre 
de $Oc(E_7)$, isomorphe \`a l'alg\`ebre de graphe $D_{10}$. Nous appelons ``$D_{10}$'' le sous-espace engendr\'e 
par les \'el\'ements de type $\ud{a}$. L'alg\`ebre de graphe de $E_7$ n'apparait 
pas comme une sous-alg\`ebre de $Oc(E_7)$, mais comme un quotient. Nous appelons ``$E_{7}$'' le sous-espace engendr\'e 
par les \'el\'ements de type $(\ud{x})$. Nous donnons dans la Tab. \ref{D10-E7_mult} la multiplication 
des \'el\'ements de $Oc(E_7)$ par ses g\'en\'erateurs.

\begin{table}[hhh] 
$$ 
\begin{array}{|c|c|} 
\hline 
``D_{10}" & \ud1 \\ 
\hline 
\ud0 & \ud1 \\ 
\ud1 & \ud0 + \ud2 \\ 
\ud2 & \ud1 + \ud3 \\ 
\ud3 & \ud2 + \ud4 \\ 
\ud4 & \ud3 + \ud5 \\ 
\ud5 & \ud4 + \ud6 \\ 
\ud6 & \ud5 + \ud7 \\ 
\ud7 & \ud6 + \ud8 + \ud{8^{'}}\\ 
\ud8 & \ud7 \\ 
\ud{8^{'}} & \ud7 \\ 
\hline 
\hline 
``E_7" & \ud1 \\ 
\hline 
\ud{(0)} & \ud{(1)} \\ 
\ud{(1)} & \ud{(0)} + \ud{(2)} \\ 
\ud{(2)} & \ud{(1)} + \ud{(3)} \\ 
\ud{(3)} & \ud{(2)} + \ud{(4)} + \ud{(6)} \\ 
\ud{(4)} & \ud{(3)} \\ 
\ud{(6)} & \ud{(3)} + \ud{(5)} \\ 
\ud{(5)} & \ud{(6)} \\ 
\hline 
\end{array} 
\qquad \qquad \qquad \qquad 
\begin{array}{|c|c|} 
\hline 
``D_{10}" & \ud{(0)} \\ 
\hline 
\ud0 & \ud{(0)} \\ 
\ud{(0)} & \ud0 + \ud8 \\ 
\ud8 & \ud{(0)} + \ud{(4)} \\ 
\ud{(4)} & \ud8 + \ud4 \\ 
\ud4 & \ud{(4)} + \ud{(6)} \\ 
\ud{(6)} & \ud4 + \ud6 \\ 
\ud6 & \ud{(6)} + \ud{(2)} \\ 
\ud{(2)} & \ud2 + \ud6 + \ud{8^{'}}\\ 
\ud2 & \ud{(2)} \\ 
\ud8^{'} & \ud{(2)} \\ 
\hline 
\hline 
``E_7" & \ud{(0)} \\ 
\hline 
\ud1 & \ud{(1)} \\ 
\ud{(1)} & \ud1 + \ud7 \\ 
\ud7 & \ud{(1)} + \ud{(3)} \\ 
\ud{(3)} & \ud3+\ud5+\ud7 \\ 
\ud3 & \ud{(3)} \\ 
\ud5 & \ud{(3)} + \ud{(5)} \\ 
\ud{(5)} & \ud5 \\ 
\hline 
\end{array} 
$$ 
\caption{Multiplication des \'el\'ements de l'alg\`ebre d'Ocneanu de $E_7$ par ses g\'en\'erateurs.} 
\label{D10-E7_mult} 
\end{table} 
Nous pouvons voir que la multiplication des  \'el\'ements de l'alg\`ebre d'Ocneanu de $E_7$ par ses g\'en\'erateurs 
est en effet cod\'ee par le graphe d'Ocneanu de $E_7$, illustr\'e \`a la Fig. \ref{grOc_E7}. 
 
\begin{figure}[hhh] 
\unitlength 0.9mm 
\begin{center} 
\begin{picture}(50,90) 
\put(25,5){\circle*{2}} 
\multiput(25,15)(0,10){2}{\circle{2}} 
\put(25,35){\circle*{2}} 
\put(25,45){\circle{2}} 
\put(25,75){\circle*{2}} 
\put(25,85){\circle{2}} 
\multiput(5,10)(0,20){2}{\circle*{2}} 
\multiput(5,60)(0,20){2}{\circle*{2}} 
\multiput(45,10)(0,20){2}{\circle*{2}} 
\multiput(45,60)(0,20){2}{\circle*{2}} 
\put(20,60){\circle{2}} 
\put(30,60){\circle{2}} 
 
\thicklines 
\put(5,80){\line(4,1){20}} 
\put(5,80){\line(5,-4){25}} 
\put(5,60){\line(1,0){15}} 
\put(5,60){\line(4,-3){20}} 
\dashline[100]{8}(5,60)(25,25) 
\put(5,30){\line(5,6){25}} 
\put(5,30){\line(4,-3){20}} 
\put(5,10){\line(4,3){20}} 
\put(5,10){\line(4,1){20}} 
 
\thinlines 
\put(45,80){\line(-4,-1){20}} 
\put(45,60){\line(-4,3){20}} 
\put(45,60){\line(-4,-5){20}} 
\put(45,30){\line(-4,1){20}} 
\put(45,10){\line(-4,5){20}} 
\put(45,10){\line(-4,-1){20}}

\thicklines 
\dashline[50]{1}(45,80)(25,85) 
\dashline[50]{1}(45,80)(20,60) 
\dashline[50]{1}(45,60)(30,60) 
\dashline[50]{1}(45,60)(25,45) 
\dashline[50]{1}(45,60)(25,25) 
\dashline[50]{1}(45,30)(20,60) 
\dashline[50]{1}(45,30)(25,15) 
\dashline[50]{1}(45,10)(25,25) 
\dashline[50]{1}(45,10)(25,15) 
 
\thinlines 
\dashline[50]{1}(5,80)(25,75) 
\dashline[50]{1}(5,60)(25,75) 
\dashline[50]{1}(5,60)(25,35) 
\dashline[50]{1}(5,30)(25,35) 
\dashline[50]{1}(5,10)(25,35) 
\dashline[50]{1}(5,10)(25,5)

\scriptsize 
\put(25,88){\makebox(0,0){$\ud{0}$}} 
\put(25,78){\makebox(0,0){$(\ud{1})$}} 
\put(25,49.5){\makebox(0,0){$\ud{8^{'}}$}} 
\put(25,39.5){\makebox(0,0){$(\ud{3})$}} 
\put(25,28){\makebox(0,0){$\ud{6}$}} 
\put(25,18){\makebox(0,0){$\ud{4}$}} 
\put(25,8){\makebox(0,0){$(\ud{5})$}} 
 
\put(1,10){\makebox(0,0){$\ud{5}$}} 
\put(1,30){\makebox(0,0){$\ud{3}$}} 
\put(1,60){\makebox(0,0){$\ud{7}$}} 
\put(1,80){\makebox(0,0){$\ud{1}$}} 
 
\put(49,10){\makebox(0,0){$(\ud{6})$}} 
\put(49,30){\makebox(0,0){$(\ud{4})$}} 
\put(49,60){\makebox(0,0){$(\ud{2})$}} 
\put(49,80){\makebox(0,0){$(\ud{0})$}} 
 
\put(31,63){\makebox(0,0){$\ud{2}$}} 
\put(19,63){\makebox(0,0){$\ud{8}$}} 
 
\normalsize 
\end{picture} 
\caption{Le graphe d'Ocneanu de $E_7$} 
\label{grOc_E7} 
\end{center} 
\end{figure}

La table de multiplication compl\`ete de l'alg\`ebre $Oc(E_7)$ poss\`ede la structure suivante: 
$$ 
\begin{array}{cccccc} 
``D_{10}" \times ``D_{10}" &\rightarrow& ``D_{10}" \qquad \qquad &``E_{7}"\times ``D_{10}"&\rightarrow& ``E_{7}"\\   
``D_{10}" \times ``E_{7}" &\rightarrow& ``E_7" \qquad \qquad & ``E_{7}" \times ``E_{7}" &\rightarrow& ``D_{10}"   
\end{array} 
$$  
Appelons $\ud{a}, \ud{b},\ud{c}$ des \'el\'ements de la partie ``$D_{10}$'', et $(\ud{x}),(\ud{y})$ des \'el\'ements 
de la partie ``$E_7$''.  
La structure multiplicative de la partie $``D_{10}" \times ``D_{10}" \rightarrow ``D_{10}"$ est cod\'ee 
par les matrices de fusion $G^{D_{10}}$ de $D_{10}$, et la structure compl\`ete est donn\'ee par: 
$$ 
\begin{array}{cccccc} 
\ud{a} . \ud{b}     &=& \displaystyle \sum_{c} (G_a^{D_{10}})_{bc}\; \ud{c} \qquad \qquad &  
(\ud{x}) . \ud{a}   &=& \displaystyle \sum_{y} (s'_x)_{ay}\; (\ud{y})\\   
\ud{a} . (\ud{x})   &=& \displaystyle \sum_{y} (s_a)_{xy}\; (\ud{y})  \qquad \qquad &  
(\ud{x}) . (\ud{y}) &=& \displaystyle \sum_{a} (s''_x)_{ya}\; \ud{a}\\   
\end{array} 
$$  
o\`u les 10 matrices $7 \times 7$ $s_a$ codent l'action de $D_{10}$ sur $E_7$, et sont des combinaisons lin\'eaires des matrices de fusion  
de $E_7$\footnote{Bien que les matrices de fusion $G$ de $E_7$ 
poss\`edent des coefficients n\'egatifs, les matrices $s_a$ d\'efinies comme des combinaisons lin\'eaires de telles  
matrices sont \`a coefficients entiers non-n\'egatifs.}: 
$$ 
\begin{array}{lclclclcl} 
s_0 = G_0 &\quad& s_2 = G_2 &\quad& s_4 = G_4 +G_6 &\quad& s_6 = G_6 + G_2 &\quad& s_8 = G_0 + G_4 \\  
s_1 = G_1 &\quad& s_3 = G_3 &\quad& s_5 = G_5 +G_3 &\quad& s_7 = G_3+G_1 &\quad& s_9 = G_2 \\  
\end{array} 
$$ 
Les matrices $s'_x$ et $s''_x$ sont des matrices rectangulaires $10 \times 7$ et $7 \times 10$ d\'efinies par: 
$$ 
(s'_x)_{ay} = (s''_x)_{ya} = (s_a)_{xy} \qquad \qquad s''=s^T 
$$ 
La structure multiplicative compl\`ete de $Oc(E_7)$ est cod\'ee par les matrices $O_x$, qui,  dans la base 
$\{x\} = \{\ud{a},(\ud{y})\}, x = 0,2,\cdots 16$, sont donn\'ees par: 
$$ 
O_x = \left\lbrace 
\begin{array}{cc} 
\left( \begin{array}{c|c} G_x^{D_{10}} & . \\ \hline  . & s_x \end{array} \right)  &  
\mathrm{pour \ } x= 0,1,\cdots , 9 \\ 
{ } & { } \\ 
\left( \begin{array}{c|c} . & s'_{x-10} \\ \hline  (s'_{x-10})^T & . \end{array} \right)  & 
\mathrm{pour \ } x=10,11,\cdots,16  \\ 
\end{array} 
\right. 
$$ 
Les matrices $s_a$ d\'efinissent l'action de $D_{10}$ sur $E_7$. L'action d'un \'el\'ement  
de $Oc(E_7)$ sur $E_7$ est obtenue en utilisant notre r\'ealisation alg\'ebrique de $Oc(E_7)$. 
Pour un \'el\'ement de $Oc(E_7)$ de la forme $x = \sigma_a \otimesdot \sigma_b$, avec 
$\sigma_a, \sigma_b \in D_{10}$, alors: 
\begin{equation} 
x . \sigma_c^{E_7} = \sum_{d \in E_7} (S_x)_{cd} \; \sigma_d^{E_7} \qquad \qquad S_x = s_a . s_b 
\end{equation}

\paragraph{Dimension des blocs} 
La dimension $d_i$ des blocs de la big\`ebre $\mathcal{B}E_7$ pour la loi de composition est donn\'ee par  
$d_i = \sum_{a,b} (F_i^{E_7})_{ab}$ pour $a,b \in E_7$ et $i \in A_{17}$. Nous avons explicitement: 
$$ 
d_i : (7,12,17,22,27,30,33,34,35,34,33,30,27,22,17,12,7) 
$$ 
La dimension des blocs de $\mathcal{B}(E_7)$ pour la loi de convolution est obtenue en sommant les  
\'el\'ements des matrices $S_x$, pour $x \in Oc(E_7)$: 
$$ 
d_x : (7,12,17,22,27,30,33,34,18,17,12,24,34,44,30,16,22) 
$$ 
Les r\`egles de somme lin\'eaire et quadratique sont v\'erifi\'ees: 
\begin{equation} 
\begin{array}{rclcl} 
\dim (\mathcal{B}(E_7)) = \displaystyle \sum_{i \in A_{17}} d_i^2 &=& \displaystyle \sum_{x \in Oc(E_7)} d_x^2 &=& 10 905\\  
\displaystyle \sum_{i} (d_i) &=& \displaystyle \;\; \sum_{x} d_x &=& 399 
\end{array} 
\end{equation} 
La relation de masse quantique entre $Oc(E_7)$ et $\mathcal{A}(E_7)=A_{17}$ est satisfaite: 
\begin{equation} 
m(Oc(E_7)) = \frac{m(D_{10}).m(D_{10})}{m(J)} = m(A_{17}) 
\label{masseE7} 
\end{equation} 
o\`u $m(J) = qdim^2(\sigma_0)+qdim^2(\sigma_2)+qdim^2(\sigma_4)+qdim^2(\sigma_6)+qdim^2(\sigma_8)+qdim^2(\sigma_{8'})$ 
est la masse quantique de la sous-alg\`ebre $J$ de $D_{10}$. La masse quantique de $Oc(7)$ est d\'efinie dans  
(\ref{masseE7}) en fonction de la r\'ealisation $Oc(E_7) = D_{10} \otimes_{\rho,J} D_{10}$. Notons que nous avons  
aussi: 
$$ 
\frac{m(D_{10}) . m(E_7)}{m(J')} = m(A_{17}) 
$$ 
o\`u $m(J') = qdim^2(\sigma_1)+qdim^2(\sigma_3)+qdim^2(\sigma_5)$, avec $\{\sigma_1,\sigma_3,\sigma_5\} \in E_7$. 
Cette derni\`ere relation nous sugg\`ere que l'alg\`ebre d'Ocneanu de $E_7$ puisse aussi \^etre r\'ealis\'ee 
par $D_{10} \otimes_{J'} E_7$, ce qui expliquerait mieux pourquoi $E_7$ aparaisse comme 
un quotient de $Oc(E_7)$.

\paragraph{Matrices toriques et fonctions de partition g\'en\'eralis\'ees} 
L'action de $A_{17}$ (\`a gauche et \`a droite) sur $Oc(E_7)$ est calcul\'ee -- utilisant la r\'ealisation  
$Oc(E_7) = D_{10} \otimes_{\rho} D_{10}$-- \`a travers l'action de $A_{17}$ sur $D_{10}$, cod\'ee par les  
matrices $F_i^{D_{10}}$. Pour un \'el\'ement $x = \sigma_a \otimesdot \sigma_b$ de $Oc(E_7)$: 
\begin{equation} 
\tau_i . x . \tau_j = \tau_i . (\sigma_a \otimesdot \sigma_b) . \tau_j =  
\sum_{c \in D_{10}} \sum_{d \in D_{10}} (F_i)_{ac} (F_j)_{bd} \; (\sigma_c \otimesdot \sigma_d) 
\end{equation} 
Il faut alors projeter les \'el\'ements $\sigma_c \otimesdot \sigma_d$ sur la base de $Oc(E_7)$, d'apr\`es 
les identifications (\ref{identif1}) et (\ref{identif2}). Pour obtenir l'expression de la  matrice $W_{xy}$,  
il faut calculer les termes proportionnels \`a l' 
\'el\'ement $y$ apres la projection. Il est difficile  
d'\'ecrire une formule g\'en\'erale pour la matrice $W_{xy}$, cependant, une fois le $y$ choisi, le calcul 
est simple \`a travers les identifications. Nous avons par exemple: 
\begin{eqnarray*} 
\tau_i . (\sigma_a \otimesdot \sigma_b) . \tau_j  &=& \sum_{c} \sum_{d \in J} (F_i)_{ac} (F_j)_{bd} \; 
(\sigma_c . \rho(\sigma_d) \otimesdot \sigma_0) + \sum_{c} \sum_{d \notin J} \cdots \\ 
{ } &=& \sum_{c} \sum_{d \in J} \sum_e (F_i)_{ac} (F_j)_{bd} (G_c)_{\rho(d)e} \;(\sigma_e \otimesdot \sigma_0)  
+ \sum_{c} \sum_{d \notin J} \cdots  
\end{eqnarray*} 
et les matrices toriques $W_x = W_{x0}$ et les matrices toriques g\'en\'eralis\'ees sont alors donn\'ees par: 
\begin{equation} 
\begin{array}{|c|} 
\hline 
{ } \\ 
\begin{array}{rcl} 
W_{x} &=& \displaystyle \sum_{c \in J} (F_i)_{ac} (F_j)_{b\rho(c)} \\ 
W_{xy} &=& \displaystyle \sum_{z \in Oc(E_7)} (O_x)_{yz} \; W_z  
\end{array} \\ 
{ } \\ 
\hline 
\end{array} 
\end{equation} 
En particulier, l'invariant modulaire  est \'egal \`a: 
$$ 
\mathcal{M} = W_{00} = \sum_{c \in J} (F_i)_{0c} (F_j)_{0\rho(c)}  
$$ 
Les fonctions de partition g\'en\'eralis\'ees du mod\`ele $E_7$ sont d\'efinies en fonction des caract\`eres  
$\chi_i(q)$ de $\widehat{su}(2)$ par: 
$$ 
\mathcal{Z}_{x|y} = \sum_{i \in A_{17}} \sum_{j \in A_{17}} \chi_i(q) \; (W_{xy})_{ij} \ov{\chi}_j(q) 
$$ 
Les fonctions de partition \`a une ligne de d\'efauts sont publi\'ees dans \cite{Coq_Gil-ADE}. Elles 
s'\'ecrivent de mani\`ere plus compacte en fonction des caract\`eres \'etendus du mod\`ele $D_{10}$: 
$$ 
\mathcal{Z}_{x|0} = \sum_{c \in J} \hat{\chi}_{ac} \ov{\hat{\chi}}_{b\rho(c)} 
$$ 
avec: 
$$ 
\hat{\chi}_{ab} = \sum_{c \in D_{10}} (G_a^{D_{10}})_{cb} \hat{\chi}_{c} \qquad \qquad  
\hat{\chi}_c = \sum_{i \in A_{17}} \chi_i  
$$ 
La fonction de partition invariante modulaire est: 
\begin{eqnarray*} 
\mathcal{Z}_{E_7} &=& \hxa{0} + \hxa{4} +  \hxa{6} + \hxa{8'} + (\hch{2}.\hoch{8} + \textrm{h.c.}) \\ 
{ }               &=& \xaa{0}{16} + \xaa{4}{12} + \xaa{6}{10} + \xa{8} + [(\chi_2 + \chi_{14}).\ov{\chi}_8 + \textrm{h.c.}]  
\end{eqnarray*}



\section{Quelques exemples du cas $su(3)$}
Les diagrammes de Coxeter-Dynkin des cas $su(3)$ sont publi\'es dans 
\cite{Oc-Bariloche}. Nous utilisons les m\^emes m\'ethodes que celles
introduites dans le cas $su(2)$ pour construir l'alg\`ebre d'Ocneanu des 
trois cas exceptionnels poss\'edant {\it self-fusion} du syst\`eme $su(3)$, 
et ainsi obtenir les fonctions de partition g\'en\'eralis\'ees associ\'ees.


\subsection{Le cas $\mathcal{E}_5$} 
 
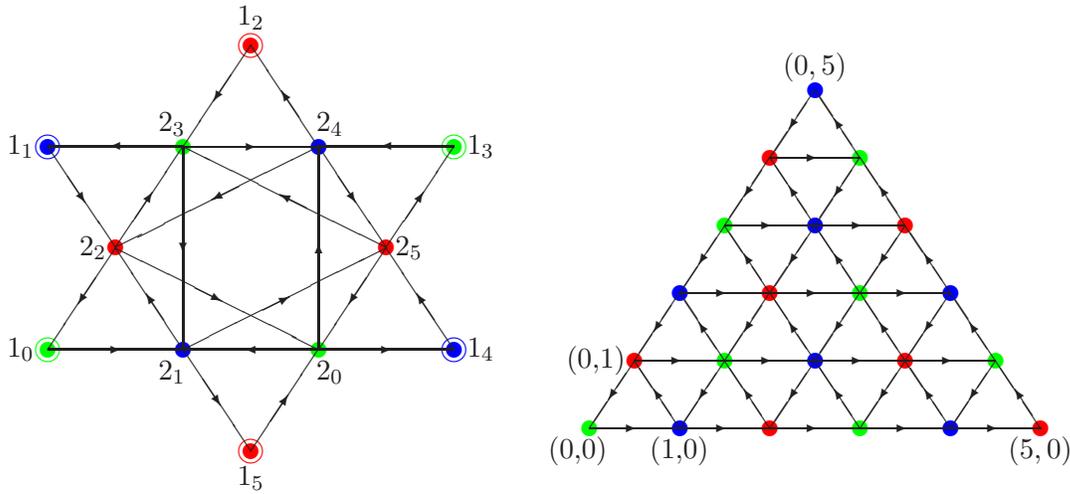
\begin{figure}[hhh] 
\begin{center} 
\unitlength 0.30mm 
 
\begin{picture}(440,200)(0,-10) 
 
\put(0,45){\begin{picture}(60,45) 
\put(0,0){\color{green} \circle*{7}} 
\put(0,0){\color{green} \circle{10}} 
\put(60,0){\color{blue} \circle*{7}} 
\put(30,45){\color{red} \circle*{7}} 
\put(0,0){\vector(1,0){32.5}} 
\put(30,0){\line(1,0){30}} 
\put(60,0){\vector(-2,3){16.5}} 
\put(45,22.5){\line(-2,3){15}} 
\put(30,45){\vector(-2,-3){16.5}} 
\put(15,22.5){\line(-2,-3){15}} 
\end{picture}} 
 
\put(120,45){\begin{picture}(60,45) 
\put(0,0){\color{green} \circle*{7}} 
\put(60,0){\color{blue} \circle{10}} 
\put(60,0){\color{blue} \circle*{7}} 
\put(30,45){\color{red} \circle*{7}} 
\put(0,0){\vector(1,0){32.5}} 
\put(30,0){\line(1,0){30}} 
\put(60,0){\vector(-2,3){16.5}} 
\put(45,22.5){\line(-2,3){15}} 
\put(30,45){\vector(-2,-3){16.5}} 
\put(15,22.5){\line(-2,-3){15}} 
\end{picture}} 
 
\put(60,135){\begin{picture}(60,45) 
\put(0,0){\color{green} \circle*{7}} 
\put(60,0){\color{blue} \circle*{7}} 
\put(30,45){\color{red} \circle*{7}} 
\put(30,45){\color{red} \circle{10}} 
\put(0,0){\vector(1,0){32.5}} 
\put(30,0){\line(1,0){30}} 
\put(60,0){\vector(-2,3){16.5}} 
\put(45,22.5){\line(-2,3){15}} 
\put(30,45){\vector(-2,-3){16.5}} 
\put(15,22.5){\line(-2,-3){15}} 
\end{picture}}

\put(0,90){\begin{picture}(60,45) 
\put(0,45){\color{blue} \circle*{7}} 
\put(0,45){\color{blue} \circle{10}} 
\put(60,45){\vector(-1,0){32.5}} 
\put(30,45){\line(-1,0){30}} 
\put(30,0){\vector(2,3){16.5}} 
\put(45,22.5){\line(2,3){15}} 
\put(0,45){\vector(2,-3){16.5}} 
\put(15,22.5){\line(2,-3){15}} 
\end{picture}} 
 
\put(120,90){\begin{picture}(60,45) 
\put(60,45){\color{green} \circle*{7}} 
\put(60,45){\color{green} \circle{10}} 
\put(60,45){\vector(-1,0){32.5}} 
\put(30,45){\line(-1,0){30}} 
\put(30,0){\vector(2,3){16.5}} 
\put(45,22.5){\line(2,3){15}} 
\put(0,45){\vector(2,-3){16.5}} 
\put(15,22.5){\line(2,-3){15}} 
\end{picture}} 
 
\put(60,0){\begin{picture}(60,45) 
\put(30,0){\color{red} \circle*{7}} 
\put(30,0){\color{red} \circle{10}} 
\put(60,45){\vector(-1,0){32.5}} 
\put(30,45){\line(-1,0){30}} 
\put(30,0){\vector(2,3){16.5}} 
\put(45,22.5){\line(2,3){15}} 
\put(0,45){\vector(2,-3){16.5}} 
\put(15,22.5){\line(2,-3){15}} 
\end{picture}} 
 
\put(60,135){\vector(0,-1){47.5}} 
\put(60,90){\line(0,-1){45}} 
\put(60,45){\vector(2,1){47.2}} 
\put(105,67.5){\line(2,1){45}} 
\put(150,90){\vector(-2,1){47.2}} 
\put(105,112.5){\line(-2,1){45}} 
 
\put(120,45){\vector(0,1){47.5}} 
\put(120,90){\line(0,1){45}} 
\put(120,135){\vector(-2,-1){47.2}} 
\put(75,112.5){\line(-2,-1){45}} 
\put(30,90){\vector(2,-1){47.2}} 
\put(75,67.5){\line(2,-1){45}}

\put(240,10){\begin{picture}(200,180) 
\put(0,0){\begin{picture}(40,40) 
\put(0,0){\color{green} \circle*{7}} 
\put(40,0){\color{blue} \circle*{7}} 
\put(20,30){\color{red} \circle*{7}} 
\put(0,0){\vector(1,0){21}} 
\put(20,0){\line(1,0){20}} 
\put(40,0){\vector(-2,3){11.5}} 
\put(30,15){\line(-2,3){10}} 
\put(20,30){\vector(-2,-3){11.5}} 
\put(10,15){\line(-2,-3){10}}\end{picture}}

\put(40,0){\begin{picture}(40,40) 
\put(40,0){\color{red} \circle*{7}} 
\put(20,30){\color{green} \circle*{7}} 
\put(0,0){\vector(1,0){21}} 
\put(20,0){\line(1,0){20}} 
\put(40,0){\vector(-2,3){11.5}} 
\put(30,15){\line(-2,3){10}} 
\put(20,30){\vector(-2,-3){11.5}} 
\put(10,15){\line(-2,-3){10}}\end{picture}} 
 
\put(80,0){\begin{picture}(40,40) 
\put(40,0){\color{green} \circle*{7}} 
\put(20,30){\color{blue} \circle*{7}} 
\put(0,0){\vector(1,0){21}} 
\put(20,0){\line(1,0){20}} 
\put(40,0){\vector(-2,3){11.5}} 
\put(30,15){\line(-2,3){10}} 
\put(20,30){\vector(-2,-3){11.5}} 
\put(10,15){\line(-2,-3){10}}\end{picture}}

\put(120,0){\begin{picture}(40,40) 
\put(40,0){\color{blue} \circle*{7}} 
\put(20,30){\color{red} \circle*{7}} 
\put(0,0){\vector(1,0){21}} 
\put(20,0){\line(1,0){20}} 
\put(40,0){\vector(-2,3){11.5}} 
\put(30,15){\line(-2,3){10}} 
\put(20,30){\vector(-2,-3){11.5}} 
\put(10,15){\line(-2,-3){10}}\end{picture}}

\put(160,0){\begin{picture}(40,40) 
\put(40,0){\color{red} \circle*{7}} 
\put(20,30){\color{green} \circle*{7}} 
\put(0,0){\vector(1,0){21}} 
\put(20,0){\line(1,0){20}} 
\put(40,0){\vector(-2,3){11.5}} 
\put(30,15){\line(-2,3){10}} 
\put(20,30){\vector(-2,-3){11.5}} 
\put(10,15){\line(-2,-3){10}}\end{picture}}

\put(20,30){\begin{picture}(40,40) 
\put(20,30){\color{blue} \circle*{7}} 
\put(0,0){\vector(1,0){21}} 
\put(20,0){\line(1,0){20}} 
\put(40,0){\vector(-2,3){11.5}} 
\put(30,15){\line(-2,3){10}} 
\put(20,30){\vector(-2,-3){11.5}} 
\put(10,15){\line(-2,-3){10}}\end{picture}} 
 
\put(60,30){\begin{picture}(40,40) 
\put(20,30){\color{red} \circle*{7}} 
\put(0,0){\vector(1,0){21}} 
\put(20,0){\line(1,0){20}} 
\put(40,0){\vector(-2,3){11.5}} 
\put(30,15){\line(-2,3){10}} 
\put(20,30){\vector(-2,-3){11.5}} 
\put(10,15){\line(-2,-3){10}}\end{picture}} 
 
\put(100,30){\begin{picture}(40,40) 
\put(20,30){\color{green} \circle*{7}} 
\put(0,0){\vector(1,0){21}} 
\put(20,0){\line(1,0){20}} 
 
\put(40,0){\vector(-2,3){11.5}} 
\put(30,15){\line(-2,3){10}} 
\put(20,30){\vector(-2,-3){11.5}} 
\put(10,15){\line(-2,-3){10}}\end{picture}}

\put(140,30){\begin{picture}(40,40) 
\put(20,30){\color{blue} \circle*{7}} 
\put(0,0){\vector(1,0){21}} 
\put(20,0){\line(1,0){20}} 
\put(40,0){\vector(-2,3){11.5}} 
\put(30,15){\line(-2,3){10}} 
\put(20,30){\vector(-2,-3){11.5}} 
\put(10,15){\line(-2,-3){10}}\end{picture}}

\put(40,60){\begin{picture}(40,40) 
\put(20,30){\color{green} \circle*{7}} 
\put(0,0){\vector(1,0){21}} 
\put(20,0){\line(1,0){20}} 
\put(40,0){\vector(-2,3){11.5}} 
\put(30,15){\line(-2,3){10}} 
\put(20,30){\vector(-2,-3){11.5}} 
\put(10,15){\line(-2,-3){10}}\end{picture}} 
 
\put(80,60){\begin{picture}(40,40) 
\put(20,30){\color{blue} \circle*{7}} 
\put(0,0){\vector(1,0){21}} 
\put(20,0){\line(1,0){20}} 
\put(40,0){\vector(-2,3){11.5}} 
\put(30,15){\line(-2,3){10}} 
\put(20,30){\vector(-2,-3){11.5}} 
\put(10,15){\line(-2,-3){10}}\end{picture}}

\put(120,60){\begin{picture}(40,40) 
\put(20,30){\color{red} \circle*{7}} 
\put(0,0){\vector(1,0){21}} 
\put(20,0){\line(1,0){20}} 
\put(40,0){\vector(-2,3){11.5}} 
\put(30,15){\line(-2,3){10}} 
\put(20,30){\vector(-2,-3){11.5}} 
\put(10,15){\line(-2,-3){10}}\end{picture}}

\put(60,90){\begin{picture}(40,40) 
\put(20,30){\color{red} \circle*{7}} 
\put(0,0){\vector(1,0){21}} 
\put(20,0){\line(1,0){20}} 
\put(40,0){\vector(-2,3){11.5}} 
\put(30,15){\line(-2,3){10}} 
\put(20,30){\vector(-2,-3){11.5}} 
\put(10,15){\line(-2,-3){10}}\end{picture}}

\put(100,90){\begin{picture}(40,40) 
\put(20,30){\color{green} \circle*{7}} 
\put(0,0){\vector(1,0){21}} 
\put(20,0){\line(1,0){20}} 
\put(40,0){\vector(-2,3){11.5}} 
\put(30,15){\line(-2,3){10}} 
\put(20,30){\vector(-2,-3){11.5}} 
\put(10,15){\line(-2,-3){10}}\end{picture}}

\put(80,120){\begin{picture}(40,40) 
\put(20,30){\color{blue} \circle*{7}} 
\put(0,0){\vector(1,0){21}} 
\put(20,0){\line(1,0){20}} 
\put(40,0){\vector(-2,3){11.5}} 
\put(30,15){\line(-2,3){10}} 
\put(20,30){\vector(-2,-3){11.5}} 
\put(10,15){\line(-2,-3){10}}\end{picture}} 
 
\put(-5,-10){\makebox(0,0){(0,0)}} 
\put(40,-10){\makebox(0,0){(1,0)}} 
\put(3,30){\makebox(0,0){(0,1)}} 
\put(200,-10){\makebox(0,0){$(5,0)$}} 
\put(100,160){\makebox(0,0){$(0,5)$}} 
\end{picture}}

\put(-12,45){\makebox(0,0){$1_0$}} 
\put(192,45){\makebox(0,0){$1_4$}} 
\put(-12,135){\makebox(0,0){$1_1$}} 
\put(192,135){\makebox(0,0){$1_3$}} 
\put(90,-12){\makebox(0,0){$1_5$}} 
\put(90,192){\makebox(0,0){$1_2$}} 
\put(55,35){\makebox(0,0){$2_1$}} 
 
\put(125,35){\makebox(0,0){$2_0$}} 
\put(55,145){\makebox(0,0){$2_3$}} 
\put(125,145){\makebox(0,0){$2_4$}} 
\put(20,90){\makebox(0,0){$2_2$}} 
\put(160,90){\makebox(0,0){$2_5$}} 
 
\end{picture} 
\label{gr_E5} 
\caption{Les diagrammes de Coxeter-Dynkin g\'en\'eralis\'es ${\cal E}_5$ et ${\cal A}_5$.}

\end{center} 
\end{figure}

\paragraph{Graphe et matrice de fusion} 
Le graphe ${\cal E}_5$ est illustr\'e \`a la Fig. \ref{gr_E5}. Le niveau est $\ell = 5$, l'altitude   
$\kappa = l+3=8$ et la norme $\beta = 1+2\cos(2  \pi /8) = 1 + \sqrt2$. L'identit\'e est $1_0$, et la 
fondamentale \`a laquelle correspond le graphe est $2_1$ (la fondamentale conjugu\'ee est $2_2$). La conjuguaison 
correspond \`a la sym\'etrie par rapport \`a l'axe qui passe par les vertex $1_0$ et $1_3$ (seules irreps r\'eelles): 
$1_0^*=1_0$,\mbox{$1_1^*=1_5$},\mbox{$1_2^*=1_4$},\mbox{$1_3^*=1_3$} ; $2_0^*=2_3,2_1^*=2_2,2_4^*=2_5$.

Le graphe $\mathcal{E}_5$ d\'etermine de mani\`ere unique son alg\`ebre de graphe, dont la table de 
multiplication commutative est donn\'ee par: 
\begin{equation} 
\begin{array}{rcl} 
1_i . 1_j & = & 1_{i+j}    \\ 
1_i . 2_j = 2_i . 1_j & = & 2_{i+j}   \\ 
2_i . 2_j & = & 2_{i+j} + 2_{i+j-3} + 1_{i+j-3} 
\end{array} 
\label{mult_E5} 
\end{equation} 
o\`u les indices $i$ et $j$ varient de 0 \`a 5$\mod 6$. Clairement le sous-ensemble $1_i$ forme une  
sous-alg\`ebre de $\mathcal{E}_5$. La matrice de fusion correspondant \`a la fondamentale $2_1$ 
est la matrice d'adjacence du graphe\footnote{Dans le livre \cite{FMS-book}, la matrice d'adjacence  
de $\mathcal{E}_5$ est incorrecte: la discussion subs\'equente n'est donc pas \`a prendre en compte.}.  
La conjuguaison se traduit au niveau matriciel par la transposition: $G_{a^*} = G_a^T$.  
\`A partir de la table de multiplication (\ref{mult_E5}), il est imm\'ediat de calculer 
les matrices de fusion $G_a$ associ\'ees aux vertex $a \in \mathcal{E}_5$.

\paragraph{Induction-restriction} 
Le graphe de la s\'erie $\mathcal{A}$ 
de m\^eme norme que $\mathcal{E}_5$ est $\mathcal{A}_5$, illustr\'e aussi \`a la Fig. \ref{gr_E5}, \`a partir de laquelle nous obtenons 
les matrices de fusion $N_{1,0}$ de la repr\'esentation fondamentale $(1,0)$, et de 
la fondamentale conjugu\'ee: \mbox{$N_{0,1} = N_{1,0}^T$.}   
Les autres matrices de fusion $N_i$ et les matrices $F_i$ codant l'action de $\mathcal{A}_5$ sur $\mathcal{E}_5$  
(pour $i = (\lambda_1,\lambda_2) \in \mathcal{A}_5$) sont 
d\'etermin\'ees par la formule de r\'ecurrence tronqu\'ee de $SU(3)$ (avec $F_{1,0} = G_{2_1}$).  
La matrice essentielle $E_{1_0}$ s'obtient par $(E_{1_0})_{ia} = (F_i)_{1_0 a}$: elle contient 21 lignes labell\'ees 
par les vertex $i \in \mathcal{A}_5$ et 12 colonnes labell\'ees par les vertex $a \in \mathcal{E}_5$. De cette 
matrice, nous lisons les r\`egles de branchement $\mathcal{A}_5 \hookrightarrow \mathcal{E}_5$ ainsi que 
les r\`egles d'induction, donn\'ees par: 
$$ 
\begin{array}{ccc} 
1_0 \hookleftarrow (0,0) , (2,2) &\qquad \qquad \qquad&  
2_0 \hookleftarrow (1,1) , (3,0) , (2,2) , (1,4) \\ 
1_1 \hookleftarrow (0,2) , (3,2) &\qquad&  
2_1 \hookleftarrow (1,0) , (2,1) , (1,3) , (3,2) \\ 
1_2 \hookleftarrow (1,2) , (5,0) &\qquad&  
2_2 \hookleftarrow (0,1) , (1,2) , (3,1) , (2,3) \\ 
1_3 \hookleftarrow (3,0) , (0,3) &\qquad&  
2_3 \hookleftarrow (1,1) , (0,3) , (2,2) , (4,1) \\ 
1_4 \hookleftarrow (2,1) , (0,5) &\qquad&  
2_4 \hookleftarrow (0,2) , (2,1) , (4,0) , (1,3) \\ 
1_5 \hookleftarrow (2,0) , (2,3) &\qquad&  
2_5 \hookleftarrow (2,0) , (1,2) , (3,1) , (0,4)  
\end{array} 
$$ 
La valeur de $\hat{T}$ sur les vertex $i = (\lambda_1, \lambda_2) \in \mathcal{A}_5$ est donn\'ee dans la Tab. \ref{T_A5} 
 
\begin{table}[hhh] 
\small 
$$ 
\begin{array}{|c||c|c|c|c|c|c|c|c|c|c|c|c|} 
\hline 
(\lambda_1,\lambda_2) & (0,0) & (1,0) &(2,0) &(3,0) &(4,0) &(5,0) 
&(1,1) &(2,1) &(3,1) &(4,1) &(2,2) &(3,2) \\ 
                        &  {}   & (0,1) &(0,2) &(0,3) &(0,4) &(0,5) & 
{}   &(1,2) &(1,3) &(1,4) & {}   &(2,3) \\ 
\hline 
\hline 
\hat{T}  & 5 & 1 & 19 & 11 & 1 & 13 & 20 & 13 & 4 & 17 & 5 & 19 \\ 
\hline 
\end{array} 
$$ 
\label{T_A5} 
\caption{Valeur de $\hat{T}$ sur les vertex $(\lambda_1, \lambda_2)$ du graphe $\mathcal{A}_5$.} 
\normalsize 
\end{table}

Nous pouvons v\'erifier que les valeurs de $\hat{T}$ sur les vertex $(\lambda_1,\lambda_2)$ dont la restriction 
donne un vertex $1_i$ sont \'egales: ceci permet de d\'efinir une valeur fixe de $\hat{T}$ aux vertex du type $1_i$.  
Par contre, pour les vertex de type $2_i$, il n'est pas possible de d\'efinir une valeur fixe de $\hat{T}$.  
Ceci nous donne une caract\'erisation de la sous-alg\`ebre $J$, engendr\'ee par les \'el\'ements $1_i$.

\begin{figure} 
\begin{center} 
\includegraphics*{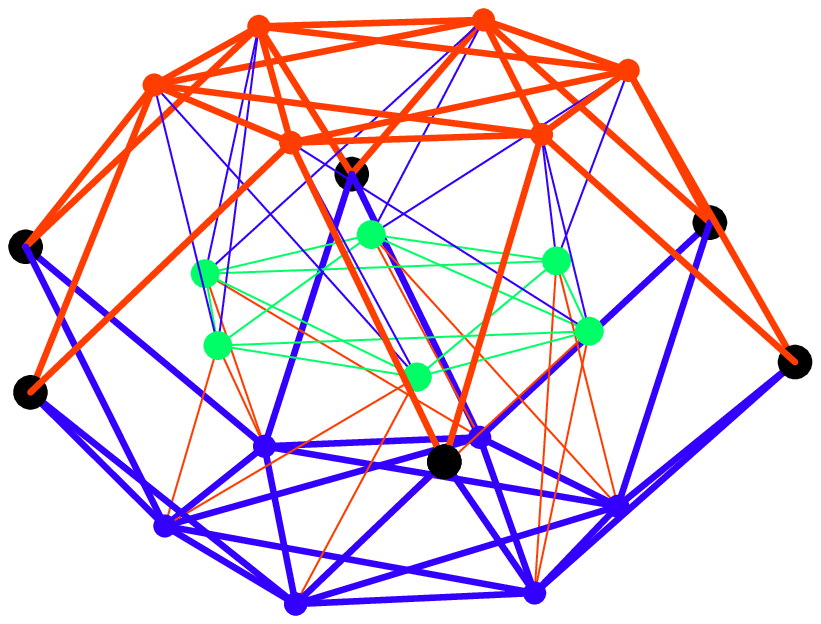} 
\caption{Graphe d'Ocneanu de ${\cal E}_{5}$} 
\label{OcGraphE5} 
\end{center} 
\end{figure}

\paragraph{Alg\`ebre d'Ocneanu} 
Nous conjecturons alors que l'alg\`ebre d'Ocneanu de $\mathcal{E}_5$ est d\'efinie par: 
\begin{equation} 
Oc(\mathcal{E}_5) = \mathcal{E}_5 \otimes_J \mathcal{E}_5 = \mathcal{E}_5 \otimesdot \mathcal{E}_5, 
\end{equation} 
o\`u nous identifions les \'el\'ements $(a \otimesdot b.c)$ avec $(a.b^* \otimesdot c)$, 
pour $b \in J = \{1_i\}$. 
L'alg\`ebre $Oc(\mathcal{E}_5)$ est de dimension $12 \times 2 = 24$, et une base est donn\'ee par les 
\'el\'ements $a \otimesdot 1_0$ et $a \otimesdot 2_0$. 
Les identifications dans $Oc(\mathcal{E}_5)$ sont donn\'ees par: 
$$ 
\begin{array}{rccclr} 
1_i \otimesdot 1_j &=& 1_{i+j^*} \otimesdot 1_0 &=& 1_0 \otimesdot 1_{i^*+j}  & A \\ 
2_i \otimesdot 1_j &=& 2_{i+j^*} \otimesdot 1_0 &=& 2_0 \otimesdot 1_{i^*+j}  & L\\ 
1_i \otimesdot 2_j  =  1_i \otimesdot 1_j . 2_0 &=& 1_{i+j^*} \otimesdot 2_0 &=& 1_0 \otimesdot 2_{i^*+j} \qquad \qquad & R\\ 
2_i \otimesdot 2_j  =  2_i \otimesdot 1_j . 2_0 &=& 2_{i+j^*} \otimesdot 2_0 &=& 2_0 \otimesdot 2_{i^*+j} & C 
\end{array} 
$$ 
La multiplication dans $Oc(\mathcal{E}_5)$ 
est d\'efinie \`a travers celle de $\mathcal{E}_5$ et les identifications pr\'ec\'edentes. 
La sous-alg\`ebre chirale gauche  
est engendr\'ee par $L = \{ a \otimesdot 1_0\}$; la sous-alg\`ebre chirale droite est engendr\'ee par  
$R = \{1_0 \otimesdot a\}$. 
Ces deux sous-alg\`ebres sont de dimension 12. Leur intersection est la sous-alg\`ebre 
ambichirale $A = \{1_i \otimesdot 1_0 = 1_0 \otimesdot 1_{i^*}\}$ de dimension 6. Le sous-espace suppl\'ementaire 
est engendr\'e par $C = \{2_i \otimesdot 2_0 = 2_{0} \otimesdot 2_{i^*}\}$, aussi de dimension 6. 
Le graphe d'Ocneanu (de multiplication par les g\'en\'erateurs gauche $2_1 \otimesdot 1_0$ et droit 
$1_0 \otimesdot 2_1$) est illustr\'e \`a la Fig. \ref{OcGraphE5}, comme deux graphes $\mathcal{E}_5$ 
superpos\'es (rouge et bleu), ayant comme vertex communs les points ext\'erieurs ambichiraux (noir), et la  
partie suppl\'ementaire \`a l'int\'erieur (vert). Les lignes bleues et rouges correspondent \`a la  
multiplication par les g\'en\'erateurs chiraux gauche et droit, les lignes vertex proviennent de la 
superposition de lignes bleus et rouges. Le graphe est orient\'e, mais nous n'indiquons pas  
l'orientation des arcs pour ne pas surcharger la figure.

En utilisant les identifications dans $\mathcal{E}_{5}$, nous pourrions alors facilement 
obtenir les matrices $O_x$ ($x \in Oc(\mathcal{E}_5)$) qui codent la multiplication dans $Oc(\mathcal{E}_5)$.   
L'action d'un \'el\'ement \mbox{$x= a \otimesdot 1_0$} (resp. $x= a \otimesdot 2_0$) de $Oc(\mathcal{E}_5)$  sur  
un \'el\'ement $b$ de $\mathcal{E}_5$  
donne l'\'el\'ement $(a.1_0.b) \in \mathcal{E}_5$ (resp. $(a.2_0.b)$). Les matrices $S_x$ qui codent cette  
action sont donc \'egales \`a: 
\begin{equation} 
S_x = \left\lbrace \begin{array}{rclc} 
S_x &=& G_a \qquad \qquad & x=a\otimesdot 1_0 \\ 
S_x &=& G_a.G_{2_0} \qquad \qquad & x=a\otimesdot 2_0 
\end{array} 
\right. 
\end{equation}

\paragraph{Dimensions des blocs} 
La dimension des blocs de la dig\`ebre $\mathcal{B}\mathcal{E}_5$ pour ses deux lois  
multiplicatives est donn\'ee par la somme des \'el\'ements des matrices $F_i$ 
et $S_x$. Les r\`egles de somme lin\'eaire et quadratique sont v\'erifi\'ees: 
$$ 
\sum_{i \in \mathcal{A}_5} d_i =  \sum_{x \in (Oc{\mathcal{E}}_5)} d_x = 720 \qquad \qquad 
\dim (\mathcal{B}\mathcal{E}_5) = \sum_{i} d_i^2 = \displaystyle \sum_{x} d_x^2 = 29376   
$$ 
Les masses quantiques de $\mathcal{E}_5$ et $\mathcal{A}_5$ sont donn\'ees par:  
$m(\mathcal{E}_5) = 12(2+\sqrt{2})$ et $m(\mathcal{A}_5) = 48(3+2\sqrt{2})$. D\'efinissons la masse  
quantique du graphe $Oc(\mathcal{E}_5)$ par $m(\mathcal{E}_5)^2 / m(J)$, o\`u $J = \{1_i\}$ et 
$m(J) = \sum_{i} qdim^2(1_i) = 6$, alors 
la relation de masse quantique est satisfaite: 
$$ 
m(Oc(\mathcal{E}_5)) = \frac{m(\mathcal{E}_5).m(\mathcal{E}_5)}{m(J)} = m(\mathcal{A}_5) = 48(3+2\sqrt{2}) 
$$ 
 
\paragraph{Matrices toriques et fonctions de partition g\'en\'eralis\'ees} 
Les matrices toriques g\'en\'eralis\'ees sont d\'efinies par l'action de $\mathcal{A}_5$ sur $Oc(\mathcal{E}_5)$. 
Pour un \'el\'ement $x \in Oc(\mathcal{E}_5)$ de la forme $a \otimesdot 1_0$, l'action \`a droite et \`a gauche  
d'\'el\'ements $i,j \in \mathcal{A}_5$ est: 
\begin{eqnarray*} 
i . (a \otimesdot 1_0) . j &=& \sum_b \sum_c (F_i)_{ab} (F_j)_{1_0 c}\; (b \otimes c) \\ 
{ } &=& \sum_b \left\lbrack \sum_{c \in J}  (F_i)_{ab} (F_j)_{1_0 c} \; (b.c^* \otimesdot 1_0)  
     +   \sum_{c \notin J}  (F_i)_{ab} (F_j)_{1_0 c} \; (b.\tilde{c}^* \otimesdot 2_0) \right\rbrack \\ 
{ } &=& \sum_b  \sum_{c \in J} \sum_d  (F_i)_{ab} (F_j)_{1_0 c} (G_b)_{c^*d} \; (d \otimesdot 1_0)   \\ 
    &+&  \sum_b  \sum_{c \notin J} \sum_d  (F_i)_{ab} (F_j)_{1_0 c} (G_b)_{\tilde{c}^*d} \; (d \otimesdot 2_0)  
\end{eqnarray*} 
o\`u $\tilde{2}_i = 1_i$. Introduisons les matrices $G'_c$ d\'efinies par: 
$$ 
(G'_c)_{ab} = (G_a)_{bc} \qquad \qquad a.b = \sum_c (G_a)_{bc} \; c = \sum_c (G'_c)_{ab} \; c 
$$ 
Nous avons alors par exemple: 
$$ 
W_{a \otimesdot 1_0,d \otimesdot 1_0} = \sum_b \sum_{c \in J} (F_i)_{ab} (F_j)_{1_0 c} (G'_d)_{bc^*} 
$$  
Les autres matrices $W_{xy}$ se calculent de la m\^eme mani\`ere, et les $24.24=576$ matrices toriques  
g\'en\'eralis\'ees de $\mathcal{E}_5$ s'obtiennent par: 
\begin{equation} 
\begin{array}{|c|} 
\hline 
{ } \\ 
W_{xy} = \left\lbrace 
\begin{array}{lc} 
\displaystyle \sum_b \sum_{c \in J} (F_i)_{ab} (F_j)_{1_0 c} (G'_d)_{bc^*} \qquad \qquad& x = a \otimesdot 1_0 , y = d \otimesdot 1_0 \\ 
\displaystyle \sum_b \sum_{c \notin J} (F_i)_{ab} (F_j)_{1_0 c} (G'_d)_{b\tilde{c}^*} & x = a \otimesdot 1_0 , y = d \otimesdot 2_0 \\ 
\displaystyle \sum_b \sum_{c \in J} (F_i)_{ab} (F_j)_{2_0 c} (G'_d)_{bc^*} & x = a \otimesdot 2_0 , y = d \otimesdot 1_0 \\ 
\displaystyle \sum_b \sum_{c \notin J} (F_i)_{ab} (F_j)_{2_0 c} (G'_d)_{b\tilde{c}^*} & x = a \otimesdot 2_0 , y = d \otimesdot 2_0  
\end{array} 
\right. \\ 
{ } \\ 
\hline 
\end{array} 
\end{equation} 
Pour $d=1_0$, nous avons $(G'_d)_{bc^*} = \delta_{bc}$, alors l'invariant modulaire $\mathcal{M}$ 
est donn\'e par: 
\begin{equation} 
\mathcal{M} = W_{1_0 \otimesdot 1_0, 1_0 \otimesdot 1_0} = \sum_{c \in J} (F_i)_{1_0 c}(F_j)_{1_0 c } 
\label{W_E5} 
\end{equation}

Les fonctions de partition g\'en\'eralis\'ees sont donn\'ees par: 
\begin{equation} 
\mathcal{Z}_{xy} = \sum_{i \in \mathcal{A}_5} \sum_{j \in \mathcal{A}_5} \chi_i (q) (W_{xy})_{ij} \ov{\chi_j}(q)  
\end{equation} 
o\`u les $\chi_i$ sont les caract\`eres de l'alg\`ebre $\widehat{su}(3)$. Introduisons les caract\`eres 
\'etendus $\hat{\chi}_a (q)$ (un pour chaque point du graphe $\mathcal{E}_5$): 
\begin{equation} 
\hat{\chi}_a (q) = \sum_i (F_i)_{1_0 a}\; \chi_i (q) = \sum_i (E_{1_0})_{ia} \;\chi_i (q)  
\end{equation} 
Ils sont enti\`erement d\'etermin\'es en fonction des caract\`eres de $\widehat{su}(3)$ par la connaissance de la matrice essentielle $E_0$: nous les pr\'esentons dans l'Annexe {\bf D}. Introduisons aussi les 
caract\`eres \'etendus g\'en\'eralis\'es: 
\begin{equation} 
\hat{\chi}_{ab} (q) = \sum_c (G_a)_{bc} \hat{\chi}_c (q) = \hat{\chi}_{ba} (q) 
\end{equation} 
qui sont des combinaisons lin\'eaires des caract\`eres \'etendus $\chi_a$. Alors, les fonctions de partition  
g\'en\'eralis\'ees du mod\`ele $\mathcal{E}_5$ sont donn\'ees par: 
\begin{equation} 
\begin{array}{|c|} 
\hline 
{ } \\ 
\mathcal{Z}_{xy} = \left\lbrace 
\begin{array}{lc} 
\displaystyle \sum_b \sum_{c \in J}\; \hat{\chi}_{ab}\; (G'_d)_{bc^*} \; \ov{\hat{\chi}}_{c 1_0}  \qquad \qquad& x = a \otimesdot 1_0 , y = d \otimesdot 1_0 \\ 
\displaystyle \sum_b \sum_{c \notin J}\; \hat{\chi}_{ab}\; (G'_d)_{bc^*} \; \ov{\hat{\chi}}_{c 2_0}  & x = a \otimesdot 1_0 , y = d \otimesdot 2_0 \\ 
\displaystyle \sum_b \sum_{c \in J}\; \hat{\chi}_{ab}\; (G'_d)_{b\tilde{c}^*} \; \ov{\hat{\chi}}_{c 1_0} & x = a \otimesdot 2_0 , y = d \otimesdot 1_0 \\ 
\displaystyle \sum_b \sum_{c \notin J}\; \hat{\chi}_{ab}\; (G'_d)_{b\tilde{c}^*} \; \ov{\hat{\chi}}_{c 2_0}& x = a \otimesdot 2_0 , y = d \otimesdot 2_0  
\end{array} 
\right.\\ 
{ } \\ 
\hline 
\end{array} 
\end{equation} 
Les fonctions de partition \`a une ligne de d\'efauts $\mathcal{Z}_{x} = \mathcal{Z}_{x,1_0 \otimesdot 1_0}$  
sont donn\'ees par: 
\begin{eqnarray} 
\mathcal{Z}_{a \otimesdot 1_0} &=& \sum_{c \in J} \; \hat{\chi}_{ac} \; \ov{\hat{\chi}}_c  
= \sum_{e} \sum_{c \in J} \; (G_a)_{ce} \; \hat{\chi}_{e}\; \ov{\hat{\chi}}_c \\ 
\mathcal{Z}_{a \otimesdot 2_0} &=& \sum_{c \notin J}\; \hat{\chi}_{ac}\; \ov{\hat{\chi}}_c  
= \sum_{e} \sum_{c \notin J}\; (G_a)_{ce}\;  \hat{\chi}_{e} \; \ov{\hat{\chi}}_c  
\end{eqnarray} 
Elles sont explicitement donn\'ees en fonction des caract\`eres \'etendus $\hat{\chi}_a$ dans l'appendice. 
La fonction de partition invariante modulaire, correspondant \`a $1_0 \otimesdot 1_0$, s'\'ecrit: 
\begin{eqnarray*} 
\mathcal{Z}_{\mathcal{E}_5} =  
\sum_{c \in J} |\hat{\chi}_{c}|^2 &=& |\chi_{(0,0)} + \chi_{(2,2)}|^2 + |\chi_{(0,2)} + \chi_{(3,2)}|^2 
 + |\chi_{(2,0)} + \chi_{(2,3)}|^2 \\ 
{ } &+& |\chi_{(2,1)} + \chi_{(0,5)}|^2 + |\chi_{(3,0)} + \chi_{(0,3)}|^2 + |\chi_{(1,2)} + \chi_{(5,0)}|^2 
 \end{eqnarray*} 
et nous pouvons v\'erifier qu'elle correspond \`a la classification de Gannon \cite{gannon-class}.


\subsection{Le cas $\mathcal{E}_{9}$} 
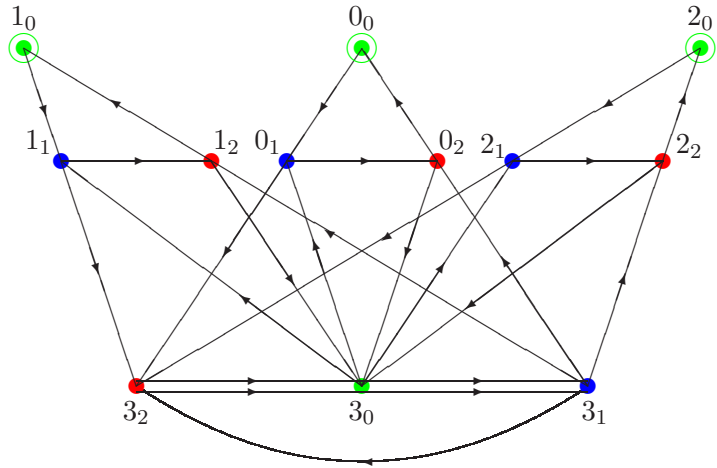
\begin{figure}[hhh] 
\unitlength=0.5mm 
\begin{center} 
 
\begin{picture}(180,130)(0,-20) 
\put(70,60){\begin{picture}(40,30) 
\put(0,0){\color{blue} \circle*{4}} 
\put(40,0){\color{red} \circle*{4}} 
\put(20,30){\color{green} \circle*{4}} 
\put(0,0){\vector(1,0){22.5}} 
\put(20,0){\line(1,0){20}} 
\put(40,0){\vector(-2,3){11.5}} 
\put(30,15){\line(-2,3){10}} 
\put(20,30){\vector(-2,-3){11.5}} 
\put(10,15){\line(-2,-3){10}} 
\end{picture}} 
 
\qbezier[920](30,0)(90,-40)(150,0) 
\put(90,-20){\vector(-1,0){0}} 
 
\put(30,0){\color{red} \circle*{4}} 
\put(90,0){\color{green} \circle*{4}} 
\put(150,0){\color{blue} \circle*{4}} 
\put(10,60){\color{blue} \circle*{4}} 
\put(50,60){\color{red} \circle*{4}} 
\put(130,60){\color{blue} \circle*{4}} 
\put(170,60){\color{red} \circle*{4}} 
\put(0,90){\color{green} \circle*{4}} 
\put(180,90){\color{green} \circle*{4}} 
 
\put(0,90){\color{green} \circle{8}} 
\put(90,90){\color{green} \circle{8}} 
\put(180,90){\color{green} \circle{8}} 
 
\put(0,90){\vector(1,-3){6.0}} 
\put(10,60){\line(-1,3){4.5}} 
\put(10,60){\vector(1,0){22.5}} 
\put(50,60){\line(-1,0){17.5}} 
 
\put(0,90){\line(5,-3){150}} 
\put(23,76.4){\vector(-3,2){0}} 
 
\put(30,0){\line(-1,3){10.8}} 
\put(10,60){\vector(1,-3){10}} 
 
\put(110,60){\vector(-1,-3){8.6}} 
\put(90,0){\line(1,3){15}} 
\put(90,0){\vector(-1,3){13}} 
\put(70,60){\line(1,-3){10}} 
 
\put(170,60){\line(-1,-3){10.8}} 
\put(150,0){\vector(1,3){10}} 
 
\put(180,90){\line(-5,-3){150}} 
\put(154,74){\vector(-3,-2){0}} 
 
\put(130,60){\vector(1,0){22.5}} 
\put(170,60){\line(-1,0){17.5}} 
\put(170,60){\vector(1,3){6.0}} 
\put(180,90){\line(-1,-3){4.5}} 
 
\put(70,60){\line(-2,-3){40}} 
\put(70,60){\vector(-2,-3){17}} 
\put(110,60){\line(2,-3){40}} 
\put(150,0){\vector(-2,3){23}} 
 
\put(30,1.5){\line(1,0){120}} 
\put(30,-1.5){\line(1,0){120}} 
\put(30,1.5){\vector(1,0){32.5}} 
\put(30,-1.5){\vector(1,0){32.5}} 
\put(90,1.5){\vector(1,0){32.5}} 
\put(90,-1.5){\vector(1,0){32.5}} 
 
\put(90,0){\line(-4,3){80}} 
\put(90,0){\vector(-4,3){32.5}} 
\put(90,0){\line(4,3){80}} 
\put(170,60){\vector(-4,-3){52.5}} 
 
\put(50,60){\line(2,-3){40}} 
\put(50,60){\vector(2,-3){22}} 
\put(90,0){\line(2,3){40}} 
\put(90,0){\vector(2,3){22}} 
 
\put(80,41.7){\vector(-3,2){0}} 
\put(95,38.7){\vector(-3,-2){0}} 
 
\put(0,98){\makebox(0,0){$1_0$}} 
\put(90,98){\makebox(0,0){$0_0$}} 
\put(180,98){\makebox(0,0){$2_0$}} 
 
\put(4,65){\makebox(0,0){$1_1$}} 
\put(65,65){\makebox(0,0){$0_1$}} 
\put(125,64){\makebox(0,0){$2_1$}} 
 
\put(54,65){\makebox(0,0){$1_2$}} 
\put(114,65){\makebox(0,0){$0_2$}} 
\put(177,64){\makebox(0,0){$2_2$}} 
 
\put(30,-7){\makebox(0,0){$3_2$}} 
\put(90,-7){\makebox(0,0){$3_0$}} 
\put(152,-7){\makebox(0,0){$3_1$}}

\end{picture} 
\end{center} 
\caption{Diagramme de Dynkin g\'en\'eralis\'e ${\cal E}_9$} 
\label{grE9} 
\end{figure}

\paragraph{Graphe et matrices de fusion} 
Le graphe $\mathcal{E}_{9}$ est illustr\'e \`a la Fig. \ref{grE9}, son niveau est $\ell=9$, son altitude 
$\kappa =9+3=12$ et sa norme $\beta = 1+2 \cos (2 \pi /12)$. L'identit\'e est $0_0$ et la
fondamentale $0_1$ ($0_2$ est la conjugu\'ee de $0_1$). Ce graphe serait mieux illustr\'e 
en 3 dimensions, en analogie avec un avion, avec le cockpit central form\'e par les vertex $0_i$ et $3_i$  
, et les deux ailes form\'ees respectivement par les vertex $1_i$ et $2_i$, pour mieux illustrer 
la sym\'etrie existant entre les deux ailes (l'indice $i=(0,1,2)$  
repr\'esente la trialit\'e des vertex).  
 
La conjuguaison correspond \`a l'axe passant par les vertex $0_0$ et $3_0$: \mbox{$0_0^*=0_0, 1_0^*=2_0, 3_0^*=3_0$}, 
$0_1^*=0_2,1_1^*=2_2,1_2^*=2_1,3_1^*=3_2$. Comme pour les cas $D_{2n}$ de $SU(2)$, le graphe  
${\cal E}_9$ ne d\'etermine pas de mani\`ere 
unique une structure alg\'ebrique associative, d\^ue \`a la sym\'etrie entre les deux ailes. 
Cependant, en {\sl imposant} que les coefficients de structure soient des entiers non-n\'egatifs, alors 
la solution est {\sl unique}. Comme la d\'etermination des matrices de fusion $G_{1_i}$ et $G_{2_i}$  
n'est pas directe, nous donnons les matrices de fusion correspondant aux vertex $1_0$ et $2_0$,  
\`a partir desquelles les autres se calculent facilement.  
Dans la base $(0_0,1_0,2_0,3_0;  
0_1,1_1,2_1,3_1; 0_2,1_2,2_2,3_2)$, elles sont donn\'ees par: 
 
\scriptsize 
$$ 
G_{{1}_{0}}  =  
\left( \begin{array}{cccccccccccc} 
. & 1 & . & . & . & . & . & . & . & . & . & .  \\ 
. & . & 1 & . & . & . & . & . & . & . & . & .  \\ 
1 & . & . & . & . & . & . & . & . & . & . & .  \\ 
. & . & . & 1 & . & . & . & . & . & . & . & .  \\ 
. & . & . & . & . & 1 & . & . & . & . & . & .  \\ 
. & . & . & . & . & . & 1 & . & . & . & . & .  \\ 
. & . & . & . & 1 & . & . & . & . & . & . & .  \\ 
. & . & . & . & . & . & . & 1 & . & . & . & .  \\ 
. & . & . & . & . & . & . & . & . & 1 & . & .  \\ 
. & . & . & . & . & . & . & . & . & . & 1 & .  \\ 
. & . & . & . & . & . & . & . & 1 & . & . & .  \\ 
. & . & . & . & . & . & . & . & . & . & . & 1  
\end{array} 
\right) 
\quad  
G_{{2}_{0}}  =  
\left( \begin{array}{cccccccccccc} 
. & . & 1 & . & . & . & . & . & . & . & . & .  \\ 
1 & . & . & . & . & . & . & . & . & . & . & .  \\ 
. & 1 & . & . & . & . & . & . & . & . & . & .  \\ 
. & . & . & 1 & . & . & . & . & . & . & . & .  \\ 
. & . & . & . & . & . & 1 & . & . & . & . & .  \\ 
. & . & . & . & 1 & . & . & . & . & . & . & .  \\ 
. & . & . & . & . & 1 & . & . & . & . & . & .  \\ 
. & . & . & . & . & . & . & 1 & . & . & . & .  \\ 
. & . & . & . & . & . & . & . & . & . & 1 & .  \\ 
. & . & . & . & . & . & . & . & 1 & . & . & .  \\ 
. & . & . & . & . & . & . & . & . & 1 & . & .  \\ 
. & . & . & . & . & . & . & . & . & . & . & 1  
\end{array} 
\right) 
$$ 
\normalsize 
 
\begin{figure} 
\begin{center} 
\includegraphics*{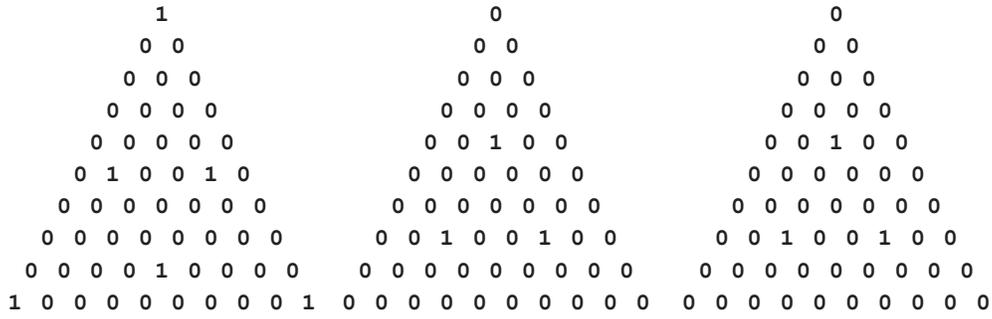} 
\caption{Induction correspondant aux vertex $0_0, 1_0$ et $2_0$ de ${\cal E}_9$.} 
\label{ColE0_E9} 
\end{center} 
\end{figure}

\paragraph{Induction-restriction} 
Le graphe de la s\'erie $\mathcal{A}$ de m\^eme norme que $\mathcal{E}_9$ est $\mathcal{A}_9$,  
poss\'edant $10 \times 11/2 = 55$ vertex.  
L'action de $\mathcal{A}_9$ sur $\mathcal{E}_9$ est cod\'ee par les matrices $F_i$, d'o\`u nous obtenons 
la matrice essentielle $E_0$, dont les colonnes correspondant aux vertex $0_0, 1_0$ et $2_0$ sont 
pr\'esent\'ees \`a la Fig. \ref{ColE0_E9}. 
Nous lisons de la Fig. \ref{ColE0_E9} l'induction $\mathcal{E}_9 \hookleftarrow \mathcal{A}_9$: 
\begin{eqnarray*} 
0_0 &\hookleftarrow  & (0,0) , (4,4) , (4,1) , (1,4) , (9,0) , (0,9) \\ 
1_0 &\hookleftarrow  & (2,2) , (5,2) , (2,5) \\ 
2_0 &\hookleftarrow  & (2,2) , (5,2) , (2,5)  
\end{eqnarray*} 
Une valeur d\'etermin\'ee de $\hat{T}$ peut \^etre d\'efinie pour les vertex $0_0, 1_0$ et $2_0$: 
$\hat{T}(0_0) = 9$ et $\hat{T}(1_0)=\hat{T}(2_0)=21$, ce qui n'est pas possible pour les autres  
vertex de $\mathcal{E}_9$. La sous-alg\`ebre $J$ est donc engendr\'ee par  
$\{0_0,1_0,2_0\}$. 
 
\paragraph{Alg\`ebre d'Ocneanu}  
Nous serions tent\'es de d\'efinir l'alg\`ebre d'Ocneanu de $\mathcal{E}_9$ par: 
\begin{equation} 
Oc(\mathcal{E}_9) = \mathcal{E}_9 \otimes_J \mathcal{E}_9 
\label{defOcE9} 
\end{equation} 
et avec cette d\'efinition, la relation de masse quantique serait satisfaite: 
$m(Oc(\mathcal{E}_9)) = m(\mathcal{E}_9).m(\mathcal{E}_9) /m(J) = m(\mathcal{A}_9) = 432(7+4\sqrt{3})$. 
Cependant, comme pour les cas $D_{2n}$ de $SU(2)$, $\hat{T}$ poss\`ede 
la m\^eme valeur sur les vertex sym\'etriques $1_0$ et $2_0$ de $J$. Dans le cas $D_{2n}$, 
cette particularit\'e conduit \`a la non-commutativit\'e de l'alg\`ebre $Oc(D_{2n})$. 
L'alg\`ebre d\'efinie en (\ref{defOcE9}) est commutative:  
nous nous atendons \`a ce que $Oc(\mathcal{E}_9)$ contienne aussi une composante  
matricielle tenant compte de la non-commutativit\'e.  
Connaissant le graphe d'Ocneanu de $D_{2n}$, nous sommes parvenus \`a donner une r\'ealisation de 
l'alg\`ebre d'Ocneanu. Cependant, dans le cas $\mathcal{E}_9$, sans la connaissance du graphe, 
nos m\'ethodes bas\'ees sur les donn\'ees provenant de l'op\'erateur modulaire $\hat{T}$ 
semblent insuffisantes pour d\'eterminer la structure compl\`ete de $Oc(\mathcal{E}_9)$, 
qui nous permettraient d'obtenir toutes les matrices toriques et fonctions de partition g\'en\'eralis\'ees. 
 
Notons toutefois que l'identit\'e de $Oc(\mathcal{E}_9)$ est $0_0 \otimesdot 0_0$, et nous pouvons  
calculer la fonction de partition associ\'ee \`a ce point.  
D\'efinissant les caract\`eres \'etendus de $\mathcal{E}_9$ suivants: 
\begin{eqnarray*} 
\hat{\chi}_0  &=& \chi_{0,0} + \chi_{0,9} + \chi_{9,0} + \chi_{1,4} + \chi_{4,1} + \chi_{4,4} \\ 
\hat{\chi}_{1_0} = \hat{\chi}_{2_0} &=& \chi_{2,2} + \chi_{2,5} + \chi_{5,2}   
\end{eqnarray*} 
nous trouvons: 
\begin{eqnarray*} 
\mathcal{Z}_{\mathcal{E}_9} =  \mathcal{Z}_{0_0 \otimesdot 0_0} = \sum_{c \in J}\, (F_i)_{0c}\, (F_j)_{0c} =  
|\hat{\chi}_{0_0}|^2 + 2 \, |\hat{\chi}_{1_0}|^2   
\end{eqnarray*} 
et cette expression correspond bien \`a la fonction de partition invariante modulaire de la  
classification de Gannon \cite{gannon-class}.


\subsection{Le cas $\mathcal{E}_{21}$} 
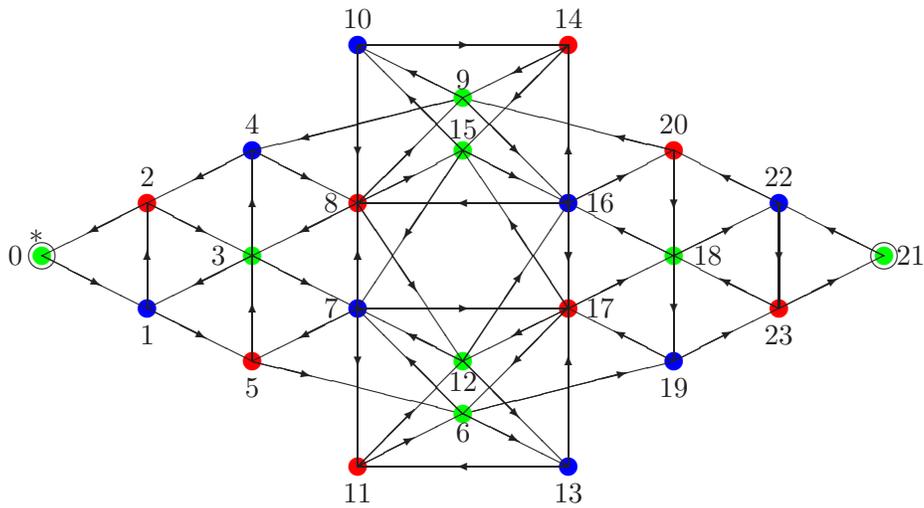
\begin{figure}[hhh] 
\begin{center} 
\begin{picture}(320,170) 
 
\put(0,60){\begin{picture}(40,40) 
\put(0,20){\color{green} \circle*{7}} 
\put(40,0){\color{blue} \circle*{7}} 
\put(40,40){\color{red} \circle*{7}} 
\put(0,20){\line(2,-1){40}} 
\put(0,20){\line(2,1){40}} 
\put(40,0){\line(0,1){40}} 
\put(40,40){\vector(-2,-1){22.5}} 
\put(0,20){\vector(2,-1){22.5}} 
\put(40,0){\vector(0,1){22.5}} 
\end{picture}} 
 
\put(40,40){\begin{picture}(40,40) 
\put(40,0){\color{red} \circle*{7}} 
\put(40,40){\color{green} \circle*{7}} 
\put(0,20){\line(2,-1){40}} 
\put(0,20){\line(2,1){40}} 
\put(40,0){\line(0,1){40}} 
\put(40,40){\vector(-2,-1){22.5}} 
\put(0,20){\vector(2,-1){22.5}} 
\put(40,0){\vector(0,1){22.5}} 
\end{picture}} 
 
\put(40,80){\begin{picture}(40,40) 
\put(40,40){\color{blue} \circle*{7}} 
\put(0,20){\line(2,-1){40}} 
\put(0,20){\line(2,1){40}} 
\put(40,0){\line(0,1){40}} 
\put(40,40){\vector(-2,-1){22.5}} 
\put(0,20){\vector(2,-1){22.5}} 
\put(40,0){\vector(0,1){22.5}} 
\end{picture}} 
 
\put(80,60){\begin{picture}(40,40) 
\put(40,0){\color{blue} \circle*{7}} 
\put(40,40){\color{red} \circle*{7}} 
\put(0,20){\line(2,-1){40}} 
\put(0,20){\line(2,1){40}} 
\put(40,0){\line(0,1){40}} 
\put(40,40){\vector(-2,-1){22.5}} 
\put(0,20){\vector(2,-1){22.5}} 
\put(40,0){\vector(0,1){22.5}} 
\end{picture}}

\put(80,40){\line(4,-1){80}} 
\put(80,40){\vector(4,-1){22.5}} 
\put(80,120){\line(4,1){80}} 
\put(160,140){\vector(-4,-1){62.5}}

\put(280,60){\begin{picture}(40,40) 
\put(0,0){\color{red} \circle*{7}} 
\put(40,20){\color{green} \circle*{7}} 
\put(0,40){\color{blue} \circle*{7}} 
\put(0,0){\line(0,1){40}} 
\put(0,0){\line(2,1){40}} 
\put(0,40){\line(2,-1){40}} 
\put(0,0){\vector(2,1){22.5}} 
\put(40,20){\vector(-2,1){22.5}} 
\put(0,40){\vector(0,-1){22.5}} 
\end{picture}} 
 
\put(240,40){\begin{picture}(40,40) 
\put(0,0){\color{blue} \circle*{7}} 
\put(0,40){\color{green} \circle*{7}} 
\put(0,0){\line(0,1){40}} 
\put(0,0){\line(2,1){40}} 
\put(0,40){\line(2,-1){40}} 
\put(0,0){\vector(2,1){22.5}} 
\put(40,20){\vector(-2,1){22.5}} 
\put(0,40){\vector(0,-1){22.5}} 
\end{picture}} 
 
\put(240,80){\begin{picture}(40,40) 
\put(0,40){\color{red} \circle*{7}} 
\put(0,0){\line(0,1){40}} 
\put(0,0){\line(2,1){40}} 
\put(0,40){\line(2,-1){40}} 
\put(0,0){\vector(2,1){22.5}} 
\put(40,20){\vector(-2,1){22.5}} 
\put(0,40){\vector(0,-1){22.5}} 
\end{picture}} 
 
\put(200,60){\begin{picture}(40,40) 
\put(0,0){\color{red} \circle*{7}} 
 
\put(0,40){\color{blue} \circle*{7}} 
\put(0,0){\line(0,1){40}} 
\put(0,0){\line(2,1){40}} 
\put(0,40){\line(2,-1){40}} 
\put(0,0){\vector(2,1){22.5}} 
\put(40,20){\vector(-2,1){22.5}} 
\put(0,40){\vector(0,-1){22.5}} 
\end{picture}} 
\put(160,20){\line(4,1){80}} 
\put(160,20){\vector(4,1){62.5}} 
 
\put(240,120){\line(-4,1){80}} 
\put(240,120){\vector(-4,1){22.5}}

\put(120,0){\color{red} \circle*{7}} 
\put(200,0){\color{blue} \circle*{7}} 
\put(160,20){\color{green} \circle*{7}} 
\put(160,40){\color{green} \circle*{7}} 
\put(160,120){\color{green} \circle*{7}} 
\put(160,140){\color{green} \circle*{7}} 
\put(120,160){\color{blue} \circle*{7}} 
\put(200,160){\color{red} \circle*{7}}

\put(200,0){\line(-1,0){80}} 
\put(200,0){\vector(-1,0){42.5}} 
\put(120,0){\line(2,1){40}} 
\put(120,0){\line(1,1){40}} 
\put(120,0){\vector(2,1){21.5}} 
\put(120,0){\vector(1,1){21.5}} 
\put(160,20){\line(2,-1){40}} 
\put(160,20){\vector(2,-1){21.5}} 
\put(160,40){\line(1,-1){40}} 
\put(160,40){\vector(1,-1){21.5}} 
\put(200,100){\line(-1,0){80}} 
\put(200,100){\vector(-1,0){42.5}} 
\put(120,100){\line(2,1){40}} 
\put(120,100){\line(1,1){40}} 
 
\put(120,100){\vector(2,1){21.5}} 
\put(120,100){\vector(1,1){21.5}} 
\put(160,120){\line(2,-1){40}} 
\put(160,120){\vector(2,-1){21.5}} 
\put(160,140){\line(1,-1){40}} 
\put(160,140){\vector(1,-1){21.5}}

\put(120,160){\line(1,0){80}} 
\put(120,160){\vector(1,0){42.5}} 
\put(200,160){\line(-2,-1){40}} 
\put(200,160){\line(-1,-1){40}} 
\put(200,160){\vector(-2,-1){21.5}} 
\put(200,160){\vector(-1,-1){21.5}} 
\put(160,140){\line(-2,1){40}} 
\put(160,140){\vector(-2,1){21.5}} 
\put(160,120){\line(-1,1){40}} 
\put(160,120){\vector(-1,1){21.5}} 
 
\put(120,60){\line(1,0){80}} 
\put(120,60){\vector(1,0){42.5}} 
\put(200,60){\line(-2,-1){40}} 
\put(200,60){\line(-1,-1){40}} 
\put(200,60){\vector(-2,-1){21.5}} 
\put(200,60){\vector(-1,-1){21.5}} 
\put(160,40){\line(-2,1){40}} 
\put(160,40){\vector(-2,1){21.5}} 
\put(160,20){\line(-1,1){40}} 
\put(160,20){\vector(-1,1){21.5}}

\put(120,0){\line(0,1){160}} 
\put(200,0){\line(0,1){160}} 
\put(120,60){\vector(0,1){22.5}} 
\put(120,60){\vector(0,-1){22.5}} 
\put(120,160){\vector(0,-1){42.5}} 
\put(200,100){\vector(0,-1){22.5}} 
\put(200,100){\vector(0,1){22.5}} 
\put(200,0){\vector(0,1){42.5}}

\put(200,60){\line(-2,3){40}} 
\put(200,60){\vector(-2,3){22.5}} 
\put(160,120){\line(-2,-3){40}} 
\put(160,120){\vector(-2,-3){22.5}}

\put(120,100){\line(2,-3){40}} 
\put(120,100){\vector(2,-3){22.5}} 
\put(160,40){\line(2,3){40}} 
\put(160,40){\vector(2,3){22.5}} 
 
\put(120,60){\line(-2,-1){40}} 
\put(120,60){\vector(-2,-1){22.5}} 
\put(80,120){\line(2,-1){40}} 
\put(80,120){\vector(2,-1){22.5}} 
 
\put(240,40){\line(-2,1){40}} 
\put(240,40){\vector(-2,1){22.5}} 
\put(200,100){\line(2,1){40}} 
\put(200,100){\vector(2,1){22.5}}

\put(-10,80){\makebox(0,0){0}} 
\put(67,80){\makebox(0,0){3}} 
\put(253,80){\makebox(0,0){18}} 
\put(330,80){\makebox(0,0){21}} 
 
\put(40,50){\makebox(0,0){1}} 
\put(40,110){\makebox(0,0){2}} 
 
\put(80,30){\makebox(0,0){5}} 
\put(80,130){\makebox(0,0){4}} 
\put(120,-10){\makebox(0,0){11}} 
\put(120,170){\makebox(0,0){10}} 
\put(200,-10){\makebox(0,0){13}} 
\put(200,170){\makebox(0,0){14}} 
\put(240,30){\makebox(0,0){19}} 
\put(240,130){\makebox(0,0){20}} 
\put(280,50){\makebox(0,0){23}} 
\put(280,110){\makebox(0,0){22}} 
\put(160,13){\makebox(0,0){6}} 
\put(160,32){\makebox(0,0){12}} 
\put(160,128){\makebox(0,0){15}} 
\put(160,147){\makebox(0,0){9}} 
\put(110,60){\makebox(0,0){7}} 
\put(110,100){\makebox(0,0){8}} 
\put(212,60){\makebox(0,0){17}} 
\put(212,100){\makebox(0,0){16}}

\put(0,80){\circle{10}} 
\put(320,80){\circle{10}}

\put(-5,85){$\ast$} 
 
\end{picture} 
\end{center} 
\caption{Le diagramme de Dynkin g\'en\'eralis\'e $\mathcal{E}_{21}$.} 
\label{grE21} 
\end{figure}

\paragraph{Graphe et matrices de fusion} 
Le graphe $\mathcal{E}_{21}$ est illustr\'e \`a la Fig. \ref{grE21}. Son niveau $\ell =21$, son altitude  
$\kappa = 21 + 3 = 24$ et sa norme $\beta = 1+ 2\cos(2 \pi / 24)$. $0$ est l'identit\'e, $1$ et $2$ sont  
les g\'en\'erateurs (l'orientation du graphe correspond au g\'en\'erateur 1). Le graphe poss\`ede 24 vertex. La conjuguaison 
correspond \`a la sym\'etrie par rapport \`a l'axe horizontal passant par les vertex 0 et 21.  
 
Le graphe $\mathcal{E}_{21}$ d\'etermine de mani\`ere unique son alg\`ebre de graphe. La multiplication par 
le g\'en\'erateur 1 est cod\'ee par la matrice d'adjacence du graphe: $G_1=\mathcal{G}_{\mathcal{E}_{21}}$. La conjuguaison se traduit 
au niveau matriciel par la transposition: $G_{i^*} = G_i^T$. La d\'etermination des matrices de fusion 
est directe jusqu'au vertex 9. Par exemple, 1.1=2+5 nous donne $G_5 = G_1.G_1 - G_2$. Le graphe est aussi 
sym\'etrique par rapport au point central (centre de l'\'etoile): nous appelons $R$ cette sym\'etrie (par exemple $0^R = 21, 1^R = 22, 
11^R=14$). 
Au niveau de la multiplication, ceci se traduit par: 
\begin{equation} 
a . b = a^R.b^R  
\label{sym_R_E21} 
\end{equation} 
Utilisant cette sym\'etrie, il est alors imm\'ediat de compl\'eter la table de multiplication et d'obtenir 
les autres matrices de fusion. Pour $a=0$, l'\'equation (\ref{sym_R_E21}) donne $b = 21. b^R$.  
En particulier $9 = 21.6$, ce qui nous donne 
$G_{21} = G_9.G_6^{-1}$. Les autres matrices $G_b$ s'obtiennent alors facilement par $G_b = G_{21}.G_{b^R}$  
(par exemple $G_{22}=G_{21}.G_1, G_{23} = G_{21}.G_2$).  
 
\begin{figure} 
\mbox{\scalebox{0.9}{\includegraphics{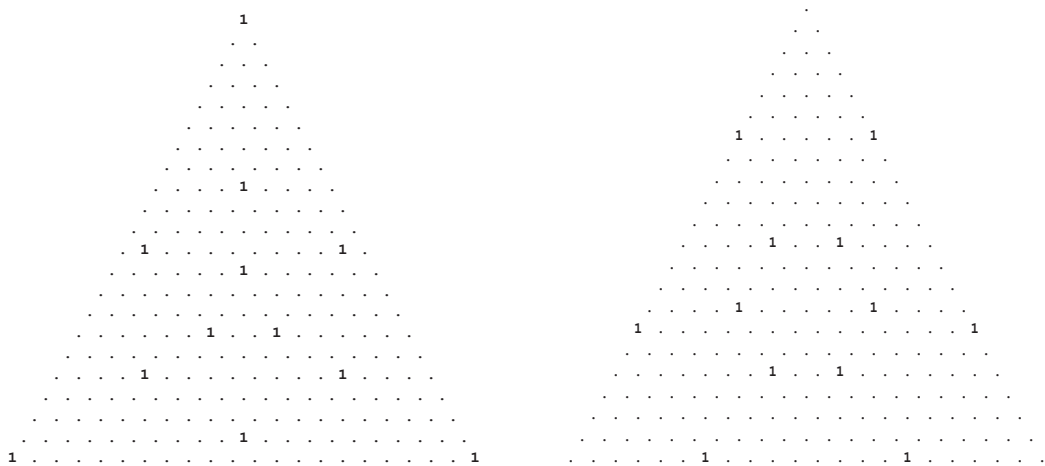}}} 
\caption{Induction correspondant aux vertex extr\'emaux 0 et 21 de ${\cal E}_{21}$} 
\label{E0E21Col_E21} 
\end{figure}

\paragraph{Induction-restriction} 
Le graphe $\mathcal{A}$ de m\^eme norme que $\mathcal{E}_{21}$ est $\mathcal{A}_{21}$, qui poss\`ede 
$22 \times 23 /2 = 253$ vertex, labell\'es par $(\lambda_1, \lambda_2)$. L'action de $\mathcal{A}_{21}$ sur 
$\mathcal{E}_{21}$ est cod\'ee par les 253 matrices $F_i = F_{(\lambda_1, \lambda_2)}$ obtenues 
par la relation de r\'ecurrence tronqu\'ee de $su(3)$, avec $F_1 = F_{(1,0)} = G_1$. Les matrices essentielles 
-- en particulier  $E_0$ -- s'obtiennent comme d'habitude: elles poss\`edent 253 lignes labell\'ees par les 
vertex de $\mathcal{A}_{21}$ et 24 colonnes labell\'ees par les vertex de $\mathcal{E}_{21}$, et $E_0$ nous 
donne l'induction-restriction entre $\mathcal{A}_{21}$ et $\mathcal{E}_{21}$. Nous illustrons 
\`a la Fig \ref{E0E21Col_E21} les lignes de $E_0$ correspondant aux vertex 0 (\`a gauche) et 21 (\`a droite).  
 
Nous pouvons v\'erifier que la valeur de l'op\'erateur modulaire $\hat{T}$ est constante pour 
les vertex de $\mathcal{A}_{21}$ dont la restriction donne le vertex 0 ($\hat{T} = 21$ sur \'el\'ements non-nuls de  
la partie gauche de la Fig. \ref{E0E21Col_E21}). Ceci est vrai aussi pour le vertex 21 ($\hat{T} = 39$).  
Par contre, $\hat{T}$ \'evalu\'e  
sur les \'el\'ements non-nuls des 22 autres colonnes de $E_0$ n'est pas constant: la sous-alg\`ebre 
$J$ est donc engendr\'ee par les deux vertex extr\'emaux: $J = \{0,21\}$.

\paragraph{Alg\`ebre d'Ocneanu} 
Nous d\'efinissons alors l'alg\`ebre d'Ocneanu $Oc(\mathcal{E}_{21})$ par: 
\begin{equation} 
Oc(\mathcal{E}_{21}) = \mathcal{E}_{21} \otimes_J \mathcal{E}_{21} = \mathcal{E}_{21} \otimesdot \mathcal{E}_{21} 
\end{equation} 
o\`u nous identifions les \'el\'ements $a \otimesdot b.c$ avec $a.b^* \otimesdot c = a.b \otimesdot c$ pour  
\mbox{$b \in J$ ($0^*=0, 21^*=21$).} 
$J$ -- ou de mani\`ere \'equivalente la r\'eflexion $R$ -- fournit une partition de $\mathcal{E}_{21}$ en classes 
d'\'equivalence \`a deux \'el\'ements $\{a, a^R\}$. Nous avons $J = \tilde{0} = \tilde{21} = \{0,21\}$,  
$\tilde{1} = \tilde{22} = \{1,22\}$, etc. Nous choisissons un repr\'esentant $\phi(a)$ dans chaque classe 
d'\'equivalence: 
\begin{equation} 
\Phi = \{\phi(b) | b \in \mathcal{E}_{21} \} = \{ 0,1,2,3,4,5,6,7,8,10,11,12\}. 
\end{equation}  
Pour $b \in \Phi$, $\phi(b)=b$. Un \'el\'ement $b \notin \Phi$ s'\'ecrit $b = 21. \phi(b)$. Introduisons  
l'application $\rho$: 
$$ 
\begin{array}{lcl} 
\rho(b) =  0  & \qquad &\textrm{si } b \in \Phi  \\ 
\rho(b) =  21 & \qquad &\textrm{si } b \notin \Phi \end{array} 
$$ 
Alors, nous avons les identifications suivantes dans l'alg\`ebre d'Ocneanu $Oc(\mathcal{E}_{21})$: 
\begin{equation} 
a \otimesdot b = a . \rho(b) \otimesdot \phi(b) 
\label{identif_E21} 
\end{equation} 
et une base de $Oc(\mathcal{E}_{21})$ est donn\'ee par les $24 \times 12 = 288$ \'el\'ements lin\'eairement  
ind\'ependants: 
\begin{equation} 
a \otimesdot b \qquad \qquad a \in \mathcal{E}_{21}, \qquad b \in \Phi. 
\end{equation} 
La sous-alg\`ebre chirale gauche est engendr\'ee par $L = \{a \otimesdot 0\}$ et la sous-alg\`ebre 
chirale droite est engendr\'ee par $R = \{ 0 \otimesdot a = \rho(a) \otimesdot \phi(a)\}$ 
toutes deux de dimension 24. 
L'intersection est la sous-alg\`ebre ambichirale, engendr\'ee par les deux \'el\'ements $\{ 0 \otimesdot 0 =  
21 \otimesdot 21\}$ et $\{ 21 \otimesdot 0 = 0 \otimesdot 21 \}$.  
Enfin, l'action d'un \'el\'ement $x = a \otimesdot b$ de $Oc(\mathcal{E}_{21})$ sur un \'el\'ement $c$ de  
$\mathcal{E}_{21}$ donne l'\'el\'ement $(a.b.c)$ de $\mathcal{E}_{21}$; et les matrices qui codent cette 
action sont \'egales \`a: 
\begin{equation} 
S_x = G_a .G_b \qquad \qquad \textrm{pour } x = a \otimesdot b \in Oc(\mathcal{E}_{21}) 
\end{equation}

\begin{figure}[H] 
\begin{center} 
\mbox{\scalebox{1.0}{\includegraphics{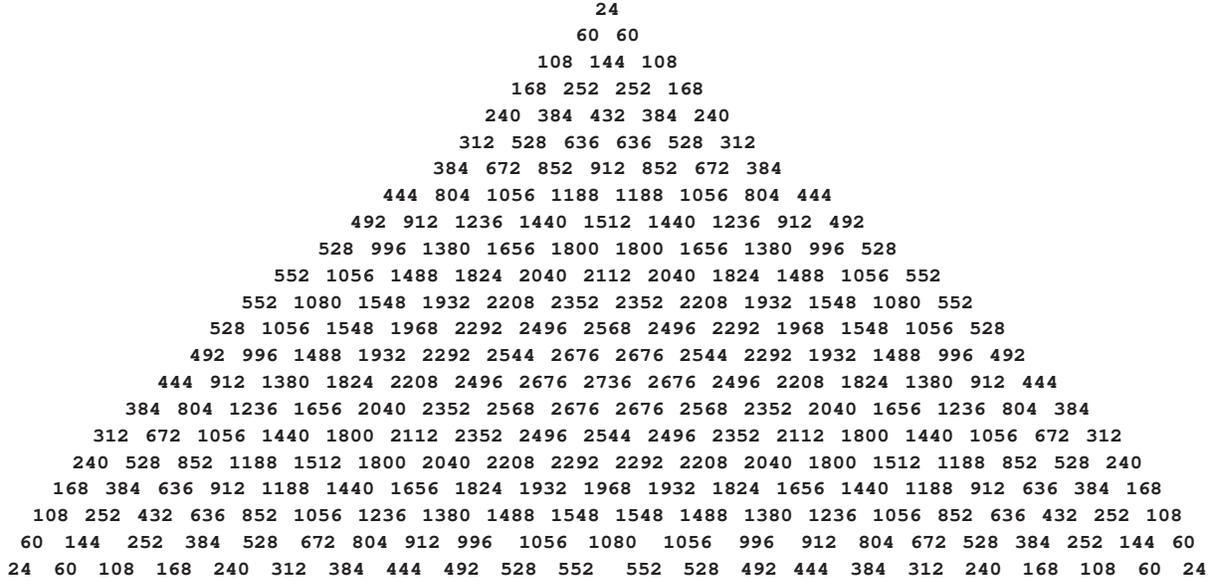}}} 
\caption{Dimensions des blocs pour la loi $\circ$ de la big\`ebre $\mathcal{BE}_{21}$.} 
\label{Dim_E21} 
\end{center} 
\end{figure}

\paragraph{Dimensions des blocs} 
Les dimensions $d_j$, pour $j = (\lambda_1, \lambda_2)$ des 253 blocs de la big\`ebre 
$\mathcal{B}(\mathcal{E}_{21})$, 
pour la loi de composition, sont obtenues en sommant les \'el\'ements de matrice de $F_j$: elles 
sont pr\'esent\'ees \`a la Fig. \ref{Dim_E21}, dispos\'ees en analogie avec le graphe $\mathcal{A}_{21}$. 
Les blocs de la deuxi\`eme structure de la big\`ebre (loi de convolution) sont labell\'es 
par les 288 points du graphe d'Ocneanu, et les dimensions $d_x$ du bloc $x$ sont obtenues en sommant les \'el\'ements 
de matrice de $S_x$, pour $x \in Oc(\mathcal{E}_{21})$. Nous trouvons (l'indice donne la multiplicit\'e de la 
dimension): 
{\small 
$$ 
\begin{array}{lcl} 
\text{Ambichirale} &:& (24)_2 \\ 
\text{Gauche (pas ambichirale)} &:& (60)_4 (108)_4 (132)_4 (144)_2 (168)_2 (216)_2 (252)_4 \\ 
\text{Droite (pas ambichirale)}&:& (60)_4 (108)_4 (132)_4 (144)_2 (168)_2 (216)_2 (252)_4 \\ 
\text{Suppl\'ement} &:&  
(168)_{8} (312)_{16} (384)_{16} (420)_{8} (492)_{8} (600)_{8} (636)_{8} (744)_{32} (804)_{8} (936)_{8} \\ 
{} & {} & (948)_{8} (996)_{8} (1080)_{2} (1188)_{8} (1236)_{8}  
(1272)_{4} (1440)_{16} (1512)_{2} (1548)_{8} \\ 
{} & {} & (1656)_{4} (1800)_{16} (1932)_{8} (1968)_{4} (2292)_{8}  
(2568)_{2} (2988)_{8} (3480)_{8} 
\end{array} 
$$} 
Les r\`egles de somme lin\'eaire et quadratiques sont v\'erifi\'ees: 
$$ 
\sum_{i \in \mathcal{A}_{21}} d_i   = \sum_{x \in Oc(\mathcal{E}_{21})} d_x   = 288\,576\, ,  \qquad \qquad 
\dim(\mathcal{B}\mathcal{E}_{21}) = \sum_i d_i^2 = \sum_x d_x^2 = 480\,701\,952. 
$$ 
Nous v\'erifions aussi la relation de masse quantique entre $Oc(\mathcal{E}_{21})$ et $\mathcal{A}_{21}$: 
$$ 
m(Oc(\mathcal{E}_{21})) = \frac{m(\mathcal{E}_{21}).m(\mathcal{E}_{21})}{m(J)} = m(\mathcal{A}_{21}) 
= 288 \left( 18 + 10 \sqrt{3} + \sqrt{6(97 + 56\sqrt{3})} \right)^2. 
$$

\paragraph{Matrices toriques et fonctions de partition g\'en\'eralis\'ees} 
Nous calculons l'action des \'el\'ements $i,j$ de $\mathcal{A}_{21}$ sur un \'el\'ement  
$x = a\otimesdot b$ de $Oc(\mathcal{E}_{21})$ en utilisant les identifications (\ref{identif_E21}): 
\begin{eqnarray*} 
i . (a \otimesdot b ) . j &=& \sum_c \sum_d (F_i)_{ac} \, (F_j)_{bd} \; (c \otimesdot d) \\ 
{ } &=&  \sum_c \sum_d (F_i)_{ac}\, (F_j)_{bd}\; (c. \rho(d) \otimesdot \phi(d)) \\ 
{ } &=&  \sum_c \sum_d \sum_e (F_i)_{ac}\, (F_j)_{bd}\, (G'_e)_{c \rho(d)}\;  (e  \otimesdot \phi(d))  
\end{eqnarray*} 
o\`u les matrices $G'_e$ sont d\'efinies par $(G'_e)_{ab} = (G_a)_{be}$. S\'eparant alors la sommation  
sur $d$ en une sommation sur les \'el\'ements de la m\^eme classe d'\'equivalence, les 
$288 \times 288 = 82\,944$ 
matrices toriques g\'en\'eralis\'ees de $\mathcal{E}_{21}$ s'obtiennent sous la forme compacte suivante: 
\begin{equation} 
\begin{array}{|c|} 
\hline 
{ } \\ 
W_{ab,ef} = \displaystyle \sum_c \sum_{d \in \tilde{f}} (F_i)_{ac} (F_j)_{bd} (G'_e)_{c \rho(d)} \\ 
{ } \\ 
\hline 
\end{array} 
\end{equation} 
o\`u la sommation sur $d$ se fait sur les \'el\'ements de la classe d'\'equivalence de $f$.  
Pour $e=0,f=0$, la sommation sur $d$ devient une sommation sur $J=\{0,21\}$, alors $\rho(d)=d$ 
et $(G'_0)_{cd} = \delta_{c,d}$: l'invariant modulaire $\mathcal{M}$ s'\'ecrit donc: 
$$ 
\mathcal{M} = W_{00,00} = \sum_{c \in J} (F_i)_{0c} \, (F_j)_{0c} 
$$  
Les fonctions de partition g\'en\'eralis\'ees s'obtiennent alors par: 
\begin{equation} 
\mathcal{Z}_{ab,ef} = \sum_{i \in \mathcal{A}_{21}} \sum_{j \in \mathcal{A}_{21}} \chi_i (W_{ab,ef})_{ij} \ov{\chi}_j   
= \sum_{c \in \mathcal{E}_{21}} \sum_{d \in \tilde{f}}\, \hat{\chi}_{ac}\, (G'_e)_{c \rho(d)}\, \ov{\hat{\chi}}_{bd}\, 
\end{equation} 
o\`u les $\chi_i$ sont les caract\`eres de l'alg\`ebre $\widehat{su}(3)$ et les $\hat{\chi}_{ab}$ sont les 
caract\`eres 
\'etendus g\'en\'eralis\'es de $\mathcal{E}_{21}$, combinaisons lin\'eaires des caract\`eres \'etendus $\hat{\chi}_a$: 
\begin{equation} 
\hat{\chi}_{ab} = \sum_{i \in \mathcal{A}_{21}} (F_i)_{ab} \; \chi_i \; = \;  
\sum_c (G_a)_{bc} \; \hat{\chi}_c \qquad \qquad 
\hat{\chi}_{c}  = \sum_i (F_i)_{a0} \; \chi_i  
\end{equation} 
Les 288 fonctions de partition \`a une ligne de d\'efaut $\mathcal{Z}_{ab} = \mathcal{Z}_{ab,00}$ s'\'ecrivent: 
\begin{equation} 
\mathcal{Z}_{ab} = \displaystyle \sum_{c \in J}\;  \hat{\chi}_{ac} \, \ov{\hat{\chi}}_{bc}
\end{equation} 
Les caract\`eres \'etendus $\hat{\chi}_c$ pour $c \in J$ sont: 
\footnotesize 
\begin{eqnarray*} 
\hat{\chi}_0   &=& \chi_{(0,0)} +  \chi_{(4,4)} + \chi_{(6,6)} + \chi_{(10,10)} + \chi_{(0,21)} + \chi_{(21,0)} +  \chi_{(1,10)} + \chi_{(10,1)} + \chi_{(4,13)} + \chi_{(13,4)} +  \chi_{(6,9)} + \chi_{(9,6)} \\ 
\hat{\chi}_{21}&=& \chi_{(0,6)} +  \chi_{(6,0)} + \chi_{(0,15)} + \chi_{(15,0)} + \chi_{(4,7)} + \chi_{(7,4)} +   
\chi_{(4,10)} + \chi_{(10,4)} + \chi_{(6,15)} + \chi_{(15,6)} + \chi_{(7,10)} + \chi_{(10,7)}    
\end{eqnarray*} 
\normalsize 
et la fonction de partition de $\mathcal{E}_{21}$ \'ecrite en fonction des caract\`eres de $\widehat{su}(3)$ correspond 
\`a l'expression obtenues dans la classification de Gannon \cite{gannon-class}: 
\begin{equation} 
\mathcal{Z}_{\mathcal{E}_{21}} = \sum_{c \in J} |\hat{\chi}_c|^2 = |\hat{\chi}_0|^2 + |\hat{\chi}_{21}|^2 
\end{equation}


\chapter*{Conclusion et perspectives} 
\addcontentsline{toc}{chapter}{Conclusion et perspectives} 
\markboth{\uppercase{Conclusion et perspectives}} 
{\uppercase{Conclusion et perspectives}}
\thispagestyle{empty}

Dans cette th\`ese nous avons pr\'esent\'e les profondes relations qui existent
entre les classifications des th\'eories conformes \`a deux dimensions dans divers
environnements et les graphes codant les diff\'erentes structures 
d'une alg\`ebre de Hopf faible: celle-ci appara\^{\i}t ainsi comme la sym\'etrie quantique
naturelle associ\'ee \`a ces mod\`eles conformes.\\

Cette alg\`ebre de Hopf faible a \'et\'e introduite par A. Ocneanu dans \cite{Oc-Marseille, Oc-paths}.
Plus pr\'ecisemment, Ocneanu associe \`a chaque diagramme de Dynkin $G$ de type $ADE$ une {\sl dig\`ebre} 
$\mathcal{B}(G)$: elle est constitu\'ee de l'espace vectoriel des endomorphismes gradu\'es de chemins
essentiels sur $G$ muni de deux lois multiplicatives $\circ$ et $\odot$ donnant lieu \`a deux autres 
graphes, appel\'es respectivement $\mathcal{A}(G)$ et $Oc(G)$. Ces derniers poss\`edent toujours
une structure
multiplicative (alg\`ebre de graphe): nous obtenons deux alg\`ebres not\'ees par le m\^eme symbole
que le graphe lui-m\^eme. 
L'alg\`ebre $Oc(G)$ est aussi appel\'ee l'alg\`ebre des sym\'etries quantiques de $G$. 
Il est assez remarquable que ces trois graphes ($G$, $\mathcal{A}(G)$ et $Oc(G)$) avec leur structure
alg\'ebrique codent les informations sur les diff\'erents coefficients permettant de d\'efinir les 
fonctions de partition du mod\`ele conforme $\widehat{su}(2)$ associ\'e au graphe $G$.

Le travail central de cette th\`ese a \'et\'e la pr\'esentation d'une r\'ealisation de l'alg\`ebre $Oc(G)$
construite comme un certain quotient du carr\'e tensoriel de l'alg\`ebre du graphe $G$. 
Cette r\'ealisation permet d'obtenir un algorithme tr\`es simple pour le calcul   
des divers coefficients d\'efinissant les fonctions de partition  du mod\`ele 
conforme $\widehat{su}(2)$ associ\'e au graphe $G$ \cite{Coq_Gil-ADE}. 
Par la suite, nous avons observ\'e que l'alg\`ebre d'Ocneanu peut -- dans la plupart des cas -- \^etre
reconstruite d'apr\`es les propri\'et\'es modulaires du graphe $G$. Nous avons alors utilis\'e
cette observation pour \'etudier plusieurs exemples appartenant au mod\`ele conforme $\widehat{su}(3)$
et obtenir ainsi les expressions des fonctions de partition associ\'ees \cite{Coq_Gil-Tmod}.

Les mod\`eles minimaux (``usuels'') -- construits par des irreps de l'alg\`ebre de Virasoro -- 
sont reli\'es aux mod\`eles $\widehat{su}(2)$. L'alg\`ebre de Virasoro est un cas particulier
d'alg\`ebre $\mathcal{W}_n$: $Vir=\mathcal{W}_2$. Les mod\`eles $\mathcal{W}_n$-minimaux, construits
par des irreps de l'alg\`ebre $\mathcal{W}_n$ g\'en\'eralisant l'alg\`ebre de Virasoro, sont eux reli\'es aux
mod\`eles $\widehat{su}(n)$. Les expressions des fonctions de partition associ\'ees aux mod\`eles 
$\widehat{su}(3)$ obtenues dans le chapitre {\bf 4} permettent ainsi l'\'etude des mod\`eles
$\widehat{W}_3$-minimaux (voir \cite{Coq_Marina}). \\

Les graphes d'Ocneanu ont \'et\'e conceptuellement d\'efinis par Ocneanu comme provenant de la 
diagonalisation de $\mathcal{B}(G)$ pour la loi $\odot$, mais il est int\'eressant de noter qu'ils
n'ont -- \`a notre connaissance -- jamais \'et\'e obtenus de cette mani\`ere $\ldots$ Un autre axe de 
recherche de cette th\`ese est l'\'etude approfondie des structures de la dig\`ebre $\mathcal{B}(G)$:
nous voulons d'une part v\'erifier que  $\mathcal{B}(G)$ satisfait {\it de facto} les 
divers axiomes d\'efinissant une alg\`ebre de Hopf faible, et d'autre part construire explicitement
le graphe d'Ocneanu \`a partir de la diagonalisation de $\mathcal{B}(G)$. Les premiers r\'esultats
obtenus se limitent pour l'instant aux graphes de la s\'erie $A$ \cite{Coq_Gil-bigebra}. \\

Les relations entre les classifications des mod\`eles conformes $\widehat{su}(2)$ et les graphes 
correspondants semblent de nos jours bien connues. Cependant, les g\'en\'eralisations de diverses
structures \`a des mod\`eles $\widehat{su}(n)$ restent encore \`a \^etre formul\'ees de mani\`ere 
pr\'ecise. 
Nous avons grand espoir qu'une meilleure compr\'ehension de ces g\'en\'eralisations
fassent appara\^{\i}tre de nouvelles structures math\'ematiques, permettant \`a leur tout une meilleure
compr\'ehension des mod\`eles physiques sous-jascents. \\

\noindent Pour conclure, citons divers probl\`emes ouverts qui devraient \^etre mieux compris:

\begin{itemize}

\item[$\bullet$] Peut-on d\'efinir une matrice R pour l'alg\`ebre de Hopf faible et obtenir une \'equation
de Yang-Baxter (g\'en\'eralis\'ee)? Quels seraient les mod\`eles int\'egrables associ\'es?
\item[$\bullet$]  Trouver une d\'efinition simple et directe -- valable dans tous les cas -- pour le 
produit de convolution $\odot$. Peut-on obtenir le produit de convolution \`a partir du carr\'e tensoriel 
d'un produit $\star$ d\'efini directement sur l'espace des chemins essentiels?  
\item[$\bullet$]  Quelle est l'origine de la r\`egle de somme lin\'eaire et de la r\`egle de masse 
quantique?
\item[$\bullet$]  La g\'en\'eralisation des diagrammes $ADE$ pour les cas $su(3)$ et $su(4)$ est connue,
mais une d\'efinition rigoureuse des ``diagrammes g\'en\'eralis\'es de Coxeter-Dynkin'' devrait 
\^etre formalis\'ee (et publi\'ee). 
\item[$\bullet$] La structure alg\'ebrique associ\'ee \`a un diagramme $ADE$ est l'alg\`ebre de Lie.
Quelles seraient les structures alg\'ebriques (g\'en\'eralisant la notion d'alg\`ebre de Lie) associ\'ees
\`a des diagrammes g\'en\'eralis\'es?  
\item[$\bullet$]  Nous avons d\'efini les fonctions de partition pour les mod\`eles affines d\'efinis sur le
tore, mais une g\'en\'eralisation dans diverses directions est envisageable. En particulier,
que se passe-t-il pour des syst\`emes d\'efinis sur des surfaces
de genre plus \'elev\'e que le tore?

\end{itemize}


\appendix


\chapter{Diagrammes de Dynkin $ADE$ et $ADE^{(1)}$} 
 \thispagestyle{empty}

\unitlength=0.06cm

\begin{table}[hhh] 
$$ 
\begin{array}{|c|ccl|} 
\hline
{ } & { } & { }   & { }  \\ 
{ } & \textrm{diagramme} & \kappa & \textrm{exposants} \\ 
{ } & { } & { }   & { }  \\ 
\hline 
\quad A_n \quad & \begin{picture}(100,15)(0,15) 
\put(5,15){\begin{picture}(0,0) 
\thicklines 
\multiput(0,0)(15,0){3}{\circle*{2}} 
\multiput(60,0)(15,0){3}{\circle*{2}} 
\multiput(30,0)(5,0){6}{\line(1,0){2}} 
\put(0,0){\line(1,0){30}} 
\put(60,0){\line(1,0){30}} 
\put(0,-5){\makebox(0,0){$\tau_0$}} 
\put(15,-5){\makebox(0,0){$\tau_1$}} 
\put(30,-5){\makebox(0,0){$\tau_2$}} 
\put(60,-5){\makebox(0,0){$\tau_{n-3}$}} 
\put(75,-5){\makebox(0,0){$\tau_{n-2}$}} 
\put(90,-5){\makebox(0,0){$\tau_{n-1}$}} 
\end{picture}} 
\end{picture} & n+1   & 1,2,\ldots,n \\  
D_{n} & \begin{picture}(100,30)(0,15) 
\put(5,15){\begin{picture}(0,0) 
\multiput(0,0)(15,0){3}{\circle*{2}} 
\multiput(45,0)(15,0){2}{\circle*{2}} 
\put(70,10){\circle*{2}} 
\put(70,-10){\circle*{2}} 
\thicklines 
\put(0,0){\line(1,0){30}} 
\put(45,0){\line(1,0){15}} 
\put(60,0){\line(1,1){10}} 
\put(60,0){\line(1,-1){10}} 
\multiput(30,0)(3,0){6}{\line(1,0){1}} 
\put(0,-5){\makebox(0,0){$\sigma_0$}} 
\put(15,-5){\makebox(0,0){$\sigma_1$}} 
\put(30,-5){\makebox(0,0){$\sigma_2$}} 
\put(44,-5){\makebox(0,0){$\sigma_{n-4}$}} 
\put(58,-5){\makebox(0,0){$\sigma_{n-3}$}} 
\put(77,10.5){\makebox(0,0){$\sigma_{n-2}$}} 
\put(77,-9.5){\makebox(0,0){$\sigma_{n-2}^{'}$}} 
\end{picture}} 
\end{picture} &  2(n-1)   & 1,3,\ldots,2n-5,2n-3;n-1 \\  
E_6 & \begin{picture}(100,30)(0,15) 
\put(5,15){\begin{picture}(0,0) 
\thicklines 
\multiput(0,0)(15,0){5}{\circle*{2}} 
\put(30,15){\circle*{2}} 
\put(0,0){\line(1,0){60}} 
\put(30,0){\line(0,1){15}} 
\put(0,-5){\makebox(0,0){$\sigma_0$}} 
\put(15,-5){\makebox(0,0){$\sigma_1$}} 
\put(30,-5){\makebox(0,0){$\sigma_2$}} 
\put(45,-5){\makebox(0,0){$\sigma_5$}} 
\put(60,-5){\makebox(0,0){$\sigma_4$}} 
\put(35,15){\makebox(0,0){$\sigma_3$}} 
\end{picture}} 
\end{picture} & 12   & 1,4,5,7,8,11 \\  
E_7 & \begin{picture}(100,30)(0,15) 
\put(5,15){\begin{picture}(0,0) 
\thicklines 
\multiput(0,0)(15,0){6}{\circle*{2}} 
\put(30,15){\circle*{2}} 
\thicklines 
\put(0,0){\line(1,0){75}} 
\put(30,0){\line(0,1){15}} 
\put(0,-5){\makebox(0,0){$\sigma_0$}} 
\put(15,-5){\makebox(0,0){$\sigma_1$}} 
\put(30,-5){\makebox(0,0){$\sigma_2$}} 
\put(45,-5){\makebox(0,0){$\sigma_6$}} 
\put(60,-5){\makebox(0,0){$\sigma_5$}} 
\put(75,-5){\makebox(0,0){$\sigma_4$}} 
\put(35,15){\makebox(0,0){$\sigma_3$}} 
\end{picture}} 
\end{picture} & 18   & 1,5,7,9,11,13,17 \\  
E_8 & \begin{picture}(100,30)(0,15) 
\put(5,15){\begin{picture}(0,0) 
\thicklines 
\multiput(0,0)(15,0){7}{\circle*{2}} 
\put(30,15){\circle*{2}} 
\put(0,0){\line(1,0){90}} 
\put(30,0){\line(0,1){15}} 
\put(0,-5){\makebox(0,0){$\sigma_0$}} 
\put(15,-5){\makebox(0,0){$\sigma_1$}} 
\put(30,-5){\makebox(0,0){$\sigma_2$}} 
\put(45,-5){\makebox(0,0){$\sigma_7$}} 
\put(60,-5){\makebox(0,0){$\sigma_6$}} 
\put(75,-5){\makebox(0,0){$\sigma_5$}} 
\put(90,-5){\makebox(0,0){$\sigma_4$}} 
\put(35,15){\makebox(0,0){$\sigma_3$}} 
\end{picture}} 
\end{picture} & 30   & 1,7,11,13,17,19,23,29 \\ 
{ } & { } & { }   & { }  \\ 
\hline
\end{array} 
$$ 
\caption{Diagrammes de Dynkin $ADE$, leur nombre de Coxeter $\kappa$ et leurs exposants de Coxeter $m^C$} 
\end{table}

\begin{table}[hhh] 
$$ 
\begin{array}{|c|cc|}
\hline  
{ } & { } & { }     \\ 
{ } & \textrm{diagramme} & \Gamma  \\ 
{ } & { } & { }    \\ 
\hline 
\quad A_n^{(1)} \quad& \begin{picture}(130,30)(0,20) 
\put(5,20){\begin{picture}(0,0) 
\multiput(0,0)(15,0){3}{\circle*{2}} 
\multiput(60,0)(15,0){3}{\circle*{2}} 
\put(45,15){\circle*{2}} 
\thicklines 
\multiput(30,0)(5,0){6}{\line(1,0){2}} 
\put(0,0){\line(3,1){45}} 
\put(90,0){\line(-3,1){45}} 
\put(0,0){\line(1,0){30}} 
\put(60,0){\line(1,0){30}} 
\put(0,-5){\makebox(0,0){1}} 
\put(15,-5){\makebox(0,0){1}} 
\put(30,-5){\makebox(0,0){1}} 
\put(60,-5){\makebox(0,0){1}} 
\put(75,-5){\makebox(0,0){1}} 
\put(90,-5){\makebox(0,0){1}} 
\put(45,20){\makebox(0,0){1}} 
\end{picture}} 
\end{picture} & \quad \mathcal{C}_n \quad    \\  
D_{n}^{(1)} & \begin{picture}(130,40)(0,20) 
\put(5,20){\begin{picture}(0,0) 
\multiput(15,0)(15,0){2}{\circle*{2}} 
\multiput(60,0)(15,0){2}{\circle*{2}} 
\multiput(30,0)(5,0){6}{\line(1,0){2}} 
\put(5,10){\circle*{2}} 
\put(5,-10){\circle*{2}} 
\put(85,10){\circle*{2}} 
\put(85,-10){\circle*{2}} 
\thicklines 
\put(15,0){\line(1,0){15}} 
\put(60,0){\line(1,0){15}} 
\put(15,0){\line(-1,1){10}} 
\put(15,0){\line(-1,-1){10}} 
\put(75,0){\line(1,1){10}} 
\put(75,0){\line(1,-1){10}} 
\put(0,10){\makebox(0,0){1}} 
\put(0,-10){\makebox(0,0){1}} 
\put(90,10){\makebox(0,0){1}} 
\put(90,-10){\makebox(0,0){1}} 
\put(16,-5){\makebox(0,0){2}} 
\put(74,-5){\makebox(0,0){2}} 
\put(30,-5){\makebox(0,0){2}} 
\put(60,-5){\makebox(0,0){2}} 
\end{picture}} 
\end{picture} &   \mathcal{D}_{n-2} \\  
E_6^{(1)} & \begin{picture}(130,40)(0,20) 
\put(20,20){\begin{picture}(0,0) 
\multiput(0,0)(15,0){5}{\circle*{2}} 
\multiput(30,10)(0,10){2}{\circle*{2}} 
\thicklines 
\put(0,0){\line(1,0){60}} 
\put(30,0){\line(0,1){20}} 
\put(0,-5){\makebox(0,0){1}} 
\put(15,-5){\makebox(0,0){2}} 
\put(30,-5){\makebox(0,0){3}} 
\put(45,-5){\makebox(0,0){2}} 
\put(60,-5){\makebox(0,0){1}} 
\put(35,10){\makebox(0,0){2}} 
\put(35,20){\makebox(0,0){1}} 
\end{picture}} 
\end{picture}    & \mathcal{T} \\  
E_7^{(1)} & \begin{picture}(130,40)(0,20) 
\put(5,20){\begin{picture}(0,0) 
\multiput(0,0)(15,0){7}{\circle*{2}} 
\put(60,10){\circle*{2}} 
\thicklines 
\put(0,0){\line(1,0){90}} 
\put(60,0){\line(0,1){10}} 
\put(0,-5){\makebox(0,0){1}} 
\put(15,-5){\makebox(0,0){2}} 
\put(30,-5){\makebox(0,0){3}} 
\put(45,-5){\makebox(0,0){4}} 
\put(60,-5){\makebox(0,0){3}} 
\put(75,-5){\makebox(0,0){2}} 
\put(90,-5){\makebox(0,0){1}} 
\put(60,15){\makebox(0,0){2}} 
\end{picture}} 
\end{picture}    & \mathcal{O} \\  
E_8^{(1)} & \begin{picture}(130,40)(0,20) 
\put(5,20){\begin{picture}(0,0) 
\multiput(0,0)(15,0){8}{\circle*{2}} 
\put(75,10){\circle*{2}} 
\thicklines 
\put(0,0){\line(1,0){105}} 
\put(75,0){\line(0,1){10}} 
\put(0,-5){\makebox(0,0){1}} 
\put(15,-5){\makebox(0,0){2}} 
\put(30,-5){\makebox(0,0){3}} 
\put(45,-5){\makebox(0,0){4}} 
\put(60,-5){\makebox(0,0){5}} 
\put(75,-5){\makebox(0,0){6}} 
\put(90,-5){\makebox(0,0){4}} 
\put(105,-5){\makebox(0,0){2}} 
\put(75,15){\makebox(0,0){3}} 
\end{picture}} 
\end{picture}   & \mathcal{I} \\ 
{ } & { } & { }     \\ 
{ } & { } & { }     \\ 
{ } & { } & { }     \\
\hline
 
\end{array} 
$$ 
\caption{Diagrammes de Dynkin $ADE^{(1)}$ et le sous-groupe $\Gamma \subset SU(2)$ correspondant.} 
\end{table}


\chapter{Correspondance de McKay classique et quantique} 
\thispagestyle{empty}

\section{Correspondance de McKay classique et graphes $ADE^{(1)}$ } 
 
\subsection{Le groupe $SU(2)$ classique} 
Consid\'erons le groupe $SU(2)$ et ses repr\'esentations irr\'eductibles{\footnote{Nous \'ecrirons par 
la suite ``irrep''  
comme un abr\'eg\'e ``franglais'' pour repr\'esentation irr\'eductible.}. Ce groupe poss\`ede des irreps 
de dimension $n$, not\'ees $(n)$, pour tout entier positif $n \geq 1$. L'irrep $(1)$ est l'identit\'e, 
l'irrep $(2)$ est la fondamentale. Nous pouvons former le produit tensoriel  
$(i) \otimes (j)$, et d\'ecomposer le r\'esultat en somme directe d'irreps $(k)$: 
\begin{equation} 
(i) \otimes (j) = \bigoplus_k \mathcal{N}_{ij}^{k} \; (k), \qquad \qquad \qquad \mathcal{N}_{ij}^{k} \in 
\mathbb{N},
\label{SU2} 
\end{equation} 
o\`u $\mathcal{N}_{ij}^{k}$ est la multiplicit\'e de $(k)$ dans $(i) \otimes (j)$. 
Un r\'esultat bien connu est la d\'ecomposition du produit tensoriel d'une irrep $(n)$ par la fondamentale: 
\begin{equation} 
(2) \otimes (n) = (n-1) \oplus (n+1).
\label{couplage1/2} 
\end{equation}

\begin{remarq} 
En adoptant le langage de spin pour des particules de type fermionique, bien connu des physiciens, 
l'irrep de spin $j$, $j$ demi-entier, est de dimension $n=2j+1$. La repr\'esentation fondamentale, de 
dimension deux, correspond au spin $1/2$. Alors, la formule (\ref{couplage1/2}) correspond simplement 
au couplage 
entre une particule de spin $1/2$ et une particule de spin $j$, qui donne la somme de particules 
de spins $(j-1/2)$ et $(j+1/2)$. 
\end{remarq} 
Tout irrep peut \^etre obtenue \`a partir d'une certaine puissance du produit tensoriel de la 
repr\'esentation fondamentale. En effet, en \'ecrivant: 
$$ 
(2)^{\otimes^n} = \underbrace{(2) \otimes (2) \otimes \cdots \otimes (2)}_{n}, 
$$ 
et utilisant (\ref{couplage1/2}), nous avons: 
\begin{eqnarray*} 
(2)^{\otimes^2} &=& (1) \oplus \ud{(3)} \\ 
(2)^{\otimes^3} &=& (2) \oplus (2) \oplus \ud{(4)} \\ 
(2)^{\otimes^4} &=& (1) \oplus (1) \oplus (3) \oplus (3) \oplus (3) \oplus \ud{(5)} 
\end{eqnarray*} 
En utilisant l'associativit\'e de $\otimes$ et en effectuant les calculs dans l'espace virtuel des 
repr\'esentations  
(c.\`a.d. en admettant des signes n\'egatifs dans l'\'etape interm\'ediaire des calculs), nous
 pouvons, \`a partir 
de la donn\'ee de (\ref{couplage1/2}), calculer la d\'ecomposition de $(i) \otimes (j)$. 
Par exemple, $(2)^{\otimes^2} = (1) \oplus (3)$, donc, en \'ecrivant $(3) = (2)^{\otimes^2} - (1)$, 
nous pouvons 
calculer la d\'ecomposition de $(3) \otimes (n)$, pour tout irrep $(n)$. Un exemple: 
\begin{eqnarray*} 
(3) \otimes (3) &=& \left( (2)^{\otimes^2} - (1) \right) \otimes (3)    \\ 
{ }             &=& (2) \otimes (2) \otimes (3) - (1) \otimes (3) \\ 
{ }             &=& (1) \oplus (3) \oplus (3) \oplus (5) - (3) \\ 
{ }             &=& (1) \oplus (3) \oplus (5) 
\end{eqnarray*} 
\`A la fin du calcul, nous retrouvons seulement des signes $\oplus$. 
Par la m\^eme m\'ethode, nous pouvons ainsi calculer toute d\'ecomposition $(i) \otimes (j)$, et obtenir 
ainsi les coefficients entiers non-n\'egatifs $\mathcal{N}_{ij}^{k}$ de (\ref{SU2}).

\subsection{Formulation matricielle et graphe $A_{\infty}$} 
Le r\'esultat de la d\'ecomposition (\ref{couplage1/2}) peut \^etre cod\'e par le graphe 
$A_{\infty}$ de $SU(2)$:
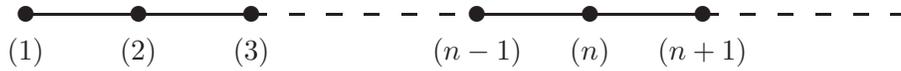
\begin{figure}[hhh] 
\unitlength=0.1cm 
\begin{center} 
\begin{picture}(130,10)(0,5) 
\thinlines 
\multiput(5,10)(15,0){3}{\circle*{2}} 
\multiput(65,10)(15,0){3}{\circle*{2}} 
\thicklines 
\multiput(35,10)(5,0){6}{\line(1,0){2}} 
\put(5,10){\line(1,0){30}} 
\put(65,10){\line(1,0){30}} 
\multiput(95,10)(5,0){6}{\line(1,0){2}} 
\put(5,5){\makebox(0,0){$(1)$}} 
\put(20,5){\makebox(0,0){$(2)$}} 
\put(35,5){\makebox(0,0){$(3)$}} 
\put(65,5){\makebox(0,0){$(n-1)$}} 
\put(80,5){\makebox(0,0){$(n)$}} 
\put(95,5){\makebox(0,0){$(n+1)$}} 
\end{picture} 
\end{center} 
\caption{Graphe $A_{\infty}$ de $SU(2)$.} 
\end{figure}

Les vertex du graphe $A_{\infty}$ sont labell\'es par les irreps $(i)$ de $SU(2)$, et  
$(2) \otimes (i)$ se d\'ecompose en la somme directe de irreps $(j)$, tel que 
$(j)$ soit voisin de $(i)$ sur le graphe (voisin dans le sens qu'un arc 
relie $(i)$ \`a $(j)$)\footnote{Il y a un arc reliant $(n)$ \`a $(n-1)$ car  
$(2) \otimes (n) = (n-1) \oplus (n+1)$, mais il y a aussi l'arc inverse reliant 
$(n-1)$ \`a $(n)$ car $(2) \otimes (n-1) = (n-2) \oplus (n)$: le graphe $A_{\infty}$ 
est donc bi-orient\'e.}.   
Soit $\mathcal{G}$ la matrice d'adjacence du graphe $A_{\infty}$, 
c.\`a.d. une matrice carr\'ee infinie telle que $(\mathcal{G})_{ij}=1$ si $(i)$ et $(j)$ 
sont voisins sur le graphe, et $(\mathcal{G})_{ij}=0$ sinon. Alors, la formule 
(\ref{couplage1/2}) s'\'ecrit aussi: 
\begin{equation}
(2) \otimes (i) = \bigoplus_j \; (\mathcal{G})_{ij} \; (j). 
\end{equation}
D'une mani\`ere g\'en\'erale, nous pouvons coder matriciellement le r\'esultat de la 
d\'ecomposition (\ref{SU2}): pour chaque irrep $(i)$, introduisons une matrice carr\'ee 
infinie $N_i$ telle que $(N_i)_{jk} = \mathcal{N}_{ij}^k$. Alors, (\ref{SU2}) s'\'ecrit aussi: 
\begin{equation}
(i) \otimes (j) = \bigoplus_k \; (N_i)_{jk} \; (k). 
\end{equation}
Nous avons $N_1 = \munite$ et $ N_2 = \mathcal{G}$. Les autres matrices $N_i$ s'obtiennent 
\`a partir de la connaissance de $N_1$ et $N_2$ par la {\bf formule de r\'ecurrence pour $SU(2)$}: 
\begin{equation} 
N_i = N_2 . N_{i-1} - N_{i-2}, \qquad \qquad \qquad \forall i \geq 3. 
\label{NSU2} 
\end{equation}  
 
\begin{conclu} 
Le graphe $A_{\infty}$ (ou sa matrice d'adjacence $\mathcal{G}=N_2$) code 
la d\'ecomposition de $(2) \otimes (i)$ en somme directe de irreps $(j)$. 
\`A partir de ces donn\'ees, nous sommes en mesure de calculer facilement toutes les d\'ecompositions 
$(i) \otimes (j)$. 
\end{conclu}

\subsection{Sous-groupes $\Gamma$ de $SU(2)$ et graphes $ADE^{(1)}$} 
Les sous-groupes finis $\Gamma$ de $SU(2)$ ont \'et\'e classifi\'es il y a plus d'un 
si\`ecle\footnote{D'une certaine fa\c{c}on cette classification a \'et\'e r\'ealis\'ee \`a l'\'epoque 
de Platon, il y a donc plus de 2000 ans!} par Felix Klein \cite{Klein}. Ils forment deux s\'eries 
infinies: $\mathcal{C}_n$ (le groupe cyclique d'ordre $n$) et 
$\mathcal{D}_n$ 
(le groupe binaire dih\'edrique d'ordre $4n$); et trois cas exceptionnels: $\mathcal{T}, \mathcal{O}$ et 
$\mathcal{I}$ 
(respectivement le groupe binaire t\'etrah\'edrique, octah\'edrique et icosah\'edrique, d'ordre 24, 48 et 
120). 
L'image $(T,O,I) \subset SO(3,\mathbb{R})$ de ces trois groupes exceptionnels sont les groupes de 
sym\'etries des cinq solides 
platoniques ($T$ pour le t\'etrah\`edre, $O$ pour l'octah\`edre et son dual le cube, et $I$ pour 
l'icosah\`edre 
et son dual le dod\'ecah\`edre). De m\^eme, l'image $(C_n, D_n) \subset SO(3,\mathbb{R})$ des groupes 
cycliques 
et binaires dih\'edriques peuvent \^etre vus comme les groupes de sym\'etrie respectivement d'une 
pyramide et d'un prisme, \`a $n$ faces. 
Pour ces sous-groupes finis $\Gamma \subset SU(2)$, nous pouvons aussi consid\'erer leurs repr\'esentations 
irr\'eductibles, qui forment maintenant un ensemble fini, not\'e $Irr(\Gamma)$, et calculer la 
d\'ecomposition 
de tout produit tensoriel: 
\begin{equation} 
(\sigma_i) \otimes (\sigma_j) = \bigoplus_k \; \mathcal{N}_{ij}^k \; (\sigma_k), \qquad \qquad \qquad 
\forall (\sigma_i),(\sigma_j),(\sigma_k) \in Irr(\Gamma). 
\label{Gamma} 
\end{equation} 
De mani\`ere parall\`ele au cas $SU(2)$, formons le produit tensoriel $(2) \otimes (\sigma_i)$, o\`u $(2)$ 
est la repr\'esentation fondamentale induite de $SU(2)$\footnote{$(2) \in Irr(\Gamma)$ pour tous les 
sous-groupes finis de $SU(2)$, \`a l'exception de $\mathcal{C}_n$, pour lequel $(2)$ est  
r\'eductible.} et d\'ecomposons le r\'esultat en somme directe de $(\sigma_j)$, o\`u $(\sigma_i)$ et 
$(\sigma_j)$ $\in Irr(\Gamma)$. 
Le r\'esultat suivant est d\^u \`a J. McKay \cite{McKay} et est connu sous le nom de 
{\bf correspondance de McKay}: 
 
\begin{theo}[McKay]
Pour tout sous-groupe fini $\Gamma \subset SU(2)$, la d\'ecomposition en irreps $(\sigma_j)$ du produit 
tensoriel 
$(2) \otimes (\sigma_i)$, o\`u $(2)$ est la repr\'esentation fondamentale induite de $SU(2)$ et 
$(\sigma_i), (\sigma_j) \in Irr(\Gamma)$, est donn\'ee par: 
\begin{equation} 
(2) \otimes (\sigma_i) = \bigoplus_j \; (\mathcal{G}_{\Gamma})_{ij} \; (\sigma_j),
\label{deuxGamma} 
\end{equation} 
o\`u $\mathcal{G}_{\Gamma}$ est la matrice d'adjacence\footnote{Il faut choisir un ordre pour repr\'esenter 
la base 
des irreps $\sigma_i$: nous prendrons toujours comme premier \'el\'ement la repr\'esentation identit\'e 
$(\sigma_1)$ et comme deuxi\`eme \'el\'ement la repr\'esentation fondamentale $(\sigma_2$).} d'un 
diagramme de Dynkin affine $ADE^{(1)}$. 
\end{theo} 
 
Les diagrammes $ADE^{(1)}$ sont illustr\'es dans l'Annexe {\bf A}. Les vertex de ces graphes 
sont labell\'es  
par les irreps $(\sigma_i)$ du sous-groupe fini $\Gamma \subset SU(2)$ correspondant.  
Pour tout $\Gamma$, la plus grande valeur propre de sa matrice  
d'adjacence $\mathcal{G}_{\Gamma}$, not\'ee  $\beta$, est \'egale \`a 2. Elle correspond \`a la 
dimension de la repr\'esentation fondamentale $(\sigma_2)$. 
Le vecteur-propre normalis\'e correspondant \`a $\beta$ est appel\'e vecteur-propre de Perron-Frobenius, 
not\'e $P$. La normalisation est telle $P(1) = 1 = \textrm{dim}(\sigma_1)$, et les composantes de ce 
vecteur donnent les dimensions des irreps correspondantes: $P(i) =$ dim$(\sigma_i)$. 
 
Nous connaissons la d\'ecomposition de $(\sigma_1) \otimes (\sigma_i)$ (l'identit\'e) 
et de $(\sigma_2) \otimes (\sigma_i)$ (par la donn\'ee du graphe ou de (\ref{deuxGamma})).  
En utilisant l'associativit\'e de $\otimes$,  
par des calculs similaires \`a ceux effectu\'es pour le cas $SU(2)$, nous pouvons alors calculer toute 
d\'ecomposition  
$(\sigma_i) \otimes (\sigma_j)$ (il faut parfois utiliser des arguments de sym\'etrie pour compl\'eter 
la table de tensorialisation: voir l'exemple du  cas $E_7^{(1)}$). Nous obtenons ainsi les entiers 
non-n\'egatifs 
${\mathcal{N}_{ij}^k}$ de (\ref{Gamma}).  
 
Soit $r$ le nombre de irreps de $\Gamma$: \`a chaque irrep $\sigma_i$ nous lui associons une matrice 
$r \times r$ $N_i$, telle que $(N_i)_{jk} = \mathcal{N}_{ij}^k$. Nous avons $N_1 = \munite_{r \times r}$ 
et $N_2 = \mathcal{G}_{\Gamma}$.  
Ayant calcul\'e la d\'ecomposition de $(\sigma_i) \otimes (\sigma_j)$, nous obtenons directement les autres  
matrices $N_i$. 
Ces $r$ matrices $N_i$ ainsi obtenues commutent toutes entre-elles: elles peuvent donc \^etre 
simultan\'ement  
diagonalis\'ees, \`a l'aide d'une matrice que nous appelerons $S$.  
Un fait remarquable est que cette matrice $S$, proprement ordonn\'ee, repr\'esente  
la table des caract\`eres de $\Gamma$ \cite{McKay-gener, Konstant}. Nous l'obtenons sans faire 
appel \`a la connaissance des classes de conjuguaison 
ni \`a la donn\'ee explicite des caract\`eres: ce r\'esultat est connu sous le nom de 
{\bf correspondance de McKay g\'en\'eralis\'ee}.

\begin{conclu} 
Il existe une correspondance entre les irreps $(\sigma_i)$ des sous-groupes finis $\Gamma$ de $SU(2)$ et 
les vertex des  
graphes $ADE^{(1)}$. Ces graphes codent la d\'ecomposition du produit tensoriel $(2) \otimes (\sigma_i)$, 
o\`u 
$(2)$ est la repr\'esentation fondamentale induite de $SU(2)$. La simple donn\'ee du graphe nous permet 
alors de  
d\'eterminer les dimensions des repr\'esentations (\`a travers les composantes du vecteur de 
Perron-Frobenius), de reconstruire la table de tensorialisation $(\sigma_i) \otimes (\sigma_j)$, et 
d'obtenir la table des caract\`eres de $\Gamma$. 
\end{conclu}

\subsection{Exemple: le groupe binaire octah\'edrique $\mathcal{O}$ et le graphe $E_7^{(1)}$} 
L'exemple du groupe binaire t\'etrah\'edrique $\mathcal{T}$ est largement trait\'e dans 
\cite{Coq-classtetra} (voir aussi \cite{Coq-classqpoly}); 
nous traiterons ici bri\`evement l'exemple du groupe binaire octah\'edrique $\mathcal{O}$. 
Consid\'erons le groupe de sym\'etrie $O$ du cube, ou de son dual l'octah\`edre. $O$ est isomorphe\footnote{Attention, 
nous parlons ici du groupe
$O$ d\'efini comme sous-groupe de $SO(3)$ (toutes les transformations ont un d\'eterminant \'egal \`a 1), et non
pas du groupe complet des sym\'etries du cube (ou de l'octah\`edre), qui est un sous-groupe de $O(3)$,
de dimension 48 (on inclut les r\'eflexions). Le groupe binaire octah\'edrique $\mathcal{O}$ que nous consid\'erons 
ici est \'egalement d'ordre 48, mais il n'est pas isomorphe au pr\'ec\'edent; de plus, c'est un sous-groupe de $SU(2)$,
non de $O(3)$.} au 
groupe sym\'etrique de permutations \`a 4 \'el\'ements, $S_4$, d'ordre $4!=24$. 
La th\'eorie des repr\'esentations des groupes 
sym\'etriques de permutation est bien connue: $S_4$ poss\`ede 5 irreps not\'ees $(\sigma_{1},\sigma_{1'},
\sigma_{2''},\sigma_{3},\sigma_{3'})$, respectivement de dimension: (1,1,2,3,3). Nous avons bien s\^ur: 
$Ordre(O) = 1^2 + 1^2 + 2^2 + 3^2 + 3^2 = 24$.
Comme sous-groupe de $SO(3)$, $O$ poss\`ede 5 irreps: ce sont aussi des irreps de sa  
pr\'e-image dans $SU(2)$, le groupe binaire octah\'edrique $\mathcal{O}$. Mais celui-ci en poss\`ede trois autres,
de dimension 2, 4 et 2.   
En tout, $\mathcal{O}$ poss\`ede 8 irreps, not\'ees $\{\sigma_1, \sigma_{1'}, \sigma_2 = f, \sigma_{2'}, 
\sigma_{2''},\sigma_3, \sigma_{3'}, \sigma_4 \}$. On v\'erifie que $Ordre(\mathcal{O})=(1^2+1^2+2^2+3^2+3^2)+(2^2+4^2+2^2)
=48$. 
Formons le graphe ayant comme vertex les irreps $\sigma_i \in Irr(\mathcal{O})$, et tel que 
$\sigma_j$ soit voisin de $\sigma_i$ si $\sigma_j$ apparait dans la d\'ecomposition de  $f \otimes 
\sigma_i$. 
Nous obtenons alors le graphe suivant, qui est le diagramme de Dynkin affin $E_7^{(1)}$ (correspondance 
de McKay): 
\begin{figure}[hhh] 
\begin{center} 
\unitlength=0.1cm 
\begin{picture}(90,20)(0,-5) 
\thinlines 
\multiput(0,0)(15,0){7}{\circle*{2}} 
\put(45,10){\circle*{2}} 
\thicklines 
\put(0,0){\line(1,0){90}} 
\put(45,0){\line(0,1){10}} 
\put(0,-5){\makebox(0,0){$\sigma_1$}} 
\put(15,-5){\makebox(0,0){$\sigma_2 = f$}} 
\put(30,-5){\makebox(0,0){$\sigma_3$}} 
\put(45,-5){\makebox(0,0){$\sigma_4$}} 
\put(60,-5){\makebox(0,0){$\sigma_{3'}$}} 
\put(75,-5){\makebox(0,0){$\sigma_{2'}$}} 
\put(90,-5){\makebox(0,0){$\sigma_{1'}$}} 
\put(45,15){\makebox(0,0){$\sigma_{2''}$}} 
\end{picture} 
\caption{Graphe $E_7^{(1)}$} 
\end{center} 
\end{figure} 
 
Par exemple, du graphe, nous lisons: 
$f \otimes \sigma_4 = \sigma_3 \oplus \sigma_{3'} \oplus \sigma_{2''}$, $f \otimes \sigma_{2''}=\sigma_4$. 
Choisissons comme ordre des irreps: $\{ \sigma_{1},\sigma_{1'},\sigma_{2},\sigma_{2'}, 
\sigma_{2''},\sigma_{3},\sigma_{3'},\sigma_{4}\}$, alors la matrice d'adjacence du graphe dans cette base 
s'\'ecrit: 
$$ 
\mathcal{G} =  
\left( 
\begin{array}{cccccccc} 
. & . & 1 & . & . & . & . & . \\ 
. & . & . & 1 & . & . & . & . \\ 
1 & . & . & . & . & 1 & . & . \\ 
. & 1 & . & . & . & . & 1 & . \\ 
. & . & . & . & . & . & . & 1 \\ 
. & . & 1 & . & . & . & . & 1 \\ 
. & . & . & 1 & . & . & . & 1 \\ 
. & . & . & . & 1 & 1 & 1 & .  
\end{array} 
\right) 
$$ 
La norme du graphe est d\'efinie comme \'etant \'egale \`a sa plus grande valeur propre, et vaut 
$\beta = 2$. Les composantes du vecteur-propre correspondant (Perron-Frobenius), normalis\'ees 
telles que $P(\sigma_1)=1$ sont: 
$$ 
P = (1,1,2,2,2,3,3,4) 
$$ 
Nous reconnaissons les dimensions des irreps. 
Par des calculs similaires \`a celui effectu\'e pour le cas $SU(2)$, nous pouvons 
remplir la table de tensorialisation des irreps. Mais il faut aussi utiliser la sym\'etrie  
$\mathbb{Z}_2$ du graphe par rapport au vertex $\sigma_4$ pour compl\'eter la table. 
Appelons $\theta$ la r\'eflexion par rapport au vertex $\sigma_4$. Alors, $\theta (\sigma_i) = \sigma_{i'}$ 
pour $i \in \{1,2,3\}$, et $\theta (\sigma_4) = \sigma_4$, $\theta(\sigma_{2''}) = \sigma_{2''}$, et 
nous avons: 
$$ 
\sigma_i \otimes \theta (\sigma_j) = \theta (\sigma_i) \otimes \sigma_j,  \qquad \qquad 
\sigma_i \otimes \sigma_j = \theta (\sigma_i) \otimes \theta (\sigma_j).  
$$  
Nous pouvons alors compl\'eter la table de tensorialisation, pr\'esent\'ee dans la Tab.\ref{tabbinocta} 
(pour une meilleure 
visualisation, 
les irreps $\sigma_i$ sont indiqu\'ees simplement par leur indice $i$, et $\oplus$ est remplac\'e par le 
signe +): 
 
\begin{table}[hhh] 
\scriptsize 
$$ 
\begin{array}{|c||ccccc|ccc|} 
\hline 
\otimes & 1 & 1' & 2'' & 3 & 3' & 2 & 2' & 4 \\ 
\hline 
\hline 
1 & 1 & 1' & 2'' & 3 & 3' & 2 & 2' & 4  \\ 
1' & 1' & 1 & 2'' & 3' & 3 & 2' & 2 & 4 \\ 
2'' & 2'' & 2'' & 1+1'+2'' & 3+3' & 3+3' & 4 & 4 & 2+2'+4 \\ 
3 & 3 & 3' & 3+3'& 1+2''+3+3' & 1'+2''+3+3' & 2+4 & 2'+4 & 2+2'+4+4 \\ 
3' & 3' & 3 & 3+3'& 1'+2''+3+3' & 1+2''+3+3' & 2'+4 & 2+4 & 2+2'+4+4 \\ 
\hline 
2 & 2 & 2' & 4 & 2+4 & 2'+4 & 1+3 & 1'+3' & 2''+3+3' \\ 
2' & 2' & 2 & 4 & 2'+4 & 2+4 & 1'+3' & 1+3 & 2''+3+3' \\ 
4 & 4 & 4 & 2+2'+4 & 2+2'+4+4 & 2+2'+4+4 & 2''+3+3' & 2''+3+3' & 1+1'+2''+3+3+3'+3' \\ 
\hline 
\end{array} 
$$ 
\caption{D\'ecomposition du produit tensoriel $\sigma_i \otimes \sigma_j$ pour les irreps $\sigma$ du groupe 
binaire octah\'edrique $\mathcal{O}$.} 
\label{tabbinocta}
\end{table} 
\normalsize 
Dans cette table, nous avons visuellement s\'epar\'e des trois autres les irreps $(\sigma_1, \sigma_{1'}, \sigma_{2''}, 
\sigma_3, 
\sigma_{3'})$ qui sont aussi des irreps de $O$, et qui se d\'ecomposent entre-elles.   
\`A partir de cette table, nous obtenons imm\'ediatement les huit matrices $8\times 8$ $N_i$ qui codent la  
d\'ecomposition dans $\mathcal{O}$ (et aussi les cinq matrices $5 \times 5$ codant la d\'ecomposition  
dans $O$). Proprement ordonn\'ees, les matrices $S(O)$ et $S(\mathcal{O})$ qui 
diagonalisent ces matrices sont donn\'ees par: 
$$ 
S(O) =  
\left( 
\begin{array}{rrrrr} 
 1 & 1 & 1 & 1 & 1 \\ 
 1 & -1 & 1 & 1 & -1 \\ 
 2 & 0 & 2 & -1 & 0 \\ 
 3 & 1 & -1 & 0 & -1 \\ 
 3 & -1 & -1 & 0 & 1  
\end{array} 
\right) 
\qquad  
S(\mathcal{O}) =  
\left( 
\begin{array}{rrrrrrrr} 
1 & 1 & 1 & 1 & 1 & 1 & 1 & 1 \\ 
1 & 1 & -1 & 1 & 1 & 1 & -1 & -1 \\ 
2 & 2 & 0 & -1 & 2 & -1 & 0 & 0 \\ 
2 & -2 & 0 & -1  & 0 & 1 & \sqrt2 & -\sqrt2 \\ 
2 & -2 & 0 & -1 & 0 & 1 & -\sqrt2 & \sqrt2 \\ 
3 & 3 & -1 & 0 & -1 & 0 & 1 & 1 \\ 
3 & 3 & 1 & 0 & -1 & 0 & -1 & -1 \\ 
4 & -4 & 0 & 1 & 0 & -1 & 0 & 0  
\end{array} 
\right) 
$$ 
Nous pouvons v\'erifier que $S(O)$ repr\'esente la table des caract\`eres de $O$ \cite{IMPA} et 
$S(\mathcal{O})$ 
repr\'esente celle de $\mathcal{O}$ \cite{Frame, McKay-gener}. Nous les obtenons facilement 
sans faire appel aux classes 
de conjuguaison ou \`a la donn\'ee explicite des caract\`eres. 
 
\subsection{$\widehat{\Gamma}$ comme module sur $\widehat{SU}(2)$ et r\`egles de branchement 
$SU(2) \hookrightarrow \Gamma$} 
Soient $(i)$ une irrep de $SU(2)$ et $(\sigma_j)$ une irrep de $\Gamma \subset SU(2)$. Nous avons vu que 
$(i)$ peut \^etre obtenue \`a partir de l'irrep fondamentale (2) de $SU(2)$. D'autre part, le graphe 
$ADE^{(1)}$ correspondant \`a $\Gamma$ code la d\'ecomposition de $(2) \otimes (\sigma_i)$ en irreps 
$(\sigma_j)$ de $\Gamma$. Nous sommes donc en mesure de calculer facilement toute d\'ecomposition de la 
forme: 
\begin{equation} 
(i) \otimes (\sigma_j) = \bigoplus_k \; \mathcal{F}_{ij}^k \; (\sigma_k), 
\label{SU2_Gamma} 
\end{equation} 
o\`u $\mathcal{F}_{ij}^k $ est la multiplicit\'e de $(\sigma_k)$ dans $(i) \otimes (\sigma_j)$. 
En d'autres termes, nous dirons que les irreps $(\sigma_i)$ de $\Gamma$ forment un module sous l'action 
(\ref{SU2_Gamma}) des irreps $(i)$ de $SU(2)$.  
 
Nous savons tensorialiser les irreps $(\sigma_i)$ de $\Gamma$ entre-elles: l'action (\ref{SU2_Gamma}) de 
$SU(2)$ sur $\Gamma$ peut aussi \^etre obtenue par les r\`egles de branchement $SU(2) \hookrightarrow 
\Gamma$. 
La tensorialisation par $(1)$ et $(2)$ est imm\'ediate et nous donne les r\`egles de branchement $(1) 
\hookrightarrow(\sigma_1)$ et $(2) \hookrightarrow (\sigma_2)$. 
Pour obtenir les autres, il suffit de comparer les puissances des repr\'esentations 
fondamentales $(2)$ et $(\sigma_2)$.  
Prenons l'exemple de $E_7^{(1)}$. Nous avons: 
\small 
$$ 
\begin{array}{cclccl} 
(2)^{\otimes 2} &=& (1) \oplus (3)  & (\sigma_2)^{\otimes 2} &=& (\sigma_1) \oplus (\sigma_3)   \\ 
(2)^{\otimes 3} &=& (2) \oplus (2) \oplus (4) & (\sigma_2)^{\otimes 3} &=& (\sigma_2) \oplus (\sigma_2) 
\oplus (\sigma_4) \\ 
(2)^{\otimes 4} &=& (1) \oplus (1) \oplus (3) \oplus (3) \oplus (3) \oplus (5)  \;\;  & (\sigma_2)^{\otimes 
4} &=&  
(\sigma1) \oplus (\sigma_1) \oplus (\sigma_3) \oplus (\sigma_3) \oplus (\sigma_{2''}) \oplus 
(\sigma_{3'})    
\end{array} 
$$ 
\normalsize 
Nous en d\'eduisons les r\`egles de branchement suivantes: 
$$ 
(3) \hookrightarrow (\sigma_3), \qquad \qquad 
(4) \hookrightarrow (\sigma_4), \qquad \qquad 
(5) \hookrightarrow (\sigma_{2''}) \oplus (\sigma_{3'}).  
$$ 
\`A partir de la connaissance des r\`egles de branchement $SU(2) \hookrightarrow \Gamma$ et de la 
tensorialisation 
des irreps de $\Gamma$, nous pouvons alors calculer 
toute d\'ecomposition $(i) \otimes (\sigma_k)$ en irreps $\sigma_l$. Si $(i) \hookrightarrow \oplus_j\, 
(\sigma_j)$,  
alors $(i) \otimes (\sigma_k) = \oplus_j (\sigma_j) \otimes (\sigma_k)$. 
\begin{conclu} 
Soient $(i)$ une irrep de $SU(2)$ et $(\sigma_j)$ une irrep de $\Gamma \subset SU(2)$. 
Les r\`egles de branchement $SU(2) \hookrightarrow \Gamma$ nous permettent de calculer la d\'ecomposition de 
$(i) \otimes (\sigma_j)$ en irreps $\sigma_k$. Nous dirons par la suite que les irreps de $\Gamma$ forment 
un module sous l'action des irreps de $SU(2)$, ou plus simplement que $\widehat{\Gamma}$ est un module 
sur $\widehat{SU}(2)$. 
\end{conclu} 
 
\section{Correspondance de McKay quantique et graphes $ADE$} 
Nous voulons maintenant g\'en\'eraliser les r\'esultats pr\'ec\'edents au cas ``quantique''.

\subsection{Le groupe quantique $U_q(sl(2))$} 
L'alg\`ebre $U_q(sl(2))$ est l'alg\`ebre engendr\'ee par les \'el\'ements $\{ K, K^{-1}, X_{\pm}\}$ 
satisfaisant les  
relations suivantes: 
\begin{eqnarray*} 
K \cdot K^{-1}  &=& K^{-1} \cdot K = 1  \\ 
K \cdot X_{\pm} &=& q^{\pm 2} X_{\pm} \cdot K \\ 
\left[ X_+, X_- \right]   &=& \frac{K - K^{-1}}{q - q^{-1}}   
\end{eqnarray*} 
D\'efinissant la comultiplication ($\Delta$), la counit\'e ($\epsilon$) et l'antipode ($S$) sur les  
g\'en\'erateurs par: 
$$ 
\begin{array}{cclccl} 
\Delta (X_+) &=& X_+ \otimes 1 + K \otimes X_+ \qquad \qquad & \Delta (K) &=& K \otimes K \\ 
\Delta (X_-) &=& X_- \otimes K^{-1} + 1 \otimes X_- \qquad \qquad & \Delta (K^{-1}) &=& K^{-1} \otimes 
K^{-1} \\ 
S (X_+) &=& -K^{-1} \cdot X_+  \qquad \qquad & S (K) &=& K^{-1}  \\ 
S (X_-) &=& -X_{-} \cdot K  \qquad \qquad & S (K^{-1}) &=& K  \\ 
\epsilon (X_{\pm}) &=& 0 \qquad \qquad & \epsilon (K^{\pm 1}) &=& 1   
\end{array} 
$$ 
munissent $U_q(sl(2))$ d'une structure d'alg\`ebre de Hopf \cite{kassel} (pour une d\'efinition 
g\'en\'erale des relations  
d\'efinissant une alg\`ebre de Hopf, voir l'Annexe {\bf C}). Nous dirons aussi par la suite que $U_q(sl(2))$ 
est un 
groupe quantique. 
Une mani\`ere \'el\'egante d'obtenir les relations pr\'ec\'edentes est de voir $U_q(sl(2))$ comme le dual 
de $Fun_q(SL(2))$. Consid\'erons deux \'el\'ements $x$ et $y$, satisfaisant la relation suivante, appel\'ee  
{\bf relation de plan quantique}: 
\begin{xalignat}{2} 
xy &= qyx,  &\qquad  q &\in \mathbb{C}.  
\label{plan_q} 
\end{xalignat}   
Alors $Fun_q(SL(2))$ est l'alg\`ebre des transformations de coordonn\'ees, de d\'eterminant \'egal \`a 1, 
qui pr\'eservent 
la relation (\ref{plan_q}). Les relations d\'efinissant la structure d'alg\`ebre de Hopf de $Fun_q(SL(2))$ 
sont d\'efinies 
de mani\`ere naturelle sur ses g\'en\'erateurs. $Fun_q(SL(2))$ et $U_q(sl(2))$ sont duales (suivant la 
notion de 
dualit\'e introduite par M. Takeushi \cite{takeu}) dans le sens qu'il existe une forme bilin\'eaire  
$\scal{}{} \rightarrow \mathbb{C}$, appel\'ee {\it pairing}, reliant leurs structures. La donn\'ee du 
{\it pairing} entre  
$U_q(sl(2))$ et $Fun_q(SL(2))$ et des relations de $Fun_q(SL(2))$ permettent de retrouver de mani\`ere 
\'el\'egante 
les relations d\'efinissant $U_q(sl(2))$ \cite{DEA}. 
Notons que ces deux groupes quantiques -- $U_q(sl(2))$ et $Fun_q(SL(2))$ -- sont de dimension infinie.

\subsection{Quotient de $U_q(sl(2))$ et graphe $A_n$} 
Consid\'erons maintenant la relation de plan quantique dans le cas o\`u $q$ est une 
racine N$^{\text{i\`eme}}$ 
de l'unit\'e: $q^{N} = 1, q \neq 1$.  Le {\bf plan quantique r\'eduit} est d\'efini par la relation 
(\ref{plan_q}) 
et les relations suivantes: 
\begin{xalignat}{2} 
x^{N} &= 1, &\qquad y^{N} &= 1. \label{plqr} 
\end{xalignat} 
L'alg\`ebre des transformations de coordonn\'ees, de d\'eterminant \'egal \`a 1, qui pr\'eservent 
les relations de plan quantique r\'eduit est appel\'ee $\mathcal{F}$. $\mathcal{F}$ est obtenue en 
quotientant 
$Fun_q(SL(2))$ par des id\'eaux bilat\`eres, qui sont des id\'eaux de Hopf. $\mathcal{F}$ est donc aussi  
un groupe quantique, mais de dimension finie: $dim(\mathcal{F}) = N^3$.  
Le dual de $\mathcal{F}$, appel\'e $\mathcal{H}$ (et souvent not\'e $u_q(sl(2))$ dans la litt\'erature), 
est aussi un groupe quantique de dimension finie. Il est 
obtenu en quotientant $U_q(sl(2))$ par les id\'eaux bilat\`eres engendr\'es par les relations suivantes 
\footnote{Ces relations de quotient sont strictement valables pour le cas $N$ impair. Pour le cas N pair,  
elles d\'ependent de la parit\'e de $\frac{N}{2}$, et la discussion est un peu plus d\'elicate (pour une  
discussion de ce probl\`eme, voir \cite{glusch1}).}: 
\begin{equation} 
K^{N}=\munite, \qquad \qquad X_+^{N}=0, \qquad  \qquad X_-^{N}=0. 
\end{equation} 
L'\'etude de la repr\'esentation r\'eguli\`ere (\`a gauche) de $U_q(sl(2))$ , avec q racine de  
l'unit\'e, est pr\'esent\'ee dans \cite{glusch1}.  
Une autre mani\`ere d'\'etudier les repr\'esentations de $\mathcal{H}$ est pr\'esent\'ee 
dans \cite{DEA,Coq_Gil-Lodz}, \`a travers un isomorphisme entre $\mathcal{H}$ et un 
groupe quantique construit \`a partir  
de variables anticommutantes \cite{oleg} (variables de Grassmann). 
Cet isomorphisme a \'et\'e explicitement d\'ecrit dans \cite{m3m21} pour le cas N=3 (voir aussi \cite{Coq_Gil-Lodz}), 
et de fa\c{c}on plus g\'en\'erale dans une des sections de \cite{oleg}.

\subsubsection{Exemple: cas N=5} 
Soit $\mathcal{H}$ le quotient de $U_q(sl(2))$, pour $q^5=1$. Alors, $\mathcal{H}$ est isomorphe 
\`a \cite{oleg}: 
\begin{equation*} 
\mathcal{H} \cong M_5 \oplus \Mlo{4 | 1}  \oplus \Mlo{3 | 2} = \Mie{\Lambda}{5} 
\end{equation*}     
Un \'el\'ement $h$ de \Mie{\Lambda}{5} est de la forme suivante: 
\begin{equation*} 
h =   
\bmatr \ast & \ast & \ast & \ast & \ast \\ 
       \ast & \ast & \ast & \ast & \ast \\ 
       \ast & \ast & \ast & \ast & \ast \\ 
       \ast & \ast & \ast & \ast & \ast \\ 
       \ast & \ast & \ast & \ast & \ast   \ematr 
\oplus 
\bmatr \bullet & \bullet & \bullet & \bullet & \circ \\ 
       \bullet & \bullet & \bullet & \bullet & \circ \\ 
       \bullet & \bullet & \bullet & \bullet & \circ \\ 
       \bullet & \bullet & \bullet & \bullet & \circ \\ 
       \circ   & \circ   & \circ   & \circ   & \bullet     \ematr 
\oplus 
\bmatr \bullet & \bullet & \bullet & \circ   & \circ \\ 
       \bullet & \bullet & \bullet & \circ   & \circ \\ 
       \bullet & \bullet & \bullet & \circ   & \circ \\ 
       \circ   & \circ   & \circ   & \bullet & \bullet \\ 
       \circ   & \circ   & \circ   & \bullet & \bullet     \ematr 
\end{equation*} 
o\`u nous avons introduit les notations suivantes: 
\begin{itemize} 
\item[-] $\ast$ pour un \'el\'ement de $\mathbb{C}$ 
\item[-] $\bullet$ pour un \'el\'ement de la forme $\alpha + \beta \theta_1\theta_2$ 
$\qquad \quad \alpha, \beta \in \mathbb{C}$  
\item[-] $\circ$ pour un \'el\'ement de la forme $\gamma \theta_1 + \delta \theta_2$ 
$\qquad \quad \gamma, \delta \in \mathbb{C}$ 
\end{itemize} 
et o\`u $\theta_1$ et $\theta_2$ sont deux \'el\'ements (variables de Grassmann) qui v\'erifient les 
relations suivantes: 
\begin{xalignat}{2} 
\theta_1^{2} &= \theta_2^{2}  &\qquad  \theta_1 \theta_2 &= - \theta_2 \theta_1 
\end{xalignat} 
Cet isomorphisme permet de construire les repr\'esentations de $\mathcal{H}$, et notamment d'obtenir ses 
repr\'esentations irr\'eductibles. Si nous n\'egligeons les repr\'esentations
de q-dimension nulle (ce sont les repr\'esentations projectives ind\'ecomposables) et que nous
dessinions le diagramme de tensorialisation par la repr\'esentation de dimension 2, nous obtenons
alors le graphe $A_4$ \cite{Coq-qtetra}. Ce r\'esultat se g\'en\'eralise du moins pour le cas $N$ impair, 
et nous obtenons ainsi le graphe $A_{N-1}$.
 
\subsection{``Sous-groupes'' finis de $U_q(sl(2))$ et graphes $ADE$} 
Le groupe quantique $U_q(sl(2))$ poss\`ede, pour $q^N=1$, des quotients de Hopf de 
dimension finie: ces quotients sont des alg\`ebres non semi-simples mais poss\`edent des 
repr\'esentations irr\'eductibles en nombre fini. L'une d'entre-elles, de dimension classique $N$,
est de q-dimension nulle, les autres sont de q-dimension non nulle. 
De la m\^eme mani\`ere que les irreps de $SU(2)$ sont labell\'ees par les vertex du graphe $A_{\infty}$, 
le quotient de $U_q(sl(2))$ pour $q^N=1$ poss\`ede des irreps de q-dimension non nulle labell\'ees par 
les vertex du graphe 
$A_{N-1}$, not\'es $(\tau_0, \tau_1, \cdots, \tau_{N-2})$: ce graphe code la tensorialisation  
des irreps $\tau_i$ par la fondamentale $\tau_1$, de q-dimension 2. 
Nous avons donc un analogue quantique du cas $SU(2)$, les graphes $A_{N-1}$ pouvant \^etre vus comme  
la correspondance quantique du graphe $A_{\infty}$. 
 
Nous voudrions, d'une mani\`ere analogue \`a ce qui a \'et\'e vu pour le cas du groupe $SU(2)$, classifier 
les ``sous-groupes'' finis de $U_q(sl(2))$. \'Evidemment, le probl\`eme ici est tr\`es diff\'erent,  
$U_q(sl(2))$ n'\'etant pas un groupe, il faut trouver une formulation ad\'equate de ce probl\`eme. 
Une pr\'esentation de la correspondance de McKay quantique, utilisant le langage 
des cat\'egories, est pr\'esent\'ee dans \cite{qMcKay}.  \\

\noindent Rappelons ici les r\'esultats \'enonc\'es dans le chapitre {\bf 3}): 

\begin{itemize} 
\item[$\bullet$] Les repr\'esentations irr\'eductibles $\sigma$ des ``sous-groupes'' finis de 
$U_q(sl(2))$ sont labell\'ees 
par les vertex des diagrammes de Dynkin $A_n$, $D_{2n}$, $E_6$ et $E_8$. Les irreps $\sigma$ peuvent \^etre 
tensorialis\'ees entre-elles, nous dirons que les cas en question poss\`edent la propri\'et\'e
de {\it self-fusion}.

\item[$\bullet$] \`A c\^ot\'e de ces ``sous-groupes'', il existe aussi des ``modules'', pour lesquels 
les irreps peuvent \^etre tensorialis\'ees par les irreps du quotient de $U_q(sl(2))$ 
correspondant, mais qui ne peuvent pas \^etre tensorialis\'ees entre-elles. 
Les ``modules'' qui ne sont pas des ``sous-groupes'' sont d\'ecrits par les diagrammes $D_{2n+1}$ et $E_7$. 
\end{itemize}

\noindent Les diagrammes $ADE$, apparaissant ainsi dans l'analogue quantique de la correspondance de McKay, 
sont illustr\'es dans l'Annexe {\bf A}.  \\

Aucun lien direct entre les r\'esultats que nous venons de rappeler (li\'es aux groupes quantiques aux racines
de l'unit\'e) et les big\`ebres $\mathcal{B}(G)$ (li\'ees \`a des chemins sur des diagrammes de Coxeter-Dynkin)
n'est connu \`a ce jour. C'est une direction qu'il serait int\'eressant d'explorer.


\chapter{Quelques d\'efinitions alg\'ebriques}
\thispagestyle{empty}
 
\section{Alg\`ebre de Hopf} 
Une alg\`ebre $(A,+,\cdot ;k)$ sur un corps $k$ est 
un espace vectoriel sur $k$, muni d'un produit 
associant \`a tout couple $(a,b) \in A \times A$ un \'el\'ement $a \cdot b$
de $A$, de mani\`ere compatible avec l'addition:
\begin{equation*}
a \cdot (b+c) = (a\cdot b) + (a\cdot c) \hspace{2cm} \forall a,b,c 
\in A
\end{equation*}
et avec l'action de $k$:
\begin{equation*}
\alpha(a\cdot b)=(\alpha a) \cdot b = a \cdot (\alpha b) \hspace{2cm}
\forall a,b \in A, \;\; \forall \alpha \in k
\end{equation*}
Ces deux propri\'et\'es peuvent \^etre r\'esum\'ees en imposant 
que le produit -- qui sera aussi not\'e $\mu$ -- soit une application lin\'eaire
de $A \otimes A \longrightarrow A$. Nous avons: $\mu(a \otimes b) = a \cdot b$, pour $a,b \in A$. 
De m\^eme, l'existence d'un neutre $1_A \in A$ pour le produit se traduit par
l'existence d'une application lin\'eaire $\eta :k \longrightarrow A$ telle que
$\eta (1)=1_A$. Par la suite, nous allons consid\'erer des alg\`ebres associatives 
et unitales. Le fait que le produit soit associatif 
$(a\cdot (b\cdot c)=((a \cdot b)\cdot c)$ et poss\`ede un neutre \`a gauche 
et \`a droite $(a \cdot 1_A = 1_A \cdot a=a)$ s'\'ecrit sous la forme de 
diagrammes commutants, introduits par Manin \cite{manin}.

\unitlength 0.05cm

$$
\begin{array}{lcc}
\parbox{120pt}{\begin{picture}(100,50)
\thicklines

\put(0,0){\makebox(0,0){$A \otimes A$}}
\put(100,0){\makebox(0,0){$A$}}
\put(0,50){\makebox(0,0){$A \otimes A \otimes A$}}
\put(100,50){\makebox(0,0){$A \otimes A$}}

\put(30,0){\vector(1,0){40}}
\put(30,50){\vector(1,0){40}}
\put(0,40){\vector(0,-1){30}}
\put(100,40){\vector(0,-1){30}}

\put(50,55){\makebox(0,0){$\mu \otimes \id$}}
\put(50,-5){\makebox(0,0){$\mu$}}
\put(-15,27){\makebox(0,0){$\id \otimes \mu$}}
\put(107,27){\makebox(0,0){$\mu$}}
\end{picture}}
& 
\qquad \qquad \qquad \qquad & \textrm{associativit\'e} \\[2cm]
\end{array}
$$
$$
\begin{array}{lcc}

\parbox{120pt}{\begin{picture}(100,55)
\thicklines

\put(50,0){\makebox(0,0){$A$}}
\put(50,50){\makebox(0,0){$A \otimes A$}}
\put(0,50){\makebox(0,0){$k \otimes A$}}
\put(100,50){\makebox(0,0){$A \otimes k$}}

\put(15,50){\vector(1,0){20}}
\put(85,50){\vector(-1,0){20}}
\put(50,42){\vector(0,-1){34}}
\put(8,42){\vector(1,-1){35}}
\put(92,42){\vector(-1,-1){35}}

\put(75,57){\makebox(0,0){$\id \otimes \eta$}}
\put(25,57){\makebox(0,0){$\eta \otimes \id$}}
\put(55,27){\makebox(0,0){$\mu$}}
\put(20,20){\makebox(0,0){$\cong$}}
\put(79,20){\makebox(0,0){$\cong$}}
\end{picture}}
& 
\qquad \qquad \qquad \qquad & \qquad \textrm{unit\'e} \qquad \\[1.5cm]
\end{array}
$$
L'\'ecriture de ces diagrammes a \'et\'e faite pour faciliter l'introduction  
de la notion de cog\`ebre.
Une cog\`ebre est un espace vectoriel, muni d'applications 
lin\'eaires $\Delta : A \longrightarrow A \otimes A$ et 
$\epsilon :A \longrightarrow k$, qui v\'erifient les propri\'et\'es de 
coassociativit\'e et de counit\'e obtenues en inversant le sens des 
fl\`eches dans les diagrammes pr\'ec\'edents.\\

$$
\begin{array}{lcc}
\parbox{120pt}{\begin{picture}(100,50)
\thicklines

\put(0,0){\makebox(0,0){$A \otimes A$}}
\put(100,0){\makebox(0,0){$A\otimes A \otimes A$}}
\put(0,50){\makebox(0,0){$A $}}
\put(100,50){\makebox(0,0){$A \otimes A$}}

\put(30,0){\vector(1,0){40}}
\put(30,50){\vector(1,0){40}}
\put(0,40){\vector(0,-1){30}}
\put(100,40){\vector(0,-1){30}}

\put(50,55){\makebox(0,0){$\Delta$}}
\put(50,-5){\makebox(0,0){$\id \otimes \Delta$}}
\put(-7,27){\makebox(0,0){$\Delta$}}
\put(115,27){\makebox(0,0){$\Delta \otimes \id$}}
\end{picture}}
& 
\qquad \qquad \qquad \qquad & \textrm{coassociativit\'e} \\[2.6cm]

\parbox{120pt}{\begin{picture}(100,50)
\thicklines

\put(50,0){\makebox(0,0){$A$}}
\put(50,50){\makebox(0,0){$A \otimes A$}}
\put(0,50){\makebox(0,0){$k \otimes A$}}
\put(100,50){\makebox(0,0){$A \otimes k$}}

\put(35,50){\vector(-1,0){20}}
\put(65,50){\vector(1,0){20}}
\put(50,8){\vector(0,1){34}}
\put(42,8){\vector(-1,1){34}}
\put(58,8){\vector(1,1){34}}

\put(75,57){\makebox(0,0){$\id \otimes \epsilon$}}
\put(25,57){\makebox(0,0){$\epsilon \otimes \id$}}
\put(55,27){\makebox(0,0){$\Delta$}}
\put(20,20){\makebox(0,0){$\cong$}}
\put(79,20){\makebox(0,0){$\cong$}}
\end{picture}}
& 
\qquad \qquad \qquad \qquad & \textrm{counit\'e} \\[1.5cm] 
\end{array}
$$

\noindent Nous pouvons maintenant \'enoncer les d\'efinitions d'alg\`ebre et de cog\`ebre de mani\`ere concise:

\begin{defin} 
Une {\bfseries alg\`ebre} est un triplet $(A,\mu,\eta)$, o\`u $A$ est  
un espace vectoriel, et: 
$$ 
\mu:A \otimes A \longrightarrow A \; ,\hspace{2cm} \eta: k \longrightarrow A \; ,
$$ 
sont des applications lin\'eaires, appel\'ees produit et unit\'e,  
satisfaisant les relations suivantes: 
$$
\begin{array}{rclcl} 
\mu \circ (\mu \otimes \id\,) &=& \mu \circ (\id \otimes \mu)  & & \\ 
\mu \circ (\eta \otimes \id\,) &=& \mu \circ (\id \otimes \eta)  &=& \id 
\end{array} 
$$
\end{defin} 
 
\begin{defin} 
Une {\bfseries cog\`ebre} est un triplet $(A,\Delta,\epsilon)$,  
o\`u $A$ est un espace vectoriel, et: 
$$ 
\Delta:A \longrightarrow A \otimes A \; ,  \hspace{2cm} \epsilon: A \longrightarrow k  \; ,
$$ 
sont des applications lin\'eaires, appel\'ees coproduit et counit\'e,  
satisfaisant les relations suivantes: 
$$
\begin{array}{rclcl} 
(\Delta \otimes  \id \,)\circ \Delta &=&(\id \otimes \Delta)  \circ   
\Delta & & \\ 
(\epsilon \otimes  \id \,) \circ \Delta &=& (\id \otimes  \, \epsilon \,) \circ   
\Delta &=& \id
\end{array}
$$ 
\end{defin} 
 
\noindent {\bf Notation de Sweedler}: nous introduisons une convention de notation, 
introduite par Sweedler \cite{sweedler}, tr\`es utile pour la clart\'e des calculs. Soit $A$
une cog\`ebre et un \'el\'ement $a \in A$. Nous avons $\Delta(a) \in A \otimes A$, que nous notons:
\begin{eqnarray*}
\Delta(a) &=& \sum_i a_{(1)_{(i)}} \otimes a_{(2)_{(i)}} \\
          &=& a_{(1)} \otimes a_{(2)}  \hspace{2cm} \textrm{sommation implicite}
\end{eqnarray*}

\begin{defin}
Le produit tensoriel de deux alg\`ebres $A_1$ et $A_2$
est une alg\`ebre dont l'espace vectoriel est le produit tensoriel des espaces
vectoriels de $A_1$ et de $A_2$, et dont le produit $\mu': (A \otimes A) \otimes (A \otimes A) \longrightarrow A \otimes A$ 
est une application lin\'eaire telle que:
$$
\mu' = (\mu \circ \mu) \circ (\id \otimes \tau \otimes \id)\; ,
$$
o\`u $\tau$ est l'op\'erateur de flip: $\tau(a \otimes b) = b \otimes a$.
Nous pouvons \'ecrire le produit plus simplement comme:
$$
(a \otimes b) \cdot (c \otimes d) = (a \cdot c) \otimes (c \cdot d) \; ,
$$

\end{defin}

\begin{defin}
Le produit tensoriel de deux cog\`ebres $A_1$ et $A_2$ est une 
cog\`ebre dont l'espace vectoriel est le produit tensoriel des 
espaces vectoriels de $A_1$ et de $A_2$, et dont le coproduit $\Delta': A \otimes A \longrightarrow 
A \otimes A \otimes A \otimes A$ est une application lin\'eaire telle que:
\begin{eqnarray*}
\Delta' &=& (\id \otimes \tau \otimes \id) \circ (\Delta \otimes \Delta) \\
i.e. \qquad \qquad \qquad  \Delta'(a \otimes b) &=& a_{(1)} \otimes b_{(1)} \otimes a_{(2)} \otimes b_{(2)} 
\qquad \qquad \qquad\end{eqnarray*}
\end{defin}

\noindent Par abus de langage, nous noterons souvent le produit et le coproduit $(\mu', \Delta')$ aussi par 
$(\mu, \Delta)$.  

\begin{theo} 
Supposons que A poss\`ede une structure d'alg\`ebre ($A, \mu, \eta$) et une structure 
de cog\`ebre ($A, \Delta, \epsilon$). Alors, les deux assertions suivantes sont  
\'equivalentes \cite{kassel}: 
\begin{itemize} 
\item $\mu$ et $\eta$ sont des morphismes de cog\`ebres. 
\item $\Delta$ et $\epsilon$ sont des morphismes d'alg\`ebres. 
\end{itemize} 
\end{theo} 

\noindent Par exemple, pour satisfaire la deuxi\`eme assertion, il faut v\'erifier que:
$$
\begin{array}{rclcrcl}
\Delta(a \cdot b) &=& \Delta(a) \cdot \Delta(b)  &\qquad \qquad \qquad&  \Delta (1_A) &=& 1_A \otimes 1_A \\
\epsilon \,(a \cdot b) &=& \; \epsilon \,(a) \cdot \epsilon \,(b) &\qquad& \epsilon \, (1_A) &=& 1   
\end{array}
$$

\begin{defin} 
Une {\bfseries big\`ebre} est un quintuple $(A,\mu,\eta,\Delta, 
\epsilon)$, o\`u $(A,\mu,\eta)$ est une alg\`ebre,  
$(A,\Delta,\epsilon)$ est une cog\`ebre, et qui  
v\'erifie une des deux conditions \'equivalentes pr\'ec\'edentes. 
\end{defin} 
Etant donn\'ees une alg\`ebre $(B,\mu,\eta)$ et une cog\`ebre  
$(C,\Delta,\epsilon)$, consid\'erons deux applications lin\'eaires 
f et g de $C$ vers $B$. Nous d\'efinissons alors la {\bfseries convolution} 
$f \star g$ qui est la composition des applications suivantes: 
$C \stackrel{\Delta}{\longrightarrow} C \otimes C  \stackrel{f\otimes g}{\longrightarrow} B \otimes B  
\stackrel{\mu}{\longrightarrow} B$, qui s'\'ecrit: 
\begin{equation*} 
f \star g = \mu \circ (f \pv g) \circ \Delta 
\end{equation*} 
Lorsque nous avons une big\`ebre \mbox{$(A,\mu,\eta,\Delta,\epsilon)$}, nous pouvons  
consid\'erer le cas \mbox{$B=C=A$} et d\'efinir la convolution sur  
l'espace vectoriel des endomorphismes de $A$: 
\begin{defin} 
Un endomorphisme $S:A \longrightarrow A$ est appel\'e {\bfseries antipode}, si: 
\begin{equation*} 
\begin{array}{crclcl} 
{} & S \star \id &=& \id \star S &=& \eta \circ \epsilon\\ 
i.e.  \qquad \qquad & \mu \circ (S \otimes \id) \circ \Delta &=& \mu \circ
(\id \otimes S) \circ \Delta &=& \eta \circ \epsilon \qquad \qquad
\end{array} 
\end{equation*} 
\end{defin} 

\noindent Une big\`ebre avec antipode est une big\`ebre de Hopf, comun\'ement appel\'ee {\bfseries alg\`ebre de Hopf}. 

 
\begin{defin} 
Le {\bfseries dual} d'une alg\`ebre de Hopf $(A, \mu, \eta, \Delta, \epsilon)$ est l'espace dual 
$\widehat{A} = Hom_k(A,k)$ muni 
des structures $(\widehat{\mu},\widehat{\eta},\widehat{\Delta},\widehat{\epsilon},\widehat{S})$ d\'efinies \`a partir 
de celles de A par la donn\'ee d'un {\it pairing} $\langle \, , \, \rangle : \widehat{A} \times A \rightarrow k$ 
tel que, pour $x,y \in A$ et $\psi, \phi \in \widehat{A}$: 
\begin{eqnarray*} 
\langle \widehat{\mu}(\psi \otimes \phi) , x \rangle &=& \langle \psi \pv \phi , \Delta(x)\rangle \\ 
\langle \widehat{\Delta}(\psi), x \pv y \rangle &=& \langle \psi , \mu(x \otimes y) \rangle \\ 
\langle \widehat{\eta} (1),x \rangle &=& \epsilon (x) \\ 
\widehat{\epsilon}(\psi) &=& \langle \psi , \eta(1) \rangle \\ 
\langle \widehat{S}(\psi), x \rangle &=& \langle \psi , S(x) \rangle 
\end{eqnarray*}   
\end{defin}

\section{Alg\`ebre de Hopf faible} 
Nous pr\'esentons ici les axiomes d'une alg\`ebre de Hopf faible, tels qu'ils ont \'et\'e  
pr\'esent\'es dans \cite{Bohm-WHA}. 
\begin{defin} 
Une alg\`ebre de Hopf faible (WHA) est un sextuple $(A,\mu,\eta,\Delta,\epsilon,S)$ satisfaisant 
les axiomes 1 \`a 4 suivants: 
\begin{axiome} 
$(A,\mu,\eta)$ est une alg\`ebre: 
$$
\begin{array}{rclcl} 
\mu \circ (\mu \otimes \id) &=& \mu \circ (\id \otimes \mu) & &\\ 
\mu \circ (\eta \otimes \id) &=& \mu \circ (\id \otimes \eta) &=& \id 
\end{array} 
$$
\end{axiome}   
\begin{axiome} 
$(A,\Delta,\epsilon)$ est une cog\`ebre: 
$$
\begin{array}{rclcl}
(\Delta \otimes  \id) \circ \Delta &=& (\id \otimes  \Delta)  \circ \Delta &{ }& \\ 
(\epsilon \otimes  \id) \circ \Delta &=&(\id \otimes  \epsilon) \circ \Delta  &=& \id
\end{array}
$$ 
\end{axiome} 
\begin{axiome} 
Les deux structures sont compatibles selon: 
\begin{itemize}
\item[(i)] $\Delta$ est multiplicatif:
\begin{eqnarray*}
\Delta \circ \mu &=& (\mu  \circ \mu)  \circ (\id \otimes \tau \otimes \id) \circ (\Delta \otimes \Delta) \\
i.e., \qquad \qquad \qquad \Delta(x \cdot y) &=& \Delta(x) \cdot \Delta(y) 
\end{eqnarray*} 
\item[(ii)] $\epsilon$ est faiblement multiplicatif:
\begin{eqnarray*}
(\epsilon \otimes \epsilon) \circ (\mu \otimes \mu) \circ (\id \otimes \Delta \otimes \id) &=& \epsilon \circ \mu \circ (\mu \otimes \id) \\
(\epsilon \otimes \epsilon) \circ (\mu \otimes \mu) \circ (\id \otimes \Delta^{\textrm{op}} \otimes \id) &=& 
\epsilon \circ \mu \circ (\mu \otimes \id)
\end{eqnarray*}
o\`u $\Delta^{\textrm{op}} = \tau \circ \Delta$. Utilisant la convention de Sweedler, ces deux \'equations s'\'ecrivent
plus simplement:
\begin{eqnarray*}
\epsilon (x \cdot y \cdot z) &=& \epsilon (x \cdot y_{(1)}) \cdot \epsilon (y_{(2)} \cdot z) \\ 
\epsilon (x \cdot y \cdot z) &=&\epsilon (x \cdot y_{(2)})  \cdot \epsilon (y_{(1)} \cdot z) 
\end{eqnarray*}
\item[(iii)] $\eta$ est faiblement multiplicatif:
\begin{eqnarray*}
\Delta^2(1_A) &=& (\Delta(1_A) \pv 1_A) \cdot (1_A \pv \Delta(1_A)) \\ 
\Delta^2(1_A) &=&(1_A \pv \Delta(1_A)) \cdot (\Delta(1_A) \pv 1_A)  
\end{eqnarray*} 
o\`u $\Delta^2 = (\Delta \otimes \id) \circ \Delta = (\id \otimes \Delta) \circ \Delta $
\end{itemize}
\end{axiome} 
\begin{axiome} 
Existence d'une antipode $S$ satisfaisant: 
\begin{align} 
S(x) &= S(x_{(1)}) \cdot x_{(2)} \cdot S(x_{(3)}) \\ 
x_{(1)} \cdot S(x_{(2)}) &=  \epsilon (1_{(1)} \cdot x ) \cdot 1_{(2)} \\ 
S(x_{(1)}) \cdot x_{(2)} &= 1_{(1)} \cdot \epsilon (x 1_{(2)})  
\end{align} 
\end{axiome} 
\end{defin} 
 
Une WHA devient une alg\`ebre de Hopf (usuelle) si l'une des conditions suivantes est satisfaite:
\begin{itemize} 
\item[$\bullet$] $\displaystyle \Delta(1_A) = 1_A \pv 1_A$ 
\item[$\bullet$] $\displaystyle \epsilon (x \cdot y) = \epsilon (x) \cdot \epsilon (y)$
\end{itemize}

\section{Divers} 
Consid\'erons le groupe de permutations $S_n$ sur $n$ objets, engendr\'e par les transpositions $t_i$ ($1 \leq i \leq n-1$),
qui permutent les objets $i$ et $i+1$. En consid\'erant une combinaison lin\'eaire dans 
$\mathbb{C}$ de tels \'el\'ements, nous obtenons l'alg\`ebre de permutations $\mathbb{C}S_n$: 
\begin{defin} 
Soit un entier $n \geq 1$. L'alg\`ebre du groupe de permutations $\mathbb{C}S_n$ est l'alg\`ebre associative 
unitale engendr\'ee par $n$ g\'en\'erateurs $(1,t_1,t_2,\ldots,t_{n-1})$ satisfaisant aux relations suivantes: 
$$ 
\begin{array}{crcl} 
(i)   \qquad & t_i^2                       &=& 1 \\ 
(ii)  \qquad & t_i \,  t_j                 &=& t_j  \, t_i \qquad \qquad \qquad \text{pour } |i-j|\geq 2   \\ 
(iii) \qquad & t_{i+1} \, t_{i} \, t_{i+1} &=& t_{i} \, t_{i+1} \, t_{i}  
\end{array} 
$$ 
\end{defin}

\begin{defin} 
Soit un entier $n \geq 1$. L'alg\`ebre du groupe des tresses $\mathbb{C}B_n$ est l'alg\`ebre associative 
unitale engendr\'ee par $n$ g\'en\'erateurs $(1,b_1,b_2,\ldots,b_{n-1})$ satisfaisant aux relations suivantes: 
$$ 
\begin{array}{crcl} 
(i)  \qquad & b_i \,  b_j                 &=& b_j  \, b_i \qquad \qquad \qquad \text{pour } |i-j|\geq 2   \\ 
(ii) \qquad & b_{i+1} \, b_{i} \, b_{i+1} &=& b_{i} \, b_{i+1} \, b_{i}  
\end{array} 
$$ 
\end{defin}

\begin{defin} 
Soit un entier $n \geq 1$ et un param\`etre $q \in \mathbb{C}$. L' 
alg\`ebre de Hecke $H_n(q)$ est l'alg\`ebre associative 
unitale engendr\'ee par $n$ g\'en\'erateurs $(1,g_1,g_2,\ldots,g_{n-1})$ satisfaisant aux relations suivantes: 
$$ 
\begin{array}{crcl} 
(i)   \qquad & g_i^2                       &=& (q-1)g_i + q \\ 
(ii)  \qquad & g_i \,  g_j                 &=& g_j  \, g_i \qquad \qquad \qquad \text{pour } |i-j|\geq 2   \\ 
(iii) \qquad & g_{i+1} \, g_{i} \, g_{i+1} &=& g_{i} \, g_{i+1} \, g_{i}  
\end{array} 
$$ 
\end{defin} 
Pour $q \rightarrow 1$, l'alg\`ebre de Hecke $H_n(q)$ se r\'eduit \`a $\mathbb{C}S_n$. 
Nous pouvons donc voir l'alg\`ebre de Hecke comme une d\'eformation de l'alg\`ebre du groupe des permutations. \\ 
Dans $H_n(q)$, effectuons le chagement de base suivant: 
$$ 
g_i = (\hat{q}^2+1) \hat{e}_i - 1 \qquad \qquad \text{o\`u } \hat{q}^2 = q 
$$ 
Dans cette nouvelle base de $n$ g\'en\'erateurs $1, \hat{e}_1, \ldots, \hat{e}_{n-1}$, les relations
s'\'ecrivent: 
$$ 
\begin{array}{crcl} 
(i)   \qquad & \hat{e}_i^2                       &=& \hat{e}_i \\ 
(ii)  \qquad & \hat{e}_i \,  \hat{e}_j                 &=& \hat{e}_j  \, \hat{e}_i \qquad \qquad \qquad \text{pour } |i-j|\geq 2   \\ 
(iii) \qquad & \displaystyle \hat{e}_{i+1} \, \hat{e}_{i} \, \hat{e}_{i+1} - \frac{\hat{q}^2}{(1+\hat{q}^2)^2} \hat{e}_{i+1} &=&  
\displaystyle  \hat{e}_{i} \, \hat{e}_{i+1} \, \hat{e}_{i}  -  \frac{\hat{q}^2}{(1+\hat{q}^2)^2} \hat{e}_i 
\end{array} 
$$ 
Jusqu'ici, nous n'avons fait que reformuler la d\'efinition de l'alg\`ebre de Hecke $H_n(q)$. \\ 
Maintenant, imposons la {\bf relation de Jones}: 
$$ 
\hat{e}_{i+1} \, \hat{e}_{i} \, \hat{e}_{i+1} - \tau  \hat{e}_{i+1} =0 
$$ 
o\`u $\tau = \frac{1}{\beta^2} = \frac{\hat{q}^2}{(1+\hat{q}^2)^2}$ est le param\`etre de Jones. Nous obtenons alors  
l'alg\`ebre de Temperley-Lieb $T_n(\hat{q})$: 
\begin{defin} 
Soit un entier $n \geq 1$ et un param\`etre $\tau \in \mathbb{C}$. L'alg\`ebre de Temperley-Lieb $T_n(\tau)$  
est l'alg\`ebre associative unitale engendr\'ee par $n$ g\'en\'erateurs $(1,e_1,e_2,\ldots,e_{n-1})$ satisfaisant  
aux relations suivantes: 
$$ 
\begin{array}{crcl} 
(i)   \qquad & e_i^2                       &=& e_i \\ 
(ii)  \qquad & e_i \,  e_j                 &=& e_j  \, e_i \qquad \qquad \qquad \text{pour } |i-j|\geq 2   \\ 
(iii) \qquad & e_{i} \, e_{i \pm 1} \, e_{i} &=& \tau e_{i}  
\end{array} 
$$ 
\end{defin} 
Nous voudrions obtenir une $C^{*}-$alg\`ebre en imposant la {\bf condition de Jones}: 
\begin{equation} 
e_i^* = e_i 
\label{reljones} 
\end{equation} 
\begin{theo}[Jones] 
Il n'est possible d'imposer la relation (\ref{reljones}) que pour les valeurs suivantes de $\beta$: 
\begin{itemize} 
\item $\beta \geq 2$ 
\item $\displaystyle \beta = 2 \cos(\frac{\pi}{N})$ pour un entier $N>2$ 
\end{itemize} 
\end{theo}


\chapter{Fonctions de partition g\'en\'eralis\'ees} 
\thispagestyle{empty}
 
\section{Cas $\widehat{su}(2)$} 

\subsection{Le cas $A_4$}

\footnotesize
\begin{table}[H] 
$$ 
\begin{array}{|rclcrcl|} 
\hline  
{ } & { } & { } & { } & { } & { } & { }  \\ 
 { } & { } & { } &\qquad& \mathcal{Z}_{11} &=& \mathcal{Z}_{00} + \mathcal{Z}_{20} \\ 
{ } & { } & { } & { } & { } & { } & { }  \\ 
\mathcal{Z}_{A_4} = \mathcal{Z}_{00} &=&  \sum_{i=0}^{3} |\chi_i|^2 
&\qquad& \mathcal{Z}_{12} = \mathcal{Z}_{21} &=& \mathcal{Z}_{10} + \mathcal{Z}_{30} \\ 
{ } & { } & { } & { } & { } & { } & { }  \\ 
\mathcal{Z}_{10} = \mathcal{Z}_{01} &=& (\chi_0 \ov{\chi}_1 + \chi_1 \ov{\chi}_2 + \chi_2 \ov{\chi}_3 ) + \textrm{h.c.} 
&\qquad& \mathcal{Z}_{13} = \mathcal{Z}_{31} &=& \mathcal{Z}_{20} \\ 
{ } & { } & { } & { } & { } & { } & { }  \\ 
\mathcal{Z}_{20} = \mathcal{Z}_{02} &=& |\chi_1|^2 + |\chi_2|^2 + \left[ (\chi_0 \ov{\chi}_2 + \chi_1 \ov{\chi}_3) + \textrm{h.c.}\right]  
&\qquad& \mathcal{Z}_{22} &=& \mathcal{Z}_{00} + \mathcal{Z}_{20} \\ 
{ } & { } & { } & { } & { } & { } & { }  \\ 
\mathcal{Z}_{30} = \mathcal{Z}_{03} &=& (\chi_0 \ov{\chi}_3 + \chi_1 \ov{\chi}_2) + \textrm{h.c.}  
&\qquad& \mathcal{Z}_{23} &=& \mathcal{Z}_{10} \\ 
{ } & { } & { } & { } & { } & { } & { }  \\ 
{ } & { } & { } &\qquad& \mathcal{Z}_{33} &=& \mathcal{Z}_{00}\\ 
{ } & { } & { } & { } & { } & { } & { }  \\ 
\hline 
\end{array}
$$ 
\caption{Fonctions de partition g\'en\'eralis\'ees du mod\`ele $A_4$.} 
\end{table} 
\normalsize

 
\subsection{Le cas $E_6$}

\footnotesize
\begin{table}[H] 
$$ 
\begin{array}{|rclcrcl|} 
\hline 
{ } & { } & { } & { } & { } & { } & { } \\ 
\hat{\chi}_0 &=& \chi_0 + \chi_6  &\qquad& \hat{\chi}_3 &=& \chi_3 + \chi_7 \\ 
{ } & { } & { } & { } & { } & { } & { } \\ 
\hat{\chi}_1 &=& \chi_1 + \chi_5 + \chi_7    &\qquad& \hat{\chi}_4 &=& \chi_4 + \chi_{10} \\ 
{ } & { } & { } & { } & { } & { } & { } \\ 
\hat{\chi}_2 &=& \chi_2 + \chi_4 + \chi_6 + \chi_8    &\qquad& \hat{\chi}_5 &=& \chi_3 + \chi_5 + \chi_9 \\ 
{ } & { } & { } & { } & { } & { } & { } \\ 
\hline 
\end{array} 
$$ 
\caption{Caract\`eres \'etendus du mod\`ele $E_6$ en fonction des caract\`eres de $A_{11}$.} 
\end{table} 
\normalsize

\begin{table}[H] 
$$ 
\begin{array}{|rclcrcl|} 
\hline 
{ } & { } &{ } &{ } &{ } &{ } &{ } \\ 
\mathcal{Z}_{E_6} =  \mathcal{Z}_{\ud{0}} &=& \hxa{0} + \hxa{3} + \hxa{4}  &\qquad \qquad &  
\mathcal{Z}_{\ud{11'}} &=&  \hxa{1} + \hxa{2} + \hxa{5} \\ 
{ } & { } &{ } &{ } &{ } &{ } &{ } \\ 
\mathcal{Z}_{\ud{3}} &=& (\hch{0} + \hch{4})\hoch{3} + \hch{3} (\hoch{0} + \hoch{4}) &\qquad \qquad & \mathcal{Z}_{\ud{21'}} &=&  (\hch{1} + \hch{5})\hoch{2} + \hch{2} (\hoch{1} + \hoch{5}) \quad \\ 
{ } & { } &{ } &{ } &{ } &{ } &{ } \\ 
\mathcal{Z}_{\ud{4}} &=& \hxa{3} + \hxx{0}{4} + \hxx{4}{0}   &\qquad \qquad & \mathcal{Z}_{\ud{51'}} &=&   \hxa{2} + \hxx{1}{5} + \hxx{5}{1}  \\ 
{ } & { } &{ } &{ } &{ } &{ } &{ } \\ 
\mathcal{Z}_{\ud{1}} &=& \hxx{1}{0} + \hxx{2}{3} + \hxx{5}{4}   &\qquad \qquad & \mathcal{Z}_{\ud{1'}} &=& \mathcal{Z}_{\ud{1}}^* \\  
{ } & { } &{ } &{ } &{ } &{ } &{ } \\ 
\mathcal{Z}_{\ud{2}} &=& \hch{2} (\hoch{0} + \hoch{4}) + (\hch{1} + \hch{5}) \hoch{3}  &\qquad \qquad & \mathcal{Z}_{\ud{31'}} &=&  \mathcal{Z}_{\ud{2}}^* \\ 
{ } & { } &{ } &{ } &{ } &{ } &{ } \\ 
\mathcal{Z}_{\ud{5}} &=& \hxx{1}{4} + \hxx{2}{3}+ \hxx{5}{0}  &\qquad \qquad & \mathcal{Z}_{\ud{41'}} &=&   
\mathcal{Z}_{\ud{5}}^* \\ 
{ } & { } &{ } &{ } &{ } &{ } &{ } \\ 
\hline 
\end{array} 
$$ 
\caption{Fonctions de partition (\`a une ligne de d\'efauts) du mod\`ele $E_6$.} 
\end{table}

\subsection{Le cas $E_8$} 
\begin{table}[H] 
$$ 
\begin{array}{|rcl|} 
\hline 
{ } & { } & { } \\  
\quad \hat{\chi}_0 &=& \chi_0 + \chi_{10} + \chi_{18} + \chi_{28} \\ 
{ } & { } & { } \\  
\hat{\chi}_1 &=& \chi_1 + \chi_{9} + \chi_{11} + \chi_{17} + \chi_{19} + \chi_{27}\\ 
{ } & { } & { } \\  
\hat{\chi}_2 &=& \chi_2 + \chi_{8} + \chi_{10} + \chi_{12} + \chi_{16} + \chi_{18} + \chi_{20} + \chi_{26} \\ 
{ } & { } & { } \\  
\hat{\chi}_3 &=& \chi_3 + \chi_{7} + \chi_{9} + \chi_{11} + \chi_{13} + \chi_{15} + \chi_{17} + \chi_{19} + \chi_{21} + \chi_{25}\\ 
{ } & { } & { } \\  
\hat{\chi}_4 &=& \chi_4 + \chi_{6} + \chi_{8} + \chi_{10} + \chi_{12} + 2 \chi_{14} + \chi_{16} + \chi_{18} + \chi_{20} + \chi_{22} + \chi_{24} \quad  \\ 
{ } & { } & { } \\  
\hat{\chi}_5 &=& \chi_5 + \chi_{9} + \chi_{13} + \chi_{15} + \chi_{19} + \chi_{23} \\ 
{ } & { } & { } \\  
\hat{\chi}_6 &=& \chi_6 + \chi_{12} + \chi_{16} + \chi_{22} \\ 
{ } & { } & { } \\  
\hat{\chi}_7 &=& \chi_5 + \chi_{7} + \chi_{11} + \chi_{13} + \chi_{15} + \chi_{17} + \chi_{21} + \chi_{23} \\ 
{ } & { } & { } \\  
\hline 
\end{array} 
$$ 
\caption{Caract\`eres \'etendus du mod\`ele $E_8$ en fonction des caract\`eres de $A_{29}$.} 
\end{table}

\scriptsize
\begin{table}[H] 
$$ 
\begin{array}{|rclcrclcl|} 
\hline 
{ } & { } & { } & { } & { } & { } & { } & { } & { }  \\ 
{ } & { } & { } & \qquad & \mathcal{Z}_{10} &=& \hch{1}.\hoch{0} + \hch{7}.\hoch{6} &=& \mathcal{Z}_{01}^* \\ 
{ } & { } & { } & { } & { } & { } & { } & { } & { }  \\ 
{ } & { } & { }  &{ }& 
\mathcal{Z}_{70} &=& \hch{7}.\hoch{0} + (\hch{1} + \hch{7}).\hoch{6} &=& \mathcal{Z}_{07}^* = \mathcal{Z}_{61}^* \\ 
{ } & { } & { } & { } & { } & { } & { } & { } & { }  \\ 
\quad \mathcal{Z}_{E_8} = \mathcal{Z}_{00} &=& \hxa{0} + \hxa{6} &{ }& 
\mathcal{Z}_{20} &=& \hch{2}.\hoch{0} + \hch{4}.\hoch{6} &=& \mathcal{Z}_{02}^* \\ 
{ } & { } & { } & { } & { } & { } & { } & { } & { }  \\ 
\mathcal{Z}_{60} &=& \hxa{6} + \hch{0}.\hoch{6} + \hch{6}.\hoch{0} & { }& 
\mathcal{Z}_{40} &=& \hch{4}.\hoch{0} + (\hch{2} + \hch{4}).\hoch{6}  &=& \mathcal{Z}_{04}^* = \mathcal{Z}_{62}^* \\  
{ } & { } & { } & { } & { } & { } & { } & { } & { }  \\ 
\mathcal{Z}_{11} &=& \hxa{1} + \hxa{7} & { } &  
\mathcal{Z}_{50} &=& \hch{5}.\hoch{0} + \hch{3}.\hoch{6} &=& \mathcal{Z}_{05}^* \\ 
{ } & { } & { } & { } & { } & { } & { } & { } & { }  \\ 
\mathcal{Z}_{71} &=& \hxa{7} + \hch{1}.\hoch{7} + \hch{7}.\hoch{1} & { } & 
\mathcal{Z}_{30} &=& \hch{3}.\hoch{0} + (\hch{3}+\hch{5}).\hoch{6}  &=& \mathcal{Z}_{03}^* = \mathcal{Z}_{65}^* \\ 
{ } & { } & { } & { } & { } & { } & { } & { } & { }  \\ 
\mathcal{Z}_{22} &=& \hxa{2} + \hxa{4} & { } & 
\mathcal{Z}_{21} &=& \hch{2}.\hoch{1} + \hch{4}.\hoch{7} &=& \mathcal{Z}_{12}^* \\ 
{ } & { } & { } & { } & { } & { } & { } & { } & { }  \\ 
\mathcal{Z}_{42} &=& \hxa{4} + \hch{2}.\hoch{4} + \hch{4}.\hoch{2} & { } & 
\mathcal{Z}_{41} &=& \hch{4}.\hoch{1} + (\hch{2}+\hch{4}).\hoch{7}   &=& \mathcal{Z}_{14}^* = \mathcal{Z}_{72}^* \\ 
{ } & { } & { } & { } & { } & { } & { } & { } & { }  \\ 
\mathcal{Z}_{55} &=& \hxa{3} + \hxa{5} & { } &  
\mathcal{Z}_{52} &=& \hch{5}.\hoch{2} + \hch{3}.\hoch{4} &=& \mathcal{Z}_{25}^* \\ 
{ } & { } & { } & { } & { } & { } & { } & { } & { }  \\ 
\mathcal{Z}_{35} &=& \hxa{3} + \hch{5}.\hoch{3} + \hch{3}.\hoch{5} & { } &  
\mathcal{Z}_{32} &=& \hch{3}.\hoch{2} + (\hch{3}+\hch{5}).\hoch{4}  &=& \mathcal{Z}_{23}^* = \mathcal{Z}_{45}^* \\ 
{ } & { } & { } & { } & { } & { } & { } & { } & { }  \\ 
{ } & { } & { } & \qquad & \mathcal{Z}_{15} &=& \hch{1}.\hoch{5} + \hch{7}.\hoch{3} &=& \mathcal{Z}_{51}^* \\ 
{ } & { } & { } & { } & { } & { } & { } & { } & { }  \\ 
{ } & { } & { } & \qquad & \mathcal{Z}_{75} &=& \hch{7}.\hoch{5} + (\hch{1}+\hch{7}).\hoch{3}  &=& \mathcal{Z}_{57}^* = \mathcal{Z}_{31}^* \quad \\ 
 
{ } & { } & { } & { } & { } & { } & { } & { } & { }  \\ 
\hline 
\end{array} 
$$ 
\caption{Fonctions de partition (\`a une ligne de d\'efauts) du mod\`ele $E_8$.} 
\end{table} 
\normalsize

\subsection{Le cas $D_4$} 

\footnotesize 
\begin{table}[H] 
$$ 
\begin{array}{|ccccc|} 
\hline 
{ } & { } & { } & { } & { } \\ 
\hat{\chi}_0 = \chi_0 + \chi_4  &\qquad& \hat{\chi}_1 = \chi_1 + \chi_3 &\qquad& \hat{\chi}_2 = \hat{\chi}_{2'} = \chi_2 \\  
{ } & { } & { } & { } & { }  \\ 
\hline 
\end{array} 
$$ 
\caption{Caract\`eres \'etendus du mod\`ele $D_4$ en fonction des caract\`eres de $A_5$.} 
\end{table} 
\normalsize

\footnotesize
\begin{table}[H] 
$$ 
\begin{array}{|rclcrcl|} 
\hline 
{ } & { } & { } & { } & { } & { } & { } \\ 
\mathcal{Z}_{D_4} = \mathcal{Z}_{0,+} &=& \hxa{0}+\hxa{2}+\hxa{2'}  &\qquad& \mathcal{Z}_{0,-} &=& \hxa{1} \\ 
{ } & { } & { } & { } & { } & { } & { } \\ 
\mathcal{Z}_{1,+} &=& \hch{1}.(\hoch{0} + \hoch{2} + \hoch{2'})    &\qquad& \mathcal{Z}_{1,-} &=& \mathcal{Z}_{1,+}^* \\ 
{ } & { } & { } & { } & { } & { } & { } \\ 
\mathcal{Z}_{2,+} &=& \hch{0}.\hoch{2}+\hch{2}.\hoch{2'}+\hch{2'}.\hoch{0} &\qquad& \mathcal{Z}_{2,-} &=& \hxa{1} \\ 
{ } & { } & { } & { } & { } & { } & { } \\ 
\mathcal{Z}_{2',+} &=&  \hch{0}.\hoch{2'}+\hch{2'}.\hoch{2}+\hch{2}.\hoch{0} = \mathcal{Z}_{2,+}  &\qquad& \mathcal{Z}_{2',-} &=& \hxa{1} \\ 
{ } & { } & { } & { } & { } & { } & { } \\ 
\hline 
\end{array} 
$$ 
\caption{Fonctions de partition (\`a une ligne de d\'efauts) du mod\`ele $D_4$.} 
\end{table} 
\normalsize

\subsection{Le cas $D_6$} 
 
\footnotesize
\begin{table}[H] 
$$ 
\begin{array}{|ccccc|} 
\hline 
{ } & { } & { } & { } & { } \\ 
\hat{\chi}_0 = \chi_0 + \chi_8  &\qquad& \hat{\chi}_2 = \chi_2 + \chi_6 &\qquad& \hat{\chi}_4 = \chi_4 \\  
{ } & { } & { } & { } & { }  \\ 
\hat{\chi}_1 = \chi_1 + \chi_7  &\qquad& \hat{\chi}_3 = \chi_5 + \chi_5 &\qquad& \hat{\chi}_{4'} = \chi_{4'} \\  
{ } & { } & { } & { } & { }  \\\hline 
\end{array} 
$$ 
\caption{Caract\`eres \'etendus du mod\`ele $D_6$ en fonction des caract\`eres de $A_{9}$.} 
\end{table} 
\normalsize
 
\footnotesize
\begin{table}[H] 
$$ 
\begin{array}{|rclcrcl|} 
\hline 
{ } & { } & { } & { } & { } & { } & { } \\ 
\mathcal{Z}_{D_6} = \mathcal{Z}_{0,+} &=& \hxa{0}+\hxa{2}+\hxa{4}+\hxa{4'}  &\quad&  
\mathcal{Z}_{0,-} &=& \hxa{1} + \hxa{3} \\ 
{ } & { } & { } & { } & { } & { } & { } \\ 
\mathcal{Z}_{1,+} &=& \hch{1}.(\hoch{0} + \hoch{2}) + \hch{3}.(\hoch{2} + \hoch{4} + \hoch{4'})    &\quad&  
\mathcal{Z}_{1,-} &=& \mathcal{Z}_{1,+}^* \\ 
{ } & { } & { } & { } & { } & { } & { } \\ 
\mathcal{Z}_{2,+} &=& \hxa{2}+ (\hch{0}.\hoch{2} + \hch{2}.\hoch{4} + \hch{2}.\hoch{4'} + \hch{4}.\hoch{4'}  + \textrm{h.c.}) &\quad&  
\mathcal{Z}_{2,-} &=& \hxaa{1}{3} + \hxa{3}\\ 
{ } & { } & { } & { } & { } & { } & { } \\ 
\mathcal{Z}_{3,+} &=&  \hch{1}.(\hoch{2}+\hoch{4}+\hoch{4'}) + \hch{3}.(\hoch{0}+ 2 \hoch{2}+\hoch{4}+\hoch{4'})    &\quad&  
\mathcal{Z}_{3,-} &=&  \mathcal{Z}_{3,+}^* \\ 
{ } & { } & { } & { } & { } & { } & { } \\ 
\mathcal{Z}_{4,+} &=&  \hxa{2} + \hxa{4} + (\hch{0}.\hoch{4} + \hch{2}.\hoch{4'} + \textrm{h.c.}) &\quad&  
\mathcal{Z}_{4,-} &=& \hxa{3} + (\hch{1}.\hoch{3} + \textrm{h.c.}) \\ 
{ } & { } & { } & { } & { } & { } & { } \\ 
\mathcal{Z}_{4',+} &=&  \hxa{2} + \hxa{4'} + (\hch{0}.\hoch{4'} + \hch{2}.\hoch{4} + \textrm{h.c.}) = \mathcal{Z}_{4,+} &\quad&  
\mathcal{Z}_{4',-} &=& \mathcal{Z}_{4,-} \\ 
{ } & { } & { } & { } & { } & { } & { } \\ 
\hline 
\end{array} 
$$ 
\caption{Fonctions de partition (\`a une ligne de d\'efauts) du mod\`ele $D_6$.} 
\end{table} 
\normalsize

\subsection{Le cas $D_5$} 
 
\scriptsize
\begin{table}[H]  
$$ 
\begin{array}{|rcl|} 
\hline 
{ }  & { } &  { }  \\ 
\mathcal{Z}_{D_5} = \mathcal{Z}_0  &=& \xa{0} + \xa{2} + \xa{3} + \xa{4} + \xa{6} + (\xx{1}{5} + \textrm{h.c.}) \\ 
{ }  & { } &  { }  \\ 
\mathcal{Z}_1  &=& (\ch{0} + \ch{2}).\och{1} + \ch{4}.(\och{0} + \och{2}) + 
\ch{1} .(\och{4} + \och{6})+ (\ch{4} + \ch{6}).\och{5} + [(\xx{2}{3} 
+ \xx{3}{4}) + \textrm{h.c.}]\\ 
{ }  & { } &  { }  \\ 
\mathcal{Z}_2  &=& \xaa{2}{4} + \xa{3} + [( \xx{0}{2} + \xx{1}{3} + \xx{1}{5} + 
\xx{3}{5} + \xx{4}{6}) + \textrm{h.c.}] \\ 
{ }  & { } &  { }  \\ 
\mathcal{Z}_3  &=& [(\ch{0} + \ch{2} + \ch{4} + \ch{6}).\och{3} + (\ch{1} + \ch{5}). 
(\och{2} + \ch{4})  + \textrm{h.c.}]  \\ 
{ }  & { } &  { }  \\ 
\mathcal{Z}_4  &=& \xa{1} + \xa{3} + \xa{5} + \xaa{2}{4} + [ (\xx{0}{4} + \xx{1}{3} 
       + \xx{2}{6} + \xx{3}{5} ) + \textrm{h.c.} ] \\ 
{ }  & { } &  { }  \\ 
\mathcal{Z}_5  &=&  \mathcal{Z}_{1}^* \\ 
{ }  & { } &  { }  \\ 
\mathcal{Z}_6  &=&  \xa{1} + \xa{3} + \xa{5} + [(\xx{0}{6} + \xx{2}{4}) + \textrm{h.c.}] \\ 
{ }  & { } &  { }  \\ 
\hline 
\end{array} 
$$  
\caption{Fonctions de partition (\`a une ligne de d\'efauts) du mod\`ele $D_5$.} 
\end{table} 
\normalsize
 
\subsection{Le cas $E_7$}

\footnotesize
\begin{table}[H] 
$$ 
\begin{array}{|ccccccccc|} 
\hline 
{ } & { } & { } & { } & { } & { } & { } & { } & { } \\ 
\hat{\chi}_0 = \chi_0 + \chi_{16}  &\quad& \hat{\chi}_2 = \chi_2 + \chi_{14} &\quad&  
\hat{\chi}_4 = \chi_4 + \chi_{12}  &\quad& \hat{\chi}_6 = \chi_6 + \chi_{10} &\quad& \hat{\chi}_8 = \chi_8 \\  
{ } & { } & { } & { } & { } & { } & { } & { } & { } \\ 
\hat{\chi}_1 = \chi_1 + \chi_{15}  &\quad& \hat{\chi}_3 = \chi_3 + \chi_{13} &\quad&  
\hat{\chi}_5 = \chi_5 + \chi_{11}  &\quad& \hat{\chi}_7 = \chi_7 + \chi_9    &\quad& \hat{\chi}_{8'} = \chi_8  \\ 
{ } & { } & { } & { } & { } & { } & { } & { } & { } \\ 
\hline 
\end{array} 
$$ 
\caption{Caract\`eres \'etendus du mod\`ele $D_{10}$ (et $E_7$) en fonction des caract\`eres de $A_{17}$.} 
\end{table} 
\normalsize
 
\footnotesize
\begin{table}[H] 
$$ 
\begin{array}{|rcl|} 
\hline 
{ } & { } & { }  \\ 
\mathcal{Z}_{E_7} =  \mathcal{Z}_{0} &=& \hxa{0} + \hxa{4} +  \hxa{6} + \hxa{8'} + (\hch{2}.\hoch{8} + \textrm{h.c.}) \\ 
{ } & { } & { }  \\ 
\mathcal{Z}_{1}= \mathcal{Z}_{(0)}^* &=& \hch{1}.(\hoch{0} + \hoch{8}) + \hch{3}.(\hoch{4} + \hoch{8}) +  
\hch{5}.(\hoch{4} + \hoch{6}) + \hch{7}.(\hoch{2} + \hoch{6} + \hoch{8'})   \\ 
{ } & { } & { }  \\ 
\mathcal{Z}_{2} = \mathcal{Z}_{8}^* &=& \hxaa{4}{6} + \hch{2}.(\hoch{0} + \hoch{4}) + \hch{8}.(\hoch{6} + \hoch{8'}) 
+ (\hch{0} + \hch{4}).\hoch{8} + (\hch{6} + \hch{8'}).\hoch{2} \\ 
{ } & { } & { }  \\ 
{ } &+& \hch{2}.\hoch{8} + (\hch{6}.\hoch{8'} + \textrm{h.c.})\\ 
{ } & { } & { }  \\ 
\mathcal{Z}_{3} = \mathcal{Z}_{(4)}^* &=& \hch{1}.(\hoch{4} + \hoch{8}) + \hch{3}.(\hoch{0} + \hoch{4} + \hoch{6} + \hoch{8}) +  
\hch{5}.(\hoch{2} + \hoch{4} + \hoch{6} + \hoch{8} + \hoch{8'})\\ 
{ } & { } & { }  \\ 
{ } &+& \hch{7}.(\hoch{2} + \hoch{4} + 2\, \hoch{6} + \hoch{8'})  \\ 
{ } & { } & { }  \\ 
\mathcal{Z}_{4} &=& \hxaa{4}{6} + \hxa{6} + \hxa{8'} + \lbrack \hch{0}.\hoch{4} + \hch{2}.(\hoch{4}+ \hoch{6}+ \hoch{8}) 
+(\hch{4}+\hch{6}).(\hoch{8}+\hoch{8'}) + \textrm{h.c.} \rbrack \\ 
{ } & { } & { }  \\ 
\mathcal{Z}_{5} = \mathcal{Z}_{(6)}^* &=& \hch{1}.(\hoch{4} + \hoch{6}) + \hch{3}.(\hoch{2} + \hoch{4} + \hoch{6} + \hoch{8} + \hoch{8'}) +  
\hch{5}.(\hoch{0} + \hoch{2} + \hoch{4} + 2\, \hoch{6} + \hoch{8} + \hoch{8'})\\ 
{ } & { } & { }  \\ 
{ } &+& \hch{7}.(\hoch{2} + 2\, \hoch{4} + 2\, \hoch{6} + \hoch{8} + \hoch{8'})  \\ 
{ } & { } & { }  \\ 
\mathcal{Z}_{6} &=& |\hch{2}+\hch{4}+\hch{6}|^2 + \hxa{6} + \hxa{8} + \left\lbrack \hch{0}.\hoch{6} + \hch{2}.\hoch{8'}  
+ \hch{4}.\hoch{6} + (\hch{4}+\hch{6}).(\hoch{8}+\hoch{8'}) + \textrm{h.c.} \right\rbrack\\ 
{ } & { } & { }  \\ 
\mathcal{Z}_{7} = \mathcal{Z}_{(2)}^* &=& \hch{1}.(\hoch{2} + \hoch{6}+ \hoch{8'}) + \hch{3}.(\hoch{2} + \hoch{4} + 2\,  \hoch{6} + \hoch{8'}) +  
\hch{5}.(\hoch{2} + 2\, \hoch{4} + 2\, \hoch{6} + \hoch{8} + \hoch{8'})\\ 
{ } & { } & { }  \\ 
{ } &+& \hch{7}.(\hoch{0}+\hoch{2} + 2\, \hoch{4} + 2\, \hoch{6} + 2\, \hoch{8} + \hoch{8'})  \\ 
{ } & { } & { }  \\ 
\mathcal{Z}_{8'} &=& \hxa{2} + \hxaa{4}{6} + \hxa{8} + \hxa{8'} + [ (\hch{0}+\hch{4}).\hoch{8'} +  
(\hch{2}+\hch{8}).\hoch{6} + \textrm{h.c.}] \\ 
{ } & { } & { }  \\ 
\mathcal{Z}_{(1)} &=& \hxaa{1}{7} + \hxaaa{3}{5}{7} + \hxa{5}  \\ 
{ } & { } & { }  \\ 
\mathcal{Z}_{(3)} &=& \hxaaa{3}{5}{7}  + \hxaa{5}{7} + \hxa{7} + [\hch{1}.(\hoch{3}+\hoch{5}+\hoch{7}) + 
\hch{3}.(\hoch{5}+\hoch{7}) + \hch{5}.\hoch{7} + \textrm{h.c.}] \\  
{ } & { } & { }  \\ 
\mathcal{Z}_{(5)} &=& \hxaa{3}{7} + \hxa{5} + [\hch{1}.\hoch{5} + \hch{5}.\hoch{7} +  + \textrm{h.c.}] \\ 
{ } & { } & { }  \\ 
\hline 
\end{array} 
$$ 
\caption{Fonctions de partition (\`a une ligne de d\'efauts) du mod\`ele $E_7$.} 
\end{table} 
\normalsize

\section{Cas $\widehat{su}(3)$: $\mathcal{E}_5$} 
 
 
 

\footnotesize 
\begin{table}[H] 
$$ 
\begin{array}{|ccccccc|} 
\hline 
{ } & { } & { } & { } & { } & { } & { }  \\ 
\hat{\chi}_{1_0} = \chi_{0,0} + \chi_{2,2}  &\quad&  
\hat{\chi}_{1_3} = \chi_{0,3} + \chi_{3,0}  &\quad&  
\hat{\chi}_{2_0} =  \chi_{1,1} + \chi_{1,4} + \chi_{2,2} + \chi_{3,0} &\quad& 
\hat{\chi}_{2_3} =  \chi_{1,1} + \chi_{1,4} + \chi_{2,2} + \chi_{3,0} \\  
{ } & { } & { } & { } & { } & { } & { }  \\ 
\hat{\chi}_{1_1} = \chi_{0,2} + \chi_{3,2}  &\quad&  
\hat{\chi}_{1_5} = \chi_{2,0} + \chi_{2,3}  &\quad&  
\hat{\chi}_{2_1} =  \chi_{1,0} + \chi_{1,3} + \chi_{2,1} + \chi_{3,2} &\quad&  
\hat{\chi}_{2_2} =  \chi_{0,1} + \chi_{3,1} + \chi_{1,2} + \chi_{2,3} \\  
{ } & { } & { } & { } & { } & { } & { }  \\ 
\hat{\chi}_{1_2}  = \chi_{1,2} + \chi_{5,0}  &\quad&  
\hat{\chi}_{1_4}  = \chi_{2,1} + \chi_{0,5}  &\quad&  
\hat{\chi}_{2_4} =  \chi_{0,2} + \chi_{1,3} + \chi_{2,1} + \chi_{4,0} &\quad& 
\hat{\chi}_{2_5} =  \chi_{2,0} + \chi_{3,1} + \chi_{2,1} + \chi_{3,2} \\ 
{ } & { } & { } & { } & { } & { } & { }  \\  
\hline 
\end{array} 
$$ 
\caption{Caract\`eres \'etendus du mod\`ele $\mathcal{E}_5$ en fonction des caract\`eres de 
$\mathcal{A}_{5}$.} 
\end{table} 
\normalsize

\scriptsize
\begin{table}[H] 
$$ 
\begin{array}{|rclrl|} 
\hline 
 { } & { } & { } & { } & { } \\ 
\quad \mathcal{Z}_{\mathcal{E}_5} = \mathcal{Z}_{1_0 \otimesdot 1_0} &=& \hxa{1_0} + \hxa{1_1} + \hxa{1_2} + \hxa{1_3} + \hxa{1_4} + \hxa{1_5} & { } & { }\\ 
 { } & { } & { } & { } & { } \\ 
\mathcal{Z}_{1_3 \otimesdot 1_0} &=& \hxx{1_0}{1_3} + \hxx{1_1}{1_4} + \hxx{1_2}{1_5} + \textrm{h.c. } & { } & { }\\ 
 { } & { } & { } & { } & { } \\ 
\mathcal{Z}_{1_1 \otimesdot 1_0} &=& \hxx{1_0}{1_1} + \hxx{1_1}{1_2} + \hxx{1_2}{1_3} + \hxx{1_3}{1_4} +  
\hxx{1_4}{1_5} + \hxx{1_5}{1_0} &\qquad = & \mathcal{Z}_{1_5 \otimesdot 1_0}^* \quad \\ 
 { } & { } & { } & { } & { } \\ 
\mathcal{Z}_{1_2 \otimesdot 1_0} &=& \hxx{1_0}{1_2} + \hxx{1_1}{1_3} + \hxx{1_2}{1_4} + \hxx{1_3}{1_5} +  
\hxx{1_4}{1_0} + \hxx{1_5}{1_1} &\qquad = & \mathcal{Z}_{1_4 \otimesdot 1_0}^*\\ 
 { } & { } & { } & { } & { } \\ 
\mathcal{Z}_{2_0 \otimesdot 1_0} &=& \hxx{2_0}{1_3} + \hxx{2_1}{1_4} + \hxx{2_2}{1_5} + \hxx{2_3}{1_0} + \hxx{2_4}{1_1} + \hxx{2_5}{1_2} &\qquad = & \mathcal{Z}_{1_0 \otimesdot 2_0}^*\\ 
 { } & { } & { } & { } & { } \\ 
\mathcal{Z}_{2_1 \otimesdot 1_0} &=& \hxx{2_0}{1_4} + \hxx{2_1}{1_5} + \hxx{2_2}{1_0} + \hxx{2_3}{1_1} + \hxx{2_4}{1_2} + \hxx{2_5}{1_3} &\qquad = & \mathcal{Z}_{1_5 \otimesdot 2_0}^* \\ 
 { } & { } & { } & { } & { } \\ 
\mathcal{Z}_{2_2 \otimesdot 1_0} &=& \hxx{2_0}{1_5} + \hxx{2_1}{1_0} + \hxx{2_2}{1_1} + \hxx{2_3}{1_2} + \hxx{2_4}{1_3} + \hxx{2_5}{1_4} &\qquad = & \mathcal{Z}_{1_4 \otimesdot 2_0}^* \\ 
 { } & { } & { } & { } & { } \\ 
\mathcal{Z}_{2_3 \otimesdot 1_0} &=& \hxx{2_0}{1_0} + \hxx{2_1}{1_1} + \hxx{2_2}{1_2} + \hxx{2_3}{1_3} + \hxx{2_4}{1_4} + \hxx{2_5}{1_5} &\qquad = & \mathcal{Z}_{1_3 \otimesdot 2_0}^*  \\ 
 { } & { } & { } & { } & { } \\ 
\mathcal{Z}_{2_4 \otimesdot 1_0} &=& \hxx{2_0}{1_1} + \hxx{2_1}{1_2} + \hxx{2_2}{1_3} + \hxx{2_3}{1_4} + \hxx{2_4}{1_5} + \hxx{2_5}{1_0} &\qquad = & \mathcal{Z}_{1_2 \otimesdot 2_0}^*  \\ 
 { } & { } & { } & { } & { } \\ 
\mathcal{Z}_{2_5 \otimesdot 1_0} &=& \hxx{2_0}{1_2} + \hxx{2_1}{1_3} + \hxx{2_2}{1_4} + \hxx{2_3}{1_5} + \hxx{2_4}{1_0} + \hxx{2_5}{1_1}  &\qquad = & \mathcal{Z}_{1_1 \otimesdot 2_0}^* \\ 
{ } & { } & { } & { } & { } \\ 
\mathcal{Z}_{2_0 \otimesdot 2_0} &=& \hxa{2_0} + \hxa{2_1} + \hxa{2_2} + \hxa{2_3} + \hxa{2_4} + \hxa{2_5}& { } & { } \\ 
 { } & { } & { } & { } & { } \\ 
\mathcal{Z}_{2_3 \otimesdot 2_0} &=& \hxx{2_0}{2_3} + \hxx{2_1}{2_4} + \hxx{2_2}{2_5} + \textrm{h.c. } & { } & { }\\ 
 { } & { } & { } & { } & { } \\ 
\mathcal{Z}_{2_1 \otimesdot 2_0} &=& \hxx{2_0}{2_1} + \hxx{2_1}{2_2} + \hxx{2_2}{2_3} + \hxx{2_3}{2_4} +  
\hxx{2_4}{2_5} + \hxx{2_5}{2_0} &\qquad = & \mathcal{Z}_{2_5 \otimesdot 2_0}^* \\ 
 { } & { } & { } & { } & { } \\ 
\mathcal{Z}_{2_2 \otimesdot 2_0} &=& \hxx{2_0}{2_2} + \hxx{2_1}{2_3} + \hxx{2_2}{2_4} + \hxx{2_3}{2_5} +  
\hxx{2_4}{2_0} + \hxx{2_5}{2_1} &\qquad = & \mathcal{Z}_{2_4 \otimesdot 2_0}^* \\ 
 { } & { } & { } & { } & { } \\ 
\hline 
\end{array} 
$$ 
\caption{Fonctions de partition (\`a une ligne de d\'efauts) du mod\`ele $\mathcal{E}_5$.} 
\end{table}
\normalsize





\begin{thebibliography}{99} 
\thispagestyle{empty}

\addcontentsline{toc}{chapter}{Bibliographie} 

\bibitem{glusch1} A. Yu. Alekseev, D. V. Gluschenkov, A. V. Lyakhovskaya, {\em Regular representation of 
the quantum group $Sl_q(2)$ (q is a root of unity)}, Algebra i Analiz {\bf 6} (1994) 88--125.

\bibitem{Bannai_Ito} E. Bannai, T. Ito, {\em Algebraic combinatorics I: Association schemes}, 
Benjamin/Cummings, 1984.

\bibitem{behrend} R. E. Behrend, P. A. Pearce, V. Petkova, J.-B. Zuber, {\em On the classification of Bulk 
and Boundary Conformal Field Theories}, Phys. Lett. {\bf B444} (1998) 163--166.

\bibitem{behrend-BCFT} R. E. Behrend, P. A. Pearce, V. Petkova, J.-B. Zuber, {\em Boundary Counditions in  
Rational Conformal Field Theories}, Nucl. Phys. {\bf B579} (2000) 707--773.

\bibitem{BPZ-Ising} A.A. Belavin, A.M. Polyakov, A.B. Zamolodchikov, {\em Infinite conformal symmetry 
in two-dimensional quantum field theory}, Nucl. Phys. {\bf B241} (1984) 333--380.

\bibitem{Bock_Evans} J. B\"ockenhauer, D. E. Evans, {\em Modular invariants, 
graphs and $\alpha$-induction for nets of subfactors I}, 
Commun. Math. Phys. {\bf 197} (1998) 361--386; {\em II}, Commun. Math. Phys. {\bf 200} (1999) 57--103;  
{\em III}, Commun. Math. Phys. {\bf 205} (1999) 183--200. 
 
\bibitem{Bock_Evans_Kawa} J. B\"ockenhauer, D. E. Evans, Y. Kawahigashi, {\em On $\alpha$-induction, 
chiral generators and modular invariants for subfactors}, Commun. Math. Phys. {\bf 208} (1999) 429--487.

\bibitem{Bohm} G. B\"{o}hm, K. Szlach\'anyi, {\em A coassociative $C^{\star}$ quantum group with 
non-integral dimensions}, Lett. Math. Phys. {\bf 38} (1996) 437--456.

\bibitem{Bohm-WHA} G. B\"{o}hm, F. Nill, K. Szlach\'anyi, {\em Weak Hopf Algebras I. Integral theory and  
$C^{\star}$ structure}, J. Algebra {\bf 221} (1999) 385--438.  

\bibitem{CIZ-class1} A. Cappelli, C. Itzykson, J.-B. Zuber, {\em Modular invariant partition 
functions in two dimensions},  Nucl. Phys. {\bf B280} (1987) 445--465. 
 
\bibitem{CIZ-class2} A. Cappelli, C. Itzykson, J.-B. Zuber, {\em The ADE classification of 
minimal and $A_1^{(1)}$ conformal invariant theories}, Commun. Math. Phys. {\bf 113} (1987) 1--26.

\bibitem{cardy-bord1} J. L. Cardy, {\em Conformal invariance and surface critical behavior},  
Nucl. Phys. {\bf B240} (1984) 514--532. 
 
\bibitem{cardy-invmod} J. L. Cardy, {\em Operator content of two-dimensional conformally 
invariant theories}, Nucl. Phys. {\bf B270} (1986) 186--204. 
 
\bibitem{cardy-bord2} J. L. Cardy, {\em Effect of boundary conditions on the operator content of  
two-dimensional conformally invariant theories}, Nucl. Phys. {\bf B275} (1986) 200--218. 
 
\bibitem{cardy-eq} J. L. Cardy, {\em Boundary conditions, fusions rules and the Verlinde formula},  
Nucl. Phys. {\bf B324} (1989) 581--596. 


\bibitem{cartier} P. Cartier, {\em Andr\'e Weil (1906-1998): adieu \`a un ami}, S\'eminaire de Philosophie 
et de Math\'ematiques, \'Ecole Normale Sup\'erieure (1998).

\bibitem{Chui} C.H.O. Chui, C. Mercat, P. Pearce, {\em Integrable and conformal twisted boundary 
conditions for $sl(2)$ A-D-E lattice models}, J. Phys. {\bf A36} (2003) 2623--2662.


\bibitem{con_krei-1} A. Connes, D. Kreimer, {\em Lessons from Quantum Field Theory -- Hopf Algebras and 
Spacetime Geometries}, Lett. Math. Phys. {\bf 48} (1999) 85--96; 
{\em Hopf algebras, renormalization and non-commutative geometry}, Commun. Math. Phys. {\bf 199} 
(1998) 203--242.

\bibitem{con_krei-2} A. Connes, D. Kreimer, {\em Renormalization in quantum field theory and the 
Riemann-Hilbert problem I: the Hopf algebra structure of graphs and the main theorem}, Commun. 
Math. Phys. {\bf 210} (2000) 249--273; 
{\em Renormalization in quantum field theory and the Riemann-Hilbert problem II: the $\beta$ function, 
diffeomorphisms and the renormalization group}, Commun. Math. Phys. {\bf 216} (2001) 215--241.


\bibitem{Coq-classtetra} R. Coquereaux, {\em Notes on the classical and quantum tetrahedron}, unpublished.

\bibitem{Coq-classqpoly} R. Coquereaux, {\em Classical and quantum polyhedra: A fusion graph algebra   
point of view}, Lectures given at the Karpacz Winter School 2001, AIP Conf. Proc. {\bf 589} (2001) 181--203.  
  
\bibitem{m3m21} R. Coquereaux, {\em On the finite dimensional quantum group  
$M_3 \oplus (M_{2|1}(\Lambda^{2}))_0$}, Lett. Math. Phys. {\bf 42}, 
(1997) 309--328.  
  
\bibitem{Coq-qtetra} R. Coquereaux, {\em Notes on the quantum tetrahedron}, 
Moscow Math. J. {\bf 2}, no.1 (2002) 41--80.

\bibitem{Coq_Gil-Lodz} R. Coquereaux, G. Schieber, {\em Action of a finite quantum group on the algebra 
of complex $N \times N$ matrices}, Particles, Fields and Gravitation, Lodz Conference, AIP Conf. Proc. 
{\bf 453} (1998) 9--23. 

\bibitem{Coq_Gil-ADE} R. Coquereaux, G. Schieber, {\em Twisted partition 
functions for $ADE$  
boundary conformal field theories and Ocneanu algebra of quantum symmetries}, J. of Geom. 
and Phys. {\bf 781} (2002) 1--43.

\bibitem{Coq_Gil-Tmod} R. Coquereaux, G. Schieber, {\em Determination of quantum symmetries 
for higher $ADE$ systems from the modular $T$ matrix}, J. of Math. Physics {\bf 44} (2003) 3809--3837.

\bibitem{Coq_Gil-bigebra} R. Coquereaux, G. Schieber, R. Trinchero, {\em Coxeter-Dynkin diagrams, Ocneanu bigebras 
and quantum groupoids}, in preparation.

\bibitem{Coq_Marina} R. Coquereaux, M. Huerta, {\em Torus structure on graphs and twisted partition 
functions for minimal and affine models}, hep-th/0301215, to appear in J. of Geom. and Phys.

\bibitem{Coq-private} R. Coquereaux, {\em About cells}, unpublished.  

\bibitem{DiFran} P. Di Francesco, {\em Integrable lattice models, graphs and modular invariant conformal 
field theories }, Int. J. of Mod. Phys. {\bf A7}, no. 3 (1992) 407--500.

\bibitem{DiFZub} F. Di Francesco, J.-B. Zuber, {\em SU(N) Lattice integrable  
models associated with graphs}, Nucl. Phys {\bf B338} (1990) 602--646. 

\bibitem{DiFran_Zub} P. Di Francesco, J.-B. Zuber, {\em $SU(N)$ Lattice Integrable Models and Modular 
Invariance}, Recents 
Developments in Conformal Field Theories, Trieste Conference (1989), S. Randjbar-Daemi, E. Sezgin, J.-B. 
Zuber 
eds., World Scientific (1990).

\bibitem{FMS-book} P. Di Francesco, P. Matthieu, D. Senechal, {\em Conformal Field Theory}, 
Springer, 1997.

\bibitem{dijkgraaf} R. Dijkgraaf, E. Verlinde, {\em Modular invariance and the fusion algebra}, 
Nucl. Phys. (Proc. Suppl.) {\bf 5B} (1998) 87--97.


\bibitem{Drinfeld} V. G. Drinfel'd, {\em Quantum groups}, Proc. of the Intern. Congress of Mathematicians,
Berkeley, A. M. Gleason ed. (1986) 798--820.

\bibitem{dyson} F. J. Dyson, {\em Missed opportunities}, Bull. Amer. Math. Soc. {\bf 78} (1972) 635--652.

\bibitem{kawa} D. E. Evans, Y. Kawahigashi, {\em Quantum symmetries on operator algebras}, Clarendon Press, Oxford, 1998.

\bibitem{QISM} L. D. Faddeev, E. K. Sklyanin, Takhtajan, {\em Quantum inverse problem method}, Theor. Math.
Phys. {\bf 40} (1979) 194--220.


\bibitem{Fendley}  P. Fendley, P. Ginsparg, {\em Non-critical orbifolds}, Nucl. Phys. {\bf B324} 
(1989) 549--580.\\ 
P. Fendley, {\em New exactly solvable models}, J. Phys. {\bf A22} (1989) 4633--4642. 
 
\bibitem{Fuchs2} B. L. Feigin, D. B. Fuchs, {\em Skew-symmetric differential operators on the 
line and Verma modules over the Virasoro algebra}, Funct. Anal. and Appl. {\bf 16} (1982) 114--126. 

\bibitem{Frame} J. S. Frame, {\em Charasteristic vectors for a product of $n$ reflections}, Duke Math. J. 
{\bf 18} (1951) 783--785.

\bibitem{Fuchs1} J. Fuchs, {\em Affine Lie Algebras and Quantum Groups}, Cambridge University 
Press, 1992.  

\bibitem{garland} H. Garland, {\em Arithmetic theory of loop algebras}, J. Algebra {\bf 53} (1978) 480--551.

\bibitem{gannon-class} T. Gannon, {\em The classification of affine su(3) modular invariants}, Commun. Math. 
Phys. {\bf 161} (1994) 233--263.

\bibitem{ginsparg} P. Ginsparg, {\em Applied conformal field theory}, Les Houches, session XLIX, 
Champs, cordes et ph\'enom\`enes critiques, E. Br\'ezin, J. Zinn-Justin eds., Elsevier, New York, 1989.

\bibitem{goddard-coset1} P. Goddard, A. Kent, D. Olive, {\em Virasoro algebras and 
coset space models}, Phys. Lett. {\bf 152B} (1985) 88--92. 
 
\bibitem{goddard-coset2} P. Goddard, A. Kent, D. Olive, {\em Unitary representations of the 
Virasoro and super-Virasoro algebras}, Commun. Math. Phys. {\bf 103} (1986) 105--119. 

\bibitem{Jones-book} F.M. Goodman, P. de la Harpe and V.F.R Jones,   
{\em Coxeter graphs and towers of algebras}, MSRI publications {\bf 14}, Springer, 1989. 

\bibitem{ItzZub-torus} C. Itzykson, J.-B. Zuber, {\em Two-dimensional conformal invariant 
theories on a torus}, Nucl. Phys. {\bf B275} (1986) 580--616. 

\bibitem{Kac1} V. G. Kac, {\em Contravariant form for infinite dimensional Lie algebras and superalgebras}, 
Lect. Notes in Phys. {\bf 94} (1979) 441--445.  
 
\bibitem{Kac2} V. G. Kac, {\em Infinite dimensional algebras}, Cambridge University Press, 1990.

\bibitem{kassel} C. Kassel, {\em Quantum Groups}, Springer-Verlag, 1995, Graduate Texts in Mathematics. 
 
\bibitem{ketov} S. V. Ketov, {\em Conformal field theory}, World Scientific, Singapore, 1994. 

\bibitem{qMcKay} A. Kirillov Jr., V. Ostrik, {\em On a q-analog of the McKay correspondence and the 
$ADE$ classification 
of $\widehat{sl}_2$ conformal field theories}, Adv. Math. {\bf 171} (2002) 183--227.

\bibitem{Klein} F. Klein, {\em Lectures on the Icosahedron and the solution of the equation of the 
fifth degree}, Dover Publ.,  New York, 1956.

\bibitem{sunder} V. Kodiyalam, V. S. Sunder, {\em Flatness and fusion coefficients}, Pacific J. of 
Math. {\bf 201} (2001).  

\bibitem{Konstant} B. Konstant, {\em The McKay correspondence, the Coxeter element and representation 
theory}, The Mathematical heritage of \'Elie Cartan, Ast\'erisque (1985) 209--255.


\bibitem{kreimer} D. Kreimer, {\em On the Hopf algebra structure of perturbative quantum 
field theories}, Adv. Theor. Math. Phys. {\bf 2} (1998) 303--334.


\bibitem{manin} Yu. I. Manin, {\em Quantum groups and non-commutative geometry}, Preprint Montreal University, CRM-1561, 1988.

\bibitem{McKay} J. McKay, {\em Graphs, singularities and finite groups}, Proc Symp. Pure Math., {\bf 37}   
(1980) 183--186. 

\bibitem{McKay-gener} J. McKay, {\em Representations and Coxeter graphs}, The Geometric Vein,
Springer-Verlag (1982) 549--554.

\bibitem{moore1} G. Moore, N. Seiberg, {\em Naturality in conformal field theory}, Nucl. Phys. 
{\bf B313} (1989) 16--40. 
 
\bibitem{moore2} G. Moore, N. Seiberg, {\em Classical and quantum conformal field theory},
 Commun. Math. Phys. {\bf 123} (1989) 177--254.

\bibitem{nahm} W. Nahm, {\em Conformal field theory: a bridge over troubled waters}, in 
Quantum Field Theory -- A Twentieth Century Profile, Hindustani Book Agency and Indian National Science 
Academy (2000) 571--604.

\bibitem{Oc-Guadeloupe} A. Ocneanu, {\em Quantum symmetries, operator algebras and invariant for
manifolds}, Talk given at the First Caribbean Spring School of Mathematical and Theoretical 
Physics, Saint-Fran\c{c}ois-Guadeloupe, 1993.

\bibitem{Oc-Marseille} A. Ocneanu, {\em Paths on Coxeter 
diagrams: from Platonic solids and singularities to minimal models 
and subfactors}, Talks given at the Centre de Physique Th\'eorique, 
Luminy, Marseille, 1995. 
 
\bibitem{Oc-paths} A. Ocneanu, {\em Paths on Coxeter diagrams: from 
Platonic solids and singularities to minimal models and subfactors}, 
Notes taken by S. Goto, AMS Fields Institute Monographs {\bf 13} (1999), Rajarama Bhat et al eds. 
 
\bibitem{Oc-string} A. Ocneanu, {\em Quantized groups, string algebras and Galois theory for   
algebras}, Operator algebras  
and Appl., Vol. 2, London Math. Soc. Lecture Notes Ser., {\bf 136}, Cambridge Univ. Press (1988) 
119--172.  
  
\bibitem{Oc-qsym} A. Ocneanu, (Lecture Notes written by Y. Kawahigashi), {\em Quantum Symmetry,   
Differential Geometry of Finite Graphs and Classification  
of Subfactors}, Univ. of Tokyo Seminar Notes (1990).  
  
\bibitem{Oc-MSRI} A. Ocneanu, {\em Higher Coxeter systems}, Talk given at  
MSRI, http://www.msri.org/publications/ln/msri/2000/subfactors/ocneanu. 

\bibitem{Oc-Bariloche} A. Ocneanu, {\em The Classification of  
subgroups of quantum 
SU(N)}, Lectures at Bariloche Summer School, Argentina, Jan. 2000, AMS Contemp. 
Math. {\bf 294}, R. Coquereaux, A. Garc\'{\i}a and R. Trinchero eds.
 
\bibitem{oleg} O. Ogievetsky, {\em Uses of quantum spaces}, 
Lectures at Bariloche Summer School, Argentina, Jan. 2000, AMS Contemp. 
Math. {\bf 294}, R. Coquereaux, A. Garc\'{\i}a and R. Trinchero eds.

\bibitem{Pasquier} V. Pasquier, {\em Two-dimensional critical
systems labelled by Dynkin diagrams}, Nucl.Phys. {\bf B285} (1987) 162--172.

\bibitem{pasquier-ADE} V. Pasquier, {\em Operator content of the ADE lattice models}, J. Phys. 
{\bf A 20} (1987) 5707--5717. 

\bibitem{Pet_Zub-1996} V.B. Petkova, J.-B. Zuber, {\em From CFT's to Graphs}, Nucl Phys. {\bf B463} 
(1996) 161--193.

\bibitem{Pet_Zub-1997} V.B. Petkova, J.-B. Zuber, {\em Conformal field theory and graphs}, Talk given at 
the 21st Intern. Coll. on Group Theor. Methods in Physics, Goslar, Germany, July 1996, hep-th/9701103.

\bibitem{Pet_Zub-gener} V.B. Petkova, J.-B. Zuber, {\em Generalised twisted partition 
functions}, Phys. Lett. {\bf B504}  (2001) 157--164. 
 
\bibitem{Pet_Zub-Oc} V.B. Petkova, J.B. Zuber, {\em The many faces of Ocneanu cells}, 
Nucl. Phys. {\bf B603} (2001) 449--496. 
 
\bibitem{Pet_Zub-CFT} V. Petkova, J.-B. Zuber, {\em Conformal field theories, 
graphs and quantum algebras}, hep-th/0108236. 
 
\bibitem{Roche-OcCell} P. Roche, {\em Ocneanu cell calculus and integrable lattice models},   
Commun. Math. Phys. {\bf 127} (1990) 395--424. 


 
\bibitem{IMPA} N. C. Saldanha and C. Tomei, {\em Spectra of semi-regular polytopes}, Informes de 
Matem\'atica, 
S\'erie A-109-Julho/94, IMPA.

\bibitem{DEA} G. Schieber, {\em Action d'un groupe quantique de dimension finie sur l'espace des 
matrices complexes},  
M\'emoire de DEA, Facult\'e de Sciences de Luminy, Marseille, 1998. 

\bibitem{sweedler} M. Sweedler, {\em Hopf Algebras}, W.A. Benjamin, 1969. 
 
\bibitem{takeu} M. Takeuchi, {\em Matched pairs of groups and bismash products of Hopf Algebras},  
Commun. Algebra {\bf 9} (1981) 841--882.

\bibitem{Roberto-cell} R. Trinchero, private notes.  

\bibitem{trinchero} R. Trinchero, private communication.  

\bibitem{todorov} I. T. Todorov, {\em Two-dimensional conformal field theory and beyond. 
Lessons from a continuing fashion}, International School for Advanced Studies (SISSA), Trieste (1999).

\bibitem{verlinde} E. Verlinde, {\em Fusion rules and modular transformations in 2-D conformal 
field theory}, 
Nucl. Phys. {\bf B300} (1988) 360--376.

\bibitem{zuber-CFT} J.-B. Zuber, {\em CFT, BCFT, ADE and all that}, Lectures at Bariloche Summer School, 
Argentina, Jan. 2000, AMS Contemp. Math. {\bf 294}, R. Coquereaux, A. Garc\'{\i}a and R. Trinchero eds.
 
 
\end{thebibliography}
\end{document}